\documentclass[11pt,twoside,final,letterpaper]{p11thesis-fr}

\usepackage{psfig,epsfig,bm,epsf,float,calc}
\usepackage{verbatim}
\usepackage{subfigure}
\usepackage{colortbl}
\usepackage{xspace}
\usepackage{amsmath,amssymb,amsfonts,amsbsy}
\usepackage{mathrsfs}
\usepackage{latexsym}
\usepackage{chicago}

\usepackage{deluxetable}
\usepackage{rotate}
\usepackage{rotating}

\usepackage{graphicx}
\usepackage{graphics}
\usepackage{tabularx}

\usepackage{minitoc}

\usepackage{ae}
\usepackage{aecompl}
\usepackage[cm]{aeguill}
\usepackage[latin1]{inputenc}
\usepackage[T1]{fontenc}
\usepackage[english,francais]{babel}

\usepackage{hhline}                 
\usepackage{multirow}               
\usepackage{array}                  
\usepackage{dcolumn}                
\usepackage{booktabs}

\usepackage{url}

\begin{document}

\newcommand{\dspaceon}{\renewcommand{\baselinestretch}{1.25}\large\normalsize}
\newcommand{\dspaceoff}{\renewcommand{\baselinestretch}{1}\large\normalsize}
\newcommand{\footon}{\dspaceoff\footnotesize}
\newcommand{\bk}{\!\!\!\!}
\def\etal{et~al.~}
\def\cad{c'est-\`a-dire~}
\def\ie{{\em i.e.~}}


\def\ks{km s$^{-1}$~}
\def\d{$^\circ$}
\def\m{$^\prime$}
\def\s{$^{\prime\prime}$}
\def\hh{$^{\mathrm h}$}
\def\mm{$^{\mathrm m}$}
\def\ss{$^{\mathrm s}$}
\def\ha{H$\alpha$~}
\def\nh{N(H)~}

\newcommand{\un}[1]{\;\ensuremath{\mathrm{#1}}}
\newcommand{\lsol}{L$_{\odot}$\xspace}
\newcommand{\tsol}{$\tau_{\odot}$\xspace}
\newcommand{\msun}{\;M$_{\odot}$\xspace}
\newcommand{\msol}{\msun}
\newcommand{\msunperyear}{\;M$_{\odot}$\,$\rm yr^{-1}$\xspace}
\newcommand{\msolperyear}{\msunperyear}
\newcommand{\ergpersec}{\un{erg\,s^{-1}}\xspace}
\newcommand{\ergpercmsec}{\un{erg\,cm^{-2}\,s^{-1}}\xspace}
\newcommand{\percmsec}{\un{cm^{-2}\,s^{-1}}\xspace}
\newcommand{\phpercmseckev}{\un{ph\;cm^{-2}\,s^{-1}\,keV^{-1}}\xspace}
\newcommand{\phpercmsec}{\un{ph\;cm^{-2}\,s^{-1}}\xspace}
\newcommand{\rs}{$\,r_{\rm s}$\xspace}
\newcommand{\rlso}{$\,r_{\rm lso}$\xspace}
\newcommand{\rmso}{$\,r_{\rm mso}$\xspace}


\newcommand{\spitzer}{{\em Spitzer\/}\xspace}
\newcommand{\cgro}{{\em CGRO\/}\xspace}
\newcommand{\comptel}{COMPTEL\xspace}
\newcommand{\egret}{EGRET\xspace}
\newcommand{\osse}{OSSE\xspace}
\newcommand{\integ}{{\em INTEGRAL\/}\xspace}
\newcommand{\ibis}{IBIS\xspace}
\newcommand{\isgri}{ISGRI\xspace}
\newcommand{\ibisgri}{IBIS/ISGRI\xspace}
\newcommand{\picsit}{PICSIT\xspace}
\newcommand{\spi}{SPI\xspace}
\newcommand{\irem}{IREM\xspace}
\newcommand{\jemx}{JEM-X\xspace}
\newcommand{\omc}{OMC\xspace}
\newcommand{\hegra}{HEGRA\xspace}
\newcommand{\hess}{HESS\xspace}
\newcommand{\veritas}{VERITAS\xspace}
\newcommand{\magic}{MAGIC\xspace}
\newcommand{\glast}{GLAST\xspace}
\newcommand{\rosat}{ROSAT\xspace}
\newcommand{\simbolx}{Simbol-X\xspace}
\newcommand{\sax}{BeppoSAX\xspace}
\newcommand{\iso}{ISO\xspace}

\newcommand{\heaoone}{{\em HEAO-1}\xspace}
\newcommand{\heaothree}{{\em HEAO-3}\xspace}
\newcommand{\smm}{{\em SMM}\xspace}
\newcommand{\sastwo}{{\em SAS-2}\xspace}
\newcommand{\cosb}{{\em COS-B}\xspace}
\newcommand{\einstein}{{\em Einstein}\xspace}
\newcommand{\spacelab}{{\em Spacelab\;2}\xspace}
\newcommand{\xrt}{XRT\xspace}
\newcommand{\granat}{{\em Granat}\xspace}
\newcommand{\artp}{ART-P\xspace}
\newcommand{\sigm}{SIGMA\xspace}
\newcommand{\ginga}{{\em Ginga}\xspace}
\newcommand{\asca}{{\em ASCA}\xspace}
\newcommand{\rxte}{RXTE\xspace}
\newcommand{\chandra}{{\em Chandra}\xspace}
\newcommand{\xmm}{{\em XMM-Newton}\xspace}
\newcommand{\epic}{EPIC\xspace}
\newcommand{\mos}{MOS\xspace}
\newcommand{\mosone}{MOS1\xspace}
\newcommand{\mostwo}{MOS2\xspace}
\newcommand{\pn}{PN\xspace}

\newcommand{\hst}{{\em HST\/}\xspace}
\newcommand{\hubble}{{\em Hubble Space Telescope\/}\xspace}
\newcommand{\keck}{{\it Keck}\xspace}
\newcommand{\cobe}{{\em COBE\/}\xspace}
\newcommand{\vla}{{\em VLA\/}\xspace}
\newcommand{\atca}{{\em ATCA\/}\xspace}
\newcommand{\vlt}{{\em VLT\/}\xspace}

\newcommand{\sgra}{Sgr\,A$^{*}$\xspace}
\newcommand{\sgraeast}{Sgr\,A\;East\xspace}
\newcommand{\sgrawest}{Sgr\;A\;West\xspace}
\newcommand{\igr}{IGR\,J17456--2901\xspace}
\newcommand{\ax}{AX\,J17456--2901\xspace}
\newcommand{\cxogc}{CXOGC\,J174540.0--290031\xspace}
\newcommand{\xmmu}{XMMU\,J174554.4--285456\xspace}
\newcommand{\eg}{3EG\,J1746--2851\xspace}
\newcommand{\hessgc}{HESS\,J1745--290\xspace}


\newcommand{\hel}[1]{$^{4}{}${#1}}
\newcommand{\lit}[1]{$^{7}{}${#1}}
\newcommand{\ber}[1]{$^{9}{}${#1}}
\newcommand{\carb}[1]{$^{12}{}${#1}}
\newcommand{\oxy}[1]{$^{16}{}${#1}}
\newcommand{\neon}[1]{$^{20}{}${#1}}
\newcommand{\na}[1]{$^{22}{}${#1}}
\newcommand{\mg}[1]{$^{24}{}${#1}}
\newcommand{\al}[1]{$^{26}{}${#1}}
\newcommand{\si}[1]{$^{28}{}${#1}}
\newcommand{\souf}[1]{$^{32}{}${#1}}
\newcommand{\argon}[1]{$^{36}{}${#1}}
\newcommand{\ca}[1]{$^{40}{}${#1}}
\newcommand{\ti}[1]{$^{44}{}${#1}}
\newcommand{\cro}[1]{$^{48}{}${#1}}
\newcommand{\nic}[1]{$^{56}{}${#1}}
\newcommand{\co}[1]{$^{57}{}${#1}}
\newcommand{\nicnew}[1]{$^{58}{}${#1}}
\newcommand{\fe}[1]{$^{60}{}${#1}}
\newcommand{\fenew}[1]{$^{61}{}${#1}}
\newcommand{\feter}[1]{$^{62}{}${#1}}
\newcommand{\ger}[1]{$^{64}{}${#1}}

\newcommand{\gammarays}{$\gamma$-rays\xspace}
\newcommand{\gammaray}{$\gamma$-ray\xspace}
\newcommand{\xray}{X-ray\xspace}
\newcommand{\xrays}{X-rays\xspace}
\newcommand{\fwhm}{FWHM\xspace}
\newcommand{\gti}{GTI\xspace}
\newcommand{\gtis}{GTIs\xspace}
\newcommand{\psla}{PSLA\xspace}
\newcommand{\arf}{ARF\xspace}
\newcommand{\pwn}{PWN\xspace}
\newcommand{\sn}{SN\xspace}
\newcommand{\snr}{SNR\xspace}
\newcommand{\snrs}{SNRs\xspace}
\newcommand{\gc}{Galactic centre\xspace}
\newcommand{\gd}{Galactic disk\xspace}
\newcommand{\gn}{Galactic nucleus\xspace}
\newcommand{\psf}{PSF\xspace}
\newcommand{\spsf}{SPSF\xspace}
\newcommand{\go}{GO\xspace}
\newcommand{\xspec}{XSPEC\xspace}
\newcommand{\glimpse}{GLIMPSE\xspace}
\newcommand{\ir}{IR\xspace}
\newcommand{\nir}{near-IR\xspace}
\newcommand{\uv}{UV\xspace}
\newcommand{\bh}{BH\xspace}
\newcommand{\osa}{OSA\xspace}
\newcommand{\isdc}{ISDC\xspace}
\newcommand{\ccd}{CCD\xspace}
\newcommand{\ccds}{CCDs\xspace}
\newcommand{\lmxb}{LMXB\xspace}
\newcommand{\lmxbs}{LMXBs\xspace}
\newcommand{\ism}{ISM\xspace}
\newcommand{\gmc}{GMC\xspace}
\newcommand{\gmcs}{GMCs\xspace}
\newcommand{\cmz}{CMZ\xspace}
\newcommand{\cnd}{CND\xspace}
\newcommand{\imf}{IMF\xspace}
\newcommand{\ura}{URA\xspace}
\newcommand{\mura}{MURA\xspace}
\newcommand{\ssc}{SSC\xspace}
\newcommand{\lso}{LSO\xspace}
\newcommand{\mso}{LSO\xspace}
\newcommand{\xrb}{XRB\xspace}
\newcommand{\xrbs}{XRBs\xspace}

\newcommand{\hone}{H\,{\sc i\/}\xspace}
\newcommand{\htwo}{H\,{\sc ii\/}\xspace}
\newcommand{\neontwo}{Ne\,{\sc ii\/}\xspace}

\newcommand\aj{{AJ}}%
\newcommand\araa{{ARA\&A}}%
\newcommand\apj{{ApJ}}%
\newcommand\apjl{{ApJ}}%
\newcommand\apjs{{ApJS}}%
\newcommand\ao{{Appl.~Opt.}}%
\newcommand\apss{{Ap\&SS}}%
\newcommand\aap{{A\&A}}%
\newcommand\aapr{{A\&A~Rev.}}%
\newcommand\aaps{{A\&AS}}%
\newcommand\azh{{AZh}}%
\newcommand\baas{{BAAS}}%
\newcommand\jrasc{{JRASC}}%
\newcommand\memras{{MmRAS}}%
\newcommand\mnras{{MNRAS}}%
\newcommand\pra{{Phys.~Rev.~A}}%
\newcommand\prb{{Phys.~Rev.~B}}%
\newcommand\prc{{Phys.~Rev.~C}}%
\newcommand\prd{{Phys.~Rev.~D}}%
\newcommand\pre{{Phys.~Rev.~E}}%
\newcommand\prl{{Phys.~Rev.~Lett.}}%
\newcommand\pasp{{PASP}}%
\newcommand\pasj{{PASJ}}%
\newcommand\qjras{{QJRAS}}%
\newcommand\skytel{{S\&T}}%
\newcommand\solphys{{Sol.~Phys.}}%
\newcommand\sovast{{Soviet~Ast.}}%
\newcommand\ssr{{Space~Sci.~Rev.}}%
\newcommand\zap{{ZAp}}%
\newcommand\nat{{Nature}}%
\newcommand\iaucirc{{IAU~Circ.}}%
\newcommand\aplett{{Astrophys.~Lett.}}%
\newcommand\apspr{{Astrophys.~Space~Phys.~Res.}}%
\newcommand\bain{{Bull.~Astron.~Inst.~Netherlands}}%
\newcommand\fcp{{Fund.~Cosmic~Phys.}}%
\newcommand\gca{{Geochim.~Cosmochim.~Acta}}%
\newcommand\grl{{Geophys.~Res.~Lett.}}%
\newcommand\jcp{{J.~Chem.~Phys.}}%
\newcommand\jgr{{J.~Geophys.~Res.}}%
\newcommand\jqsrt{{J.~Quant.~Spec.~Radiat.~Transf.}}%
\newcommand\memsai{{Mem.~Soc.~Astron.~Italiana}}%
\newcommand\nphysa{{Nucl.~Phys.~A}}%
\newcommand\physrep{{Phys.~Rep.}}%
\newcommand\physscr{{Phys.~Scr}}%
\newcommand\planss{{Planet.~Space~Sci.}}%
\newcommand\procspie{{Proc.~SPIE}}%

\title{\'Etalonnage d'un nouveau type de d\'etecteur bolom\'etrique
pour l'instrument PACS de l'Observatoire Spatial Herschel}

\author{Nicolas BILLOT}

\degreeday{19} \degreemonth{D\'ecembre} \degreeyear{2007}

\degree{Docteur en sciences physiques de l'UNIVERSIT\'E PARIS-SUD 11}
\field{Astronomie et Instrumentations Associ\'ees} \department{Service
d'Astrophysique, CEA-Saclay}

\advisor{Dr. Marc Sauvage} \president{Pr Guillaume Pineau-des-For\^ets}
\refereeone{Dr. Jean-Philippe Bernard} \refereetwo{Dr. Bruno Maffei}
\examinerone{Dr. Albrecht Poglitsch}
\examinertwo{Dr. Fran\c{c}ois-Xavier D\'esert}

\maketitle

\dspaceon

\begin{dedication}

\vspace{5cm} \hspace{5cm} \textit{\`a Delphine, Noah, ma famille ...}

\vspace{15cm} \hspace{5cm} \textit{Le doute est le premier pas vers la
science ou la v\'erit\'e ;}

\hspace{5cm} \textit{celui qui ne discute rien ne s'assure de rien ;}

\hspace{5cm} \textit{celui qui ne doute de rien ne d\'ecouvre rien.}

\vspace{0.5cm} \hspace{5cm} Denis Diderot, l'Encyclop\'edie.

\clearpage{\pagestyle{empty}\cleardoublepage}

\end{dedication}

\begin{abstract}

\vspace{10mm}

La mission Herschel est un des projets phare du programme scientifique
de l'agence spatiale europ\'eenne (ESA). Son objectif est d'explorer
le ciel dans l'une des r\'egions du spectre \'electromagn\'etique les
moins connues \`a ce jour~: l'infrarouge lointain. Sa r\'esolution, sa
sensibilit\'e mais aussi son domaine spectral font de Herschel un
observatoire unique et parfaitement adapt\'e \`a l'\'etude des
m\'ecanismes de formation d'\'etoiles et d'\'evolution des galaxies.
Parmi les autres th\`emes scientifiques qui b\'en\'eficieront des
observations Herschel se trouvent les noyaux actifs de galaxie, les
disques circumstellaires ou encore les com\`etes de notre syst\`eme
solaire.  

De nombreux instituts de recherche ont particip\'e \`a l'\'elaboration
de ce projet ambitieux, notamment le CEA qui a d\'evelopp\'e un
nouveau type de d\'etecteur bolom\'etrique pour le photom\`etre de
l'instrument Herschel/PACS.

Ce manuscrit rend compte du travail de recherche que j'ai effectu\'e
au Service d'Astrophysique du CEA dans le cadre de ma th\`ese de
doctorat. Ma t\^ache a consist\'e d'une part \`a d\'evelopper une
proc\'edure de caract\'erisation adapt\'ee aux nouvelles matrices de
bolom\`etres du CEA, et d'autre part \`a r\'ealiser l'\'etalonnage du
photom\`etre PACS et \`a optimiser ses performances dans les
diff\'erents modes d'observation ouverts \`a la communaut\'e
astronomique.

Dans ce manuscrit, je pr\'esenterai les grandes lignes de l'astronomie
infrarouge de la d\'ecouverte du rayonnement infrarouge par William
Herschel \`a la r\'ealisation de l'Observatoire Spatial Herschel. Je
d\'ecrirai \'egalement les d\'eveloppements d'hier et d'aujourd'hui
dans le domaine de la bolom\'etrie refroidie afin de mettre en
perspective les innovations apport\'ees par le CEA, \`a savoir la
fabrication collective de bolom\`etres, la thermom\'etrie haute
imp\'edance, le multiplexage \`a froid et l'absorption du rayonnement
par cavit\'e r\'esonante. J'exposerai ensuite le principe de
fonctionnement des matrices de bolom\`etres, \'etape n\'ecessaire pour
comprendre la probl\'ematique de la proc\'edure de caract\'erisation
que nous avons mise au point. Puis je pr\'esenterai et analyserai en
d\'etail les r\'esultats obtenus lors de la campagne d'\'etalonnage du
Photom\`etre PACS qui s'est achev\'ee en Juin~2007. Enfin, je
traduirai les mesures r\'ealis\'ees en laboratoire en terme de
performances \og observationnelles \fg du Photom\`etre PACS.




\vspace{15mm}

\newpage

{\centerline{\bf {\LARGE Abstract}}}

\vspace{15mm}

The Herschel mission is a major project at the core of the European
Space Agency (ESA) scientific program. The space telescope will
perform observations of the Universe in the far-infrared regime of the
electromagnetic spectrum, which still remains little-known today. With
its spatial resolution, sensitivity and spectral range (60 to
670~$\mu$m) Herschel will provide astronomers with unique
opportunities to decipher many aspects of star formation mecanisms and
galaxy evolution. Various fields in astrophysics such as active
galatic nuclei, circumstellar disks or solar system objects will also
benefit from Herschel observations.

Among the many research institutes involved in the development and
exploitation of this challenging observatory, the CEA designed a novel
type of bolometric detectors to equip the photometer of the PACS
instrument on-board the Herschel satellite.

In this thesis, I will report on the work I produced during my
doctorate at the \emph{Service d'Astrophysique du CEA}. My task was
twofold, I developed a characterisation procedure that takes advantage
of unique features of CEA filled bolometer arrays and I applied it to
calibrate the PACS photometer and optimize its performances in the
various observing modes open to the scientific community.

In this manuscript, I will present the basics of infrared astronomy
from its very beginning in 1800 to the European Space Agency's
Herschel Space Observatory. I will then describe past and present
developments in cryogenic bolometers, emphasising new concepts
introduced by CEA, that is to say the collective manufacturing of
bolometer arrays, the high impedance thermometers, the cold
multiplexing and the use of a resonant cavity to optimize absorption
of electromagnetic radiation. I will follow with an explanation of the
working principles of CEA bolometer arrays, a prerequisite to grasp
the strategy of the characterisation procedure that we developed. I
will then expose and analyse thoroughly the results that we obtained
during the calibration campaign of the PACS photometer. Finally, I
will express detector performances in terms of \og observational \fg
performances that future PACS users can comprehend.

\end{abstract}

\begin{acknowledgments}

\vspace{10mm}

\indent Durant ces trois ann\'ees pass\'ees au Service d'Astrophysique
du CEA, j'ai eu la chance de c\^otoyer et de travailler avec de
nombreuses personnes, toutes avec des personnalit\'es hors du commun
et bien souvent une grande gentillesse. Ces rencontres ont \'et\'e
incroyablement enrichissantes et instructives. Je suis
particuli\`erement reconnaissant envers mes trois chefs, Marc Sauvage
le premier, car il a pris sur lui la responsabilit\'e de m'amener
vivant jusqu'au doctorat.  Merci Marc pour ta perspicacit\'e \`a toute
\'epreuve, pour ton \'ecoute et ta confiance. Je reste dans le projet
PACS, nous continuerons donc \`a travailler et, je l'esp\`ere, \`a
publier ensemble. Merci \`a Olivier Boulade pour son esprit purement
cart\'esien, son grand app\'etit pour les bonnes choses de la vie (tes
gargantuesques repas bolom\`etres me manquent d\'ej\`a !), et aussi
pour nos maintes discussions sur la technique et la physique des
d\'etecteurs. Merci \`a Louis Rodriguez pour sa patience et son
ouverture d'esprit (et aussi son l\'egendaire romantisme
scientifique). Tu es et tu resteras \`a mes yeux le grand gourou des
bolom\`etres, tu m'as tout appris sur les bolos PACS et je t'en suis
vraiment reconnaissant. Koryo Okumura m'a aussi tellement
appris. C'est avec toi que j'ai le plus souvent travaill\'e pendant ma
th\`ese et notre collaboration a \'et\'e sans nulle doute la plus
productive que j'ai jamais eu. Merci pour tes maintes histoires des
sciences, pour les nombreux restaurants et weissbier que nous avons
partag\'es \`a Munich. Merci pour tout. Merci \'egalement aux
rapporteurs de ma th\`ese, Bruno Maffei et Jean-Phillipe Bernard, pour
avoir examin\'e attentivement et avec pertinence le manuscrit, et aux
autres membres du Jury, Xavier D\'esert, Albrecht Poglitsh et
Guillaume Pineau-des-For\^ets, pour leur participation active \`a la
soutenance.  \\ \indent Merci aussi \`a Ren\'e Gastaud, un personnage
unique et tr\`es attachant. Merci \`a Fr\`ed\'erique Motte pour sa
gentillesse et ses multiples conseils scientifiques et
pratiques. Merci \`a Laurent Vigroux pour ces nombreuses lettres de
recommendation et son soutien pour le postdoc \`a IPAC. Merci \`a
Herv\'e Aussel pour ses conseils informatiques qui souvent me
d\'ebloqu\`erent.  Merci \`a Daniel Dang, Mister D, pour ses
pr\'ecieuses imagettes et son esprit al\'eatoire tout \`a fait
indiscernable. Il est toutefois regrettable qu'il est d\^u quitter
notre \'equipe aussi brutalement. \\ \indent Merci aux acteurs de
l'ICC PACS qui nous ont aid\'e lors des tests au MPE : Ekki W., Thomas
M., Helmut F., Diego C., Alessandra C., Gerd J., Dieter L., Albrecht
P., Michael W., Pierre R., Bart V., Babar A., Dave F., Bruno A. et
Roland V.. \\ \indent Merci aussi aux gens du labo avec qui j'ai
beaucoup appr\'eci\'e travailler et papoter : Eric Doumayrou, Beno\^it
Horeau, J\'er\^ome Martignac, Yannick Le Pennec, Norma Hurtado (on se
retrouve au Chili...), Thierry Orduna, Michel Lortholary, Christelle
Clou\'e, G\'erard, Manu (j'esp\`ere que tu as retrouv\'e tes deux bras
et ta bonne humeur), Cyril et Fran\c{c}ois. \\ \indent Sans oublier
tous les autres permanents et volatiles du CEA et d'ailleurs avec qui
j'ai partag\'e des moments agr\'eables ces trois derni\`eres ann\'ees,
la liste est longue et dans le d\'esordre : Suzanne, Olivier L.,
Phillipe F., little Bobby, Ludo, Matthieu, Guillaume, Mickey, Pierrot,
Marie- Lyse, Robert, Ang\'elie, Nico, Isabelle LeM., Jean F.,
Pierre-Olivier L., Eric P., Alain G., Marion, Michel T., Matthias,
Fabio, Vincent M., Vincent R., Claire Z., Didier D., Dimitra K., Fred
M., Jeff S., J\'er\'emite B., Pac\^ome D., Faustine F., David E.,
Julio R., M\'ed\'eric B., Joel B., Savita M., Sergei A., Estelle D.,
Pascal G., Sacha H., Pascal L., Nestor H., Pascale P., Nico S. et
Aur\'elie, Nico M. (futur docteur ?)  et Anne-Po, Trioux, Burns,
Fabricio et bien d'autres encore...  \\ \indent Merci \`a Lady de
Nantes, une grande dame qui a fait mourir de rire mes coll\`egues \`a
plusieurs reprises. Merci \`a la plateforme de l'escalier de secours
de la petite aile du premier \'etage pour m'avoir support\'e pendant
mes innombrables pauses clopes.\\

\indent Enfin, je voudrais remercier chaleureusement ma famille qui m'a
toujours soutenu et sans qui je ne serais jamais arriv\'e jusque
l\`a. Et par dessus tout, je remercie Delphine, ma moiti\'e, qui
m'encourage et me supporte depuis pr\`es de 10 ans maintenant, et qui
m'a r\'ecemment donn\'e le plus beau b\'eb\'e de mon univers
observable.

\vspace{15mm}

\indent Merci.

\end{acknowledgments}

\begin{preambule}

\vspace{15mm}

\indent Ce manuscrit repr\'esente l'aboutissement de mon travail de
doctorat, trois ann\'ees d\'edi\'ees \`a l'\'etalonnage du
Photom\`etre HERSCHEL/PACS. Il s'agit donc d'une th\`ese
instrumentale.  J'insiste. Le travail des instrumentalistes, lorsqu'il
est bien fait, reste peu visible par la communaut\'e astronomique ;
pourtant, l'instrumentation pour l'astronomie est un sujet
tout-\`a-fait passionnant.  D'une part parce que les outils
d'observation sont aujourd'hui indispensables et qu'ils permettent aux
astrophysiciens de confronter les pr\'evisions de leurs mod\`eles aux
observables de notre Univers (les avanc\'ees techniques sont
d'ailleurs souvent suivies de d\'ecouvertes scientifiques majeures),
et d'autre part parce que c'est un domaine de recherche tr\`es
stimulant et comp\'etitif qui se trouve g\'en\'eralement \`a la pointe
de notre connaissance technologique. Les instruments modernes de
l'astronomie sont des syst\`emes de plus en plus complexes, surtout
lorsqu'il s'agit d'instruments embarqu\'es sur satellite, et leur
r\'ealisation n\'ecessite g\'en\'eralement l'implication de plusieurs
centaines de personnes sur plus d'une dizaine d'ann\'ees. C'est le cas
par exemple de l'instrument PACS sur lequel j'ai eu la chance de
travailler pendant ma th\`ese.  \\

\indent Bien que le travail pr\'esent\'e dans ce manuscrit soit
original et personnel, j'ai d\'ecid\'e de r\'ediger ma th\`ese \`a la
premi\`ere personne du pluriel car de nombreuses personnes ont
contribu\'e \'a son succ\`es et plus g\'en\'eralement au succ\`es de
la campagne d'\'etalonnage du Photom\`etre PACS.  Il serait donc
pr\'esomptueux de m'attribuer tout le m\'erite du travail accompli. Le
\emph{nous} que j'utiliserai tout au long de la narration regroupe
donc l'\'equipe bolom\`etre de Saclay ainsi que le groupe ICC PACS. Je
pr\'eciserai d'ailleurs le nom des personnes avec qui j'ai collabor\'e
au d\'ebut de chaque chapitre. Toutefois, par souci d'exactitude, je
me dois de pr\'eciser bri\`evement quelle a \'et\'e ma contribution au
projet PACS PhFPU. J'ai personnellement r\'ealis\'e les mesures de
performances des matrices de bolom\`etres au niveau d\'etecteur \`a
Saclay et au niveau instrument au MPE. Avec l'aide de Koryo Okumura,
j'ai r\'eduit et analys\'e les donn\'ees collect\'ees. Avec Louis
Rodriguez et Olivier Boulade, j'ai mis au point une proc\'edure
d'\'etalonnage adapt\'ee au fonctionnement des matrices. J'ai
\'egalement d\'evelopp\'e un programme, en IDL et en Jython, qui
permet de pr\'edire le r\'eglage des d\'etecteurs afin de minimiser la
saturation de l'\'electronique de lecture. Enfin, r\'esultats \og
observables \fg par la communaut\'e astronomique, j'ai traduit les
performances instrumentales mesur\'ees en laboratoire en termes de
performances observationnelles pour le Photom\`etre PACS. Cependant,
je n'ai pas contribu\'e \`a la conception des matrices de bolom\`etres
du PhFPU, ni aux tests de vibration ou d'irradiation. Je n'ai pas non
plus particip\'e \`a la manipulation des d\'etecteurs, c'est-\`a-dire
\`a l'aspect cryog\'enique des bancs de test ou \`a l'int\'egration
des d\'etecteurs. \\

\indent \`A propos des figures utilis\'ees dans ce manuscrit, je
citerai syst\'ematiquement la source des images emprunt\'ees ;
l'absence de source signifiera donc que je suis l'auteur de la
figure. Par ailleurs, le lecteur recontrera deux types de graphique
tout \`a fait reconnaissable : les courbes de type IDL, qui
s'appliquent g\'en\'eralement aux donn\'ees obtenues \`a Saclay lors
des tests r\'ealis\'es au niveau d\'etecteur, ainsi que les courbes de
type IA\footnote{\emph{Interactive Analysis} est le nom donn\'e au
logiciel d\'evelopp\'e par les groupes ICC pour manipuler et analyser
les donn\'ees des instruments de l'observatoire Herschel.}, qui se
r\'ef\`erent uniquement aux r\'esultats obtenus \`a Garching lors de
l'\'etalonnage du Photom\`etre PACS. J'ai de plus \'evit\'e l'usage
d'anglicismes dans le texte, mais cet exercice s'est av\'er\'e
relativemement difficile puisque nous travaillons tous les jours avec
des termes anglosaxons qui, une fois traduits en fran\c{c}ais, perdent
g\'en\'eralement leur sens technique. \\

\indent Enfin, je tiens \`a mentionner mon int\'er\^et grandissant
pour l'histoire des sciences car son \'etude permet de mettre en
perspective les connaissances actuelles. Durant ma th\`ese, je me suis
en particulier document\'e sur le travail de Sir William Herschel et
de Samuel Langley, deux grands hommes de science du
\textsc{XVIII}\ieme et \textsc{XIX}\ieme si\`ecle respectivement. Le
lecteur trouvera dans les premiers chapitres de ce manuscrit quelques
descriptions de leurs travaux dont le sujet pr\'esente un int\'er\^et
pour la suite de la th\`ese.

\vspace{15mm}

\indent Bonne lecture...

\vfill

\end{preambule}

\tableofcontents
\listoffigures
\listoftables

\begin{acronymes}

\chapter*{Glossaire}

\begin{tabular}{lll}

    \textbf{ADC}       & \emph{\textbf{A}nalog-to-\textbf{D}igital \textbf{C}onverter } &\\
    \textbf{BFP}       & \emph{\textbf{B}olometer \textbf{F}ocal \textbf{P}lane } &\\
    \textbf{BOLC}      & \emph{\textbf{BOL}ometer \textbf{C}ontrol (\'electronique chaude du PhFPU)} &\\
    \textbf{BLIP}      & \emph{\textbf{B}ackground \textbf{L}imited \textbf{I}nfrared \textbf{P}hotodetector} &\\
    \textbf{BU}       & \emph{\textbf{B}uffer \textbf{U}nit } &\\
    \textbf{CEA}       & \emph{\textbf{C}ommissariat \`a l'\textbf{\'E}nergie \textbf{A}tomique } &\\
    \textbf{CD}       & \emph{\textbf{C}ircuit de \textbf{D}\'etection } &\\
    \textbf{CL}       & \emph{\textbf{C}ircuit de \textbf{L}ecture } &\\
    \textbf{CNES}       & \emph{\textbf{C}entre \textbf{N}ational \textbf{\'E}tudes \textbf{S}patiales } &\\
    \textbf{DAPNIA}      & \emph{\textbf{D}\'epartement d'\textbf{A}strophysique, de physique des \textbf{P}articules,} \\
                         & \emph{de physique \textbf{N}ucl\'eaire et de l'\textbf{I}nstrumentation \textbf{A}ssoci\'ee} &\\
    \textbf{DDCS}      & \emph{\textbf{D}ouble \textbf{D}ifferential \textbf{C}orrelated \textbf{S}ampling} &\\
    \textbf{ESA}       & \emph{\textbf{E}uropean \textbf{S}pace \textbf{A}gency } &\\
    \textbf{FPU}       & \emph{\textbf{F}ocal \textbf{P}lane \textbf{U}nit} &\\
    \textbf{HIFI}      & \emph{\textbf{H}eterodyne \textbf{I}nstrument for the \textbf{F}ar-\textbf{I}nfrared} &\\
    \textbf{HSC}       & \emph{\textbf{H}erschel \textbf{S}cience \textbf{C}enter } &\\
    \textbf{Jy}        & \emph{\textbf{J}ansky}, 1~Jy = 10$^{-26}$~W/m$^2$/Hz &\\
    \textbf{LETI}      & \emph{\textbf{L}aboratoire d'\textbf{\'e}lectronique et de  \textbf{T}echnologie de l'\textbf{I}nformation } &\\
    \textbf{LIR}       &  \emph{\textbf{L}aboratoire \textbf{I}nfra\textbf{R}ouge } &\\
    \textbf{MPE}       & \emph{\textbf{M}ax-\textbf{P}lanck-institut f\"ur \textbf{E}xtraterrestrische Physik } &\\
    \textbf{NHSC}       & \emph{\textbf{N}ASA \textbf{H}erschel \textbf{S}cience \textbf{C}enter } &\\
    \textbf{OGSE}      & \emph{\textbf{O}ptical \textbf{G}round \textbf{S}egment \textbf{E}quipment} &\\
    \textbf{PACS}      & \emph{\textbf{P}hotodetector \textbf{A}rray \textbf{C}amera and \textbf{S}pectrometer} &\\
    \textbf{PAH}      & \emph{\textbf{P}olycyclic \textbf{A}romatic \textbf{H}ydrocarbons} &\\
    \textbf{PEL}      & \emph{\textbf{P}oint \textbf{E}l\'ementaire de \textbf{L}ecture} &\\
    \textbf{PhFPU}     & \emph{\textbf{Ph}otometer \textbf{F}ocal \textbf{P}lane \textbf{U}nit} &\\
    \textbf{PSF}     & \emph{\textbf{P}oint \textbf{S}pread \textbf{F}unction } &\\
    \textbf{SAp}     & \emph{\textbf{S}ervice d'\textbf{A}stro\textbf{p}hysique du CEA/DSM/DAPNIA} &\\
    \textbf{SED}     & \emph{\textbf{S}pectral \textbf{E}nergy \textbf{D}istribution } &\\
    \textbf{SPIRE}      & \emph{\textbf{S}pectral and \textbf{P}hotometric \textbf{I}maging and \textbf{RE}ceiver} &\\
    \textbf{ULIRG}      & \emph{\textbf{U}ltra \textbf{L}uminous \textbf{I}nfra\textbf{R}ed \textbf{G}alaxy }  &\\

\end{tabular}




\end{acronymes}


\chapter{Une br\`eve histoire de l'astronomie infrarouge~: de Herschel \`a Herschel}
\label{chap:intro_astroIR}

\pagenumbering{arabic}

\begin{center}
\begin{minipage}{0.85\textwidth}

\small Ce chapitre d'introduction est destin\'e \`a donner le contexte
de l'astronomie infrarouge et submillim\'etrique telle que nous la
connaissons aujourd'hui. La d\'ecouverte par William Herschel du
rayonnement thermique infrarouge sera notre point de d\'epart. Nous
donnerons ensuite une br\`eve description de notre Univers vu dans
l'infrarouge lointain en insistant sur les ph\'enom\`enes physiques
\'emetteurs de rayonnement, ainsi que sur les diff\'erents types de
d\'etecteurs et m\'ethodes d'observation utilis\'es par les astronomes
pour scruter le ciel dans ce domaine du spectre
\'electromagn\'etique. Enfin, nous pr\'esenterons le projet Herschel,
grand observatoire de l'agence spatiale europ\'eenne, qui abrite
l'instrument que nous avons \'etalonn\'e, Herschel/PACS, et qui fait
l'objet de ce manuscrit.


\end{minipage}
\end{center}

\section{La naissance de l'astronomie infrarouge}
\label{sec:intro_astroIR_univers_debut}

Les mots \emph{infrarouge}, \emph{rayons~X}, \emph{radio} ou encore
\emph{micro-onde} sont aujourd'hui rentr\'es dans le vocabulaire
courant, et il semble \'evident que la lumi\`ere ne s'arr\^ete pas au
domaine visible du spectre \'electromagn\'etique mais s'\'etend
au-del\`a et en-de\c{c}a de ce que nos yeux nous permettent de
voir. La premi\`ere personne qui montra l'existence d'un rayonnement
invisible \`a l'\oe il mais qui se comporte pourtant comme la
lumi\`ere visible s'appelle William Herschel. Il fit cette
d\'ecouverte en 1800 en \'etudiant le \og pouvoir des couleurs
prismatiques \`a illuminer et \`a chauffer les objets \fg
\shortcite{herschel1800a}. Je tiens \`a commencer ce manuscrit par ce
qui repr\'esente la naissance de l'astronomie infrarouge~; le but
\'etant de mettre en perspective l'\'evolution de notre connaissance
du monde infrarouge tel que nous le voyons aujourd'hui, mais aussi de
montrer \`a quel point il est fascinant de voir comment un homme
arm\'e d'un simple thermom\`etre, d'un prisme et d'une bonne dose de
curiosit\'e a pu ouvrir une nouvelle voie d'investigation pour les
sciences du XIX\textsuperscript{e} si\`ecle.\\

\begin{figure}
  \begin{center}
    \begin{tabular}{ll}
      \includegraphics[width=0.45\textwidth,angle=0]{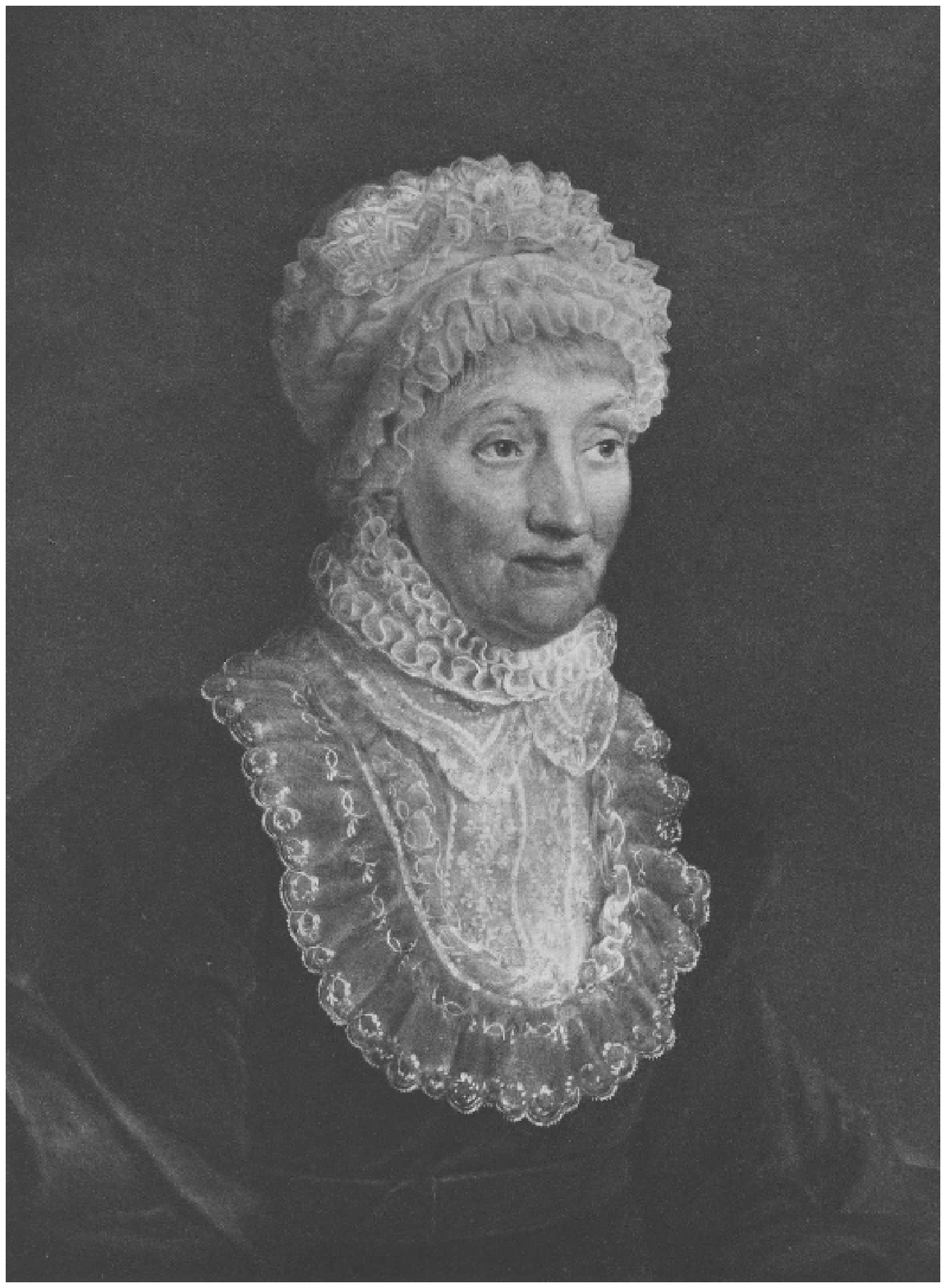}
      &
      \includegraphics[width=0.455\textwidth,angle=0]{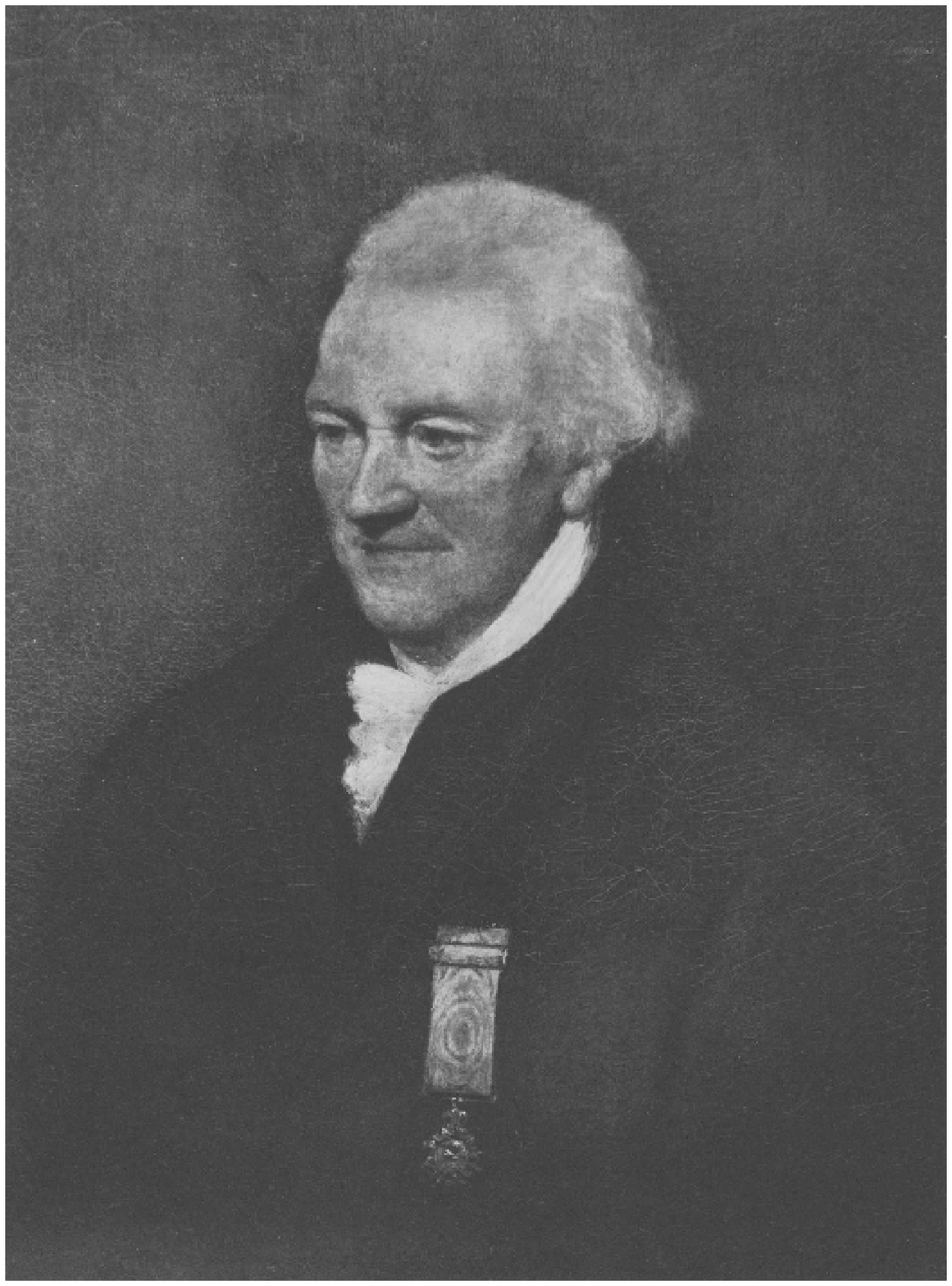}
    \end{tabular}
  \end{center}
  \caption[Caroline et William Herschel]{Portraits de Caroline (1829)
  et William Herschel (1819) extraits de~\shortciteN{dreyer1912}. Les
  peintures sont de Tielemann (\`a gauche) et Artaud (\`a droite).}
  \label{fig:intro_astroIR_univers_debut_portraits}
\end{figure}

Friederich Wilhelm Herschel est n\'e en 1738 \`a Hannover en Allemagne
(il fut naturalis\'e anglais en 1793 et pris le nom de William
Herschel). Il quitta son pays \`a l'age de 18~ans pour rejoindre
l'Angleterre o\`u il enseigna la musique dans la ville de Bath. Il
\'etait joueur-compositeur de haut-bois et vivait exclusivement de sa
musique. En 1773, Herschel s'int\'eressa \`a l'astronomie et acheta
les outils qui lui permirent de fabriquer son premier t\'elescope. Il
consacra les ann\'ees qui suivirent \`a la construction de
t\'elescopes et \`a l'observation du ciel nocturne. Herschel est
aujourd'hui connu pour son t\'elescope de 12~m de longueur focale (le
plus grand \`a l'\'epoque), pour sa d\'ecouverte de la plan\`ete
Uranus\footnote{Herschel donna initialement le nom de \emph{Georgium
Sidius} \`a la plan\`ete qu'il venait de d\'ecouvrir en l'honneur de
son Roi George III.}, pour son catalogue d'\'etoiles doubles et bien
s\^ur pour sa d\'ecouverte du rayonnement infrarouge. Pendant toute sa
carri\`ere d'astronome au service du Roi George III d'Angleterre, il
travailla avec sa s\oe ur, Caroline Herschel, qui \'etait son
assistante et l'aidait dans son exploration syst\'ematique du
ciel. Caroline Herschel \'etait \'egalement une observatrice
chevronn\'ee~; elle d\'ecouvrit de nombreuses com\`etes et publia elle
aussi un catalogue d'\'etoiles doubles. La
figure~\ref{fig:intro_astroIR_univers_debut_portraits} montre les
portraits de Caroline et William Herschel.

La contribution de Sir William Herschel en astronomie a
\'et\'e consid\'erable dans de nombreux domaines, en atteste la
collection impressionante de papiers rassembl\'es dans l'ouvrage
pr\'esent\'e par~\shortciteN{dreyer1912}. Je vais maintenant me
concentrer sur une courte p\'eriode de sa vie, quelques mois de
l'ann\'ee 1800 pour \^etre plus pr\'ecis, durant laquelle il a
\'etudi\'e le spectre solaire et a d\'ecouvert l'existence du
rayonnement infrarouge qu'il appelle la chaleur radiative, ou
\emph{radiant heat} en anglais.
\begin{center}
\begin{minipage}{0.85\textwidth}
\vspace{1cm} \small \noindent \textit{\og Radiant heat will at least,
if not chiefly, consist, if I may be permitted the expression, of
invisible light ; that is to say, of rays coming from the sun, that
have such a momentum as to be unfit for vision. And, admitting, as is
highly probable, that the organs of sight are only adapted to receive
impressions from particles of a certain momentum, it explains why the
maximum of illumination should be in the middle of the refrangible
rays ; as those which have greater or less momenta, are likely to
become equally unfit for impressions of sight.\fg }\\
\noindent Extrait du texte original de William Herschel dans les
\emph{Philosophical Transactions} (1800).\\
\end{minipage}
\end{center}

\begin{figure}
  \begin{center} \begin{tabular}{c}
    \includegraphics[width=0.7\textwidth,angle=0]{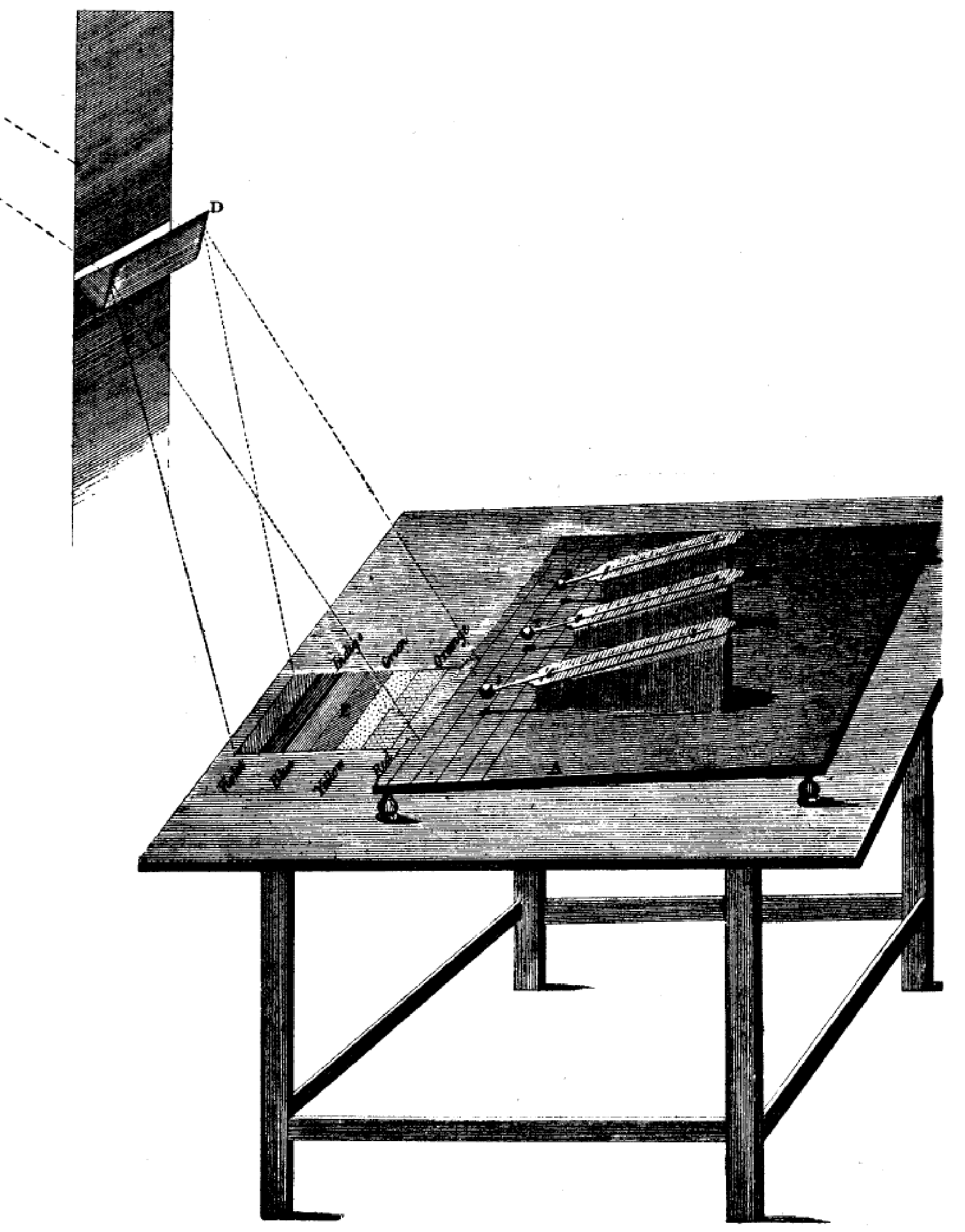}\\
    \includegraphics[width=0.9\textwidth,angle=0]{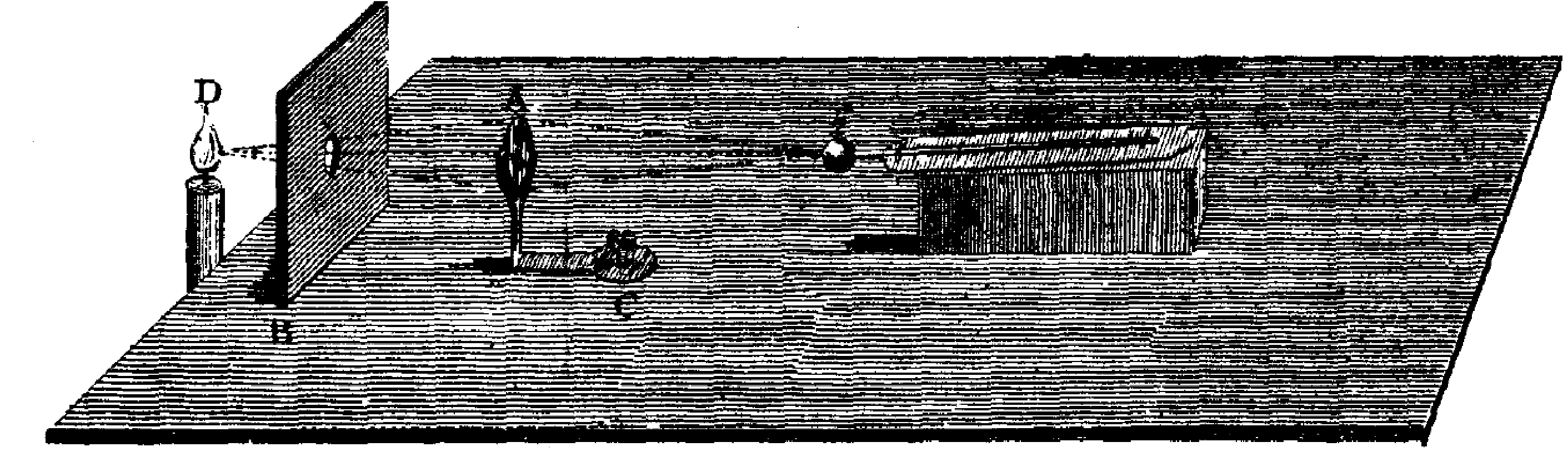}
    \end{tabular} \end{center} \caption[D\'ecouverte de l'infrarouge:
    les couleurs calorifiques]{Gravures extraites
    de~\citeN{herschel1800b} et~\citeN{herschel1800c} (p.~71
    et~87). En haut: rayonnement solaire diffract\'e par un prisme en
    verre. Le spectre est projet\'e sur la table et les thermom\`etres
    mesurent la temp\'erature de chaque couleur et mettent en
    \'evidence l'existence de la couleur calorifique, l'infrarouge. En
    bas: Mat\'eriel exp\'erimental qu'Herschel utilisa pour montrer
    que les rayons calorifiques invisibles ob\'eissent aux lois de la
    r\'efraction, tout comme les rayons de lumi\`ere visible. Lorsque
    la lentille est plac\'ee entre la flame et la pointe du
    thermom\`etre, Herschel mesure une \'el\'evation de temp\'erature
    de plus de 2\textdegree. Il montra \'egalement lors d'une autre
    exp\'erience que le foyer d'une lentille est plus \'eloign\'e de
    la lentille dans l'infrarouge que dans la lumi\`ere visible.}
    \label{fig:intro_astroIR_univers_debut_manip_herschel}
\end{figure}

Herschel utilise un prisme de verre pour diffracter la lumi\`ere du
Soleil et place trois thermom\`etres sur une table sur laquelle se
proj\`ete le spectre solaire (cf gravures de la
figure~\ref{fig:intro_astroIR_univers_debut_manip_herschel}). Deux de
ces thermom\`etres sont utilis\'es comme \'etalons pour mesurer les
variations de la temp\'erature ambiante\footnote{Herschel r\'ealise ce
que nous appelerions aujoud'hui des mesures diff\'erentielles
indispensables pour les observations infrarouges.}. Il d\'eplace le
troisi\`eme thermom\`etre pour mesurer l'\'elevation de temp\'erature
associ\'ee \`a chacune des \og couleurs prismatiques \fg. Il
s'aper\c{c}oit alors que la temp\'erature continue d'augmenter alors
que le thermom\`etre a d\'ej\`a d\'epass\'e le spectre visible du
c\^ot\'e de la couleur rouge. Il vient de d\'etecter pour la
premi\`ere fois le rayonnement infrarouge \'emis par le Soleil.

Herschel pense que la chaleur radiative doit \^etre r\'efract\'ee par
le prisme de la m\^eme mani\`ere que la lumi\`ere visible, mais que
notre \oe il n'est pas capable de d\'etecter ce rayonnement
calorifique (il suppose que les particules de lumi\`ere infrarouge ont
trop, ou pas assez, d'\'energie pour impressionner notre r\'etine). Il
poursuit donc ses recherches sur la r\'efrangibilit\'e des \og rayons
solaires et terrestres qui occasionnent de la chaleur \fg. Il utilise
toujours son prisme de verre mais r\'ealise \'egalement des
exp\'eriences avec des miroirs et des lentilles pour tester les
propri\'et\'es optiques de ces rayons invisibles. La gravure
pr\'esent\'ee en bas de la
figure~\ref{fig:intro_astroIR_univers_debut_manip_herschel} montre une
de ses exp\'eriences o\`u il teste l'aptitude d'une lentille \`a
concentrer le rayonnement calorifique sur la boule noircie de son
thermom\`etre. Les nombreuses exp\'eriences qu'il r\'ealise sont
ing\'enieuses et le lecteur avis\'e pourra consulter le livre de
\shortciteN{dreyer1912} pour de plus amples d\'etails. 

\begin{figure}
  \begin{center}
    \includegraphics[width=0.8\textwidth,angle=0]{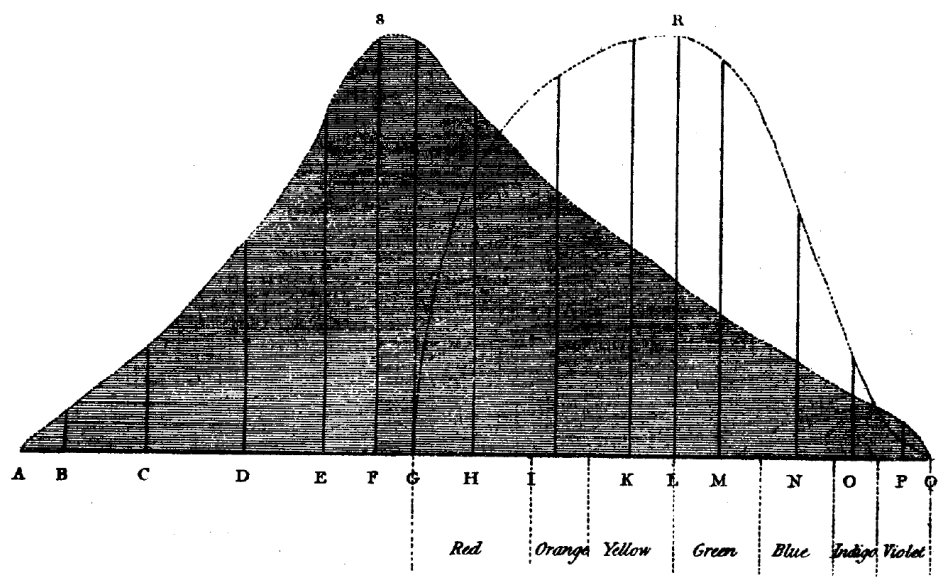}
  \end{center}
  \caption[Spectre du rayonnement calorifique mesur\'e par Herschel en
  1800]{Spectres extraits de~\citeN{herschel1800c} (p.~99). La zone
  sombre d\'elimit\'ee par le p\'erim\`etre $ASQA$ correspond au
  rayonnement invisible absorb\'e par le thermom\`etre d'Herschel en
  fonction de la couleur calorifique, i.e. de l'angle de
  r\'efraction. La zone vide comprise dans le p\'erim\`etre $GRQG$
  repr\'esente le pouvoir d'illumination de la lumi\`ere visible,
  c'est d'apr\`es Herschel la capacit\'e qu'une couleur a d'illuminer
  les objets.}
  \label{fig:intro_astroIR_univers_debut_blackbody}
\end{figure}

Apr\`es quelques mois pass\'es \`a tester les propri\'et\'es du
rayonnement infrarouge, Herschel publie deux courbes qui pourraient
aujourd'hui s'apparenter \`a des distributions spectrales
d'\'energie. Ces deux courbes sont pr\'esent\'ees dans la
figure~\ref{fig:intro_astroIR_univers_debut_blackbody}. L'axe des
abscisses correspondrait \`a la longueur d'onde observ\'ee ($\lambda$
augmente vers la gauche). La courbe de droite (courbe vide)
repr\'esente l'aptitude d'une couleur prismatique \`a illuminer un
objet. En pratique, Herschel utilise chacune des couleurs du spectre
solaire pour \'eclairer un objet et \'evaluer le contraste de la
sc\`ene qu'il observe. Sur la
figure~\ref{fig:intro_astroIR_univers_debut_blackbody} nous voyons que
le contraste est maximum pour la couleur jaune. Cette courbe est sans
aucun doute reli\'ee au spectre de corps noir du Soleil, \cad que le
maximum d'\'energie se trouve bien dans le jaune, mais elle fait aussi
intervenir l'organe de la vision (le pic d'absorption de la r\'etine
se trouve autour de 0.5~$\mu$m d'apr\`es \shortciteNP{sliney}) de
sorte que sa forme ne correspond pas exactement au spectre de corps
noir du Soleil. En ce qui concerne la courbe de gauche (courbe
pleine), elle montre l'\'evolution de la temp\'erature associ\'ee \`a
chacun des angles de r\'efraction des rayons calorifiques. Cette
courbe pique manifestement dans l'infrarouge proche, ce qui semble
contradictoire avec ce que nous savons de l'\'emission solaire. Notez
cependant que cette courbe est le produit du spectre solaire, de la
transmission de l'atmosph\`ere et du prisme, et de l'absorption de la
boule noircie du thermom\`etre. Elle ne repr\'esente donc pas
directement l'\'emission infrarouge du Soleil. Mais elle a quand
m\^eme le m\'erite d'\^etre la toute premi\`ere d\'etection de
rayonnement non-visible \`a l'\oe il nu.

\section{L'Univers infrarouge et (sub-)millim\'etrique}
\label{sec:intro_astroIR_univers}

Depuis Herschel, la d\'efinition de la \og chaleur radiative \fg a
naturellement \'evolu\'e, et pour \'eviter toute confusion dans la
suite du manuscrit, il est utile de pr\'eciser la terminologie
utilis\'ee pour d\'esigner les diff\'erents domaines du spectre
\'electromagn\'etique. Le terme \emph{infrarouge} (IR) se rapporte de
mani\`ere g\'en\'erale \`a un rayonnement \'electromagn\'etique dont
la longueur d'onde est comprise entre $\sim$1~$\mu$m et $\sim$1~mm,
\cad une fr\'equence allant de $3\times10^{14}$ \`a
$3\times10^{11}$~Hz. Dans cette gamme de longueur d'onde, nous
distinguons quatre subdivisions avec les conventions
suivantes\footnote{Ces conventions ne sont pas universelles et le
lecteur pourra en trouver d'autres d\'efinitions dans la
litt\'erature.}~: l'infrarouge proche s'\'etend de~1 \`a 5~$\mu$m,
l'infrarouge moyen de~5 \`a 30~$\mu$m (aussi appel\'e infrarouge
thermique), l'infrarouge lointain de~30 \`a 200~$\mu$m, et enfin le
sub-millim\'etrique de~200
\`a 1000~$\mu$m. Au-del\`a de 1~mm de longueur d'onde se trouve le 
domaine millim\'etrique.  Nous utiliserons \'egalement le terme
(sub-)millim\'etrique pour d\'esigner les fen\^etres atmosph\'eriques
qui se trouvent \`a cheval entre le sub-millim\'etrique et le
millim\'etrique (850~$\mu$m et 1.3~mm, cf
section~\ref{sec:intro_astro_william_obs}).

\subsection{Les \'emetteurs de rayonnement}
\label{sec:intro_astroIR_univers_emetteur}

Depuis une cinquantaine d'ann\'ees, et plus particuli\`erement depuis
les ann\'ees~80 avec l'apparition d'observatoires spatiaux (cf
section~\ref{sec:intro_astro_william_obs}) et de d\'etecteurs
infrarouges tr\`es performants (cf
section~\ref{sec:intro_astro_IR_detecteur}), notre compr\'ehension de
l'Univers infrarouge s'est consid\'erablement accrue
\shortcite{walker,low_book}. Nous savons aujourd'hui que les \'emetteurs de 
rayonnement infrarouge sont de nature tr\`es vari\'ee et qu'ils sont
pr\'esents \`a toutes les \'echelles de l'Univers, de notre syst\`eme
solaire \shortcite{muller05} aux galaxies les plus lointaines
\shortcite{sanders,lagache}, en passant par le milieu
interstellaire Galactique \shortcite{mckee} et extragalactique
\shortcite{kennicutt}.

Le rayonnement infrarouge est un vecteur d'information tr\`es riche
qui nous permet de d\'evoiler nombre d'objets enfouis au c\oe ur de
nuages de poussi\`ere opaques \`a la lumi\`ere visible. Il nous plonge
par ailleurs dans le royaume des objets froids. En effet, un photon
infrarouge poss\`ede une \'energie comprise entre~1 et 10$^{-3}$~eV,
soit une temp\'erature de~10 \`a $\sim$1000~K, o\`u nous avons
simplement utilis\'e la formule
$\mbox{E}=\mbox{h}\,\nu=\mbox{k}_{\mbox{\tiny B}}\mbox{T}$ pour
obtenir ces temp\'eratures. Parmi les principales th\'ematiques
scientifiques abord\'ees en astronomie infrarouge figurent l'\'etude
des atmosph\`eres plan\'etaires, du milieu interstellaire galactique
et extragalactique (cirrus, nuages mol\'eculaires et r\'egions de
formation d'\'etoiles) ainsi que de la formation et de l'\'evolution
des galaxies. Bien qu'ils soient souvent associ\'es au domaine radio,
l'\'emission synchrotron et l'\'emission free-free \shortcite{condon}
ainsi que le rayonnement fossile (voir plus bas dans le texte) sont
\'egalement des ph\'enom\`enes observables dans le
(sub-)millim\'etrique.

Il serait trop ambitieux dans ce manuscrit de faire un \'etat des lieux
exhaustif de nos connaissances actuelles, et plut\^ot que de faire un
inventaire des objets c\'elestes et de leurs propri\'et\'es, nous
proposons une br\`eve description de quelques processus physiques
\'emetteurs de rayonnement infrarouge et (sub-)millim\'etrique en les
repla\c{c}ant dans leur contexte astrophysique.

\subsubsection{La poussi\`ere}

\begin{figure}
  \begin{center}
    \includegraphics[width=0.95\textwidth,angle=0]{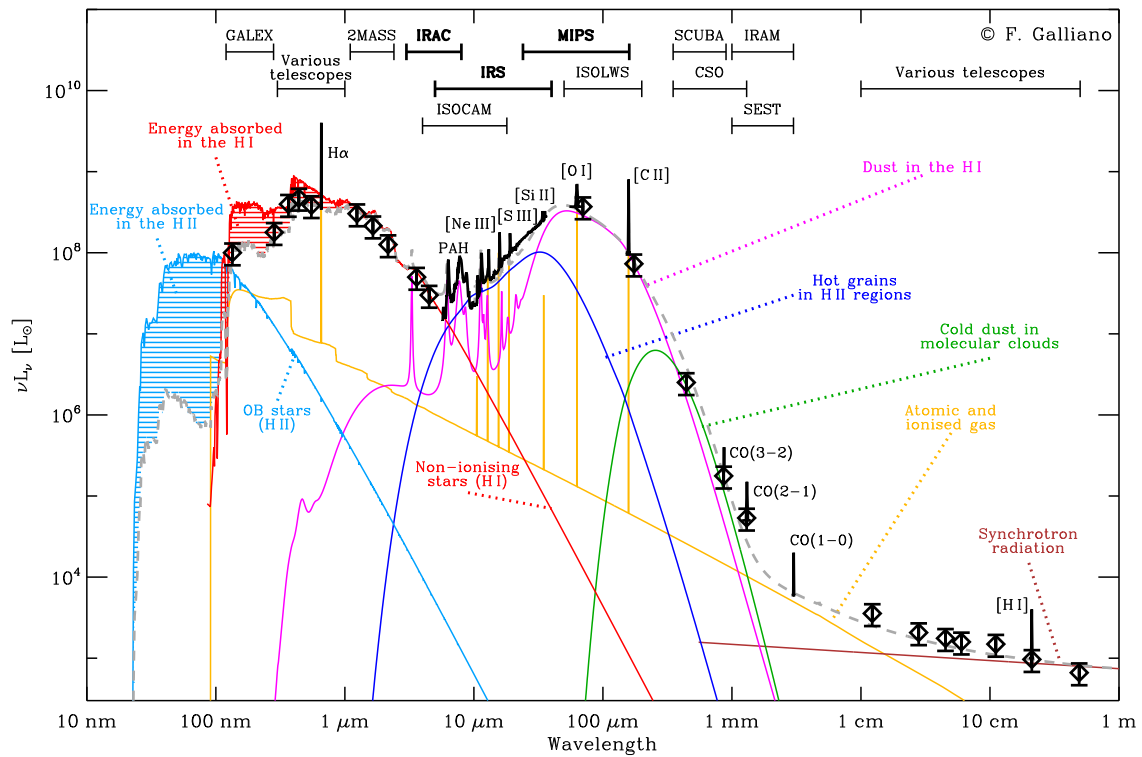}
    \end{center} \caption[Densit\'e spectrale d'\'energie d'une
    galaxie]{Cette figure montre la densit\'e spectrale d'\'energie
    d'une galaxie naine (NGC1569) de l'ultraviolet au domaine
    radio. La contribution des diff\'erents \'emetteurs de rayonnement
    visibles sur la figure a \'et\'e mod\'elis\'ee par
    \shortciteN{galliano}. De 10~$\mu$m \`a 1~mm, ce sont les grains
    de poussi\`ere chauff\'es par le rayonnement (inter-)stellaire
    (nuages mol\'eculaires et r\'egions de photodissociation) qui
    rayonnent la majorit\'e de la luminosit\'e infrarouge. Ces grains
    sont \'egalement responsables de l'absorption du rayonnement
    \'energ\'etique des jeunes \'etoiles (partie bleue hachur\'ee aux
    courtes longueurs d'onde). Image reproduite avec l'aimable
    autorisation de Fr\'ed\'eric Galliano.}  \label{fig:SED_galliano}
\end{figure}

\begin{figure}
  \begin{center}
    \includegraphics[width=1.\textwidth,angle=0]{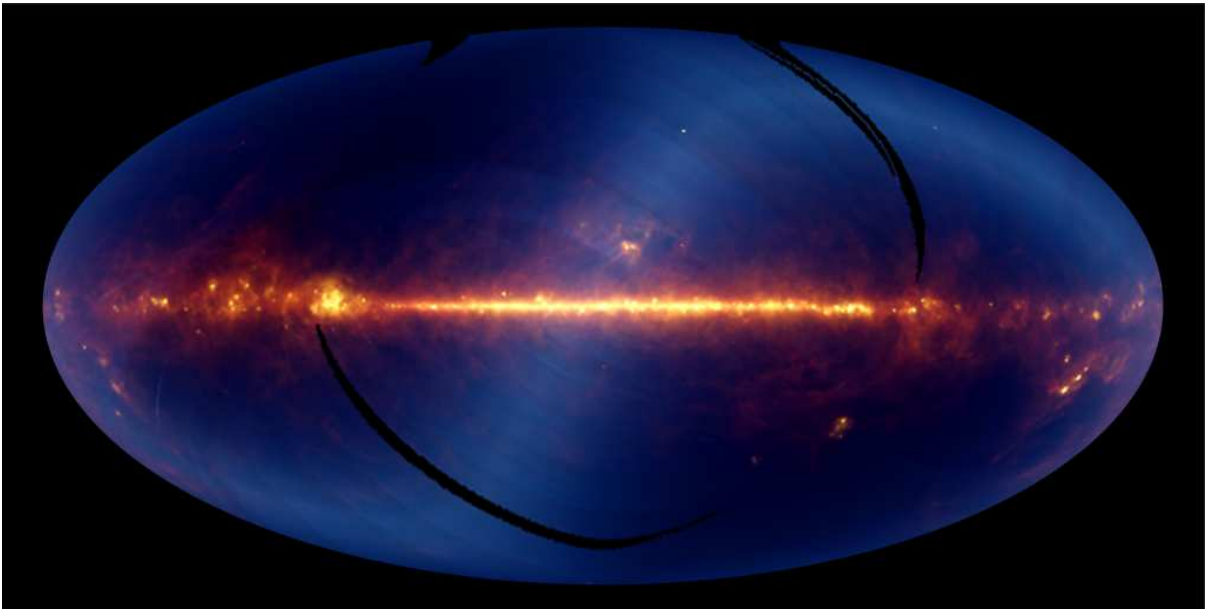}
    \end{center} \caption[Le ciel infrarouge vu par IRAS]{Carte du
    ciel infrarouge (entre 10~et 120~$\mu$m) observ\'e par le
    satellite IRAS. Le plan Galactique contient une grande quantit\'e
    de poussi\`ere qui r\'e\'emet le rayonnement interstellaire dans
    le domaine infrarouge. D'apr\`es \shortciteN{dasyra}, la masse de
    poussi\`ere s'\'el\`eve \`a $\sim10^7$~$\mbox{M}_{\odot}$ dans les
    galaxies spirales typiques. L'\'emission zodiacale produite la
    poussi\`ere interplan\'etaire, en forme de S bleut\'ee sur la
    figure, indique la position de l'\'ecliptique~; elle est
    relativement intense dans les bandes IRAS.  (lien
    www.ipac.caltech.edu)} \label{fig:allsky_IRAS}
\end{figure}

L'\'emission continuum des galaxies observ\'ee entre 10~$\mu$m et 1~mm
est domin\'ee par l'\'emission de petits grains de poussi\`ere (cf
figure~\ref{fig:SED_galliano}).  En effet, la poussi\`ere
interstellaire, dont la pr\'esence est connue par le biais des effets
d'extinction \shortcite{calzetti}, dans l'ultraviolet (UV) notamment,
absorbe le rayonnement interstellaire pour le r\'e\'emettre dans le
domaine infrarouge (cf figure~\ref{fig:allsky_IRAS}).  Cette
conversion d'\'energie des courtes vers les grandes longueurs d'onde
se manifeste particuli\`erement dans les r\'egions de formation
d'\'etoiles o\`u les jeunes astres, puissants
\'emetteurs de rayonnement UV, sont encore profond\'ement enfouis dans
le nuage de gaz et de poussi\`ere qui leur a donn\'e
naissance. L'\'emission infrarouge des galaxies est donc intimement
li\'ee \`a l'activit\'e de formation d'\'etoiles
\shortcite{kennicutt}. Dans le cas extr\^eme des ULIRGs (\emph{Ultra
Luminous Infrared Galaxy}) pour lesquelles L$_{\mbox{\tiny
IR}}\sim10^{13}$~L$_{\odot}$, la luminosit\'e infrarouge repr\'esente
la quasi-totalit\'e de la luminosit\'e bolom\'etrique
\shortcite{lonsdale}~; cela est d\^u \`a l'intense activit\'e de
formation d'\'etoiles d\'eclanch\'ee par la collision/fusion de deux
galaxies\footnote{L'\'eventuelle pr\'esence d'un noyau actif enfoui
dans un tore de poussi\`ere peut \'egalement contribuer \`a la forte
luminosit\'e infrarouge.}. Notez \'egalement que les galaxies dites
\og normales \fg \shortcite{sauvage}, \cad qui ne contiennent ni noyau
actif ni flamb\'ee d'\'etoiles, produisent tout de m\^eme un tiers de
leur luminosit\'e totale dans l'infrarouge. La poussi\`ere joue donc
un r\^ole tr\`es important dans l'\'equilibre \'energ\'etique des
galaxies. Elle tient \'egalement une place centrale dans la chimie du
milieu interstellaire, avec notamment la formation d'hydrog\`ene
mol\'eculaire \`a la surface des grains \shortcite{vidali}, et dans
l'\'etude des m\'ecanismes de formation d'\'etoiles
\shortcite{andre_coeur}, v\'eritable moteur de l'\'evolution des
galaxies.\\

D'apr\`es \shortciteN{draine2003}, les grains de poussi\`ere sont en
partie produits dans les enveloppes de g\'eantes rouges, d'\'etoiles
carbon\'ees ou encore de n\'ebuleuses plan\'etaires. Des observations
sub-millim\'etriques ont m\^eme montr\'e que les supernovae pouvaient
produire de grandes quantit\'es de poussi\`ere
\shortcite{dunne}. Toutefois, \shortciteN{draine2003} sugg\`ere que
la plupart des grains observ\'es ne serait pas simplement de la \og
poussi\`ere d'\'etoile\fg, la composition des grains interstellaires
serait plut\^ot d\'etermin\'ee par les processus physiques qui ont
lieu dans le milieu interstellaire comme par exemple les collisions
grain-grain, la coagulation, l'\'evaporation, le rayonnement cosmique,
etc... Quoiqu'il en soit, les observations montrent que les grains
sont majoritairement compos\'es de silicates (oxyde de silice
amalgam\'e avec du magn\'esium ou du fer) et de mat\'eriaux carbon\'es
(graphite, diamant, hydrocarbure polycyclique aromatique, etc...). Du
carbure de silicium ainsi que des carbonates sont aussi pr\'esents
dans le milieu interstellaire mais en moindre quantit\'e (5 et 1\%
respectivement). Dans certains environnements, les grains peuvent
\'egalement \^etre recouverts d'un manteau de glace (eau, \'ethanol,
m\'ethane, etc...) et de mat\'eriaux organiques
\shortcite{dartois,ehrenfreund}.\\

\begin{figure}
  \begin{center}
      \includegraphics[width=0.7\textwidth]{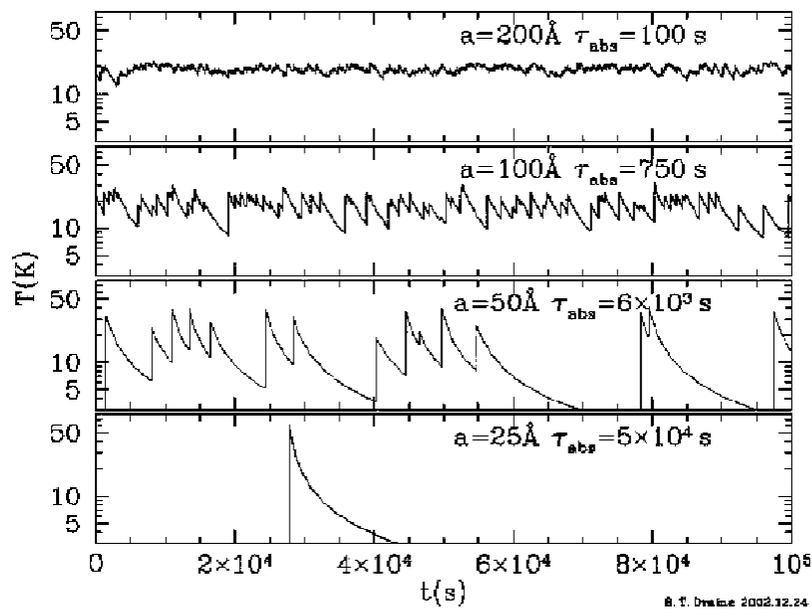}
  \end{center} \caption[Chauffage de la poussi\`ere
  interstellaire]{Chauffage de grains de poussi\`ere interstellaires
  dans un champ de radiation moyen. Nous voyons l'\'evolution de la
  temp\'erature de quatre grains carbon\'es de taille diff\'erente
  (entre~25 et 200~\r{A}) au long d'une journ\'ee. $\tau_{\mbox{abs}}$
  est le temps moyen entre deux absorptions successives. Les gros
  grains sont thermalis\'es alors que les petits subissent un
  chauffage stochastique. Figure extraite de \shortciteN{draine2003}.}
  \label{fig:dust_grain_heating}
\end{figure}

\shortciteN{mathis} ont montr\'e que l'absorption interstellaire moyenne 
\'etait parfaitement reproduite par un mod\`ele de grains qui
contient deux composantes, des graphites et des
silicates. L'ajustement des courbes d'extinction avec ce mod\`ele
indique que la distribution en taille de ces deux types de grains suit
une loi de puissance de la forme $dn/da\propto a^{-3.5}$ avec une
taille comprise entre $a_-\sim50$~\r{A} et
$a_+\sim2500$~\r{A}. \shortciteN{draine1984} ont ensuite \'etendu ce
mod\`ele \`a l'infrarouge en calculant les propri\'et\'es optiques des
grains de silicate et de graphite jusqu'\`a 1~mm de longueur d'onde.
Pour rendre compte des caract\'eristiques spectrales de la poussi\`ere
interstellaire, la mod\'elisation des grains int\`egre \'egalement
l'existence des hydrocarbures polycyliques aromatiques dans le milieu
interstellaire, aussi appel\'es PAH pour \emph{Polycylic Aromatic
Hydrocarbons} \shortcite{leger}. Ces grains peuvent \^etre aussi
petits que quelques angstr\"oms et ne contenir que quelques dizaines
d'atomes. Leur mode de chauffage par les photons UV est particulier,
il est qualifi\'e de stochastique \shortcite{draine2001} car leur
temp\'erature \'evolue plus rapidement que le temps qui s\'epare deux
absorptions cons\'ecutives (cf courbe du bas sur la
figure~\ref{fig:dust_grain_heating}). \`A l'inverse, les gros grains
sont \`a l'\'equilibre thermique avec le champ de radiation ambiant
(cf courbe du haut sur la
figure~\ref{fig:dust_grain_heating}). L'\'energie absorb\'ee par les
gros grains est convertie en \'energie interne dans le solide, les
grains thermalis\'es r\'e\'emettent alors cette \'energie avec un
spectre de corps gris. La temp\'erature associ\'ee \`a cette
\'emission d\'epend de l'environnement imm\'ediat~; elle est 
g\'en\'eralement comprise entre 10 et 50~K, ce qui correspond \`a des
longueurs d'onde de 300 et 60~$\mu$m. La figure~\ref{fig:SED_galliano}
montre la contribution \`a la SED de NGC1569 de l'\'emission des
grains de poussi\`ere contenus dans les r\'egions de photodissociation
ou bien dans les nuages d'hydrog\`ene atomique et mol\'eculaire, les
bandes PAH sont \'egalement visibles dans l'infrarouge proche et
moyen. Nous voyons sur cette figure que les points de mesure dans
l'infrarouge lointain sont d\'eterminants pour contraindre la position
du pic d'\'emission des grains de poussi\`ere et ainsi calculer leur
temp\'erature et la masse int\'egr\'ee sur la ligne de vis\'ee.

\subsubsection{Le gaz}

\begin{figure}
  \begin{center} 
       \includegraphics[width=1.\textwidth,angle=0]{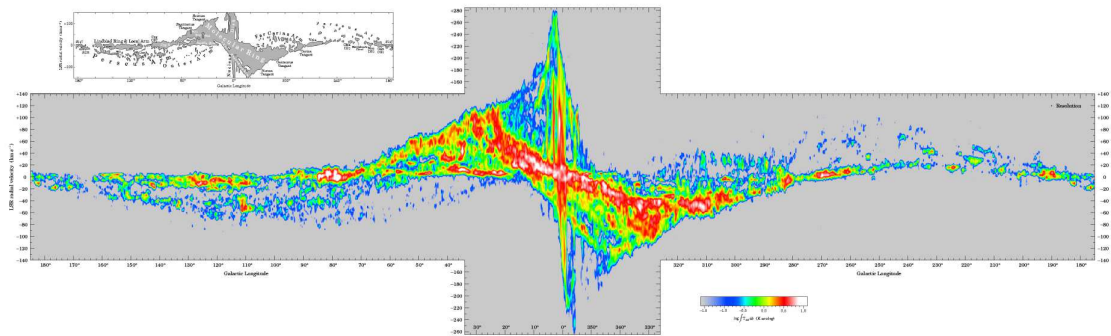}
  \end{center} \caption[Carte de vitesse de la Galaxie en
    CO]{Observations spectroscopiques du monoxyde de carbone (115~GHz)
    dans la Galaxie qui trace la dynamique des nuages d'hydrog\`ene
    mol\'eculaire. Carte de vitesse de l'\'emission CO int\'egr\'ee
    dans une bande de $\sim$4\textdegree de large en latitude
    centr\'ee sur le plan galactique. Plusieurs bras spiraux tournant
    \`a diff\'erentes vitesses sont visibles sur la figure. Le fort
    potentiel gravitationnel au centre de la Galaxie est responsable
    des grandes vitesses observ\'ees dans cette r\'egion. Figure
    extraite de \shortciteN{dame} } \label{fig:CO_MW_dame01}
\end{figure}

Dans le domaine infrarouge, le spectre des galaxies rec\`ele de
nombreuses raies atomiques, ioniques et mol\'eculaires qui ne
souffrent pas ou tr\`es peu d'extinction~; celles-ci permettent par
cons\'equent de sonder les milieux denses et poussi\`ereux tels que
les r\'egions de formation d'\'etoiles. Par ailleurs, les raies
d'\'emission du gaz sont essentielles pour l'\'etude \'energ\'etique,
dynamique et chimique du milieu interstellaire. Par exemple, les
observations spectroscopiques permettent de calculer le champ de
radiation UV des r\'egions de photodissociation, la densit\'e ou
encore la temp\'erature qui r\`egne dans le milieu interstellaire et
circumstellaire \shortcite{hollenbach}. Les abondances d'\'el\'ements
neutres ou ionis\'es peuvent aussi \^etre d\'etermin\'ees par ce
biais. De plus, certains \'el\'ements jouent un r\^ole central dans le
refroidissement des nuages mol\'eculaires, et donc dans la
condensation des c\oe urs pr\'e-stellaires, comme par exemple le
carbone et l'oxyg\`ene~; ces derniers poss\`edent en effet une grande
section efficace dans l'ultraviolet et r\'e\'emettent tr\`es
efficacement l'\'energie absorb\'ee dans l'infrarouge lointain,
notamment la raie de [C\small II \normalsize] \`a 158~$\mu$m et de
[O\small I \normalsize] \`a 63 et 146~$\mu$m. La haute r\'esolution
spectrale est un autre outils utilis\'e par les astrophysiciens pour
\'etudier entre autre la dynamique des nuages d'hydrog\`ene
mol\'eculaire, principalement gr\^ace \`a la mol\'ecule de CO qui est
un excellent traceur de H$_2$. La figure~\ref{fig:CO_MW_dame01} montre
une carte de vitesse de la Galaxie obtenue dans la raie 1-0 du CO \`a
115~GHz
\shortcite{dame}.

\subsubsection{Le rayonnement fossile}

Depuis sa d\'ecouverte en 1965 \shortcite{penzias}, le rayonnement
fossile, \'egalement appel\'e fond diffu cosmologique ou CMB pour
\emph{Cosmic Microwave Background}, repr\'esente l'un des piliers les
plus solides sur lequel s'appuie la th\'eorie du \emph{Big Bang}. Ce
rayonnement trouve son origine au moment du d\'ecouplage
mati\`ere-rayonnement environ $380\,000$~ans apr\`es la naissance de
l'Univers, lorsque la temp\'erature \'etait suffisamment basse pour
permettre aux \'electrons libres de se combiner avec les protons et
neutrons pr\'esents dans la soupe cosmique laissant ainsi la lumi\`ere
se propager librement dans l'espace. La temp\'erature d'\'equilibre
entre la mati\`ere et le rayonnement \'etait alors d'environ
3000~K. Depuis, l'expansion et le refroidissement de l'Univers ont
d\'ecal\'e vers les grandes longueurs d'onde les premiers photons
lib\'er\'es lors de la (re)combinaison\footnote{Le terme recombinaison
est g\'en\'eralement utilis\'e, toutefois le terme combinaison serait
plus appropri\'e puisque c'est la premi\`ere fois que les photons
peuvent se lier aux autres particules.}, et le spectre du rayonnement
fossile est aujourd'hui celui d'un corps noir dont la temp\'erature
est de $2.726\pm0.002$~K \shortcite{mather_CMB}. Le pic d'intensit\'e
du CMB se trouvant aux alentours de 1~mm de longueur d'onde, les
cosmologistes l'observent dans le domaine (sub-)millim\'etrique afin
de mieux contraindre son spectre. Notez que nous donnerons quelques
exemples d'observatoires d\'edi\'es \`a l'\'etude du CMB dans les
sections~\ref{sec:intro_astro_william_obs_obs}
et~\ref{sec:intro_bolometrie_bolo}.

\begin{figure}
  \begin{center}
    \includegraphics[width=0.8\textwidth,angle=0]{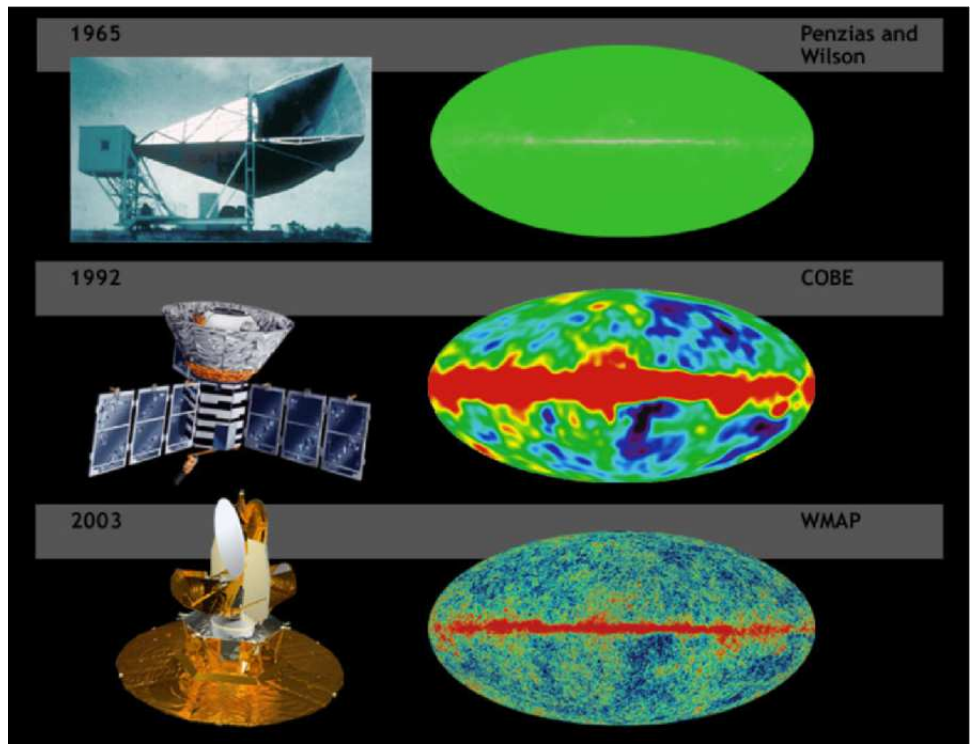}
    \end{center} \caption[\'Evolution de notre vision du CMB (Penzias
    \& Wilson, COBE et WMAP)]{La carte du haut est une simulation du
    ciel tel que l'antenne de Penzias et Wilson l'aurait observ\'e
    s'il avait \'et\'e possible d'effectuer un relev\'e complet du
    ciel avec celle-ci. Le CMB est isotrope et poss\`ede un spectre de
    corps noir \`a 2.726~K. La carte du milieu a \'et\'e obtenue par
    le satellite COBE \`a 53~GHz, elle montre la Galaxie et les
    anisotropies du CMB apr\`es soustraction du dip\^ole (effet
    doppler d\^u au mouvement de la Terre par rapport \`a la surface
    d'\'emission du CMB). L'\'echelle de couleur de cette carte montre
    des fluctuations en temp\'erature de seulement 10~$\mu$K. La carte
    du bas montre les fluctuations du CMB vues par WMAP avec une
    r\'esolution et une pr\'ecision 20~fois meilleure que COBE. (lien
    http$\!$://fr.wikipedia.org/wiki/Fond\_diffus\_cosmologique) }
    \label{fig:intro_astroIR_univers_emetteurCMB}
\end{figure}

Ce rideau de lumi\`ere cosmique qu'est le rayonnement fossile est
unique. Il rec\`ele de pr\'ecieuses informations \`a propos de
l'origine et du devenir de notre Univers. L'\'etude de ses
anisotropies (cf figure~\ref{fig:intro_astroIR_univers_emetteurCMB})
peut \'egalement nous renseigner sur les fluctuations de la
distribution de masse dans l'Univers primordial au moment de la
recombinaison. Par ailleurs, des mesures pr\'ecises de la polarisation
de ce rayonnement pourraient \^etre utilis\'ees pour d\'etecter des
ondes gravitationnelles \'emises pendant la phase d'inflation post-Big
Bang. Enfin, en suppl\'ement d'observations en rayons~X,
l'effet~Sunyaev-Zel'dovich\footnote{Effet~SZ~: Lorsqu'un photon
faiblement \'energ\'etique du CMB traverse un amas de galaxies, le
plasma d'\'electrons qui baigne l'amas lui communique une certaine
quantit\'e d'\'energie par l'effet Compton inverse modifiant ainsi le
spectre du CMB dans la direction de l'amas.} \shortcite{SZ} permet
d'\'etudier les structures gravitationnellement li\'ees les plus
massives de l'Univers, \`a savoir les amas de galaxies, et de calculer
quelques param\`etres physiques de ces amas tels que leur masse ou
leur vitesse.

\subsection{Les d\'etecteurs de rayonnement}
\label{sec:intro_astro_IR_detecteur}

La photod\'etection est un domaine tr\`es vaste dans lequel les
applications sont multiples et tr\`es diversifi\'ees. L'utilisateur
averti choisira donc le d\'etecteur le plus appropri\'e au type de
mesures qu'il compte effectuer connaissant les contraintes et
limitations de son syst\`eme d'\'etude et du d\'etecteur en
question. En astronomie par exemple, la limitation est souvent fix\'ee
par les flux extr\^emement faibles des sources observ\'ees
($\sim10^{-26}$~W/m$^2$/Hz dans l'IR). Les cam\'eras astronomiques
modernes poss\`edent g\'en\'eralement des performances qui approchent
les limites fondamentales de la photod\'etection, nous parlons alors
de d\'etecteurs BLIP pour \emph{Background Limited Infrared
Photodetector} (bruit d\'etecteur $\ll$ bruit de photon, cf
section~\ref{sec:intro_bolometrie_thermo_principe_bruit}). Le champ de
vue des instruments modernes a \'egalement tendance \`a s'\'elargir,
augmentant ainsi la vitesse de cartographie de grandes r\'egions du
ciel~; par exemple les relev\'es 2MASS
\shortcite{skrutskie}, UKIDSS \shortcite{lawrence}, CFHT Legacy Survey
\shortcite{cfhtls} , SDSS \shortcite{york},
etc... \shortciteN{bahcall} d\'efinissent une quantit\'e qu'ils
appellent le \og potentiel astronomique \fg et qui est proportionnelle
\`a~:
\begin{equation}
\left(\mbox{nombre de pixel}\right)\times\left(\mbox{sensibilit\'e du
pixel}\right)^2
\label{eq:bahcall_potentiel_astro}
\end{equation}
Cette expression indique le nombre de positions dans le ciel qui
peuvent \^etre observ\'ees en un temps donn\'e \`a une limite de
d\'etection donn\'ee. D'apr\`es \shortciteN{rieke}, l'astronomie
infrarouge moderne est fond\'ee sur le d\'eveloppement des
d\'etecteurs mono-pixel de la p\'eriode 1965~-~1985~; et les
v\'eritables matrices de d\'etecteurs comprenant un nombre raisonnable
de pixels ($\sim32\times32$) sont apparues il y a environ
20~ans. Depuis cette \'epoque, le potentiel astronomique au sens de
l'\'equation~(\ref{eq:bahcall_potentiel_astro}) a doubl\'e tous les
7~mois environ dans le domaine de la d\'etection infrarouge proche.

Les avanc\'ees fulgurantes de ces derni\`eres d\'ecennies dans le
domaine de l'imagerie thermique (de 1~\`a~30~$\mu$m) ont
\'et\'e principalement d\'evelopp\'ees par et pour les militaires. Ils
ont en effet \'etabli les bases de l'architecture et des processus de
fabrication des matrices de d\'etecteurs haute-performance. Quelques
industriels comme Raytheon Vision Systems ou Teledyne Imaging Sensors
(anciennement Rockwell Scientific Company) ont poursuivi les
d\'eveloppements et sont aujourd'hui les principaux fournisseurs de
d\'etecteurs infrarouge proche et moyen pour les grands
observatoires. La situation est sensiblement diff\'erente dans le
domaine de l'infrarouge lointain et du sub-millim\'etrique (de
30~\`a~1000~$\mu$m), en effet les astronomes repr\'esentent la
majorit\'e des utilisateurs de ce type de d\'etecteurs et ils ont
jou\'e un r\^ole majeur dans le d\'eveloppement des technologies
utilis\'ees aujourd'hui (bolom\'etrie infrarouge et d\'etecteurs
TeraHertz).

Nous donnons maintenant une br\`eve description du principe de
fonctionnement des d\'etecteurs les plus r\'epandus dans le domaine de
l'astronomie infrarouge~; et nous faisons la distinction entre les
d\'etecteurs quadratiques, qui sont sensibles uniquement au carr\'e de
l'amplitude du signal incident (i.e. la puissance), et les
d\'etecteurs coh\'erents, qui mesurent \`a la fois l'amplitude et la
phase du champ de radiation.

\subsubsection{Les d\'etecteurs large bande : Photoconducteurs et Bolom\`etres}
\label{sec:intro_astro_IR_detecteur_largebande}

La photod\'etection dans le r\'egime infrarouge proche et moyen repose
sur le principe de la photoconductivit\'e. L'\'el\'ement de base d'un
photoconducteur est un petit bloc de mat\'eriau semiconducteur
connect\'e de part et d'autre par deux \'electrodes qui \'etablissent
un champ \'electrique dans le volume du semiconducteur. Lorsqu'un
photon interagit avec le mat\'eriau, il peut \'eventuellement
lib\'erer des porteurs de charges qui migrent vers les \'electrodes
produisant ainsi un faible courant qui est ensuite mesur\'e par des
amplificateurs haute performance. Si l'\'energie du photon est
sup\'erieure \`a l'\'energie de liaison des \'electrons dans le
semiconducteur, alors les porteurs de charges sont lib\'er\'es en
brisant une liaison du cristal semiconducteur~; nous parlons alors de
\emph{photoconductivit\'e intrins\`eque}. Dans le jargon de la
physique du solide, nous dirions qu'un \'electron est promu de la
bande de valence vers la bande de conduction. Les d\'etecteurs bas\'es
sur l'absorption intrins\`eque peuvent fonctionner jusqu'\`a une
longueur d'onde de coupure $\lambda_C$ qui correspond \`a l'\'energie
de la bande interdite du mat\'eriau~:
\begin{equation}
\lambda_C=\frac{hc}{E_g}=\frac{1.24\,\mu\mbox{m}}{E_g\,(\mbox{eV})}
\label{eq:photoconduc_lambdaC}
\end{equation}
o\`u $h$ est la constante de Planck, $c$ la c\'el\'erit\'e de la
lumi\`ere et $E_g$ l'\'energie de la bande interdite ($E_g$ = gap
energy). Le silicium, par exemple, a une bande interdite de
$\sim1.12$~eV, ce qui correspond \`a une longueur d'onde d'absorption
maximale $\lambda_C\sim1.1$~$\mu$m.

Pour observer dans l'infrarouge moyen et lointain, il est possible de
r\'eduire significativement $E_g$ en introduisant des impuret\'es dans
le mat\'eriau semiconducteur. Ces impuret\'es sont des atomes
exog\`enes qui se logent dans le r\'eseau cristallin et dont les
niveaux \'energ\'etiques se trouvent entre la bande de valence et la
bande de conduction du semiconducteur. Nous parlons alors de
\emph{photoconductivit\'e extrins\`eque}. Il est par ailleurs possible
d'am\'eliorer les performances d'un photoconducteur en s\'eparant les
fonctions optiques (cr\'eation des porteurs de charges) des fonctions
\'electriques (transport des porteurs de charges). C'est par exemple
le cas des photodiodes \`a base de InSb avec $\lambda_C=5.5$~$\mu$m
\shortcite{hoffman}, ou bien \`a base de Hg$_{(1-x)}$Cd$_x$Te. Le
grand avantage de ce dernier exemple est que $E_g$ peut \^etre
modifi\'e en changeant la composition \'el\'ementaire du mat\'eriau
$x$ \shortcite{garnett}, $\lambda_C$ peut atteindre
$\sim$15~$\mu$m. Une autre approche de s\'eparation des fonctions
optiques et \'electriques d'un photoconducteur est \`a l'oeuvre dans
les d\'etecteurs \emph{IBC}, \emph{Impurity band conduction}. Ils
offrent d'excellentes performances jusqu'\`a 18~$\mu$m pour un
semiconducteur de type Si$:$Ga \shortcite{hogue}, 28~$\mu$m pour le
Si$:$As qui a \'et\'e utilis\'e pour les trois instruments de Spitzer,
ou encore 40~$\mu$m pour le Si$:$Sb \shortcite{huffman}. Au del\`a de
40~$\mu$m, aucun semiconducteur bas\'e sur les technologies silicium
n'offre des performances suffisamment bonnes pour l'instrumentation
astronomique. Les d\'etecteurs infrarouge lointain utilisent plut\^ot
des photoconducteur \`a base de germanium comme par exemple les
d\'etecteurs du satellite AKARI \shortcite{fujiwara} qui couvrent le
r\'egime spectral de 50~\`a 180~$\mu$m, ou bien le spectrom\`etre de
l'instrument Herschel/PACS \shortcite{poglitsch2003}. Pour atteindre
des \'energies aussi faibles, de l'ordre de $6\times 10^{-3}$~eV, des
tensions m\'ecaniques doivent \^etre appliqu\'ees aux mat\'eriaux
semiconducteurs pour modifier physiquement la taille du r\'eseau
cristallin et ainsi rapprocher la bande de conduction de la bande
d'impuret\'e. Par exemple les d\'etecteurs du spectrom\`etre PACS (cf
section~\ref{sec:detect_observatoire_sat_instru}) sont press\'es par
un \og \'etau \fg qui applique une pression uniaxiale de plus de
700~N/mm$^2$~; le plan focal de l'instrument est alors tr\`es
encombr\'e et le nombre de pixels est limit\'e \`a quelques milliers
($16\times25$ pour PACS). D'autre part, l'efficacit\'e quantique de ce
type de d\'etecteur chute avec la longueur d'onde et ne d\'epasse pas
40~\% au del\`a de 200~$\mu$m.\\

Pour les plus grandes longueur d'onde, les photoconducteurs sont
supplant\'es par les bolom\`etres. Ce sont en effet les d\'etecteurs
qui offrent les meilleures performances en termes de sensibilit\'e
pour observer dans l'infrarouge lointain et le
sub-millim\'etrique. Notez que le principe de d\'etection d'un
bolom\`etre est tr\`es diff\'erent de celui d'un photoconducteur, il
repose en fait sur l'absorption du rayonnement par un petit
\'el\'ement isol\'e thermiquement de la structure du d\'etecteur,
l'\'energie radiative est alors transform\'ee en chaleur et
l'\'el\'evation de temp\'erature induite est mesur\'ee par un
thermom\`etre. Le senseur thermique est en g\'en\'eral une
r\'esistance dont l'imp\'edance varie fortement avec la
temp\'erature. Nous d\'ecrirons en d\'etails le fonctionnement d'un
bolom\`etre dans la section~\ref{sec:intro_bolometrie_thermo}. Par
ailleurs, \shortciteN{richards} pr\'esente une excellente revue sur
les d\'etecteurs bolom\'etriques pour l'infrarouge et le
millim\'etrique.\\

Photoconducteurs et bolom\`etres sont qualifi\'es de d\'etecteurs
large bande car ils peuvent absorber le rayonnement
\'electromagn\'etique sur des plages de longueur d'onde relativement
\'etendues.  Pour les photoconducteurs, cette plage est limit\'ee aux
hautes \'energies par la faible section efficace des particules
incidentes par rapport \`a un semiconducteur d'\'epaisseur donn\'ee~;
aux basses fr\'equences la limite est fix\'ee comme nous l'avons vu
par l'\'energie n\'ecessaire aux photons pour lib\'erer des porteurs
de charge. Pour les bolom\`etres, l'absorption du rayonnement ne
d\'epend pas de la longueur d'onde, et il serait en th\'eorie possible
de fabriquer des bolom\`etres pour d\'etecter le rayonnement
\'electromagn\'etique des ondes radio aux rayons~$\gamma$. Toutefois,
les performances des bolom\`etres ne sont aujourd'hui comp\'etitives
que pour la d\'etection du rayonnement infrarouge
lointain/(sub-)millim\'etrique et les rayons~X.

\subsubsection{Les d\'etecteurs h\'et\'erodynes}
\label{sec:intro_astro_IR_detecteur_heterodyne}

Le principe de la d\'etection coh\'erente repose sur la mesure
simultan\'ee de l'amplitude et de la phase du champ de radiation
incident, \cad que le signal \'electrique d\'etect\'e par le
r\'ecepteur pulse \`a la m\^eme fr\'equence que l'onde
\'electromagn\'etique. La technologie utilis\'ee pour ces r\'ecepteurs
est \`a base de FET (\emph{Field Effect Transistors}) ou de HEMT
(\emph{High Electron Mobility Transistors}). Pour des fr\'equences
sup\'erieures au GHz, il est en pratique tr\`es difficile d'amplifier
le signal et de le transporter sur de grandes longueurs de c\^able~;
il est alors n\'ecessaire de ramener la fr\'equence du signal
astronomique \`a une fr\'equence interm\'ediaire plus basse et
d'effectuer l'amplification \`a cette fr\'equence l\`a. Ces
d\'etecteurs sont qualifi\'es de h\'et\'erodyne \`a cause de la
transposition de fr\'equence du signal utile. Cette op\'eration est
r\'ealis\'ee par un composant \'electronique non-lin\'eaire, le
m\'elangeur, et d'un oscillateur local qui fournit une onde
sinuso\"idale de fr\'equence constante assez proche de celle du signal
astronomique.

Plusieurs types de m\'elangeurs sont disponibles. Les diodes Shottky
sont principalement utilis\'ees aux grandes longueurs d'ondes
($\lambda>1$~cm), elles sont peu on\'ereuses et peuvent fonctionner
\`a temp\'erature ambiante. Dans le domaine millim\'etrique, les
m\'elangeurs SIS (\emph{Supraconducteur-Isolant-Supraconducteur}) sont
plus sensibles et donc plus int\'eressants pour les applications en
astronomie. L'\'el\'ement non-lin\'eaire se pr\'esente sous la forme
d'une jonction dans laquelle une couche d'isolant tr\`es fine (environ
30~Angstroms) s\'epare deux \'electrodes supraconductrices typiquement
en niobium. Un tel empilement s'appelle une jonction de Josephson
\shortcite{josephson}. 
Dans le domaine sub-millim\'etrique, les \'el\'ements m\'elangeurs les
plus sensibles sont des HEB (\emph{Hot Electron Bolometers}). Les
bolom\`etres sont en effet des composants non-lin\'eaires, mais la
plupart d'entre eux n'est pas suffisamment rapide pour \^etre
utilis\'e comme des m\'elangeurs dans le r\'egime
TeraHertz. Toutefois, pour les HEB, les \'electrons du mat\'eriau
poss\`edent une temp\'erature sup\'erieure \`a celle des phonons,
donnant ainsi une constante de temps compatible avec les hautes
fr\'equences du domaine sub-millim\'etrique. Avec le d\'eveloppement
r\'ecent de ces HEB, nous rentrons v\'eritablement dans l'\`ere de
l'astronomie TeraHertz\footnote{Le terme TeraHertz est utilis\'e pour
faire r\'ef\'erence au mode de d\'etection h\'et\'erodyne en
opposition au r\'egime sub-millim\'etrique qui fait r\'ef\'erence \`a
un mode de d\'etection incoh\'erent.}. Par exemple,
\shortciteN{cherednichenko} pr\'esentent des m\'elangeurs HEB \`a base
de NbN fonctionnant \`a 2.5~THz~; \shortciteN{hubers} utilisent le
m\^eme mat\'eriau et poussent la limite jusqu'\`a 5.2~THz.

Le grand avantage de la d\'etection coh\'erente est qu'elle permet une
r\'esolution spectrale virtuellement infinie gr\^ace aux
autocorr\'elateurs qui poss\`edent des dixaines de milliers de
voies. L'instrument HERSCHEL/HIFI poss\`ede par exemple une
r\'esolution $R=\frac{\nu}{\Delta\nu}\sim 10^7$
\shortcite{degraauw}. Notez toutefois que la bande passante des
d\'etecteurs h\'et\'erodynes est relativement limit\'ee
($\sim$4~GHz)~; ces d\'etecteurs sont donc d\'edi\'es \`a l'\'etude
des raies spectrales contrairement aux d\'etecteurs large bande qui
sont plus adapt\'es \`a l'observation du continuum. Une des faiblesses
des d\'etecteurs h\'et\'erodynes est qu'ils ne poss\`edent
g\'en\'eralement qu'un seul pixel ce qui rend la cartographie du ciel
extr\`emement longue~; cependant, depuis une d\'ecennie, nous voyons
appara\^itre des matrices de d\'etecteurs comme par exemple CHAMP$^+$
qui fonctionne \`a 450~et 350~$\mu$m sur le t\'elescope APEX et qui
poss\`ede 8 pixels \shortcite{kasemann}. L'instrument SUPERCAM
fonctionne \`a 870~$\mu$m et devrait \^etre install\'e fin~2007 sur le
t\'elescope Heinrich Hertz en Arizona, il contiendra 64~pixels
\shortcite{groppi}. Enfin, un dernier avantage notable de la
d\'etection coh\'erente est qu'elle permet d'obtenir des r\'esolutions
spatiales inf\'erieures \`a la milliseconde d'arc par la technique
d'interf\'erom\'etrie et de synth\`ese d'ouverture.





\subsection{Observations du ciel dans l'infrarouge lointain et le (sub-)millim\'etrique}
\label{sec:intro_astro_william_obs}


Dans l'infrarouge lointain, l'atmosph\`ere terrestre est un \'ecran
opaque, brillant et turbulent. Il existe toutefois quelques
\emph{fen\^etres atmosph\'eriques} \`a travers lesquelles les
astronomes peuvent observer le ciel avec des t\'elescopes au sol.

\subsubsection{Atmosph\`ere et vapeur d'eau}
\label{sec:intro_astro_william_obs_atmo}

\begin{figure}
  \begin{center}
    \includegraphics[width=0.7\textwidth,angle=0]{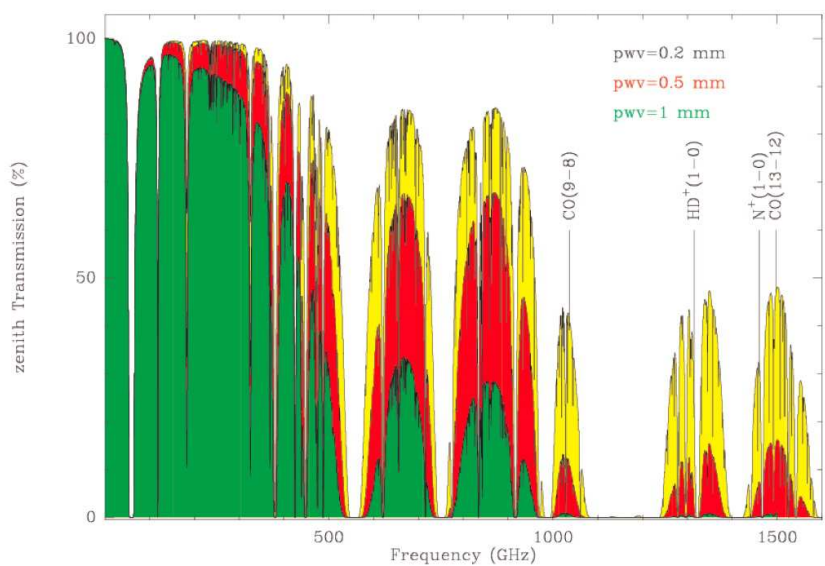}
  \end{center}
  \caption[Courbes de transmisison de l'atmosph\`ere au-dessus du
  Cerro Chajnantor]{Courbes de transmission de l'atmosph\`ere au
  dessus du Cerro Chajnantor. Chaque couleur correspond \`a une
  quantit\'e de vapeur d'eau pr\'ecipitable donn\'ee (0.2, 0.5 et
  1~mm). La transmission dans les fen\^etres \`a 350~et 450~$\mu$m
  (850~et 660~GHz) est tr\`es d\'ependente des conditions
  atmosph\'eriques. Une fen\^etre \`a 200~$\mu$m (1500~GHz) s'ouvre
  p\'eniblement pour des PWV inf\'erieures 0.2~mm. Cette courbe est
  extraite de \shortciteN{reveret}.}
  \label{fig:intro_astro_william_obs_atmo_chajnantor}
\end{figure}

L'absorption du rayonnement sub-millim\'etrique est principalement
caus\'ee par les transitions mol\'eculaires de type rotationnelle des
mol\'ecules de vapeur d'eau pr\'esentes dans les couches inf\'erieures
de l'atmosph\`ere \`a environ 2~km d'altitude. Des \'el\'ements tels
que l'oxyg\`ene ou l'ozone contribuent \'egalement, mais plus
faiblement,
\`a l'absorption du rayonnement. La
figure~\ref{fig:intro_astro_william_obs_atmo_chajnantor} montre la
courbe de transmission de l'atmosph\`ere jusqu'\`a 1500~GHz,
\cad $\lambda>200$~$\mu$m, au-dessus du site de Chajnantor dans le
d\'esert d'Atacama. Dans le sub-millim\'etrique, nous trouvons trois
fen\^etres partiellement ouvertes \`a 350, 450~et 870~$\mu$m. Dans le
millim\'etrique, les fen\^etres \`a 1.3, 2~et 3~mm sont plus larges et
offrent une bien meilleure transmission.

Les astronomes quantifient la transmission atmosph\'erique de deux
mani\`eres diff\'erentes. La plus r\'epandue consiste \`a effectuer
des \emph{skydips}, ou profils d'atmosph\`ere~; \cad qu'ils mesurent
l'\'emission de l'atmosph\`ere en fonction de l'\'el\'evation
\`a l'aide d'un radiom\`etre, puis ils calculent l'opacit\'e
atmosph\'erique gr\^ace \`a un mod\`ele de transfert
radiatif. L'opacit\'e \'etant habituellement mesur\'ee \`a une
fr\'equence de 225~GHz, les conditions atmosph\'eriques sont donc
repr\'esent\'ees par la grandeur $\tau_{225}$. L'autre mani\`ere
consiste \`a \'evaluer la quantit\'e de vapeur d'eau pr\'esente
au-dessus du t\'elescope. Cette quantit\'e est souvent donn\'ee en
millim\`etre d'eau pr\'ecipitable (\emph{PWV: Precipitable Water
Vapor} en anglais). Les meilleurs sites au monde offrent des
\emph{PWV} de l'ordre de 0.1-0.2~mm pour seulement quelques dizaines
de nuits par an. Notez que les bandes spectrales \`a 350~et 450~$\mu$m
sont particuli\`erement sensibles \`a l'humidit\'e de l'atmosph\`ere
(cf
figure~\ref{fig:intro_astro_william_obs_atmo_chajnantor}). \shortciteN{thomas}
proposent une relation entre l'opacit\'e \`a 225~GHz et la quantit\'e
de vapeur d'eau pr\'ecipitable, relation utile pour transposer les
conditions d'observations du millim\'etrique vers l'infrarouge
lointain.

En plus d'absorber le rayonnement submillim\'etrique, l'atmosph\`ere
est environ dix mille fois plus brillante que la plupart des sources
astrophysiques. Elle se comporte en effet comme un corps gris \`a
$\sim$250~K dont l'\'emissivit\'e d\'epend de la quantit\'e de vapeur
d'eau pr\'esente dans l'atmosph\`ere.  Le contraste des images
sub-millim\'etriques (ou rapport signal-\`a-bruit) peut \^etre
astucieusement augment\'e par le biais de modes d'observation
particuliers et en co-additionant un grand nombre d'images
individuelles.

D'autre part, la vapeur d'eau n'est pas distribu\'ee de fa\c{c}on
homog\`ene dans l'atmosph\`ere, elle est plut\^ot condens\'ee dans des
cellules de taille variable qui se d\'eplacent au gr\'e des
vents. L'intensit\'e lumineuse re\c{c}ue au niveau du d\'etecteur
varie donc au rythme du passage de la vapeur d'eau dans le champ de
vue du t\'elescope. Ces lentes fluctuations introduisent du bruit
basse fr\'equence que l'on appelle \emph{sky noise} en anglais. Il est
possible de s'affranchir de ce bruit en modulant le signal \`a une
fr\'equence plus rapide que la vitesse de d\'efilement des
cellules. Nous parlons alors d'observations chopp\'ees (du mot anglais
\emph{chopper}). Un autre effet ind\'esirable d\^u au d\'eplacement
des cellules est l'\'equivalent du ph\'enom\`ene de
\emph{seeing} dans le domaine optique. En effet, la distribution de
vapeur d'eau le long du trajet optique \'evolue avec le temps,
l'indice de r\'efraction change \'egalement avec le d\'eplacement des
cellules et peut introduire une incertitude non-n\'egligeable sur le
pointage du t\'elescope.

\subsubsection{Les grands observatoires}
\label{sec:intro_astro_william_obs_obs}

Les observations dans le domaine sub-millim\'etrique sont relativement
difficiles \`a r\'ealiser du fait de la pr\'esence de l'atmosph\`ere
et du contraste extr\`emement faible des sources astrophysiques par
rapport \`a l'\'emission d'avant-plan (atmosph\`ere, t\'elescope et
int\'erieur de l'instrument). Pour explorer le ciel dans l'une des
r\'egions les moins \'etudi\'ees du spectre \'electromagn\'etique, les
astronomes ont construits de nombreux observatoires sp\'ecialement
d\'edi\'es aux r\'egimes infrarouge lointain et
(sub-)millim\'etrique.

Pour minimiser les effets n\'efastes de l'atmosph\`ere, il faut
r\'eduire au maximum la colonne d'air humide qui se trouve entre le
t\'elescope et la source. Cela implique de construire les grands
observatoires en altitude dans des endroits extr\`emement secs. Les
trois principaux sites qui autorisent des observations de bonnes
qualit\'es dans l'infrarouge lointain sont~:
\begin{itemize}
\item Le Mauna Kea \`a Hawaii (4200~m) o\`u se trouvent trois
observatoires majeurs qui sont le CSO (Caltech Submillimeter
Observatory, 10~m de diam\`etre) avec son optique active unique
\shortcite{leong}, le JCMT (James Clerk Maxwell Telescope, 15~m) et le
SMA (SubMillimeter Array, 8~antennes de 6~m).
\item Le Cerro Chajnantor dans le d\'esert d'Atacama au Chili
(5200~m). Les t\'elescopes APEX (Atacama Pathfinder EXperiment, 12~m)
et NANTEN2 (nanten signifie ciel austral en japonais, 4~m) y sont
d\'ej\`a install\'es. Ce site accueillera \'egalement le grand
interf\'erom\`etre ALMA (Atacama Large Millimeter Array, environ
50~antennes de 12~m) et tr\`es probablement le t\'elescope CCAT
(Cornell Caltech Atacama Telescope, 25~m).
\item Le plateau antarctique (3000~m) poss\`ede d\'ej\`a plusieurs
t\'elescopes am\'ericains sur leur base du p\^ole sud avec entre autre
le SPT (South Pole Telescope, 10~m) et AST/RO (Antarctic Submillimeter
Telescope and Remote Observatory, 1.7~m). L'atmosph\`ere au-dessus du
continent est tr\`es s\`eche \`a cause des tr\`es basses
temp\'eratures qui y r\`egnent. De nombreux auteurs s'accordent \`a
dire que l'antarctique est potentiellement le meilleur endroit sur
Terre pour observer le ciel du visible au sub-millim\'etrique
\shortcite{burton,sironi,mould,minier}.
\end{itemize}
Les observations (sub-)millim\'etriques \'etant moins exigentes en
terme de PWV, les t\'elescopes sont plus nombreux et se trouvent en
g\'en\'eral \`a des endroits plus faciles d'acc\`es mais toujours en
altitude. Par exemple nous trouvons le t\'elescope de 30~m de l'IRAM
au pico Veleta en Espagne \`a 2850~m, l'interf\'erom\`etre CARMA
(Combined Array for Millimeter Wave Astronomy) en Californie \`a
2200~m ou encore le tr\`es r\'ecent LMT (Large Millimeter Telescope)
de 50~m de diam\`etre au Mexique \`a 4600~m d'altitude.\\

Pour pouvoir observer en-dehors des fen\^etres atmosph\'eriques, il
est n\'ecessaire de monter encore plus haut en altitude. L'avion KAO
(Kuiper Airborne Observatory) par exemple volait \`a environ 12~km
d'altitude et observait le ciel entre 1~et 500~$\mu$m avec un
t\'elescope embarqu\'e de 90~cm. Son successeur SOFIA (Stratospheric
Observatory for Infrared Astronomy) devrait abriter un t\'elescope de
2.5~m de diam\`etre ainsi que 9~instruments. 

Les ballons stratosph\'eriques sont aussi des options int\'eressantes
pour s'affranchir de l'atmosph\`ere terrestre, ils permettent en effet
de s'\'elever \`a environ 38~km d'altitude. Par exemple, lors de son
troisi\`eme vol fin 2006, l'exp\'erience BLAST
\shortcite{devlin} a observ\'e le ciel sub-millim\'etrique avec son
t\'elescope de 2~m de diam\`etre pendant plus de 10~jours alors qu'il
tournait autour du continent antarctique port\'e par les vents
circumpolaires. En 2000 et 2001, les exp\'eriences en ballon MAXIMA
\shortcite{hanany_maxima} et BOOMERANG \shortcite{lange} ont
r\'ecolt\'e des donn\'ees tr\`es importantes qui ont permis de montrer
que l'Univers \'etait \og plat \fg, au sens topologique du terme.
Notez \'egalement les exp\'eriences franco-europ\'eennes PRONAOS
\shortcite{serra}, Archeops \shortcite{benoit} et le futur ballon
PILOT \shortcite{bernard} qui devrait mesurer entre 2010 et 2012 la
polarisation du milieu interstellaire \`a 240~et 550~$\mu$m avec des
matrices de bolom\`etres du type HERSCHEL/PACS \shortcite{billot}. 

La derni\`ere \'etape, la plus on\'ereuse mais \'egalement la plus
efficace, consiste \`a satelliser le t\'elescope. Il existe certes de
lourdes contraintes sur la charge utile embarqu\'ee notamment sur la
taille du r\'eflecteur, mais dans l'espace la transmission est de
100~\% \`a toutes les longueurs d'onde, l'environnement est tr\`es
stable, les \'el\'ements optiques sont plus froids et donc moins
brillants dans l'infrarouge lointain. La sensibilit\'e des
t\'elescopes spatiaux est sans commune mesure avec celle des meilleurs
observatoires terrestres ou m\^eme stratosph\'eriques. Une \'etape
importante en terme de seuil de d\'etection ($\sim$1~Jy) a en effet
\'et\'e franchie en 1983 avec le lancement du satellite IRAS
\shortcite{neugebauer}. En 10~mois d'op\'eration, il a observ\'e 96~\%
du ciel entre 10~et 120~$\mu$m (cf figure~\ref{fig:allsky_IRAS}) et a
d\'etect\'e environ 350000 sources dont 20000 galaxies. La NASA a
ensuite lanc\'e le satellite COBE
\shortcite{boggess} pour
\'etudier le fond diffu cosmologique ainsi que les avant-plans
astrophysiques. Puis le satellite europ\'een ISO \shortcite{kessler} a
observ\'e le ciel entre 2.5~et 240~$\mu$m pendant deux ans et demi, et
a r\'ev\'el\'e entre autre l'omnipr\'esence de la poussi\`ere froide
Galactique et extra-galactique ind\'etectable par IRAS. En 2001,
l'agence spatiale am\'ericaine lance le t\'elescope WMAP
\shortcite{bennett} pour mesurer les anisotropies du CMB avec une 
meilleure r\'esolution que COBE. Puis vinrent les satellites Spitzer
\shortcite{werner} et AKARI \shortcite{murakami}, les dignes successeurs 
de ISO et IRAS respectivement.
Les satellites de l'ESA, Planck \shortcite{tauber} et
Herschel \shortcite{pilbratt}, devraient \^etre lanc\'es fin 2008 (cf
section~\ref{sec:detect_observatoire_FIRST}). Enfin, dans un futur
plus lointain, des projets ambiteux tels que le JWST
\shortcite{clampin} avec son t\'elescope d\'epliable de 6~m de
diam\`etre, l'observatoire japonais SPICA \shortcite{nakagawa} avec
son miroir de 3.5~m refroidi \`a 4.5~K, ou encore SAFIR
\shortcite{lester} et son miroir de 8-10~m \`a 5~K devraient voir le
jour pour pousser encore plus loin les limites de notre connaissance,
particuli\`erement en ce qui concerne la d\'etection de plan\`etes
extrasolaires, la formation des \'etoiles et des galaxies, et de
mani\`ere plus g\'en\'erale l'origine de notre Univers.

\section{L'Observatoire Spatial Herschel}
\label{sec:detect_observatoire_FIRST}

\subsection{La mission Herschel et ses objectifs scientifiques}
\label{sec:detect_observatoire_FIRST_drivers}

La mission spatiale Herschel est un projet de tr\`es grande envergure,
c'est la quatri\`eme \og pierre angulaire\footnote{Les
\emph{cornerstone missions} sont les plus gros projets organis\'es par
l'agence, leur budget d\'epasse bien souvent le milliard
d'euros. Notez que Herschel \'etait initialement pr\'evu pour \^etre
la troisi\`eme pierre angulaire de l'ESA mais elle a \'et\'e
d\'ecal\'ee et remplac\'ee par la mission ROSETTA \shortcite{rosetta}
car le voyage de la sonde inter-plan\'etaire doit durer plus de 10~ans
pour rejoindre son objectif~: la com\`ete Churyumov-Gerasimenko.} \fg
du programme scientifique \emph{Horizon 2000} de l'Agence Spatiale
Europ\'eenne (ESA). Son but est d'observer l'Univers dans une des
r\'egions du spectre \'electromagn\'etique les moins \'etudi\'ees \`a
ce jour, entre 60~et 670~$\mu$m, en s'affranchissant de tout effet
atmosph\'erique, \cad faible \'emission de fond et acc\`es \`a toutes
les longueurs d'onde du spectre. Par rapport aux missions spatiales
pr\'ec\'edentes, Herschel permettra d'observer des objets plus froids
et plus lointains avec une meilleure r\'esolution spatiale et
spectrale~; son \og potentiel d\'ecouverte \fg est consid\'erable.

L'observatoire poss\`ede un t\'elescope de 3.5~m de diam\`etre et
trois instruments scientifiques (photom\'etrie et spectroscopie). Le
satellite embarque une quantit\'e d'h\'elium suffisante pour assurer le
fonctionnement des instruments sur une dur\'ee minimum de 3~ans. Le
lancement est actuellement pr\'evu pour la fin de l'ann\'ee~2008.

\subsubsection{Les motivations premi\`eres}
\label{sec:detect_observatoire_FIRST_drivers_1}

\begin{figure}
  \begin{center}
    \begin{tabular}[t]{ll}
      \includegraphics[height=0.6\textheight]{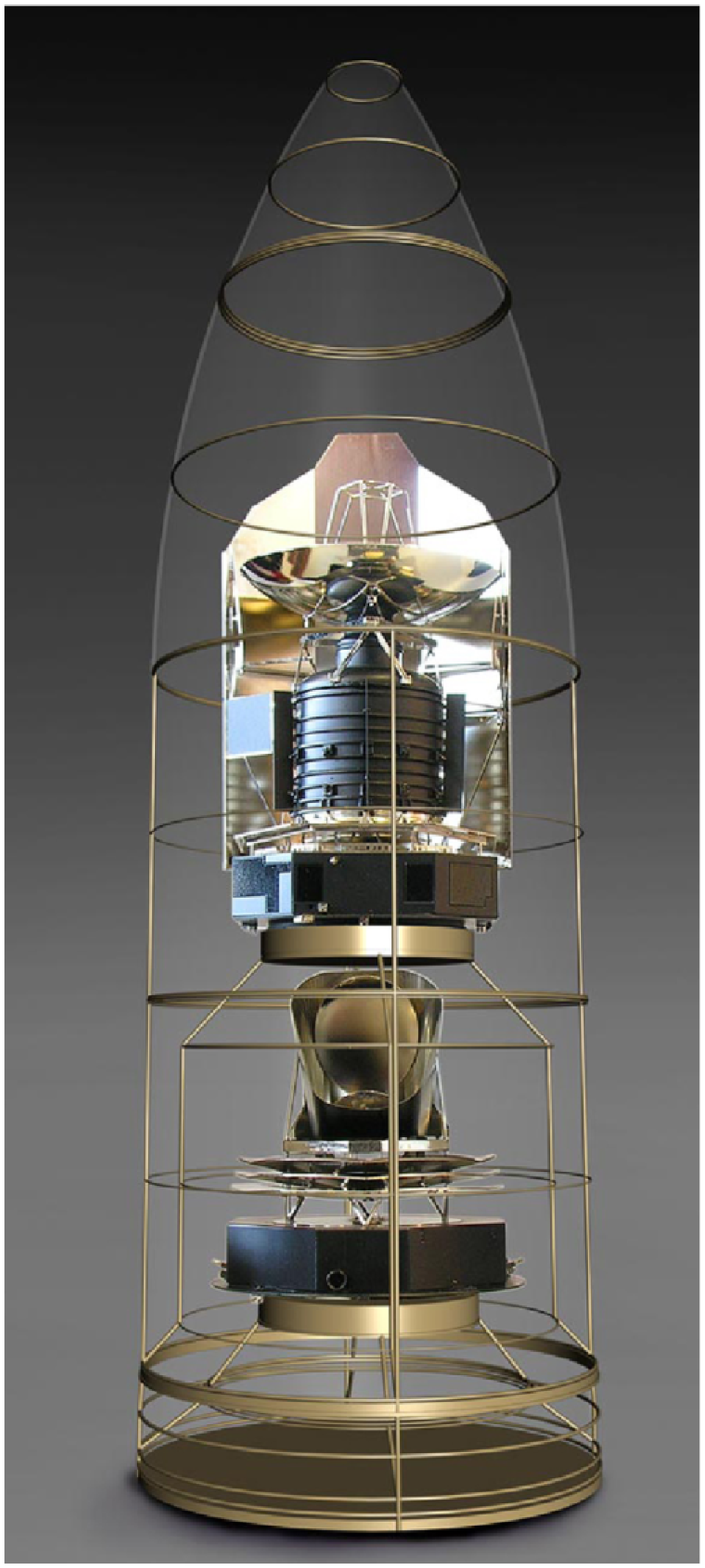} & \includegraphics[height=0.6\textheight]{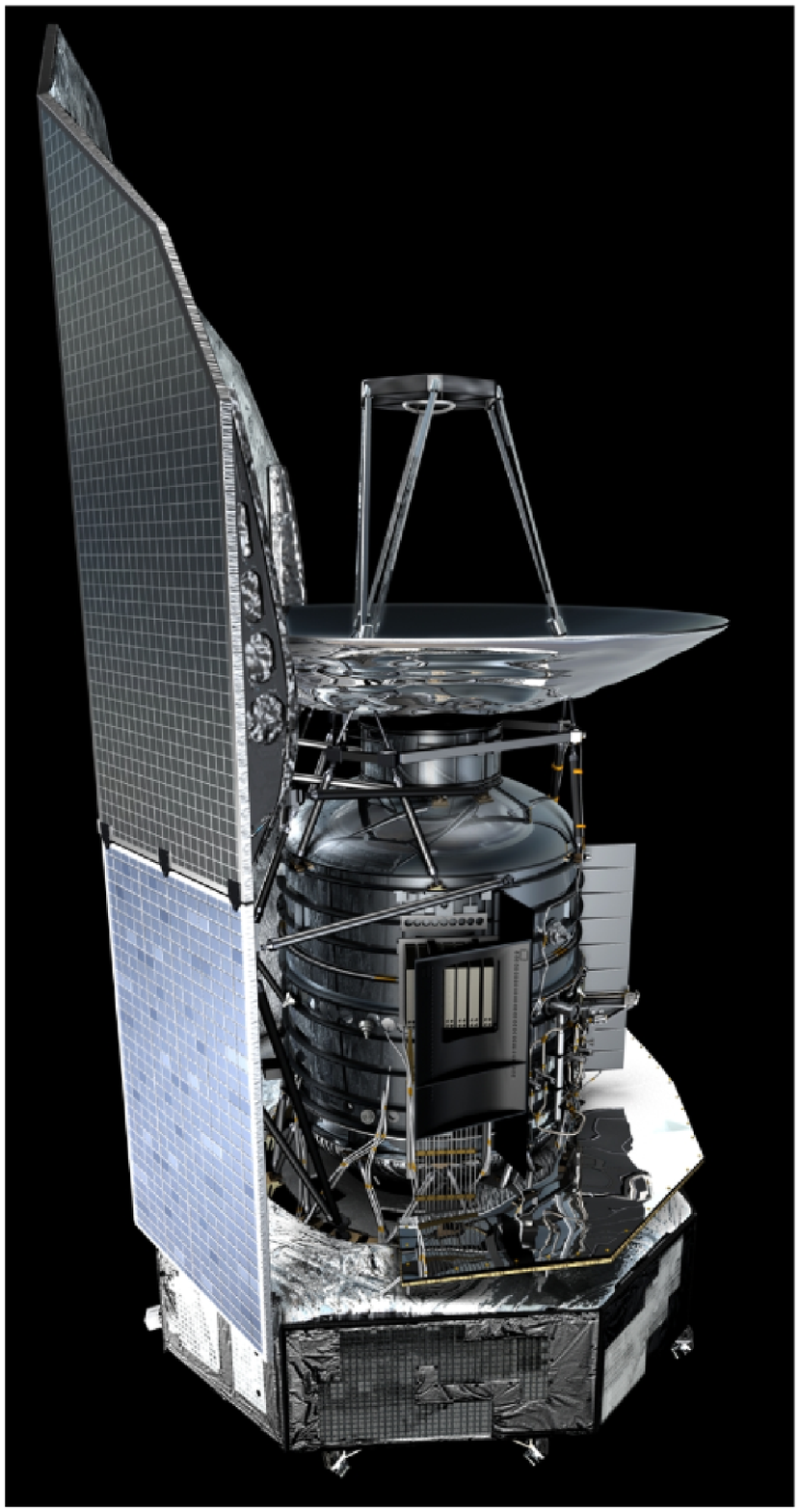} 
    \end{tabular}
  \end{center}
  \caption[L'Observatoire Herschel (+Planck) sous la coiffe de
Ariane~5]{\`A gauche: Les deux satellites de l'ESA Herschel (en haut)
et Planck (en bas) seront lanc\'es en 2008 par une fus\'ee Ariane~5 de
la base de Kourou en Guyane fran\c{c}aise. \`A droite, une vue
d'artiste de l'Observatoire Spatial Herschel (HSO). Herschel mesure
plus de 9~m de haut pour un diam\`etre de 4~m. (lien http$\!$://esamultimedia.esa.int/)
  \label{fig:detect_observatoire_herschel_ariane_HSO}}
\end{figure}

D'apr\`es G\"oran Pilbratt, le responsable scientifique de la mission,
l'id\'ee de spatialiser un t\'elescope pour \'etudier le ciel dans
l'infrarouge lointain a \'emerg\'e dans les ann\'ees~70. C'est ensuite
en~1980 que l'ESA a re\c{c}ut une proposition de la part de la
communaut\'e scientifique europ\'eenne pour r\'ealiser un tel
projet. L'ESA lance donc une \'etude de faisabilit\'e qui conclue que
ce projet est r\'ealisable moyennant quelques avanc\'ees
technologiques, notamment \`a propos de m\'elangeurs SIS fonctionnant
\`a 1.5~THz (il s'est av\'er\'e par la suite que cette technologie
n'est pas adapt\'ee \`a de si hautes fr\'equences). L'observatoire fut
provisoirement appel\'e FIRST (\emph{Far InfraRed Space
Telescope}). Dans les ann\'ees qui suivirent, le design de FIRST
\'evolua vers un t\'elescope de 3.5~m refroidit passivement avec trois
instruments scientifiques \`a son bord. En 1997, l'ESA lance un appel
\`a proposition pour la construction des 3~instruments
scientifiques. Le projet regroupe aujourd'hui plusieurs dizaines
d'instituts de recherche europ\'eens et am\'ericains~; de nombreux
industriels europ\'eens ont \'egalement \'et\'e contract\'es pour
r\'ealiser le satellite, notamment Alcatel Space, Astrium GmbH et
Aliena Spazio. De plus, la mission Herschel repr\'esente un
v\'eritable d\'efi technologique~; par exemple les matrices de
bolom\`etres du photom\`etre courte longueur d'onde sont les premiers
d\'etecteurs bolom\'etriques de ce type, l'instrument h\'et\'erodyne
poss\`ede \'egalement des m\'elangeurs fonctionnant dans le domaine du
TeraHertz qui est un champ d'\'etude encore nouveau, et le
t\'elescope de l'Observatoire Herschel est aussi une premi\`ere
puisqu'il sera le plus grand miroir jamais envoy\'e dans l'espace.
En~2000, \`a la conf\'erence de Toledo \shortcite{pilbratt_toledo},
FIRST est renomm\'e en l'honneur de la personne qui a d\'ecouvert le
rayonnement infrarouge et qui a construit les plus grands t\'elescopes
de son \'epoque~; l'observatoire spatial de l'ESA prend alors le nom
de Herschel.

\`A la m\^eme \'epoque, l'ESA pr\'eparait une autre mission, de taille
moyenne celle-ci (de type M3), dont l'objectif principal est de
cartographier les anisotropies du CMB sur tout le ciel avec une
pr\'ecision de $\Delta T/T=2\times10^{-6}$ et une r\'esolution
angulaire de 10~minutes d'arc~: il s'agit de la mission Planck. Les
similitudes entre les missions Herschel et Planck ont pouss\'e le
comit\'e scientifique de l'ESA \`a combiner ces deux missions
\shortcite{passvogel}. Les deux satellites seront donc mis en orbite
par le m\^eme lanceur. La
figure~\ref{fig:detect_observatoire_herschel_ariane_HSO} montre la
configuration choisie pour installer Herschel et Planck sous la coiffe
de la fus\'ee Ariane~5.

\subsubsection{Les objectifs scientifiques}
\label{sec:detect_observatoire_FIRST_objectifs}

\begin{figure}
  \begin{center}
    \includegraphics[width=0.7\textwidth,angle=0]{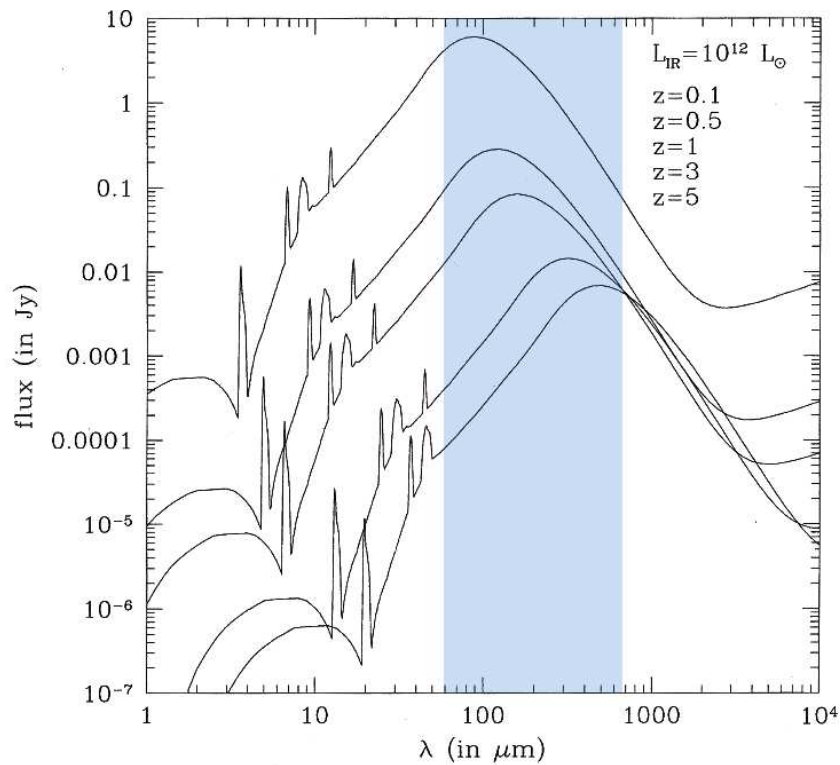}
  \end{center}
\caption[Couverture spectrale de Herschel et spectres
d'ULIRGs]{Spectres mod\'elis\'es d'une galaxie de type ULIRG en
fonction de son redshift dans le r\'ef\'erentiel de l'observateur. La
r\'egion surlign\'ee correspond \`a la couverture spectrale du
t\'elescope Herschel. Le pic d'\'emission de la galaxie dans cette
gamme de $\lambda$ est d\^u \`a l'\'emission des grains de
poussi\`eres chauff\'es par les \'episodes violents de formation
d'\'etoiles massives qui ont lieu au sein de la galaxie. Notez qu'aux
grandes longueurs d'onde le flux apparent d\'epend faiblement du
redshift (le d\'ecalage du pic est compens\'e par l'\'eloignement de
la source). Cette figure est extraite de \shortciteN{guiderdoni}.
\label{fig:detect_observatoire_FIRST_objectifs_guiderdoni}}
\end{figure}

La cible du t\'elescope Herschel est l'Univers froid. En effet, dans
sa gamme de longueur d'onde se trouve le pic d'\'emission de corps
noirs dont la temp\'erature est comprise entre 5~et 50~K, ainsi que
les raies d'\'emission atomiques et mol\'eculaires les plus fortes
pour des gaz dont la temp\'erature est comprise entre 10~K et quelques
centaines de~K. Nous retrouvons ces conditions physiques de notre
syst\`eme solaire jusqu'aux confins de l'Univers. Les objets
observables par Herschel sont donc de nature tr\`es vari\'ee, ils
vont des ast\'ero\"ides aux AGN en passant par les nuages
mol\'eculaires, les supernovae, les disques proto-plan\'etaires ou
encore les galaxies \`a flamb\'ee d'\'etoiles. Toutefois, la dur\'ee
de la mission \'etant limit\'ee, les grandes th\'ematiques porteront
en priorit\'e sur l'\'evolution des galaxies, la formation des
\'etoiles et la relation qui existe entre ces deux ph\'enom\`enes~;
le but \'etant de comprendre pourquoi et comment notre Univers est
devenu ce qu'il est aujourd'hui. Remarquez que le domaine spectral de
Herschel est particuli\`erement bien adapt\'e \`a l'\'etude de ces
m\'ecanismes. Par exemple, la
figure~\ref{fig:detect_observatoire_FIRST_objectifs_guiderdoni} montre
un mod\`ele de densit\'e spectrale d'\'energie pour une galaxie de
type ULIRG et son \'evolution en fonction du redshift
\shortcite{guiderdoni}. La quasi-totalit\'e de l'\'energie contenue
dans ce spectre provient de grains de poussi\`ere r\'echauff\'es par
le rayonnement \'energ\'etique des \'etoiles massives nouvellement
form\'ees.  Ce type d'objets donne le maximum de son \'energie dans la
bande Herschel jusqu'\`a un redshift de~$\sim$5. D'autre part,
\shortciteN{andre} montrent que les premi\`eres \'etapes
d'effondrement de proto-\'etoiles pr\'esentent des densit\'es
spectrales d'\'energies qui piquent aux alentours de
100-300~$\mu$m. Les six bandes spectrales de Herschel r\'eparties
entre 60~et 670~$\mu$m permettent donc d'ajuster pr\'ecis\'ement les
spectres, et ainsi de bien contraindre la temp\'erature et la masse de
poussi\`ere contenue le long de la ligne de vis\'ee. Mais sans se
cantonner \`a ces deux grands axes, les th\`emes scientifiques
abord\'es seront nombreux, nous en dressons une liste non-exhaustive
inspir\'ee de l'article de \shortciteN{pilbratt_toledo}~:
\begin{itemize}
\item \textbf{Les galaxies}
\begin{itemize} 
\item Formation, \'evolution et physique des galaxies 
\item Le \emph{Cosmic Infrared Background} 
\item \'Evolution du taux de formation d'\'etoiles dans l'Univers
\item \'Evolution du taux de production des \'el\'ements lourds 
\end{itemize}
\item \textbf{La formation des \'etoiles et des syst\`emes
plan\'etaires, physique et chimie du milieu interstellaire}
\begin{itemize}
\item Structure, dynamique et composition du milieu interstellaire
\item Les c\oe urs pr\'e-stellaires et les YSO (Young Stellar Objects)
\item La fin de vie des \'etoiles
\item L'enrichissement du milieu interstellaire - Astrochimie
\item L'\'etude d\'etaill\'ee des galaxies proches r\'esolues
\end{itemize}
\item \textbf{Le syst\`eme solaire - atmosph\`eres plan\'etaires et
com\'etaires}
\begin{itemize}
\item Composition des com\`etes et objets du syst\`eme solaire externe
\item Composition de l'atmosph\`ere des plan\`etes g\'eantes
\item Origine et r\^ole de l'eau
\end{itemize}
\end{itemize}
\shortciteANP{pilbratt_toledo} note \`a juste titre que l'Observatoire
Herschel est de plus \'equip\'e pour effectuer des \emph{follow-up}
spectroscopiques sur les objets d\'ecouverts lors des grands relev\'es
du ciel et qui pr\'esenteraient un int\'er\^et particulier. En effet,
ces observations seront certainement n\'ecessaires puisque Herschel
est son propre \'eclaireur (il ouvre la voie et permet de nombreuses
d\'ecouvertes)~; il n'a pas de pr\'ed\'ecesseur sur lesquels nous
pourrions nous baser pour pr\'eparer les futures observations comme
c'\'etait le cas de ISO avec IRAS, puis de Spitzer avec ISO.

Au total, la mission Herschel offre environ 20000~heures de temps
d'observation pour les programmes scientifiques. Un tiers de ce temps
est r\'eserv\'e aux instituts qui ont fourni \`a l'ESA les instruments
scientifiques. Nous parlons de temps garanti. Les deux tiers restant
sont ouverts \`a toute la communaut\'e scientifique (pas seulement
europ\'eenne), c'est le temps ouvert. Il n'y aura que trois appels \`a
proposition, le premier concerne les projets clefs, les deux suivants
les projets standards. Un projet clef est un programme d'observation
qui n\'ecessite beaucoup de temps d'observation et qui exploite les
performances uniques de Herschel tout en traitant un sujet
scientifique int\'eressant et pertinent. Les projets clefs du temps
garanti ont d\'ej\`a \'et\'e s\'electionn\'es\footnote{Le lecteur
trouvera les programmes et les objets r\'eserv\'es pour les projets
clefs de temps garanti sur le site web du workshop tenu \`a l'ESTEC en
F\'evrier~2007~: http$:$//herschel.esac.esa.int/OT\_KP\_wkshop.shtml}.


\subsection{Le satellite}
\label{sec:detect_observatoire_sat}

\subsubsection{L'orbite}
\label{sec:detect_observatoire_sat_orbite}

\begin{figure}
  \begin{center}
    \begin{tabular}[t]{ll}
      \includegraphics[height=0.245\textheight]{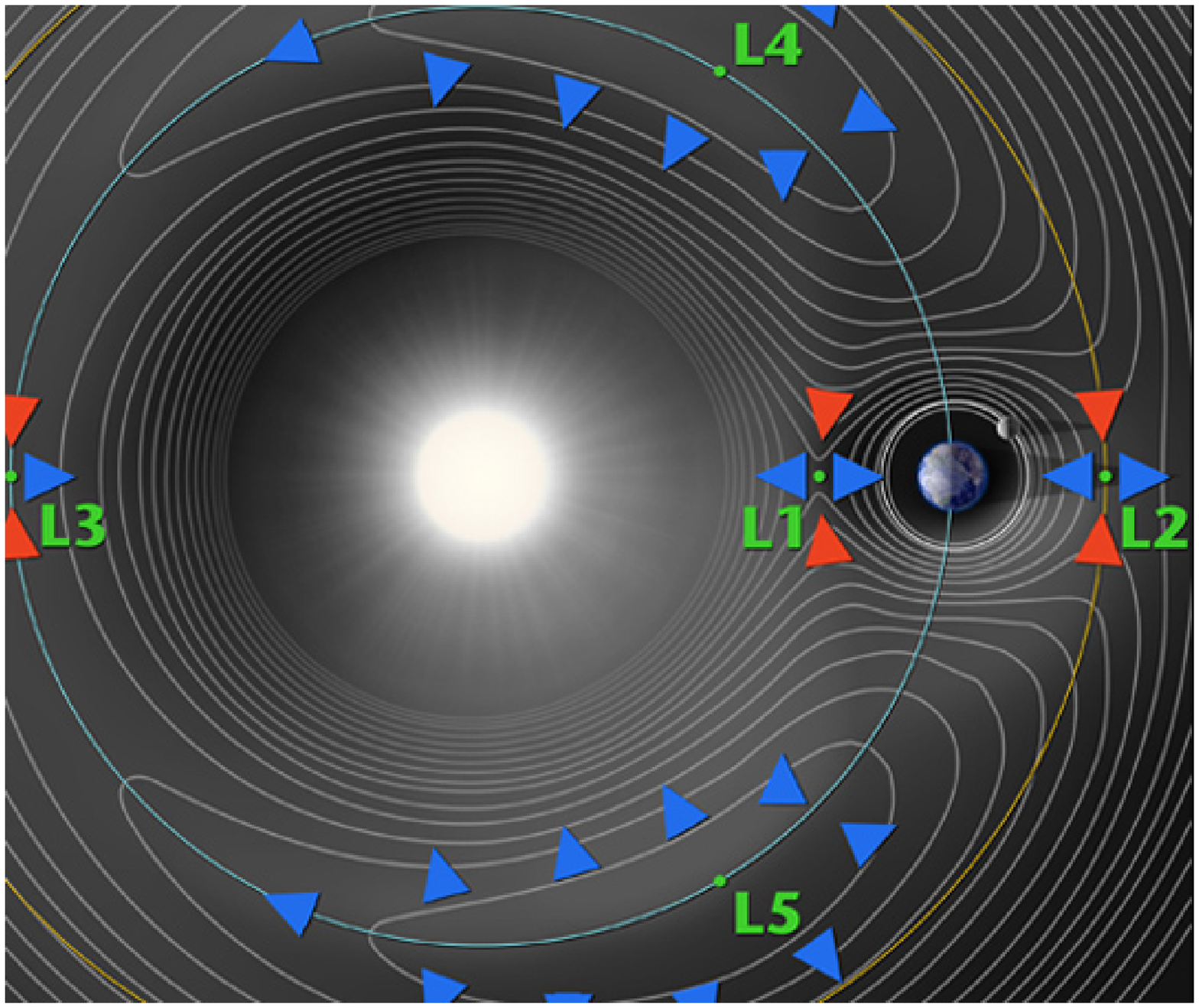} & \includegraphics[height=0.245\textheight]{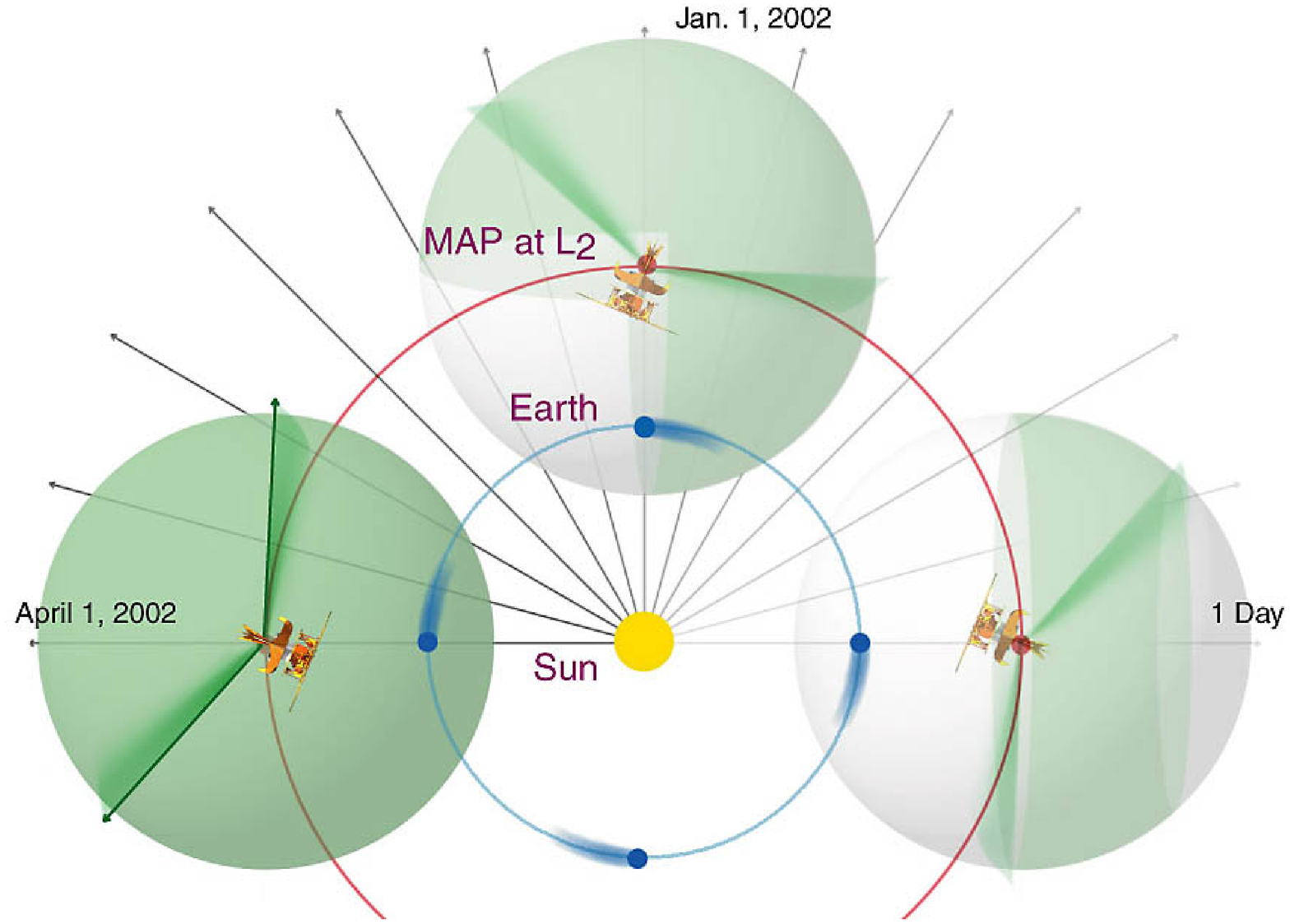} 
    \end{tabular}
  \end{center}
  \caption[Point de Lagrange~L2]{\`A gauche les lignes de champ
  gravitationnel sont repr\'esent\'ees pour le syst\`eme
  Terre-Soleil. Herschel orbitera autour de L2 \`a 1.5 million de km
  de la Terre. \`A droite nous voyons l'exemple du t\'elescope WMAP
  qui a d\'ej\`a orbit\'e autour de L2 (les distances ne sont pas \`a
  l'\'echelle). Notez que le miroir pointe toujours dans une direction
  oppos\'ee \`a celle du Soleil. (lien http$\!$://www.iscap.columbia.edu/ 
  et http$\!$://map.gsfc.nasa.gov/)
  \label{fig:detect_observatoire_sat_orbite_lagrange}}
\end{figure}

Le satellite Herschel doit \^etre satellis\'e sur une orbite de
Lissajou autour du point de Lagrange~L2 du syst\`eme Terre-Soleil \`a
environ 1.5 millions de km de la Terre. L2~est un point de l'espace
o\`u la d\'eriv\'ee du champ de gravitation est nulle (cf
figure~\ref{fig:detect_observatoire_sat_orbite_lagrange})~; c'est un
point m\'etastable, le satellite n\'ecessite donc quotidiennement de
petits ajustements de trajectoire. Cette orbite est particuli\`erement
int\'eressante car elle offre un environnement thermique tr\`es stable
pour le t\'elescope. L'environnement
\'electromagn\'etique est \'egalement tr\`es stable, le satellite ne
traverse pas les ceintures de radiation qui entourent la Terre comme
c'\'etait le cas d'ISO sur son orbite HEO (\emph{Highly Elliptical
Orbit}).

Le transfert du satellite de la Terre vers L2 dure environ 6~mois
durant lesquels les instruments seront test\'es dans des conditions
repr\'esentatives des conditions d'op\'eration. La phase de
v\'erification des performances ainsi que l'optimisation des modes
d'observation auront lieu pendant le transfert, ceci afin de s'assurer
que les trois ann\'ees d'op\'eration \`a~L2 seront efficaces d\`es le
premier jour.

\subsubsection{Le cryostat}
\label{sec:detect_observatoire_sat_cryo}

\begin{figure}
  \begin{center}
    \begin{tabular}{c}
    \includegraphics[width=0.9\textwidth,angle=0]{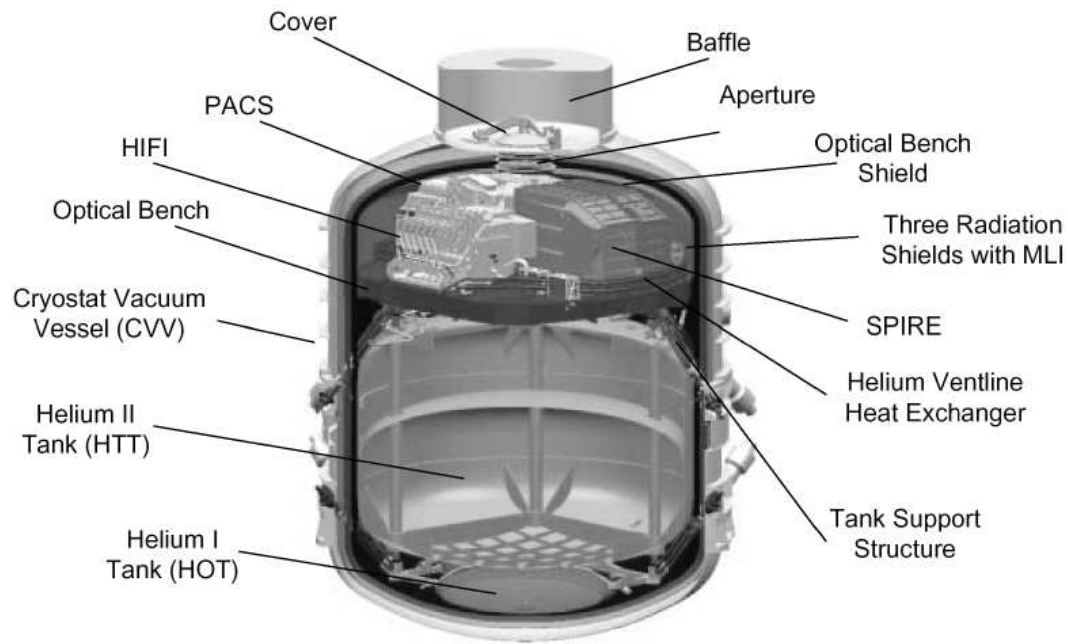}\\
    \includegraphics[width=0.7\textwidth,angle=0]{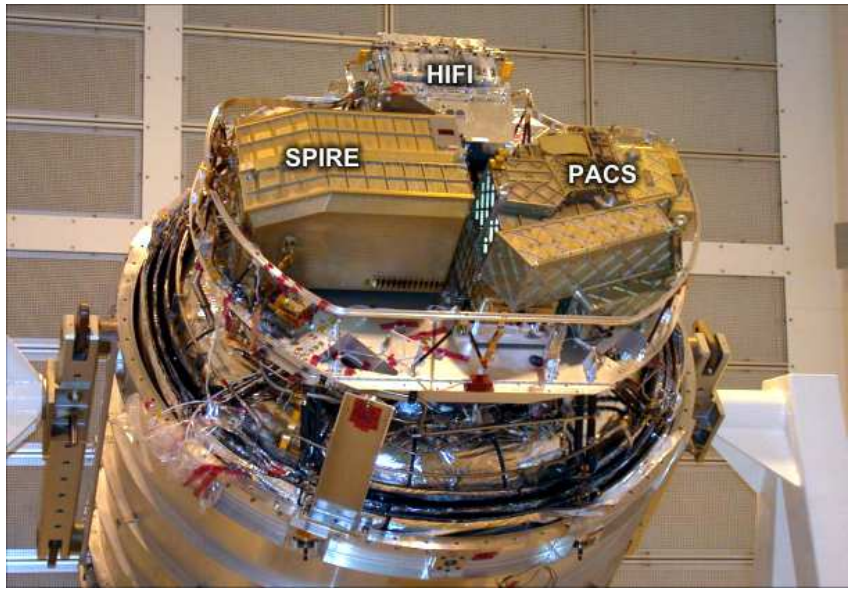}\\
    \end{tabular}
  \end{center}
\caption[Les instruments du satellite Herschel dans le cryostat]{Vue
en coupe du cryostat du satellite Herschel, il peut contenir
3500~litres d'h\'elium superfluide. Photographie des trois instruments
int\'egr\'es sur le plateau optique \`a l'int\'erieur du cryostat. Ces
instruments sont les mod\`eles de qualification. (lien http$\!$://esamultimedia.esa.int/)
\label{fig:detect_observatoire_sat_instru_tous}}
\end{figure}

Tous les instruments scientifiques embarqu\'es sur Herschel
n\'ecessitent des temp\'eratures de fonctionnement cryog\'eniques. Le
satellite embarque donc un cryostat qui contient une grande quantit\'e
d'h\'elium superfluide. Il fournit aux instruments deux niveaux de
temp\'erature, le \emph{level~0} \`a 1.7~K (temp\'erature de
l'h\'elium pomp\'e) et le \emph{level~1} \`a environ 5~K
(temp\'erature de l'h\'elium gaz qui s'\'echappe du cryostat mais qui
est quand m\^eme utilis\'e pour refroidir les instruments avant
d'\^etre lib\'er\'e dans le vide). Remarquez que la dur\'ee de la
mission est d\'etermin\'ee par la quantit\'e d'h\'elium pr\'esente
dans le cryostat au moment du lancement~; en l'occurence les
3500~litres doivent assurer une mission de 3.5~ans (6~mois de
transfert et 3~ans d'observations
\`a L2). Les trois instruments sont install\'es \`a l'int\'erieur du
cryostat comme le montre la photographie de la
figure~\ref{fig:detect_observatoire_sat_instru_tous}. Pour r\'eduire
les co\^uts de d\'eveloppement, ce cryostat utilise les technologies
d\'ej\`a d\'evelopp\'ees pour le cryostat de ISO
\shortcite{juillet}.

\subsubsection{Le module de service}
\label{sec:detect_observatoire_sat_SVM}

Le r\^ole du module de service (SVM) est de contr\^oler tout ce qui se
passe \`a bord du satellite, en particulier le contr\^ole d'attitude et
d'orbite. Le SVM est \'egalement appel\'e plateforme car il sert de
support \`a toute la structure, il est visible sur la
figure~\ref{fig:detect_observatoire_herschel_ariane_HSO} sous le
cryostat. Il contient les boitiers d'\'electronique chaude de chacun
des instruments scientifiques et assure les communications entre la
Terre et les instruments. Notez que la bande passante est limit\'ee
\`a 130~kbps (kbits/seconde).


\subsubsection{Le t\'elescope}
\label{sec:detect_observatoire_sat_telescope}

Le t\'elescope Herschel est de type Ritchey-Chr\'etien avec un miroir
primaire de 3.5~m de diam\`etre et un miroir secondaire l\'eg\`erement
sous-dimensionn\'e~; le diam\`etre effectif du t\'elescope est alors
de 3.3~m, ce qui en fait le t\'elescope le plus grand jamais envoy\'e
dans l'espace. La pr\'ecision de surface (\emph{WaveFront Error})
mesur\'ee \`a 70~K est de 5.5~$\mu$m. Le miroir est constitu\'e \`a
90~\% de carbure de silicium (SiC) et p\`ese seulement 315~kg.

Au-del\`a de la r\'esolution spatiale impos\'ee par la taille du
miroir, le t\'elescope joue un r\^ole crucial sur les performances
finales des instruments. En effet, la sensibilit\'e des photom\`etres
d\'epend significativement du flux incident, ce dernier \'etant
impos\'e par la temp\'erature et l'\'emissivit\'e des miroirs primaire
et secondaire. Pour plus de d\'etails, le lecteur peut se reporter \`a
l'article de \shortciteN{fischer} qui pr\'esente des mesures
d'\'emissivit\'e en fonction de la temp\'erature du miroir et de sa
\og propret\'e \fg (en terme de d\'epots de poussi\`eres). Pour un
t\'elescope refroidi passivement \`a $\sim$85~K,
\shortciteANP{fischer} donnent l'\'emissivit\'e~:
\begin{equation}
\epsilon(\lambda)=3.36\times10^{-5}\lambda^{-1/2}+2.73\times10^{-7}\lambda^{-1}
\label{eq:emissivite_fischer}
\end{equation}
Pour ne pas perturber l'\'equilibre thermique du miroir, le
t\'elescope est en permanence \`a l'ombre du Soleil derri\`ere les
panneaux solaires. La temp\'erature du primaire devrait donc \^etre
extr\`emement homog\`ene. Les pr\'edictions donnent un gradient de
temp\'erature d'environ 0.2~K dans la direction du soleil et de
$\sim$0~K dans la direction perpendiculaire. La diff\'erence de
temp\'erature entre les miroirs primaire et secondaire devrait \^etre
de l'ordre de 2~K. D'autre part, pour que les panneaux solaires
pointent en permanence vers le Soleil, les r\'egions du ciel
accessibles \`a Herschel \`a une \'epoque donn\'ee sont limit\'ees. En
effet le satellite peut tourner sur 360\textdegree~autour de l'axe
Terre-Soleil mais il est restreint \`a $\pm$30\textdegree dans la
direction perpendiculaire. La programmation des observations d\'epend
donc fortement de la p\'eriode de l'ann\'ee consid\'er\'ee.

\begin{figure}
  \begin{center}
     \includegraphics[width=0.65\textwidth,angle=0]{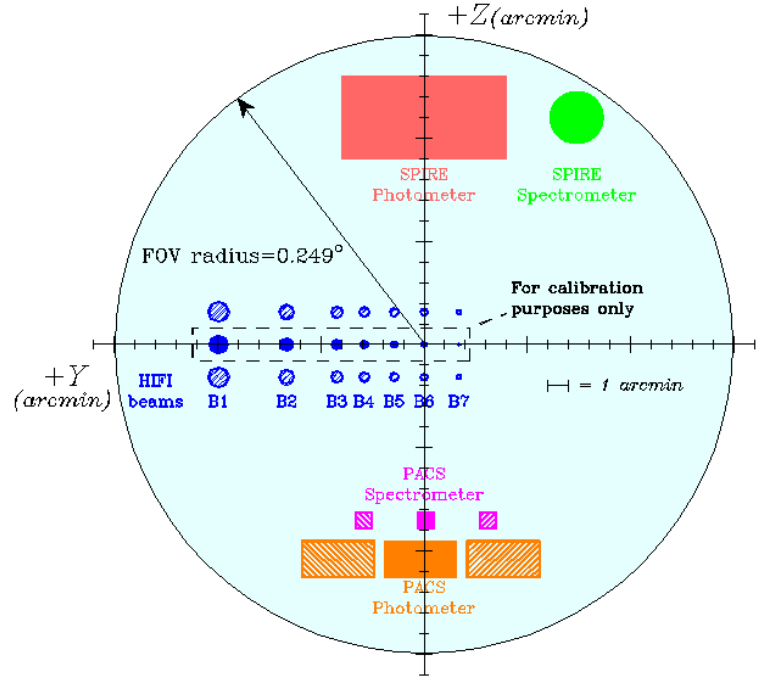}
  \end{center}
\caption[Le plan focal de l'Observatoire Herschel]{Plan focal de
l'Observatoire Herschel avec le d\'etail des champs de vue des trois
instruments scientifiques. (cr\'edits ESA)
\label{fig:detect_observatoire_sat_instru_fov}}
\end{figure}

Le plan focal du t\'elescope se trouve \`a environ 1~m sous le miroir
primaire, \`a l'int\'erieur du cryostat (cf
figures~\ref{fig:detect_observatoire_sat_instru_tous}
et~\ref{fig:detect_observatoire_sat_instru_fov}). Le plan focal est
l\'eg\`erement courb\'e ce qui entra\^ine quelques aberrations
optiques. Par ailleurs, la mise au point du t\'elescope n'est pas
r\'eglable une fois que le satellite est assembl\'e. 

Notez que pendant le premier mois apr\`es le lancement, le cryostat
restera ferm\'e et que le miroir sera chauff\'e pour permettre un
d\'egazage complet du satellite et ainsi \'eviter que le gaz n'aille
se condenser sur les parties les plus froides du satellite, i.e. les
d\'etecteurs.

\subsubsection{Les instruments scientifiques}
\label{sec:detect_observatoire_sat_instru}

L'Observatoire Spatial Herschel poss\`ede trois instruments
scientifiques. Les trois technologies que nous avons pr\'esent\'ees
dans la section~\ref{sec:intro_astro_IR_detecteur} sont
repr\'esent\'ees.\\

$\blacktriangleright$ \textbf{HIFI}\\ \indent HIFI
\shortcite{degraauw} est l'instrument h\'et\'erodyne de Herschel
(Heterodyne Instrument for the Far Infrared). C'est un spectrom\`etre
\`a tr\`es haute r\'esolution ($\lambda/\Delta\lambda>10^6$) qui
couvre 7~bandes spectrales entre 157~et 625~$\mu$m (1910-480~GHz),
avec un trou entre 212~et 240~$\mu$m. Les cinq bandes \og basse
fr\'equence \fg utilisent des m\'elangeurs SIS, les deux autres bandes
utilisent des HEB. Deux polarisations sont mesur\'ees pour chaque
bande. Les d\'etecteurs de HIFI sont mono-pixel, la r\'esolution
spatiale va de 12~\`a 40$''$.  Avec une r\'esolution en vitesse
comprise entre 0.3~et 300~km/s, HIFI est id\'eal pour \'etudier
pr\'ecis\'ement la dynamique du gaz dans des objets de nature tr\`es
diff\'erentes (galaxies, Galaxie, c\oe urs denses, jets, etc...). De
plus, il observera dans un r\'egime de longueur d'onde encore
inexplor\'e depuis l'espace, il pourrait donc d\'ecouvrir de nouvelles
mol\'ecules dans le milieu interstellaire ou dans les atmosph\`eres
plan\'etaires et com\'etaires. HIFI a \'egalement acc\`es \`a de
nombreuses raies spectrales de l'eau pour des densit\'es et niveaux
d'excitation tr\`es vari\'es, et l'\'etude de cet \'el\'ement sera
centrale dans le programme scientifique de l'instrument.

$\blacktriangleright$ \textbf{SPIRE}\\ \indent SPIRE
\shortcite{griffin} est le spectro-imageur grande longueur d'onde de
Herschel (Spectral and Photometric Imaging receiver). Le photom\`etre
observe simultan\'ement dans trois bandes spectrales centr\'ees \`a
250, 350 et 500~$\mu$m. Le champ de vue est de $4'\times8'$ et la
taille du faisceau (PSF) est de 18, 25 et 36$''$ respectivement. Le
spectrom\`etre est de type Mach-Zender, c'est un spectrom\`etre
imageur \`a transform\'ee de Fourier. Son champ de vue est de
2.6$'$. Sa r\'esolution spectrale est comprise entre 20~et 1000 \`a
250~$\mu$m. SPIRE utilise 5~matrices de bolom\`etres (3~pour le
photom\`etre, 2~pour le spectrom\`etre) pour un total de 326
bolom\`etres individuels de type \emph{spiderweb} avec thermom\`etres
NTD~Ge. Les bolom\`etres fonctionnent \`a 300~mK et sont coupl\'es
avec le t\'elescope par des c\^ones de Winston de diam\`etre
2F$\lambda$. SPIRE poss\`ede un miroir orientable tr\`es rapide qui
permet d'effectuer des observations adapt\'ees \`a ce type de plan
focal (\emph{jiggling} pour echantillonner le ciel \`a Nyquist, cf
section~\ref{sec:intro_bolometrie_bolo_matrice}).

$\blacktriangleright$ \textbf{PACS}\\ \indent PACS
\shortcite{poglitsch} est le spectro-imageur qui couvre la gamme
60-210~$\mu$m (Photodetector Array Camera and Spectrometer). Il
utilise des photoconducteurs Ge$:$Ga pour la spectrom\'etrie
($2\times16\times25$~pixels) et des matrices de bolom\`etres pour la
photom\'etrie ($16\times32$ et $32\times64$~pixels). Pour le
spectrom\`etre, le champ de vue de $47''\times47''$ est d\'ecoup\'e
sur un quadrillage de $5\times5$ puis r\'e-imag\'e le long d'une fente
pour \^etre ensuite dispers\'e par un r\'eseau et imag\'e sur les
d\'etecteurs. C'est un spectrom\`etre int\'egral de champ. La
r\'esolution spectrale est comprise entre 1000~et 2000. Le
photom\`etre sera d\'ecrit en d\'etail dans la
section~\ref{sec:herschel_oservatoire_phfpu}. Nous distinguons deux
sous-syst\`emes pour PACS~: le FPU (Focal Plane Unit) qui se trouve
dans le cryostat et l'\'electronique chaude qui est situ\'ee dans le
module de service. PACS poss\`ede deux sources de calibration internes
situ\'ees de part et d'autre du champ de vue ainsi qu'un chopper
ultra-rapide \shortcite{krause}. L'instrument au complet p\`ese 130~kg
(75~kg pour le FPU) et n\'ecessite une puissance de 100~W pour
fonctionner.

\subsection{Le photom\`etre de PACS}
\label{sec:herschel_oservatoire_phfpu}

Nous revenons dans cette section sur le photom\`etre de l'instrument
PACS, le PhFPU (\emph{Photometer Focal Plane Unit}), qui a \'et\'e
con\c{c}u par le CEA et qui abrite les 10~matrices de bolom\`etres sur
lesquelles j'ai effectu\'e mon travail de th\`ese. Nous pr\'esentons
maintenant le PhFPU de mani\`ere assez g\'en\'erale pour mettre en
avant l'environnement imm\'ediat des d\'etecteurs~; nous d\'ecrirons
en d\'etail le principe de fonctionnement des matrices dans le
chapitre~\ref{chap:detect_bolocea}.



\subsubsection{Description g\'en\'erale}
\label{sec:detect_observatoire_phfpu_description}

\begin{figure}
  \begin{center} \begin{tabular}{ll}
    \includegraphics[width=0.5\textwidth,angle=0]{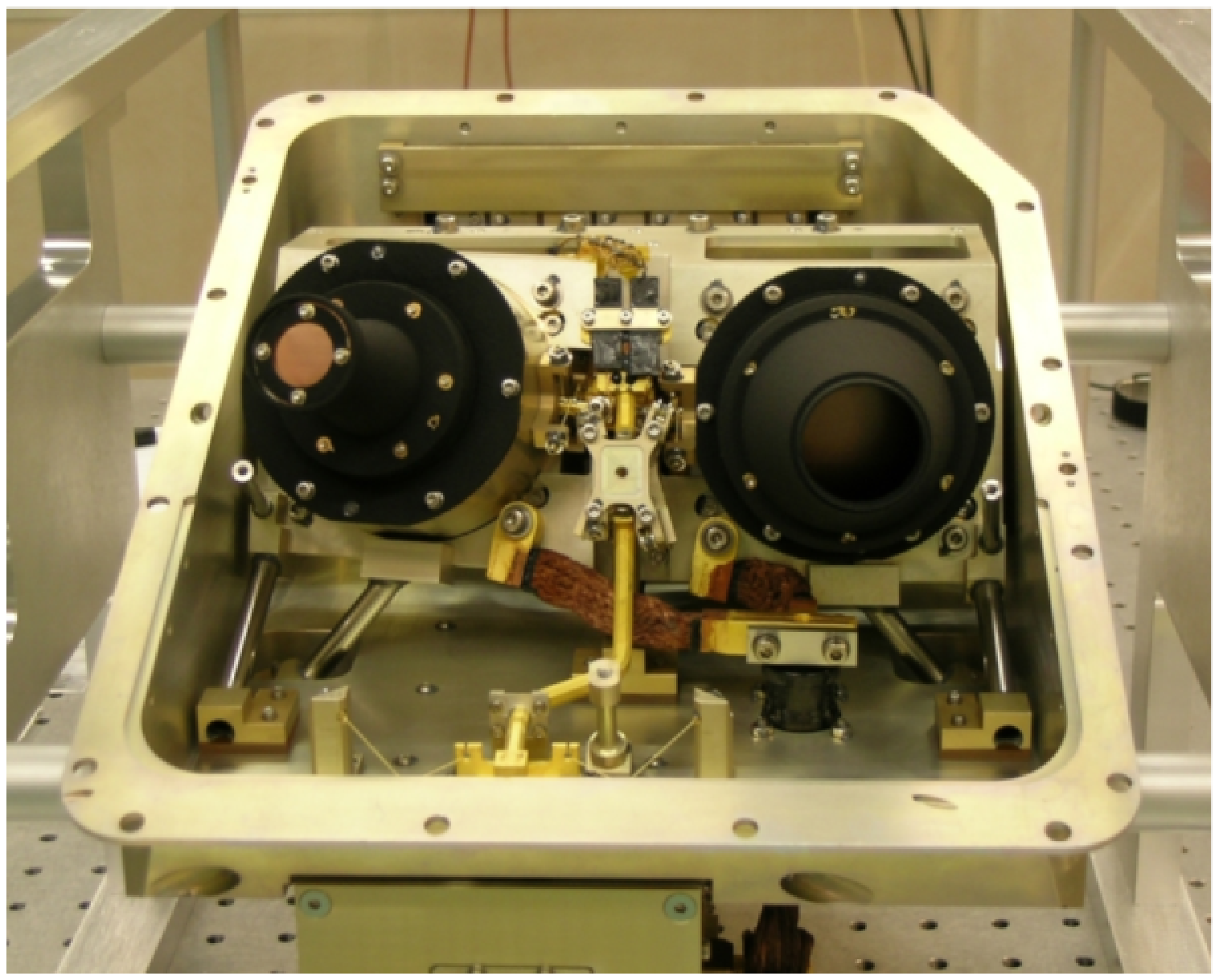}
    \includegraphics[width=0.5\textwidth,angle=0]{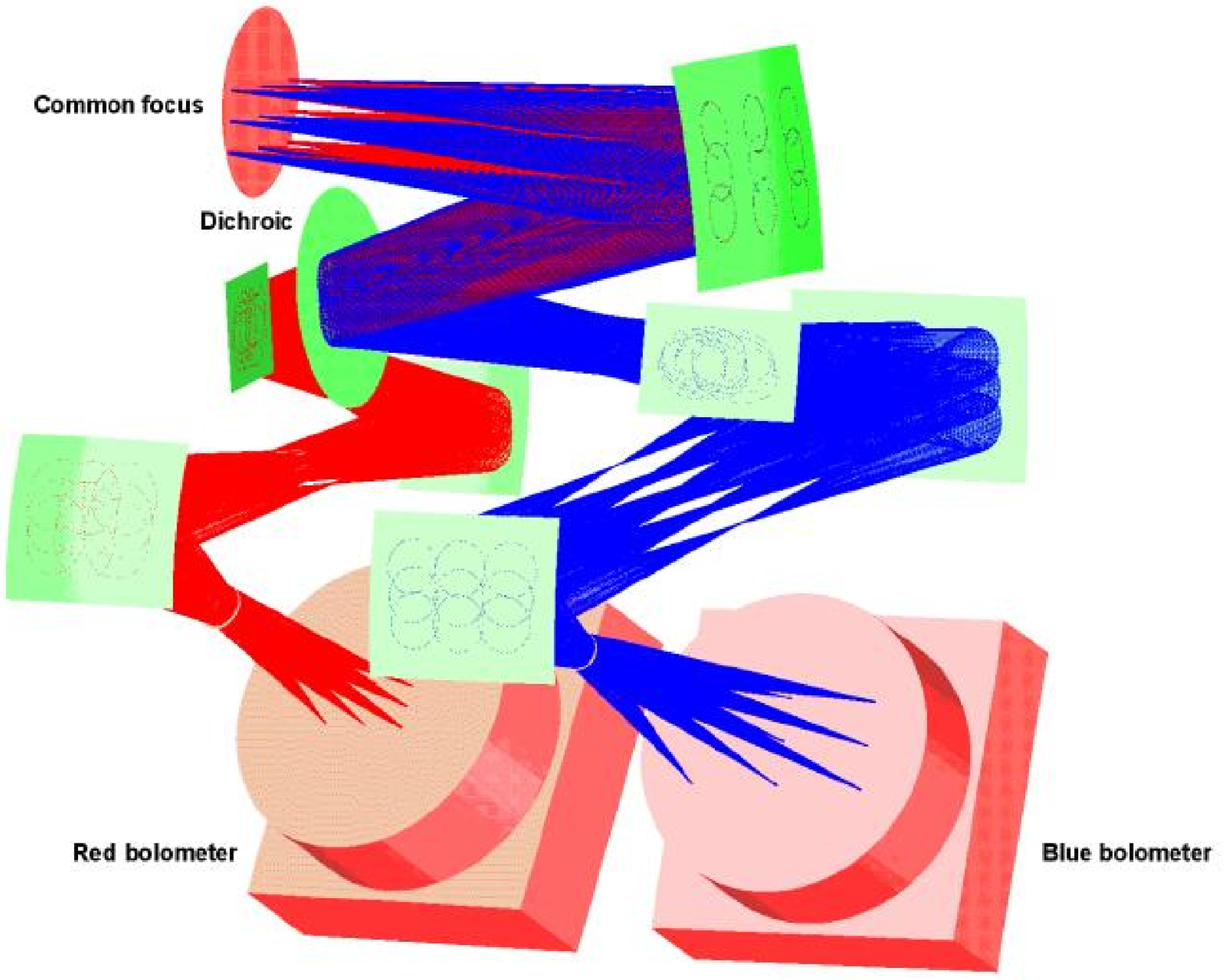}
    \end{tabular} \end{center} \caption[Les voies bleue et rouge du
    PhFPU]{Photographie du mod\`ele de vol du PhFPU lors de son
    assemblage. Les d\'etecteurs sont situ\'es \`a la base des deux
    c\^ones qui servent \`a d\'efinir le champ de vue de la cam\'era
    et \`a bloquer la lumi\`ere parasite. La voie rouge est \`a
    droite, la bleue \`a gauche. La barre m\'etallique jaune qui se
    trouve entre les deux plans focaux fournit les 300~mK
    n\'ecessaires au fonctionnement des bolom\`etres. Le
    cryo-r\'efrig\'erateur se situe sous le PhFPU. Les dimensions du
    photom\`etre sont de $260\times348.5\times216$~mm, il p\`ese
    8.2~kg. Le sch\'ema de droite montre le dichro\"ique qui s\'epare
    la lumi\`ere entre la voie bleue et rouge. (cr\'edit CEA)}
    \label{fig:detect_observatoire_phfpu_description_dichro}
\end{figure}

\begin{figure}
  \begin{center}
      \includegraphics[width=0.7\textwidth,angle=0]{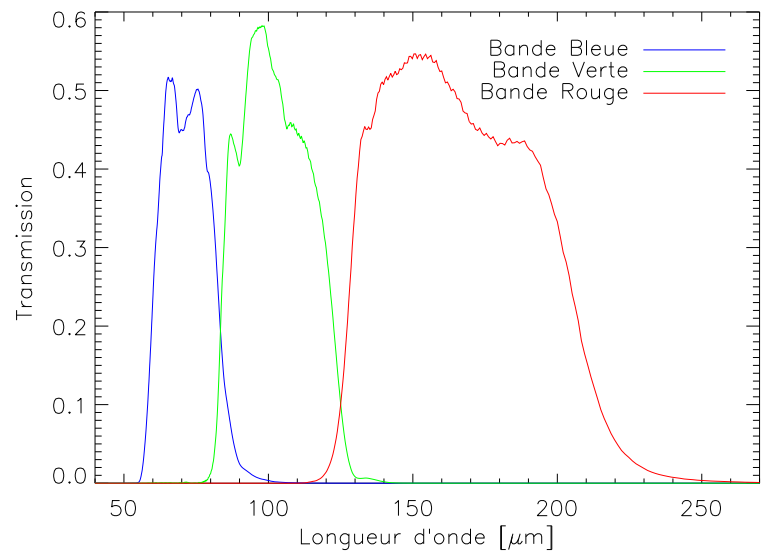}
  \end{center}
  \caption[Courbes de transmission pour le Photom\`etre PACS]{Courbes
  de transmission globale des filtres et dichro\"ique du Photom\`etre
  PACS dans chacune des bandes. La transmittance a \'et\'e mesur\'ee
  pour des filtres thermalis\'es \`a environ 5~K et un dichro\"ique
  \`a temp\'erature ambiante. L'efficacit\'e d'absorption des
  bolom\`etres n'intervient pas dans le calcul de ces courbes.}
  \label{fig:detect_observatoire_phfpu_description_filtre}
\end{figure}

Le PhFPU poss\`ede deux voies d'imagerie~: la voie \og bleue \fg qui
couvre le domaine spectral de 60~\`a 130~$\mu$m, et la voie \og rouge
\fg de 130~\`a 210~$\mu$m. Les deux voies sont illumin\'ees
simultan\'ement gr\^ace \`a un miroir dichro\"ique situ\'e sur le
chemin optique, ce miroir est r\'efl\'echissant en-dessous de
130~$\mu$m et transparent au-del\`a (cf
figure~\ref{fig:detect_observatoire_phfpu_description_dichro}).  La
voie bleue est de plus scind\'ee en deux bandes spectrales par le
biais d'une roue \`a filtre situ\'ee entre le miroir dichro\"ique et
le d\'etecteur (la roue \`a filtre n'est pas repr\'esent\'ee sur la
figure). Nous distinguons donc la bande bleue courte longueur d'onde
de 60 \`a 85~$\mu$m et la bande bleue grande longueur d'onde, aussi
appel\'ee bande verte, de 85~\`a 130~$\mu$m. Le PhFPU observe donc
dans deux bandes simultan\'ement~: les bandes bleue et rouge ou bien
verte et rouge. La
figure~\ref{fig:detect_observatoire_phfpu_description_filtre} montre
la transmission du syst\`eme optique dans chacune de ses trois bandes
spectrales, elle est d'environ~50~\%. Notez que les bolom\`etres
\'etant des d\'etecteurs large bande, il est n\'ecessaire de filtrer
le rayonnement infrarouge proche et moyen pour \'eviter une surcharge
optique sur les d\'etecteurs (augmentation ind\'esirable du bruit de
photon et de la puissance incidente). Le syst\`eme optique est
limit\'e par la diffraction \`a~$\sim$100~$\mu$m.

\begin{figure}
  \begin{center}
    \begin{tabular}[t]{ll}
      \includegraphics[height=0.25\textheight]{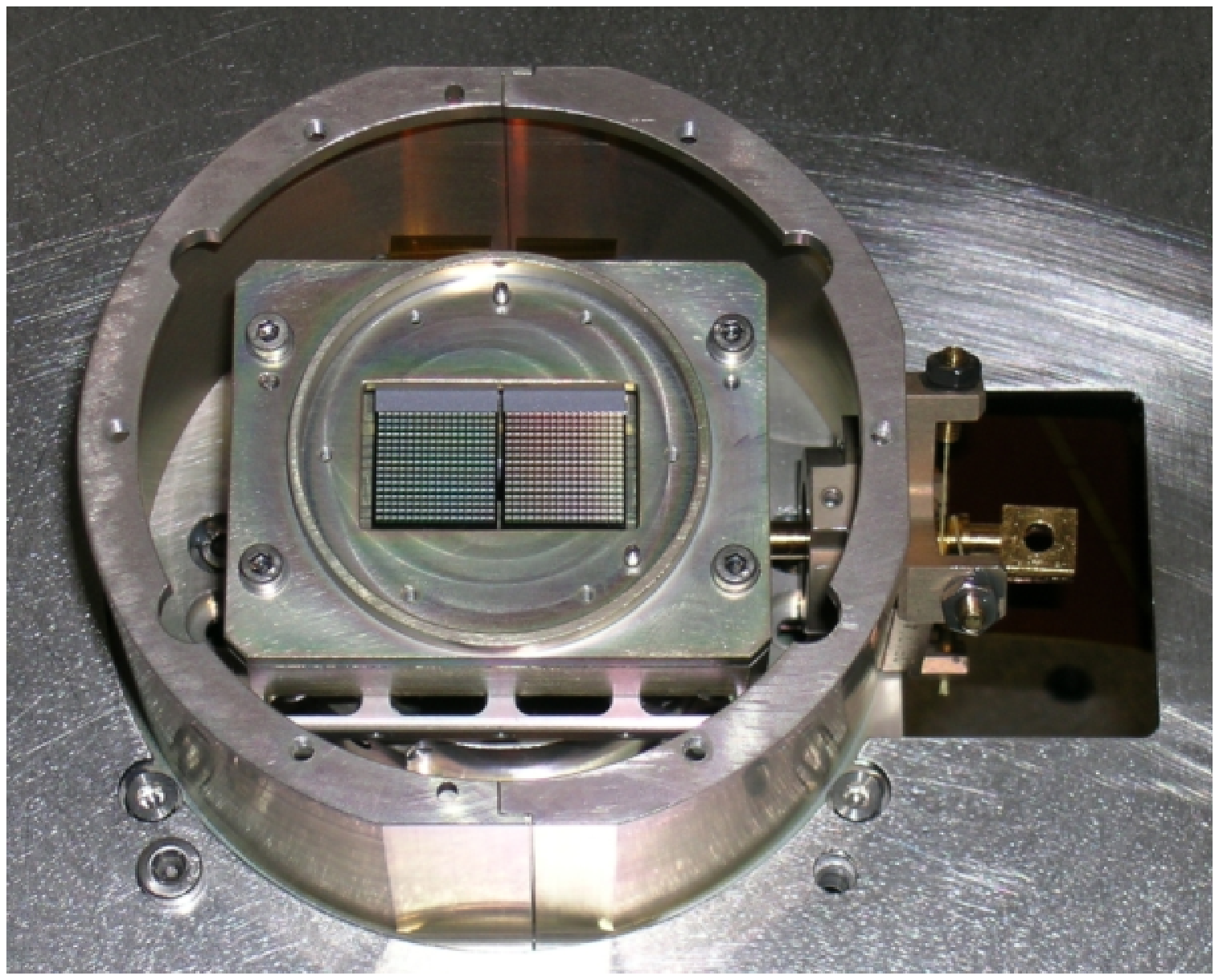} & \includegraphics[height=0.25\textheight]{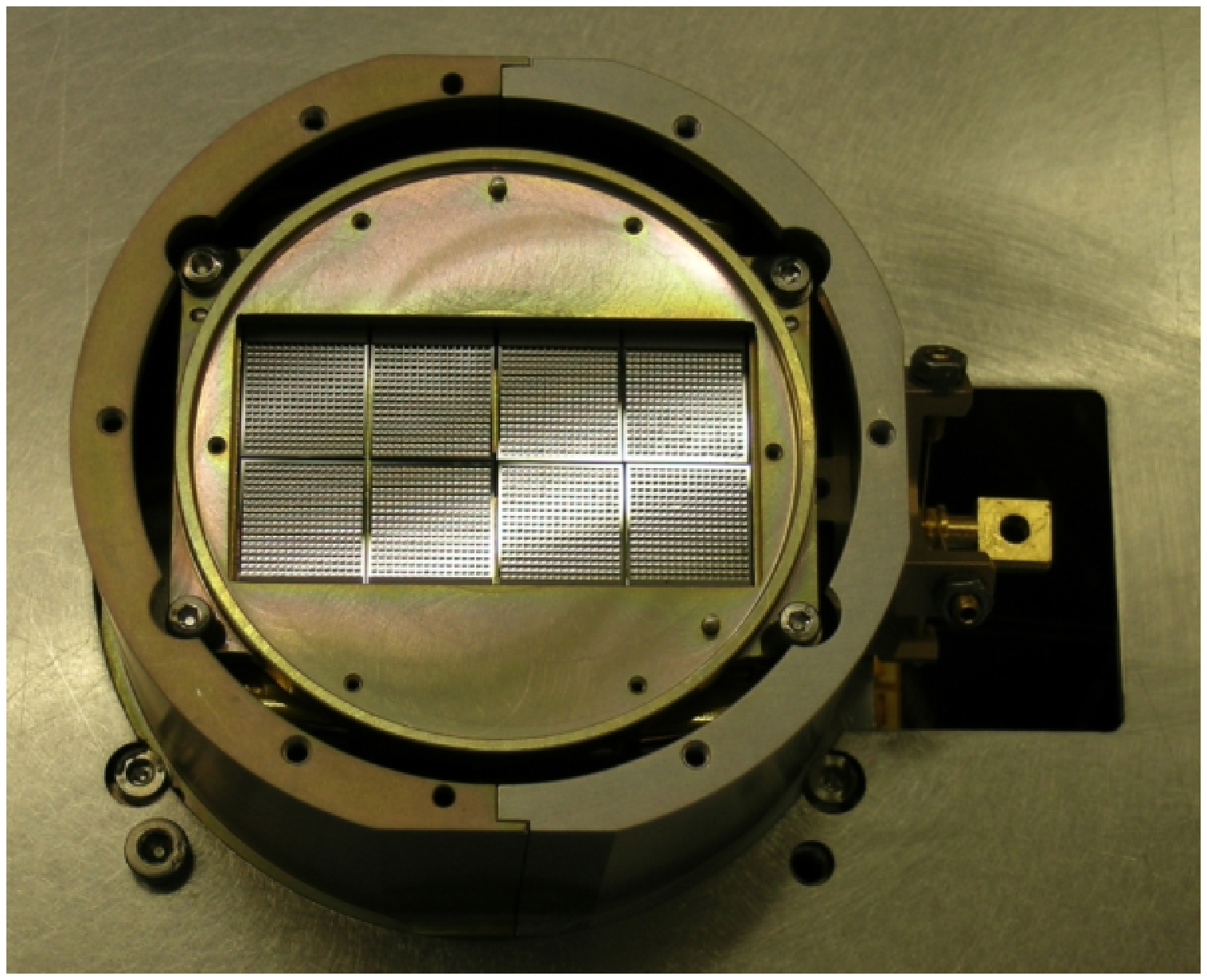} 
    \end{tabular}
  \end{center}
  \caption[Photographie des deux BFP du PhFPU]{Photographie des deux
  BFP du PhFPU avant int\'egration dans le photom\`etre. \`A gauche le
  BFP rouge contient 2~matrices de $16\times16$~bolom\`etres
  chacune. \`A droite, le BFP bleu est compos\'e d'une mosa\"ique de
  8~matrices pour un total de 2048~bolom\`etres. C'est \`a ce jour le
  d\'etecteur qui contient le plus grand nombre de bolom\`etres. 
  (cr\'edit CEA)
  \label{fig:detect_observatoire_phfpu_description_bfps}}
\end{figure}

Les plans focaux bleu et rouge (BFP, \emph{Bolometer Focal Plane})
sont compos\'es de mosa\"iques de matrices de bolom\`etres~: la voie
bleue compte 2048~pixels ($4\times2$~matrices de
$16\times16$~bolom\`etres chacune) et la voie rouge 512~pixels
(2~matrices). Le champ de vue des deux BFP est de
$3.5'\times1.75'$. Un pixel bleu projet\'e sur le ciel repr\'esente un
carr\'e de $3.2''\times3.2''$, les pixels rouges sont deux fois plus
grands (les pixels bleus et rouges ont la m\^eme taille physique mais
la longueur d'onde est deux fois plus grande sur la voie rouge).  La
figure~\ref{fig:detect_observatoire_phfpu_description_bfps} montre les
deux BFP avant int\'egration dans le PhFPU. La structure circulaire
qui entoure les d\'etecteurs est maintenue \`a 2~K alors que la
structure centrale qui supporte les bolom\`etres est thermalis\'ee \`a
300~mK. Cette derni\`ere est attach\'ee \`a la structure 2~K par un
syst\`eme de poulies et de fils de kevlar qui isole thermiquement les
deux \'el\'ements. Cette architecture est extr\^emement solide, le
PhFPU a d'ailleurs pass\'e avec succ\`es la campagne de qualification
spatiale, avec entre autre des tests de vibrations \`a froid.

\begin{figure}
  \begin{center}
      \includegraphics[width=0.8\textwidth,angle=0]{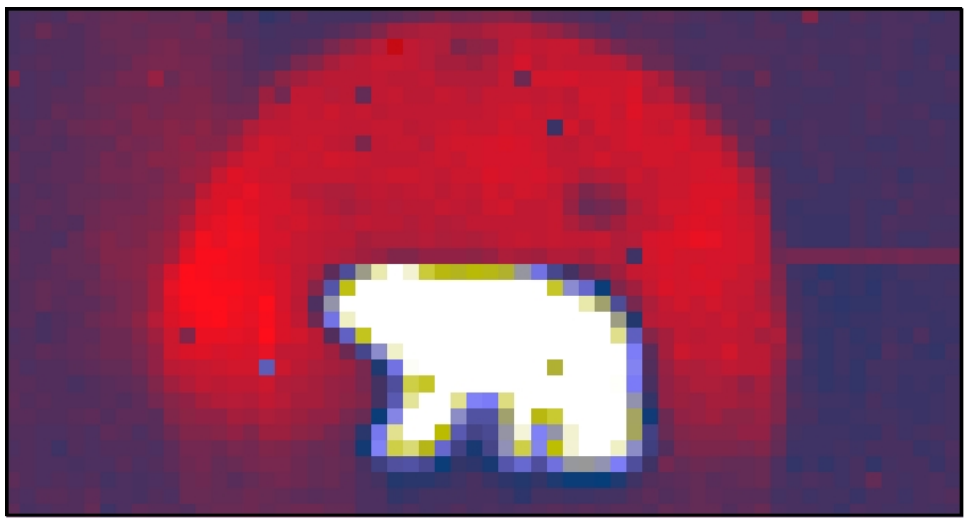}
  \end{center}
  \caption[Observation de Ursa Major avec PACS!]{Observation de Ursa
  Major avec PACS! Cette image est en fait extraite d'une simulation
  d'observation r\'ealis\'ee avec le XY-Stage du banc de test de
  Garching. Le soleil couchant (ou plut\^ot disque rouge en fond)
  correspond \`a des r\'eflexions internes sur la fen\^etre d'entr\'ee
  du cryostat. Cette image est maintenant le logo de l'instrument
  PACS.}
  \label{fig:detect_observatoire_phfpu_description_mascot}
\end{figure}

Pendant la campagne d'\'etalonnage men\'ee au MPE (Garching,
Allemagne), l'\'equipe a test\'e un \'equipement que nous appelons le
XY-Stage avec une source tout \`a fait remarquable. En fait, le
XY-Stage est un bo\^itier externe au cryostat de test qui permet
d'illuminer le FPU avec des sources g\'en\'eralement ponctuelles et de
les d\'eplacer dans le champ de vue pour simuler diff\'erentes
techniques d'observation. D'autre part, la mascotte de PACS \'etant un
ours polaire, ils ont d\'ecoup\'e la forme d'un ours dans une plaque
m\'etallique et ont utilis\'e le XY-stage pour \og simuler \fg
l'observation de Ursa Major avec le photom\`etre PACS! Une des images
du balayage est devenue le logo de l'instrument PACS, elle est
pr\'esent\'ee dans la
figure~\ref{fig:detect_observatoire_phfpu_description_mascot}. Notez
de plus que ce \og test \fg a permis de mettre en \'evidence un
cross-talk \'electrique sur les matrices du BFP rouge (cf
section~\ref{sec:calib_perflabo_ajuste_crosstalk}).

L'\'electronique chaude n\'ecessaire pour lire le signal
bolom\'etrique se situe dans le module de service du satellite, \cad
\`a une distance de 7~m et \`a une temp\'erature 10~fois plus
\'elv\'ee que le PhFPU. L'\'electronique chaude se compose de
plusieurs \'el\'ements, en partant du PhFPU vers la plateforme du
satellite nous trouvons~:
\begin{description}
\item[BOLC]: C'est le bo\^itier \'electronique qui contr\^ole les
matrices de bolom\`etres et le cryo-r\'efrig\'erateur (cf
section~\ref{sec:detect_observatoire_phfpu_cryocooler}). Il p\`ese un
peu plus de 18~kg, consomme environ 44~W en mode de fonctionnement
nominal\footnote{Il ne consomme que 6.9~W durant le recyclage du
cryo-r\'efrig\'erateur et 6.0~W en \emph{standby}.}, \cad en mode
d'observation, et fonctionne \`a une temp\'erature d'environ
300~K. Une description d\'etaill\'ee est pr\'esent\'ee dans la
section~\ref{sec:detect_bolocea_elec_lecture}. Notez simplement que le
d\'ebit de donn\'ees g\'en\'er\'ees par les deux plans
focaux\footnote{Au total 2560~pixels \'echantillonn\'es \`a 40~Hz dont
le signal de chaque pixel est cod\'e sur 16~bits.} du PhFPU est de
l'ordre de 1600~kbits/s. Les \emph{House Keeping} sont
\'echantillonn\'es \`a 0.5~Hz et occupent environ 1~\% de la bande
passante de la t\'el\'em\'etrie.
\item[DECMEC]: Ce bo\^itier contr\^ole les photoconducteurs du
spectrom\`etre ainsi que toutes les parties m\'ecaniques (roue \`a
filtres, chopper, r\'eseau) et les sources de calibration de PACS. Il
transmet \'egalement le signal brut vers le SPU (\emph{Software
Processing Unit}). Du point de vue de BOLC, DECMEC est un fil
\'electrique qui le connecte au SPU.
\item[SPU]: Du fait de la faible bande passante disponible pour
communiquer avec la Terre (130~kbps), le SPU effectue un
pr\'e-traitement du signal (photom\`etre et spectrom\`etre) ainsi
qu'une compression des donn\'ees. Pour le photom\`etre, il est
n\'ecessaire de r\'eduire le d\'ebit de donn\'ees par un
facteur~16. Un facteur de compression l\'eg\`erement inf\'erieur \`a 4
est r\'ealis\'e sans perte de donn\'ees\footnote{L'efficacit\'e de cet
algorithme d\'epend des propri\'et\'es du signal.}, et le facteur~4
restant est atteint en moyennant 4~images successives. L'avantage de
ce type de
\og compression \fg est que le rapport signal-\`a-bruit des donn\'ees
transmises est augment\'e d'un facteur~2.
\item[DPU]: C'est un bo\^itier \'electronique num\'erique qui est \`a
l'interface entre PACS et le satellite. Il g\`ere notamment l'aspect
t\'el\'em\'etrie/t\'el\'ecommande.
\end{description}
Tous les bo\^itiers de la cha\^ine \'electronique de PACS poss\`edent
des voies nominales et redondantes, ceci afin d'assurer une bonne
fiabilit\'e du syst\`eme.





\subsubsection{Le cryo-r\'efrig\'erateur}
\label{sec:detect_observatoire_phfpu_cryocooler}

Le satellite fournit au PhFPU une temp\'erature minimum de
1.7~K. Cependant, les matrices de bolom\`etres requi\`erent d'\^etre
refroidies \`a 300~mK pour fonctionner. C'est donc le r\^ole du
cryo-r\'efrig\'erateur d'abaisser la temp\'erature des BFP.  Le
cryo-r\'efrig\'erateur \`a adsorption de PACS a \'et\'e con\c{c}u par
le service des basses temp\'eratures du CEA (CEA/SBT), celui utilis\'e
pour l'instrument Herschel/SPIRE en est une copie quasi-conforme. Le
principe de fonctionnement de ce cryo-r\'efrig\'erateur est
relativement simple et efficace, il ne contient aucune partie mobile
(tr\`es adapt\'e aux contraintes spatiales), nous allons donc
bri\`evement d\'ecrire son concept et pr\'esenter ses performances.\\

\begin{figure}
  \begin{center}
    \begin{tabular}{ll}
      \includegraphics[width=0.31\textwidth,angle=0]{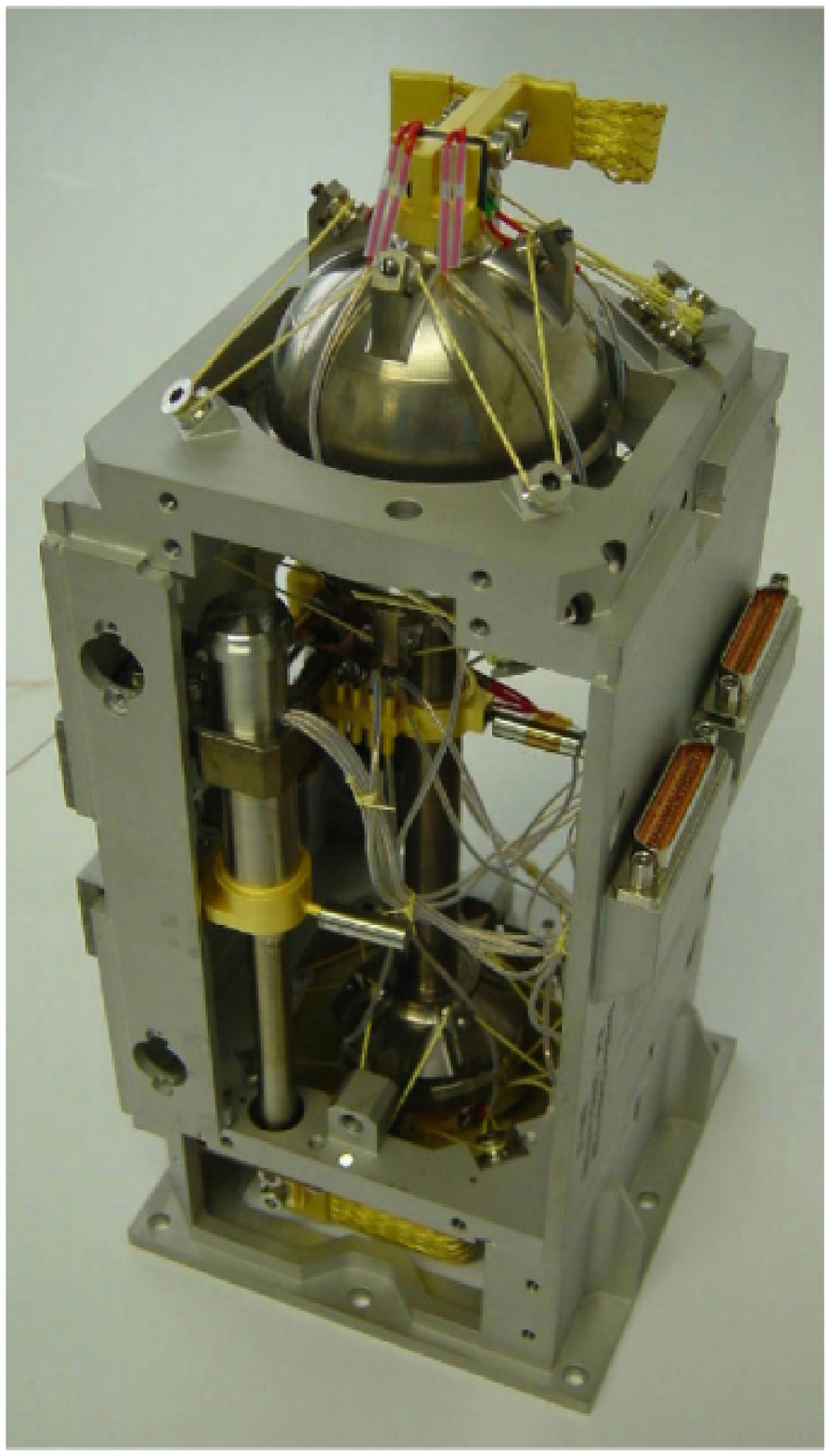}&
      \includegraphics[width=0.65\textwidth,angle=0]{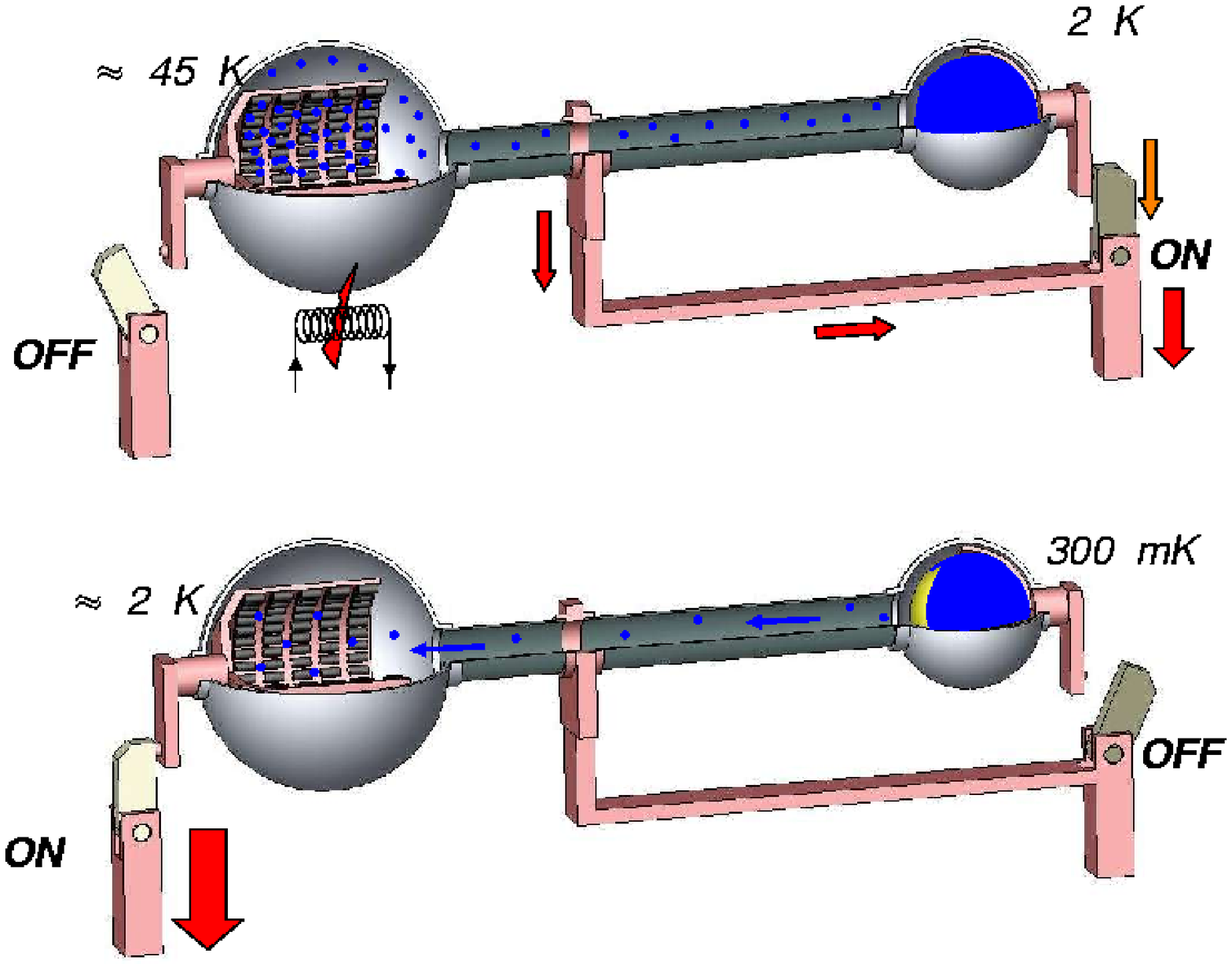}
    \end{tabular}
  \end{center}
  \caption[Le cryo-r\'efrig\'erateur de PACS]{Photographie du
  cryo-r\'efrig\'erateur de PACS. Nous voyons la sph\`ere qui est
  maintenue en suspension par des fils de kevlar. Le sch\'ema de
  droite pr\'esente le fonctionnement du cryo-r\'efrig\'erateur. En
  haut, la pompe est chauff\'ee pour d\'esorber le gaz qui va se
  condenser dans l'\'evaporateur (sph\`ere de droite). En bas, la
  pompe est connect\'ee au \emph{level~0} et les charbons actifs
  pompent l'h\'elium liquide, la temp\'erature du liquide en
  \'ebullition chute \`a 300~mK. Ce sch\'ema est extrait de la
  pr\'esentation de Laurent Clerc pour la conf\'erence \og Cryog\'enie
  spatiale \fg en avril~2005. (cr\'edit CEA)}
  \label{fig:detect_observatoire_phfpu_cryocooler_cooler}
\end{figure}

Le fonctionnement du cryo-r\'efrig\'erateur \`a adsorption repose sur
l'\'evaporation d'une petite quantit\'e de $^3$He qui permet
d'abaisser la temp\'erature de la phase liquide jusqu'\`a
250-300~mK. Sch\'ematiquement, le cryo-r\'efrig\'erateur se pr\'esente
sous la forme de deux sph\`eres en titane reli\'ees par un tube. Dans
l'une de ces sph\`eres que nous appelons la \emph{pompe} se trouve
quelques grammes de charbons actifs. Ces charbons sont tr\`es poreux
et poss\`edent une capacit\'e extraordinaire \`a adsorber le gaz (un
gramme de charbon actif repr\'esente une surface sp\'ecifique
d'environ 2000~m$^2$). La pompe est reli\'ee au \emph{level~0}
($\sim2$~K) par un interrupteur thermique. Une chaufferette situ\'ee
sur la sph\`ere permet de changer la temp\'erature des charbons, et
ainsi de contr\^oler leur vitesse d'adsorption et de
d\'esorption. L'autre sph\`ere s'appelle l'\emph{\'evaporateur}, elle
contient une mousse qui permet de retenir le liquide par capillarit\'e
en l'abscence de pesanteur. Elle est \'egalement li\'ee au
\emph{level~0} par un interrupteur thermique ce qui permet de
condenser l'h\'elium pendant la phase de recyclage. Le tube qui relie
les deux sph\`eres est thermalis\'e \`a 2~K pour isoler thermiquement
la partie chaude de la partie froide, \cad la pompe de
l'\'evaporateur. La structure pompe/\'evaporateur/tube est maintenue
dans un chassis \`a 4~K par des fils de kevlar pour assurer une bonne
isolation thermique. La
figure~\ref{fig:detect_observatoire_phfpu_cryocooler_cooler}
pr\'esente une photographie du cryo-r\'efrig\'erateur de PACS et un
sch\'ema explicatif de son fonctionnement.

\begin{figure}
  \begin{center}
    \begin{tabular}{ll}
      \includegraphics[width=0.48\textwidth,angle=0]{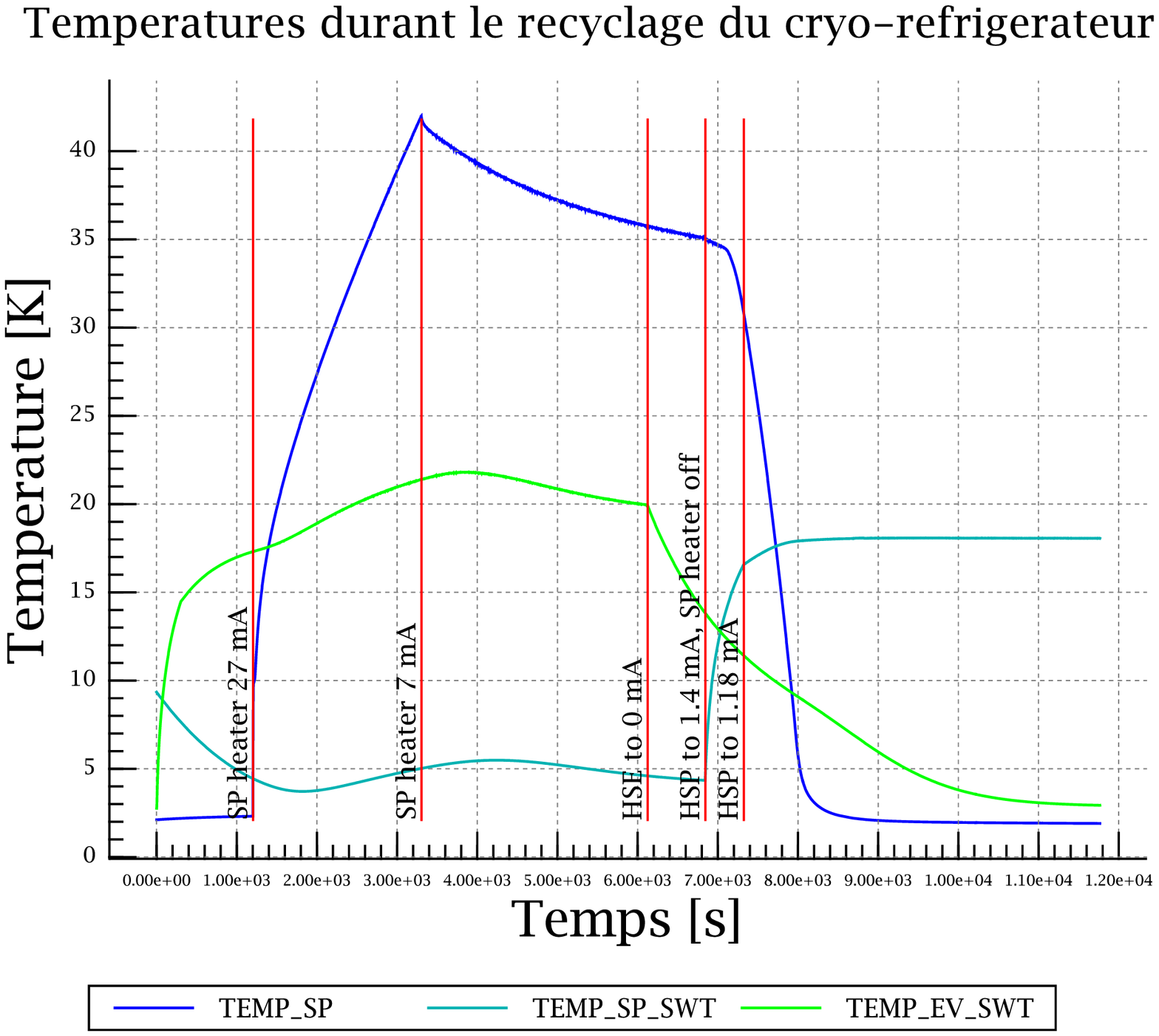}&
      \includegraphics[width=0.48\textwidth,angle=0]{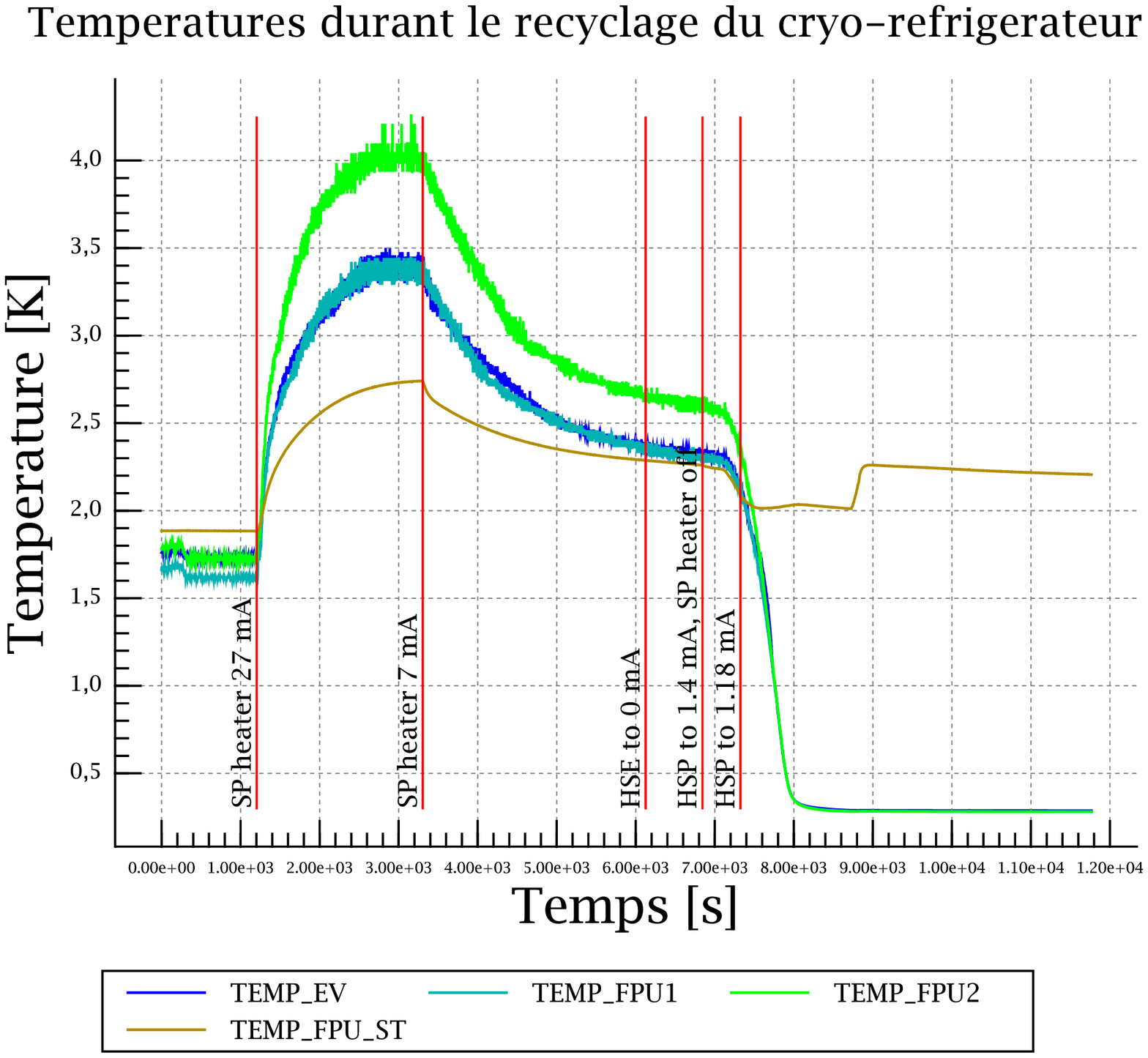}
    \end{tabular}
  \end{center}
  \caption[Courbes de recyclage du cryo-r\'efrig\'erateur]{\'Evolution
  de la temp\'erature des diff\'erents \'el\'ements du
  cryo-r\'efrig\'erateur (\`a gauche) et du PhFPU (\`a droite) durant
  un recyclage. TEMP\_SP et TEMP\_EV sont les temp\'eratures de la
  pompe et de l'\'evaporateur respectivement. TEMP\_SP\_SWT et
  TEMP\_EV\_SWT sont les temp\'eratures des interrupteurs thermiques
  de la pompe et de l'\'evaporateur respectivement (ils sont passants
  quand ils sont chauds). TEMP\_FPU1 est la temp\'erature du BFP
  rouge, TEMP\_FPU2 celle du BFP bleu et TEMP\_FPU\_ST celle de la
  structure du PhFPU. Ce recyclage a dur\'e environ 2~heures.}
  \label{fig:detect_observatoire_phfpu_cryocooler_recyclage}
\end{figure}

L'\'etape de recyclage consiste \`a chauffer la pompe avec la
chaufferette jusqu'\`a environ 45~K. Le gaz contenu dans les charbons
actifs est d\'esorb\'e et va condenser dans l'\'evaporateur qui est
maintenu \`a 2~K. Lorsque tout l'h\'elium est en phase liquide, le
chauffage est arr\^et\'e, la pompe est connect\'ee au \emph{level~0}
par l'interrupteur thermique et l'\'evaporateur est d\'econnect\'e du
\emph{level~0} en ouvrant son interrupteur thermique. La temp\'erature
des charbons passe \`a 2~K augmentant ainsi consid\'erablement leur
pouvoir d'adsorption, pour suivre la chute de pression dans
l'\'evaporateur l'h\'elium liquide se met \`a bouillir et sa
temp\'erature chute alors \`a $\sim$260~mK. Le cryo-r\'efrig\'erateur
maintient cette temp\'erature au niveau de l'\'evaporateur tant qu'il
reste du liquide \`a \'evaporer. La
figure~\ref{fig:detect_observatoire_phfpu_cryocooler_recyclage} montre
l'\'evolution des temp\'eratures du cryo-r\'efrig\'erateur durant un
recyclage.

Lors des tests effectu\'es \`a Saclay, nous avons obtenus une
autonomie de 59~heures avec un recyclage de 2~heures et un
\emph{level~0} \`a 1.6~K. Durant la campagne d'\'etalonnage de PACS au
MPE, un recyclage de 2~heures ne donnait qu'une autonomie de
$\sim$35-40~heures (le \emph{level~0} du cryostat PACS est sup\'erieur
\`a 1.6~K). N\'eanmoins, des recyclages de 3~heures nous ont permis
d'obtenir une autonomie de 45-50~heures (il faut attendre que
l'\'evaporateur passe en dessous des 2~K avant de le connecter aux
BFP). Bien que l'autonomie du cryo-r\'efrig\'erateur d\'epende du
\emph{level~0}, la temp\'erature de l'\'evaporateur est extr\^emement
reproductible quel que soit le \emph{level~0}.\\

La description du photom\`etre PACS a \'et\'e inspir\'ee de l'article
publi\'e dans les proceedings de la conf\'erence SPIE~2006 d'Orlando
o\`u j'ai pr\'esent\'e \`a l'oral les performances du PhFPU. Le
lecteur pourra consulter cet article dans l'annexe~\ref{a:publi}.

\newpage{\pagestyle{empty}\cleardoublepage}
\chapter{La bolom\'etrie}
\label{chap:intro_bolometrie}

\begin{center}
\begin{minipage}{0.85\textwidth}

\small Ce chapitre introduit les concepts fondamentaux qui r\'egissent
le fonctionnement d'un bolom\`etre. Nous commencerons par d\'ecrire le
travail de S.P.~Langley qui fabriqua le tout premier bolom\`etre de
l'histoire \`a la fin du XIX\textsuperscript{e} si\`ecle. Nous verrons
ensuite la description d\'etaill\'ee du comportement thermique de ce
type de d\'etecteur ainsi que les diff\'erentes sources de bruit
inh\'erentes \`a l'utilisation de bolom\`etres. Enfin, nous ferons le
tour des instruments actuellement en op\'eration sur les grands
t\'elescopes (sub-)millim\'etriques en insistant sur les aspects
thermom\'etriques et optiques de la d\'etection bolom\'etrique.
\end{minipage}
\end{center}

\section{Des d\'etecteurs thermiques}
\label{sec:intro_bolometrie_thermo}

\begin{center}
\begin{minipage}{0.85\textwidth}

\small \emph{\og I call the instrument provisionally the ``Bolometer''
($\beta o \lambda\acute{\eta},\,\mu\acute{\epsilon}\tau\rho o\nu$), or
``Actinic Balance'', because it measures radiations and acts by the
method of the ``bridge'' or ``balance'', there being always two arms,
usually in juxtaposition, and exposed alike to every similar change of
temperature arising from surrounding objects, air-currents, etc., so
that the needle is (in theory at least) only affected when radiant
heat, from which one balance-arm is shielded, falls on the other.\fg}\\
\noindent Extrait de \shortciteN{langley1881}, p.349.\\

\end{minipage}
\end{center}

\subsection{De la thermopile \`a l'invention du bolom\`etre}
\label{sec:intro_bolometrie_thermo_langley}

Le bolom\`etre fut invent\'e en 1880 par un ing\'enieur am\'ericain du
nom de Samuel Pierpont Langley \shortcite{langley1881}. Il fabriqua
cet instrument de mesure dans le but d'\'etudier la \og distribution
de chaleur\fg dans le spectre du Soleil. \`A cette \'epoque, les
physiciens spectroscopistes utilisaient des prismes en verre pour
r\'efracter la lumi\`ere du Soleil et ainsi s\'eparer la lumi\`ere
visible et le rayonnement thermique, comme l'avait fait Herschel au
d\'ebut du XIX\textsuperscript{e} si\`ecle (cf
section~\ref{sec:intro_astroIR_univers_debut}). Cependant Langley
savait que les lois qui r\'egissent la r\'efraction de \og
l'ultra-rouge\fg dans un prisme d\'ependent des propri\'et\'es
physiques du mat\'eriau. Donc, en plus du verre, il travaillait avec
des prismes en sel cristallin (\emph{rock-salt} en anglais) ou en
fluorite. Il a d'ailleurs mesur\'e l'indice de r\'efraction du sel
cristallin jusqu'\`a 10~$\mu$m \shortcite{langley_1900} en s'appuyant
sur les r\'esultats de \shortciteN{paschen94} et
\shortciteN{keeler}. Il a donc pu d\'eduire une relation entre l'angle
de r\'efraction et la longueur d'onde, relation qu'il utilisait
ensuite pour \'etalonner ses \og bolographes\fg (spectres) en longueur
d'onde.\\ Avant 1880, Langley mesurait le flux d'\'energie IR
r\'efract\'e par le prisme avec une thermopile\footnote{Une thermopile
est une association en s\'erie de thermocouples qui utilisent l'effet
thermo-\'electrique, ou effet Seebeck, pour g\'en\'erer une
diff\'erence de potentiel \`a partir d'un gradient de temp\'erature le
long d'un mat\'eriau conducteur.}. Mais apr\`es plusieurs ann\'ees
pass\'ees \`a travailler avec ces thermopiles, il en est arriv\'e \`a
la conclusion que ce type de d\'etecteur n'\'etait qu'un simple
indicateur de rayonnement et non un instrument de mesure suffisamment
fiable pour \'etudier le spectre solaire. Il se lance en 1879 dans la
fabrication d'un nouvel instrument qu'il d\'ecide d'appeller
\emph{bolom\`etre}:\\
\begin{center}
  \begin{minipage}{0.85\textwidth}
    \emph{\og The earliest design was to have two strips of thin
      metal, virtually forming arms of a Wheatstone's Bridge, placed
      side by side in as nearly as possible identical conditions as to
      environment, of which one could be exposed at pleasure to the
      source of radiation. As it was warmed by this radiation and its
      electric resistance proportionally increased over that of the
      other, this increased resistance to the flow of the current from
      a battery would be measured (by the disturbance of the equality
      of the ``bridge'' currents) by means of a galvanometer.\fg}
      Extrait de \shortciteNP{langley1881}, page 343.\\
  \end{minipage}
\end{center}
\begin{figure}
  \begin{center}
    \begin{tabular}{cc}
      \includegraphics[width=0.35\textwidth,angle=0]{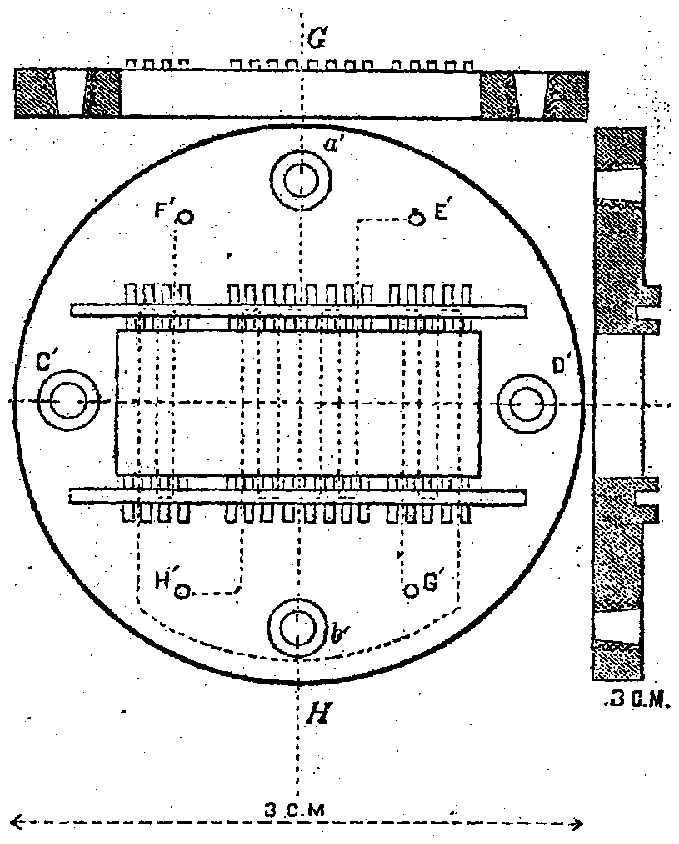}&\includegraphics[width=0.50\textwidth,angle=0]{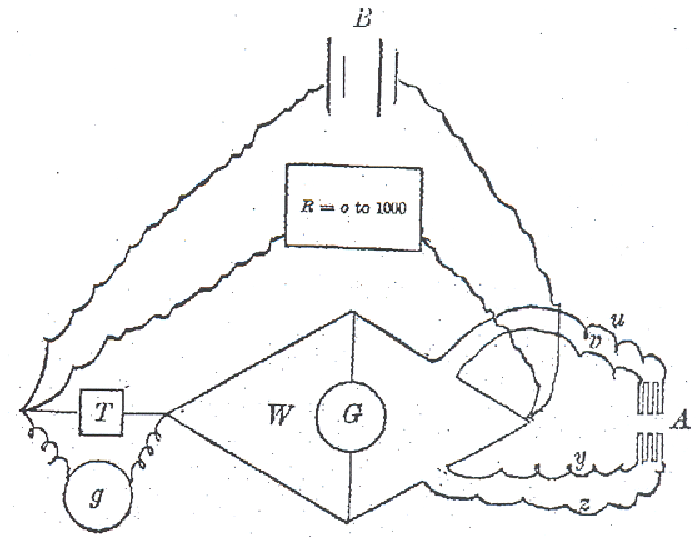}\\
      \multicolumn{2}{l}{\includegraphics[width=1.\textwidth,angle=0]{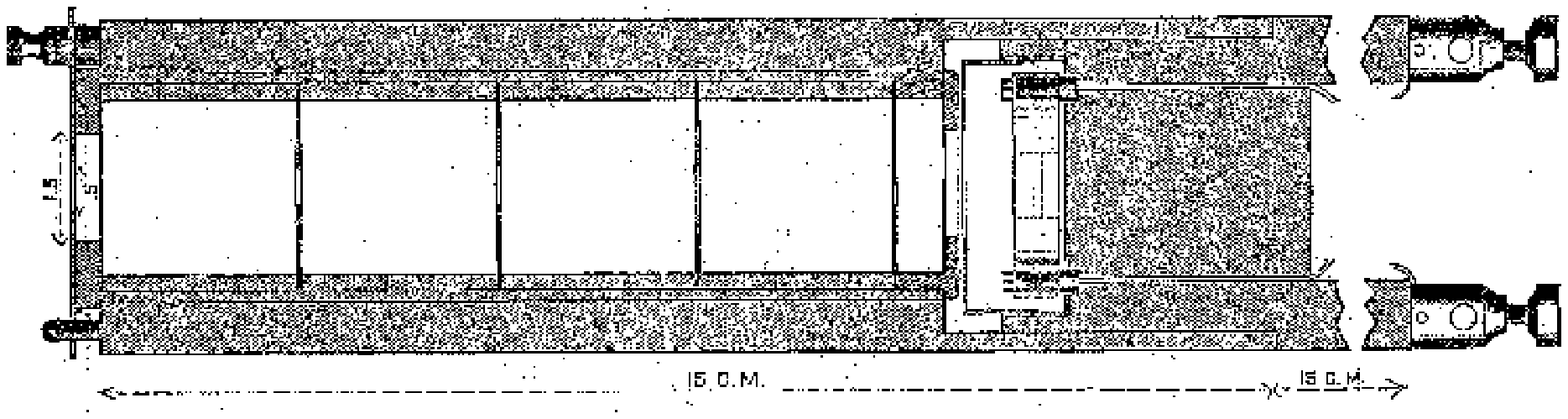}}
    \end{tabular}
  \end{center}
  \caption[Sch\'ema du premier bolom\`etre (1881)]{Sch\'ema du premier
  bolom\`etre con\c{c}u par Samuel Langley en 1881 avec lequel il
  mesura le spectre solaire dans l'infrarouge proche. \emph{En bas}~:
  Enceinte du bolom\`etre. La lumi\`ere rentre par la gauche, passe
  les 4~diaphragmes qui minimisent la circulation d'air dans
  l'instrument (donc les fluctuations de temp\'erature), puis elle est
  absorb\'ee par les bandelettes de platine qui sont reli\'ees
  \'electriquement \`a l'ext\'erieur par deux fils connecteurs
  (visibles en bas \`a droite). \emph{En haut \`a gauche}~: Vues de
  dessus et en coupe du d\'etecteur. Les bandelettes de platine sont
  suspendues et plac\'ees l'une \`a la suite de l'autre en
  s\'erie. \emph{En haut \`a droite}~: Sch\'ema \'electrique du pont
  de Wheaston utilis\'e par Langley pour mesurer les variations
  d'imp\'edance de son bolom\`etre. Il mesure le d\'es\'equilibre du
  pont avec le galvanom\`etre $G$. Le potentiom\`etre $R$ est
  utilis\'e pour r\'eguler le courant qui alimente le pont
  bolom\'etrique.
  \label{fig:intro_bolometrie_thermo_langley_wheatstone}}
\end{figure}
Les premiers travaux de Langley sur les bolom\`etres sont consign\'es
dans un article tr\`es int\'eressant publi\'e en 1881
\shortcite{langley1881} dans les \og Proceedings of the American
Academy of Arts and Sciences\fg. La
figure~\ref{fig:intro_bolometrie_thermo_langley_wheatstone} contient
trois illustrations extraites de cet article. Nous trouvons en bas un
dessin de l'enceinte qui abrite le bolom\`etre, en haut \`a gauche se
trouve un dessin de la partie sensible \`a la lumi\`ere infrarouge,
\cad le bolom\`etre lui-m\^eme, et en bas \`a droite nous voyons le
sch\'ema \'electrique, en configuration de pont de Wheatstone, que
Langley utilisait pour lire le signal aux bornes de la
r\'esistance. La r\'egion d\'enomm\'e $A$ sur la droite du sch\'ema
constitue la partie sensible du d\'etecteur, une r\'esistance est
expos\'ee \`a la radiation thermique du Soleil alors que l'autre ne
l'est pas. Le changement de temp\'erature d\^u au flux incident change
l'imp\'edance de la r\'esistance expos\'ee et des\'equilibre le pont
de Wheatstone. Un galvanom\`etre ($G$ sur la figure) mesure la
variation du courant qui circule dans le bolom\`etre et d\'etecte
ainsi l'\'energie absorb\'ee. Langley a fabriqu\'e de nombreux
bolom\`etres avec des r\'esistances en fer, en acier, en carbone mais
le mat\'eriau qui donne les meilleurs r\'esultats reste le platine (ce
qui n'est vrai qu'\`a temp\'erature ambiante, le platine est en plus
tr\`es pratique puisque c'est un mat\'eriau relativement
mal\'eable). Il fabrique lui-m\^eme ces r\'esistances en forme de
fines bandelettes de 2~\`a~10~$\mu$m d'\'epaisseur, de 1~cm de long et
de 1~mm de large. Ces bandelettes de platine ont une imp\'edance de
quelques Ohms, et les variations de temp\'erature typiques,
lorsqu'elles sont expos\'ees \`a la lumi\`ere r\'efract\'ee du Soleil,
sont de l'ordre de 7~degr\'es. D'apr\`es son article de 1881, cet
instrument de mesure est capable de d\'etecter un \'echauffement de la
r\'esistance inf\'erieur \`a \og $\frac{1}{10000}$\fg degr\'es
Celsius! Les performances du bolom\`etre de Samuel Langley sont
exceptionnelles, elles sont bien meilleures que celles de la plus
sensible des thermopiles.\\ Cependant, pour Samuel Langley,
l'invention du bolom\`etre n'\'etait pas une finalit\'e mais plut\^ot
un moyen pour \'etudier le Soleil dans un domaine de longueur d'onde
encore inexplor\'e. Il mesure le spectre solaire dans l'infrarouge
proche\footnote{Il nomme en fait cette r\'egion du spectre
\emph{lowest infrared} plut\^ot que \emph{near infrared} comme nous le
faisons aujourd'hui.} en 1881 et d\'ecouvre des bandes d'absorption
\`a 0.92 et 1.1~$\mu$m qu'il attribue \`a la vapeur d'eau contenue
dans l'atmosph\`ere terrestre. En 1900, alors qu'il est directeur du
Smithsonian Astrophysical Observatory qu'il a lui m\^eme fond\'e,
Langley publie ses r\'esultats sur les diverses observations du
Soleil, ainsi que de quelques sources terrestres, avec son
spectro-bolom\`etre \shortcite{langley_1900}. La
figure~\ref{fig:intro_bolometrie_thermo_langley_specSun} montre le
spectre solaire qu'il a obtenu en 1896 au Mount Whitney en
Californie.\\
\begin{figure}
  \begin{center}
      \includegraphics[width=0.6\textwidth,angle=0]{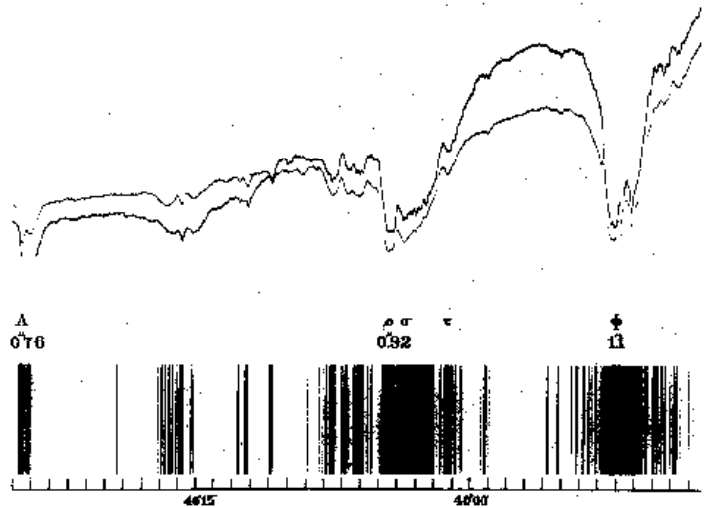}
  \end{center}
  \caption[Spectre du Soleil dans l'infrarouge proche (1896)]{Spectre
  du Soleil dans l'infrarouge proche mesur\'e avec le
  spectro-bolom\`etre de Samuel Langley. Ce \og bolographe\fg a
  \'et\'e obtenu au Mount Whitney en 1896 avec un prisme en sel
  cristallin. Langley attribue les deux larges raies d'absorption \`a
  0.92 et 1.1~$\mu$m \`a l'absorption de la vapeur d'eau
  atmosph\'erique.
  \label{fig:intro_bolometrie_thermo_langley_specSun}}
\end{figure}
En 1881 d\'ej\`a, Langley avait mesur\'e le spectre du Soleil de
l'ultra-violet \`a \og l'ultra-rouge\fg et affirmait que le pic
d'\'emission se situe dans l'orange, autour de 0.6~$\mu$m, et non dans
la partie \og non-lumineuse du spectre\fg comme le pensaient les
scientifiques de l'\'epoque. De plus, il a montr\'e que l'\'emission
thermique des sources terrestres pique \`a de plus grandes longueurs
d'onde que celle du Soleil, et que plus la temp\'erature de la source
augmente et plus le pic se d\'eplace vers les courtes longueurs
d'onde. D'apr\`es les travaux de \shortciteN{paschen97} sur
l'\'emission de corps noir, il mesure pr\'ecisemment la temp\'erature
de quelques sources chauff\'ees \`a la bougie. L'\'emission du corps
noir sera expliqu\'ee quelques ann\'ees plus tard par Max Planck qui
introduisit le quantum d'\'energie et ouvrit la voie \`a la
m\'ecanique quantique.








\subsection{G\'en\'eralit\'es et principes de fonctionnement}
\label{sec:intro_bolometrie_thermo_principe}

Depuis la d\'ecouverte de l'infrarouge par Herschel, les d\'etecteurs
thermiques sont devenus les d\'etecteurs de pr\'edilection pour
l'\'etude du rayonnement \'electromagn\'etique dans le r\'egime
infrarouge lointain (100~$\mu$m \`a 1~mm). Ils ont bien s\^ur trouv\'e
de nombreuses applications en astronomie, mais la spectroscopie a
\'egalement jou\'e un r\^ole moteur dans le d\'eveloppement et
l'am\'elioration de ces d\'etecteurs au cours du XX\textsuperscript{e}
si\`ecle. Il existe diff\'erentes sortes de d\'etecteurs thermiques,
mais tous ont un point commun~: ils utilisent la temp\'erature comme
vecteur d'information. Ce qui les distingue toutefois l'un de l'autre
est la mani\`ere dont les excursions en temp\'erature sont
d\'etect\'ees. Par exemple, pour les cellules pneumatiques de Golay,
le rayonnement est absorb\'e par une fine couche de m\'etal qui est en
contact thermique avec une cellule contenant un petit volume de
gaz. La chaleur transf\'er\'ee du m\'etal vers le gaz produit une
augmentation de pression et d\'eforme une extr\'emit\'e de la cellule
sur laquelle se trouve une membrane r\'efl\'echissante. Les variations
de temp\'erature, \cad variations de forme, peuvent alors \^etre
mesur\'ees par un syst\`eme optique. Les cellules de Golay sont
utilis\'ees pour la spectroscopie infrarouge depuis les ann\'ees~60,
elles sont relativement sensibles et facile d'utilisation car elles
fonctionnent \`a temp\'erature ambiante. D'autres d\'etecteurs
thermiques relativement r\'epandus tels que les d\'etecteurs
pyro\'electriques ou les thermopiles sont actuellement utilis\'es dans
le domaine de la s\'ecurit\'e, de la surveillance, de la d\'etection
de flammes, de la mesure de temp\'erature, etc... Mais dans le domaine
de l'astronomie, les sensibilit\'es \`a atteindre sont tellement
extr\^emes que les d\'etecteurs doivent \^etre refroidis \`a tr\`es
basse temp\'erature pour modifier leurs propri\'et\'es physiques et
les adapter aux applications astronomiques. Le plupart des
bolom\`etres modernes sont op\'er\'es \`a des temp\'eratures
inf\'erieures \`a 0.3~K. Leur mise en \oe uvre est donc plus
contraignante que celle des d\'etecteurs \`a temp\'erature ambiante.\\

\begin{figure}
  \begin{center}
    \begin{tabular}[t]{ll}
      \includegraphics[width=0.50\textwidth,angle=0]{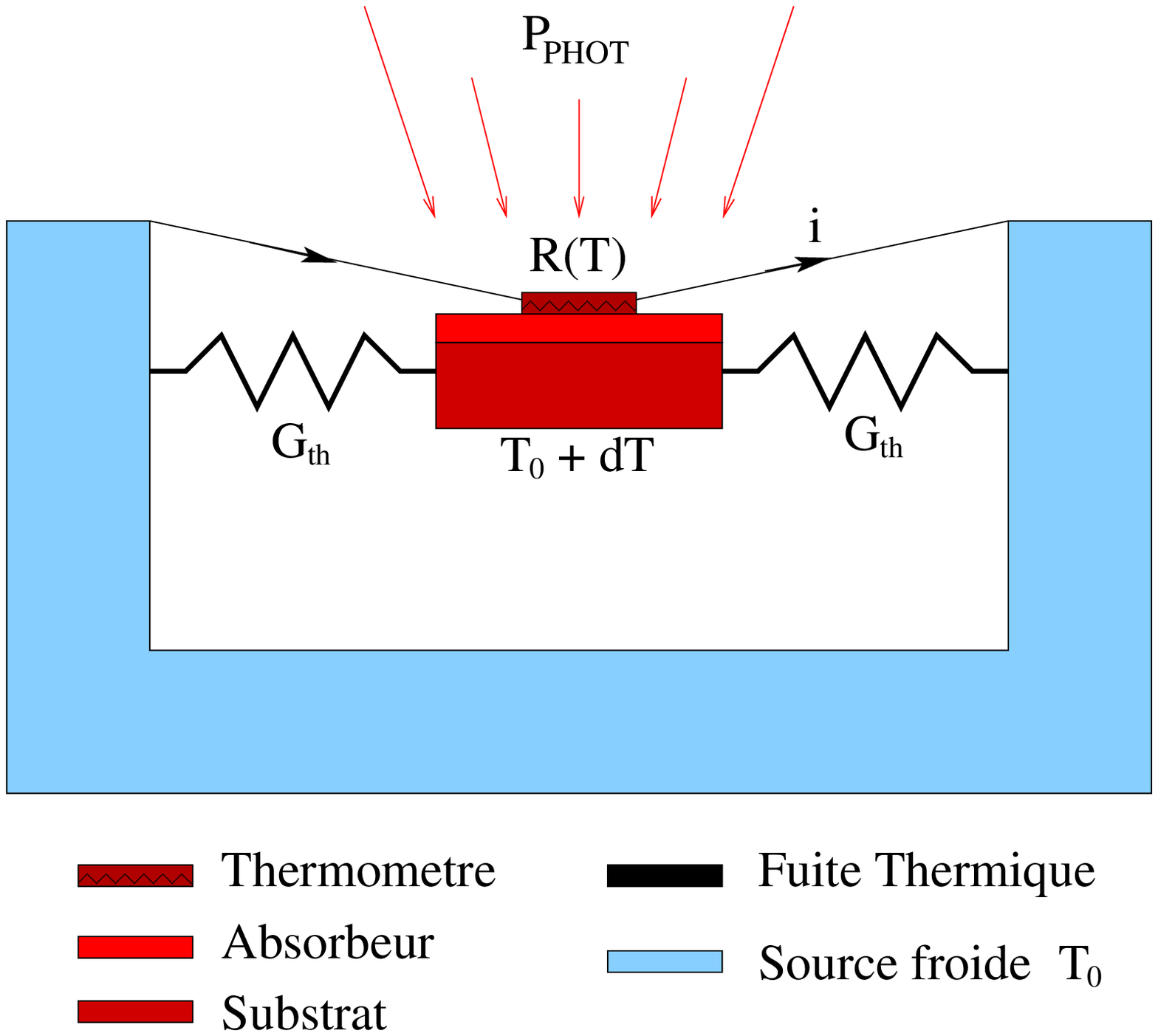}
      &
      \includegraphics[width=0.45\textwidth,angle=0]{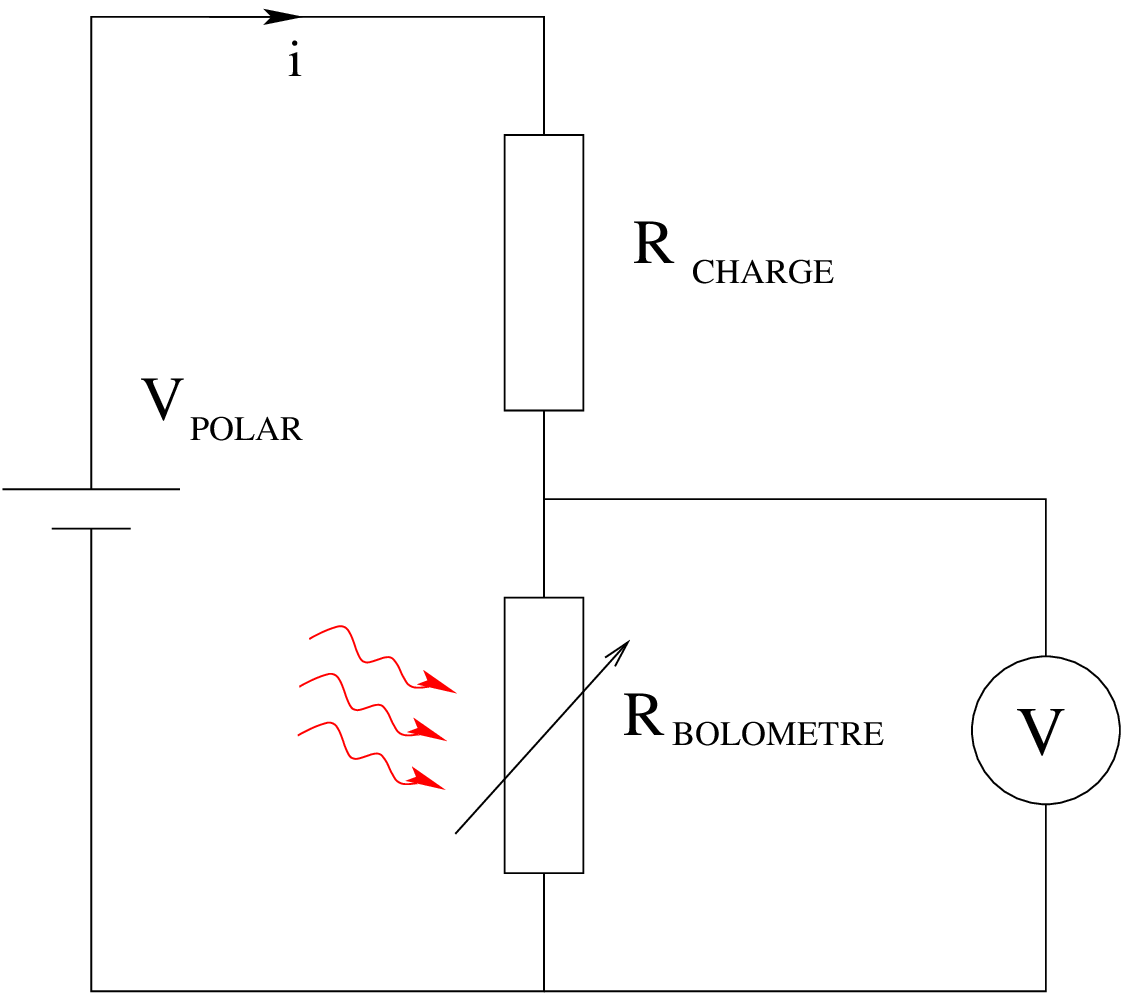}
    \end{tabular}
  \end{center}
  \caption[Sch\'ema \'electrique et thermique d'un bolom\`etre
  typique]{Un bolom\`etre est un syst\`eme opto-\'electro-thermique et
  son \'etude fait appel \`a divers domaines de la physique. Le
  sch\'ema de gauche (inspir\'e de~\shortciteNP{buzzi_these}) montre
  le bolom\`etre suspendu par des poutres fines reliant thermiquement
  la source froide (temp\'erature $T_0$) et l'absorbeur (temp\'erature
  $T_0+dT$). Le thermom\`etre est \'echauff\'e par la puissance
  photonique $P_{PHOT}$ et par la dissipation \'electrique
  $P_{Joule}=Ri^2$. La chaleur accumul\'ee dans le composant de
  capacit\'e calorifique $C_{th}$ est \'evacu\'ee par la fuite
  thermique de r\'esistance thermique $R_{th}=1/G_{th}$.  Le sch\'ema
  de droite montre le montage \'electrique, \cad un pont diviseur de
  tension, typiquement utilis\'e pour mesurer le signal aux bornes de
  la thermistance. L'imp\'edance varie fortement avec la
  temp\'erature.
  \label{fig:intro_bolometrie_thermo_principe_bolo1}}
\end{figure}

Un bolom\`etre est essentiellement compos\'e de trois briques de
bases~: un \emph{absorbeur} de rayonnement, un \emph{senseur
thermique} et une \emph{fuite thermique} qui connecte l'absorbeur \`a
une source froide. Historiquement, les premiers bolom\`etres \'etaient
compos\'es d'un seul \'el\'ement qui combinait les fonctions
d'absorbeur et de thermom\`etre~; par exemple, le bolom\`etre d\'ecrit
par \citeN{low} est compos\'e d'une r\'esistance de Germanium dop\'e
recouverte de peinture noire, et cette r\'esistance est reli\'ee au
bain thermique seulement par ses deux fils connecteurs. Aujourd'hui,
ces fonctions de bases sont remplies par des \'el\'ements distincts,
il est ainsi possible d'optimiser s\'epar\'ement les performances de
chacun des \'el\'ements constitutifs du bolom\`etre. Nous parlons
alors de \emph{bolom\`etres composites}. La
figure~\ref{fig:intro_bolometrie_thermo_principe_bolo1} montre
l'empilement typique des \'el\'ements constituant un bolom\`etre
composite pour la d\'etection FIR/submm~: l'absorbeur est une couche
mince de m\'etal d\'epos\'ee sur un substrat dont la capacit\'e
calorifique $C_{th}$ est tr\`es faible, le thermom\`etre est en
contact thermique avec l'absorbeur et est reli\'e \'electriquement \`a
la structure par deux connecteurs. Le courant $i$ qui circule dans la
r\'esistance $R(T)$ dissipe une puissance \'electrique, dite puissance
Joule $P_J$, dans l'absorbeur. Une puissance photonique $P_{PHOT}$ est
\'egalement absorb\'ee par le bolom\`etre et contribue \`a
l'\'el\'evation de temp\'erature de la partie suspendue. Le
bolom\`etre est reli\'e m\'ecaniquement et thermiquement \`a une
source froide maintenue \`a la temp\'erature $T_0$ par des poutres de
faible conductance thermique $G_{th}$. Le sch\'ema de droite dans la
figure~\ref{fig:intro_bolometrie_thermo_principe_bolo1} pr\'esente le
montage typiquement utilis\'e pour lire le signal \'electrique aux
bornes du thermom\`etre. Il s'agit en fait d'un simple montage en pont
diviseur de tension.

D'apr\`es \citeN{jones}, et de mani\`ere g\'en\'erale, nous pouvons
d\'efinir le bolom\`etre comme \'etant un d\'etecteur de rayonnement
dont le mode d'op\'eration se r\'esume de la mani\`ere suivante:
\begin{itemize}
\item le rayonnement incident sur le d\'etecteur est absorb\'e, ce qui
entra\^ine un \'echauffement du bolom\`etre,
\item l'\'el\'evation de temp\'erature modifie la r\'esistance
\'electrique du bolom\`etre,
\item une batterie et une r\'esistance de charge mont\'ees en s\'erie
avec le bolom\`etre sont alors utilis\'ees pour convertir ce
changement de r\'esistance en un changement de potentiel \'electrique,
\item l'\'energie stock\'ee dans le bolom\`etre est \'evacu\'ee vers
une source froide par un faible lien thermique.
\end{itemize}
Jones pr\'ecise \'egalement dans son article qu'un bolom\`etre n'est
pas un transducteur, il ne g\'en\`ere pas de signal \'electrique, il
module simplement le courant \'electrique qui le traverse. Un
thermocouple, par contre, est un transducteur, il convertit une
puissance radiative en une puissance \'electrique. D'autre part, un
photoconducteur n'est pas un bolom\`etre car le changement de
r\'esistivit\'e n'est pas la cons\'equence d'un changement de
temp\'erature mais plut\^ot le passage d'un certain nombre
d'\'electrons dans la bande de conduction du mat\'eriau
semiconducteur.

Un bolom\`etre est un syst\`eme particuli\`erement int\'eressant pour
un physicien car, malgr\'e son principe de fonctionnement relativement
simple \`a premi\`ere vue, il se trouve \`a la crois\'ee de plusieurs
domaines de la physique~; et pour \'etudier exhaustivement son
comportement, il faut faire appel \`a la thermodynamique, \`a
l'\'electronique, \`a l'optique, \`a la m\'ecanique, \`a la physique
du solide, aux techniques de cryog\'enie, \`a la micro-technologie,
etc... Pour d\'ebuter, je propose une analyse thermodynamique simple
du fonctionnement et des performances d'un bolom\`etre. Notez qu'au
fil du manuscrit je donnerai des descriptions de plus en plus
d\'etaill\'ees sur les diff\'erents aspects du comportement d'un
bolom\`etre.\\

Pour une quantit\'e donn\'ee d'\'energie absorb\'ee $\delta Q$, la
sensibilit\'e d'un bolom\`etre est d'autant plus grande que son
\'el\'evation de temp\'erature $\delta T$ est grande. Ces deux
quantit\'es sont reli\'ees par la capacit\'e calorifique $C_{th}$ du
bolom\`etre par l'\'equation suivante~:
\begin{equation}
\delta T = \frac{\delta Q}{C_{th}}
\label{eq:fig:intro_bolometrie_thermo_principe_capacitecalorific}
\end{equation}
Un bolom\`etre tr\`es sensible aura donc une capacit\'e calorifique
tr\`es faible. Notez de plus que la capacit\'e calorifique est le
produit de la chaleur sp\'ecifique du mat\'eriau constituant
l'absorbeur~$s$ et de sa masse~$m$.
$$C_{th}=m\times s$$ L'optimisation des performances d'un bolom\`etre
n\'ecessite donc de r\'eduire la masse de l'\'el\'ement absorbeur. Il
existe toutefois une limite \`a la taille minimale que peut avoir un
bolom\`etre. En effet l'absorption du rayonnement est inefficace si
l'absorbeur est plus petit que la longueur d'onde $\lambda$ du
rayonnement \`a d\'etecter, ceci est d\^u au ph\'enom\`ene de
diffraction qui devient non-n\'egligeable. L'absorbeur a donc l'aspect
d'une plaque tr\`es fine, avec une \'epaisseur de quelques $\mu$m
impos\'ee par les contraintes m\'ecaniques, dont la taille
caract\'eristique est approximativement $\lambda\times \lambda$. Une
fois la taille fix\'ee, il est encore possible de r\'eduire la
quantit\'e de mati\`ere en \og creusant \fg des ouvertures dans la
plaque. Si ces trous sont plus petits que la longueur d'onde
$\lambda$, alors l'onde incidente \og voit \fg le bolom\`etre comme
une plaque pleine et l'efficacit\'e d'absorption reste
inchang\'ee. Nous verrons dans la
section~\ref{sec:detect_bolocea_grille} que le pixel des matrices de
bolom\`etres du CEA ressemble en effet \`a un fin squelette de
mati\`ere et que sa masse est de l'ordre du $\mu$g. De plus, en
r\'eduisant la surface de l'absorbeur, sa section efficace diminue et
les int\'eractions avec les rayons cosmiques sont minimis\'ees.

D'autre part, la capacit\'e calorifique peut \^etre consid\'erablement
r\'eduite en abaissant la temp\'erature de fonctionnement d'un
bolom\`etre. En effet, en-dessous de la temp\'erature de Debye, la
chaleur sp\'ecifique des semiconducteurs utilis\'es pour la
fabrication des substrats ou des thermom\`etres varie comme
$T^3$. Pour les m\'etaux, l'absorbeur \`a proprement parler, la
chaleur sp\'ecifique varie lin\'eairement avec la temp\'erature $T$.
Notez que les bolom\`etres sont les d\'etecteurs qui n\'ecessitent les
temp\'eratures d'op\'eration les plus basses en astronomie, elles sont
en g\'en\'eral inf\'erieures \`a 0.3~K et nous obtenons des
capacit\'es calorifiques de l'ordre de 10$^{-13}$-10$^{-14}$~J/K. Pour
la d\'etection dans l'infrarouge lointain, il faut \'egalement
refroidir l'environnement imm\'ediat du bolom\`etre pour s'assurer par
exemple que l'\'electronique de lecture n'\'emette pas dans la bande
de d\'etection (\`a 300~mK le bolom\`etre \'emet dans le domaine
centim\'etrique). Un autre avantage de travailler \`a des
temp\'eratures cryog\'eniques est que le bruit g\'en\'er\'e par le
bolom\`etre est significativement r\'eduit. Les diff\'erentes sources
de bruit intrins\`eques \`a l'utilisation d'un bolom\`etre sont
d\'etaill\'ees dans la
section~\ref{sec:intro_bolometrie_thermo_principe_bruit}.\\

Les variations de temp\'erature d'un bolom\`etre sont mesur\'ees par
une thermistance, \cad une r\'esistance dont l'imp\'edance varie
fortement avec sa temp\'erature. Cette r\'esistance est en principe
coll\'ee sur l'absorbeur avec une colle Epoxy qui assure un tr\`es bon
contact thermique~; la r\'esistance doit \^etre parfaitement
thermalis\'ee avec l'absorbeur. La plupart des bolom\`etres modernes
utilise des thermistances r\'esistives, mais dans quelques ann\'ees
les thermistances supraconductrices seront certainement beaucoup plus
r\'epandues car elles permettent en th\'eorie une meilleure
sensibilit\'e. Ces deux types de bolom\`etres seront pr\'esent\'es
plus en d\'etails dans les
sections~\ref{sec:intro_bolometrie_bolo_resistif},
\ref{sec:intro_bolometrie_bolo_supra}~
et~\ref{sec:detect_bolocea_fabrication_thermo}.\\

Les poutres qui supportent m\'ecaniquement l'absorbeur et le
connectent \`a la source froide jouent un r\^ole d\'eterminant dans
l'\'equilibre thermique du bolom\`etre. Elles transportent en effet la
chaleur absorb\'ee par le bolom\`etre vers une monture massive plus
froide. Cette monture poss\`ede une grande capacit\'e calorifique de
sorte que sa temp\'erature n'est pas influenc\'ee par l'\'energie
provenant du bolom\`etre (comme indiqu\'e par
l'\'equation~\ref{eq:fig:intro_bolometrie_thermo_principe_capacitecalorific}). Le
puits de chaleur est g\'en\'eralement le doigt froid d'un
cryo-r\'efrig\'erateur qui fournit \`a tout le d\'etecteur une
temp\'erature constante T$_0$. La description du
cryo-r\'efrig\'erateur utilis\'e pour le Photom\`etre PACS est
donn\'ee dans la
section~\ref{sec:detect_observatoire_phfpu_cryocooler}. La conductance
thermique des poutres est un param\`etre critique dans la conception
d'un bolom\`etre~; elle est controll\'ee par leur longueur et leur
diam\`etre. La diffusion de chaleur le long de ces poutres suit
l'\'equation suivante~:
\begin{equation}
\frac{dQ}{dt}=G_{th}\,(T-T_0)
\end{equation}
o\`u $dQ$ est l'\'energie transf\'er\'ee en un temps $dt$ le long des
poutres, $G_{th}$ est la conductance thermique ([W/K]), $T_0$ est la
temp\'erature du puits de chaleur et $T$ est la temp\'erature de
l'absorbeur.  Pour faire le bilan \'energ\'etique du bolom\`etre, il
faut prendre en compte les sources d'\'energie pr\'esentes au niveau
de l'absorbeur, \cad la puissance radiative absorb\'ee et la puissance
Joule dissip\'ee par la thermistance. Toutefois nous ne faisons pour
le moment aucune distinction entre ces deux sources d'\'energie, et
nous appelons $P$ la puissance totale absorb\'ee. Nous pouvons alors
\'ecrire que la variation d'\'energie contenue dans l'absorbeur est
\'egale \`a la puissance entrant dans le syst\`eme moins la puissance
sortante~:
\begin{equation}
C_{th}\,\frac{dT}{dt}=P-G_{th}\,(T-T_0)
\label{eq:intro_bolometrie_thermo_principe_cdTdt}
\end{equation}
\`A l'\'equilibre, la temp\'erature de l'absorbeur est constante et
vaut $T=T_0+P/G_{th}$. Il vient alors~:
\begin{equation}
\frac{dT}{dP}=\frac{1}{G_{th}}
\end{equation}
 o\`u $\frac{dT}{dP}$ repr\'esente la r\'eponse thermique du
d\'etecteur, \cad que pour un flux incident donn\'e, les excursions en
temp\'erature sont d'autant plus grandes que la conductance thermique
est petite. Nous avons donc int\'er\^et \`a minimiser $G_{th}$ pour
isoler thermiquement le bolom\`etre et ainsi augmenter sa
sensibilit\'e. Cependant, une autre propri\'et\'e importante du
bolom\`etre d\'epend \'egalement de $G_{th}$, c'est la constante de
temps.\\

Si le d\'etecteur est en op\'eration et \`a l'\'equilibre thermique
\`a la temp\'erature $T_1$, et que la source d'\'energie s'arr\`ete
brusquement, alors la temp\'erature du bolom\`etre chute jusqu'\`a
$T_0$ en suivant l'\'equation~:
\begin{equation}
C_{th}\,\frac{dT}{dt}=-G_{th}\,(T-T_0)
\end{equation}
La solution de cette \'equation diff\'erentielle du premier ordre est~:
\begin{equation}
T=T_0+(T_1-T_0)\,exp(-t/\tau_{th})
\end{equation}
o\`u $\tau_{th}=C_{th}/G_{th}$ est la constante de temps thermique du
bolom\`etre. Plus les poutres sont aptes \`a transf\'erer de la
chaleur, \cad $G_{th}$ est grand, plus la chaleur contenue dans
l'absorbeur est \'evacu\'ee rapidement. D'autre part, plus la
capacit\'e calorifique $C_{th}$ est grande, plus la quantit\'e de
chaleur stock\'ee dans l'absorbeur est grande, il faut donc plus
longtemps pour retourner \`a l'\'equilibre thermique.

Prenons maintenant une puissance incidente qui oscille \`a la
fr\'equence angulaire $\omega$. Nous faisons l'hypoth\`ese que cette
puissance peut \^etre d\'ecompos\'ee en une partie constante
d'amplitude r\'eelle $P_C$, et d'une partie p\'eriodique d'amplitude
complexe $\bar{P}$. La puissance instantan\'ee s'\'ecrit alors
$P(t)=P_C+\bar{P}\,exp(\boldsymbol{i}\omega t)$. Comme une oscillation
forc\'ee, la temp\'erature du bolom\`etre s'\'etablit dans un r\'egime
similaire avec une partie constante et une partie p\'eriodique~:
$T(t)=T_C+\bar{T}\,exp(\boldsymbol{i}\omega t)$. Le d\'ephase entre
$P(t)$ et $T(t)$ est contenu dans l'amplitude complexe $\bar{T}$. En
rempla\c{c}ant $P(t)$ et $T(t)$ dans
l'\'equation~\ref{eq:intro_bolometrie_thermo_principe_cdTdt} et en
\'egalisant les termes oscillants, nous obtenons~:
\begin{equation}
\boldsymbol{i}\omega C_{th} \bar{T} = \bar{P} - G_{th} \bar{T}
\end{equation}
et l'amplitude de la r\'eponse en temp\'erature vaut alors~:
\begin{equation}
|\bar{T}|=\frac{|\bar{P}|}{G_{th}}\frac{1}{\sqrt{1+\omega^2\tau_{th}^2}}
\label{eq:intro_bolometrie_thermo_principe_respomega}
\end{equation}
Cette formule nous indique que pour une m\^eme amplitude en puissance,
l'amplitude des oscillations de la temp\'erature diminue avec la
fr\'equence de modulation du signal. En fait, l'\'energie stock\'ee
dans l'absorbeur et la faible fuite thermique se combinent pour
cr\'eer un filtre passe-bas du premier ordre. Nous verrons les
cons\'equences de cette constante de temps sur les observations
astronomiques dans la section~\ref{sec:calib_perfobs_scan}.

Nous arr\^etons ici l'analyse, nous la reprendrons dans la
section~\ref{sec:detect_outils_loadcurves} o\`u nous utiliserons une
approche \'electro-thermique pour d\'ecrire la m\'ethode standard de
caract\'erisation des bolom\`etres. Nous nous int\'eressons maintenant
aux diff\'erentes sources de bruit inh\'erentes \`a l'utilisation d'un
bolom\`etre.




\subsection{Les sources de bruit}
\label{sec:intro_bolometrie_thermo_principe_bruit}

Il existe une limitation fondamentale dans le processus de d\'etection
du rayonnement \'electromagn\'etique~: les fluctuations quantiques du
flux de photons incident sur le d\'etecteur~; et l'objectif de tout
instrument destin\'e \`a l'astronomie est d'atteindre cette limite de
sensibilit\'e en r\'eduisant les bruits intrins\`eques g\'en\'er\'es
par le d\'etecteur ou son \'electronique de lecture. Dans le domaine
infrarouge au sens large du terme, \cad de 1~$\mu$m \`a 1~mm,
lorsqu'un d\'etecteur poss\`ede effectivement un bruit intrins\`eque
inf\'erieur au bruit de photon, il est alors qualifi\'e de \emph{BLIP}
(\emph{B}ackground \emph{L}imited \emph{I}nfrared
\emph{P}hotodetector). Dans cette section, nous pr\'esentons les
diff\'erentes sources de bruit rencontr\'ees en bolom\'etrie, et nous
donnons \'egalement, lorsque cela est possible, une formule analytique
qui permet de calculer leur contribution au bruit total du
d\'etecteur. Nous utiliserons souvent le terme de \emph{NEP} pour
quantifier un niveau de bruit. Notez simplement que cette quantit\'e
repr\'esente la plus petite puissance optique d\'etectable dans une
bande passante de 1~Hz avec un signal-\`a-bruit de~1, elle sera
d\'ecrite plus en d\'etails dans les
sections~\ref{sec:calib_perflabo_sensibilite}
et~\ref{sec:calib_perfobs_nep}.

\subsubsection{Le bruit de photon}
\label{sec:intro_bolometrie_thermo_principe_bruit_photon}

Nous avons vu dans la section~\ref{sec:intro_astro_william_obs} que la
principale contribution au flux incident sur un d\'etecteur infrarouge
lointain est l'\'emission du t\'elescope et de l'atmosph\`ere. M\^eme
dans le cas d'un observatoire spatial comme Herschel o\`u
l'atmosph\`ere n'est plus un probl\`eme et que le t\'elescope est
refroidi \`a environ 80~K, la source d'\'emission la plus puissante
n'est pas le ciel mais plut\^ot le t\'elescope lui-m\^eme. Puisque la
quasi totalit\'e des photons d\'etect\'es par un bolom\`etre provient
de son environnement, nous allons d\'evelopper le formalisme
n\'ecessaire au calcul du bruit de photon pour une \'emission
d'avant-plan de type corps gris, \cad un corps noir avec une
\'emissivit\'e $\epsilon (\nu)$ qui d\'epend de la fr\'equence du
rayonnement. D'apr\`es~\citeN{rohlfs}, la fluctuation \emph{r.m.s.} du
nombre de photons \'emis \`a la fr\'equence~$\nu$ est~:
\begin{equation}
(\Delta n_{r.m.s.})^2=n\times(n+1)
\label{eq:fluctuation_stat}
\end{equation}
o\`u $n$ est le nombre d'occupation des photons dans le mode
consid\'er\'e. \\
\noindent Pour un corps noir \`a l'\'equilibre \`a la temp\'erature
$T$, la distribution de Bose-Einstein donne
$n=[exp(h\nu/kT)-1]^{-1}$. Aux courtes longueurs d'onde, \cad jusque
dans l'infrarouge proche o\`u $h\nu/kT\gg 1$, nous avons $n\ll 1$ de
sorte que les photons suivent une statistique de Poisson, ils arrivent
sur le d\'etecteur de fa\c{c}on al\'eatoire, et $(\Delta
n_{r.m.s.})^2=n$. Par contre, dans le r\'egime radio o\`u $n\gg 1$,
les photons arrivent par groupe et interf\`erent, le bruit vaut alors
$(\Delta n_{r.m.s.})^2=n^2$ \shortcite{richards,zmuidzinas}. Notez que
dans l'infrarouge lointain aucun de ces deux termes n'est
n\'egligeable, nous les gardons donc dans le reste de nos calculs.\\
\noindent Pour d\'eterminer la NEP d\'etecteur, nous calculons le
nombre de photons transmis par le syst\`eme optique et absorb\'e par
le bolom\`etre. Nous devons donc consid\'erer la quantit\'e~:
\begin{equation}
n=\eta_{det}\,\epsilon(\nu)\,t(\nu)\,\frac{1}{e^{\frac{h\nu}{kT}}-1}
\end{equation}
o\`u $\eta_{det}$ est l'efficacit\'e du d\'etecteur pour convertir un
photon en signal, $\epsilon(\nu)$ est l'\'emissivit\'e du corps noir
et $t(\nu)$ est la transmission du syst\`eme optique. Plut\^ot que
d'exprimer le bruit de photon en unit\'e de [nombre de photon], nous
pouvons calculer la NEP$_{photon}$ en multipliant
l'\'equation~(\ref{eq:fluctuation_stat}) par l'\'etendue de faisceau
du syst\`eme optique ($A\Omega$), par la densit\'e d'\'etats du mode
consid\'er\'e ($2h\nu^3/c^2$), et par un facteur~2 pour prendre en
compte les 2~\'etats de polarisation possibles. Il faut ensuite
int\'egrer sur les fr\'equences et nous obtenons finalement la formule
g\'en\'erale de la NEP photonique~:
\begin{equation}
NEP_{photon}=2\,A\Omega \,\frac{h}{c}\,\sqrt{\int^{+\infty}_0\!\nu^4\,\frac{\eta_{det}\epsilon(\nu)t(\nu)}{e^{\frac{h\nu}{kT}}-1}\,\left[1+\frac{\eta_{det}\epsilon(\nu)t(\nu)}{e^{\frac{h\nu}{kT}}-1}\right]\,d\nu}
\label{eq:NEPphoton}
\end{equation}

\subsubsection{Le bruit thermique ou bruit de phonon}
\label{sec:intro_bolometrie_thermo_principe_bruit_phonon}

Un phonon d\'esigne un quantum de vibration dans un solide cristallin.
La chaleur accumul\'ee dans un bolom\`etre s'\'evacue via les poutres
de fuite thermique (cf
figure~\ref{fig:intro_bolometrie_thermo_principe_bolo1}) par le biais
de phonons, et nous pouvons exprimer les fluctuations d'\'energie
li\'ees \`a ce transfert de chaleur de la fa\c{c}on suivante~:
\begin{equation}
<(\Delta U)^2>=kT_0^2C_{th}
\end{equation}
o\`u $\Delta U$ repr\'esente la fluctuation d'\'ebergie, $k$ est la
constante de Boltzmann, $T_0$ est la temp\'erature de la source froide
et $C_{th}$ est la capacit\'e calorifique de l'absorbeur. Ce bruit se
traduit en fluctuations de puissance dissip\'ee dans le d\'etecteur
tel que~:
\begin{equation}
<(\Delta P)^2>=4kT_0^2G_{th}
\end{equation}
o\`u $G_{th}$ est la conductance thermique des poutres reliant
l'absorbeur \`a la source froide. Notez que ce bruit est ind\'ependant
de la fr\'equence, c'est un bruit blanc, et qu'il peut \^etre
consid\'erablement r\'eduit si le d\'etecteur est utilis\'e \`a tr\`es
basse temp\'erature .\\
\noindent D'autre part, cette derni\`ere expression suppose que le
bolom\`etre soit \`a la m\^eme temp\'erature que le bain thermique. En
pratique, la puissance photonique absorb\'ee et la puissance
\'electrique dissip\'ee \'el\`event l\'eg\`erement sa temp\'erature,
typiquement de l'ordre de 10~\% d'apr\`es \shortciteN{low}, de sorte
que la conductance thermique est en r\'ealit\'e un peu plus grande. Ce
ph\'enom\`ene est qualifi\'e de contre-r\'eaction
thermique. Toutefois, ce ph\'enom\`ene est rarement pris en compte, si
bien que la plupart des auteurs se contentent de remplacer $T_0$ par
la temp\'erature effective du bolom\`etre $T$, et $G_{th}$ par la
conductance dynamique $(G_{th})_d=\frac{\partial P}{\partial T}$, il
vient~:
\begin{equation}
<(\Delta P)^2>=4kT^2(G_{th})_d
\label{eq:phonon}
\end{equation}
Dans sa th\'eorie du bruit d'un bolom\`etre hors-\'equilibre,
\shortciteN{mather82} a montr\'e que la contre-r\'eaction apporte un
terme correctif \`a cette expression de l'ordre de 30~\% (le bruit
r\'eel devrait \^etre 30~\% inf\'erieur). 

\subsubsection{Le bruit Johnson}
\label{sec:intro_bolometrie_thermo_principe_bruit_johnson}

Le bruit Johnson \shortcite{johnson} est un bruit fondamental
pr\'esent aux bornes de toute r\'esistance, m\^eme en l'absense de
polarisation. En effet, l'agitation thermique des \'electrons
pr\'esents dans une r\'esistance cr\'ee un faible courant \`a moyenne
nulle mais dont la variance vaut~:
\begin{equation}
<(\Delta V)^2>=4kTR
\end{equation}
o\`u $R$ est la valeur de la r\'esistance. Pour un pont
bolom\'etrique, la r\'esistance \`a consid\'erer dans le calcul du
bruit est la r\'esistance \'equivalente du circuit. Du point de vue du
circuit de lecture, les deux r\'esistances du pont bolom\'etrique sont
mont\'ees en parall\`ele, et nous trouvons~:
\begin{equation}
\sigma_{Jonhson}=\sqrt{4kTR_e}
\,\,\,\,\,\,\,\,\,\,\,\,\,\,\,\,\mbox{et}\,\,\,\,\,\,\,\,\,\,\,\,\,\,\,\,R_e=\frac{R_{bolo}R_{charge}}{R_{bolo}+R_{charge}}
\label{eq:johnson}
\end{equation}
Le bruit Jonhson est \'egalement un bruit blanc.

\subsubsection{Le bruit de lecture}
\label{sec:intro_bolometrie_thermo_principe_bruit_lecture}

Le signal \'electrique au niveau du pont bolom\'etrique n'est en
principe pas directement exploitable, soit parce qu'il est trop faible
pour \^etre mesur\'e avec un simple voltm\`etre
(quelques~10$^{-9}$~V), soit parce que le bolom\`etre est trop
r\'esistif et qu'il faut adapter le circuit en imp\'edance. Quoiqu'il
en soit, il est souvent n\'ecessaire d'avoir une \'electronique de
lecture plus ou moins complexe pour pouvoir lire le signal du pont
bolom\'etrique, et le bruit g\'en\'er\'e par cette \'electronique
contribue bien s\^ur au bruit total du d\'etecteur.\\ 
\noindent Nous verrons par exemple comment le circuit de multiplexage
(section~\ref{sec:calib_procedure_explore_imped}) ou bien le mode de
lecture diff\'erentielle (section~\ref{sec:calib_perflabo_compare})
peuvent changer sensiblement le comportement des matrices de
bolom\`etres du CEA, ce qui se traduit le plus souvent par une
augmentation du niveau de bruit.

\subsubsection{Autres sources de bruit}
\label{sec:intro_bolometrie_thermo_principe_bruit_autre}

$\blacktriangleright$ \textbf{Le bruit basse fr\'equence ou Flicker
noise}\\ \indent Le bruit basse fr\'equence, souvent appel\'e bruit en
1/f \`a cause de sa d\'ependance en fr\'equence dans l'espace de
Fourier, est omnipr\'esent dans la nature~; nous le retrouvons par
exemple dans les battements cardiaques~\shortcite{kobayashi}, les
march\'es financiers~\shortcite{bonanno}, la musique et la
parole~\shortcite{voss}, les r\'esistances
\'electriques~\shortcite{voss76} ou encore les transistors
MOSFET~\shortcite{zhu}. L'origine de ce bruit d\'epend bien s\^ur du
syst\`eme consid\'er\'e mais il s'exprime toujours de la m\^eme
fa\c{c}on, par une lente d\'erive du signal qui ressort en un exc\`es
de bruit aux basses fr\'equences dans la densit\'e spectrale de
bruit.\\ Les bolom\`etres ne d\'erogent pas \`a la r\`egle, et m\^eme
s'il n'y a pas aujourd'hui de th\'eorie commun\'ement accept\'ee, le
bruit en 1/f a \'et\'e mod\'elis\'e par de nombreux auteurs
\shortcite{hooge,shklovskii,damico} dans le but de comprendre puis de
ma\^itriser la fabrication des bolom\`etres. Par exemple,
\shortciteN{buzzi_these} pr\'esente des simulations num\'eriques du
bruit en percolation dans les milieux d\'esordonn\'es o\`u la
conduction s'effectue par sauts entre impuret\'es comme c'est le cas
pour les matrices de bolom\`etres du CEA. Ce bruit peut
significativement alt\'erer les performances d'un instrument, et nous
verrons que les modes d'observation d'un t\'elescope sont souvent
adapt\'es aux caract\'eristiques du bruit en 1/f pour pouvoir corriger
les d\'erives basses fr\'equences des bolom\`etres (cf
section~\ref{sec:calib_perfobs_oof}).\\

$\blacktriangleright$ \textbf{Le bruit de courant}\\ \indent Lorsque
les contacts \'electriques entre deux composants ne sont pas de bonne
qualit\'e, le passage des \'electrons au-dessus des barri\`eres de
potentiel cr\'e\'ees aux interfaces g\'en\`ere un bruit de courant, ou
\emph{shot noise} en anglais, qui s'exprime de la fa\c{c}on suivante
d'apr\`es~\shortciteN{buzzi_these}~:
\begin{equation}
i^2=\sqrt{2qI}
\end{equation}
o\`u $q$ est la charge \'electronique et $I$ est le courant de
polarisation du d\'etecteur. Ce bruit de courant est converti en
bruit de tension aux bornes de la thermistance. Le bruit de tension
est d'autant plus grand que l'imp\'edance de la thermistance est
grande.\\

$\blacktriangleright$ \textbf{Le bruit microphonique}\\ \indent La
microphonie est g\'en\'er\'ee par la vibration m\'ecanique des
conducteurs \'electriques. Ces vibrations sont souvent dues aux pompes
utilis\'ees pour la cryog\'enie. En effet, lorsque deux fils proches
se d\'eplacent l'un par rapport \`a l'autre, des variations de
capacit\'e \'electrique apparaissent et se transforment en
fluctuations de tension ou de courant. Pour limiter les probl\`emes de
microphonie, il faut r\'eduire au maximum la longueur des c\^ables et
s'assurer qu'ils sont correctement fix\'es \`a la structure de
l'instrument.\\

$\blacktriangleright$ \textbf{Le bruit li\'e aux d\'erives de
temp\'erature du cryostat}\\ \indent Les bolom\`etres \'etant des
d\'etecteurs thermiques, ils sont tr\`es sensibles aux fluctuations de
temp\'erature de leur environnement. Or, la temp\'erature du bain
cryog\'enique peut \'eventuellement fluctuer, si un \'el\'ement de
l'instrument est mal thermalis\'e par exemple, et ceci se traduit par
des d\'erives basses fr\'equences du signal bolom\'etrique. Le bruit
li\'e aux d\'erives de temp\'erature est souvent fortement
corr\'el\'e. Une solution \'el\'egante pour monitorer et corriger ces
d\'erives en temp\'erature consiste \`a utiliser un bolom\`etre
aveugle, \cad que ce bolom\`etre est d\'edi\'e \`a la mesure de la
temp\'erature du plan focal, ind\'ependamment du flux incident. Nous
verrons dans la section~\ref{sec:detect_bolocea_elec_lecture} que le
concept initial des matrices de bolom\`etres du CEA contenait
plusieurs pixels aveugles que nous n'avons pas pu garder pour les
mod\`eles de vol du Photom\`etre PACS.\\

$\blacktriangleright$ \textbf{Le bruit li\'e aux perturbations
\'electromagn\'etiques}\\ \indent La question des perturbations
\'electromagn\'etiques et surtout des ondes radio-\'electriques est un
probl\`eme g\'en\'eral en d\'etection, notamment si l'on a des boucles
\`a haute imp\'edance comme en bolom\'etrie. Les fils reliant les
bolom\`etres aux pr\'e-amplificateurs doivent donc \^etre blind\'es et
les plus courts possible. Pour le Photom\`etre PACS, ce probl\`eme est
d'autant plus important que les panneaux solaires du satellite
Herschel sont de puissants \'emetteurs de rayonnement magn\'etique.

\section{L'av\`enement des bolom\`etres sur les grands t\'elescopes}
\label{sec:intro_bolometrie_bolo}

Nous nous int\'eressons dans cette section aux instruments
bolom\'etriques d'hier et d'aujourd'hui qui ont marqu\'e l'histoire de
l'astronomie (sub-)millim\'etrique soit par les innombrables
d\'ecouvertes scientifiques qu'ils ont permises, soit par leur r\^ole
novateur dans le domaine de la bolom\'etrie infrarouge. Nous
distinguerons deux familles de senseurs thermiques~: les bolom\`etres
r\'esistifs qui exploitent la transition m\'etal-isolant de
semiconducteurs dop\'es, et les bolom\`etres supraconducteurs qui
profitent de la forte d\'ependance en temp\'erature de certains
m\'etaux lorsqu'ils passent de l'\'etat de conducteur \`a celui de
supraconducteur. Enfin nous aborderons le sujet du couplage de
rayonnement entre le t\'elescope et les bolom\`etres en insistant sur
le potentiel observationnel des matrices de bolom\`etres qui
poss\`edent de bons facteurs de remplissage.

\subsection{Les bolom\`etres r\'esistifs}
\label{sec:intro_bolometrie_bolo_resistif}


L'histoire moderne des bolom\`etres a v\'eritablement commenc\'e avec
le d\'etecteur de \shortciteN{boyle_rodgers} qui utilisait une
r\'esistance en carbone comme thermom\`etre. L'avantage de cette
technologie est qu'elle est relativement peu on\'ereuse, facile \`a
fabriquer et que la capacit\'e calorifique du composant chute
significativement \`a basse temp\'erature. La r\'esistance jouait le
r\^ole d'absorbeur et de senseur thermique. Cependant ces
r\'esistances poss\`edent un fort bruit basse fr\'equence qui limite
les performances du bolom\`etre~; de plus le carbone est un mat\'eriau
amorphe qui ne permet pas d'atteindre des capacit\'es calorifiques
aussi faibles que celles des mat\'eriaux cristallins. Le stade suivant
du d\'eveloppement des bolom\`etres est donc l'invention de
thermom\`etres basse temp\'erature \`a base de germanium cristallin
fortement dop\'e et compens\'e. L'article de \shortciteN{low}
pr\'esente les performances sup\'erieures d'un tel d\'etecteur. Il
mesure une NEP de $5\times10^{-13}$~W/$\sqrt{\mbox{Hz}}$ \`a 2~K et
pr\'edit une NEP de $10^{-15}$~W/$\sqrt{\mbox{Hz}}$ \`a 0.5~K avec une
constante de temps de 1~ms. \`A noter \'egalement le travail de
\shortciteN{downey} qui a introduit le concept de bolom\`etres
monolithiques \`a base de silicium o\`u le thermom\`etre est
directement implant\'e dans le substrat par implantation ionique
compens\'ee. Il obtient une NEP de
$4\times10^{-16}$~W/$\sqrt{\mbox{Hz}}$ \`a 350~mK et une constante de
temps de 20~ms. Les bolom\`etres de \shortciteANP{downey} sont
d'autant plus int\'eressants que leur conception est assez proche de
celle des matrices de bolom\`etres du CEA que nous d\'ecrirons en
d\'etail dans la section~\ref{sec:detect_bolocea_principe}.\\

Aujourd'hui, la grande majorit\'e des bolom\`etres composites utilise
des thermom\`etres de type NTD~Ge (\emph{Neutron-Transmutation-Doped
Germanium}) car ils pr\'esentent une faible capacit\'e calorifique et
des niveaux de bruit peu \'elev\'es \shortcite{lange83}. La m\'ethode
de fabrication de ces thermom\`etres consiste \`a placer le germanium
dans un flux de neutrons provenant d'un r\'eacteur nucl\'eaire pendant
une dur\'ee bien pr\'ecise. Certains noyaux de $^{70}$Ge vont alors
subir une d\'ecroissance b\'eta pour devenir du $^{71}$Ge. Cet isotope
a une demi-vie de 11~jours et se transmute finalement en $^{71}$Ga qui
est un accepteur d'\'electron par rapport au germanium. D'autres
isotopes du $^{70}$Ge se transmutent en $^{75}$As qui, lui, est un
donneur d'\'electron. Le rapport des isotopes est tel que le cristal
finit par contenir un exc\`es d'accepteur d'\'electrons ce qui rend le
mat\'eriau conducteur. Le m\'ecanisme de conduction dans ce type de
mat\'eriau (semiconducteur dop\'e et compens\'e) sera d\'ecrit dans la
section~\ref{sec:detect_bolocea_fabrication_thermo}. Cette technique a
l'avantage d'\^etre reproductible et permet d'obtenir de grandes
quantit\'es de mat\'eriau dop\'e de fa\c{c}on tr\`es homog\`ene, ce
qui conf\`ere aux thermom\`etres un bruit basse fr\'equence
relativement faible. Toutefois, le taux de compensation (rapport
accepteurs/donneurs) n'est pas contr\^olable, il est en effet impos\'e
par le rapport des isotopes de germanium. De plus, ces thermom\`etres
doivent ensuite \^etre assembl\'es individuellement sur les
substrats. Cette technologie est tr\`es d\'elicate et ne permet pas la
fabrication collective de bolom\`etres.\\

La cam\'era SCUBA \shortcite{holland} poss\'edait par exemple des
thermom\`etres NTD~Ge fonctionnant \`a 100~mK. Elle \'etait
install\'ee \`a Hawaii sur le JCMT et observait dans les fen\^etres
\`a 350, 450, 750 et 850~$\mu$m. Elle a \'et\'e r\'ealis\'ee \`a
Edinburgh (UK) par la m\^eme \'equipe qui a construit UKT14
\shortcite{duncan}, le pr\'ecurseur des cam\'eras sub-millim\'etriques
actuelles. Alors que UKT14 ne poss\'edait qu'un seul pixel, SCUBA
comptait 37~bolom\`etres dans les deux bandes grande longueur d'onde
et 91~dans les deux autres. SCUBA fut le premier instrument \`a offrir
une r\'eelle capacit\'e d'imagerie dans le
sub-millim\'etrique. D'apr\`es \shortciteANP{holland}, SCUBA \'etait
dix mille fois plus rapide que UKT14 pour cartographier le
ciel. L'arriv\'ee de SCUBA a \'et\'e une v\'eritable r\'evolution pour
la communaut\'e sub-millim\'etrique et a permis de nombreuses
d\'ecouvertes. Par exemple, \shortciteN{hughes} ont observ\'e le
Hubble Deep Field pendant 51 heures et ont d\'ecouvert une population
de galaxies distantes extr\^emement brillantes. L'\'etude des
m\'ecanismes de formation d'\'etoiles \shortcite{kirk} ou de disques
de poussi\`ere \shortcite{wyatt} a \'egalement profit\'e des
performances de SCUBA.\\

Au milieu des ann\'ees~90, l'\'equipe du GSFC (Goddard Space Flight
Center) a d\'evelopp\'e des d\'etecteurs bolom\'etriques de type
\emph{pop-up} pour la cam\'era SHARC \shortcite{wang}. Elle \'etait
install\'ee sur le t\'elescope CSO et observait \`a 350~et
450~$\mu$m. La thermom\'etrie est r\'ealis\'ee par silicium implant\'e
phosphore et compens\'e bore, comme pour les matrices de bolom\`etres
du CEA (cf section~\ref{sec:detect_bolocea_fabrication_thermo} pour
plus de d\'etails). La concentration de dopant est choisie de sorte
que l'on obtienne une imp\'edance de 10~M$\Omega$ \`a 300~mK. Le plan
focal est compos\'e de bolom\`etres monolithiques planaires qui se
pr\'esentent sous la forme d'une barrette lin\'eaire de
24~pixels. Tous ces pixels sont manufactur\'es simultan\'ement \`a
partir d'une m\^eme plaque de silicium par des techniques de
micro-lithographie. L'absorption est r\'ealis\'ee par une fine couche
de bismuth d\'epos\'ee sur le substrat en silicium. La NEP est de
5$\times10^{-15}$~W/$\sqrt{\mbox{Hz}}$. En 2002, SHARC est remplac\'e
par SHARC~II \shortcite{dowell,silverberg} qui repose sur la m\^eme
m\'ethode de fabrication mais qui contient 12 barrettes de 32~pixels
(85~\% des pixels sont fonctionnels). L'id\'ee originale du GSFC pour
pallier au probl\`eme d'encombrement des longues poutres qui relient
l'absorbeur \`a la structure a \'et\'e de les plier
perpendiculairement au plan des bolom\`etres. De cette fa\c{c}on les
barrettes ont pu \^etre assembl\'ees de mani\`ere contig\"ue pour
obtenir un bon facteur de remplissage du plan focal. De plus, les
bolom\`etres de SHARC~II sont plac\'es dans une cavit\'e quart-d'onde
pour augmenter l'absorption du rayonnement (ce syst\`eme d'absorption
a \'et\'e introduit par le CEA pour ses matrices de bolom\`etres, nous
y reviendrons dans la
section~\ref{sec:detect_bolocea_fabrication_cavite}). Bas\'e sur la
m\^eme technologie, le GSFC a construit HAWC, un photom\`etre
optimis\'e pour observer de 50 \`a 250~$\mu$m, il est destin\'e \`a
l'observatoire SOFIA \shortcite{harper} qui devrait bient\^ot prendre
son envol.\\

Le groupe du Max-Planck-Institut f\"ur Radioastronomie de Bonn
poss\`ede \'egalement une longue exp\'erience dans la fabrication de
bolom\`etres infrarouges. Ils ont entre autre fourni \`a l'IRAM la
cam\'era MAMBO \shortcite{kreysa98} qui poss\`ede 117~bolom\`etres et
observe \`a 1.3~mm au Pico Veleta. Tr\`es r\'ecemment, le groupe a
effectu\'e avec succ\`es la v\'erification scientifique de leur
dernier instrument LABOCA \shortcite{kreysa} sur le t\'elescope
APEX. Cette cam\'era contient 295~bolom\`etres, dont 16~\% sont
d\'efectueux, elle fonctionne \`a 300~mK et utilise des thermom\`etres
NTD~Ge. Sa sensibilit\'e \`a 870~$\mu$m d\'epend du mode d'observation
mais vaut environ 75~mJy/$\sqrt{\mbox{Hz}}$.\\

Du c\^ot\'e fran\c{c}ais, le CEA s'est lanc\'e en 1995 dans le
d\'eveloppement de matrices de bolom\`etres pour l'infrarouge
lointain. Ces matrices sont innovantes sur de nombreux points,
notamment sur la fabrication collective des bolom\`etres, le circuit
de lecture multiplex\'e ou encore l'utilisation de cavit\'es
r\'esonantes pour optimiser l'absorption du rayonnement. Nous
reviendrons en d\'etails sur ces d\'etecteurs dans le
chapitre~\ref{chap:detect_bolocea}. Notez \'egalement la mise en place
d'une collaboration \`a l'\'echelle nationale, le DCMB
(\emph{D\'eveloppement Concert\'e de Matrices de Bolom\`etres}), dont
l'objectif est d'exploiter les comp\'etences de plusieurs laboratoires
de recherche fran\c{c}ais et de d\'evelopper conjointement des
matrices de bolom\`etres d'une nouvelle g\'en\'eration pour
l'astronomie (sub-)millim\'etrique. Le groupe est men\'e par Alain
Beno\^it, il rassemble dix laboratoires situ\'es \`a Grenoble, en
r\'egion parisienne et \`a Toulouse. Les travaux de recherche sont
financ\'es par le CNRS et le CNES. Les th\'ematiques scientifiques
vis\'ees par le DCMB concernent principalement la mesure de la
polarisation du CMB, avec le projet spatial SAMPAN/Bpol
\shortcite{desert}, ainsi que la mesure de l'effet Sunyaev-Zeldovich
pour le projet stratosph\'erique Olimpo \shortcite{masi}. Le groupe a
d\'ej\`a r\'ealis\'e des matrices prototypes contenant jusqu'\`a 200
pixels, les thermom\`etres employ\'es peuvent \^etre de type
r\'esistif \`a tr\`es haute imp\'edance, ou de type
supraconducteur. Les thermom\`etres sont r\'ealis\'es par d\'ep\^ot de
couches minces de NbSi. Ce mat\'eriau se comporte comme un isolant
d'Anderson ou comme un supraconducteur suivant la stochiom\'etrie des
deux \'el\'ements. L'\'electronique de lecture est alors adapt\'ee au
type de thermom\`etres utilis\'es, des SQUID pour les TES et des
QPC-HEMT pour les r\'esistifs.\\


\begin{figure}
  \begin{center}
    \begin{tabular}[t]{ll}
      \includegraphics[height=0.26\textheight]{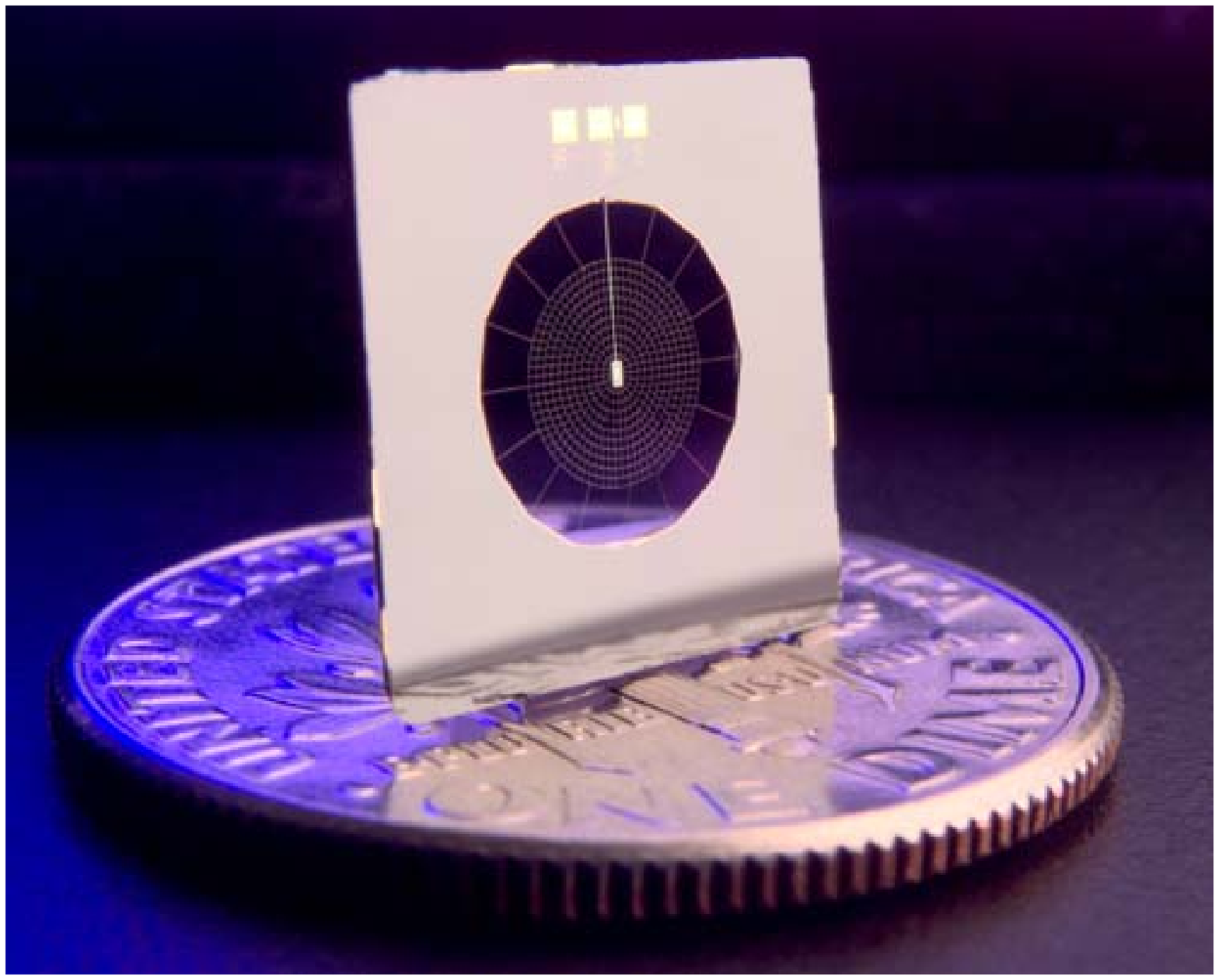} & \includegraphics[height=0.26\textheight]{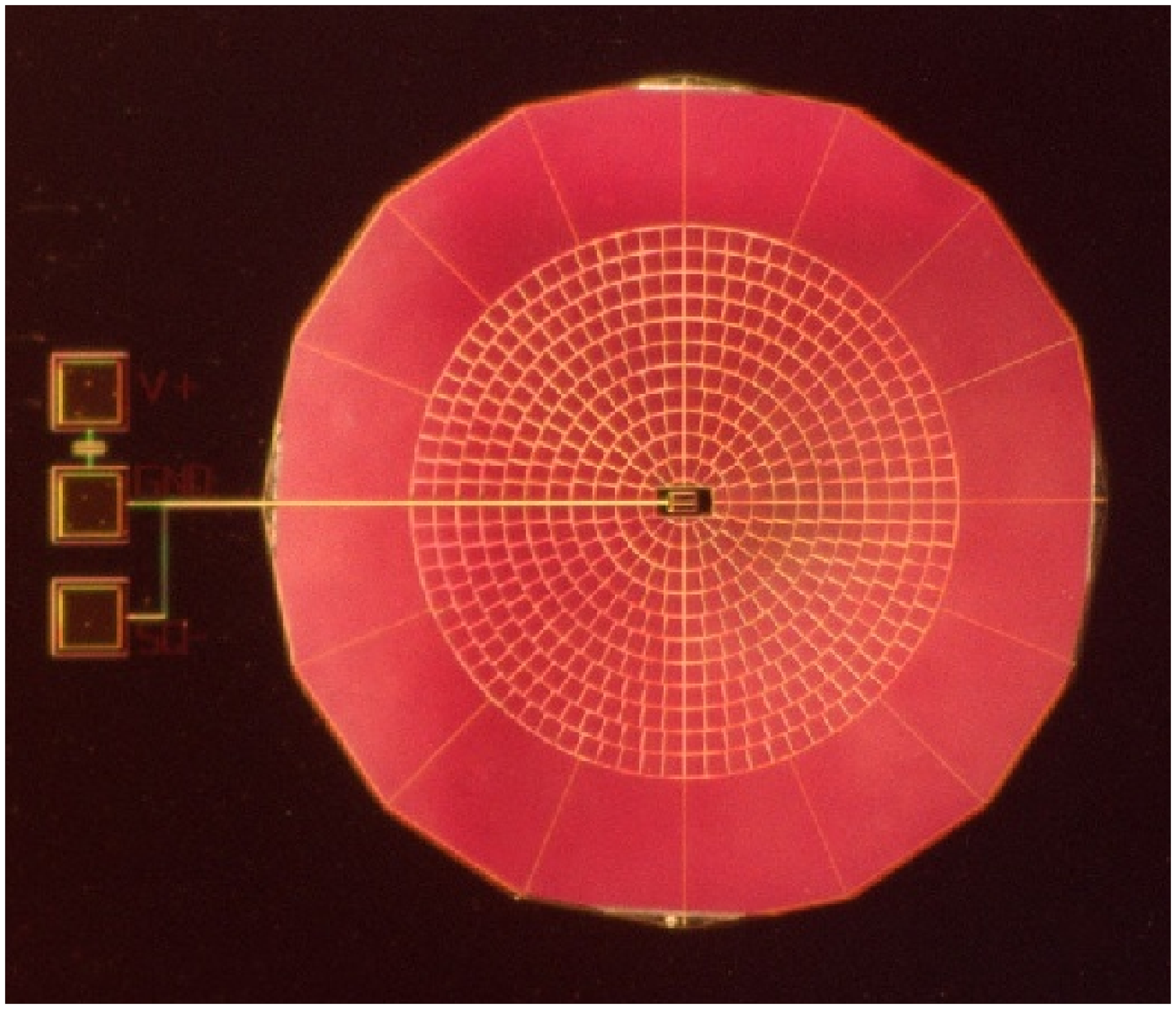} 
    \end{tabular}
  \end{center}
  \caption[Bolom\`etres \emph{spiderweb} du JPL]{Photographies de
  bolom\`etres de type \emph{spiderweb} con\c{c}us par le JPL. La
  masse suspendue est r\'eduite au maximum abaissant ainsi la
  capacit\'e calorifique de la grille absorbante. Le thermom\`etre est
  coll\'e au centre de la \emph{toile} et est connect\'e
  \'electriquement au circuit de lecture par une piste m\'etallis\'ee
  visible sur les deux photos. (lien www.planck.fr/ et
  www2.jpl.nasa.gov/)
  \label{fig:intro_bolometrie_bolo_resistif_spiderweb}}
\end{figure}

Enfin, nous terminons ce tour d'horizon non-exhaustif des bolom\`etres
r\'esistifs par les d\'etecteurs les plus r\'epandus actuellement en
bolom\'etrie infrarouge. Il s'agit des bolom\`etres \emph{spiderweb}
du JPL/Caltech (\emph{Jet Propulsion Laboratory/California Institute
of Technology}), initialement con\c{c}us par J. Bock
\shortcite{bock,turner}. La
figure~\ref{fig:intro_bolometrie_bolo_resistif_spiderweb} montre deux
photographies de ces bolom\`etres. Le pixel est constitu\'e d'une
grille \`a g\'eom\'etrie circulaire, en forme de toile d'araign\'ee,
sur laquelle est d\'epos\'ee une fine couche de bismuth pour absorber
le rayonnement (sub-)millim\'etrique. La structure de la grille
pr\'esente de nombreux avantages, notamment une faible masse (donc une
faible capacit\'e calorifique), une faible conductance thermique,
ainsi qu'une faible section efficace g\'eom\'etrique aux particules
cosmiques qui perturbent les signaux bolom\'etriques. La grille est
r\'ealis\'ee en Si$_3$N$_4$ par des techniques de micro-usinage
d\'ej\`a utilis\'ees en micro-\'electronique. Un thermom\`etre NTD~Ge
est plac\'e au centre de la grille. Initialement coll\'es par \'epoxy,
les thermom\`etres sont maintenant soud\'es \`a l'indium sur un
d\'ep\^ot Cr-Au. Les connections \'electriques sont r\'ealis\'ees par
des fils en NbTi qui contribuent \`a la conductance thermique. Les
performances de ces d\'etecteurs en terme de sensibilit\'e sont
excellentes~: NEP$=1.2\times10^{-17}$~W/$\sqrt{\mbox{Hz}}$ \`a 300~mK
et $2\times10^{-18}$~W/$\sqrt{\mbox{Hz}}$ \`a 100~mK. Ces bolom\`etres
\'equipent de nombreux instruments comme par exemple BOLOCAM sur le
CSO qui observe \`a 1.1, 1.4 et 2.1~mm avec ses 144~pixels
\shortcite{glenn}. Caltech devrait livrer une cam\'era quasi-identique
\`a BOLOCAM pour le LMT au Mexique. Les projets ACBAR au p\^ole sud,
BOOMERANG, MAXIMA et ARCHEOPS en ballon, Planck/HFI
\shortcite{lamarre} et Herschel/SPIRE \shortcite{griffin} dans
l'espace utilisent \'egalement des bolom\`etres spiderweb.

\subsection{Les bolom\`etres supraconducteurs}
\label{sec:intro_bolometrie_bolo_supra}

Les premi\`eres r\'ealisations de bolom\`etres supraconducteurs datent
des ann\'ees~40. Par exemple \shortciteN{andrews} de l'Universit\'e
Johns Hopkins dans le Maryland ont fabriqu\'e un bolom\`etre
supraconducteur en utilisant une fine bandelette de CbN maintenue \`a
$\sim$14.36~K autour de sa phase de transition. Malgr\'e de bonnes
performances en terme de sensibilit\'e et de rapidit\'e, ce type de
d\'etecteur ne s'est pas vraiment d\'evelopp\'e \`a cette \'epoque car
il \'etait relativement difficile \`a fabriquer et \`a maintenir dans
sa zone de transition. D'apr\`es \shortciteN{zwerdling} les
bolom\`etres en germanium d\'ecrit par \shortciteN{low} \'etaient bien
plus pratiques que les bolom\`etres supraconducteurs et suffisamment
performants. \c{C}a n'est que tr\`es r\'ecemment que les mat\'eriaux
supraconducteurs ont repris le pas sur les thermom\`etres NTD~Ge en
terme de sensibilit\'e et de rapidit\'e. Les bolom\`etres dont la
thermom\`etrie est bas\'ee sur des mat\'eriaux supraconducteurs sont
appel\'es des TES (\emph{Transition-Edge Sensors}).\\

La r\'eponse d'un bolom\`etre est bien souvent exprim\'ee par le
coefficient de temp\'erature $\alpha$~; nous avons trouv\'e plusieurs
d\'efinitions dans la litt\'erature, mais nous adoptons pour le moment
la convention $\alpha=\frac{1}{R}\frac{dR}{dT}$. Notez que ce
param\`etre renseigne en effet sur la capacit\'e qu'a un thermom\`etre
\`a changer d'imp\'edance lorsque sa temp\'erature \'evolue.  Les
m\'etaux supraconducteurs pr\'esentent des coefficients~$\alpha$ qui
peuvent d\'epasser de plusieurs ordres de grandeur ceux des
bolom\`etres r\'esistifs lorsqu'ils passent de l'\'etat normal, \cad
conducteur, \`a celui de supraconducteur. Cependant, la plage de
temp\'erature sur laquelle le param\`etre~$\alpha$ est
particuli\`erement int\'eressant est g\'en\'eralement tr\`es
\'etroite. D'apr\`es \shortciteN{richards}, cette plage vaut environ
l'inverse du coefficient~$\alpha$ tel que d\'efinit
pr\'ec\'edemment. Il est donc en th\'eorie possible d'obtenir des
$\alpha$ tr\`es grands, i.e. des bolom\`etres tr\`es sensibles, \`a
condition de pouvoir maintenir le m\'etal supraconducteur dans sa
phase de transition. Les bolom\`etres modernes utilisent le principe
de contre-r\'eaction \'electro-thermique pour garder le mat\'eriau \`a
sa temp\'erature de transition. Le principe de cette contre-r\'eaction
est de r\'eduire la dissipation \'electrique dans le thermom\`etre
lorsque sa temp\'erature s'\'el\`eve. Cette condition est r\'ealis\'ee
pour des thermom\`etres polaris\'es en tension, nous parlons alors de
d\'etecteurs VSB (\emph{Voltage-biased Supraconductor Bolometers}). En
effet, lorsque le flux incident chauffe le bolom\`etre, le
supraconducteur s'\'eloigne de sa transition et l'imp\'edance augmente
tr\`es rapidement, la dissipation Joule $P=V^2/R$ diminue et
contre-balance l'apport d'\'energie radiative. Le bolom\`etre retourne
alors rapidement \`a sa temp\'erature d'\'equilibre autour de sa
transition de phase. Les fortes contre-r\'eactions permettent en plus
d'obtenir des bolom\`etres tr\`es rapides. Par exemple, en
rempla\c{c}ant le traditionnel thermom\`etre NTD~Ge par un film de
titane sur un bolom\`etre de type spiderweb, \shortciteN{lee97} ont
mesur\'e des constantes de temps de 5~ms \`a 300~mK pour une NEP de
$1.1\times10^{-17}$~W/$\sqrt{\mbox{Hz}}$.

Dans leur r\'egime de fonctionnement, les thermom\`etres TES
poss\`edent des imp\'edances bien inf\'erieures \`a 1~$\Omega$ ce qui
rend les d\'etecteurs inadapt\'es aux circuits de lecture
conventionnels \`a base de transistors FET. Pour lire le signal
bolom\'etrique, le TES est mont\'e en s\'erie avec une source de
tension et une bobine supraconductrice~; lorsque l'imp\'edance du TES
change, le courant dans la bobine varie et g\'en\`ere un faible champ
magn\'etique qui est d\'etect\'e par un SQUID (\emph{Superconducting
Quantum Interference Device}). Un SQUID est en effet un
magn\'etom\`etre extr\^emement sensible qui fonctionne \`a tr\`es
basse temp\'erature et qui dissipe tr\`es peu d'\'energie. Tous les
bolom\`etres TES utilisent des SQUID pour leur circuit de lecture
\shortcite{irwin}.\\

Il existe aujourd'hui un v\'eritable engouement pour les bolom\`etres
supraconducteurs car ils offrent d'excellentes performances, mais
\'egalement parce qu'ils sont compatibles avec des m\'ethodes de
fabrication collective. C'est d'ailleurs l'objectif que s'est fix\'ee
l'\'equipe d'Edinburgh, en collaboration avec d'autres laboratoires
britanniques, canadiens et am\'ericains, de fabriquer une cam\'era
sub-milim\'etrique \`a base de bolom\`etres TES contenant pas moins de
10000~pixels. Ce projet colossal et ambitieux porte le nom de SCUBA2
\shortcite{holland06}, digne successeur de SCUBA. Cet instrument
devrait \^etre install\'e sur le t\'elescope JCMT pour observer \`a
450 et 850~$\mu$m. Les bolom\`etres, fournis par le NIST, sont
refroidis \`a 100~mK et poss\`edent une NEP de
$\sim1\times10^{-16}$~W/$\sqrt{\mbox{Hz}}$ \`a 450~$\mu$m et
$\sim3\times10^{-17}$~W/$\sqrt{\mbox{Hz}}$ \`a 850~$\mu$m. Les TES
sont form\'es d'un film bi-couche de Mo/Cu dont la transition
s'\'etale sur seulement 1-2~mK \shortcite{audley} pour permettre une
grande sensibilit\'e. Les bolom\`etres sont hybrid\'es au circuit de
lecture par billes d'indium tout comme les photoconducteurs d\'ecrits
dans la section~\ref{sec:intro_astro_IR_detecteur}. La lecture du
signal est multiplex\'ee par des SQUID qui se trouvent juste sous les
bolom\`etres. Cette cam\'era est tr\`es attendue par la communaut\'e
des astronomes car elle devrait \^etre 1000~fois plus rapide que SCUBA
pour cartographier le ciel. Toutefois, comme tout projet ambitieux, la
livraison de l'instrument a d\'ej\`a pris du retard (probl\`eme de
blindage magn\'etique, \'evolution du design des d\'etecteurs,
etc...), la mise en service de l'instrument complet aura certainement
lieu courant~2008.

Le NIST d\'eveloppe \'egalement d'autres d\'etecteurs en collaboration
avec le GSFC, les BUG (\emph{Backshort-Under-Grids arrays}), qui
reposent aussi sur une fabrication collective de milliers de
bolom\`etres multiplex\'es par des SQUID
\shortcite{allen}. L'instrument GISMO par exemple est destin\'e au
t\'elescope de 30~m de l'IRAM, il poss\`ede $8\times16$ bolom\`etres
TES qui fonctionnent \`a 2~mm et offre une NEP de
$4\times10^{-17}$~W/$\sqrt{\mbox{Hz}}$
\shortcite{staguhn}. \shortciteN{benford} montrent les premi\`eres
images astronomiques obtenues avec un d\'etecteur bolom\`etrique
supraconducteur multiplex\'e. Ils pr\'esentent des observations \`a
350~$\mu$m de Jupiter, Venus et d'une r\'egion de formation d'\'etoile
G34.3+0.2. L'instrument est install\'e au CSO, il s'appelle FIBRE et a
\'egalement \'et\'e construit par la m\^eme \'equipe NIST/GSFC.

Le groupe de Berkeley men\'e par Paul Richards et Adrian Lee est
\'egalement tr\`es prolifique en mati\`ere de bolom\`etres TES. C'est
d'ailleurs l'\'equipe qui a mis au point la th\'eorie des VSB
\shortcite{lee96,lee98}. Ils ont de plus d\'evelopp\'e la technique de
multiplexage fr\'equentiel \shortcite{lanting}. L'id\'ee est
d'associer \`a chaque TES une bobine dont l'inductance varie
l\'eg\`erement d'un pixel \`a l'autre, chaque bolom\`etre poss\`ede
alors une fr\'equence propre diff\'erente de celle des autres
pixels. Ces bolom\`etres sont mont\'es en s\'erie, \cad sur un seul
fil, ils sont aliment\'es par une onde porteuse qui contient la
fr\'equence de tous les pixels, ce signal est ensuite lu par un seul
SQUID et d\'emodul\'e par une \'electronique chaude pour extraire le
courant qui circule dans chacun des bolom\`etres. Cette technique
permet \`a pr\'esent de multiplexer 8~pixels~; la limite aux basses
fr\'equences est fix\'ee par la constante de temps des bolom\`etres,
celle aux hautes fr\'equences est d\'etermin\'ee par la constante de
temps des c\^ables qui relient les bolom\`etres \`a l'\'electronique
chaude. L'equipe de Berkeley a r\'ecemment r\'ealis\'e avec succ\`es
la v\'erification scientifique de leur cam\'era APEX-SZ au Chili. Elle
contient 300~bolom\`etres de type spiderweb avec des thermom\`etres
TES fonctionnant \`a $\sim$300~mK \shortcite{dobbs}. L'objectif de cet
instrument est de d\'ecouvrir et d'\'etudier des amas de galaxies dans
le domaine millim\'etrique en utilisant l'effet Sunyaev Zel'dovich. Un
instrument similaire, mais contenant pr\`es de 1000~bolom\`etres, a
\'egalement \'et\'e install\'e sur l'antenne de 10~m du SPT.\\
\begin{figure}
  \begin{center}
      \includegraphics[width=1.\textwidth,angle=0]{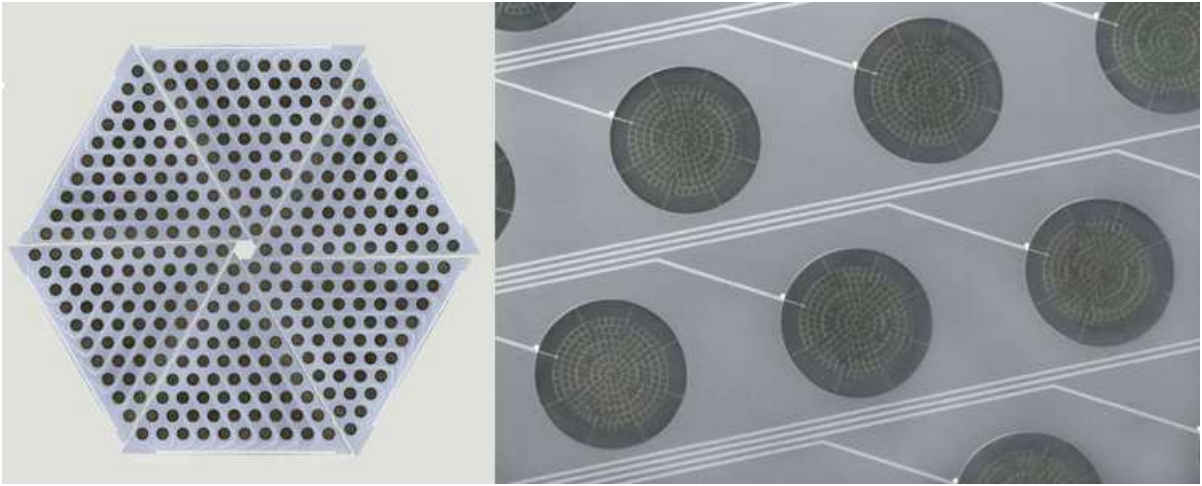}
  \end{center}
  \caption[Bolom\`etres piderweb avec thermom\`etre TES pour
  APEX-SZ]{Ces deux photographies montrent le plan focal de APEX-SZ
  ainsi qu'une vue rapporch\'ee d'un des pixels. Nous voyons
  300~bolom\`etres \`a thermom\'etrie TES dont le signal est
  multiplex\'e dans le domaine fr\'equentiel (8 -> 1). \`A droite,
  nous retrouvons la g\'eom\'etrie spiderweb de la grille suspendue et
  les lignes \'electriques grav\'ees dans le silicium. (lien
  http$\!$://bolo.berkeley.edu/)
  \label{fig:intro_bolometrie_bolo_supra}}
\end{figure}

\subsection{Couplage optique des bolom\`etres et remplissage du plan focal}
\label{sec:intro_bolometrie_bolo_matrice}

\begin{figure}
  \begin{center}
    \begin{tabular}[t]{ll}
      \includegraphics[height=0.215\textheight]{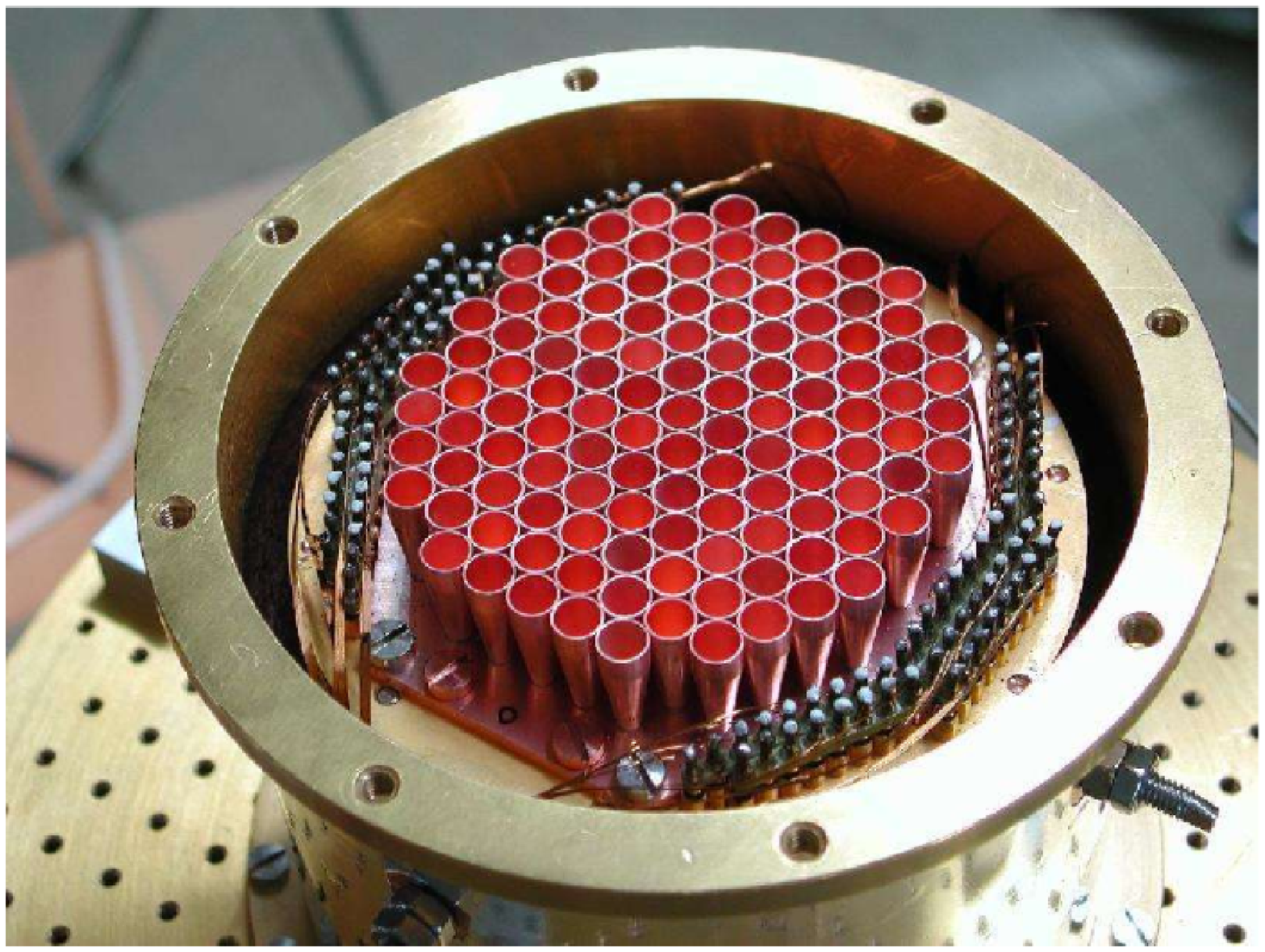} & \includegraphics[height=0.215\textheight]{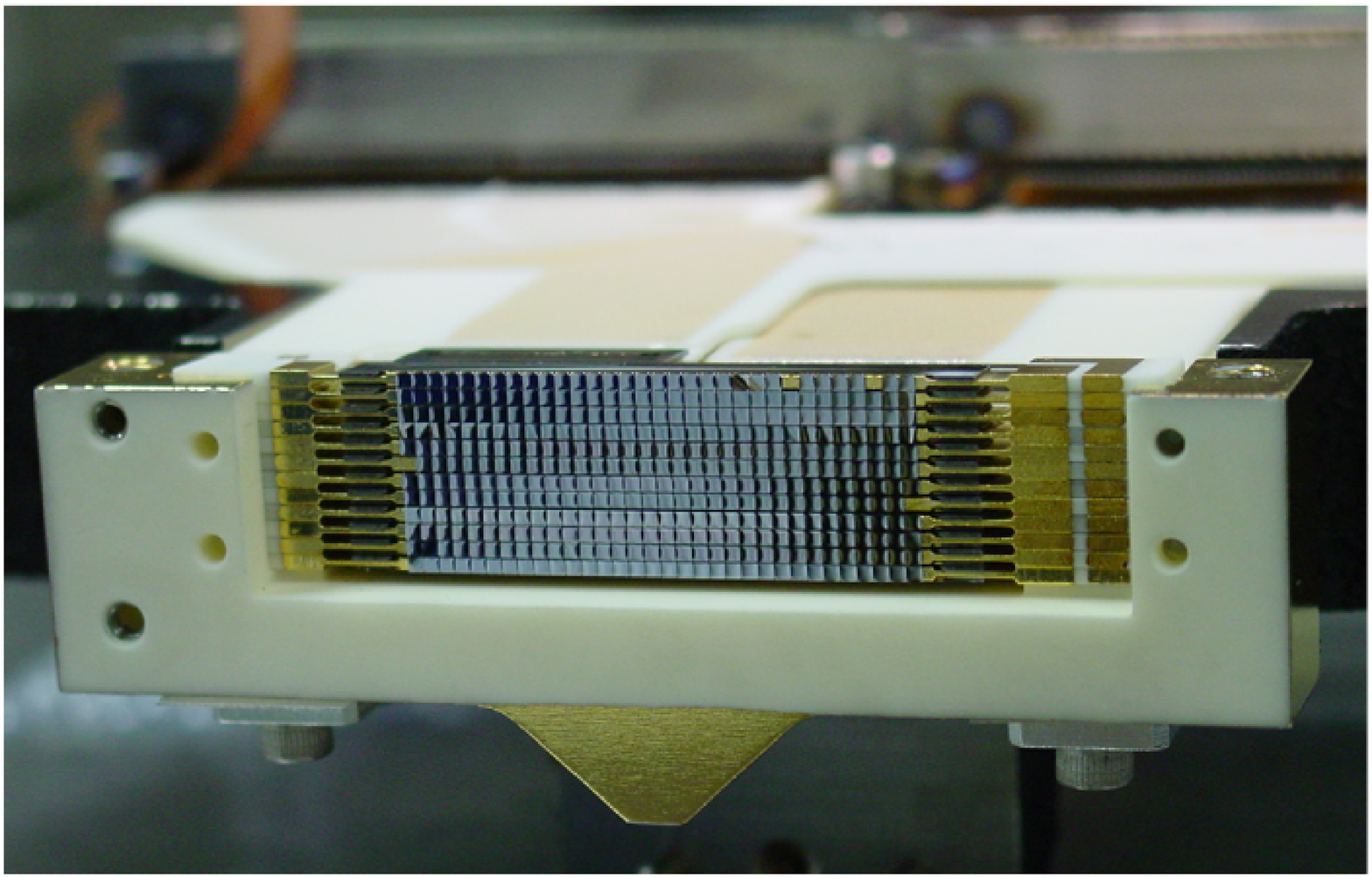} 
    \end{tabular}
  \end{center}
  \caption[Bolom\`etres avec et sans c\^ones de Winston]{Photographies
  des plans focaux de MAMBO-II, \`a gauche, et de SHARC-II, \`a
  droite. Les c\^ones de Winston sont dispos\'es de fa\c{c}on
  hexagonale, deux pixels adjacents sont s\'epar\'es de 2F$\lambda$
  pour MAMBO-II. Les matrices de bolom\`etres nus poss\`edent en
  revanche des pixels carr\'es, contigus et espac\'es de
  0.65F$\lambda$ pour SHARC-II~; ils offrent un facteur de remplissage
  proche de~1. (lien www.mpifr-bonn.mpg.de et www.submm.caltech.edu)
  \label{fig:intro_bolometrie_bolo_matrice_cones}}
\end{figure}

Une fois les performances \'electro-thermiques d'un bolom\`etre
fix\'ees, un autre aspect de la d\'etection rentre en jeu~: c'est
l'absorption du rayonnement \'electromagn\'etique. Pour les
bolom\`etres modernes dont le substrat est g\'en\'eralement du
silicium transparent dans l'infrarouge lointain, il est n\'ecessaire
de d\'eposer une couche de m\'etal sur ce substrat pour absorber
l'\'energie radiative incidente. L'imp\'edance de cet absorbeur doit
\^etre adapt\'ee \`a l'imp\'edance du vide pour \'eviter les
r\'eflections ind\'esirables. Dans ce cas, l'absorption peut en
th\'eorie atteindre~50~\%. Il existe plusieurs mani\`eres d'optimiser
l'absorption du rayonnement. La plus r\'epandue est celle des
concentrateurs de lumi\`ere. En effet, \`a l'exception des
bolom\`etres popup du GSFC, tous les d\'etecteurs actuellement en
op\'eration sur les grands t\'elescopes (sub-)millim\'etriques
utilisent ces concentrateurs pour optimiser le couplage du rayonnement
entre les bolom\`etres et le t\'elescope. Ils se composent de deux
\'el\'ements~: un cornet et une sph\`ere int\'egratrice. Le cornet
(\emph{feedhorn} en anglais), souvent appel\'e c\^one de Winston du
nom de la personne qui les a con\c{c}u \shortcite{winston}, joue le
r\^ole d'un guide d'onde. Il d\'efinit le champ de vue du d\'etecteur,
\cad que les rayons lumineux dont l'angle d'incidence est sup\'erieur
\`a une valeur fix\'ee par la g\'eom\'etrie du c\^one n'atteignent pas
le bolom\`etre. Le diagramme de rayonnement d'un c\^one est
approximativement gaussien. Les autres rayons sont transmis vers une
sph\`ere qui se situe \`a la pointe du c\^one et qui contient le
bolom\`etre. Le r\^ole de cette sph\`ere int\'egratrice est de
pi\'eger les rayons luminueux afin de multiplier le nombre
d'interactions entre le bolom\`etre et le champ de radiation. Les
parois internes de la sph\`ere sont r\'efl\'echissantes, les rayons
qui y p\'en\`etrent ne peuvent plus en ressortir, ils sont
in\'evitablement absorb\'es par le bolom\`etre. L'utilisation de ces
c\^ones pour les d\'etecteurs infrarouges et (sub-)millim\'etriques a
\'et\'e \'etudi\'ee en d\'etail par
\shortciteN{hildebrand82}. L'efficacit\'e des c\^ones de Winston
d\'epend principalement de leur diam\`etre. \shortciteN{griffin02}
d\'efinissent l'efficacit\'e d'ouverture comme \'etant le rapport
entre la puissance de la PSF (\emph{Point Spread Function}) et la
puissance intercept\'ee par un bolom\`etre. Ils trouvent que les
c\^ones de Winston poss\`edent un maximum d'efficacit\'e de~0.7 pour
un diam\`etre de 2F$\lambda$, o\`u $\lambda$ est la longueur d'onde et
F est le rapport focal de l'instrument (F=f/D avec f la longueur
focale et D le diam\`etre du t\'elescope). Un diam\`etre de
2F$\lambda$ permet donc un couplage optimal entre le t\'elescope et le
bolom\`etre, la plupart des d\'etecteurs poss\`ede d'ailleurs des
c\^ones de cette taille l\`a. D'apr\`es \shortciteANP{griffin02}, les
principaux avantages \`a utiliser des c\^ones avec sph\`eres
int\'egratrices sont~:
\begin{itemize}
\item Les c\^ones \`a 2F$\lambda$ offrent les meilleures performances
pour d\'etecter des sources ponctuelles dont la position est connue,
\cad que presque toute la puissance d'une source est collect\'ee sur
un m\^eme d\'etecteur.
\item Les propri\'et\'es sont bien comprises et la technique de
fabrication est ma\^itris\'ee.
\item Le champ de vue des bolom\`etres est parfaitement d\'efini,
les c\^ones offrent donc une tr\`es bonne r\'ejection de la lumi\`ere
parasite.
\item La structure c\^one+sph\`ere agit comme une cage de Faraday, les
bolom\`etres sont donc peu susceptibles aux perturbations
\'electromagn\'etiques.
\end{itemize}
Les principaux inconv\'enients sont~:
\begin{itemize}
\item Jusqu'\`a 70~\% de la puissance radiative incidente au niveau du
plan focal n'est pas absorb\'ee par les bolom\`etres, soit parce
qu'elle est r\'efl\'echie par les c\^ones et repart vers l'entr\'ee de
l'instrument, soit parce qu'elle passe entre des c\^ones adjacents
(facteur de remplissage modeste, cf
figure~\ref{fig:intro_bolometrie_bolo_matrice_cones}). En particulier,
l'efficacit\'e d'un c\^one \`a 2F$\lambda$ peut consid\'erablement
chuter pour une source ponctuelle si sa position n'est pas connue~; la
PSF peut en effet tomber entre deux c\^ones et une large fraction de
la puissance lumineuse serait alors perdue.
\item L'efficacit\'e de cartographie est limit\'ee par l'utilisation
de c\^ones de Winston. Nous reviendrons sur ce point de fa\c{c}on plus
quantitative dans la suite du texte.
\item Pour des diam\`etres de c\^ones sup\'erieurs \`a 0.5F$\lambda$,
les bolom\`etres sont trop espac\'es et ne permettent pas
d'\'echantillonner le ciel au crit\`ere de Nyquist. Des techniques
d'observation peu efficaces sont alors n\'ecessaires pour
reconstruire des cartes du ciel correctement \'echantillonn\'ees. Par
exemple pour des c\^ones \`a 2F$\lambda$, il faut effectuer
16~pointages diff\'erents, m\^eme pour cartographier une r\'egion plus
petite que le champ de vue de la cam\'era, nous parlons
d'observations en mode \emph{jiggling}.  La cartographie du ciel par
balayage permet \'egalement d'obtenir des images correctement
\'echantillonn\'ees \`a condition de choisir un angle de balayage
appropri\'e \`a la g\'eom\'etrie du plan focal.
\end{itemize}
J'ajouterai aux inconv\'enients list\'es par \shortciteANP{griffin02}
que l'utilisation de c\^ones de Winston n'est pas compatible avec la
fabrication de grands plans focaux contenant des milliers de
bolom\`etres. Il est en effet m\'ecaniquement et thermiquement
difficile de fabriquer un instrument qui contiendrait des milliers de
c\^ones refroidis \`a 300 voire 100~mK. De plus, les c\^ones \`a
2F$\lambda$ \'etant relativement volumineux, aucun plan focal ne
pourrait accueillir un d\'etecteur de $\sim$500F$\lambda$ de
c\^ot\'e.\\

\begin{figure}
  \begin{center}
      \includegraphics[height=0.3\textheight]{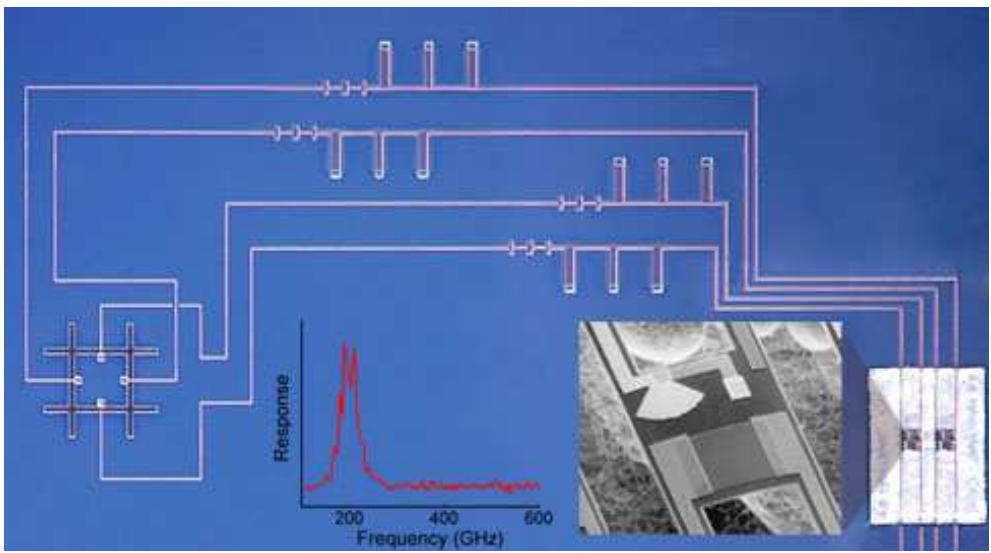}
  \end{center}
  \caption[Bolom\`etres \`a antennes de Berkeley pour l'instrument
  PolarBear]{Exemple de bolom\`etre \`a antennes. Pixel prototype pour
  l'instrument PolarBear. L'antenne est le motif en forme de \# en bas
  \`a gauche de l'image. Les quatres branches de l'antenne sont
  connect\'ees \`a deux bolom\`etres (en bas \`a droite) par des
  microstrips. Les microstrips poss\`edent des filtres (barres
  perpendiculaires en haut de l'image) qui s\'electionnent la
  fr\'equence du signal. La courbe au centre montre la transmission du
  syst\`eme. L'encadr\'e en bas \`a droite montre une vue rapproch\'ee
  du bolom\`etre qui mesure la polarisation horizontale. (lien
  http$\!$://bolo.berkeley.edu/)
  \label{fig:intro_bolometrie_bolo_matrice_antenne}}
\end{figure}

Remarquez n\'eanmoins qu'il est aujourd'hui possible de se passer de
c\^ones de Winston pour coupler le t\'elescope aux bolom\`etres et
d'optimiser l'absorption du rayonnement. Deux voies technologiques ont
\'et\'e explor\'ees et ont montr\'e d'excellents r\'esultats~: les
cavit\'es r\'esonantes et les bolom\`etres \`a antennes. Le principe
de cavit\'e r\'esonante a \'et\'e introduit par le CEA au milieu des
ann\'ees~90, il est aujourd'hui utilis\'e par SHARC-II et offre des
absorptions proches de 100~\%. Les instruments PACS, ARTEMIS, HAWC ou
SCUBA-2 exploiteront \'egalement les performances des cavit\'es
r\'esonantes. Le but de ces cavit\'es est de s\'electionner la bande
de longueur d'onde qui est efficacement absorb\'ee par le
bolom\`etre. Toutefois, elles n'offrent aucune directivit\'e, le
rayonnement est absorb\'e sur 2$\pi$ st\'eradians. Il est donc
indispensable de placer un diaphragme refroidi (\emph{cold stop})
au-dessus du plan focal pour d\'efinir le champ de vue des
bolom\`etres. Le diagramme de rayonnement d'un tel dispositif est
approximativement une fonction porte, \cad une transmission de~1 en
direction du t\'elescope et~0 ailleurs. Compar\'e au profil gaussien
des c\^ones, les bolom\`etres nus (\emph{bare arrays} dans
\shortciteANP{griffin02}) sont plus efficacement illumin\'es par la
partie externe du t\'elescope~; c'est la raison pour laquelle, \`a une
taille donn\'ee, l'efficacit\'e d'ouverture des \emph{bare arrays}
avec pixels carr\'es est toujours sup\'erieure \`a celle des
\emph{feedhorn-coupled arrays} (cf fig.2 dans
\shortciteANP{griffin02}). Nous d\'ecrirons en d\'etail le principe
d'absorption des cavit\'es r\'esonantes dans la
section~\ref{sec:detect_bolocea_fabrication_cavite}. D'autre part, les
bolom\`etres \`a antennes d\'etectent le rayonnement
\'electromagn\'etique de mani\`ere similaire aux d\'etecteurs
h\'et\'erodynes, \cad que les \'electrons libres \`a l'int\'erieur
d'une antenne sont excit\'es \`a la m\^eme fr\'equence que le champ
\'electrique de l'onde incidente (cf
section~\ref{sec:intro_astro_IR_detecteur}). La g\'eom\'etrie de
l'antenne d\'efinit le champ de vue du pixel et permet en plus
d'effectuer des mesures de polarisation. Les fluctuations de courant
induites par le champ de radiation sont transmises le long de lignes
\'electriques appel\'ees \emph{microstrips}\footnote{Les microstrips
sont des sortes de c\^ables coaxiaux \`a 2-dimensions grav\'es sur
circuit imprim\'e.} \shortcite{dunlop} vers un \'el\'ement dispersif,
g\'en\'eralement une r\'esistance. L'\'energie lib\'er\'ee est alors
mesur\'ee par un bolom\`etre (voir
figure~\ref{fig:intro_bolometrie_bolo_matrice_antenne}). De plus, il
est possible d'ajouter des filtres le long de la ligne de transmission
et ainsi de d\'etecter plusieurs bandes de fr\'equences avec un seul
pixel, \cad une seule antenne (mais un bolom\`etre est tout de m\^eme
n\'ecessaire pour lire chaque bande spectrale). Nous parlons alors de
pixels multi-couleurs. Notez \'egalement le travail de
\shortciteN{perera} qui repose sur un principe diff\'erent
d'absorption mais dont le but est \'egalement de mesurer une puissance
radiative \`a plusieurs longueurs d'onde avec un seul pixel. Mais
c'est encore une fois le groupe de Berkeley qui montre une large
avance sur le d\'eveloppement de ce genre de technologie~; ils ont en
effet d\'ej\`a r\'ealis\'e de nombreux prototypes de bolom\`etres \`a
antennes et se sont r\'ecemment lanc\'e dans la fabrication de
l'instrument PolarBear qui est destin\'e \`a l'\'etude de la
polarisation du CMB \shortcite{myers04,myers}. La
figure~\ref{fig:intro_bolometrie_bolo_matrice_antenne} montre un pixel
prototype de PolarBear contenant deux bolom\`etres, chacun mesure une
polarisation du rayonnement. Elles ne sont pas visibles sur la figure
mais de petites lentilles plan-concaves en silicium sont d\'epos\'ees
sur chacun des pixels pour am\'eliorer le couplage entre le
t\'elescope et l'antenne.

Le grand avantage des bolom\`etres \`a antennes ou des cavit\'es
r\'esonantes r\'eside dans le fait qu'ils sont compatibles avec des
m\'ethodes de fabrication collective. En effet ces d\'etecteurs
peuvent \^etre r\'ealis\'es par des techniques couramment utilis\'ees
en micro-\'electronique, les fonctions de couplage optique et
d'absorption du rayonnement sont alors int\'egr\'ees dans le circuit
de d\'etection des bolom\`etres. Ceci ouvre la voie \`a la fabrication
de tr\`es grands plans focaux, et par extension au d\'eveloppement de
l'astronomie sub-millim\'etrique grand champ comme par exemple les
projets ARTEMIS ou SCUBA-2.

D'autre part, l'abandon des c\^ones de Winston permet de r\'ealiser
des matrices de bolom\`etres tr\`es compactes avec des facteurs de
remplissage proches de~1 (cf
figure~\ref{fig:intro_bolometrie_bolo_matrice_cones} pour SHARC-II ou
section~\ref{sec:detect_bolocea_grille} pour PACS). Les bolom\`etres
peuvent \'egalement \^etre dimensionn\'es de sorte qu'ils
\'echantillonnent le ciel instantan\'ement au crit\`ere de Nyquist,
\cad les bolom\`etres sont espac\'es d'au plus 0.5F$\lambda$, tout en
conservant un couplage efficace avec le t\'elescope et une absorption
optimale. Les m\'ethodes d'observations utilis\'ees dans l'infrarouge
proche, qui sont plus efficaces que le traditionnel mode
\emph{jiggling}, peuvent alors \^etre adopt\'ees pour ce type de
d\'etecteurs. Par ailleurs, \shortciteN{griffin02} calculent la
vitesse de cartographie de bolom\`etres \`a cornets et de bolom\`etres
nus~; moyennant quelques hypoth\`eses raisonnables sur le syst\`eme
optique et les performances des d\'etecteurs (BLIP), ils trouvent
qu'une matrice de bolom\`etres nus \`a 0.5F$\lambda$ est jusqu'\`a
3.5~fois plus rapide qu'un assemblage de bolom\`etres \`a cornets de
2F$\lambda$. Il faut cependant noter que les bolom\`etres nus \`a
0.5F$\lambda$ re\c{c}oivent $\sim$4~fois moins de flux que les
bolom\`etres \`a cornets \`a 2F$\lambda$. Par cons\'equent, pour
obtenir les m\^emes performances en terme de sensibilit\'e BLIP, les
petits pixels doivent avoir une NEP $\sim$2~fois meilleure que celle
des pixels \`a 2F$\lambda$.

\vfill

\cleardoublepage

\chapter{Les matrices de bolom\`etres du Photom\`etre Herschel/PACS}
\label{chap:detect_bolocea}

\begin{center}
\begin{minipage}{0.85\textwidth}

\small Ce chapitre est enti\`erement d\'edi\'e \`a la description des
matrices de bolom\`etres du CEA. Nous introduirons dans un premier
temps le principe de fonctionnement de ces d\'etecteurs de nouvelle
g\'en\'eration en insistant sur leur pertinence pour l'astronomie
infrarouge et sub-millim\'etrique. Dans un deuxi\`eme temps, nous
pr\'esenterons les propri\'et\'es physiques des matrices de
bolom\`etres, notamment la thermom\'etrie haute imp\'edance et la
cavit\'e r\'esonante qui sont deux concepts originaux introduits par
le CEA. Enfin, nous donnerons une description d\'etaill\'ee de
l'\'electronique de lecture des matrices, \'etape n\'ecessaire pour
comprendre la probl\'ematique de la proc\'edure d'\'etalonnage.
\end{minipage}
\end{center}

\section{Les matrices de bolom\`etres du CEA~: principes novateurs}
\label{sec:detect_bolocea_principe}

\subsection{Contexte et motivations}
\label{sec:detect_bolocea_printpe_contexte}

\begin{figure}[!htb]
  \begin{center}
    \includegraphics[width=0.8\textwidth,angle=0]{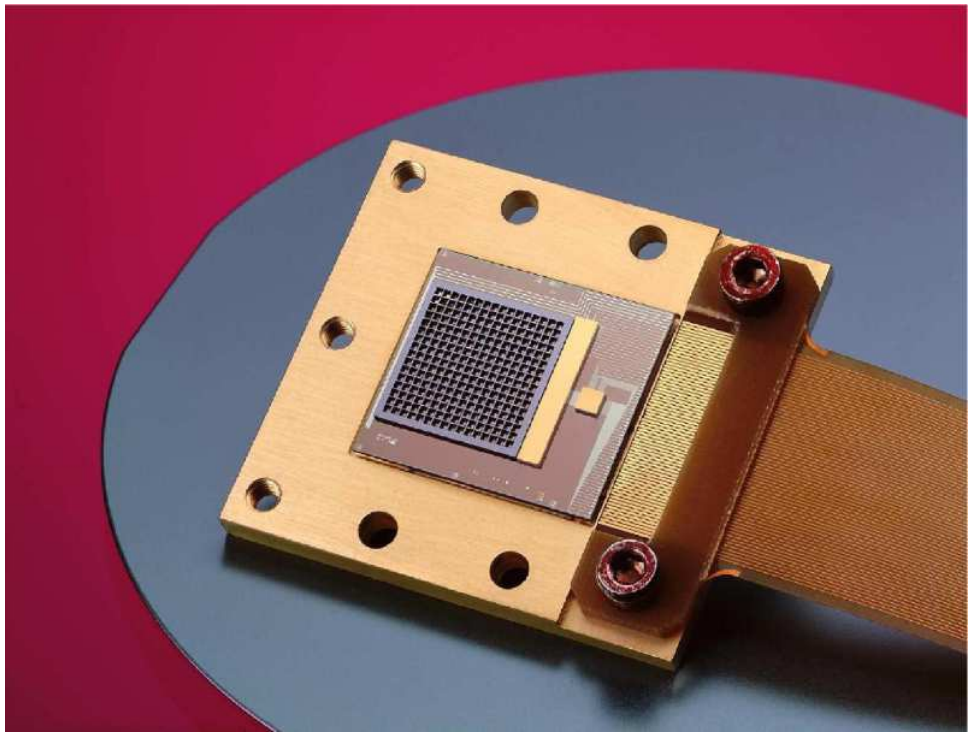}
  \end{center}
\caption[Premi\`ere matrice de bolom\`etres fonctionnelle en
  1999.]{Photographie de la premi\`ere matrice de bolom\`etres
  fonctionnelle $16\times16$~pixels r\'ealis\'ee par le LETI en
  1999. Sur ce mod\`ele le circuit de lecture multiplex\'e est
  situ\'e \`a c\^ot\'e du circuit de d\'etection, et non en-dessous
  comme c'est le cas pour les mod\`eles actuels. (cr\'edit CEA)
\label{fig:detect_bolocea_principe_motive_matrice}}
\end{figure}

L'histoire des matrices de bolom\`etres du CEA commence en 1995 au
Laboratoire Infrarouge du CEA/LETI \`a Grenoble. C'est Patrick
Agn\`ese, organisateur de cette premi\`ere r\'eunion, qui rassemble
les responsables de quelques grands laboratoires de recherche~:
Laurent Vigroux pour le CEA/SAp, Jean-Michel Lamarre pour l'IAS et
Bernard Lazareff pour l'IRAM. L'objectif de cette rencontre est de
discuter du potentiel du LETI pour la bolom\'etrie en astronomie
submillim\'etrique, particuli\`erement pour l'instrument SPIRE du
satellite FIRST. En effet, le LETI a d\'evelopp\'e des technologies et
des savoir-faire pour l'instrumentation spatiale lors du projet ISOCAM
en collaboration avec le CEA/SAp \shortcite{mottier,cesarsky}, et ses
multiples comp\'etences en micro-\'electronique, lithographie,
etc... pourraient permettre de franchir une barri\`ere dans la
fabrication de bolom\`etres submillim\'etriques. Deux id\'ees majeures
propos\'ees par le LETI \shortcite{buzzi_these} ont \'et\'e retenues
lors de cette r\'eunion~:
\begin{itemize}
\item la fabrication collective de bolom\`etres devrait assurer un
haut rendement de production et ouvrir la voie \`a l'astronomie grand
champ,
\item l'introduction de grilles r\'esonantes pour l'absorption du
rayonnement pourrait conduire \`a une v\'eritable r\'evolution pour
les matrices de bolom\`etres.
\end{itemize}
De plus, les possibles applications d'un tel composant sont multiples,
la premi\`ere \'etant bien \'evidemment pour l'instrumentation des
t\'elescopes au sol pour les plus grandes longueurs d'onde, mais la
conception de ces d\'etecteurs pourrait \'eventuellement \^etre
adapt\'ee \`a des longueurs d'onde plus courtes pour la surveillance
et la s\'ecurit\'e (d\'etection de mines anti-personnelles enfouies
sous terre par exemple).\\ En Avril~1996, la phase de pr\'e-\'etude
des matrices de bolom\`etres d\'emarre sur financement du Service
d'Astrophysique. En 1997, le CEA r\'epond \`a l'appel \`a proposition
de l'ESA pour l'instrument SPIRE, et c'est en 1999 que le LETI produit
la premi\`ere matrice fonctionnelle de bolom\`etres
submillim\'etriques $16\times16$~pixels avec un circuit de lecture
multiplex\'e \`a 300~mK \shortcite{agnese99}. Une photographie de
cette matrice est repr\'esent\'ee sur la
figure~\ref{fig:detect_bolocea_principe_motive_matrice}.  Le projet du
CEA est tellement ambitieux que le consortium SPIRE refuse d'utiliser
les matrices de bolom\`etres car il juge le concept trop risqu\'e pour
un projet spatial, et c'est finalement les bolom\`etres \emph{spider
web} du JPL, dont la technologie est connue depuis plus de 10~ans, qui
sont s\'electionn\'es pour l'instrument SPIRE. C'est en 2000 que les
bolom\`etres du CEA se retrouvent \`a nouveau dans le projet Herschel,
mais cette fois pour l'instrument PACS. En effet, le design optique de
l'instrument PACS \'etant tr\`es complexe, le consortium d\'ecide de
remplacer les photoconducteurs initialement pr\'evus pour le
photom\`etre par les matrices de bolom\`etres du CEA. Les \'equipes du
LETI et du SAp ont ensuite r\'eagi rapidement pour apporter aux
matrices les modifications n\'ecessaires pour adapter leur domaine
d'absorption \`a la gamme de longueur d'onde de PACS. Lorsque ma
th\`ese a d\'ebut\'e, la campagne d'\'etalonnage des mod\`eles CQM
(Cryogenic Qualification Model) venait tout juste de se terminer, et
les premi\`eres matrices de type FM (Flight Model) commen\c{c}aient
\`a arriver au SAp pour y \^etre test\'ees et s\'electionn\'ees pour
constituer le plan focal de vol et le plan focal de rechange (Flight
Spare).

\subsection{Pertinence du concept et innovations technologiques}
\label{sec:detect_bolocea_printpe_innovation}

Les matrices de bolom\`etres du CEA sont v\'eritablement des
d\'etecteurs d'une nouvelle g\'en\'eration. L'introduction de
cavit\'es r\'esonantes pour optimiser l'absorption du champ de
radiation incident est effectivement un concept r\'evolutionnaire en
bolom\'etrie, et la plupart des bolom\`etres modernes a d'ailleurs
adopt\'e cette technologie brevet\'ee par le CEA/LETI. Nous
d\'ecrirons bri\`evement le principe d'absorption de rayonnement dans
la section~\ref{sec:detect_bolocea_fabrication_cavite}. Notons
premi\`erement que la cavit\'e r\'esonante permet, entre autre,
d'abandoner l'usage des c\^ones de Winston traditionnels et des
sph\`eres int\'egratrices qui leur sont g\'en\'eralement associ\'ees
(cf section~\ref{sec:intro_bolometrie_bolo_matrice})~; ceci pr\'esente
trois int\'er\^ets majeurs:
\begin{itemize}
\item les c\^ones de Winston sont volumineux et \og encombrent \fg le
plan focal (faibles facteurs de remplissage), alors que le concept de
cavit\'e r\'esonante permet de fabriquer des pixels contigus sans
perdre de place et ainsi atteindre des facteurs de remplissage proche
de 1.
\item Les pixels \'etant contigus, ils peuvent \^etre dimensionn\'es
pour satisfaire le crit\`ere d'\'echantillonnage de Nyquist sans
affecter le couplage optique avec le t\'elescope.
\item Les c\^ones de Winston \'etant relativement massifs, s'en
affranchir relaxe des contraintes thermiques et m\'ecaniques au niveau
du plan focal de sorte que le nombre de pixels peut maintenant \^etre
consid\'erablement augment\'e.
\end{itemize}
Pour mettre \`a profit ce dernier point, le LETI a mis en \oe uvre une
proc\'edure de fabrication collective qui permet de r\'ealiser des
matrices de bolom\`etres contenant 256~pixels \`a partir d'un m\^eme
wafer\footnote{Un wafer est le nom donn\'e aux plaques de silicium
utilis\'ees en micro-\'electronique, c'est le produit de d\'epart de
tout circuit int\'egr\'e.}. Les bolom\`etres d'une m\^eme matrice
pr\'esentent alors des propri\'et\'es physiques tr\`es similaires, ce
qui est difficile \`a r\'ealiser et \'eventuellement couteux pour les
bolom\`etres traditionnels fabriqu\'es de fa\c{c}on individuelle. La
conception des matrices permet en plus de les abouter sur trois
c\^ot\'es pour former des plans focaux de plusieurs milliers de
pixels. La cam\'era ARTEMIS par exemple comptera plus de
4000~bolom\`etres op\'erant \`a 450~$\mu$m \shortcite{talvard}. Ce
nombre est \`a comparer avec les cam\'eras bolom\'etriques
actuellement en op\'eration o\`u les bolom\`etres sont assembl\'es un
par un pour obtenir quelques centaines de pixels au maximum dans le
plan focal. La fabrication collective ouvre assur\'ement la voie vers
la production de grands plans focaux pour l'astronomie grand champ
dans les domaines de l'infrarouge lointain et du submillim\'etrique. 

D'autre part, l'\'electronique de lecture n\'ecessaire pour
lire les milliers de pixels op\'erant \`a 300~mK repr\'esente
\'egalement une premi\`ere technologique dans le domaine de la
bolom\'etrie refroidie. Cette \'electronique multiplex\'ee utilise des
transistors MOS\footnote{Les bolom\`etres r\'esistifs sont
habituellement lus par des transistors JFET car leurs performances en
terme de bruit sont tr\`es bonnes ($\sim$nV). Cependant, ce type de
composant ne fonctionne pas \`a 300~mK, la technologie MOS a donc
\'et\'e choisie pour le multiplexage \`a froid malgr\'e son niveau de
bruit relativement \'elev\'e.} \`a 300~mK. Nous la d\'ecrirons en
d\'etail dans la section~\ref{sec:detect_bolocea_elec}. Notez
toutefois que pour minimiser les co\^uts de d\'eveloppement,
l'architecture du circuit de lecture a \'et\'e largement inspir\'ee de
celle de la cam\'era ISOCAM \shortcite{cesarsky} qui fut \'egalement
con\c{c}ue par le CEA/LETI.

Pour atteindre notre objectif, \cad une NEP de l'ordre de $1\times
10^{-16}$~W/$\sqrt{\mbox{Hz}}$, il est n\'ecessaire de compenser le
bruit intrins\`eque relativement \'elev\'e de ces transistors (une
fraction de $\mu$V) par une r\'eponse bien sup\'erieure \`a celle des
bolom\`etres traditionnels. Il faut en effet atteindre une r\'eponse
de quelques $10^{10}$~V/W. C'est la raison pour laquelle les
thermom\`etres fonctionnent dans un r\'egime o\`u leur imp\'edance
d\'epasse les $10^{12}$~$\Omega$.

En bref, les innovations technologiques majeures d\'evelopp\'ees pour
la conception des matrices de bolom\`etres sont les suivantes~:
\begin{itemize}
\item Cavit\'e r\'esonante pour optimiser l'absorption du
rayonnement,
\item Fabrication collective des bolom\`etres,
\item Multiplexage \`a froid du signal bolom\'etrique,
\item Thermom\'etrie tr\`es haute imp\'edance.
\end{itemize}
\noindent La fabrication des matrices de bolom\`etres du CEA fait donc
appel \`a plusieurs technologies de pointe allant de la
micro-\'electronique \`a la lithographie en passant par le
micro-usinage, l'implantation ionique et les techniques
d'hybridation. La production de ces matrices est un processus complexe
qui n\'ecessite plusieurs mois, ce qui repr\'esente un r\'eel d\'efi
technologique.
\subsection{Description g\'en\'erale de l'architecture des matrices}
\label{sec:detect_bolocea_printpe_description}


Les matrices de bolom\`etres du CEA sont bas\'ees sur une conception
tout-silicium qui met largement \`a profit les techniques
g\'en\'eralement utilis\'ees pour la micro-\'electronique. Les
bolom\`etres sont de type composite, \cad que chaque \'el\'ement
constitutif des matrices poss\`ede une fonction bien d\'efinie qui est
en g\'en\'eral ind\'ependante de celle des autres \'el\'ements (cf
section~\ref{sec:intro_bolometrie_thermo_principe}). Nous dressons
dans un premier temps un portrait global de la fabrication et du
fonctionnement des matrices, puis nous reviendrons plus en d\'etail
sur certains \'el\'ements clefs de ces d\'etecteurs dans les
sections~\ref{sec:detect_bolocea_fabrication}
et~\ref{sec:detect_bolocea_elec}.\\

\begin{figure}
  \begin{center}
    \includegraphics[width=0.8\textwidth,angle=0]{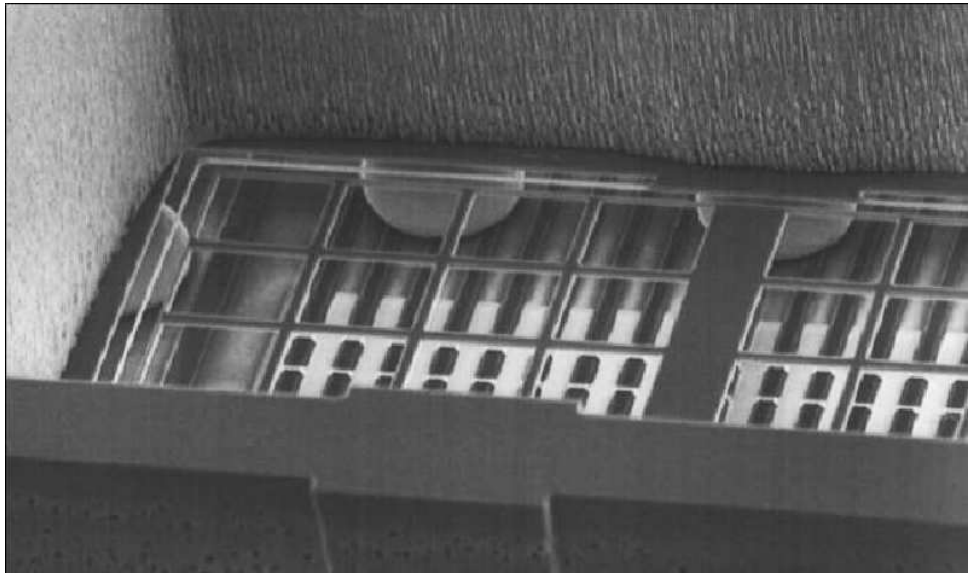}
  \end{center}
\caption[Photographie de l'hybridation par billes
d'indium]{Photographie des billes d'indium prises en sandwich entre le
circuit de d\'etection (en haut) et le circuit de lecture (en
bas). Ces billes assurent le contact \'electrique et thermique entre
les deux plaques de silicium. Elles jouent \'egalement le r\^ole de
support m\'ecanique et d\'etermine la taille de la cavit\'e
r\'esonante. (cr\'edit CEA)
\label{fig:detect_bolocea_principe_description_matriceoverview_bille}}
\end{figure}

Les matrices de bolom\`etres du CEA se pr\'esentent sous la forme de
deux fines plaques de silicium hybrid\'ees l'une sur l'autre par des
centaines de billes d'indium. L'hybridation est une technique
d'assemblage utilis\'ee en micro-\'electronique pour relier
\'electriquement deux composants. Elle est par exemple utilis\'ee lors
de la fabrication des d\'etecteurs infrarouges modernes pour connecter
le mat\'eriau photosensible \`a l'\'electronique de lecture (voir
\shortciteNP{rieke} pour une revue r\'ecente des techniques
d'hybridation pour les d\'etecteurs IR). Cette technique consiste \`a
d\'eposer de petites quantit\'es d'indium au bout de certaines lignes
\'electriques d'un circuit imprim\'e et de disposer sur celui-ci un
autre circuit de sorte que les billes d'indium pro\'eminentes se
positionnent en d\'ebut de ligne \'electrique sur le second
circuit. Nous obtenons alors deux circuits superpos\'es et connect\'es
\'electriquement dans la troisi\`eme dimension. Les billes jouent
\'egalement le r\^ole de contact thermique\footnote{\`A 300~mK,
l'indium est un supraconducteur, \cad que le mat\'eriau ne pr\'esente
aucune r\'esistance \'electrique~; cependant la conduction de la
chaleur est tr\`es inefficace dans un supraconducteur car seuls les
phonons peuvent la transporter. Ce sont en fait les \'electrons
pr\'esents dans la fine couche d'oxyde (non-supra) qui recouvre la
bille qui conduisent la chaleur et qui permettent de thermaliser les
deux composants.} et de support m\'ecanique entre les deux plaques. La
figure~\ref{fig:detect_bolocea_principe_description_matriceoverview_bille}
pr\'esente une photographie prise au microscope \'electronique de
quelques billes d'indium prises en sandwich entre les deux plaques de
silicium micro-usin\'ees. Notez que, mis \`a part le principe de
d\'etection, la conception des matrices de bolom\`etres se rapproche
de celle des d\'etecteurs IR modernes. Les modes d'observation du
Photom\`etre PACS (cf section~\ref{sec:calib_perfobs_oof}) sont
eux-aussi tr\`es similaires aux modes d'observation utilis\'es pour
les photoconducteurs de l'infrarouge proche et moyen. La technique
d'hybridation pour la bolom\'etrie refroidie est relativement
r\'ecente, et elle s'annonce prometteuse pour le d\'eveloppement des
futures cam\'eras grand champ dans le domaine sub-millim\'etrique
\shortcite{agnese99,allen}.\\

\begin{figure}
  \begin{center}
    \begin{tabular}[t]{c}
      \includegraphics[width=0.6\textwidth,angle=0]{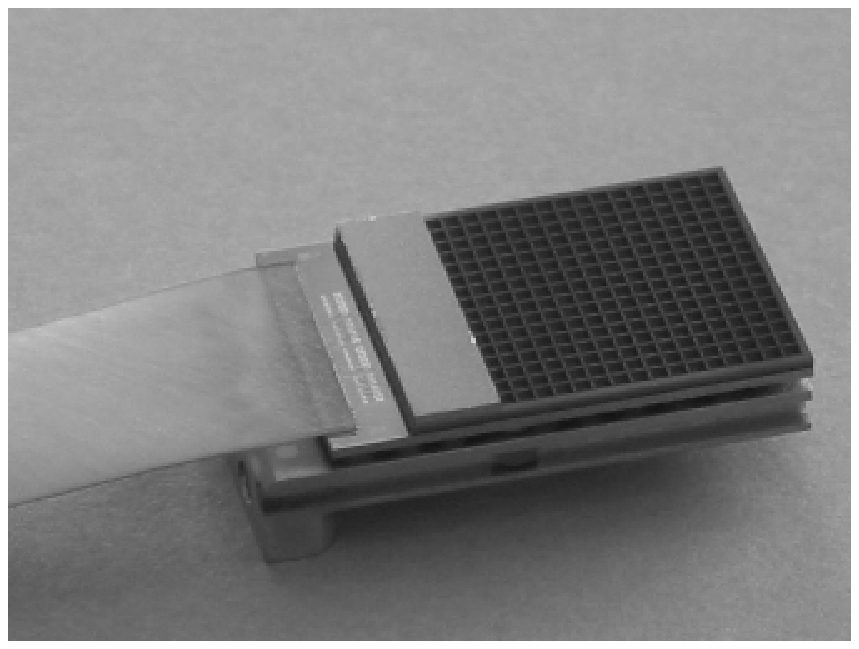} \\ \includegraphics[width=1.\textwidth,angle=0]{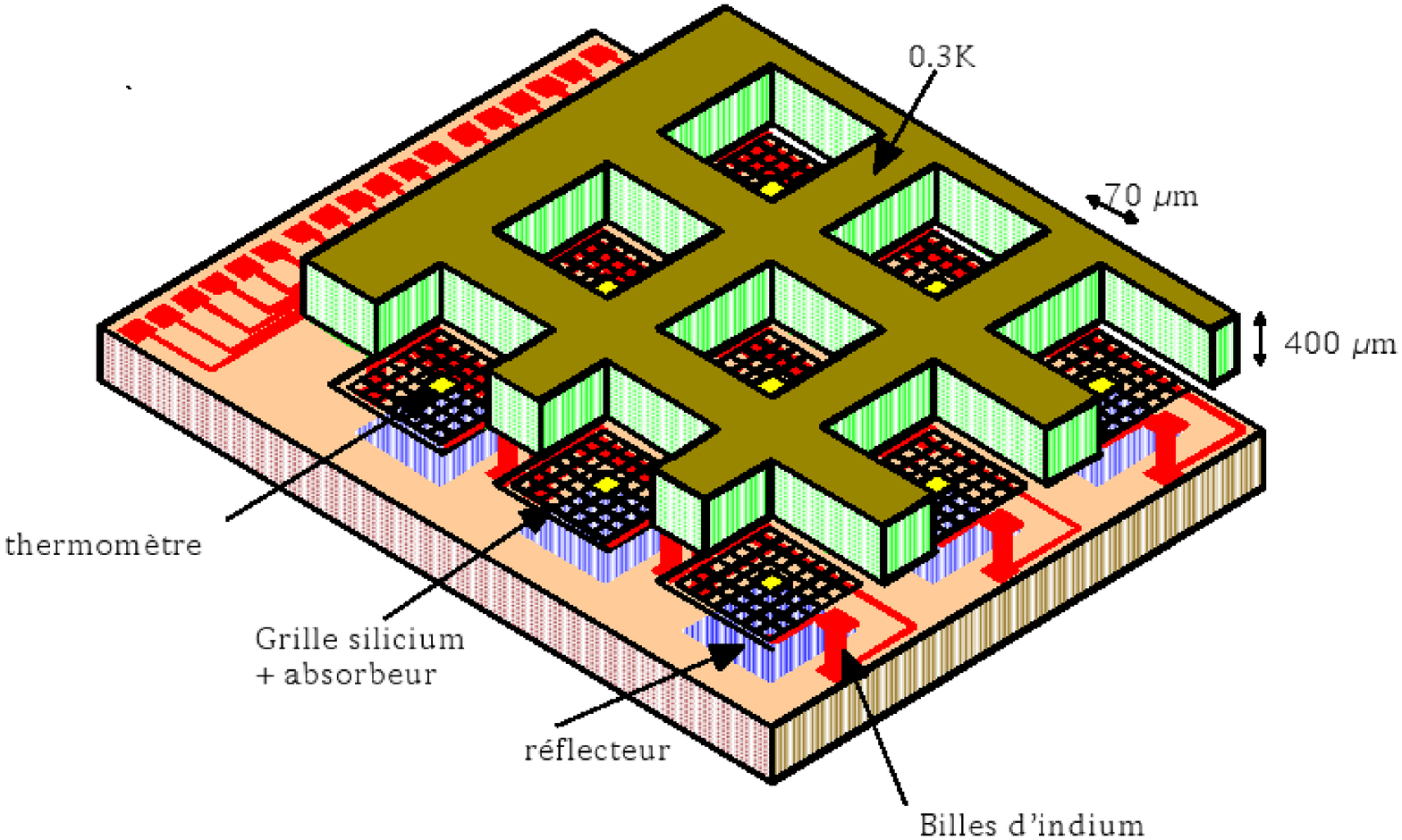} 
    \end{tabular}
  \end{center}
  \caption[Photographie et sch\'ema explicatif d'une matrice de
  bolom\`etres]{La photographie du haut pr\'esente une matrice de
  bolom\`etres assembl\'ee. Le circuit de d\'etection (\emph{CD}) est
  expos\'e au rayonnement \'electromagn\'etique et contient
  256~bolom\`etres. Juste en-dessous se trouve le circuit de lecture
  multiplex\'e reli\'e au reste de l'\'electronique par un c\^able
  flex en kapton. Sur le sch\'ema du bas, nous retrouvons le \emph{CD}
  hybrid\'e au \emph{CL} par les billes d'indium. Chaque grille
  suspendue poss\`ede une thermistance pour mesurer les variations de
  temp\'erature dues au flux incident. Sous chaque pixel se trouve un
  r\'eflecteur pour former une cavit\'e r\'esonante et ainsi optimiser
  l'absorption du rayonnement. (cr\'edit CEA)
  \label{fig:detect_bolocea_principe_description_matriceoverview_photo}}
\end{figure}

La
figure~\ref{fig:detect_bolocea_principe_description_matriceoverview_photo}
montre la photographie d'une matrice assembl\'ee par hybridation ainsi
qu'un sch\'ema explicatif de son architecture. La plaque expos\'ee au
champ de radiation constitue ce que nous appellons le \emph{Circuit de
D\'etection}, ou \emph{CD}, c'est elle qui contient les bolom\`etres
\`a proprement parler. La plaque inf\'erieure constitue
l'\'electronique de lecture multiplex\'ee que nous appellons
\emph{Circuit de Lecture}, ou \emph{CL}. Ces deux plaques de silicium
sont manufactur\'ees s\'epar\'ement et de fa\c{c}on totalement
ind\'ependante, il est donc possible de leur donner des fonctions
tr\`es sp\'ecialis\'ees et dissoci\'ees l'une de l'autre. Le \emph{CD}
en l'occurence est une plaque de silicium d'environ 1~cm de c\^ot\'e
et de 400~$\mu$m d'\'epaisseur qui est micro-usin\'ee en profondeur
pour obtenir 256~grilles suspendues de 5~$\mu$m d'\'epaisseur. Ces
pixels carr\'es sont dispos\'es selon une grille cart\'esienne et sont
s\'epar\'es par des murs interpixels, ce qui donne au \emph{CD} un
aspect gauffr\'e (cf
figures~\ref{fig:detect_bolocea_principe_description_matriceoverview_photo}
et~\ref{fig:detect_bolocea_grille_photopix}). Les murs interpixels
\'etant beaucoup plus massifs que les grilles suspendues, ils ont une
grande capacit\'e calorifique et servent de fuites thermiques pour les
bolom\`etres. D'autre part le silicium est transparent dans
l'infrarouge lointain, les grilles sont donc recouvertes de m\'etal
pour absorber le rayonnement \'electromagn\'etique. Deux
thermom\`etres de tr\`es haute imp\'edance sont implant\'es sur chaque
pixel, un au centre de la grille suspendue et l'autre sur le mur
interpixel adjacent (cf
section~\ref{sec:detect_bolocea_fabrication_thermo}). Celui qui se
trouve sur le mur est thermalis\'e \`a la temp\'erature du bain
thermique alors que l'autre se trouve \`a la temp\'erature de la
grille absorbante qui, elle, varie avec le flux incident. Cette
structure de silicium est maintenue au-dessus du circuit de lecture
par les billes d'indium. Des r\'eflecteurs en or sont dispos\'es sous
chaque pixel pour former des cavit\'es r\'esonantes et ainsi optimiser
l'absorption du rayonnement dans une certaine gamme de longueur
d'onde. L'absorption d\'epasse g\'en\'eralement les 95~\%. Nous
traiterons ce sujet plus en d\'etail dans la
section~\ref{sec:detect_bolocea_fabrication_cavite}. \\

\begin{figure}
  \begin{center}
    \includegraphics[width=0.9\textwidth,angle=0]{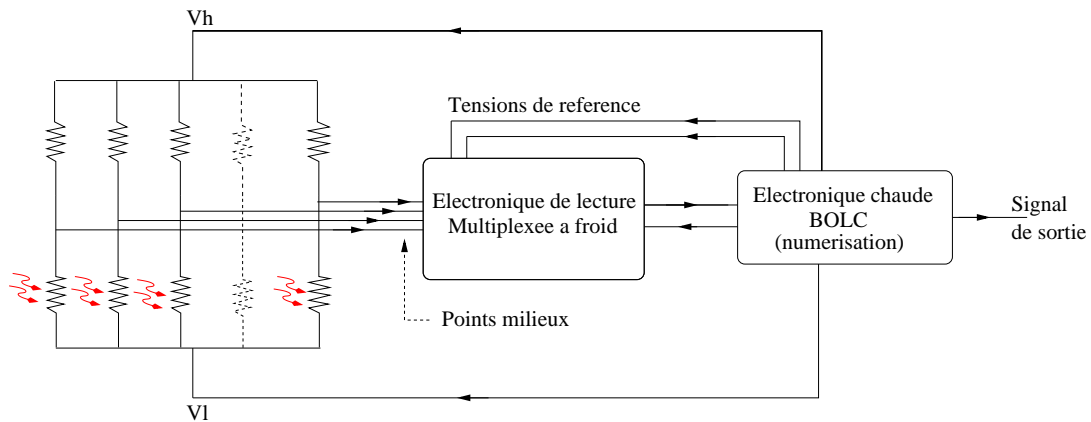}
  \end{center}
\caption[Diagramme \'electrique d'une matrice de
bolom\`etres]{Diagramme \'electrique d'une matrice de
bolom\`etres. Les ponts bolom\'etriques sont repr\'esent\'es \`a
gauche. Les thermistances situ\'ees sur les grilles sont soumises aux
variations de temp\'erature induites par le flux radiatif incident
(fl\`eches rouges). Le signal utile est la tension mesur\'ee au milieu
des ponts diviseur de tension, nous appelons ces tensions les
\emph{points milieux}. Les ponts sont mont\'es en
parall\`ele. L'\'electronique froide permet de multiplexer la lecture
des points milieux et de r\'eduire le nombre de lignes \'electriques
d'un facteur~16. L'\'electronique chaude alimente en tension le
circuit de lecture et les bolom\`etres eux-m\^eme via les tensions
$V_h$ et $V_l$. Deux tensions de r\'ef\'erence sont inject\'ees au
niveau de l'\'electronique froide.
\label{fig:detect_bolocea_principe_description_matriceoverview_schema}}
\end{figure}

Le \emph{CL} est un circuit imprim\'e \`a base de transistors MOSFET,
il en contient plus de~4000. Son r\^ole est de lire la tension aux
bornes des thermom\`etres et de transporter cette information vers
l'\'electronique chaude. Cependant, une matrice comprend 256~pixels
fonctionnant \`a 300~mK, et il n'est pas possible de faire sortir
256~fils connecteurs du plan focal pour chaque matrice du Photom\`etre
PACS~; le budget thermique est en effet extr\'emement restreint \`a
ces temp\'eratures. Il est donc n\'ecessaire de multiplexer le signal,
\cad lire plusieurs pixels avec le m\^eme circuit \'electrique. Pour
PACS, le \emph{CL} est con\c{c}u pour r\'eduire le nombre de fils
sortant des matrices par un facteur 16 (1~circuit lit une colonne
enti\`ere de 16~pixels), ce qui rend la dissipation \'electrique du
circuit de lecture compatible avec le budget thermique disponible \`a
300~mK. L'autre raison qui a motiv\'e l'implantation du circuit de
lecture \`a proximit\'e des bolom\`etres est la tr\`es haute
imp\'edance des thermom\`etres. Il faut en effet minimiser la
capacit\'e \'electrique des c\^ables (elle est proportionnelle \`a leurs
longueurs) qui transmettent le signal pour minimiser la constante de
temps \'electrique du circuit de lecture, et ainsi s'assurer que le
temps de r\'eponse des bolom\`etres reste compatible avec la
fr\'equence de lecture du signal. Gr\^ace \`a la technique
d'hybridation, il est possible d'implanter un premier \'etage
d'adaptation d'imp\'edance \`a seulement quelques millim\`etres des
thermom\`etres. Le \emph{CL} est un \'el\'ement clef des matrices de
bolom\`etres et son r\'eglage est critique pour les performances
finales du d\'etecteur. Nous en donnerons une description beaucoup
plus d\'etaill\'ee dans la section~\ref{sec:detect_bolocea_elec}. La
figure~\ref{fig:detect_bolocea_principe_description_matriceoverview_schema}
montre le sch\'ema \'electrique simplifi\'e d'une matrice de
bolom\`etres pour le Photom\`etre PACS. Chaque pixel contient deux
thermistances mont\'ees en pont diviseur de tension, et tous les ponts
d'une m\^eme matrice sont mont\'es en parall\`ele. Nous appelons
\emph{point milieu} la tension \'electrique \`a la sortie du pont
bolom\'etrique comme indiqu\'e sur la figure. L'\'electronique froide
\'echantillonne le point milieu de chaque pixel et ram\`ene
l'imp\'edance du circuit de lecture vers des valeurs compatibles avec
des lignes de transmissions longues de plus de 7~m~; l'imp\'edance
passe de quelques $10^{12}-10^{13}$~$\Omega$ \`a environ 10~k$\Omega$
sur les 10~premiers centim\`etres du circuit. L'\'electronique chaude
porte le nom de \emph{BOLC} dans l'instrument PACS, son r\^ole est de
num\'eriser le signal bolom\'etrique et de fournir les tensions
\'electriques \`a tout le circuit de lecture et aux bolom\`etres
eux-m\^emes (via les tensions $V_h$ et $V_l$). Une particularit\'e
int\'eressante de BOLC est qu'il permet d'injecter des tensions de
r\'ef\'erence au niveau de l'\'electronique froide~; cela autorise une
lecture diff\'erentielle du signal bolom\'etrique et donc une
meilleure r\'ejection des d\'erives basses fr\'equences de
l'\'electronique de lecture et des diverses perturbations
(microphonie, pickup HF, environnement \'electromagn\'etique dans le
satellite). Nous verrons en plus que ces tensions de r\'ef\'erence
jouent un r\^ole central dans la proc\'edure d'\'etalonnage que nous
avons mise au point (cf section~\ref{sec:detect_outils_concept}).\\


\section{Propri\'et\'es physiques des matrices de bolom\`etres}
\label{sec:detect_bolocea_fabrication}

\subsection{La grille absorbante: le motif de la matrice}
\label{sec:detect_bolocea_grille}

\begin{figure}
  \begin{center}
    \begin{tabular}[t]{ll}
      \includegraphics[height=0.22\textheight]{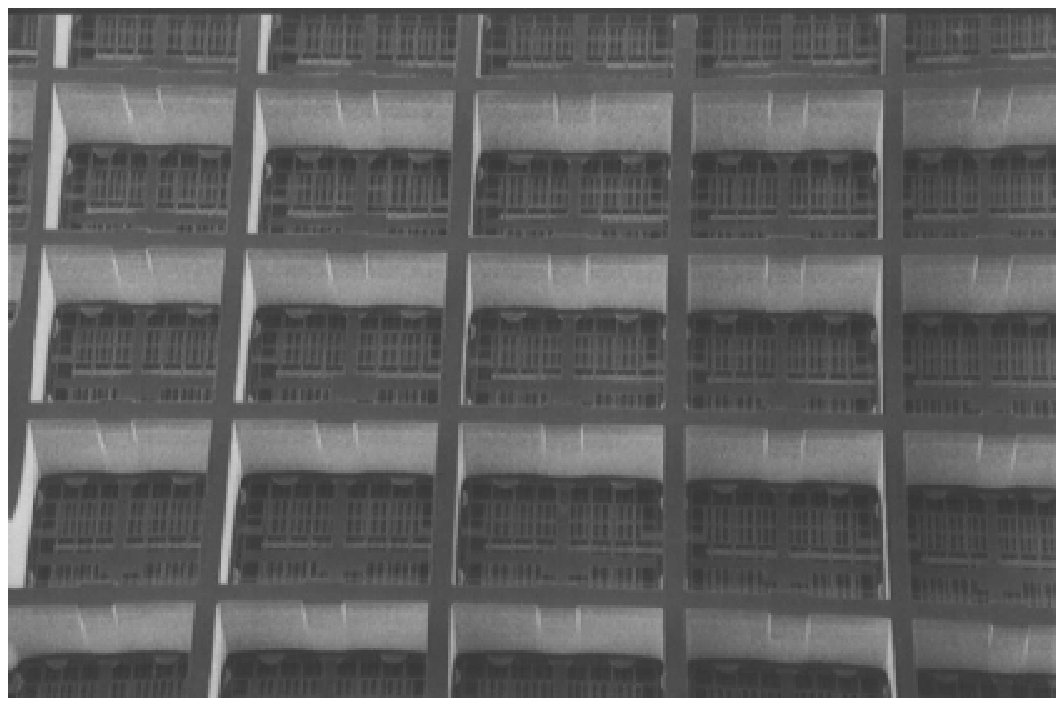} & \includegraphics[height=0.22\textheight]{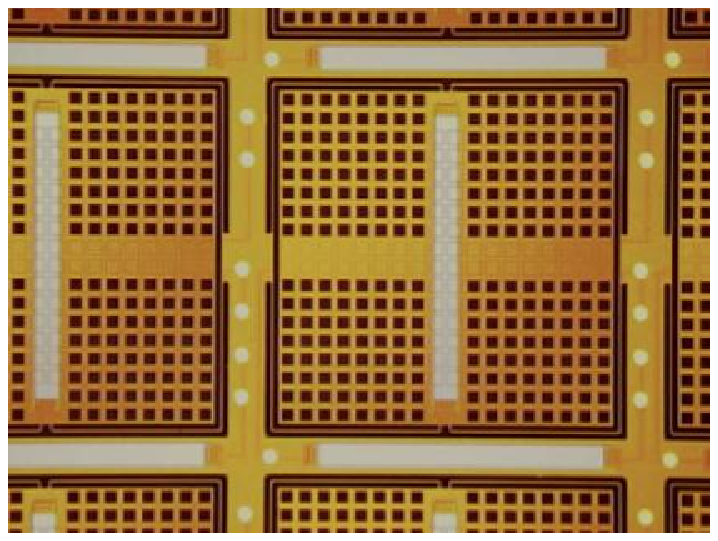} 
    \end{tabular}
  \end{center}
  \caption[Vues rapproch\'ees du circuit de d\'etection]{Vues
  rapproch\'ees du circuit de d\'etection par dessus (\`a gauche) et
  par dessous avant l'\'etape d'hybridation (\`a droite). Nous
  retrouvons les murs interpixels, les grilles suspendues, les
  thermom\`etres (longues bandes claires sur la photographie de
  droite) et les contacts pr\'evus pour les billes d'indium (ronds
  clairs situ\'es sous les murs interpixels).
  \label{fig:detect_bolocea_grille_photopix}}
\end{figure}

Nous nous int\'eressons dans un premier temps \`a la partie sensible
du d\'etecteur, \cad \`a la structure suspendue qui absorbe le
rayonnement et dont la temp\'erature \'evolue avec le flux
incident. La figure~\ref{fig:detect_bolocea_grille_photopix}
pr\'esente deux vues rapproch\'ees du circuit de d\'etection. 

\`A gauche, nous voyons le \emph{CD} par dessus, les grilles
absorbantes se situent \`a la base des murs interpixels et nous
retrouvons bien cette forme gauffr\'ee des matrices. Nous pouvons
m\^eme apercevoir les billes d'indium qui supportent le \emph{CD}. Les
murs interpixels mesurent 400~$\mu$m de haut, les grilles ne mesurent
que 5~$\mu$m dans la m\^eme dimension. Les pixels sont carr\'es et
espac\'es de 750~$\mu$m sur les mod\`eles PACS. Les grilles de
silicium font 640~$\mu$m de c\^ot\'e, ce qui donne un facteur de
remplissage g\'eom\'etrique d'environ 0.73. Cependant, l'\'epaisseur
des murs interpixels (70~$\mu$m) est inf\'erieure \`a la longueur
d'onde et le silicium est transparent dans l'infrarouge lointain~; par
cons\'equent, un photon qui tomberait exactement sur un mur serait
partiellement absorb\'e sur les deux pixels adjacents, nous parlons
alors de couplage capacitif entre pixels. Le facteur de remplissage
effectif est donc sup\'erieur \`a 73~\%, nous le confirmerons dans la
section~\ref{sec:detect_bolocea_fabrication_cavite} par des mesures
d'absorption effectu\'ees avec un spectrom\`etre \`a transform\'ee de
Fourier.

\`A droite, nous pr\'esentons une photographie du \emph{CD} vu par
dessous avant l'\'etape d'hybridation. Les ronds clairs situ\'es sous
les murs interpixels marquent l'emplacement des contacts \'electriques
pour les billes d'indium. Il y en a 7~par pixels, ce qui fait un total
d'environ 1800~billes par matrice. Les longues bandes claires visibles
sous la grille et sous les murs interpixels sont les thermom\`etres,
nous les d\'ecrirons en d\'etails dans la
section~\ref{sec:detect_bolocea_fabrication_thermo}.\\

La fabrication du circuit de d\'etection n\'ecessite plusieurs
dixaines d'\'etapes successives pour micro-usiner la plaque de
silicium, implanter les thermom\`etres et les lignes \'electriques
n\'ecessaires \`a la lecture du signal. La th\`ese de Christophe Buzzi
\shortcite{buzzi_these} pr\'esente la fabrication des premiers
mod\`eles de matrices de bolom\`etres. 
La r\'ealisation de ces matrices est complexe et d\'elicate, elle
rel\`eve presque de la joaillerie, et pour paraphraser George Rieke
\`a propos des matrices de bolom\`etres du CEA~:
\begin{center}
\begin{minipage}{0.85\textwidth}
\emph{\og The delicate construction of the detector depends on the
ability to etch exquisitely complex miniature structures in
silicon. The development of ``silicon micromachining'' has enabled
substantial advances in bolometer construction generally and is
central to making large-scale arrays\fg}\\ Extrait de
\shortciteNP{rieke}.
\end{minipage}
\end{center}

\subsubsection{Capacit\'e calorifique}
\label{sec:detect_bolocea_grille_capacite}

Pour une quantit\'e d'\'energie absorb\'ee donn\'ee, la r\'eponse en
temp\'erature d'un bolom\`etre est inversement proportionnelle \`a sa
capacit\'e calorifique (cf
section~\ref{sec:intro_bolometrie_thermo_principe}). Le \emph{CD} a
donc \'et\'e con\c{c}u pour minimiser la masse du substrat suspendu
ainsi que sa chaleur sp\'ecifique. La premi\`ere \'etape consiste \`a
rendre la grille absorbante aussi fine que possible tout en gardant
une \'epaisseur compatible avec les contraintes m\'ecaniques de la
structure (5~$\mu$m en l'occurence). Des trous carr\'es sont ensuite
usin\'es dans la grille pour r\'eduire encore la masse (cf
photographie de droite sur la
figure~\ref{fig:detect_bolocea_grille_photopix}). Ces trous sont
espac\'es de 35~$\mu$m et les structures \og inter-trous \fg sont des
poutres \`a section carr\'ee de 5~$\mu$m de c\^ot\'e. Nous obtenons
ainsi une masse suspendue de silicium de 2.02$\times10^{-6}$~g. Notez
qu'avec une si faible masse, les grilles sont tr\`es peu sensibles aux
vibrations m\'ecaniques qu'elles subiront lors du d\'ecollage de la
fus\'ee. D'autre part, la chaleur sp\'ecifique du silicium varie comme
T$^3$ aux tr\`es basses temp\'eratures~; une temp\'erature de
fonctionnement de 300~mK pour les bolom\`etres PACS permet donc de
r\'eduire consid\'erablement la chaleur sp\'ecifique. Pour calculer
pr\'ecis\'ement la capacit\'e calorifique de la partie suspendue, il
faut prendre en compte le substrat en silicium mais \'egalement le
thermom\`etre et le m\'etal absorbeur d\'epos\'e \`a la surface de la
grille. \shortciteN{reveret_these} donne la formule suivante pour la
capacit\'e calorifique totale~:
\begin{equation}
C(T)=V_{Si}\times5.7\times10^{-7}\,T^3+S_{abs}\times5.1\times10^{-11}\,exp\left(\frac{-3.17}{T}\right)+S_{th}\times6.6\times10^{-13}\,T
\end{equation}
o\`u $V_{Si}$, $S_{abs}$ et $S_{th}$ sont respectivement le volume de
la grille de silicium en cm$^3$, et les surfaces de l'absorbeur
m\'etallique et du thermom\`etre.\\ Pour les pixels du photom\`etre
PACS, nous trouvons une capacit\'e calorifique totale d'environ
$2.1\times10^{-14}$~J/K.



\subsubsection{Conductance thermique}
\label{sec:detect_bolocea_grille_conductance}

Les grilles absorbantes sont reli\'ees aux murs interpixels par
4~poutres en silicium de 600~$\mu$m de long et de section
$2\times5$~$\mu$m. Elles sont visibles sur les photographies des
figures~\ref{fig:detect_bolocea_grille_photopix}
et~\ref{fig:detect_bolocea_fabrication_thermo_photo}. Elles jouent le
r\^ole de suspension m\'ecanique et de lien thermique entre la grille
et la source froide. Deux d'entre elles sont recouvertes de m\'etal
pour connecter \'electriquement le thermom\`etre au circuit de
lecture.

La conductance thermique de ces poutres est une quantit\'e critique
pour les performances du d\'etecteur, elle contribue \`a la r\'eponse
en temp\'erature du bolom\`etre ainsi qu'\`a sa constante de temps
(cf section~\ref{sec:intro_bolometrie_thermo_principe}). Durant son
stage de DEA au CEA/LETI, Olivier Savry a mesur\'e la conductance
thermique \`a 300~mK des poutres, il trouve~:
\begin{equation}
G_{th}\approx 0.6\times10^{-11}\,\,\mbox{W/K}
\end{equation}

\subsection{Les thermom\`etres}
\label{sec:detect_bolocea_fabrication_thermo}

Les thermom\`etres utilis\'es pour les matrices de bolom\`etres du CEA
sont de type r\'esistif. Nous exploitons la forte d\'ependance en
temp\'erature de la r\'esistance du mat\'eriau autour de sa transition
m\'etal-isolant \shortcite{mott_book} pour mesurer les fluctuations de
temp\'erature de la grille absorbante. Comme le reste des matrices,
les thermom\`etres sont bas\'es sur une technologie
silicium\footnote{D'autres mat\'eriaux thermom\'etriques ont tout de
m\^eme \'et\'e envisag\'es et rapidement rejet\'es: le NTD Ge (il ne
permet pas une r\'ealisation collective des matrices), les TES
(inadapt\'es \`a une lecture par transistors MOS) et les couches
minces du type NbSi ou AuGe (\'el\'elements non-standards dans les
fili\`eres technologiques classiques).}. Toutefois, \`a 300~mK, les
porteurs de charges sont gel\'es dans la plupart des semiconducteurs,
il est alors n\'ecessaire de doper le silicium avec des impuret\'es de
type~n (donneur d'\'electrons) et de type~p (accepteur d'\'electrons)
pour permettre le transport de charges et offrir une fonction
thermom\'etrique aux bolom\`etres. Dans les mat\'eriaux faiblement
dop\'es, comme c'est le cas pour les bolom\`etres du CEA, le mode de
conduction principal est la conduction par sauts entre impuret\'es
(\og \emph{hopping conduction} \fg en anglais). Le silicium \'etant un
\'el\'ement t\'etravalent, le phosphore qui se trouve dans le groupe~V
de la table p\'eriodique joue le r\^ole de donneur d'\'electrons, et
le bore qui se trouve dans le groupe~III se comporte comme un
accepteur d'\'electrons. D'autre part, les atomes de phosphore ont des
niveaux d'\'energie proche des nombreux niveaux de la bande de
conduction dans le silicium, et la conduction \'electrique dans un
thermom\`etre silicium dop\'e phosphore s'effectue par sauts des
\'electrons entre atomes de phosphore neutres et atomes de phosphore
ionis\'es. La pr\'esence d'atomes de Bore\footnote{Une concentration
en bore deux fois moins \'elev\'ee qu'en phosphore donne les
r\'esultats les plus satisfaisants en terme de conduction
\'electrique.} est donc n\'ecessaire pour accepter les \'electrons
provenant du phosphore et ainsi permettre la conduction \'electrique
dans le mat\'eriau. Le saut des \'electrons entre deux atomes de
phosphore est activ\'e par la temp\'erature. Un grand nombre de
mod\`eles th\'eoriques ont \'et\'e d\'evelopp\'es pour expliquer les
mesures de r\'esistivit\'e \`a basses temp\'eratures, les principaux
\'etant le mod\`ele du \og \emph{Nearest Neighbor Hopping} \fg
\shortcite{miller}, celui du \og \emph{Variable Range Hopping} \fg
\shortcite{mott}, et celui du \og \emph{Coulomb gap} \fg
\shortcite{efros}. La r\'esistivit\'e du mat\'eriau doit donc
\'evoluer avec la temp\'erature comme suit~:
\begin{equation}
\rho=\rho_0\,exp\left(\frac{T_0}{T}\right)^n
\end{equation}
avec $n=1$ pour le mod\`ele de Miller-Abrahams, $n=1/4$ pour Mott et
$n=1/2$ pour Shklovskii-Efros. Les param\`etres $\rho_0$ et $T_0$
d\'ependent de la g\'eom\'etrie et du dopage du
thermom\`etre. D'apr\`es la th\`ese de \shortciteN{buzzi_these} qui
pr\'esente une analyse d\'etaill\'ee de ces trois mod\`eles, les
diff\'erents r\'egimes de conduction devraient se succ\'eder \`a
mesure que la temp\'erature diminue.

\begin{figure}
  \begin{center}
    \includegraphics[width=0.6\textwidth,angle=0]{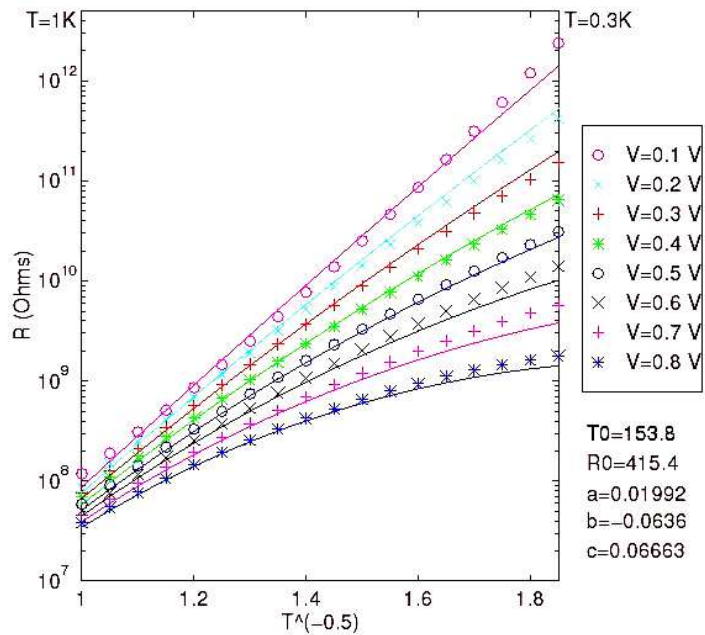}
  \end{center}
\caption[Ajustement avec loi d'Efros modifi\'ee pour les
thermom\`etres
$\mbox{Si}\!\!:\!\!\mbox{P}\!\!:\!\!\mbox{B}$]{\'Evolution de la
r\'esistance d'un thermom\`etre
$\mbox{Si}\!\!:\!\!\mbox{P}\!\!:\!\!\mbox{B}$ en fonction de sa
temp\'erature et de la tension de polarisation appliqu\'ee \`a ses
bornes. Les mesures, repr\'esent\'ees par les symboles, sont bien
ajust\'ees par le mod\`ele de Buzzi. Les coefficients \`a droite de la
courbe sont ceux qui ajuste le mieux les donn\'ees selon la loi
d'Efros modifi\'ee (\'equation~\ref{eq:efros}). Cette figure est
extraite de la th\`ese de \shortciteN{buzzi_these}.
\label{fig:detect_bolocea_fabrication_thermo_buzzieffros}}
\end{figure}

Notez toutefois que ces mod\`eles de conduction ne sont valables que
dans la limite des faibles champs \'electriques. En effet, la
probabilit\'e qu'un \'electron saute d'un atome de phosphore \`a un
autre peut \^etre modifi\'ee en appliquant un champ \'electrique \`a
travers le mat\'eriau. Les bolom\`etres du CEA \'etant lus par des
transistors MOS relativement bruyants, il est n\'ecessaire de
g\'en\'erer des forts signaux \'electriques en sortie des
bolom\`etres. Les thermom\`etres sont donc fortement polaris\'es pour
surpasser le bruit de l'\'electronique de lecture~: les effets de
champ ne sont alors plus n\'egligeables. En s'appuyant sur le mod\`ele
de \shortciteN{hill}, C.~Buzzi et P.~Agn\`ese ont r\'ealis\'e des
simulations num\'eriques de la conduction en utilisant des techniques
de percolation \shortcite{buzzi_these}, puis ils ont ajust\'e les
mesures exp\'erimentales effectu\'ees sur un thermom\`etre
$\mbox{Si}\!\!:\!\!\mbox{P}\!\!:\!\!\mbox{B}$ avec leur mod\`ele. Une
loi d'Efros modifi\'ee semble \^etre la plus repr\'esentative du
comportement des bolom\`etres. La
figure~\ref{fig:detect_bolocea_fabrication_thermo_buzzieffros} montre
l'ajustement des mesures pour diff\'erentes tensions de polarisation
et diff\'erentes temp\'eratures de fonctionnement. Pour un champ
direct, \cad parall\`ele \`a la direction de conduction, ils trouvent
que la formule analytique suivante d\'ecrit correctement l'\'evolution
de la r\'esistance $R$ du thermom\`etre en fonction de sa
temp\'erature $T$ et du champ \'electrique $E$ qui r\`egne en son
sein~:
\begin{equation}
R(T,E)=R_0\,exp\!\left(\sqrt{\frac{T_0}{T}}\right)exp\!\left(-\frac{eEL(T)}{k_BT}\right)
\label{eq:efros}
\end{equation}
o\`u $e$ est la charge \'el\'ementaire, $k_B$ est la constante de
Boltzmann et $L(T)$ est la longueur moyenne de saut entre deux atomes
de phosphore. Buzzi utilise un polyn\^ome du second ordre pour
repr\'esenter la longueur de saut~: $L(T)=aT^2+bT+c$.\\

Le CEA/LETI a fabriqu\'e de nombreux types de thermom\`etres de forme
et de dopage diff\'erents dans le but d'en extraire leurs param\`etres
physiques et de les ajuster aux besoins des d\'etecteurs PACS. Les
thermom\`etres carr\'es ou en serpentin montrent un fort effet de
champ de sorte que leur imp\'edance s'\'ecroule pour des tensions de
polarisation sup\'erieures \`a $\sim$100~mV. La g\'eom\'etrie qui
minimise les effets de champ est la g\'eom\'etrie longiligne. Les
thermom\`etres s\'electionn\'es pour les matrices de bolom\`etres sont
donc des bandes de 40~$\mu$m de large pour 600~$\mu$m de long.

\begin{figure}
  \begin{center}
    \includegraphics[width=0.8\textwidth,angle=0]{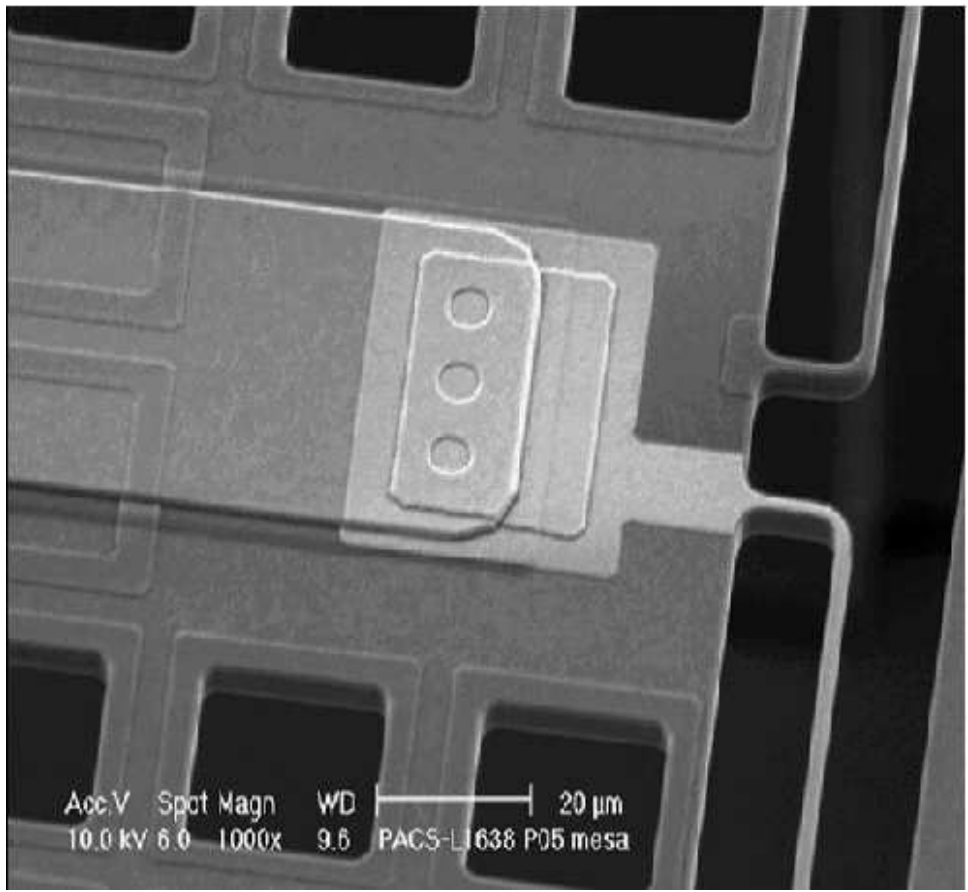}
  \end{center}
\caption[Photographie d'un thermom\`etre PACS en configuration
mesa]{Vue rapproch\'ee de l'extr\'emit\'e d'un thermom\`etre
$\mbox{Si}\!\!:\!\!\mbox{P}\!\!:\!\!\mbox{B}$ en configuration
mesa. Le thermom\`etre est bien en relief par rapport \`a la grille,
il est isol\'e du substrat par une couche de SiO$_2$ qui emp\`eche les
ions de diffuser dans la grille. La r\'egion claire \`a
l'extr\'emit\'e du thermom\`etre et sur la poutre du bas correspond au
d\'ep\^ot m\'etallique qui connecte \'electriquement la r\'esistance
au circuit de lecture. Notez la finesse des poutres de suspension
($2\times5$~$\mu$m) qui maintiennent la structure. Nous apercevons
\'egalement des motifs carr\'es qui entourent les trous perc\'es dans
la grille, ce sont des d\'ep\^ots de TiN qui servent \`a absorber le
rayonnement \'electromagn\'etique.
\label{fig:detect_bolocea_fabrication_thermo_photo}}
\end{figure}

D'autre part, la dose d'impuret\'es implant\'ees est choisie telle que
le coefficient $\alpha=\frac{\partial R}{\partial T}$ soit le plus
grand possible dans le domaine de fonctionnement des bolom\`etres
PACS. L'uniformit\'e du dopage est \'egalement un param\`etre
critique. La conductance \'electrique est en effet une fonction
exponentielle de la densit\'e d'impuret\'es, et si l'implantation
n'est pas homog\`ene, alors des courants non-uniformes peuvent se
former et engendrer des points chauds par auto-\'echauffement ou bien
g\'en\'erer un exc\`es de bruit \`a basses fr\'equences. Deux
m\'ethodes de fabrication ont \'et\'e test\'ees~:
\begin{itemize}
\item l'implantation du phosphore et du bore dans le corps du substrat
en silicium suivi d'un recuit \`a temp\'erature mod\'er\'ee pour
limiter la diffusion des impuret\'es,
\item la configuration mesa dans laquelle le thermom\`etre est
sur\'elev\'e par rapport \`a la grille et isol\'e par une fine couche
de SiO$_2$ imperm\'eable \`a la diffusion des ions, ce qui permet un
recuit \`a environ 1000\textdegree C pendant plusieurs heures.
\end{itemize}
Seule la deuxi\`eme m\'ethode autorise une bonne homog\'en\'eit\'e des
dopants dans le volume du thermom\`etre \shortcite{simoens}. Notez
toutefois que cette fa\c{c}on de proc\'eder n\'ecessite pas moins de
40~\'etapes suppl\'ementaires dans la fabrication des matrices par
rapport \`a une implantation dans le corps du substrat. La
figure~\ref{fig:detect_bolocea_fabrication_thermo_photo} montre une
vue rapproch\'ee d'un thermom\`etre en configuration mesa. Nous voyons
\'egalement un d\'ep\^ot m\'etallique sur le thermom\`etre et la
poutre qui joue le r\^ole de connecteur \'electrique entre la
r\'esistance et le circuit de lecture.

Les deux thermom\`etres qui composent le pont bolom\'etrique d'un
m\^eme pixel sont identiques en forme et en dopage (cf
figure~\ref{fig:detect_bolocea_grille_photopix}). La fabrication
collective des bolom\`etres garantit en fait que tous les
thermom\`etres d'une m\^eme matrice poss\`edent des caract\'eristiques
physiques identiques, ce qui n'est pas le cas des bolom\`etres \`a
base de germanium qui sont fabriqu\'es puis assembl\'es sur les
grilles absorbantes de fa\c{c}on individuelle.\\

Dans le cas plus sp\'ecifique des d\'etecteurs du Photom\`etre PACS
(cf section~\ref{sec:herschel_oservatoire_phfpu}), les matrices de
bolom\`etres qui \'equipent le plan focal sont de deux types. Elles se
diff\'erencient par le dopage de leurs thermom\`etres~: les
bolom\`etres du BFP bleu sont par construction plus imp\'edants que
ceux du BFP rouge. Mis \`a part cette diff\'erence, les matrices
bleues et rouges sont g\'eom\'etriquement identiques, \cad qu'elles
poss\`edent des absorbeurs, des cavit\'es r\'esonantes et des
thermom\`etres de m\^eme taille.

\begin{figure}
  \begin{center}
    \includegraphics[width=0.95\textwidth,angle=0]{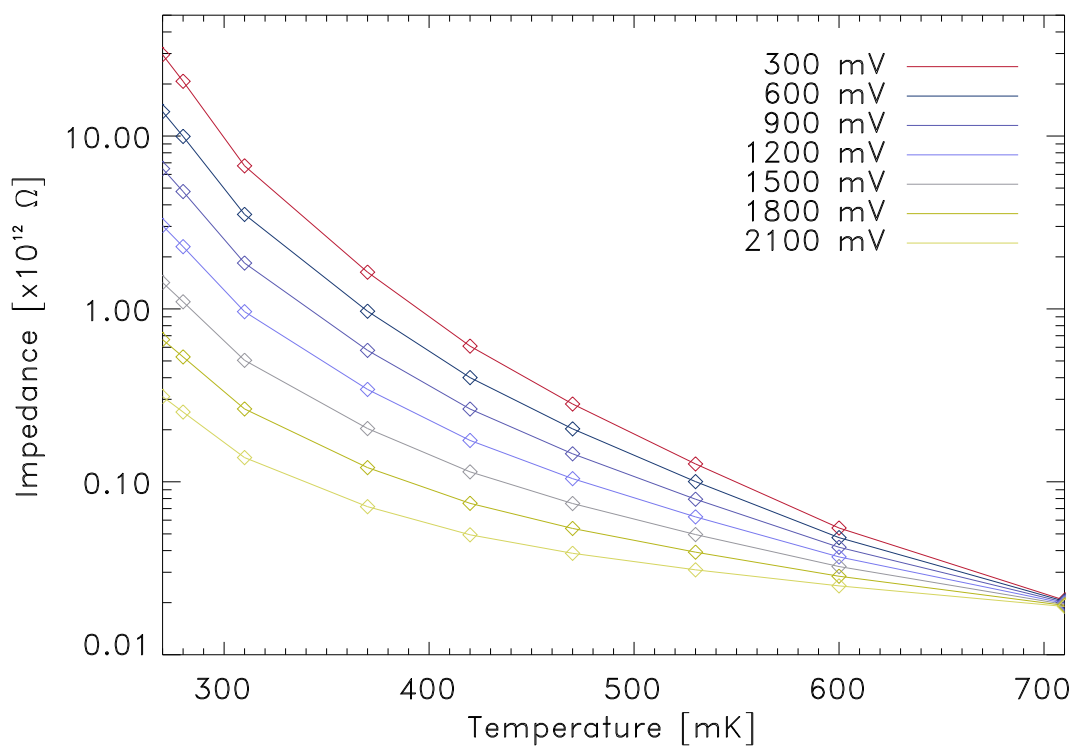}
  \end{center}
\caption[Imp\'edance des thermom\`etres PACS en fonction de la
temp\'erature et de la tension]{\'Evolution de l'imp\'edance moyenne
des thermom\`etres du mod\`ele de rechange PACS sur la voie bleue en
fonction de la temp\'erature et de la tension appliqu\'ee \`a leurs
bornes. Ces thermom\`etres sont du m\^eme type que ceux du mod\`ele de
vol rouge. L'imp\'edance est de l'ordre de quelques T$\Omega$ dans le
r\'egime de fonctionnement nominal des bolom\`etres PACS. L'effet de
champs r\'eduit consid\'erablement l'imp\'edance des thermom\`etres
lorsque la tension augmente. Remarquez que ces courbes ne
repr\'esentent pas le comportement d'un seul pixel mais celui d'un
pixel moyen calcul\'e sur deux matrices (512~bolom\`etres en
parall\`eles).
\label{fig:detect_bolocea_fabrication_thermo_R2T}}
\end{figure}

Des mesures r\'ecentes (cf section~\ref{sec:calib_perflabo_ajuste_IV}
pour plus de d\'etails) r\'ealis\'ees sur le mod\`ele de rechange
bleu\footnote{Notez que le dopage des thermom\`etres du mod\`ele de
rechange bleu est le m\^eme que celui des thermom\`etres du mod\`ele
de vol rouge.} du BFP PACS ont permis de calculer l'\'evolution de
l'imp\'edance des bolom\`etres en fonction de leur temp\'erature et de
la tension appliqu\'ee \`a leurs bornes. Les r\'esultats sont
pr\'esent\'es dans la
figure~\ref{fig:detect_bolocea_fabrication_thermo_R2T}. Nous mesurons
une imp\'edance de l'ordre de quelques T$\Omega$ dans le r\'egime de
fonctionnement des bolom\`etres PACS (cf
chapitres~\ref{chap:calib_procedure} et~\ref{chap:calib_perflabo} pour
la d\'etermination de ce r\'egime). D'autre part, nous retrouvons la
forte d\'ependance en temp\'erature n\'ecessaire aux thermom\`etres
pour atteindre l'objectif de quelques 10$^10$~V/W de r\'eponse.

\subsection{La cavit\'e r\'esonante}
\label{sec:detect_bolocea_fabrication_cavite}

Le principe d'absorption du rayonnement \'electromagn\'etique par une
cavit\'e r\'esonante a \'et\'e mis au point par les militaires durant
la deuxi\`eme guerre mondiale dans le but d'am\'eliorer la furtivit\'e
de leurs \'equipements. La surface des avions \'etaient par exemple
recouverte d'une multitude de petites structures capable d'absorber le
rayonnement \'emis par les radars. Lorsque cette technologie est
pass\'ee dans le domaine public, le CEA/LETI l'a reprise et adapt\'ee
\`a la d\'etection sub-millim\'etrique pour la bolom\'etrie
refroidie. Le concept a d'ailleurs fait l'objet d'un brevet. Ce
principe a depuis \'et\'e adopt\'e par la quasi-totalit\'e des
d\'etecteurs bolom\'etriques actuels, qu'ils soient coupl\'es au
t\'elescope par des c\^ones de Winston ou pas.

Le silicium mono-cristallin est transparent dans l'infrarouge
lointain. Il est donc n\'ecessaire de d\'eposer une fine couche d'un
mat\'eriau opaque sur le substrat en silicium pour permettre
l'absorption du rayonnement \'electromagn\'etique. Le bolom\`etre
d\'ecrit par \shortciteN{low} \'etait recouvert d'une couche de
peinture noire qui assurait une absorption proche de~100~\%, mais qui
apportait \'egalement un exc\`es de capacit\'e calorifique
r\'edhibitoire pour les performances du d\'etecteur. Les bolom\`etres
modernes sont plut\^ot recouverts de films m\'etalliques~; le bismuth
par exemple poss\`ede une tr\`es faible capacit\'e calorifique de par
sa nature de semi-m\'etal, ou bien le HgTe qui est couramment
utilis\'e en d\'etection~X \shortcite{stahle}. Le LETI utilise du
nitrure de titane (TiN) pour les matrices de bolom\`etres car c'est un
compos\'e standard en fili\`ere silicium. Ce mat\'eriau devient
supraconducteur en-dessous de $\sim$4~K, et il offre une tr\`es faible
capacit\'e calorifique ainsi qu'un bon coefficient d'absorption des
ondes sub-millim\'etriques \shortcite{buzzi_these}. Notez que c'est la
premi\`ere fois qu'un mat\'eriau supraconducteur est utilis\'e comme
absorbeur pour la bolom\'etrie.\\

La convertion d'une \'energie radiative en chaleur est d\'ecrite en
physique classique par deux ph\'enom\`enes~: la composante
\'electrique du champ de radiation g\'en\`ere le mouvement des
\'electrons libres contenus dans le m\'etal par la force de Lorentz,
et l'\'energie cin\'etique de ces \'electrons est transform\'ee en
chaleur par dissipation Joule. Une couche m\'etallique plac\'ee dans
un champ de radiation peut ainsi absorber un maximum th\'eorique de
50~\% de l'\'energie contenue dans l'onde \'electromagn\'etique. Le
principe de la cavit\'e r\'esonante est relativement simple, il
consiste \`a cr\'eer une onde stationnaire \`a l'aide d'une plaque
r\'efl\'echissante (le champ \'electrique \`a la surface d'un
conducteur parfait a une composante perpendiculaire nulle) et de
placer l'absorbeur \`a une distance de $\lambda/4$ au-dessus du
r\'eflecteur, \cad \`a l'endroit o\`u toute l'\'energie de l'onde
\'electromagn\'etique est concentr\'ee dans sa composante \'electrique
(ventre du champ \'electrique et n\oe ud du champ magn\'etique). En
pratique, cela revient \`a fabriquer une cavit\'e quart-d'onde entre
le circuit de lecture et le circuit de d\'etection, la taille de la
cavit\'e \'etant d\'etermin\'ee par le diam\`etre des billes
d'indium. Le r\'eflecteur utilis\'e pour les matrices de bolom\`etres
est une fine couche d'or d\'epos\'ee sous chacun des pixels (cf
figure~\ref{fig:detect_bolocea_principe_description_matriceoverview_photo}). Ce
type de r\'esonance est appel\'ee r\'esonance verticale. Lorsque la
r\'esistance de surface du m\'etal est adapt\'ee \`a l'imp\'edance du
vide (377~$\Omega_{/\square}$\footnote{L'unit\'e $\Omega_{/\square}$,
\og Ohm par carr\'e \fg, est utilis\'ee pour exprimer l'imp\'edance
des mat\'eriaux en couche mince~; leur \'epaisseur doit \^etre
constante et le courant doit circuler parall\`element \`a la
surface.}), elle permet une absorption th\'eorique de 100~\%. Le
profil d'absorption d'un tel syst\`eme optique est relativement
large~; il est cependant possible de le modifier l\'eg\`erement en
changeant le motif de l'absorbeur m\'etallique. Nous parlons alors de
r\'esonance horizontale. Plusieurs g\'eom\'etries de l'absorbeur ont
\'et\'e test\'ees (en grille, en croix ou en boucle, ce qui correspond
\`a des filtres passe-bas, passe-bande ou passe-haut). Le motif retenu
pour les matrices de bolom\`etres est celui de la g\'eom\'etrie en
boucle (cf figure~\ref{fig:detect_bolocea_fabrication_thermo_photo}).

L'absorption des matrices de bolom\`etres de type PACS, \cad avec des
billes d'indium de 20~$\mu$m de diam\`etre, a \'et\'e mesur\'ee par
l'\'equipe du SAp \`a l'aide d'un spectrom\`etre \`a transform\'ee de
fourier (FTS). La m\'ethode consiste \`a mesurer le spectre d'un corps
noir r\'efl\'echi sur un miroir, $S_{reference}$, puis le spectre du
m\^eme corps noir r\'efl\'echi sur la matrice \`a tester
($S_{echantillon}$). La mesure effectu\'ee sur le miroir que l'on
suppose parfaitement r\'efl\'echissant sert de r\'ef\'erence, elle
contient la contribution spectrale du syst\`eme de mesure (la source
de rayonnement, les filtres optiques, le d\'etecteur utilis\'e pour
mesurer le flux r\'efl\'echi, etc...). En faisant l'hypoth\`ese simple
que le rayonnement qui n'est pas r\'efl\'echi par la matrice est
effectivement absorb\'e, nous calculons le spectre d'absorption $A$ de
la matrice avec la formule suivante~:
\begin{equation}
A=1-\frac{S_{echantillon}}{S_{reference}}
\end{equation}

\begin{figure}
  \begin{center}
      \includegraphics[width=0.7\textwidth,angle=0]{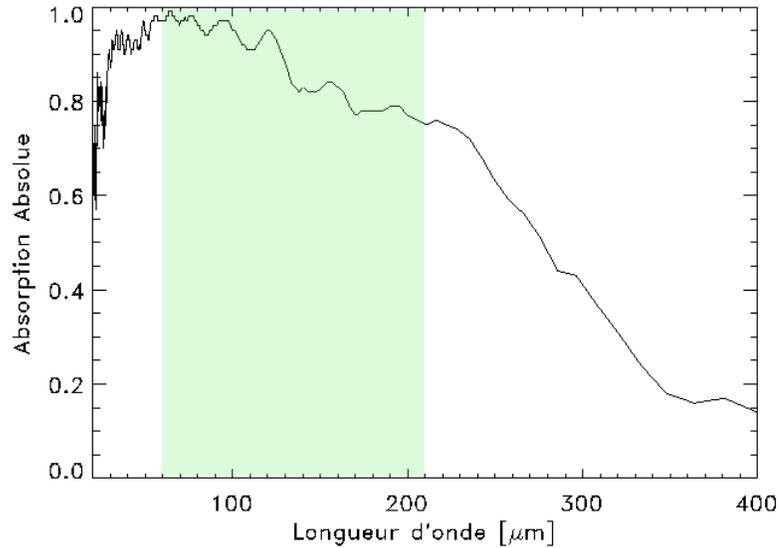}
  \end{center}
  \caption[Absorption absolue des matrices PACS]{Mesures d'absorption
  r\'ealis\'ees avec le FTS pour une matrice de type PACS, \cad avec
  une cavit\'e de 20~$\mu$m de profondeur. La bande PACS
  (60-210~$\mu$m) est sur-lign\'ee en vert sur la figure.
  \label{fig:detect_bolocea_fabrication_cavite_vincent}}
\end{figure}

Le r\'esultat de ces mesures est pr\'esent\'e dans la
figure~\ref{fig:detect_bolocea_fabrication_cavite_vincent}. Dans la
bande de PACS (60-210~$\mu$m), l'absorption est sup\'erieure \`a
75~\%.

L'absorption du rayonnement par cavit\'e r\'esonante est donc un
processus tr\`es efficace. Elle autorise de plus l'abandon des c\^ones
de Winston trop encombrants au niveau du plan focal (cf
section~\ref{sec:detect_bolocea_printpe_innovation}). Toutefois,
l'absorption n'est pas directive avec ce type de syst\`eme optique,
les d\'etecteurs sont donc tr\`es sensibles aux lumi\`eres
parasites. Il est par cons\'equent crucial de minimiser les sources
possibles de lumi\`ere parasite lors de la conception de
l'instrument. Dans le cas de PACS, les deux plans focaux sont
cloisonn\'es dans deux c\^ones dont les ouvertures d\'efinissent le
champ de vue des bolom\`etres (cf
section~\ref{sec:detect_observatoire_phfpu_description}). L'int\'erieur
des c\^ones est soigneusement noirci pour \'eviter les r\'eflections
internes.\\

\begin{figure}
  \begin{center}
    \begin{tabular}[t]{ll}
      \includegraphics[width=0.5\textwidth]{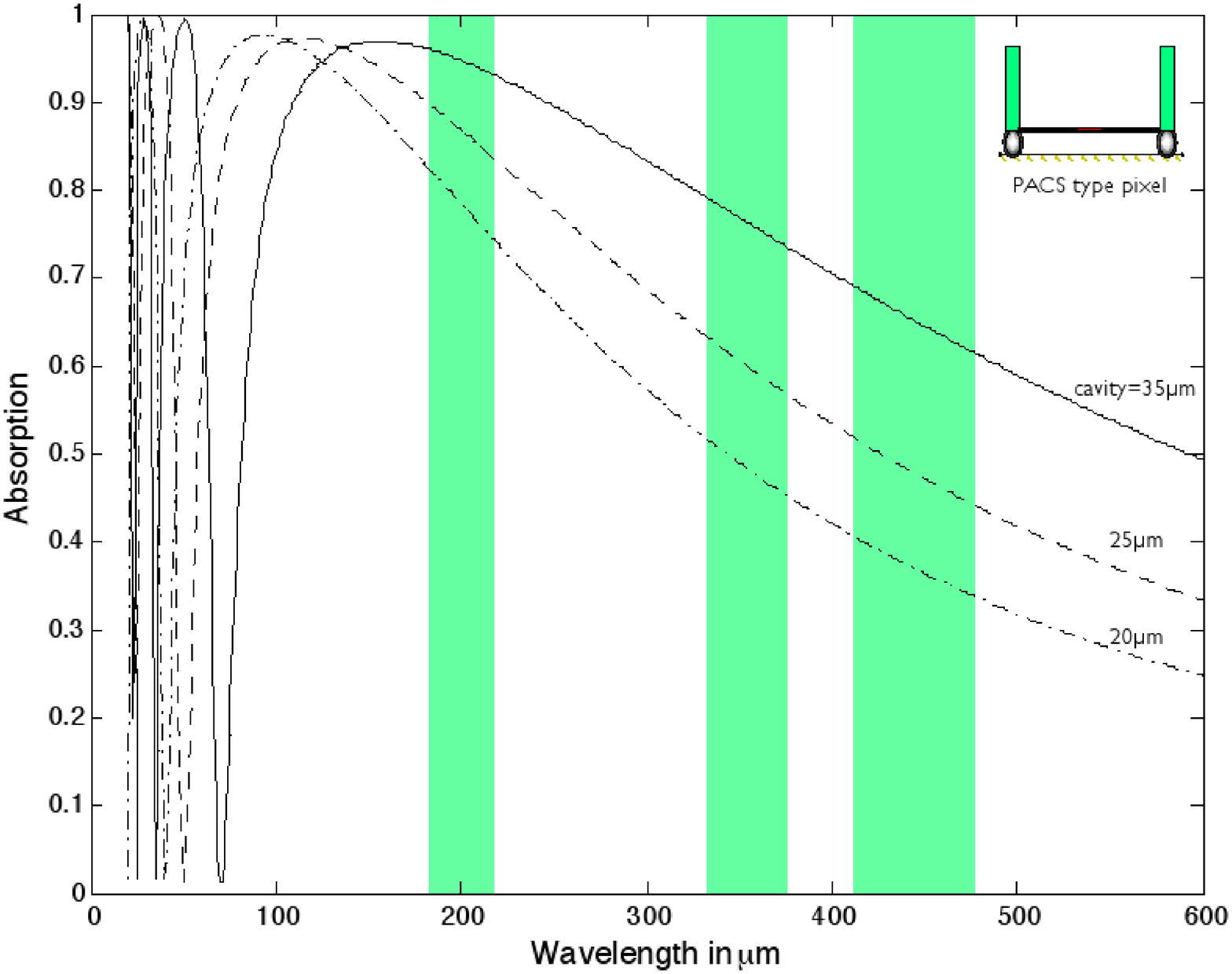} & \includegraphics[width=0.5\textwidth]{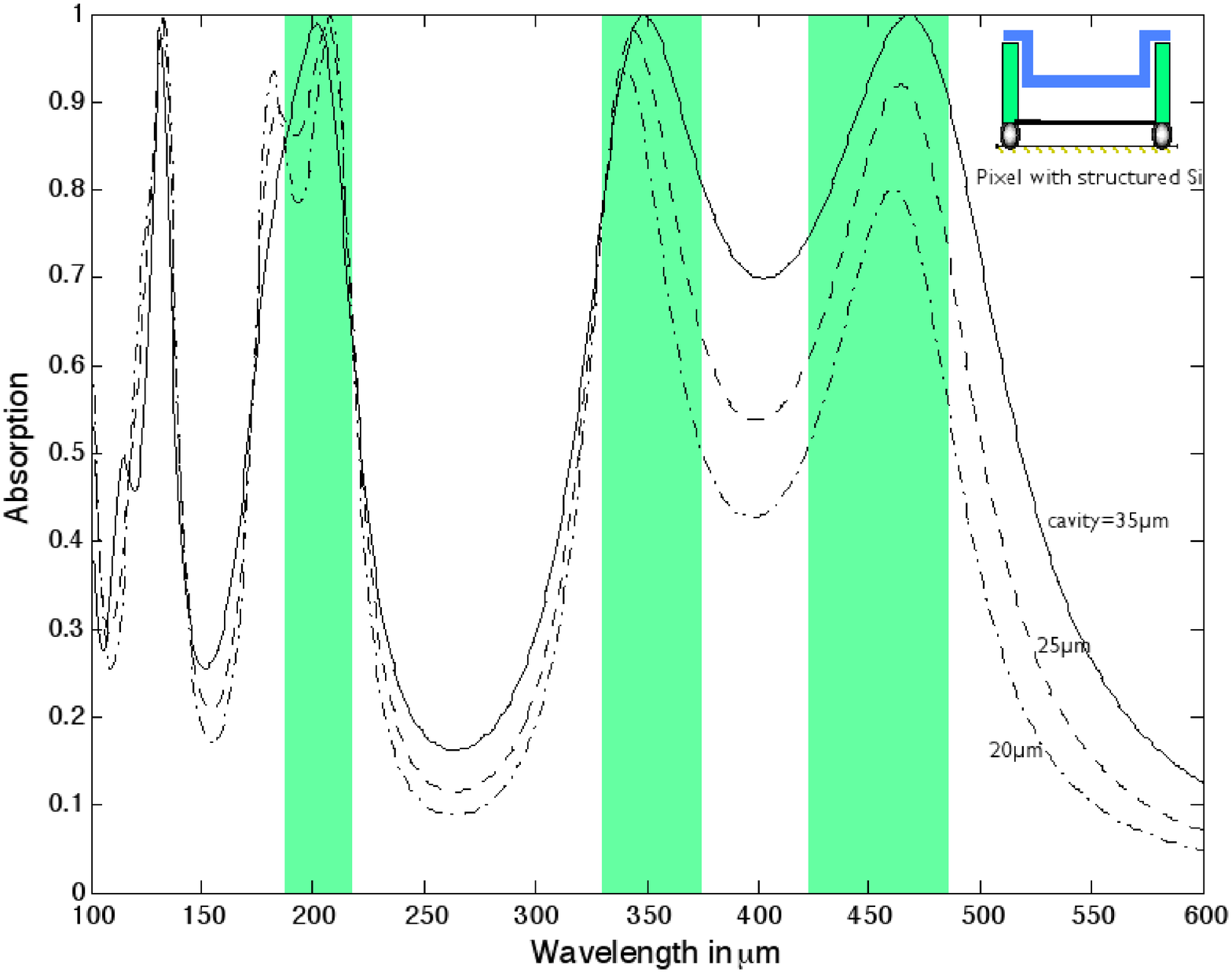} 
    \end{tabular}
  \end{center}
  \caption[Absorption des matrices de bolom\`etres
  modifi\'ees]{Simulations pr\'esentant l'absorption absolue des
  matrices de bolom\`etres du CEA pour trois \'epaisseurs de la
  cavit\'e r\'esonante. \`A gauche la matrice est de type PACS. \`A
  droite elle est recouverte d'une couche de di\'electrique
  structur\'ee qui permet d'obtenir une excellente absorption dans les
  trois bandes atmosph\'eriques \`a 200, 350 et 450~$\mu$m. Cette
  figure est extraite de \shortciteN{reveret}.
  \label{fig:detect_bolocea_fabrication_cavite_absVincent}}
\end{figure}

En principe, il est possible de d\'ecaler le domaine spectral
d'absorption des matrices de bolom\`etres en adaptant la taille de la
cavit\'e. \`A plus basses longueurs d'ondes que la bande PACS,
l'int\'er\^et pour l'astronomie est discutable puisque les
photoconducteurs traditionnels pr\'esentent de meilleures performances
en termes de sensibilit\'e. Cependant, le LETI s'est lanc\'e dans le
d\'eveloppement de cam\'eras bolom\'etriques non-cryog\'eniques
fonctionnant \`a plus courtes longueurs d'ondes pour l'industrie de la
surveillance. Aux plus grandes longueurs d'ondes, il est important de
mentioner le travail de th\`ese de Vincent Rev\'eret sur l'adaptation
des matrices de bolom\`etres pour la fen\^etre atmosph\'eriques \`a
1300~$\mu$m \shortcite{reveret_these}. \`A une telle longueur d'onde,
il n'est pas envisageable d'augmenter la taille des billes d'indium
jusqu'\`a 325~$\mu$m alors que les murs interpixels ne font que
50~$\mu$m de large~; la cavit\'e serait alors pleine d'indium et
l'efficacit\'e d'absorption chuterait radicalement. D'autre part, il
faut alt\'erer le moins possible le design des matrices PACS pour
minimiser le co\^ut de d\'eveloppement. Une autre solution technique a
\'et\'e mise au point pour \og pi\'eger \fg la composante \'electrique
de l'onde incidente dans la cavit\'e r\'esonante~: il s'agit de
cr\'eer une couche anti-reflet sur chacun des pixels \`a l'aide d'une
couche de di\'electrique. Cette couche est plane ou structur\'ee selon
la taille de l'airgap n\'ecessaire (l'airgap est le volume
d\'elimit\'e par les murs interpixels, la couche de di\'electrique et
la grille suspendue). En se basant sur le formalisme des couches
minces, V. Rev\'eret a montr\'e qu'un empilement de type [\emph{couche
di\'electrique - airgap - substrat silicium - absorbeur - cavit\'e -
r\'eflecteur}] permettait d'obtenir une absorption efficace jusqu'\`a
1300~$\mu$m. La
figure~\ref{fig:detect_bolocea_fabrication_cavite_absVincent}
pr\'esente des courbes d'absorption calcul\'ees pour diff\'erentes
tailles de bille avec et sans couche di\'electrique. Nous voyons qu'il
est possible d'obtenir des matrices de bolom\`etres adapt\'ees aux
trois fen\^etres atmosph\'eriques \`a 200, 350 et 450~$\mu$m avec une
seule couche structur\'ee de di\'electrique \shortcite{reveret}. Cette
technologie a \'et\'e test\'ee sur le t\'elescope APEX pour une
cam\'era prototype d\'evelopp\'ee par le CEA~: P-ARTEMIS. Quelques
plan\`etes du syst\`eme solaire ainsi qu'une r\'egion de formation
d'\'etoiles tr\`es brillante ont \'et\'e observ\'ees avec succ\`es \`a
450~$\mu$m par P-ARTEMIS. L'instrument ARTEMIS devrait voir le jour
courant 2009, il contiendra 16 matrices de 288~pixels chacune. Une
description d\'etaill\'ee de la cam\'era est donn\'ee dans l'article
de \shortciteN{talvard}.

\section{L'\'electronique de lecture}
\label{sec:detect_bolocea_elec}

Les bolom\`etres sont mont\'es en ponts diviseurs de tension comme
indiqu\'e sur la
figure~\ref{fig:detect_bolocea_principe_description_matriceoverview_schema}. Ils
sont aliment\'es par les tensions $V_h$ et $V_l$. L'\'equilibre
\'electrique du pont \'evolue avec la temp\'erature du
thermom\`etre. Le signal utile se situe donc au milieu des ponts
bolom\'etriques. Le r\^ole de l'\'electronique de lecture est de
transporter la tension des points milieux de chaque pixel jusqu'\`a
l'ext\'erieur du cryostat puis de la convertir en valeur num\'erique
pour la stocker.

\begin{sidewaysfigure}
  \begin{center}
      \includegraphics[width=1.0\textwidth,angle=0]{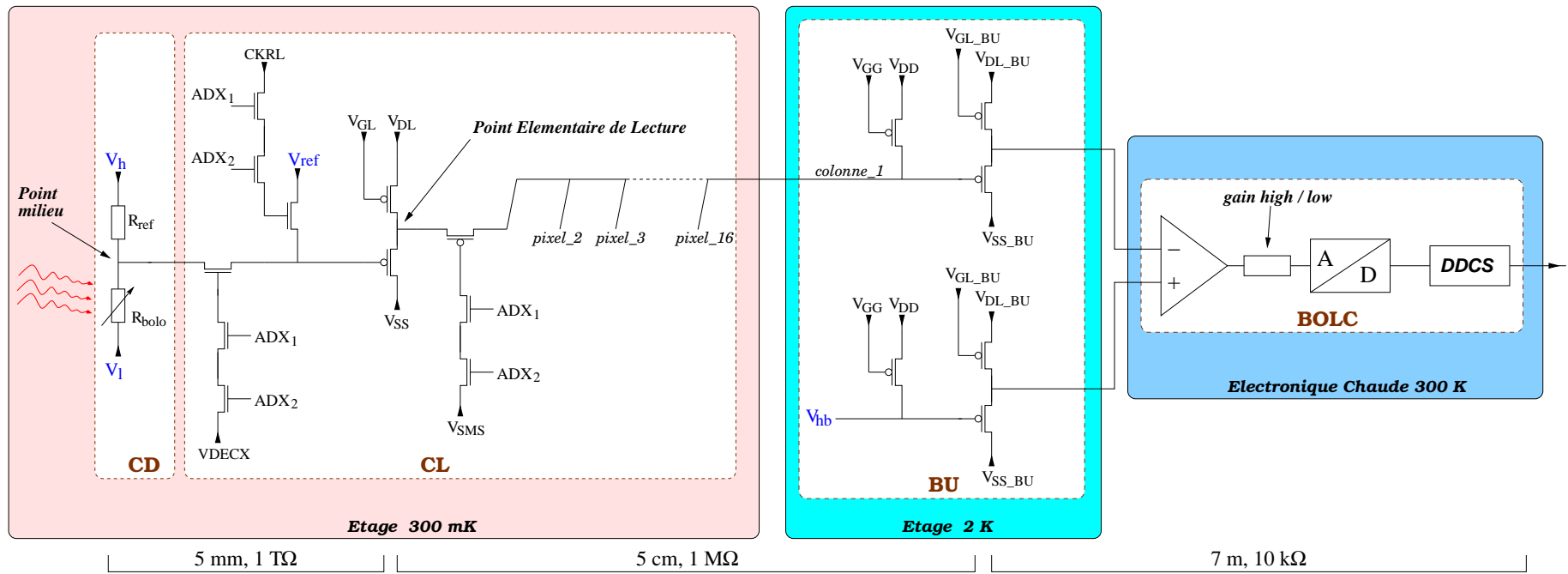}
  \end{center}
  \caption[Sch\'ema detaill\'e de l'\'electronique de lecture du
  Photom\`etre PACS]{Sch\'ema \'electronique du circuit de lecture des
  matrices de bolom\`etres. L'\'electronique se compose de trois
  \'etages d'adaptation d'imp\'edance maintenus \`a des temp\'eratures
  allant de 300~mK \`a 300~K. L'imp\'edance passe de quelques
  $10^{12}$~$\Omega$ au niveau du Circuit de D\'etection (\emph{CD})
  \`a environ $10^4$~$\Omega$ au niveau de l'\'electronique chaude
  (\emph{BOLC}). L'adaptation d'imp\'edance est r\'ealis\'ee par les
  deux circuits suiveurs command\'es par les tensions $V_{SS}$ et
  $V_{SS\_BU}$ sur la figure. Les deux transistors MOS aliment\'es par
  la tension $VDECX$ permettent de couper ou de fermer la ligne qui
  relie le bolom\`etre au circuit de lecture en aval. En fait certains
  transistors sont utilis\'es comme interrupteur. De la m\^eme
  fa\c{c}on les deux transistors MOS aliment\'es par $CKRL$ permettent
  d'injecter la tension de r\'ef\'erence $V_{ref}$ dans le circuit de
  lecture.  Les deux transistors MOS aliment\'es par la tension
  $V_{SMS}$ assurent le multiplexage des pixels (une seul pixel est
  montr\'e sur la figure). L'\'etage \`a 2~K (\emph{BU}, Buffer Unit)
  est compos\'e d'un circuit suiveur qui abaisse une seconde fois
  l'imp\'edance de la ligne pour pouvoir transmettre le signal
  \'electrique jusqu'\`a l'\'electronique chaude (BOLC) situ\'ee \`a
  plus de 7~m. La tension de r\'ef\'erence $V_{hb}$ est inject\'ee au
  niveau du \emph{BU}. Voir le texte pour une explication plus
  d\'etaill\'ee des fonctions qui composent ce circuit de lecture.
  \label{fig:detect_bolocea_elec_froide_principe}}
\end{sidewaysfigure}

Le circuit de lecture des matrices est un sous-syst\`eme critique des
d\'etecteurs, et son r\'eglage est d\'eterminant pour les performances
des bolom\`etres. Il est utile \`a ce stade du manuscrit de donner une
explication d\'etaill\'ee de son fonctionnement pour saisir les
subtilit\'es que nous avons mises \`a jour durant la calibration des
d\'etecteurs (chapitres~\ref{chap:calib_procedure}
et~\ref{chap:calib_perflabo}). Le circuit de lecture \'etant
relativement complexe, je vais prendre un soin particulier \`a rendre
sa description la plus claire et intelligible possible. Par
cons\'equent, je n'insisterai pas sur l'aspect technologique de
l'\'electronique mais plut\^ot sur la description fonctionnelle des
divers \'el\'ements du circuit. De plus, je me r\'ef\`ererai
r\'eguli\`erement au sch\'ema de la
figure~\ref{fig:detect_bolocea_elec_froide_principe} pour \'etayer mes
explications. Cette figure donne en effet une vue synth\'etique de
toute l'\'electronique de lecture.

\subsection{L'\'electronique froide}
\label{sec:detect_bolocea_elec_froide}

Pour aller \`a l'essentiel, nous pourrions r\'esumer la fonction de
l'\'electronique froide en deux points~: adaptation d'imp\'edance et
multiplexage.\\
\noindent Les thermom\`etres poss\`edent en effet des imp\'edances
gigantesques, de l'ordre de $10^{12}-10^{13}$~$\Omega$, et une si
grande imp\'edance n'est pas compatible avec des c\^ables
\'electriques long de plusieurs m\`etres. La difficult\'e majeure pour
transporter une tension \'electrique le long d'un circuit tr\`es haute
imp\'edance consiste \`a pouvoir r\'eduire suffisamment la capacit\'e
\'electrique r\'esiduelle des composants ainsi que celle des lignes
\'electriques (la capacit\'e des lignes est proportionnelle \`a leur
longueur). En effet, une r\'esistance et un condensateur s'associent
en s\'erie pour former un filtre passe-bas qui peut \'eventuellement
filtrer le signal utile et ainsi mener \`a une perte
d'information. L'adaptation d'imp\'edance se fait en deux \'etapes. La
premi\`ere a lieu au niveau du \emph{CL} \`a 300~mK, \`a seulement
quelques millim\`etres des bolom\`etres. Le but est d'abaisser
l'imp\'edance de $\sim$10$^{12}$~$\Omega$ \`a $\sim$10$^6$~$\Omega$
pour obtenir une fr\'equence de coupure de 1.5~kHz compatible avec la
fr\'equence d'\'echantillonnage de l'\'electronique de lecture. La
seconde a lieu \`a 2~K au niveau du \emph{Buffer Unit} (\emph{BU} sur
la figure~\ref{fig:detect_bolocea_elec_froide_principe}) pour abaisser
l'imp\'edance \`a $\sim$10$^3$~$\Omega$ et ainsi permettre au signal
de parcourir les 7~m de c\^able n\'ecessaire pour rejoindre
l'\'electronique chaude (\emph{BOLC}) qui se situe dans le module de
service du satellite. Ces deux \'etages d'adaptation d'imp\'edance ont
lieu sur deux circuits diff\'erents (\emph{CL} et \emph{BU}) pour des
raisons li\'ees aux budgets thermiques disponibles \`a 300~mK et \`a
2~K. Le composant \'electronique utilis\'e pour abaisser l'imp\'edance
est un circuit suiveur de tension compos\'e de deux transistors MOS en
s\'erie. Un suiveur de tension poss\`ede une imp\'edance infinie en
entr\'ee et un gain unit\'e. Il isole, en terme d'imp\'edance, le
circuit en amont du circuit en aval. Le circuit suiveur du \emph{CL}
est control\'e par les tensions $V_{GL}$, $V_{DL}$ et $V_{SS}$, celui
du \emph{BU} par $V_{GL\_BU}$, $V_{DL\_BU}$ et $V_{SS\_BU}$. La
figure~\ref{fig:detect_bolocea_elec_froide_principe} montre
l'emplacement des MOS suiveurs ainsi que leurs tensions
associ\'ees. Nous retrouvons \'egalement sur la figure la
temp\'erature, l'imp\'edance et la longueur des lignes \'electriques
de chacun des \'etages d'adaptation d'imp\'edance.

En ce qui concerne le multiplexage, nous avons d\'ej\`a mention\'e
dans la section~\ref{sec:detect_bolocea_printpe_description} sa
n\'ecessit\'e pour lire le signal de milliers de pixels fonctionnant
\`a 300~mK tout en respectant le budget thermique tr\`es serr\'e \`a
cette temp\'erature. L'\'electronique froide des matrices poss\`ede
une fonction de multiplexage qui permet de lire tous les bolom\`etres
d'une m\^eme colonne avec le m\^eme circuit de lecture. Le nombre de
fils n\'ecessaire pour extraire le signal d'une matrice passe donc
de~256 \`a~16.  Nous allons maintenant nous servir de la
figure~\ref{fig:detect_bolocea_elec_froide_principe} pour expliquer le
fonctionnement du multiplexeur. Notez cependant que, par souci de
clart\'e, la figure ne montre pas les 256 pixels du \emph{CL} mais
seulement le premier pixel de la premi\`ere colonne. De la m\^eme
mani\`ere, le \emph{BU} contient jusqu'\`a 32~entr\'ees, une seule est
visible sur la figure. Notez \'egalement que les transistors MOS qui
composent le circuit de multiplexage sont utilis\'es comme des
interrupteurs, \cad qu'ils ne peuvent prendre que deux \'etats
(passant ou bloquant) suivant la tension que nous appliquons \`a leurs
bornes. La lecture de chaque pixel s'effectue au \emph{Point
\'El\'ementaire de Lecture}, ou \emph{PEL}, qui se situe sur le
\emph{CL} \`a la sortie du premier \'etage d'adaptation d'imp\'edance,
comme indiqu\'e sur la figure. Le multiplexage se joue au niveau du
premier MOS juste apr\`es le \emph{PEL}, celui qui est control\'e par
les tensions $ADX_1$, $ADX_2$ et $V_{SMS}$. \`A tour de r\^ole, le
transistor de chacun des pixels devient passant et connecte le BU au
\emph{PEL} du pixel. Un m\^eme pixel n'est donc \og vu \fg par le
\emph{BU} qu'un seizi\`eme du temps. Les tensions $ADX_1$ et $ADX_2$
sont des tensions qui permettent d'adresser un pixel en particulier
pendant que les 15~autres sont \'electriquement d\'econnect\'es du
\emph{BU}. La tension $V_{SMS}$ sert \`a valider l'adressage.

Les premi\`eres mesures effectu\'ees sur les matrices de bolom\`etres
ont \'et\'e obtenues dans un mode particulier que nous appelons
\emph{PEL statique}. Dans ce mode, le circuit suiveur du \emph{CL} est
aliment\'e en permanence par les tensions $V_{GL}$, $V_{DL}$ et
$V_{SS}$. Cependant ce suiveur ne sert qu'un seizi\`eme du temps pour
la lecture du \emph{PEL} lorsque le pixel en question est
adress\'e. Aujourd'hui, les matrices sont utilis\'ees dans un autre
mode que nous appelons \emph{PEL commut\'e}. Dans ce mode, les
tensions $V_{GL}$ et $V_{DL}$ sont \'egales de sorte que le courant
dans le suiveur ne circule pas. Ce sont en fait les tensions
$V_{GL\_BU}$ et $V_{DL\_BU}$ qui servent d'alimentation et activent le
suiveur au moment o\`u le pixel est adress\'e uniquement, \cad un
seizi\`eme du temps. Le grand avantage du mode \emph{PEL commut\'e}
r\'eside dans le fait que la dissipation \'electrique est
consid\'erablement r\'eduite au niveau du \emph{CL}.\\


\subsection{Les modes de lecture}
\label{sec:detect_bolocea_elec_lecture}

Bien que l'\'electronique froide d\'ecrite pr\'ec\'edemment soit le
v\'eritable c\oe ur des matrices de bolom\`etres, c'est \emph{BOLC}
qui orchestre l'activit\'e \'electrique du \emph{CL} et du
\emph{BU}. Il g\'en\`ere toutes les tensions visibles sur la
figure~\ref{fig:detect_bolocea_elec_froide_principe} et coordonne
l'adressage des pixels ainsi que la conversion des signaux analogiques
en valeurs num\'eriques. Les composants qui assurent ces conversions
sont des \emph{ADC} (\emph{Analog-to-Digital Converter}) 16-bits dont
la dynamique peut \^etre ajust\'ee en changeant le gain de
l'\'electronique de \emph{BOLC}. En mode nominal de fonctionnement, le
gain de \emph{BOLC} est tel que l'\emph{ADC} \'echantillonne
correctement le bruit du signal \'electrique, \cad que le bruit
\emph{r.m.s.} doit occuper plusieurs pas codeurs pour ne pas se
retrouver limit\'e par le bruit de num\'erisation. Dans ce mode que
nous appelons \emph{gain fort}, un pas codeur vaut 5~$\mu$V, ce qui
donne une dynamique totale de 330~mV. \emph{BOLC} offre une autre
alternative, le \emph{gain faible}, pour laquelle le gain est r\'eduit
d'un facteur~4, ce qui am\`ene la dynamique totale des \emph{ADC} \`a
1.3~V. Nous verrons le grand int\'er\^et du gain faible pour
l'\'etalonnage des matrices dans le chapitre suivant. Le lecteur
pourra se r\'ef\'erer \`a l'annexe~\ref{a:dyna_BOLC} pour plus de
d\'etails sur la dynamique des convertisseurs num\'eriques.\\

\emph{BOLC} joue de plus le r\^ole d'interface entre le
photom\`etre et le reste de l'instrument PACS. En particulier, il
re\c{c}oit et ex\'ecute les t\'el\'ecommandes envoy\'ees par
l'utilisateur. Il fournit \'egalement au \emph{SPU} les conversions
num\'eriques du signal ainsi que les HK (les \og \emph{House Keeping}
\fg contiennent les informations d'\'etat du photom\`etre telles que
la temp\'erature des diff\'erentes parties de la cam\'era, les
tensions envoy\'ees vers l'\'electronique froide, etc...). 

La fr\'equence d'\'echantillonnage des points milieux est fix\'ee par
\emph{BOLC}~; sa valeur nominale est de 40~Hz. Cependant, la faible
bande passante de l'antenne \`a haut gain du satellite n\'ecessite de
moyenner 4~images successives dans le \emph{SPU} avant de les
transmettre vers la Terre (cf
section~\ref{sec:herschel_oservatoire_phfpu}). La fr\'equence
effective des donn\'ees du Photom\`etre PACS est donc de 10~Hz. Nous
avons naturellement effectu\'es la campagne d'\'etalonnage du
Photom\`etre PACS en moyennant 4~images successives pour \^etre le
plus repr\'esentatif possible des conditions de l'instrument dans
l'espace. N\'eanmoins, nous avons relax\'e cette contrainte pour tous
les tests effectu\'es \`a Saclay ce qui nous permet d'obtenir
l'information spectrale contenue dans les donn\'ees jusqu'\`a 20~Hz,
\cad la fr\'equence de Nyquist (cf
sections~\ref{sec:calib_perflabo_sensibilite_bruit},~\ref{sec:calib_perflabo_tau_fourier}
et~\ref{sec:calib_perflabo_compare_sequenceur}).\\

L'\'electronique de lecture des matrices de bolom\`etres permet
d'effectuer des mesures diff\'erentielles du signal bolom\'etrique~;
le but \'etant de corriger les d\'erives \'eventuelles du circuit de
lecture ou des bolom\`etres eux-m\^eme. Les sources de bruits
parasites peuvent \^etre de nature tr\`es diff\'erentes, comme par
exemple la d\'erive en temp\'erature du plan focal, la d\'erive des
gains et offsets des transistors, la microphonie\footnote{La
microphonie d\'ecrit le ph\'enom\`ene dans lequel les vibrations
m\'ecaniques d'un syst\`eme \'electrique se transforment en signaux
\'electriques ind\'esirables. Le terme de microphonie provient de
l'analogie avec les v\'eritables microphones pour lesquels ce
ph\'enom\`ene est intentionnel plut\^ot qu'involontaire. Notez
toutefois que les matrices de bolom\`etres sont relativement peu
sensibles aux bruits microphoniques du fait de la tr\`es bonne
qualit\'e des contacts \'electriques et de l'utilisation de c\^ables
flex en kapton.} ou bien les perturbations
\'electromagn\'etiques\footnote{L'environnement \'electromagn\'etique
du Photom\`etre PACS risque de ne pas \^etre tr\`es stable.  Les
panneaux solaires du satellite par exemple pr\'esentent une lente
oscillation du courant qu'ils g\'en\`erent produisant ainsi un flux
magn\'etique variable \`a proximit\'e du Photom\`etre. Cette \og
respiration \fg des panneaux peut potentiellement induire des courants
parasites dans l'\'electronique de lecture des matrices. Ce type de
bruit a \'et\'e largement att\'enu\'e en donnant une masse commune \`a
tous les \'el\'ements du d\'etecteur (\emph{CL}, \emph{BU} et
\emph{BOLC}) et gr\^ace aux mesures diff\'erentielles d\'ecrites dans
le corps du texte.}. Pour r\'ealiser ces mesures diff\'erentielles,
nous utilisons deux tensions de r\'ef\'erence que \emph{BOLC} injecte
au niveau de l'\'electronique froide ($V_{hb}$ et $V_{ref}$ sont
visibles en bleu sur la
figure~\ref{fig:detect_bolocea_elec_froide_principe}). Il est donc en
th\'eorie possible de corriger toutes les d\'erives qui interviennent
en aval du point d'injection des tensions de
r\'ef\'erence. \emph{BOLC} offre plusieurs modes de lecture du signal
bolom\'etrique, dont deux ont \'et\'e explor\'es de fa\c{c}on
exhaustive~: le mode \emph{direct} et une de ces variantes que nous
appelons le mode \emph{DDCS} (\emph{Double Differential Correlated
Sampling}).

\subsubsection{Le mode direct}

Le design originel des matrices contenait deux colonnes de pixels
\emph{aveugles}. Ces pixels \'etaient en tous points identiques aux
autres pixels dit \emph{actifs}, sauf qu'ils \'etaient couverts et ne
recevaient aucun flux radiatif. Il \'etait cependant possible de
changer la temp\'erature de la grille absorbante des pixels aveugles
par le biais d'une chaufferette, nous pouvions alors les placer dans
un r\'egime de fonctionnement similaire \`a celui des pixels actifs
(flux radiatif absorb\'ee~$\Longleftrightarrow$~puissance \'electrique
dissip\'ee). Leur r\^ole consistait \`a mesurer les d\'erives du
signal dues aux fluctuations de la temp\'erature du bain thermique ou
de l'\'electronique de lecture et ainsi de corriger une partie du
bruit corr\'el\'e contenu dans le signal. En pratique, il \'etait
tr\`es difficile de r\'egler les chaufferettes avec pr\'ecision~; le
signal des pixels aveugles \'etait par cons\'equent tr\`es dispers\'e
ce qui entra\^inait une s\'ev\`ere perte d'information par saturation
des convertisseurs num\'eriques. Les matrices actuelles du
Photom\`etre PACS poss\`edent toujours ces deux colonnes de pixels
aveugles mais ils sont \'electriquement d\'econnect\'es du circuit de
lecture. C'est maintenant la tension $V_{hb}$, qui est inject\'ee au
niveau du \emph{BU}, qui remplace le signal des pixels aveugles.

Le mode de lecture direct consiste \`a prendre la diff\'erence entre
le signal bolom\'etrique et la tension de r\'ef\'erence
$V_{hb}$. Cette op\'eration est r\'ealis\'ee en entr\'ee de
\emph{BOLC} par un soustracteur analogique (le point milieu rentre sur
la patte $\circleddash$ de l'amplificateur). Les deux tensions sont
soustraites de fa\c{c}on synchrone. Le r\'esultat est alors amplifi\'e
par le gain de \emph{BOLC} avant d'\^etre num\'eris\'e par les
\emph{ADC} puis envoy\'e vers le \emph{SPU}. Remarquez que le mode
direct pourrait \^etre renomm\'e mode simple diff\'erenciation pour
\^etre plus explicite. $V_{hb}$ \'etant une tension constante choisie
par l'op\'erateur, elle peut \^etre consid\'er\'ee comme un offset qui
permet de changer le niveau d'entr\'ee de l'\emph{ADC} de
\emph{BOLC}.

Dans le domaine lin\'eaire de fonctionnement de l'\'electronique de
lecture (cf section~\ref{sec:calib_procedure_vrlvhb}), nous pouvons
\'ecrire la formule analytique suivante qui relie le signal de sortie
de BOLC, $V_{sortie}$, aux autres tensions du circuit:
\begin{equation}
  V_{sortie} = G_{BOLC} \times G_{BU} \times [V_{hb} -( G_{CL}
  \times V_{entree} +O_{CL})]
\label{eq:mode_lecture_direct}
\end{equation}
o\`u chacun des termes de l'\'equation et leurs d\'ependances
vis-\`a-vis des param\`etres du syst\`eme est d\'ecrit dans la liste
suivante~:
\begin{itemize}
\item $G_{BOLC}$ est le gain de \emph{BOLC}, il peut prendre deux
valeurs (cf annexe~\ref{a:dyna_BOLC}).
\item $G_{BU}$ est le gain des transistors du \emph{BU}. Il d\'epend
de la temp\'erature du \emph{BU}, du courant qui l'alimente
$I_{VSS\_BU}$, du num\'ero de la colonne lue ainsi que de la tension
d'entr\'ee du \emph{BU}.
\item $G_{CL}$ et $O_{CL}$ sont les gains et offsets des
transistors\footnote{Les transistors sont caract\'eris\'es par leur
gains et offsets. L'offset d\'epend du seuil des transistors.} du
\emph{CL}. Ils d\'ependent de la temp\'erature du \emph{CL}, du
courant qui l'alimente $I_{VSS}$, de la position du pixel sur la
matrice ainsi que de la tension d'entr\'ee du \emph{CL}.
\item $V_{hb}$ est la tension de r\'ef\'erence fournie par \emph{BOLC}.
\item $V_{entree}$ est la tension d'entr\'ee du \emph{CL}. En mode
direct, elle vaut $V_{ptmil}$, la tension du point milieu~; mais elle
peut aussi prendre la valeur $V_{ref}$ qui est l'autre tension de
r\'ef\'erence fournie par \emph{BOLC}.
\end{itemize}
Remarquez que le signal bolom\'etrique et la tension $V_{hb}$ ne
circulent pas dans les m\^emes transistors au niveau du \emph{BU} (cf
figure~\ref{fig:detect_bolocea_elec_froide_principe})~; il faudrait
donc en toute rigueur diff\'erencier les gains et offsets des MOS qui
transportent le signal bolom\'etrique de ceux qui transportent
$V_{hb}$. Toutefois, ces MOS sont identiques par construction et sont
de surcroit aliment\'es par les m\^emes tensions g\'en\'er\'ees par
\emph{BOLC}. Nous consid\'erons alors que les gains et offsets du
\emph{BU} sont identiques~; c'est la raison pour laquelle $O_{BU}$
n'appara\^it pas dans l'equation~(\ref{eq:mode_lecture_direct}) et que
$G_{BU}$ a pu \^etre factoris\'e.

\subsubsection{Le mode DDCS}

\begin{figure}
  \begin{center}
    \includegraphics[width=1.0\textwidth,angle=0]{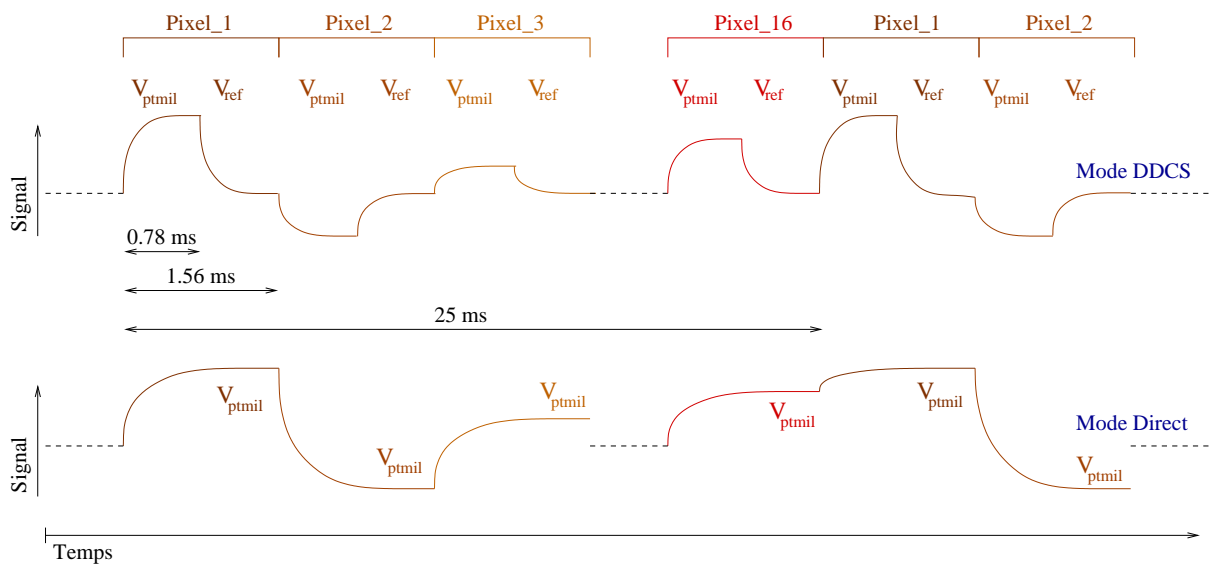}
  \end{center}
\caption[\'Evolution temporelle du signal d'entr\'ee de \emph{BOLC} en
mode DDCS]{\'Evolution temporelle du signal d'entr\'ee de \emph{BOLC}
en mode DDCS et en mode direct pour une colonne contenant
16~pixels. En mode DDCS, le signal s'\'etablit successivement au
potentiel du point milieu puis de $V_{ref}$. Le multiplexeur
s\'electionne le pixel suivant et le signal s'\'etablit \`a nouveau au
potentiel du point milieu du second pixel puis \`a celui de
$V_{ref}$. La tension $V_{ref}$ est choisie par l'op\'erateur, elle
est commune \`a tous les pixels d'un m\^eme \emph{BU}. Par contre les
points milieux sont dispers\'es \`a cause de l'inhomog\'en\'eit\'e du
processus d'implantation ionique. Les temps caract\'eristiques de
chaque \'etape du s\'equencement sont \'egalement montr\'es sur la
figure. En mode direct, les points milieux se succ\`edent. Dans les
deux modes de lecture, le temps d'\'etablissement du signal est fix\'e
par la constante de temps du circuit \`a moyenne imp\'edance.
\label{fig:detect_bolocea_elec_lecture_seq_schemaPEL}}
\end{figure}

Gr\^ace \`a la tension $V_{ref}$ qui est inject\'ee \`a l'entr\'ee du
\emph{CL}, il est possible de r\'ealiser une soustraction
suppl\'ementaire entre $V_{ptmil}$ et $V_{ref}$ qui permet de corriger
le signal bolom\'etrique des d\'erives de toute la cha\^ine
\'electronique. Le mode DDCS offre donc une lecture doublement
diff\'erentielle du signal d'o\`u son nom \emph{Double Differential
Correlated Sampling}. Notez que les tensions $V_{ptmil}$ et $V_{ref}$
\'etant ind\'ependantes, le bruit associ\'e au mode DDCS est en
th\'eorie la somme quadratique des bruits associ\'es \`a chacune de
ces tensions (cf section~\ref{sec:calib_perflabo_compare_sequenceur}
et annexe~\ref{a:seq}).

La lecture en mode DDCS consiste \`a intercaler une mesure de
$V_{ref}$ entre deux mesures successives de points milieux. Cette
fonction est remplie par les tensions $VDECX$ et $CKRL$ qui jouent le
r\^ole d'interrupteurs entre le pont bolom\'etrique et la tension
$V_{ref}$ respectivement et le PEL. En effet, leur mode de
fonctionnement est similaire \`a celui de la tension $V_{SMS}$~; \cad
que les tensions d'adressage $ADX_1$ et $ADX_2$ s\'electionnent le
pixel \`a lire, puis, selon leur valeur, $VDECX$ ouvre ou ferme la
connection \'electrique entre le \emph{CD} et le PEL, et $CKRL$ ouvre
ou ferme la connection entre $V_{ref}$ et le PEL (cf
figure~\ref{fig:detect_bolocea_elec_froide_principe}). En pratique,
\emph{BOLC} commande les tensions $VDECX$ et $CKRL$ en suivant une
s\'equence temporelle pr\'ed\'efinie par l'op\'erateur de sorte qu'il
peut \'echantillonner alternativement $V_{ptmil}$ et $V_{ref}$ sur
chacun des pixels. Nous appelons cette s\'equence temporelle le
\emph{s\'equenceur}.  

La figure~\ref{fig:detect_bolocea_elec_lecture_seq_schemaPEL} montre
l'\'evolution temporelle du signal \'electrique qui entre dans
\emph{BOLC} dans les deux modes de lecture. En mode DDCS, la
s\'equence commence avec le PEL connect\'e au pont bolom\'etrique du
pixel\_1, \cad $VDECX$ est passant~; le signal s'\'etablit au
potentiel du point milieu avec une constante de temps de l'ordre de
0.1~ms (la fr\'equence de coupure du \emph{BU} est d'environ 1500~Hz),
puis les $VDECX$ et $CKRL$ commutent pour d\'econnecter le point
milieu et relier la tension de r\'ef\'erence au PEL. Nous voyons donc
le signal s'\'etablir \`a la valeur $V_{ref}$ avec la m\^eme constante
de temps. Le multiplexeur adresse ensuite le pixel suivant et
r\'ep\`ete cette succession de $V_{ptmil}$ et $V_{ref}$. Remarquez sur
la figure que la valeur des tensions $V_{ref}$ est commune \`a tous
les pixels de la colonne\footnote{$V_{ref}$ est, comme toutes les
autres tensions envoy\'ees par \emph{BOLC}, commune \`a tous les
pixels d'un m\^eme \emph{BU}.} alors que les tensions de points
milieux sont dispers\'ees autour de $V_{ref}$ \`a cause de la
dispersion d'imp\'edance intrins\`eque des bolom\`etres. Lorsque les
16~pixels d'une m\^eme colonne ont \'et\'e lus, le multiplexeur
retourne sur le pixel\_1 de la m\^eme colonne et ainsi de suite. Le
temps qui s\'epare deux lectures successives d'un m\^eme pixel est
l'inverse de la fr\'equence d'\'echantillonnage, \cad 25~ms.

Le temps qui s\'epare la lecture de deux pixels successifs vaut un
seizi\`eme de la p\'eriode d'\'echantillonnage, \cad 1.56~ms. Enfin,
le temps qui s\'epare les mesures de points milieux et de $V_{ref}$
vaut la moiti\'e du temps pass\'e \`a adresser un m\^eme pixel, \cad
0.78~ms. Ces temps caract\'eristiques sont report\'es sur la
figure~\ref{fig:detect_bolocea_elec_lecture_seq_schemaPEL}. La partie
inf\'erieure de la figure pr\'esente la m\^eme colonne de pixels lue
cette fois-ci en mode direct. De la m\^eme mani\`ere le multiplexeur
passe d'un pixel \`a l'autre et le signal d'entr\'ee de \emph{BOLC}
est donc la succession des points milieux de chaque pixel sans passer
par la tension $V_{ref}$. En mode direct, les $VDECX$ et $CKRL$ sont
statiques.

\begin{figure}
  \begin{center}
    \includegraphics[width=0.8\textwidth,angle=0]{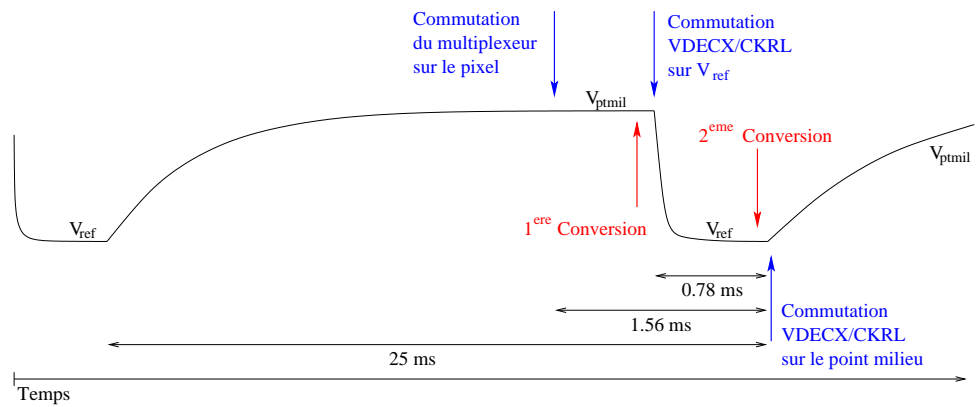}
  \end{center}
\caption[\'Evolution temporelle du signal au PEL d'un pixel en mode
DDCS]{\'Evolution temporelle du signal au niveau du PEL d'un pixel en
mode DDCS. Le PEL est connect\'e au pont bolom\'etrique pendant quinze
seizi\`eme du temps pour laisser le temps au point milieu de
s'\'etablir (la constante de temps du bolom\`etre est relativement
longue). \emph{BOLC} effectue une premi\`ere conversion num\'erique
qui contient $V{hb}-V_{ptmil}$. Apr\`es la commutation des $VDECX$ et
$CKRL$, le potentiel du PEL s'\'etablit \`a $V_{ref}$ assez rapidement
et \emph{BOLC} effectue une seconde conversion num\'erique sur
$V_{ref}$. Le signal de sortie de \emph{BOLC} vaut
$(V_{hb}-V_{ptmil})-(V_{hb}-V_{ref})$, soit $V_{ref}-V_{ptmil}$.
\label{fig:detect_bolocea_elec_lecture_seq_schemapix}}
\end{figure}

Alors qu'en mode direct la soustraction est analogique et synchrone,
la diff\'erence suppl\'ementaire du mode DDCS est num\'erique et
d\'ecal\'ee dans le temps de 780~$\mu$s. En effet, \emph{BOLC}
effectue une premi\`ere conversion num\'erique lorsque le PEL est au
potentiel du point milieu (cf
figure~\ref{fig:detect_bolocea_elec_lecture_seq_schemapix}), la valeur
de $V_{hb}-V_{ptmil}$ est alors stock\'ee dans un
r\'egistre. \emph{BOLC} effectue une seconde conversion num\'erique
sur $V_{ref}$ et stocke la valeur $V_{hb}-V_{ref}$ dans un autre
registre. En mode DDCS, \emph{BOLC} fournit la diff\'erence entre la
premi\`ere et la deuxi\`eme conversion de sorte que le signal de
sortie peut s'exprimer comme suit~:
\begin{equation}
  V_{sortie} = G_{BOLC} \times G_{BU} \times G_{CL} \times (V_{ref} -
  V_{ptmil})
\label{eq:mode_lecture_DDCS}
\end{equation}
o\`u $V_{hb}$ et les offsets des MOS \emph{CL} et \emph{BU} sont
\'elimin\'es du fait de la double diff\'erence.\\

Nous avons maintenant d\'efini les principales fonctions de
l'\'electronique de lecture, nous allons donc arr\'eter ici sa
description. Toutefois, nous reviendrons ponctuellement sur certains
aspects de son fonctionnement tout au long du manuscrit pour expliquer
et interpr\'eter le comportement des matrices de bolom\`etres.



\chapter{La proc\'edure d'\'etalonnage}
\label{chap:calib_procedure}

\begin{center}
\begin{minipage}{0.85\textwidth}

\small 

Ce chapitre pr\'esente les premi\`eres \'etapes de la proc\'edure
d'\'etalonnage que nous avons mise au point pour caract\'eriser le
Photom\`etre PACS. Nous montrerons dans un premier temps que la
m\'ethode g\'en\'eralement utilis\'ee pour \'etalonner les
bolom\`etres traditionnels n'est pas appliquable aux matrices de
bolom\`etres du CEA. Nous pr\'esenterons ensuite le d\'eroulement de
la proc\'edure de test qui nous permet aujourd'hui d'automatiser le
r\'eglage des d\'etecteurs, et qui nous fournit de surcro\^it une base
de donn\'ee compl\`ete sur laquelle nous nous appuyons pour comprendre
et pr\'edire le comportement des matrices. Ce travail est le fruit
d'une collaboration \'etroite avec Louis Rodriguez, Olivier Boulade et
Eric Doumayrou.

\end{minipage}
\end{center}

\section{Contexte et strat\'egie}
\label{sec:detect_outils_contexte}

De par leur conception originale, \`a savoir le multiplexage de
256~ponts bolom\'etriques mont\'es en parall\`ele et l'utilisation de
r\'esistances de charge sensibles aux effets de champ, les matrices de
bolom\`etres du CEA ne peuvent pas \^etre caract\'eris\'ees de la
m\^eme mani\`ere que les bolom\`etres r\'esistifs individuels
traditionnels. La raison est simple, nous n'avons pas acc\`es \`a une
des observables sur laquelle repose la quasi-totalit\'e des m\'ethodes
d'\'etalonnage. Cette (in-)observable est en fait le courant qui
circule dans chacun des bolom\`etres.

L'\'etalonnage d'une cam\'era bolom\'etrique n\'ecessite en effet de
pouvoir caract\'eriser la r\'esistance des senseurs thermiques en
mesurant des familles de \og \emph{load curves} \fg. Ces courbes
repr\'esentent le courant qui traverse un bolom\`etre en fonction de
la tension \'electrique appliqu\'ee \`a ses bornes~; en fran\c{c}ais
nous les appelons des courbes de charge ou bien des courbes~I-V.
Associ\'ees \`a un mod\`ele \'electro-thermique du bolom\`etre, elles
permettent d'extraire les param\`etres physiques du d\'etecteur tels
que par exemple la temp\'erature de l'absorbeur, la conductance
thermique des poutres ou bien la r\'eponse du bolom\`etre
\shortcite{griffin_holland}. 

Dans cette section, nous commen\c{c}ons par expliquer bri\`evement la
m\'ethode standard utilis\'ee pour \'etalonner un bolom\`etre
r\'esistif en nous appuyant sur deux articles qui pr\'esentent
clairement les diff\'erentes \'etapes de la proc\'edure
\shortcite{sudiwala,woodcraft}. Nous insisterons particuli\`erement
sur les limites de cette m\'ethode vis-\`a-vis des matrices de
bolom\`etres du CEA. Puis nous justifierons la d\'emarche qui nous a
pouss\'e \`a choisir une approche exp\'erimentale pour caract\'eriser
nos d\'etecteurs. Le d\'eroulement et les r\'esultats de la
proc\'edure de test que nous avons mis au point seront pr\'esent\'es
en d\'etails dans la suite du chapitre.

\subsection{Des outils existants inutilisables}
\label{sec:detect_outils_loadcurves}

La mesure de courbes de charge a depuis toujours \'et\'e un
\'el\'ement clef dans la caract\'erisation des bolom\`etres, et ce \`a
juste titre. En effet, il est crucial de conna\^itre parfaitement le
comportement du senseur thermique sur lequel repose le fonctionnement
d'un bolom\`etre. Ce senseur \'etant une \og simple \fg r\'esistance,
elle est caract\'eris\'ee en mesurant le courant qui la traverse en
fonction de la diff\'erence de potentiel qui lui est
appliqu\'ee. \shortciteN{jones} a d\'evelopp\'e une th\'eorie
g\'en\'erale sur les performances de bolom\`etres en se basant sur une
analyse \'electro-thermique. Le point de d\'epart de cette th\'eorie
est une courbe~I-V dont il extrait entre autre l'imp\'edance
\'electrique du bolom\`etre ainsi que sa r\'eponse. \shortciteN{low}
utilise \'egalement des courbes~I-V pour caract\'eriser un bolom\`etre
en germanium refroidi \`a 2~K. Dans l'article de
\shortciteN{zwerdling}, nous retrouvons une analyse tr\`es compl\`ete
du fonctionnement et des performances d'un bolom\`etre pour
l'infrarouge lointain. Le mod\`ele qu'il \'elabore fait aussi appel
\`a ces incontournables \og \emph{load curves} \fg. Les exemples sont
multiples \shortcite{duncan,wang,turner}. Pour illustrer l'utilisation
concr\`ete de courbes~I-V, j'ai choisi les articles plus r\'ecents de
\shortciteN{sudiwala} et de \shortciteN{woodcraft} qui pr\'esentent
conjointement la caract\'erisation d'un bolom\`etre de type spider-web
pour la bande \`a 143~GHz de l'instrument Planck/HFI.\\

D'apr\`es \shortciteN{sudiwala}, l'objectif de la proc\'edure standard
est de mesurer des courbes~I-V pour diff\'erentes temp\'eratures du
bain thermique, avec et sans flux radiatif incident~; ceci dans le but
de d\'eterminer exp\'erimentalement un jeu de param\`etres qui
d\'ecrit les propri\'et\'es physiques et le comportement des
mat\'eriaux mis en jeu dans l'\'equilibre thermique et \'electrique du
bolom\`etre. Commen\c{c}ons par introduire le formalisme et les
briques de base sur lesquelles repose le mod\`ele
\'electro-thermique. Le bolom\`etre est d\'efini de fa\c{c}on
suivante~: c'est une thermistance d'imp\'edance $R$ et de
temp\'erature $T$ qui est faiblement coupl\'ee \`a un puit de chaleur
de temp\'erature $T_0$ par une fuite thermique de conductance statique
$G_S$. Un courant $I$ traverse la thermistance cr\'eant ainsi une
diff\'erence de potentiel \`a ses bornes $V=I\times R$. Le courant est
habituellement g\'en\'er\'e par une r\'esistance de charge $R_L$
mont\'ee en s\'erie avec une source de tension. L'\'energie totale
dissip\'ee dans le bolom\`etre est $W=P+Q$ o\`u $Q$ est la puissance
radiative absorb\'ee et $P=V\times I$ est la puissance \'electrique
dissip\'ee par la thermistance elle-m\^eme. En r\'egime stationnaire,
l'\'energie \'evacu\'ee vers le puit de chaleur est
$W=G_S\times(T-T_0)$. \\
\noindent \shortciteN{sudiwala} exprime la conductance statique comme
une loi de puissance~:
\begin{equation}
G_S(\phi)=\frac{{G_S}_0}{(\beta+1)}\left(\frac{\phi^{\beta+1}-1}{\phi-1}\right)
\label{eq:gs}
\end{equation}
o\`u ${G_S}_0$ est la conductance thermique \`a $T_0$, $\beta$
l'exposant de la loi de puissance et $\phi=\frac{T}{T_0}$. Comme nous
l'avons d\'ej\`a pr\'esent\'e dans la
section~\ref{sec:detect_bolocea_fabrication_thermo}, le m\'ecanisme de
conduction dominant \`a 300~mK pour un semi-conducteur dop\'e est la
conduction par saut. La r\'esistance du bolom\`etre \'evolue alors
avec la temp\'erature et le champ \'electrique comme indiqu\'e dans
l'\'equation~(\ref{eq:efros}). Toutefois, les bolom\`etres classiques
fonctionnent avec des tensions \'electriques si faibles que l'effet de
champ peut en g\'en\'eral \^etre n\'eglig\'e. L'existence d'un
\'eventuel d\'ecouplage \'electron-phonon est \'egalement
n\'eglig\'e. L'imp\'edance de la thermistance s'\'ecrit alors~:
\begin{equation}
R(T) = R^{*} \, exp \left(\left[\frac{T_g}{T}\right]^n\right)
\label{eq:R2T}
\end{equation}
o\`u $R^*$ et $T_g$ d\'ependent du mat\'eriau et du dopage, et $n$
d\'epend du m\'ecanisme de conduction. Nous pouvons \'egalement
d\'efinir un autre param\`etre important souvent utilis\'e comme
figure de m\'erite des performances d'un bolom\`etre, c'est le
coefficient de temp\'erature~:
\begin{equation}
\alpha=\frac{1}{R}\frac{dR}{dT}=-\frac{nT_g^n}{T^{n+1}}
\end{equation}
La r\'eponse th\'eorique du bolom\`etre en fonction de la fr\'equence
de modulation du flux incident $\omega$ s'\'ecrit~:
\begin{equation}
S(\omega)=\frac{S(0)}{\sqrt{1+\omega^2\tau_e^2}}
\end{equation}
$$\mbox{avec}\,\,\,\,\,\, S(0)=\frac{dV}{dQ}=\frac{\alpha
V}{G_e}\left[\frac{R_L}{R+R_L}\right] \mbox{, }\,\,\,\,\,\,
\tau_e=C/G_e\,\, \,\,\,\,\mbox{ et }\,\,\,\,\,\, G_e=G_d-\alpha
P\left[\frac{R_L-R}{R_L+R}\right]$$
o\`u $S(0)$ est la r\'eponse \`a fr\'equence nulle, $\tau_e$ est la
constante de temps effective, $G_d=\frac{dW}{dT}$ est la conductance
thermique dynamique, et $G_e$ est la conductance thermique effective
qui rend compte de l'effet de r\'etroaction \'electro-thermique, ou
\emph{electrothermal feedback} en anglais \shortcite{mather82}. En
d\'efinissant $\delta=T_g/T_0$ et $\gamma=Q/({G_S}_0T_0)$ et en
faisant l'hypoth\`ese que $R_L\gg R$, i.e. $R_L$ est utilis\'e comme
source de courant constant, \shortciteN{sudiwala} exprime $S(0)$ en
fonction des param\`etres d\'efinis pr\'ec\'edemment~:
\begin{equation}
S(0)=\sqrt{\frac{R^*}{{G_S}_0T_0}}\left[
\frac{n\delta^n\sqrt{exp\left(\left[\frac{\delta}{\phi}\right]^n\right)
\left[\frac{\phi^{\beta+1}-1}{\beta+1}-\gamma\right]}}
{\phi^{\beta+n+1}+n\delta^n\left[\frac{\phi^{\beta+1}-1}{\beta+1}-\gamma\right]}\right]
\end{equation}
Cette formule n'est en r\'ealit\'e qu'un exemple parmi tant d'autres~;
le mod\`ele analytique permet en effet d'exprimer virtuellement toutes
les grandeurs pertinentes \`a l'utilisation d'un bolom\`etre,
notamment sa NEP, sa constante de temps ou encore la puissance
radiative incidente. L'int\'er\^et d'un tel mod\`ele est clair: il
nous permet en th\'eorie de pr\'edire les performances d'un
bolom\`etre \`a partir d'une poign\'ee de param\`etres physiques. Il
ne reste plus qu'\`a mesurer ces quelques param\`etres et \`a
v\'erifier que le mod\`ele s'applique bien au d\'etecteur dans le
domaine de fonctionnement explor\'e.\\

\begin{figure}[!htb]
  \begin{center}
      \includegraphics[width=0.6\textwidth,angle=0]{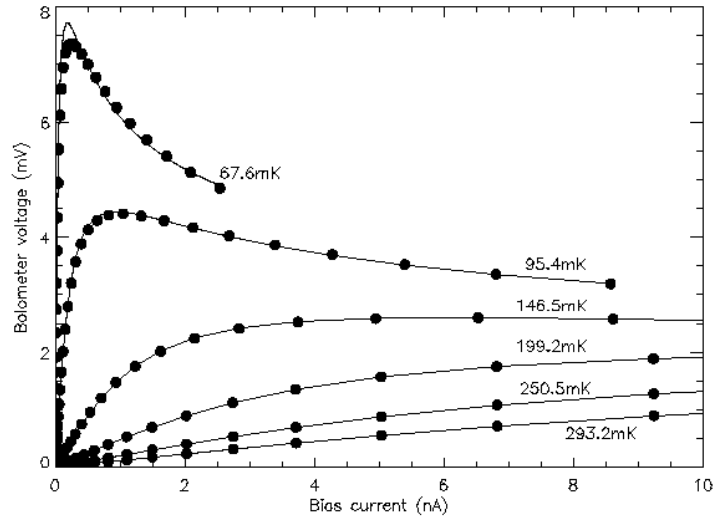}
  \end{center}
  \caption[Exemple de courbes~I-V mesur\'ees pour un bolom\`etre de
  Planck/HFI]{Cette figure est extraite de \shortciteN{woodcraft}. Les
  points repr\'esentent une famille de courbes~I-V mesur\'ees \`a
  diff\'erentes temp\'eratures du bain thermique. Les bolom\`etres ne
  sont pas illumin\'es ce qui permet d'extraire les param\`etres
  $R^*$, $T_g$ et $n$ en extrapolant vers une dissipation Joule
  nulle. Les courbes en trait plein montre l'ajustement des donn\'ees
  qui permettent d'obtenir les param\`etres ${G_S}_0$ et $\beta$. \`A
  basses temp\'eratures, le mod\`ele ne reproduit plus parfaitement le
  comportement du bolom\`etre.
  \label{fig:detect_outils_loadcurves_woodcraft}}
\end{figure}

La premi\`ere \'etape de la proc\'edure standard consiste \`a
d\'eterminer les param\`etres $R^*$, $T_g$ et $n$ \`a partir de
courbes~I-V mesur\'ees pour diff\'erentes temp\'eratures du bain
thermique $T_0$ en s'assurant qu'aucune charge optique n'est
absorb\'ee par le d\'etecteur. En effet, pour ces mesures, la
dissipation \'electrique de la thermistance doit \^etre la seule
source d'\'energie pour le bolom\`etre. L'id\'ee est de calculer
l'imp\'edance du bolom\`etre dans la limite o\`u le courant qui le
traverse tend vers 0~A. Dans ce cas, la dissipation Joule tend
vers~0~W et la temp\'erature de la grille tend vers la temp\'erature
du bain qui est connue. Nous pouvons alors calculer $R(T_0)$ pour
chacune des temp\'eratures test\'ees et ainsi caract\'eriser la
thermistance du bolom\`etre. \\
\noindent Ensuite, connaissant $R^*$, $T_g$ et $n$, il est possible
d'extraire les param\`etres ${G_S}_0$ et $\beta$ \`a partir des
m\^emes courbes~I-V en les ajustant avec le mod\`ele.  Cela
n\'ecessite de param\'etriser les grandeurs $I$ et $V$ \`a partir de
la variable $T$ (ou $\phi$). En associant la loi d'Ohm et la formule
de dissipation \'electrique nous pouvons \'ecrire $V=\sqrt{P\times
R(T)}$ et $I=\sqrt{P/R(T)}$~; d'autre part le bilan thermodynamique de
la grille suspendue nous donne~:
\begin{equation}
P(\phi)=\frac{{G_S}_0T_0}{\beta+1}(\phi^{\beta+1}-1)-Q
\end{equation}
avec $Q=0$ pour les courbes~I-V \`a ajuster. La
figure~\ref{fig:detect_outils_loadcurves_woodcraft} donne l'exemple
d'un tel ajustement pour un bolom\`etre prototype de Planck/HFI \`a
diff\'erentes temp\'eratures du bain thermique. L'ajustement du
mod\`ele est excellent au-dessus de 100~mK~; toutefois, il se
d\'egrade \`a plus basses temp\'eratures. La raison invoqu\'ee par
\shortciteN{woodcraft} est l'apparition d'une imp\'edance thermique
additionnelle qui serait due au d\'ecouplage \'electron-phonon dans la
thermistance~; il ajoute qu'un effet de champ aurait le m\^eme impact
sur les ajustements du param\`etre $\beta$.

La proc\'edure d'\'etalonnage continue par des mesures de courbes~I-V
\`a diff\'erentes temp\'eratures pour des bolom\`etres charg\'es
optiquement ($Q\neq0$). Le but principal est de mesurer l'efficacit\'e
de la cha\^ine optique, mais ces courbes permettent \'egalement de
v\'erifier la coh\'erence des param\`etres ${G_S}_0$ et $\beta$ dans
des configurations du syst\`eme proches des v\'eritables conditions
d'utilisation du d\'etecteur. \shortciteN{woodcraft} mesurent des
courbes de charge pour deux flux diff\'erents et en d\'eduit le
r\'eponse du bolom\`etre. Celle-ci est en accord avec la pr\'ediction
du mod\`ele. Cependant, \`a cause d'une d\'eg\'en\'erescence du
mod\`ele, l'ajustement de la r\'eponse ne peut en aucun cas \^etre
utilis\'e pour extraire les param\`etres physiques du probl\`eme
(plusieurs jeux de param\`etres peuvent reproduire une m\^eme courbe
de r\'eponse).\\

La pr\'esentation de ce mod\`ele montre qu'il existe des outils
standards relativement performants pour caract\'eriser les
bolom\`etres r\'esistifs. Toujours est-il que cette proc\'edure
n\'ecessite la mesure de courbes~I-V, et que nous ne pouvons pas
produire ce type de courbes pour chacun des pixels d'une matrice de
bolom\`etres. En effet, nous avons seulement acc\`es au potentiel
\'electrique des points milieux de chaque pixel,
et nous ne pouvons pas utiliser les r\'esistances de r\'ef\'erence
comme sources de courant constant car leur imp\'edance d\'epend de la
temp\'erature du plan focal et de la tension appliqu\'ee \`a leurs
bornes. Sans sources de courant connu et constant, il nous est
impossible de mesurer des courbes de charge et la proc\'edure que nous
venons de pr\'esenter devient alors inutilisable. Malgr\'e cela,
serait-il possible de mettre \`a profit le mod\`ele analytique
d\'ej\`a d\'evelopp\'e? La r\'eponse est malheureusement non.  Nos
bolom\`etres doivent en effet \^etre fortement polaris\'es pour
surmonter le bruit de l'\'electronique de lecture~; l'effet de champ
(cf section~\ref{sec:detect_bolocea_fabrication_thermo}) et le
d\'ecouplage \'electron-phonon deviennent alors des ph\'enom\`enes
non-n\'egligeables qui d\'epassent le domaine de validit\'e du
mod\`ele \'electro-thermique pr\'esent\'e. \shortciteN{grannan}
proposent une extension de ce mod\`ele pour inclure l'effet de champ
dans la pr\'ediction des performances, mais il est n\'ecessaire de
faire l'hypoth\`ese que le bolom\`etre est aliment\'e par un courant
constant, ce qui n'est pas le cas des bolom\`etres du
CEA. \shortciteN{galeazzi} proposent \'egalement un mod\`ele
analytique qui prend en compte le d\'ecouplage \'electron-phonon ainsi
que l'effet de champ, mais, encore une fois, l'utilisation de ce
mod\`ele n\'ecessite de pouvoir mesurer les param\`etres physiques du
bolom\`etre, ce que nous ne pouvons faire faute de pouvoir mesurer le
courant qui circule dans chacun des pixels individuellement.\\



\subsection{La reformulation du probl\`eme}
\label{sec:detect_outils_concept}




Les matrices de bolom\`etres du CEA sont des d\'etecteurs uniques. Il
n'est donc pas surprenant que la litt\'erature sp\'ecialis\'ee ne
puisse nous fournir d'outils adapt\'es \`a l'\'etude de leur
fonctionnement. N\'eanmoins, nous devons trouver une proc\'edure de
test compatible avec les observables auxquelles nous avons acc\`es
pour optimiser les performances des matrices. Plut\^ot que de baser
notre travail sur une mod\'elisation compl\`ete du
d\'etecteur\footnote{Notez que Vincent Rev\'eret a developp\'e un
mod\`ele num\'erique pour simuler le comportement d'un pont
bolom\'etrique tel que celui con\c{c}u pour les matrices
\shortcite{reveret_these}. Cependant, le mod\`ele ne prend pas en
compte l'\'electronique de lecture qui est un \'el\'ement clef du
d\'etecteur. Son int\'er\^et a \'et\'e relativement limit\'e au cours
de ma th\`ese puisque nous avons principalement rencontr\'e des
difficult\'es avec le circuit \'electronique.}, et compte tenu de la
complexit\'e du syst\`eme et de l'imminence de la campagne
d'\'etalonnage du Photom\`etre PACS, nous avons opt\'e pour une
approche plus pragmatique. La proc\'edure de caract\'erisation que
nous avons d\'evelopp\'ee repose en fait sur l'exploration
syst\'ematique des performances des matrices. Le but est d'obtenir un
jeu de donn\'ees le plus complet possible \`a partir duquel nous
pourrons pr\'edire le comportement des d\'etecteurs en orbite et
d\'eterminer le point de fonctionnement optimum dans les conditions
nominales d'op\'eration. La richesse des informations r\'ecolt\'ees
pourra \'egalement nous servir d'outil diagnostique pour r\'ev\'eler
d'\'eventuels dysfonctionnements ou alt\'erations du syst\`eme au
cours de la mission. D'int\'er\^et plus imm\'ediat, cette proc\'edure
nous permet de jauger le potentiel d\'ej\`a prometteur de ces nouveaux
d\'etecteurs.

Pour r\'ealiser une \'etude compl\`ete du comportement des matrices,
nous devons explorer un maximum de configurations du syst\`eme. Par
configurations, il faut comprendre combinaisons des diff\'erents
param\`etres qui peuvent potentiellement modifier le signal de sortie
de la cam\'era. Toutefois, il est en pratique impossible d'effectuer
une analyse v\'eritablement exhaustive sur tous les param\`etres
envisageables. En effet, les matrices n\'ecessitent pas moins de
19~tensions de polarisation pour alimenter les bolom\`etres et leur
circuit de lecture (seulement 15~sont visibles sur la
figure~\ref{fig:detect_bolocea_elec_froide_principe})~; et si nous
d\'ecidions de mesurer pendant 1~minute toutes les combinaisons
possibles de ces 19~param\`etres, et en ne testant que 3~valeurs
diff\'erentes pour chaque tension, il nous faudrait plus de 2000~ans
pour compl\'eter la proc\'edure de caract\'erisation! Nous devons donc
faire un choix et distinguer les param\`etres primaires que nous
explorerons effectivement de mani\`ere exhaustive, et les param\`etres
secondaires qui seront dans un premier temps fix\'es \`a une valeur
par d\'efaut. Les param\`etres primaires sont choisis pour leur
pertinence vis-\`a-vis des mesures de performances, nous en donnons la
liste et justifions pourquoi notre choix s'est port\'e sur eux~:
\begin{description}
\item[La tension de polarisation~: $\mathbf{V_{polar}=(V_h-V_l})$.]
$V_h$ et $V_l$ sont les tensions appliqu\'ees aux bornes des ponts
bolom\'etriques (cf
figure~\ref{fig:detect_bolocea_elec_froide_principe}). Elles sont
communes \`a tous les pixels d'un m\^eme BU (256 ou 512~pixels pour
les BFP rouge et bleu respectivement). La puissance \'electrique
dissip\'ee dans le bolom\`etre, \cad indirectement la temp\'erature de
l'absorbeur, ainsi que l'effet de champ d\'ecrit dans la
section~\ref{sec:detect_bolocea_fabrication_thermo} d\'ependent de
cette tension de polarisation $V_{polar}$. Le point de fonctionnement
des thermom\`etres, et donc les performances des bolom\`etres,
d\'ependent fortement de ce param\`etre.  Nous devons donc l'explorer
assez finement~; nous testons 24~valeurs de $V_{polar}$ entre 0.5~et
3.5~V pour les matrices bleues et entre 0.5~et 3~V pour les matrices
rouges (un test pr\'eliminaire a en effet indiqu\'e une tension
optimale autour de 2~V).
\item[Le flux radiatif.] Cette quantit\'e repr\'esente le flux
incident sur les bolom\`etres. Elle est exprim\'ee en pW/pixel. Il ne
faut cependant pas la confondre avec la puissance effectivement
absorb\'ee par le bolom\`etre $Q$ telle que nous l'avons d\'ecrite
pr\'ec\'edemment. Dans le cas de l'observatoire Herschel, la charge
optique est principalement d\'etermin\'ee par l'\'emission d'avant
plan, \cad le t\'elescope. En effet, la plupart des sources
astrophysiques ne devrait compter que pour un milli\`eme du flux
total. Dans un rapport interne au consortium PACS,
\shortciteN{sauvage_note} utilise la formule de \shortciteANP{fischer}
sur l'\'emissivit\'e du t\'elescope Herschel (cf
\'equation~(\ref{eq:emissivite_fischer})) pour calculer le flux
incident au niveau du plan focal pour chacune des bandes spectrales du
Photom\`etre PACS. Il trouve un flux de 2.75, 1.54 et 2.52~pW/pixel
pour les bandes \`a 85, 110 et 170~$\mu$m respectivement. D'autre
part, pour les sources tr\`es brillantes telles que les plan\`etes ou
les ast\'ero\"ides qui devraient \^etre utilis\'ees comme calibrateurs
primaires, le flux incident pourrait doubler le fond de
t\'elescope. Il est donc n\'ecessaire d'explorer ce param\`etre entre
1~et 7~pW/pixel~; ces mesures nous permettrons \'egalement de
quantifier la non-lin\'earit\'e des bolom\`etres. Remarquez de plus
que les sources d'\'etalonnage que nous utilisons au sol ne
poss\`edent pas le m\^eme spectre d'\'emission que le t\'elescope~;
nous devons alors mesurer s\'epar\'ement les flux de 1~\`a 7~pW/pixel
sur la voie rouge du photom\`etre (ce qui correspond \`a des faibles
flux sur le BFP bleu), puis de 1~\`a 7~pW/pixel sur la voie bleue (ce
qui correspond \`a des forts flux sur la voie rouge).
\item[Les tensions de r\'ef\'erence $\mathbf{V_{ref}}$ et
$\mathbf{V_{hb}}$.] Ces tensions sont utilis\'ees comme r\'ef\'erence
pour r\'ealiser les mesures diff\'erentielles d\'ecrites dans la
section~\ref{sec:detect_bolocea_elec}. Elles sont \'egalement communes
\`a tous les pixels d'un m\^eme BU. Leur r\^ole premier est de
supprimer les d\'erives basse fr\'equence du circuit de lecture, mais
elles sont aussi utilis\'ees comme des offsets pour ajuster le niveau
du signal et pour le centrer dans la dynamique des convertisseurs
num\'eriques de BOLC. Si ces deux tensions ne sont pas r\'egl\'ees
correctement, elles peuvent conduire \`a une perte d'information par
saturation des ADC. $V_{hb}$ est la tension inject\'ee au niveau du
BU. $V_{ref}$ est la tension inject\'ee en amont de l'\'electronique
froide, tr\`es proche physiquement des bolom\`etres. Du point de vue
du circuit de lecture, elle est \'equivalente \`a un point milieu dont
nous pourrions choisir la valeur. Nous explorons 125~valeurs du couple
de tension ($V_{ref}$, $V_{hb}$).
\item[Le mode de lecture.] BOLC offre deux modes de lecture (cf
section~\ref{sec:detect_bolocea_elec_lecture}). Le mode DDCS doit en
th\'eorie pr\'esenter de meilleures performances en terme de
stabilit\'e et de suceptibilit\'e \'electromagn\'etique, alors que le
mode direct doit \^etre plus int\'eressant en terme de sensibilit\'e
(la lecture en mode DDCS n\'ecessite une soustraction
suppl\'ementaire, le bruit en mode DDCS est donc la somme quadratique
des bruits des deux signaux soustraits). Puisque les deux modes sont
potentiellement int\'eressants, et que nous ne connaissons pas encore
suffisamment bien l'environnement \'electromagn\'etique de PACS, nous
devons les caract\'eriser tous les deux.
\end{description}

Remarquez que malgr\'e son r\^ole central dans le fonctionnement des
bolom\`etres, nous n'avons pas cit\'e la temp\'erature des
d\'etecteurs comme param\`etre \`a explorer. D'une part, le banc de
test que nous utilisons ne permet pas de contr\^oler la temp\'erature
du plan focal~; nous pouvons \'eventuellement la faire varier en
chauffant la pompe du cryo-r\'efrig\'erateur (cf
section~\ref{sec:detect_observatoire_phfpu_cryocooler}) mais il est
tr\`es difficile d'obtenir une temp\'erature stable sur des p\'eriodes
suffisamment longues pour effectuer des centaines de mesures. D'autre
part, le cryo-r\'efrig\'erateur fournit une temp\'erature tellement
reproductible \`a chaque recyclage que ce param\`etre est rest\'e
inchang\'e tout au long de la proc\'edure de test. Nous consid\'erons
donc qu'explorer l'influence de la temp\'erature n'est pas
n\'ecessaire pour la suite de l'\'etalonnage.

Les param\`etres secondaires sont les tensions qui alimentent le reste
du circuit de lecture, \`a savoir le multiplexeur, les circuits
suiveurs de tension et le s\'equenceur. 
A priori, ces tensions ne sont pas pertinentes pour les mesures de
performance des bolom\`etres et nous d\'ecidons de les fixer \`a des
valeurs par d\'efaut. Leur r\'eglage est bas\'e sur l'exp\'erience que
nous avons acquise durant la phase de d\'eveloppement des
matrices. Par exemple, le courant qui alimente les MOS suiveurs du CL,
$I_{VSS}$, doit \^etre le plus petit possible pour minimiser la
dissipation \'electrique \`a l'\'etage 300~mK, mais il doit \^etre
suffisamment important pour permettre le transport du signal
\'electrique vers le BU. Ce courant vaut par d\'efaut 300~nA pour un
groupe bleu qui contient 2~matrices en parall\`eles, et 150~nA pour un
groupe rouge qui n'en contient qu'une seule.\\

Jusqu'en 2005, nous r\'eglions les matrices de bolom\`etres de
fa\c{c}on empirique, \cad que l'op\'erateur devait ajuster \og \`a la
main \fg chacune des 19~tensions \`a appliquer aux d\'etecteurs, et ce
pour chacun des six groupes qui composent le PhFPU. Cette proc\'edure
de r\'eglage \'etait relativement longue, fastidieuse et
inefficace. Remarquez que pour mesurer syst\'ematiquement les
performances de la cam\'era et ainsi r\'ealiser un \'etalonnage
convenable du photom\`etre, il est n\'ecessaire de tester les matrices
dans plus d'un millier de configurations\footnote{Pour obtenir la NEP
(cf section~\ref{sec:calib_perflabo_sensibilite}), nous devons mesurer
le bruit et la r\'eponse des matrices pour 24 tensions de polarisation
dans les deux modes de lecture pour 7 flux sur le BFP bleu et 7 flux
sur le BFP rouge, ce qui donne d\'ej\`a un total de
1344~configurations \`a tester.}. L'automatisation de la proc\'edure
de r\'eglage des d\'etecteurs s'est donc av\'er\'ee indispensable.

\begin{figure}
  \begin{center}
    \includegraphics[width=0.8\textwidth,angle=0]{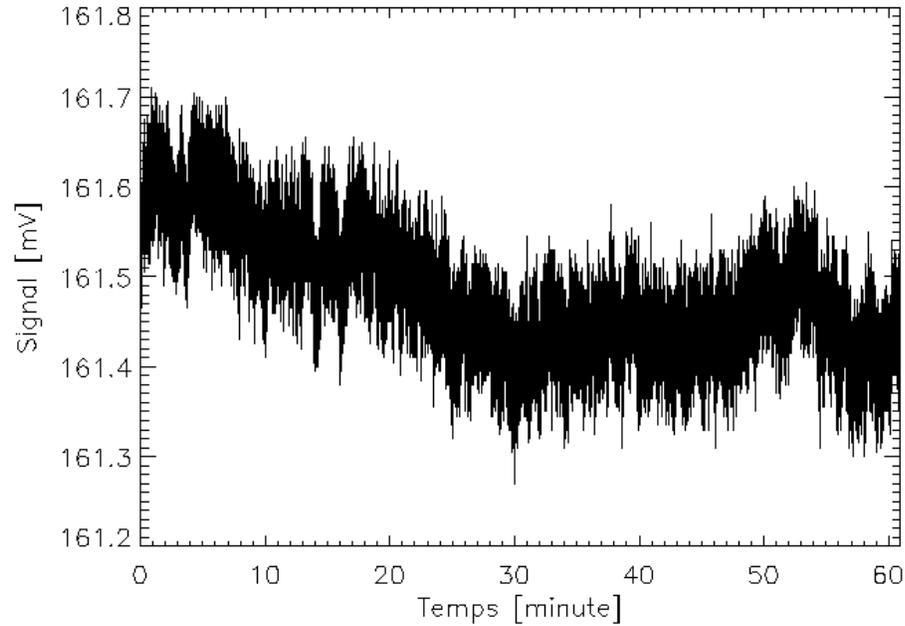}
  \end{center}
  \caption[Signal typique d'un bolom\`etre]{Signal typique d'un
  bolom\`etre mesur\'e en mode direct en sortie de BOLC pour un flux
  incident de 2~pW/pixel. L'\'echantillon contient $1.5\times10^5$
  points. La d\'eviation standard est de l'ordre de 70~$\mu$V. Durant
  cette mesure longue d'une heure, la temp\'erature a d\'eriv\'e de
  fa\c{c}on monotone sur seulement 60~$\mu$K.
  \label{fig:detect_outils_concept_typic}}
\end{figure}

Cette proc\'edure comporte trois \'etapes. La premi\`ere consiste \`a
mesurer la fonction de transfert de l'\'electronique de lecture, \cad
\`a trouver la correspondance qu'il existe entre le signal d'entr\'ee
du CL et le signal de sortie de BOLC (cf
section~\ref{sec:calib_procedure_vrlvhb}). La seconde \'etape consiste
\`a mesurer syst\'ematiquement le niveau de point milieu des matrices
pour les configurations \`a tester tout en relaxant la contrainte sur
la dynamique des ADC de BOLC (cf
section~\ref{sec:calib_procedure_explore}). Notez que l'objectif de
cette \'etape est de constituer une base de donn\'ees compl\`ete qui
nous renseigne sur l'\'etat d'\'equilibre des ponts bolom\`etriques et
qui ne d\'epend pas a priori des divers gains et offsets de la
cha\^ine \'electronique. En effet, il est important que le point
milieu soit ind\'ependant de l'\'electronique de lecture car cela nous
permet de comparer des donn\'ees obtenues avec diff\'erents r\'eglages
de l'\'electronique, \cad diff\'erents gains et offsets. Enfin, pour
la troisi\`eme \'etape, nous nous appuyons sur ces valeurs de points
milieux pour calculer les tensions \`a appliquer aux d\'etecteurs afin
de minimiser la saturation de l'\'electronique de lecture (cf
section~\ref{sec:calib_procedure_prediction}).

Remarquez que la validit\'e de cette m\'ethode d'automatisation
d\'epend de la stabilit\'e du syst\`eme. En effet, la fonction de
transfert de l'\'electronique ainsi que la base de donn\'ees des
points milieux sont obtenues \`a des intervalles de temps qui peuvent
atteindre plusieurs dizaines d'heures. Pour nous assurer que le calcul
des tensions \`a partir de ces deux \'el\'ements est assez pr\'ecis,
\cad que le r\'eglage pr\'edit n'aboutit pas \`a une saturation de
l'\'electronique chaude, nous devons comparer la d\'erive du signal
\`a la dynamique de BOLC. La
figure~\ref{fig:detect_outils_concept_typic} montre un signal temporel
typique mesur\'e en sortie de BOLC pour un flux incident constant de
2~pW/pixel. La d\'eviation standard du signal \'electrique est
d'environ 70~$\mu$V, ce qui repr\'esente environ 0.02~\% de la
dynamique de BOLC en gain fort. D'autre part, la temp\'erature du plan
focal n'a d\'eriv\'e que de 60~$\mu$K durant la mesure. Le syst\`eme
est donc suffisamment stable pour appliquer notre proc\'edure de
r\'eglage.

\section{L'\'electronique de lecture}
\label{sec:calib_procedure_vrlvhb}

La premi\`ere \'etape de la proc\'edure d'\'etalonnage des matrices de
bolom\`etres consiste \`a mesurer la fonction de transfert du circuit
de lecture. Elle sera essentielle dans la suite de la calibration lors
du calcul des points milieux. Pour effectuer cette mesure, nous
exploitons le fait que les matrices de bolom\`etres poss\`edent 2
tensions de r\'ef\'erences, $V_{ref}$ et $V_{hblind}$, qui sont
disponibles \`a chacun des deux \'etages d'adaptation
d'imp\'edance. Cette fonctionnalit\'e a \'et\'e impl\'ement\'ee pour
fournir une lecture diff\'erentielle du signal, mais il est aussi
possible de se servir de ces tensions comme \'etalons pour injecter un
signal connu dans le circuit de lecture et ainsi mesurer la fonction
de transfert. Cependant cette mesure n\'ecessite un r\'eglage
particulier des matrices de bolom\`etres que nous d\'ecrivons ici.\\
Nous isolons \'electriquement le Circuit de Lecture (CL) du Circuit de
D\'etection (CD) en imposant $V_{DecX\_h}$=$V_{DecX\_l}$=0~V et
$C\!K\!RL\_h$=$C\!K\!RL\_l$=2~V (cf
figure~\ref{fig:detect_bolocea_elec_froide_principe}). De cette
fa\c{c}on les transistors qui relient le CD et le CL deviennent
isolants, alors que ceux reliant $V_{ref}$ et le CL deviennent
passants. Du point de vue du s\'equenceur, ce r\'eglage est
\'equivalent au mode de lecture direct, mais au lieu
d'\'echantillonner le point milieu, l'\'electronique froide
\'echantillonne uniquement la tension de r\'ef\'erence
$V_{ref}$. D'autre part, les tensions secondaires qui alimentent le
circuit de lecture ne sont pas pertinentes pour les mesures de
performance, elles sont donc mises \`a leur valeurs nominales et ne
changeront pas dans la suite des tests (cf
section~\ref{sec:detect_outils_concept}). En effet, il est crucial de
garder ces m\^emes r\'eglages pour les tensions secondaires tout au
long de la proc\'edure de caract\'erisation pour s'assurer que les
gains et offsets de l'\'electronique restent inchang\'es (\`a la
d\'erive basse fr\'equence des transistors pr\`es).  Il est
\'egalement n\'ecessaire de mettre l'\'electronique chaude dans le
mode de gain faible pour explorer toute la gamme dynamique des
convertisseurs num\'eriques de BOLC et ainsi mesurer leurs limites de
saturation.

\begin{figure}
  \begin{center}
    \begin{tabular}[t]{ll}
      \includegraphics[width=0.6\textwidth,angle=0]{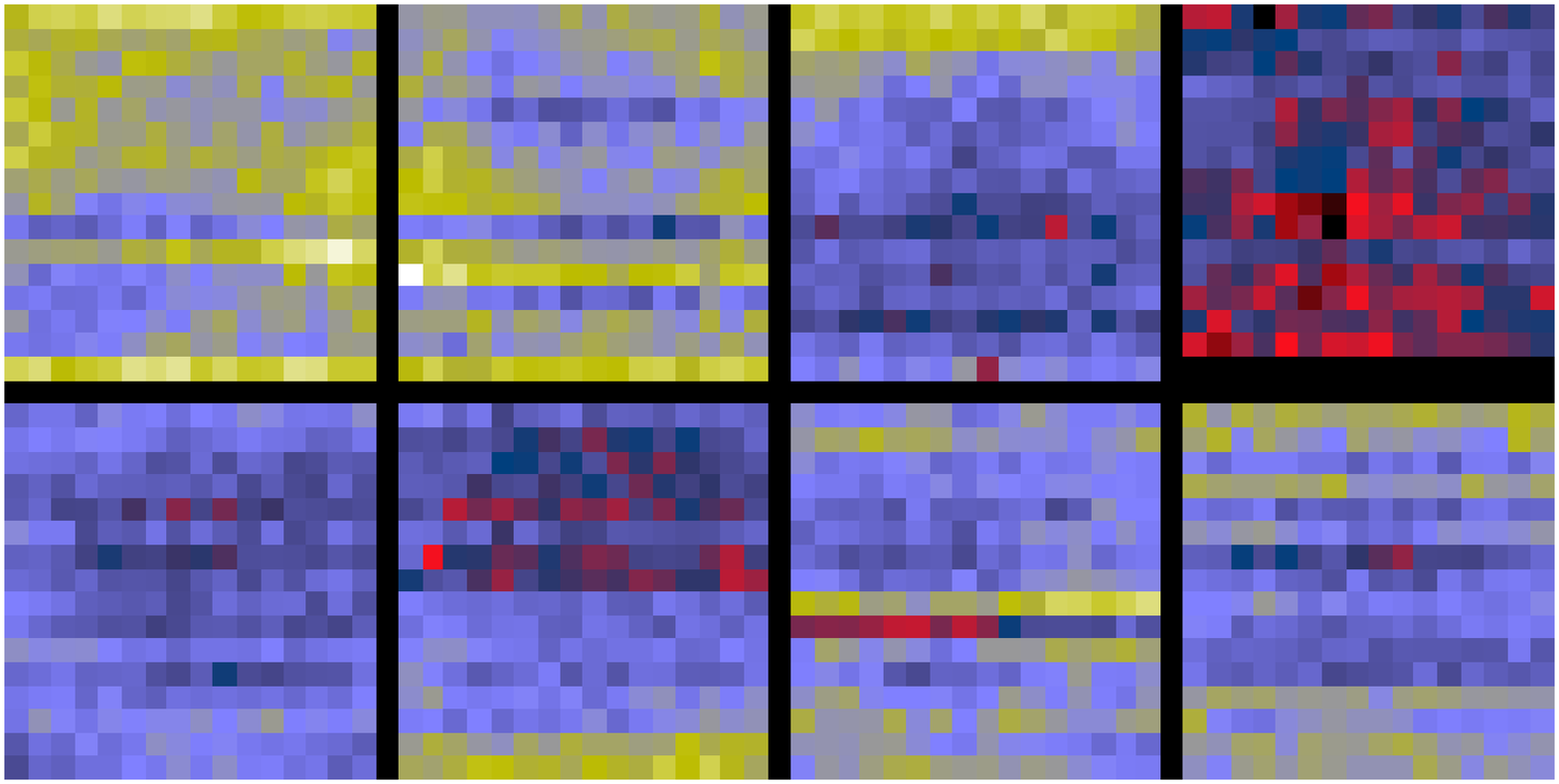} & \includegraphics[width=0.35\textwidth,angle=0]{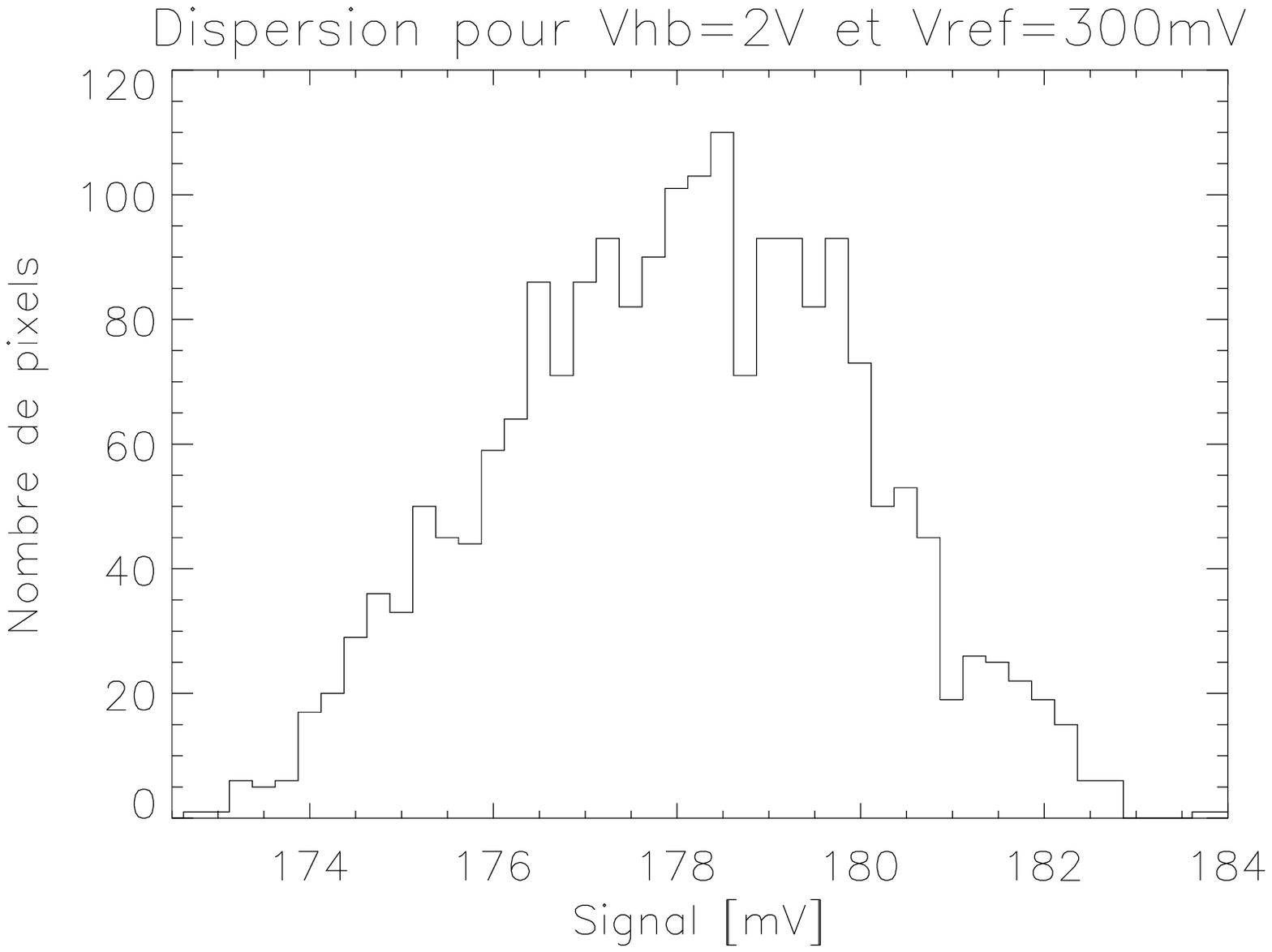} 
    \end{tabular}
  \end{center}
  \caption[Carte des $V_{ref}$: image de l'\'electronique de
  lecture]{Signal de sortie du plan focal bleu pour
  ($V_{ref}\,,\,V_{hblind}$) = (300~mV$\,,\,$2~V). Les structures
  horizontales sont dues au multiplexage. Tous les pixels sont
  fonctionnels \`a l'exception de la ligne au centre \`a droite qui a
  \'et\'e sacrifi\'ee pour pouvoir r\'ecup\'erer le reste de la
  matrice. L'histogramme montre une dispersion \emph{r.m.s.} sur cette
  carte de quelques mV.
  \label{fig:calib_procedure_vrvhb_maphisto}}
\end{figure}
\begin{figure}
  \begin{center}
    \includegraphics[width=0.7\textwidth,angle=0]{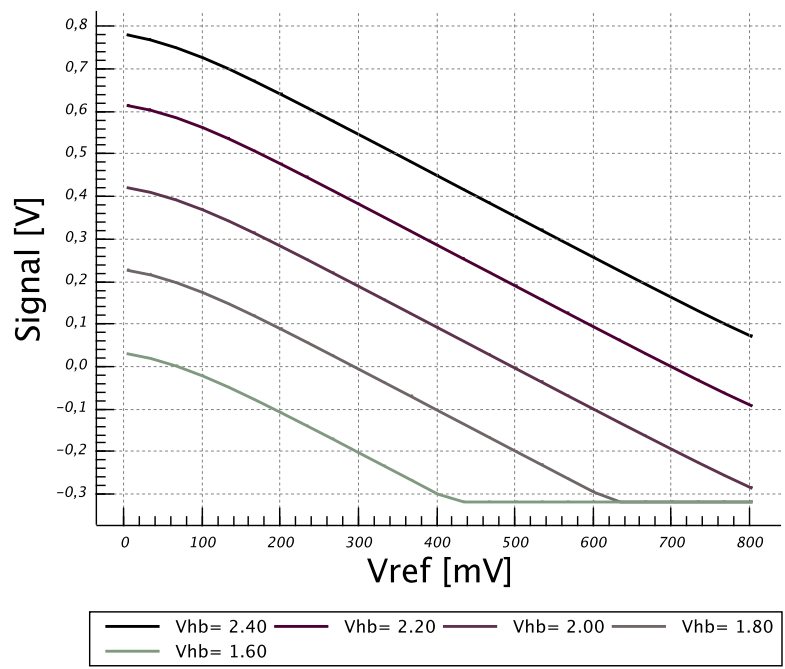}
  \end{center}
  \caption[Fonction de transfert de l'\'electronique de
  lecture]{Fonction de transfert de l'\'electronique de lecture. Elle
  repr\'esente le signal de sortie moyen d'une matrice de bolom\`etre
  PACS pour diff\'erents couples de ($V_{ref}\,,\,V_{hblind}$)
  inject\'es dans le circuit. $V_{ref}$ varie de 0 \`a 800~mV par pas
  de 33~mV et $V_{hblind}$ de 1.60 \`a 2.40~V par pas de 200~mV. BOLC
  \'etant utilis\'e avec un gain faible, le signal sature le
  convertisseur num\'erique aux alentours de -330~mV, effet visible
  aux faibles valeurs de $V_{hblind}$. Cette fonction de transfert
  sera utilis\'ee comme courbe de calibration pour le calcul des
  points milieux.
  \label{fig:calib_procedure_vrvhb_surfcal}}
\end{figure}

Nous mesurons le signal de sortie de toute la cha\^ine \'electronique
pour plusieurs couples de tension ($V_{ref}\,,\,V_{hblind}$)
inject\'es. La figure~\ref{fig:calib_procedure_vrvhb_maphisto} montre
un exemple de distribution spatiale du signal sur le BFP bleu pour un
couple ($V_{ref}\,,\,V_{hblind}$) = (300~mV$\,,\,$2~V). Cette carte ne
contient aucune contribution des bolom\`etres mais nous donne plut\^ot
une image de l'\'electronique de lecture seule. Les structures
horizontales sur chaque matrice sont dues au multiplexage et indiquent
un certain niveau de corr\'elation entre les pixels d'une m\^eme
colonne de lecture. En effet, ces colonnes poss\`edent un circuit de
lecture en commun, celui qui se trouve en aval du multiplexeur (cf
figure~\ref{fig:detect_bolocea_elec_froide_principe}). La dispersion
du signal sur tout le plan focal est d'environ 8~mV.

\begin{figure}[!htb]
  \begin{center}
    \begin{tabular}{l}
      \includegraphics[width=0.45\textwidth,angle=0]{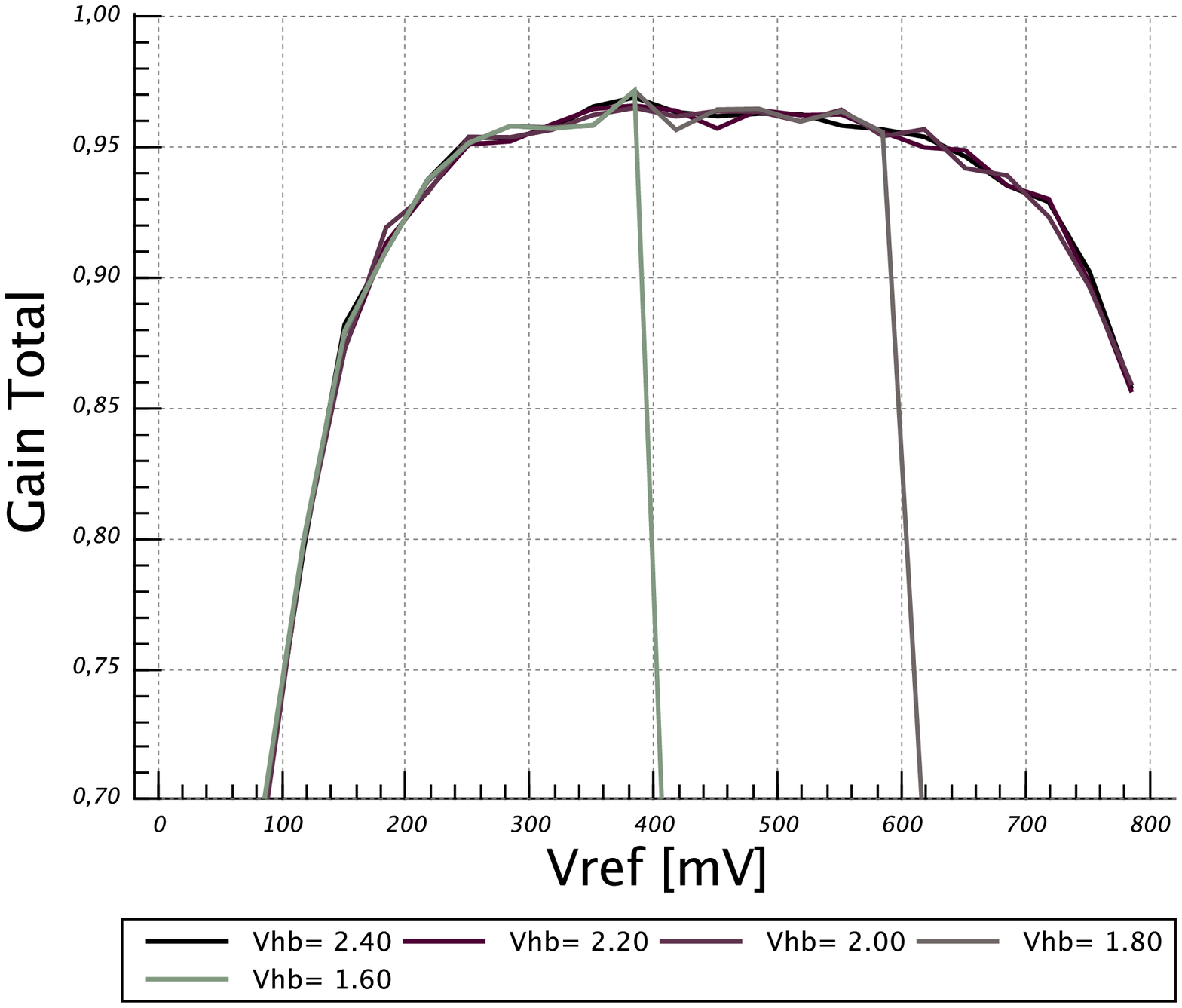}
      \includegraphics[width=0.45\textwidth,angle=0]{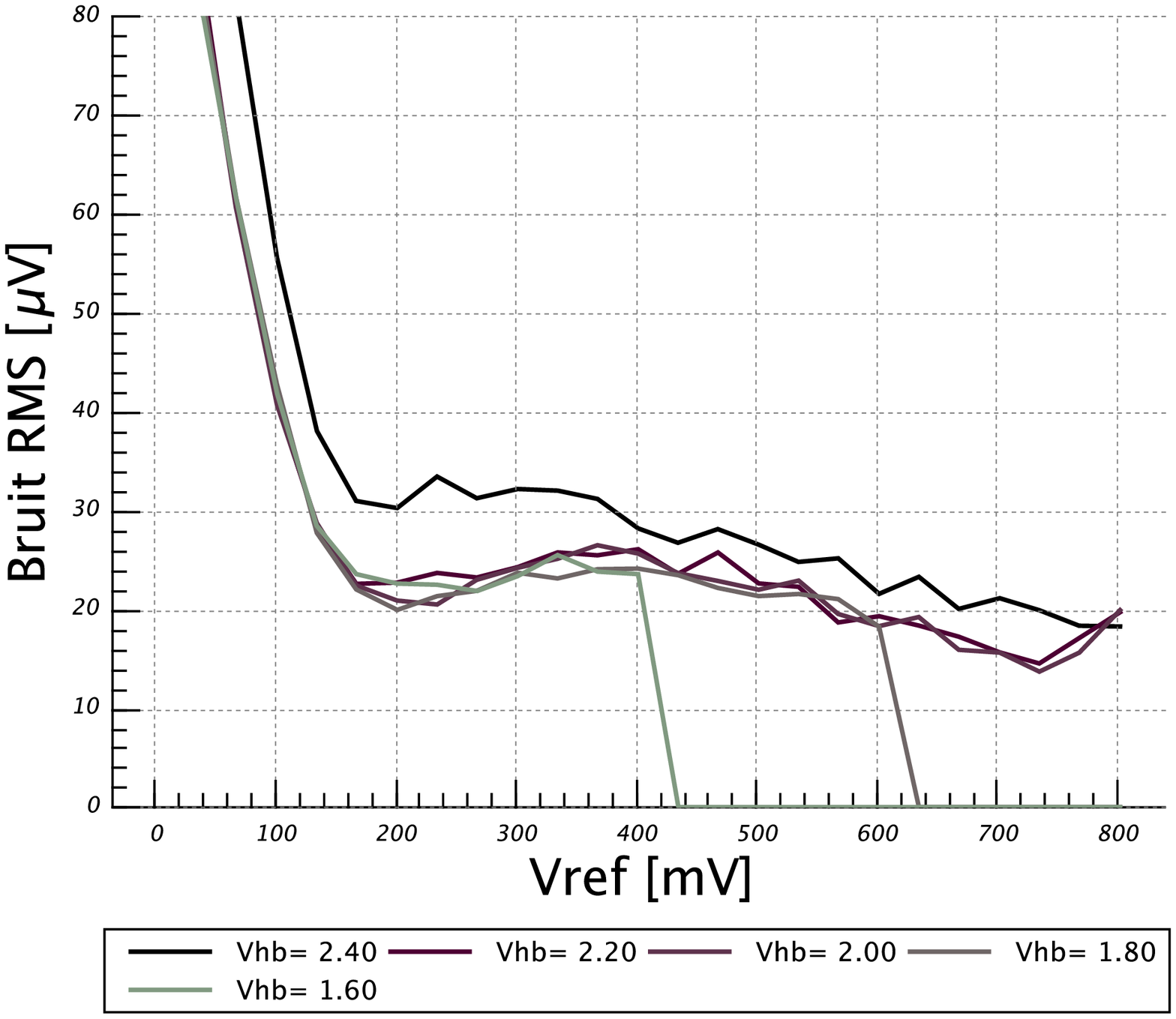}
    \end{tabular}
  \end{center}
  \caption[Gain et bruit de l'\'electronique de
  lecture]{\emph{Gauche}: Gain total de toute la cha\^ine
  \'electronique. Il est obtenu en d\'erivant la fonction de transfert
  par rapport au $V_{ref}$ qui est inject\'e en amont du circuit,
  gain=$\frac{\partial Signal}{\partial V_{ref}}$. Pour des $V_{ref}$
  compris entre 250 et 650~mV le signal est transmis avec un gain
  sup\'erieur \`a 95~\%. Les deux morceaux de courbes aberrants
  proviennent de la saturation de BOLC pour les faibles valeurs de
  $V_{hlind}$. \emph{Droite}: Bruit \emph{r.m.s.} mesur\'e pour chaque
  couple ($V_{ref}\,,\,V_{hblind}$). Pour les valeurs de $V_{ref}$
  inf\'erieures \`a 200~mV le bruit augmente fortement indiquant que
  l'\'electronique de lecture ne doit pas \^etre utilis\'ee dans ce
  r\'egime. Le bruit vaut zero lorsque le signal sature aux faibles
  valeurs de $V_{hlind}$.
  \label{fig:calib_procedure_vrvhb_gainbruit}}
\end{figure}

Afin d'\'echantillonner finement la fonction de transfert, nous
explorons $V_{ref}$ de 0 \`a 800~mV par pas de 33~mV, et $V_{hblind}$
de 1.60 \`a 2.40~V par pas de 200~mV. Le r\'esultat de ces mesures est
pr\'esent\'e dans la
figure~\ref{fig:calib_procedure_vrvhb_surfcal}. Chaque point
repr\'esente la valeur du signal de sortie moyenn\'ee sur une matrice
enti\`ere, la matrice en bas \`a gauche de la
figure~\ref{fig:calib_procedure_vrvhb_maphisto} en l'occurence. Les
mesures sont tr\`es peu bruit\'ees et l'\'electronique de lecture
semble avoir un comportement lin\'eaire sur une grande partie du
domaine explor\'e. Notez toutefois que pour les faibles valeurs de
$V_{hblind}$ le signal sature \`a -330~mV, valeur correspondant \`a la
limite inf\'erieure de la dynamique de BOLC en gain faible (cf
annexe~\ref{a:dyna_BOLC}). Nous d\'efinissons maintenant le gain total
de l'\'electronique de lecture comme \'etant la quantit\'e
$\frac{\partial Signal}{\partial V_{ref}}$. Nous le calculons
ais\'ement \`a partir de la fonction de transfert et nous le
tra\c{c}ons dans la
figure~\ref{fig:calib_procedure_vrvhb_gainbruit}. Le gain de la
cha\^ine \'electronique est sup\'erieur \`a 95~\% pour des valeurs de
$V_{ref}$ comprises entre 250 et 650~mV. Par contre, en dehors de
cette gamme, le signal entrant dans le circuit de lecture est
consid\'erablement att\'enu\'e (gain<0.5 pour $V_{ref}$<50~mV). La
partie droite de la figure~\ref{fig:calib_procedure_vrvhb_gainbruit}
montre le bruit \emph{r.m.s.}  mesur\'e pour chacun des couples
($V_{ref}\,,\,V_{hblind}$). Pour des valeurs de $V_{ref}$
sup\'erieures \`a 200~mV le bruit de lecture moyen sur une matrice est
d'environ 24~$\mu$V pour les groupes bleus et d'environ 18~$\mu$V pour
les groupes rouges\footnote{Le niveau de bruit d\'epend du courant qui
circule dans les CL. En effet, les groupes rouges contiennent une
seule matrice avec un courant CL de l'ordre de 100~nA alors que les
groupes bleus contiennent deux matrices avec un courant CL d'environ
350~nA (ce courant est distribu\'e sur les 2 CL). Les matrices des
groupes bleus sont donc probablement sur-aliment\'ees g\'en\'erant
ainsi un exces de bruit qui pourrait \^etre coup\'e en diminuant le
courant qui circule dans les CL.}. Dans les deux cas, en-dessous de
200~mV, l'\'electronique g\'en\`ere un bruit consid\'erablement plus
\'elev\'e. Nous attribuons cette augmentation de bruit \`a la
saturation des transistors control\'es par $V_{SMS}$ (adressage du
multiplexeur). Ces transistors sont \'egalement responsables de la
chute de gain pour les petites valeurs de $V_{ref}$.\\ La fonction de
transfert est un \'el\'ement clef de la proc\'edure d'\'etalonnage,
elle nous fournit l'information qui nous permettra de calculer le
niveau de point milieu de chaque pixel. Mais les mesures
pr\'esent\'ees dans cette section sont d'autant plus importantes
qu'elles d\'elimitent le domaine de fonctionnement de l'\'electronique
de lecture, et qu'elles donnent des contraintes sur le niveau de
signal qui peut, ou non, \^etre transmis jusqu'\`a BOLC. D'apr\`es la
figure~\ref{fig:calib_procedure_vrvhb_gainbruit}, nous concluons que
l'\'electronique de lecture ne fonctionne que pour des signaux entrant
compris entre 200 et 700~mV.

\section{L'exploration des ponts bolom\'etriques et de leurs points milieux}
\label{sec:calib_procedure_explore}

\subsection{Mesure, calcul et interpr\'etation des points milieux}
\label{sec:calib_procedure_explore_calcul}

Maintenant que nous avons obtenu les courbes d'\'etalonnage de
l'\'electronique de lecture, nous nous int\'eressons au comportement
des bolom\`etres eux-m\^eme. Nous explorons donc syst\'ematiquement le
niveau de point milieu de tous les pixels en fonction de quelques
param\`etres judicieusement choisis que nous avons d\'ej\`a d\'ecrit
dans la section~\ref{sec:detect_outils_concept}. L'objectif de ce test
n'est pas de mesurer les performances des matrices de bolom\`etres
mais plut\^ot de quantifier l'influence de ces quelques param\`etres
critiques sur le point milieu ainsi que sur le signal de sortie de la
cam\'era. Ces mesures sont indispensables \`a la proc\'edure
d'\'etalonnage car elles contiennent les informations n\'ecessaires
\`a la pr\'ediction du r\'eglage des d\'etecteurs pour l'\'etape
suivante qui est d\'edi\'ee aux v\'eritables mesures de performance
(cf chapitre~\ref{chap:calib_perflabo}).\\ Afin d'explorer une vaste
gamme de param\`etres de r\'eglage sans se soucier de la saturation
des convertisseurs num\'eriques, nous mettons BOLC en mode de gain
faible pour \'elargir sa dynamique \`a 1.3~V (au lieu de 330~mV en
gain fort) comme nous l'avons fait pour mesurer la fonction de
transfert de l'\'electronique. Nous fixons \'egalement les tensions
secondaires du circuit de lecture \`a leurs valeurs nominales. Le but
de ces tests \'etant de mesurer le niveau de point milieu, nous
choisissons le mode de lecture le plus simple pour interpr\'eter et
reconstruire les points milieux plus facilement, c'est-\`a-dire le
mode direct. La tension de r\'ef\'erence $V_{ref}$ n'est donc pas
utilis\'ee pour ce test. D'autre part, en mettant BOLC en mode de gain
faible, nous avons relax\'e la contrainte sur la dynamique de
l'\'electronique chaude, il n'est donc en principe pas n\'ecessaire de
modifier la tension $V_{hb}$ pour ajuster le signal dans la dynamique
de BOLC pour chaque configuration test\'ee. En pratique, pourtant,
nous avons toujours utilis\'e les valeurs de $V_{hb}$ obtenues \`a
l'occasion de tests ant\'erieurs qui centraient le signal dans la
dynamique de BOLC.

Le signal de sortie de chacune des configurations test\'ees poss\`ede
donc un offset diff\'erent qu'il faut prendre en compte pour remonter
\`a la valeur de point milieu qui a engendr\'e ce signal. Pour ce
faire, nous utilisons les courbes de calibration de l'\'electronique
de lecture (fig.~\ref{fig:calib_procedure_vrvhb_surfcal}). En effet,
la fonction de transfert est une bijection qui nous fournit un lien
direct entre le signal entrant dans le CL et celui sortant de BOLC,
moyennant que la valeur de $V_{hb}$ soit connue. En se ramenant \`a
l'entr\'ee du circuit de lecture, nous corrigeons donc les gains et
offsets de l'\'electronique de lecture. Le signal reconstruit de cette
fa\c{c}on ne d\'epend plus du r\'eglage de l'\'electronique, \`a
savoir $V_{ref}$ et $V_{hb}$, et tous les points milieux sont alors
des tensions absolues, \cad comparables entre elles. \\

\begin{figure}
  \begin{center}
    \includegraphics[width=0.6\textwidth,angle=0]{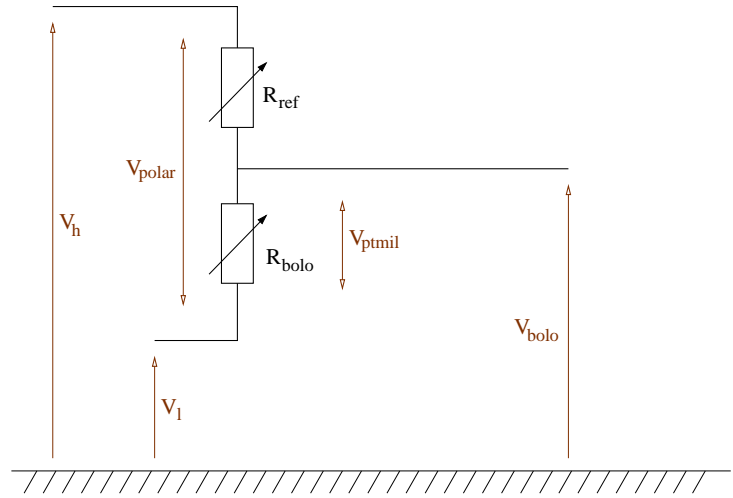}
    \caption[Sch\'ema simplifi\'e du pont bolom\'etrique]{Sch\'ema
    simplifi\'e d'un pont bolom\'etrique. Nous rappelons \'egalement
    la convention d'appellation des diff\'erentes tensions mises en
    jeu.
      \label{fig:calib_procedure_explore_schemasimple}}
  \end{center}
\end{figure}

Avant de pr\'esenter nos r\'esultats et pour faciliter leur
interpr\'etation, il est utile de rappeler que l'imp\'edance de la
r\'esistance de r\'ef\'erence $R_{ref}$ et celle du bolom\`etre
$R_{bolo}$ d\'ependent exponentiellement de leur temp\'erature et du
champ \'electrique appliqu\'e \`a leurs bornes comme d\'ecrit dans
l'\'equation~(\ref{eq:efros}). En outre, l'\'equation qui relie la
tension de polarisation du pont bolom\'etrique $V_{polar}$ \`a la
tension aux bornes de la thermistance $V_{ptmil}$ s'obtient facilement
en \'ecrivant la loi d'Ohm\footnote{Loi d'Ohm macroscopique:
$V_{tension}=R_{resistance}\times I_{courant}$} aux bornes du pont
bolom\'etrique et de la thermistance (cf
figure~\ref{fig:calib_procedure_explore_schemasimple} pour un rappel
sur les notations utilis\'ees). Nous obtenons la formule suivante en
\'egalisant les courants:
\begin{center}
\begin{equation}
V_{ptmil}=\frac{R_{bolo}}{R_{bolo}+R_{ref}}\times V_{polar}
\label{eq:midpt}
\end{equation}
\end{center}
Cette formule exprime la fonction \og pont diviseur de tension \fg du
bolom\`etre. En effet, pour une tension d'entr\'ee $V_{polar}$
donn\'ee, le rapport $\frac{R_{bolo}}{R_{bolo}+R_{ref}}$ d\'efinit
compl\`etement la tension de sortie $V_{ptmil}$ ($V_{ptmil}$ est
toujours inf\'erieure \`a $V_{polar}$). \\

\begin{figure}
  \begin{center}
    \begin{tabular}[t]{ll}
      \includegraphics[width=0.6\textwidth,angle=0]{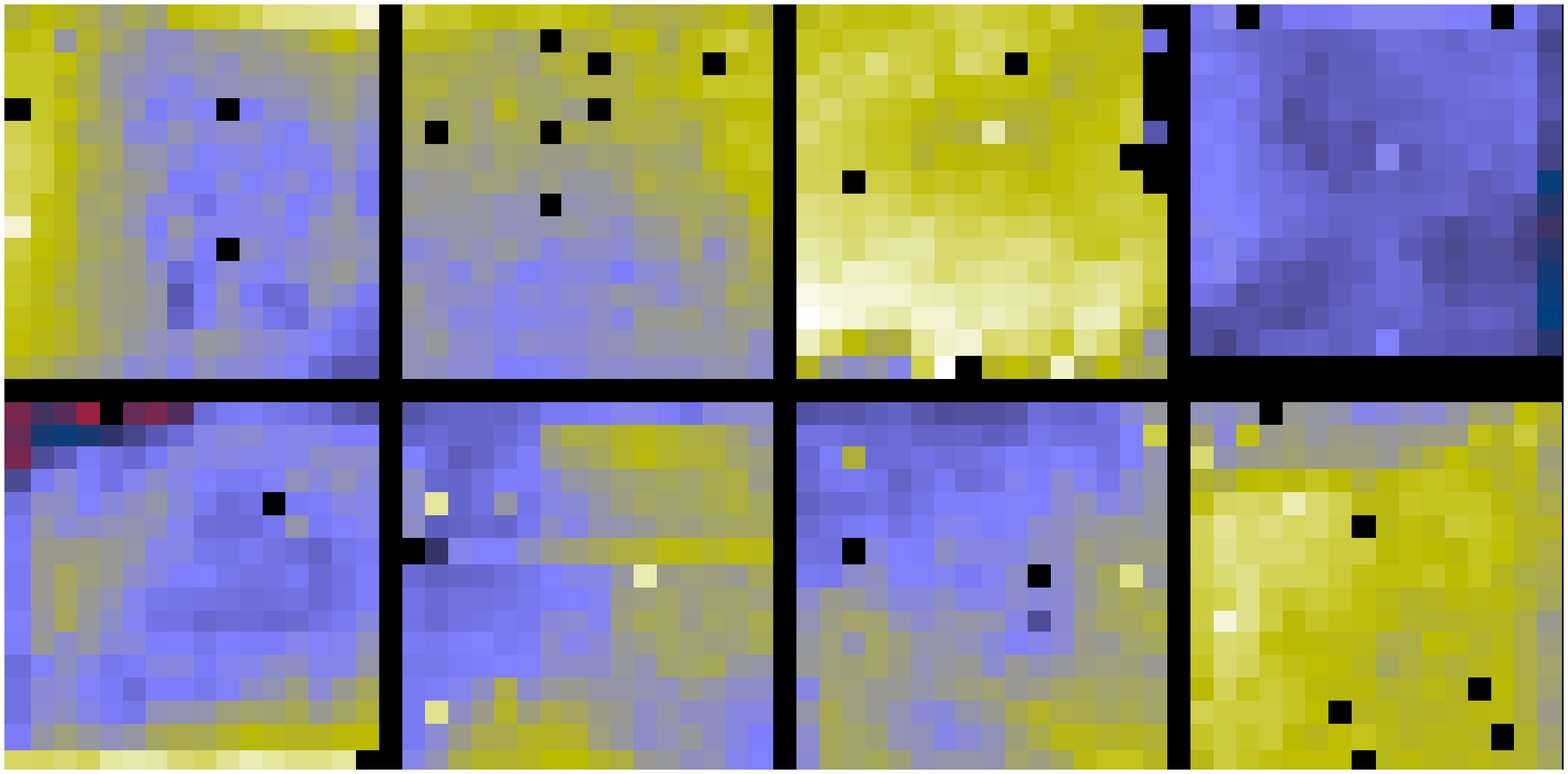} & \includegraphics[width=0.35\textwidth,angle=0]{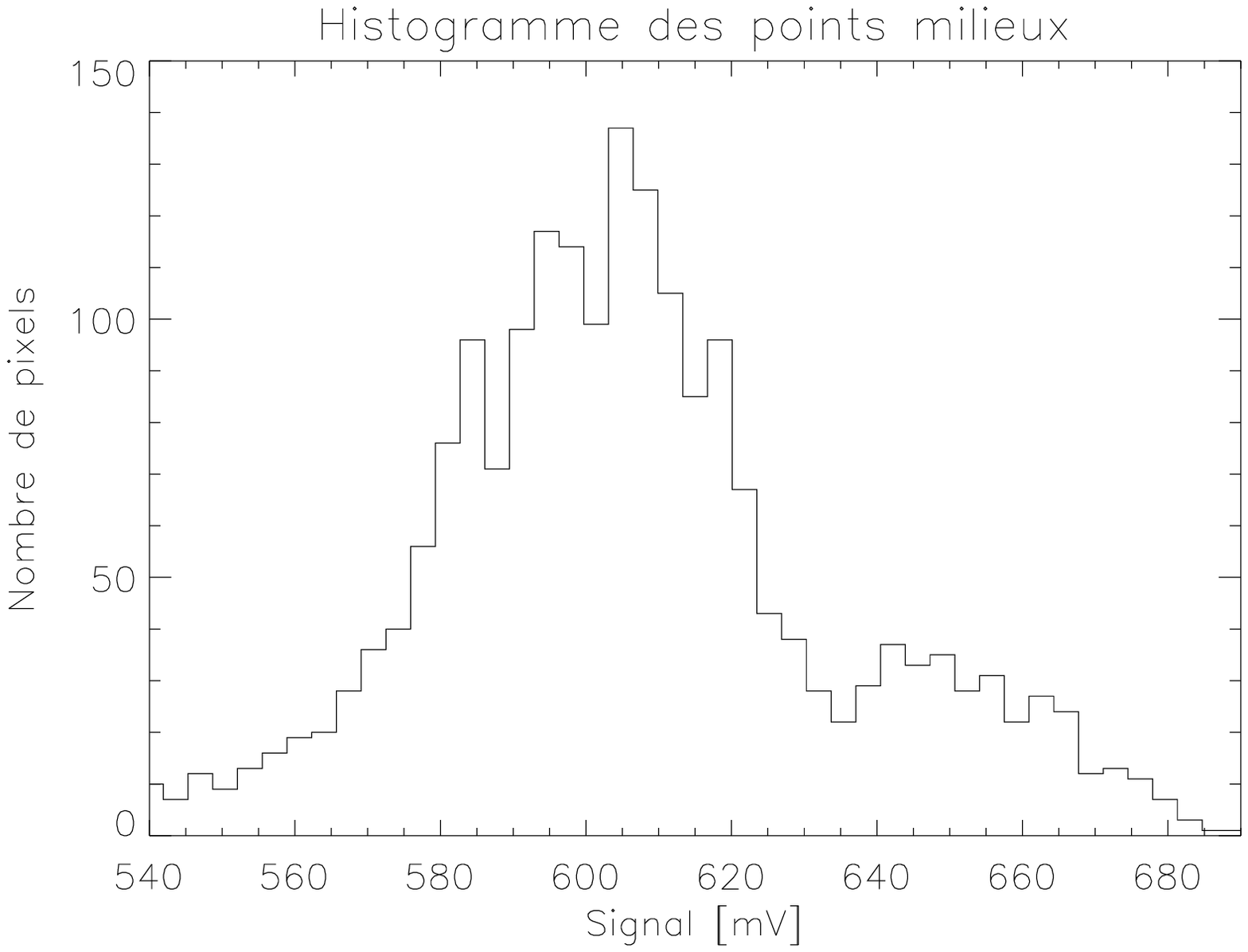} 
    \end{tabular}
  \end{center}
  \caption[Carte et dispersion de points milieux BFP
  bleu]{Distribution spatiale des points milieux du BFP bleu pour une
  tension de polarisation des ponts bolom\'etriques de 1.8~V et un
  flux incident de 2~pW/pixel. Les pixels morts repr\'esentent
  seulement 2~\% de la totalit\'e des pixels. L'histogramme de droite
  montre que la dispersion des points milieux pour cette tension de
  polarisation est de l'ordre de 150~mV.
  \label{fig:calib_procedure_explore_dispersion}}
\end{figure}

La distribution spatiale des points milieux du plan focal bleu du
PhFPU est donn\'ee dans la
figure~\ref{fig:calib_procedure_explore_dispersion} pour une tension
de polarisation de 1.8~V et un flux de 2~pW/pixel. Les espaces
inter-matrices y sont repr\'esent\'es en noir. Ils ont ici une largeur
par d\'efaut de 1~pixel mais des mesures plus pr\'ecises effectu\'ees
lors de la campagne d'\'etalonnage montrent que l'espacement est
l\'eg\`erement plus grand qu'un pixel.  Environ 2~\% des pixels ont
des points milieux aberrants, nous les appellerons par la suite les
\emph{pixels morts}. Ces pixels ne sont pas fonctionnels pour diverses
raisons. D'apr\`es les rapports de test du LETI, certains pixels
poss\`edent des poutres coll\'ees, dans ce cas la grille absorbante se
trouve thermalis\'ee \`a la temp\'erature du bain et le bolom\`etre ne
peut d\'etecter aucun flux incident, d'autres ne sont tout simplement
pas connect\'es \'electriquement au circuit de lecture, ceux-ci sont
alors totalement inutilisables. En ce qui concerne la colonne morte de
la matrice en haut \`a droite de la
figure~\ref{fig:calib_procedure_explore_dispersion} qui porte le nom
de SMD~N\textdegree05-07 (cf annexe~\ref{a:pacs_bfp}), elle a \'et\'e
sacrifi\'ee pour sauver le reste de la matrice. En effet, nous avions
s\'electionn\'e cette matrice pour ses bonnes performances, elle a
donc \'et\'e int\'egr\'ee dans le plan focal du mod\`ele de vol
d\'ebut 2005. Cependant lorsque nous avons test\'e le plan focal
complet, cette matrice n'\'etait pas fonctionnelle, la tension
$V_{SMS}$ qui contr\^ole le multiplexeur n'atteignait pas le circuit
de lecture. Le LETI a alors entrepris une op\'eration tr\'es
d\'elicate de pontage, c'est-\`a-dire qu'une ligne \'electrique
d\'edi\'ee \`a la lecture d'une colonne de pixel a \'et\'e
d\'etourn\'ee pour acheminer le signal du multiplexeur du BU vers le
CL. Le choix de la colonne s'est port\'ee sur celle qui \'etait la
plus proche physiquement de la ligne du multiplexeur \`a ponter (sinon
nous aurions choisi une colonne externe du BFP). L'op\'eration s'est
pass\'ee comme pr\'evu et nous avons r\'ecup\'er\'e les 15 autres
colonnes de la matrice SMD~N\textdegree05-07 fonctionnelles en
Septembre 2005.\\
Les pixels fonctionnels montrent une distribution relativement lisse
des points milieux sur chaque groupe. Toutefois les deux matrices du
groupe~4 (SMB~N\textdegree04-32 dans l'annexe~\ref{a:pacs_bfp}, ou
bien en haut \`a droite sur la
figure~\ref{fig:calib_procedure_explore_dispersion}) n'ont pas les
m\^eme offsets et leurs niveaux de points milieux se trouvent
d\'ecal\'es d'environ 100~mV. \'Etant donn\'e que les tensions de
r\'eglage sont appliqu\'ees \`a chaque groupe et non \`a chaque
matrice, le groupe~4 est particuli\`erement d\'elicat \`a r\'egler
puisque soit l'une, soit l'autre des matrices se retrouve proche de la
limite de saturation de BOLC. La dispersion des points milieux de la
figure~\ref{fig:calib_procedure_explore_dispersion} s'\'el\`eve \`a
environ 150~mV, mais elle peut atteindre plus de 350~mV pour les
fortes tensions de polarisation des ponts bolom\'etriques (cf
fig.~\ref{fig:calib_procedure_prediction_dispersion}). Notez que la
dynamique de BOLC en gain fort n'est que de 330~mV.\\

\begin{figure}
  \begin{center}
    \begin{tabular}[t]{ll}
      \includegraphics[width=0.48\textwidth,angle=0]{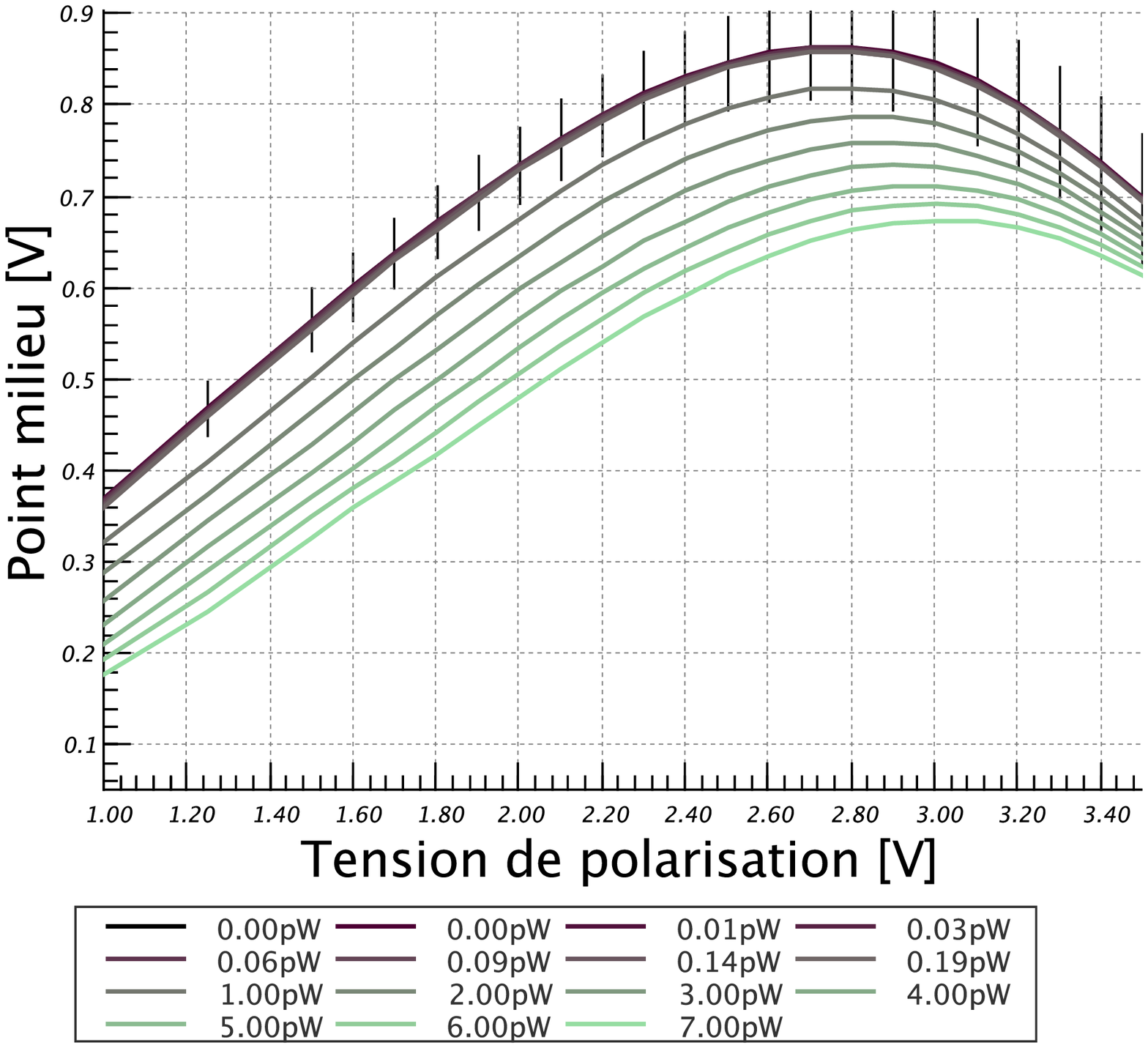} & \includegraphics[width=0.48\textwidth,angle=0]{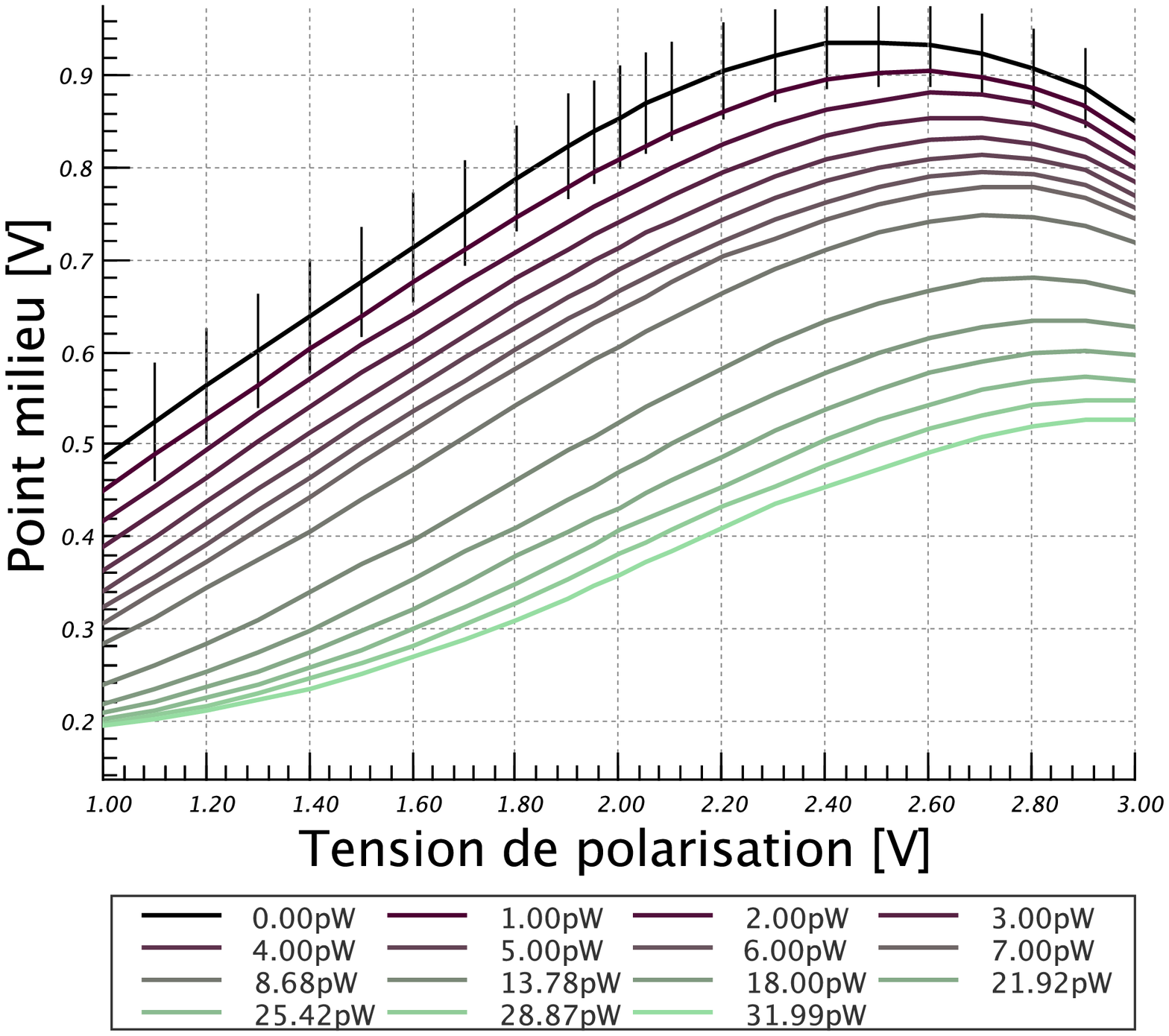} 
    \end{tabular}
  \end{center}
  \caption[\'Evolution des points milieux avec le flux et la tension
  de polarisation]{\'Evolution des points milieux moyenn\'es sur une
  matrice enti\`ere (matrice bleue \`a gauche et rouge \`a droite) en
  fonction de la tension de polarisation des ponts
  bolom\'etriques. Chaque courbe correspond \`a une temp\'erature de
  la source de calibration. Les barres d'erreur repr\'esentent la
  dispersion \emph{r.m.s.} sur la matrice enti\`ere en fonction de la
  tension de polarisation. Elles ne d\'ependent que faiblement du flux
  incident, c'est pour cela que nous ne les tra\c{c}ons que sur une
  seule courbe. Plus nous polarisons les ponts bolom\'etriques et plus
  le point milieu augmente jusqu'au point o\`u la dissipation
  \'electrique et les effets de champ font chuter l'imp\'edance des
  r\'esistances. Ces points milieux ont \'et\'e obtenus en mode de
  lecture direct avec le gain faible de BOLC.
  \label{fig:calib_procedure_explore_midpt}}
\end{figure}

Afin de tester le comportement des bolom\`etres, nous explorons 24
valeurs de $V_h$, $V_l$ et $V_{hb}$ ainsi que 16 valeurs de flux
incident. L'objectif est de tester les matrices pour des flux compris
entre 0~et 7~pW/pixel (par pas de 1~pW/pixel) sur chacun des BFP~;
mais puisque les sources de calibration du banc de test PACS n'ont pas
le m\^eme spectre d'\'emission que le t\'elescope Herschel, nous
mesurons en plus des flux de 0~\`a 0.2~pW/pixel sur le BFP bleu (ce
qui correspond \`a des flux de 0 \`a 7~pW/pixel sur le BFP rouge), et
de 8 \`a 32~pW/pixel sur le BFP rouge (ce qui correspond \`a des flux
de 0 \`a 7~pW/pixel sur le BFP bleu). Les points milieux reconstruits
\`a partir des mesures en gain faible sont pr\'esent\'es dans la
figure~\ref{fig:calib_procedure_explore_midpt}.

Chaque point de cette figure repr\'esente la moyenne spatiale des
points milieux d'une m\^eme matrice\footnote{Les pixels morts sont
masqu\'es pour \'eviter de biaiser le calcul de la moyenne.} pour une
tension de polarisation et un flux donn\'es. En utilisant la moyenne,
nous choisissons de montrer le comportement g\'en\'eral d'une matrice
enti\`ere, et nous v\'erifions que ces r\'esultats restent
repr\'esentatifs du comportement individuel de la plupart des
bolom\`etres. Chaque courbe illustre l'\'evolution des points milieux
pour un flux incident constant.

La figure~\ref{fig:calib_procedure_explore_midpt} montre que pour des
tensions de polarisation comprises entre 1~et 2~V, $V_{ptmil}$
augmente avec $V_{polar}$ (cf \'equation~\ref{eq:midpt}). Toutefois,
pour les tensions de polarisation plus \'elev\'ees, l'imp\'edance du
pont bolom\'etrique s'effondre, et ce pour deux raisons. D'une part,
la dissipation \'electrique par effet Joule \'echauffe la thermistance
et son imp\'edance chute exponentiellement en suivant
l'\'equation~(\ref{eq:efros}). D'autre part, l'effet de champ
augmente avec la tension aux bornes de la r\'esistance, ce qui
engendre \'egalement une chute exponentielle d'imp\'edance.  

Pour une tension de polarisation donn\'ee, le point milieu diminue
avec le flux incident. En effet lorsque le flux augmente, la
temp\'erature de l'absorbeur s'\'el\`eve ce qui engendre une chute de
l'imp\'edance de la thermistance, et donc une diminution de la tension
aux bornes de celle-ci. Une \'etude plus approfondie de l'influence du
flux sur l'\'evolution des points milieux est pr\'esent\'ee dans la
section~\ref{sec:calib_perflabo_nonlinear_courbe} qui est consacr\'ee
\`a la non-lin\'earit\'e des bolom\`etres. 

Ces courbes de points milieux nous donnent d\'ej\`a une premi\`ere
indication, \`a savoir que les performances des bolom\`etres se
trouvent limit\'ees aux fortes polarisations. Nous voyons en effet que
toutes les courbes semblent converger vers une m\^eme valeur de point
milieu pour les fortes polarisations quelque soit le flux
incident. C'est d'autant plus flagrant sur le BFP bleu (graphe de
gauche sur la figure~\ref{fig:calib_procedure_explore_midpt}) sur
lequel la tension de polarisation est explor\'ee jusqu'\`a 3.5~V (3~V
sur le BFP rouge). Si les courbes de point milieu se \og rapprochent
\fg, cela signifie que l'influence du flux incident diminue avec la
tension de polarisation, c'est-\`a-dire que la r\'eponse des
d\'etecteurs diminue. Nous aborderons ce sujet plus en d\'etail dans
la section~\ref{sec:calib_perflabo_sensibilite_reponse}.\\

Nous avons trouv\'e dans la section~\ref{sec:calib_procedure_vrlvhb}
la limite de fonctionnement de l'\'electronique de lecture, et il est
crucial que le signal d'entr\'ee du CL respecte cette limite. En
effet, si le r\'eglage et les conditions d'illumination des
bolom\`etres occasionnent une tension d'entr\'ee du CL inf\'erieure
\`a 250~mV, alors le signal bolom\'etrique ne sera pas transmis
correctement et le calcul de points milieux n'aura aucune
signification physique, il ne nous renseignera donc pas sur l'\'etat
du pont bolom\'etrique.  D'apr\`es la
figure~\ref{fig:calib_procedure_explore_schemasimple}, nous trouvons
la relation suivante entre le signal d'entr\'ee du CL, $V_{bolo}$, et
la tension aux bornes de la thermistance, $V_{ptmil}$:
\begin{center}
  \begin{equation}
    V_{bolo}=V_{ptmil}+V_l
  \label{eq:entreeCL}
  \end{equation}
\end{center}
Pour une valeur de $V_{ptmil}$ donn\'ee, \cad pour un flux et une
tension de polarisation donn\'es, nous pouvons choisir la valeur de
$V_l$ de sorte que la tension d'entr\'ee du CL soit sup\'erieure \`a
250~mV. Cependant, avec l'\'electronique chaude du photom\`etre PACS,
$V_l$ ne peut prendre que des valeurs n\'egatives. Dans certaines
conditions d'utilisation, le niveau de point milieu est alors trop bas
pour passer au-dessus de la barri\`ere des 250~mV, m\^eme avec un
$V_l$ maximum de 0~V. C'est par exemple le cas dans la
figure~\ref{fig:calib_procedure_explore_midpt} pour les faibles
tensions de polarisation et les forts flux du BFP rouge (graphe de
droite). De 0 \`a 7~pW/pixel, aux basses tensions de polarisation, les
courbes de points milieux sont parall\`eles~; par contre, pour les
flux sup\'erieurs \`a 10~pW/pixel, nous voyons que les courbes
convergent vers 200~mV quelque soit le flux incident. Cela signifie
que le signal d'entr\'ee du CL n'est pas suffisamment \'elev\'e dans
ces conditions d'illumination, et que les transistors du CL commencent
\`a saturer. Nous devons donc rejeter ces points milieux qui ne sont
pas repr\'esentatifs de l'\'etat des ponts bolom\'etriques. Il en est
de m\^eme pour les tensions de polarisation inf\'erieures \`a 1~V o\`u
le point milieu est le plus souvent inf\'erieur \`a 300~mV (points
non-visibles sur les graphes de points milieux).\\

Il existe par ailleurs une limitation sur la pr\'ecision des courbes
de la figure~\ref{fig:calib_procedure_explore_midpt}, et cette
incertitude est li\'ee \`a la d\'erive basse fr\'equence des
d\'etecteurs. En effet, les courbes d'\'etalonnage de l'\'electronique
permettent de corriger le signal de sortie uniquement pour les gains
et offsets du circuit de lecture \`a un instant donn\'e~; mais les
d\'erives basses fr\'equences de l'\'electronique et du bolom\`etre
lui-m\^eme ne sont pas corrig\'ees lors du calcul des points
milieux. La mesure des 168 configurations de tensions et flux
pr\'esent\'ees dans la figure~\ref{fig:calib_procedure_explore_midpt}
a n\'ecessit\'e plus de 30~heures de test~; les gains et offsets du
syst\`eme ont effectivement eu le temps de varier librement au cours
de ces mesures. En extrapolant la courbe de la
figure~\ref{fig:detect_outils_concept_typic} pour laquelle les
r\'eglages sont constants, nous estimons la d\'erive du syst\`eme \`a
un maximum de quelques~mV pour une dur\'ee de
$\sim$30~heures. Rappelons tout de m\^eme que l'objectif de ces
mesures en gain faible n'est pas de mesurer les performances des
bolom\`etres ni de quantifier leur stabilit\'e, et qu'une pr\'ecision
de quelques~mV reste tout \`a fait compatible avec notre objectif
premier d'automatiser le r\'eglage des d\'etecteurs (ces quelques~mV
sont \`a comparer avec la dynamique des ADC).

Le but de ces quelques remarques sur la stabilit\'e des d\'etecteurs
est simplement de pr\'evenir le lecteur que les d\'erives basses
fr\'equences affectent les courbes de points milieux. Nous
pr\'esentons une analyse plus d\'etaill\'ee de ces d\'erives dans les
sections~\ref{sec:calib_perflabo_nonlinear}
et~\ref{sec:calib_perfobs_oof}~; nous verrons par la suite que,
malgr\'e les lentes fluctuations de signal inh\'erentes \`a tout
instrument de mesure, la reconstruction des points milieux est un
calcul suffisament pr\'ecis pour pr\'edire le r\'eglage des
bolom\`etres, et que nous pouvons de plus extraire la r\'eponse des
bolom\`etres \`a partir des courbes de points milieux avec une
pr\'ecision tout \`a fait raisonnable de 10~\%.\\



\subsection{Les rapports d'imp\'edance}
\label{sec:calib_procedure_explore_imped}

\begin{figure}
  \begin{center}
    \begin{tabular}{l}
      \includegraphics[width=0.48\textwidth,angle=0]{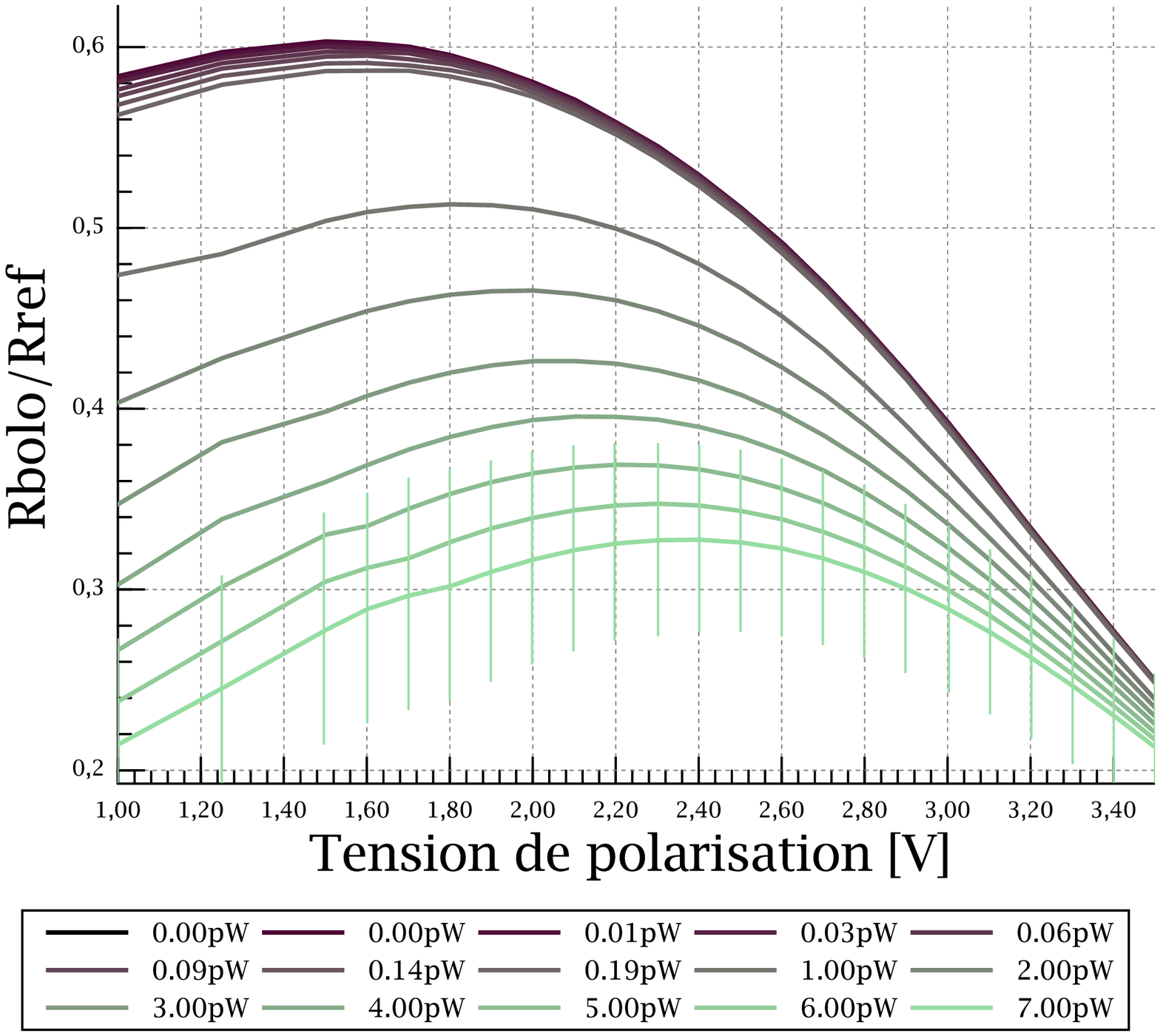}
      \includegraphics[width=0.48\textwidth,angle=0]{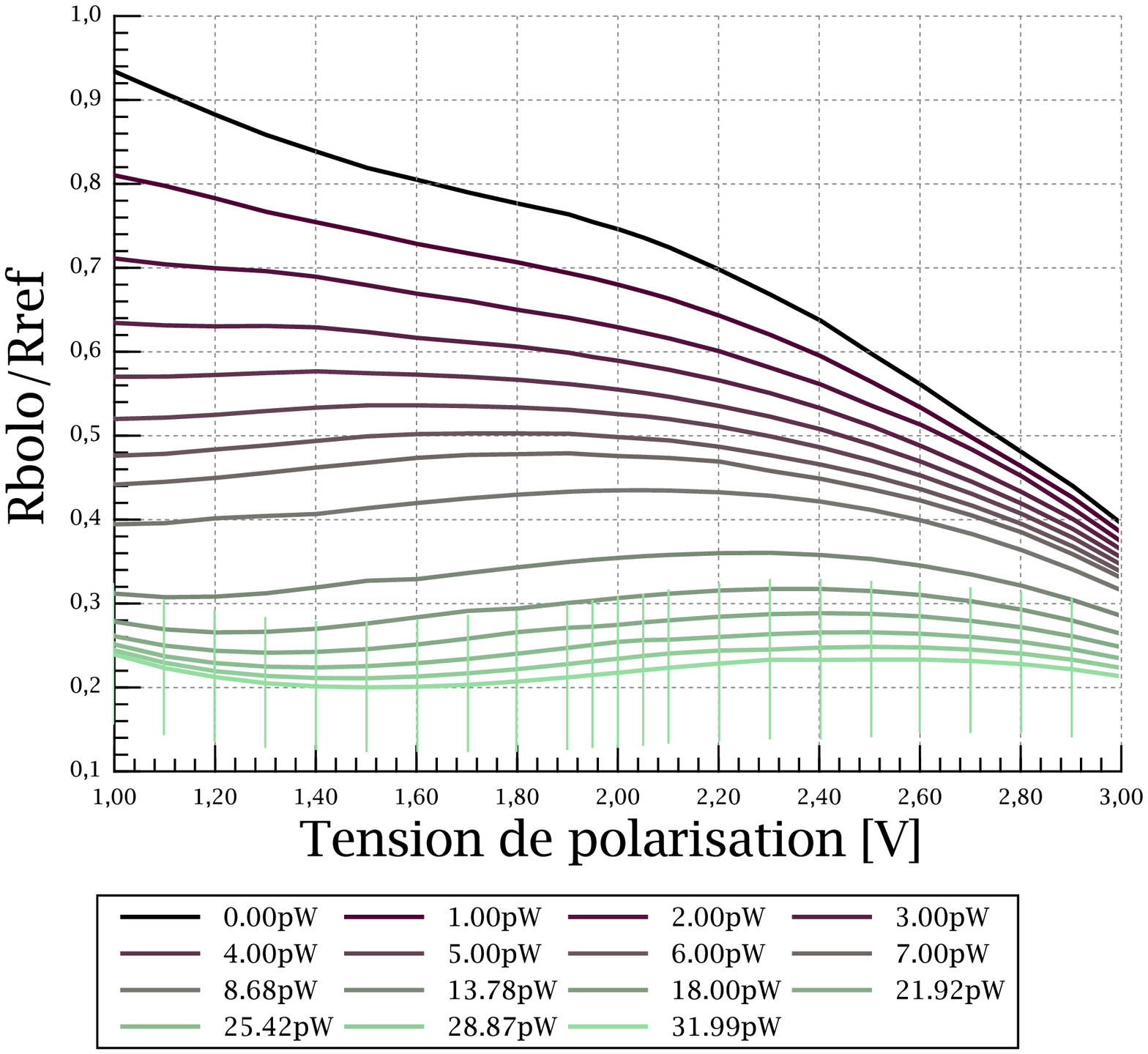}
    \end{tabular}
  \end{center}
  \caption[\'Evolution des rapports d'imp\'edance avec le flux et la
tension de polarisation]{\'Evolution des rapports d'imp\'edance
$\frac{R_{bolo}}{R_{ref}}$ moyenn\'es sur une matrice enti\`ere en
fonction de la tension de polarisation des ponts bolom\'etriques et du
flux incident. Les barres d'erreurs montrent la dispersion
\emph{r.m.s.} des rapports d'imp\'edance sur la matrice. Les courbes
de la matrice rouge (\`a droite) ont un comportement tout \`a fait
coh\'erent avec la physique du d\'etecteur, \`a savoir les rapports
tendent vers~1 pour les faibles flux et les faibles tensions. Par
contre, le fait que les courbes de la matrice bleue (\`a gauche) ne
tendent pas vers~1 mais fl\'echissent aux basses tensions indique que
pour le BFP bleu, il existe probablement un offset introduit en amont
du circuit de lecture par le multiplexeur. Le calcul de points milieux
ne prend pas en compte l'injection de charges \`a l'origine de cet
offset. Une explication plus d\'etaill\'ee est donn\'ee dans le texte.
  \label{fig:calib_procedure_explore_imped}}
\end{figure}

Rappelons que, par construction, les thermistances $R_{bolo}$ et les
r\'esistances de r\'ef\'erence $R_{ref}$ d'une m\^eme matrice sont
identiques (cf
section~\ref{sec:detect_bolocea_fabrication_thermo}). Par
cons\'equent, lorsque le flux est nul et que la tension de
polarisation tend vers 0~V (pas de dissipation Joule ni d'effet de
champ), les imp\'edances devraient tendre vers une m\^eme valeur
puisque les deux r\'esistances se trouvent thermalis\'ees \`a la
m\^eme temp\'erature, celle de la source froide. Nous allons
maintenant utiliser cette propri\'et\'e pour tester la pertinence du
calcul de points milieux. En effet, exprimer les points milieux sous
forme de rapports d'imp\'edance nous offre un diagnostique plus
quantitatif que la repr\'esentation des donn\'ees sous forme de
tensions de points milieux. \`A partir de
l'\'equation~(\ref{eq:midpt}) et des donn\'ees de la
figure~\ref{fig:calib_procedure_explore_midpt}, nous calculons le
rapport d'imp\'edance de chacun des ponts bolom\'etriques pour chaque
configuration $(tension,flux)$ avec la formule suivante~:
$$\frac{R_{bolo}}{R_{ref}}=\frac{V_{ptmil}}{V_{polar}-V_{ptmil}}$$ Le
r\'esultat de ces calculs est illustr\'e dans la
figure~\ref{fig:calib_procedure_explore_imped}. Int\'eressons-nous
dans un premier temps aux courbes du BFP rouge (graphe de droite). \\
\noindent Lorsque le flux incident sur le bolom\`etre augmente, la
temp\'erature de l'absorbeur augmente et l'imp\'edance de la
thermistance $R_{bolo}$ diminue, le rapport $R_{bolo}/R_{ref}$
d\'ecro\^it. Lorsque la tension de polarisation augmente,
l'interpr\'etation est un peu plus d\'elicate car les effets de champ
tendent \`a r\'eduire l'imp\'edance $R_{bolo}$ mais \'egalement
l'imp\'edance $R_{ref}$, et ceci est une caract\'eristique unique aux
bolom\`etres du CEA. Il est donc n\'ecessaire d'analyser l'\'evolution
du rapport $R_{bolo}/R_{ref}$ dans diff\'erents r\'egimes de
fonctionnement des bolom\`etres~:
\begin{itemize}
\item Dans un r\'egime de faibles flux et de faibles tensions
(quadrant sup\'erieur gauche), la dissipation Joule est faible et le
flux incident apporte relativement peu d'\'energie au bolom\`etre. La
temp\'erature de la thermistance est alors \`a peine plus \'elev\'ee
que celle de la r\'esistance de r\'ef\'erence, leur imp\'edance est
donc peu diff\'erente. Le rapport $R_{bolo}/R_{ref}$ doit tendre
vers~1 quand le flux et la tension diminuent.
\item Dans un r\'egime de forts flux et de faibles tensions (quadrant
inf\'erieur gauche, mais pour des flux inf\'erieurs \`a
$\sim$10~pW/pixel car au-del\`a le signal ne passe pas le circuit de
lecture), la dissipation Joule est relativement faible mais
l'absorbeur est r\'echauff\'e par le flux incident,
$R_{bolo}/R_{ref}\sim0.5$. Les deux tiers de la tension appliqu\'ee au
pont bolom\'etrique sont donc report\'es sur la r\'esistance de
r\'ef\'erence qui voit son imp\'edance chuter \`a cause de l'effet de
champ. Dans ce r\'egime de fonctionnement, \cad une r\'esistance de
r\'ef\'erence froide et fortement polaris\'ee et une thermistance
chaude et peu polaris\'ee, $R_{ref}$ chute plus rapidement que
$R_{bolo}$ (cf figure~\ref{fig:detect_bolocea_fabrication_thermo_R2T})
et le rapport $R_{bolo}/R_{ref}$ augmente l\'eg\`erement avec la
tension de polarisation.
\item Pour les fortes polarisations (moiti\'e droite du graphe), la
dissipation Joule apporte beaucoup plus d'\'energie \`a l'absorbeur
que le flux radiatif. Le comportement thermique du bolom\`etre est
alors domin\'e par la tension de polarisation quelque soit le flux~;
les courbes se rapprochent et la r\'eponse diminue (le coefficient
$\alpha=\partial R / \partial T$ s'\'ecroule, cf
figure~\ref{fig:detect_bolocea_fabrication_thermo_R2T}).
Dans ce r\'egime, \cad quelques Volts de potentiel aux bornes de la
r\'esistance de r\'ef\'erence et une thermistance significativement
plus chaude que la source froide, l'imp\'edance totale du pont
bolom\'etrique est environ 10~fois plus petite que pour une tension de
$\sim$1~V.
\end{itemize}
Les rapports d'imp\'edance du BFP rouge sont en tr\`es bon accord avec
notre compr\'ehension des bolom\`etres PACS. Par contre,
l'interpr\'etation que nous proposons n'est correcte sur le BFP bleu
que pour les forts flux et les fortes tensions de polarisation~; le
probl\`eme se trouve aux faibles tensions et faibles flux car le
rapport $R_{bolo}/R_{ref}$ ne tend pas vers~1 comme attendu. Mais pour
quelle raison les rapports d'imp\'edance des BFP bleu et rouge
seraient-ils diff\'erents? Nous avons d'abord pens\'e qu'une forte
lumi\`ere parasite sur la voie bleue du Photom\`etre \'etait
responsable du fl\'echissement des courbes aux basses tensions~; mais
aucune autre preuve ne semble indiquer l'existence d'un tel niveau de
lumi\`ere parasite (nous avons soigneusement noirci l'int\'erieur de
l'instrument pour \'eviter les r\'eflexions de lumi\`ere sur le plan
focal, et les bancs de test de Saclay et de Garching donnent
pr\'ecisemment les m\^emes r\'esultats pour un flux suppos\'e de
1~pW/pixel). Nous savons par ailleurs que les circuits de lecture des
BFP bleu et rouge sont strictement identiques, nous avons en effet
v\'erifi\'e que les courbes d'\'etalonnage de l'\'electronique
pr\'esent\'ees dans la figure~\ref{fig:calib_procedure_vrvhb_surfcal}
sont les m\^emes pour les deux BFPs. Par cons\'equent, la diff\'erence
entre les rapports d'imp\'edance bleus et rouges provient d'un
\'el\'ement qui se trouve sur le circuit de d\'etection (CD), en amont
du premier transistor de lecture. Par construction, les r\'esistances
du BFP bleu sont plus imp\'edantes que celles du BFP rouge, mais cette
diff\'erence devrait s'exprimer sur les courbes de la
figure~\ref{fig:calib_procedure_explore_imped} uniquement par une
diff\'erence de forme~; les rapports d'imp\'edance devraient quand
m\^eme tendre vers~1 pour les faibles tensions de polarisation, le
contraire ne serait pas physique. Nous sommes donc forc\'es de
constater que pour les faibles tensions de polarisation des
bolom\`etres, les points milieux que nous avons calcul\'es pour le BFP
bleu ne sont pas repr\'esentatifs de la tension aux bornes de la
thermistance.

Malgr\'e cela, la m\'ethode de calcul des points milieux telle que
nous l'avons pr\'esent\'ee dans la section pr\'ec\'edente reste
correcte. C'est l'hypoth\`ese selon laquelle la tension d'entr\'ee du
circuit de lecture est \'equivalente \`a la tension centrale du pont
bolom\'etrique qui \'echoue dans certaines configurations du
d\'etecteur. Pour lever cette hypoth\`ese, et faire une mesure plus
r\'ealiste de l'\'equilibre du pont bolom\'etrique, il nous faudrait
des abaques identiques \`a celles de la
figure~\ref{fig:calib_procedure_vrvhb_surfcal} mais qui prendraient en
compte l'\'etage haute imp\'edance des matrices de bolom\`etres. En
effet, l'\'etalonnage de la cha\^ine \'electronique consiste \`a
injecter une tension de r\'ef\'erence dans le circuit au niveau du
$V_{ref}$ (\`a ce niveau l'imp\'edance est de $\sim$5~M$\Omega$ pour
les deux BFPs)~; alors qu'il faudrait injecter une tension au niveau
du pont bolom\'etrique. Si cela \'etait possible, nous devrions voir
une diff\'erence entre les courbes d'\'etalonnage du BFP bleu et
celles du BFP rouge car leurs imp\'edances sont diff\'erentes. Pour
r\'ealiser ce type de test, il suffirait de commander $V_h=V_l$, et
d'explorer plusieurs couples $(V_l,V_{hb})$ de la m\^eme mani\`ere que
nous l'avons fait dans la section~\ref{sec:calib_procedure_vrlvhb}
pour obtenir les abaques de la
figure~\ref{fig:calib_procedure_vrvhb_surfcal}. De telles courbes
d'\'etalonnage devraient nous permettre de mesurer le v\'eritable
point milieu de chaque pont bolom\'etrique, et nous devrions trouver
un rapport d'imp\'edance qui tend vers~1 pour les faibles tensions de
polarisation. Il n'est malheureusement pas possible d'effectuer ce
test\footnote{Il serait \'eventuellement possible d'utiliser une
bo\^ite d'\'eclatement entre BOLC et le cryostat pour connecter $V_h$
\`a $V_l$ et commander \`a BOLC des excursions de $V_l$. Toutefois, le
Photom\`etre a \'et\'e int\'egr\'e dans l'instrument PACS il y a plus
d'un an, et il est maintenant hors de question de faire ce genre de
test sur un instrument spatial.} sur l'instrument PACS car BOLC ne
permet pas de commander des valeurs positives de $V_l$. 

Bien que les outils diagnostiques dont nous disposons pour le moment
soient limit\'es, nous pouvons quand m\^eme proposer un sc\'enario qui
expliquerait les diff\'erences observ\'ees entre les rapports
d'imp\'edance bleu et rouge de la
figure~\ref{fig:calib_procedure_explore_imped}. Nous pensons que
l'origine du probl\`eme se trouve au niveau du multiplexeur qui
injecterait des charges parasites dans l'\'etage haute imp\'edance de
la matrice par couplage capacitif. En effet, en mode PEL commut\'e, le
MOS de lecture (contr\^ol\'e par $V_{SMS}$,
figure~\ref{fig:detect_bolocea_elec_froide_principe}) d'un m\^eme
pixel est allum\'e pendant 1.56~ms ($\frac{1}{16\times 40}$), temps
durant lequel l'\'electronique \'echantillonne la tension \`a
l'entr\'ee du CL, puis il s'\'eteint pendant 23.44~ms jusqu'\`a ce
qu'il se rallume pour mesurer \`a nouveau le signal bolom\'etrique et
ainsi de suite (cf section~\ref{sec:detect_bolocea_elec_lecture} pour
une description d\'etaill\'ee du s\'equenceur et du multiplexeur). Le
point important ici est que les MOS de lecture sont largement mis \`a
contribution par le multiplexeur en mode PEL commut\'e~; et nous
pensons que les \'etats transitoires de ces MOS lors de l'allumage
peuvent lib\'erer des charges parasites dans le circuit \`a moyenne
imp\'edance (cf
figure~\ref{fig:detect_bolocea_elec_froide_principe}). Ces
perturbations sont ensuite communiqu\'ees \`a l'\'etage haute
imp\'edance par couplage capacitif du MOS de lecture (le mouvement des
charges dans le circuit moyenne imp\'edance induit le mouvement des
charges dans le circuit haute imp\'edance par le biais de la
capacitance du MOS de lecture, dont la fonction de transfert est
$\boldsymbol{i}C\omega$). Des perturbations \'electroniques sont donc
inject\'ees p\'eriodiquement dans l'\'etage haute imp\'edance.
Rappelons que le bolom\`etre agit comme un filtre passe-bas (cf
section~\ref{sec:intro_bolometrie_thermo_principe}), \cad qu'il est
capable de d\'etecter les pics d'un signal p\'eriodique si la
constante de temps du filtre est beaucoup plus longue que la p\'eriode
des perturbations~; dans ce cas l'injection de charges se traduit
alors par un offset au niveau du pont bolom\'etrique.

\begin{figure}
  \begin{center}
      \includegraphics[width=0.6\textwidth,angle=0]{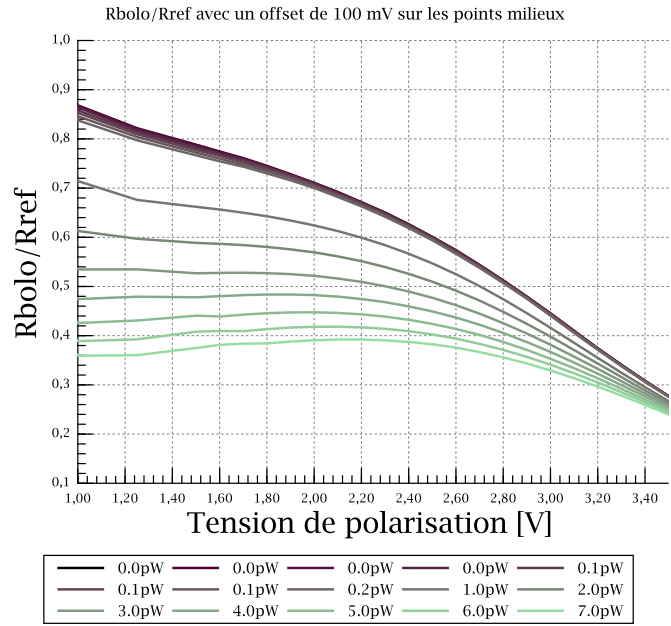}
  \end{center}
  \caption[Rapports d'imp\'edance du BFP bleu avec un offset de 100~mV
  sur les points milieux]{Rapports d'imp\'edance de la matrice~0 du
  BFP bleu pour lesquels nous avons ajout\'e syst\'ematiquement un
  offset de 100~mV \`a tous les points milieux. Bien que le
  v\'eritable offset ne soit pas a~priori constant avec la tension de
  polarisation ou le flux incident, le r\'esultat obtenu est flagrant:
  les rapports $R_{bolo}/R_{ref}$ \og modifi\'es \fg ont retrouv\'e
  une allure similaire aux rapports du BFP rouge.
  \label{fig:calib_procedure_explore_impedOffset}}
\end{figure}

Suite \`a cette analyse, nous avons essay\'e d'ajouter diff\'erentes
valeurs d'offset aux points milieux du BFP bleu pour voir si cela
pouvait r\'econcilier les rapports d'imp\'edance re-calcul\'es avec le
comportement attendu. La
figure~\ref{fig:calib_procedure_explore_impedOffset} montre
l'\'evolution du rapport $R_{bolo}/R_{ref}$ pour un offset de
100~mV. L'am\'elioration est manifeste. Les courbes deviennent
coh\'erentes avec celles du BFP rouge et nous retrouvons bien que le
rapport tend vers~1 pour les faibles tensions de polarisation. Le
ph\'enom\`ene d'injection de charges pourrait donc expliquer les
diff\'erences que nous observons entre les BFPs bleu et rouge (le BFP
bleu est plus imp\'edant donc poss\`ede une constante de temps plus
grande).
Nous verrons dans la section~\ref{sec:calib_perflabo_compare} que le
mode de lecture DDCS induit \'egalement des perturbations de ce
type.

Pour mettre \`a l'\'epreuve notre interpr\'etation, il faut pouvoir
modifier l'amplitude et la p\'eriode des injections de charges. Une
solution serait de modifier le nombre de charges inject\'ees en
changeant le courant qui alimente le MOS de lecture (PEL), \cad
$V_{gg}$ et $V_{dd}$, et de voir l'evolution des rapports
d'imp\'edance. Nous pourrions \'egalement changer la fr\'equence
d'\'echantillonnage pour que la p\'eriode des perturbations deviennent
longue devant la constante de temps des bolom\`etres, l'offset
parasite devrait alors dispara\^itre. Il faudrait aussi estimer la
valeur de cet offset lorsque les matrices fonctionnent en mode PEL
statique, ou alors en for\c{c}ant le s\'equenceur \`a
n'\'echantillonner qu'un seul pixel. Dans ce cas, le MOS de lecture
serait toujours aliment\'e et aucune charge ne devrait \^etre
inject\'ee dans le circuit haute imp\'edance. Mais, \`a nouveau, le
Photom\`etre PACS \'etant livr\'e \`a L'ESA, nous ne pouvons pas
effectuer ces tests sur le mod\`ele de vol. Cependant, nous allons
bient\^ot recevoir le mod\`ele de rechange du Photom\`etre sur lequel
nous pourrons changer le s\'equenceur et \'eventuellement confirmer la
pr\'esence d'un offset d\^u aux injections de charges du
multiplexeur.\\

Remarquez que les tests pr\'esent\'es dans cette section ne sont pas
de v\'eritables mesures de performance, leur but premier \'etait
simplement de sonder sans a priori le comportement des bolom\`etres,
mais nous avons quand m\^eme pu extraire de pr\'ecieuses informations
sur le fonctionnement physique des d\'etecteurs. Notez \'egalement que
la pr\'esence de cet offset n'a aucune incidence sur la suite de la
proc\'edure d'\'etalonnage ni sur les mesures de performance. Nous
verrons d'ailleurs dans la section~\ref{sec:calib_perflabo_nonlinear}
que les points milieux que nous avons calcul\'es et trac\'es dans la
figure~\ref{fig:calib_procedure_explore_midpt} contiennent des
informations physiques utiles \`a l'\'etalonnage des matrices de
bolom\`etres, notamment sur la non-lin\'earit\'e et la r\'eponse des
d\'etecteurs (m\^eme pour le BFP bleu). L'existence de cet offset
repr\'esente uniquement une limitation pour interpr\'eter et mesurer
les param\`etres physiques des bolom\`etres.




\section{La pr\'ediction du r\'eglage des d\'etecteurs}
\label{sec:calib_procedure_prediction}

\begin{figure}
  \begin{center}
      \includegraphics[width=0.6\textwidth,angle=0]{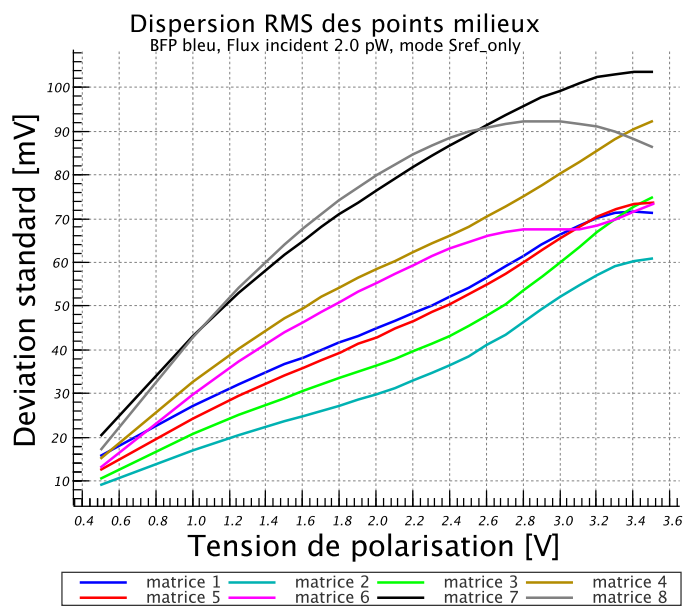}
  \end{center}
  \caption[Dispersion des points milieux en fonction de la tension de
  polarisation]{Dispersion des points mileux en fonction de la tension
  de polarisation des ponts bolom\'etriques. Cette dispersion est la
  d\'eviation standard calcul\'ee sur chaque matrice du BFP bleu pour
  un flux incident de 2~pW/pixel. Pour les fortes tensions de
  polarisation, la d\'eviation pic-\`a-pic peut d\'epasser 300~mV,
  \cad la dynamique totale des convertisseurs num\'eriques de BOLC.
  \label{fig:calib_procedure_prediction_dispersion}}
\end{figure}

Les matrices de bolom\`etres du CEA poss\`edent une dispersion
intrins\`eque des points milieux qui risque d'engendrer la saturation
d'une large fraction des d\'etecteurs lors des mesures de
performance. Cette dispersion est directement reli\'ee \`a la
pr\'ecision de fabrication des r\'esistances. Malgr\'e la grande
homog\'en\'eit\'e et la reproductibilit\'e du processus d'implantation
ionique (\`a mieux de 5~\% selon \shortciteNP{simoens}), il reste de
petites variations spatiales d'imp\'edance sur chaque matrice causant
une dispersion non-n\'egligeable sur les niveaux de points
milieux. Cette dispersion est d'autant plus amplifi\'ee que la tension
de polarisation appliqu\'ee aux bornes des ponts bolom\'etriques est
grande comme le montre la
figure~\ref{fig:calib_procedure_prediction_dispersion}. 
Dans certaines conditions d'utilisation, la dispersion pic-\`a-pic
d'une seule matrice peut \^etre sup\'erieure \`a la dynamique des
convertisseurs num\'eriques de BOLC. Le r\'eglage des tensions
$V_{ref}$ et $V_{hb}$ peut donc s'av\'erer tr\`es d\'elicat.
Au d\'ebut de ma th\`ese, ces r\'eglages \'etaient effectu\'es \og \`a
la main \fg de fa\c{c}on empirique, c'est-\`a-dire que l'op\'erateur
devait r\'egler les offsets $V_{ref}$ et $V_{hb}$ pour ajuster le
signal au milieu de la dynamique de BOLC, et ce pour chaque nouvelle
configuration du syst\`eme \`a tester. Cette m\'ethode de r\'eglage
\'etait longue et tr\`es inefficace, il \'etait donc n\'ecessaire de
l'automatiser.\\

D'une part, nous avons caract\'eris\'e le comportement des ponts
bolom\'etriques, nous connaissons donc le signal qui entre dans le
circuit de lecture quelque soient les conditions d'illumination et la
tension appliqu\'ee aux bornes des ponts bolom\'etriques. En outre, la
fonction de transfert de l'\'electronique de lecture nous permet de
calculer le signal de sortie du photom\`etre connaissant le signal
d'entr\'ee du CL. \`A partir de ces deux \'el\'ements nous pouvons
pr\'edire les valeurs de $V_{ref}$ et $V_{hb}$ \`a appliquer aux
d\'etecteurs pour que le signal de sortie soit centr\'e sur la
dynamique des convertisseurs num\'eriques de BOLC.  Notez que la
pr\'ecision avec laquelle nous pouvons r\'egler les bolom\`etres est
fix\'ee par la d\'erive basse fr\'equence du signal (thermistance +
chaine \'electronique + offset parasite). Elle est de l'ordre de
quelques~mV, ce qui ne repr\'esente qu'une fraction de la dynamique
totale de BOLC. D'autre part, l'offset que nous avons mis en
\'evidence dans la section~\ref{sec:calib_procedure_explore_imped} n'a
pas d'incidence sur la pr\'ediction du r\'eglage des bolom\`etres. En
effet, nous utilisons deux fois la courbe d'\'etalonnage de
l'\'electronique, une fois dans chaque sens\footnote{La courbe de la
figure~\ref{fig:calib_procedure_vrvhb_surfcal} est utilis\'ee une
premi\`ere fois pour convertir le signal de sortie en signal
d'entr\'ee du CL pour le calcul des points milieux ($V_{out} \to
V_{ptmil}$), et une deuxi\`eme fois pour convertir le signal
d'entr\'ee en signal de sortie ($V_{ptmil} \to V_{out}$).}, de sorte
que les erreurs introduites par cet offset s'annulent, \`a la d\'erive
pr\`es.

En pratique, nous avons d\'evelopp\'e un programme qui prend comme
variables d'entr\'ee un flux incident et une tension de polarisation
$(V_h-V_l)$, et qui g\'en\`ere en sortie les quatres tensions \`a
appliquer aux d\'etecteurs qui minimisent la saturation de BOLC et des
transistors de l'\'electronique froide. Ces tensions sont $V_h$,
$V_l$, $V_{ref}$ et $V_{hb}$. La proc\'edure de g\'en\'eration des
tensions suit la logique suivante:
\begin{itemize}
\item La premi\`ere \'etape est d'interpoler les courbes de la
figure~\ref{fig:calib_procedure_explore_midpt} pour conna\^itre le
point milieu, c'est-\`a-dire la tension aux bornes de la thermistance,
qui correspond au flux et \`a la tension de polarisation demand\'es
par l'utilisateur.
\item $V_l$ est d\'etermin\'e tel que le potentiel entrant dans le
circuit de lecture, $V_{bolo}$, soit le plus proche possible d'une
valeur cible que je fixe \`a 400~mV ($V_l=V_{cible}-V_{ptmil}$, voir
la figure~\ref{fig:calib_procedure_explore_schemasimple} et
l'\'equation~\ref{eq:entreeCL}) tout en tenant compte des
diff\'erentes contraintes li\'ees \`a l'\'electronique de lecture. En
effet, d'apr\`es les r\'esultats de la
section~\ref{sec:calib_procedure_vrlvhb} cette tension cible doit se
trouver entre 250 et 700~mV pour que le signal soit transmis jusqu'\`a
BOLC. D'autre part, la conception de BOLC et du circuit de lecture
impose que $V_l$ soit n\'egative et sup\'erieure \`a -600~mV.
\item $V_h$ est ensuite d\'eduite de la tension de polarisation
$(V_h-V_l)$ fournie par l'utilisateur et de la valeur de $V_l$ obtenue
\`a l'\'etape pr\'ec\'edente. \`A partir de l\`a, les bolom\`etres
sont polaris\'es et nous connaissons pour chacun des pixels la tension
qui entre dans le circuit de lecture.
\item $V_{ref}$ est fix\'ee \`a la valeur m\'ediane des tensions
d'entr\'ee des deux matrices d'un m\^eme groupe (m\^eme BU). Ceci
place $V_{ref}$ au centre de la distribution des signaux d'un m\^eme
BU, et dans le cas d'une mesure en mode DDCS, les matrices devraient
\^etre centr\'ees autour de z\'ero, minimisant ainsi la saturation des
convertisseurs de BOLC.
\item Puisque $V_{ref}$ repr\'esente la valeur m\'ediane des signaux
d'entr\'ee du CL, nous utilisons la
figure~\ref{sec:calib_procedure_vrlvhb} pour d\'eterminer\footnote{La
m\'ethode pour obtenir le signal de sortie connaissant le signal
d'entr\'ee est exactement l'inverse du calcul de point milieu
pr\'esent\'e dans la section pr\'ec\'edente o\`u nous avons calcul\'e
le signal entrant \`a partir du signal de sortie.} la valeur de
$V_{hb}$ qui centrera le signal de sortie sur la dynamique de BOLC. Le
centre de la dynamique d\'epend du gain de BOLC (cf
annexe~\ref{a:dyna_BOLC}), la valeur cible est de 83~mV en gain faible
et de 330~mV en gain fort.
\end{itemize}
\begin{figure}
  \begin{center}
    \begin{tabular}{ll}
      \includegraphics[width=0.49\textwidth,angle=0]{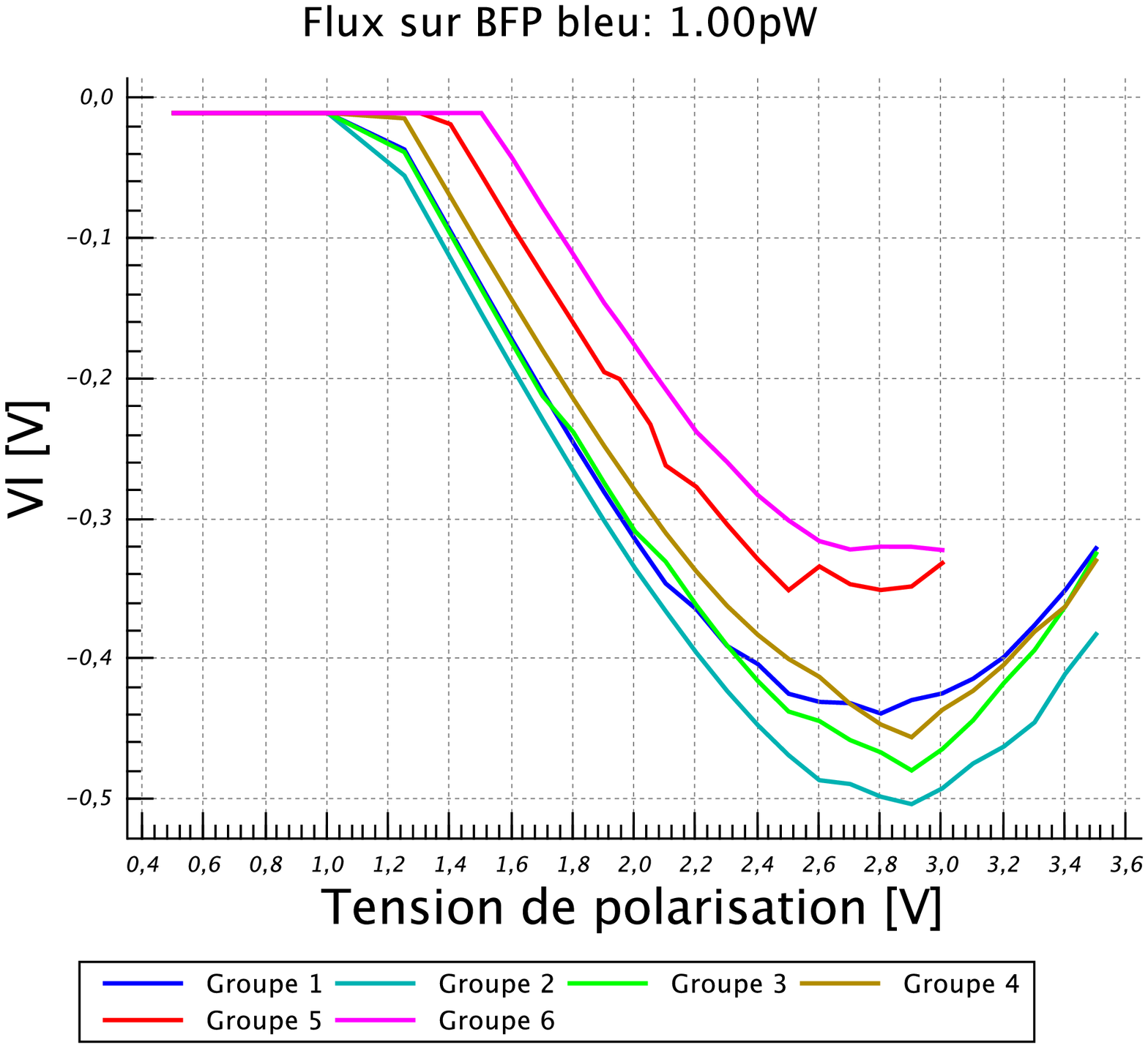}
      \includegraphics[width=0.49\textwidth,angle=0]{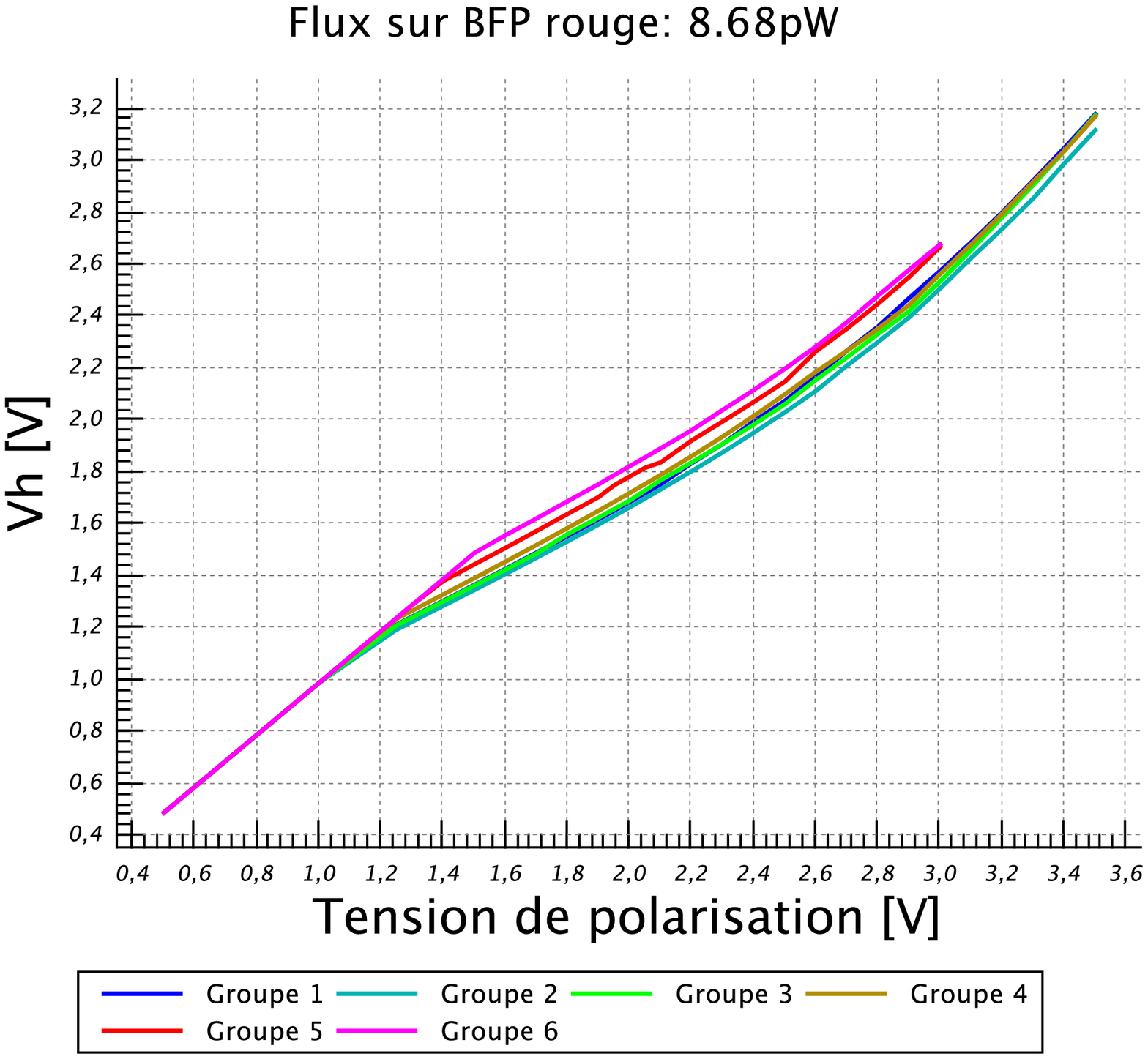}\\
      \includegraphics[width=0.49\textwidth,angle=0]{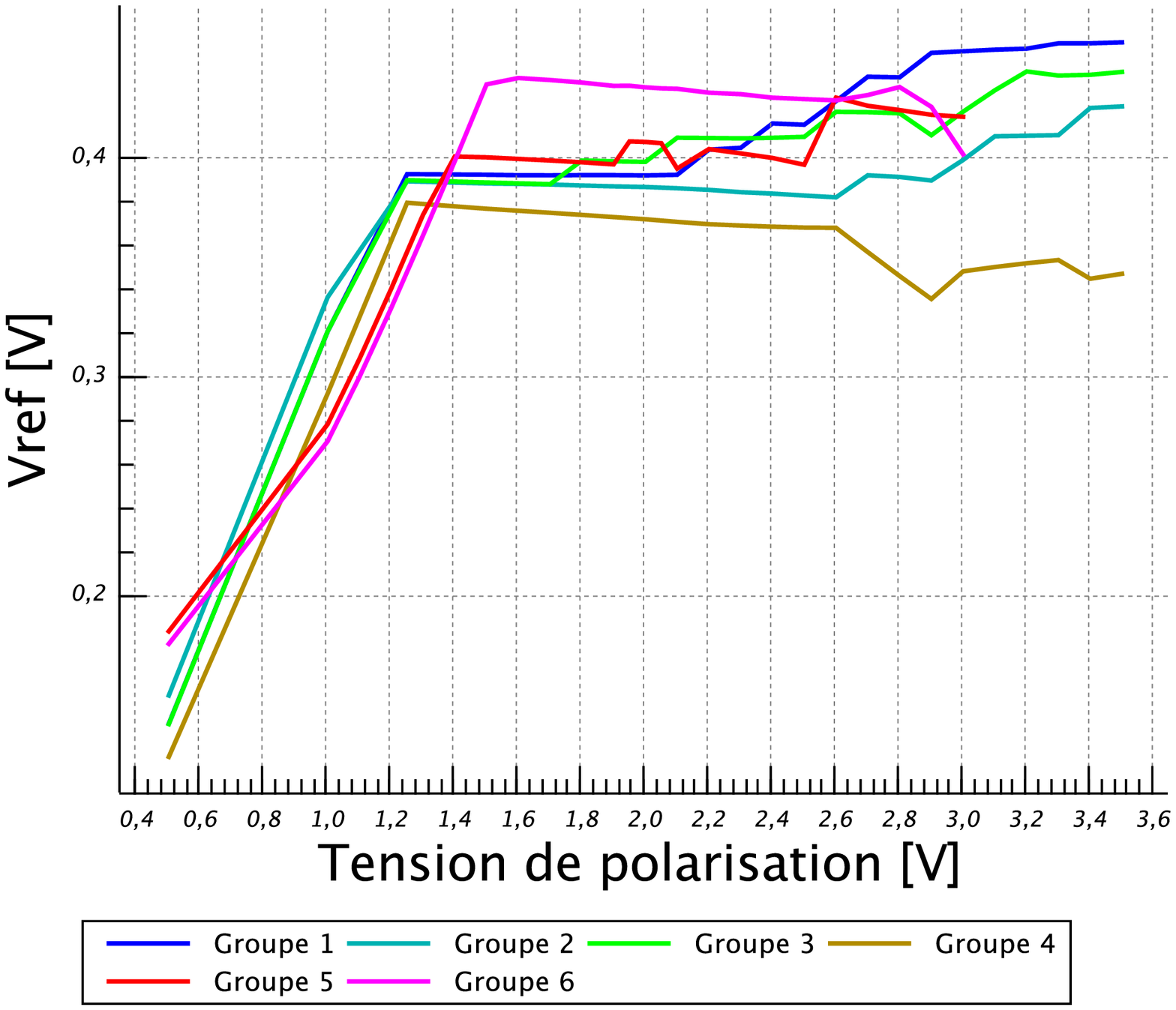}
      \includegraphics[width=0.49\textwidth,angle=0]{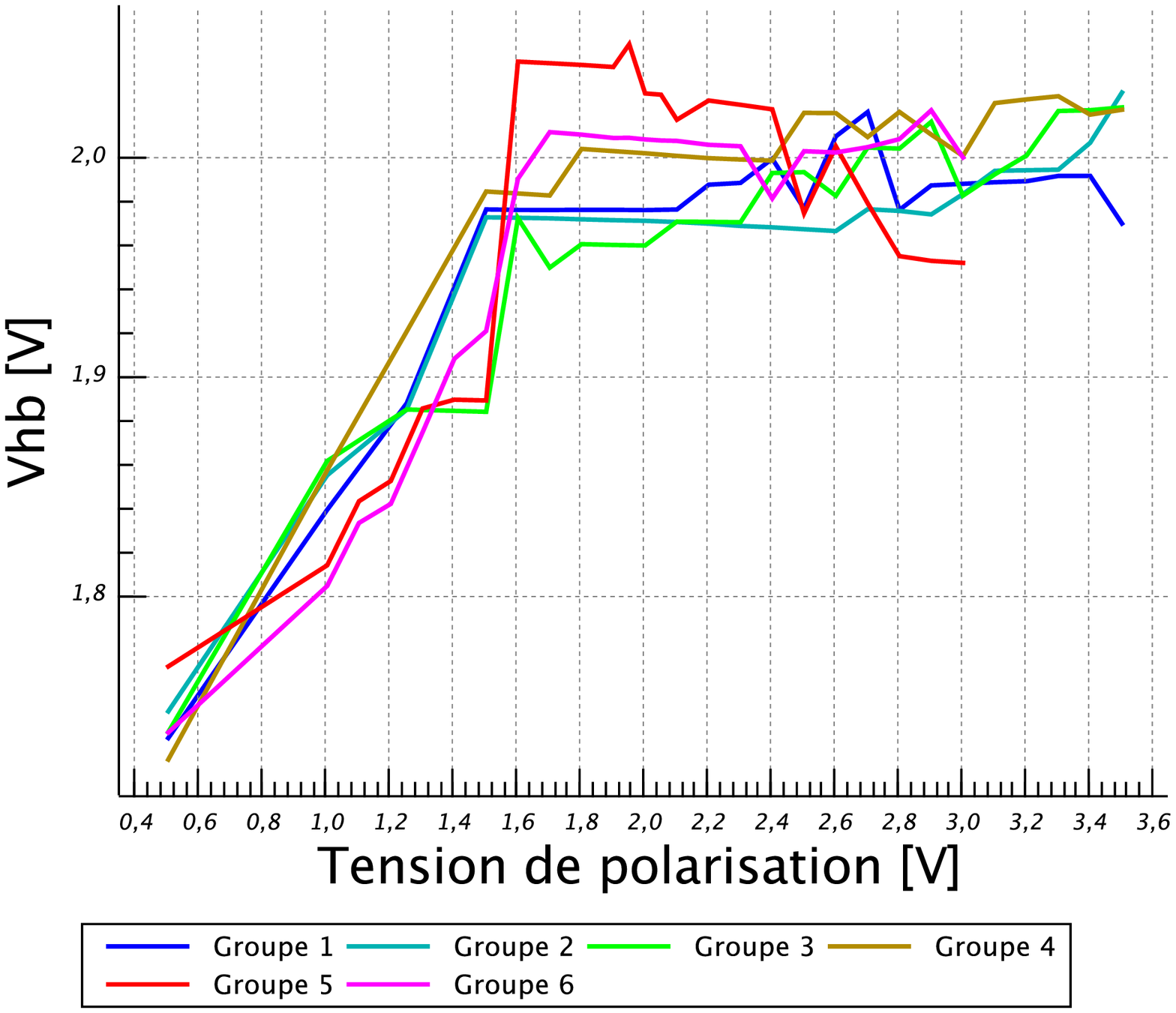}
    \end{tabular}
  \end{center}
  \caption[Pr\'ediction du r\'eglage des ponts bolom\'etriques et des
  tensions de r\'eference]{Les quatres tensions primaires $V_h$,
  $V_l$, $V_{ref}$ et $V_{hb}$ g\'en\'er\'ees par le programme de
  calcul des r\'eglages pour 24 valeurs de tensions de polarisation et
  pour un flux de 1~pW sur le BFP bleu et 8.7~pW sur le rouge. Les
  tensions sont g\'en\'er\'ees de 0.5 \`a 3.5~V sur le BFP bleu
  (Groupe 1~\`a~4) et de 0.5 \`a 3.0~V sur le rouge (Groupe
  5~et~6). La tension $V_{ref}$ est particuli\`erement int\'eressante
  puisqu'elle repr\'esente l'\'evolution du signal m\'edian qui entre
  dans le circuit de lecture pour chacun des groupes. Voir le texte
  pour une explication d\'etaill\'ee de la g\'en\'eration des tensions
  de polarisation.
  \label{fig:calib_procedure_prediction_polar}}
\end{figure}
La figure~\ref{fig:calib_procedure_prediction_polar} donne un exemple
concret de tensions g\'en\'er\'ees par ce programme pour un flux de
1~pW sur le BFP bleu et 8.7~pW sur le rouge. Pour chacun des groupes
nous obtenons 24 valeurs de $V_h$, $V_l$, $V_{ref}$ et $V_{hb}$ pour
les 24 tensions de polarisation que nous voulons tester (cf
section~\ref{sec:detect_outils_concept}). Notez qu'au-del\`a de 1.5~V,
les tensions $V_{ref}$ et $V_{hb}$ \'evoluent tr\`es peu. Cela
signifie que le signal d'entr\'ee du CL reste approximativement le
m\^eme et que les points milieux sont d\'ej\`a centr\'es sur la
dynamique de BOLC pour toutes ces configurations l\`a. En effet, pour
un couple $(tension, flux)$ donn\'e, c'est la tension $V_l$ qui
d\'etermine le niveau d'entr\'ee du CL~; et au-del\`a de 1.5~V, les
changements de $V_l$ peuvent compenser les variations de points
milieux. Par contre, pour les faibles tensions de polarisation, $V_l$
devrait prendre des valeurs positives pour permettre au niveau
d'entr\'ee du CL d'atteindre les 400~mV cible. Cependant, puisque
$V_l$ ne peut d\'epasser 0~V par construction de BOLC, le point milieu
de ces configurations n'est pas suffisamment \'elev\'e, et il ne sera
pas transmis correctement par le circuit de lecture (saturation des
MOS de lecture du CL, cf
section~\ref{sec:calib_procedure_vrlvhb}). D\`es lors, nous savons que
les tensions de polarisation inf\'erieures \`a 1~V n'offrent pas un
r\'eglage fiable des bolom\`etres.\\

Pour r\'esumer, dans ce chapitre, nous avons relax\'e les contraintes
de saturation de BOLC en travaillant en gain faible, nous avons ainsi
pu explorer et quantifier le comportement de l'\'electronique de
lecture et des ponts bolom\'etriques pour de nombreuses configurations
du syst\`eme. Nous avons \'egalement mis en \'evidence les limitations
de fonctionnement des matrices de bolom\`etres (saturation du CL et de
BOLC), ainsi que la pr\'esence tr\`es probable d'injections de charges
parasites dues au multiplexage. La richesse des mesures
syst\'ematiques que nous avons effectu\'ees nous a permis de
comprendre plus en d\'etails et de pr\'edire l'\'evolution du signal
en fonction de la tension de polarisation et du flux incident. Nous
sommes maintenant capables de g\'en\'erer des scripts de test long de
plusieurs dizaines d'heures, ce qui aurait n\'ecessit\'e des
journ\'ees enti\`eres en laboratoire pour tester toutes ces
configurations \og \`a la main \fg, et nous utilisons ces scripts pour
mesurer de mani\`ere syst\'ematique les performances des matrices de
bolom\`etres. Je tiens \`a souligner l'importance de cette \'etape
d'automatisation qui nous a apport\'e une certaine flexibilit\'e dans
l'ex\'ecution des tests d'\'etalonnage (les bolom\`etres tournent
24/24~h). Les mesures sont alors devenues tr\`es efficaces et de bonne
qualit\'e (pas de saturation), et notre compr\'ehension des
d\'etecteurs n'a cess\'e de cro\^itre depuis que nous pouvons
g\'en\'erer ces scripts de tests.



\newpage{\pagestyle{empty}\cleardoublepage}
\chapter{Les mesures de performance}
\label{chap:calib_perflabo}

\begin{center}
\begin{minipage}{0.85\textwidth}

\small Ce chapitre pr\'esente les principaux r\'esultats obtenus lors
de la campagne d'\'etalonnage du Photom\`etre PACS r\'ealis\'ee au
Max-Planck-institut f\"ur Extraterrestrische Physik (MPE) \`a Garching
en Allemagne entre Octobre 2006 et Juin 2007. Compte tenu des
plannings tr\`es serr\'es impos\'es par l'ESA et de l'exigence des
projets spatiaux, plus d'un an a \'et\'e n\'ecessaire pour mettre en
place la proc\'edure d'\'etalonnage et pour pr\'eparer efficacement
cette campagne de tests, et je tiens \`a pr\'eciser que ce chapitre ne
pr\'esente que la partie \'emerg\'ee de nos efforts. Au total nous
avons amass\'e plus d'un Terabit de donn\'ees, r\'ealis\'e des
milliers d'heures de tests et autant de temps \`a analyser les
donn\'ees. Ce travail a \'et\'e r\'ealis\'e en collaboration avec
Koryo Okumura et Marc Sauvage, il s'inscrit dans l'activit\'e du
groupe ICC (Instrument Control Center) dont je fais partie et qui est
responsable entre autre de l'\'etalonnage et du suivi en vol de
l'instrument.


\end{minipage}
\end{center}

\section{La sensibilit\'e des bolom\`etres}
\label{sec:calib_perflabo_sensibilite}

Pour la grande majorit\'e des r\'ecepteurs de rayonnement utilis\'es
pour les observations astronomiques, la sensibilit\'e du d\'etecteur
est l'un des param\`etres les plus importants pour d\'efinir les
performances globales de l'instrument. Ce param\`etre d\'etermine en
effet le temps d'int\'egration n\'ecessaire pour observer une source
donn\'ee avec un signal-\`a-bruit donn\'e (cf
section~\ref{sec:calib_perfobs_nep}). La sensibiblit\'e d'un
bolom\`etre est g\'en\'eralement exprim\'ee en terme de Puissance
Equivalente de Bruit, ou \emph{Noise Equivalent Power} en anglais
(\emph{NEP}). La \emph{NEP} est d\'efinie comme \'etant la puissance
radiative qui produit un signal \'electrique d'amplitude \'equivalente
\`a celle du bruit g\'en\'er\'e par le d\'etecteur dans une bande
passante\footnote{Pour \'eviter les confusions, il est utile de
pr\'eciser que la bande passante de 1~Hz n'est pas l'inverse d'une
longueur d'onde, mais l'inverse d'un temps d'observation.} de 1~Hz. En
d'autres termes c'est la plus petite puissance lumineuse qu'un
instrument puisse d\'etecter avec un signal-\`a-bruit de~1 ($S/N\sim
1$) en 1~seconde. La NEP est exprim\'ee en [W/$\sqrt{\mbox{Hz}}$]. \\
\noindent Une d\'efinition \'equivalente de la \emph{NEP}, mais plus
pratique \`a mettre en \oe uvre, est le rapport du bruit $\sigma$
mesur\'e dans une bande passante de 1~Hz, et de la r\'eponse $\Re$ du
d\'etecteur:
\begin{equation}
NEP=\sigma\,/\,\Re
\label{eq:NEP_br_resp}
\end{equation}
Dans cette section, nous pr\'esentons de mani\`ere g\'en\'erale et
qualitative les mesures de r\'eponse et de bruit qui sont utilis\'ees
dans le calcul de la \emph{NEP} des bolom\`etres. Nous nous appuyons
en particulier sur les mesures effectu\'ees en mode direct sur le BFP
bleu. Une analyse compar\'ee des r\'esultats entre les deux modes de
lecture et les deux BFP bleu et rouge sera donn\'ee dans la
section~\ref{sec:calib_perflabo_compare}. Les r\'esultats quantitatifs
sur les mesures de \emph{NEP} seront pr\'esent\'es dans les
sections~\ref{sec:calib_perflabo_compare}
et~\ref{sec:calib_perfobs_nep}.

\subsection{Les mesures de r\'eponse}
\label{sec:calib_perflabo_sensibilite_reponse}

Nous appelons \emph{r\'eponse} l'aptitude d'un bolom\`etre \`a
convertir une variation de flux en une variation de signal
\'electrique. Elle est exprim\'ee en [V/W].  Pour la mesurer, nous
utilisons deux sources maintenues \`a des temp\'eratures
diff\'erentes, autour de 30~K, qui \'emettent un spectre de corps
noir, ainsi qu'un chopper pour moduler le flux incident sur le plan
focal. La r\'eponse des bolom\`etres $\Re$ est obtenue en divisant
l'amplitude du signal \'electrique $\Delta{\mbox{Signal}}$ par
l'amplitude de la modulation de flux $\Delta{\mbox{flux}}$ qui a
g\'en\'er\'ee ce $\Delta{\mbox{Signal}}$:
\begin{equation}
\Re=\Delta{\mbox{Signal}}\,/\,\Delta{\mbox{flux}}\;\;\;\;\;\;\;\;\;\mbox{[V/W]}
\label{eq:resp}
\end{equation}


\noindent Nous verrons dans la
section~\ref{sec:calib_perflabo_nonlinear} que la r\'eponse des
bolom\`etres ne varie pas lin\'eairement avec le flux incident, nous
choisissons donc une modulation de faible amplitude
($\Delta{\mbox{flux}}=0.5$~pW/pixel) pour nous assurer que nous
mesurons effectivement la r\'eponse sans la sous-estimer.
Nous savons \'egalement d'apr\`es la
section~\ref{sec:intro_bolometrie_thermo_principe} que la r\'eponse
des bolom\`etres chute avec la fr\'equence de modulation. En effet,
lorsque le flux lumineux \'evolue plus vite que la constante de temps
du bolom\`etre, le pixel n'a pas le temps de se thermaliser, et
l'amplitude du signal est alors sous-estim\'ee. Par cons\'equent nous
d\'ecidons de moduler le signal \`a une fr\'equence de 0.5~Hz,
fr\'equence \`a laquelle les effets de la constante de temps sont
n\'egligeables. D'autre part, la r\'eponse varie avec le coefficient
$\alpha=\frac{1}{R}\frac{\partial{R}}{\partial{T}}$ tel que d\'efini
dans la section~\ref{sec:detect_outils_loadcurves}~; les param\`etres
susceptibles d'influencer la r\'eponse des bolom\`etres sont donc ceux
qui peuvent changer la temp\'erature de fonctionnement des
bolom\`etres ou plus g\'en\'eralement leur imp\'edance. Les deux
param\`etres principaux que nous allons explorer sont le flux incident
sur le plan focal, et la tension de polarisation qui d\'etermine
l'\'energie \'electrique dissip\'ee au niveau de l'absorbeur de
rayonnement ainsi que l'amplitude de l'effet de champ (cf
\'equation~\ref{eq:efros}).

\begin{figure}
  \begin{center}
      \includegraphics[width=0.7\textwidth,angle=0]{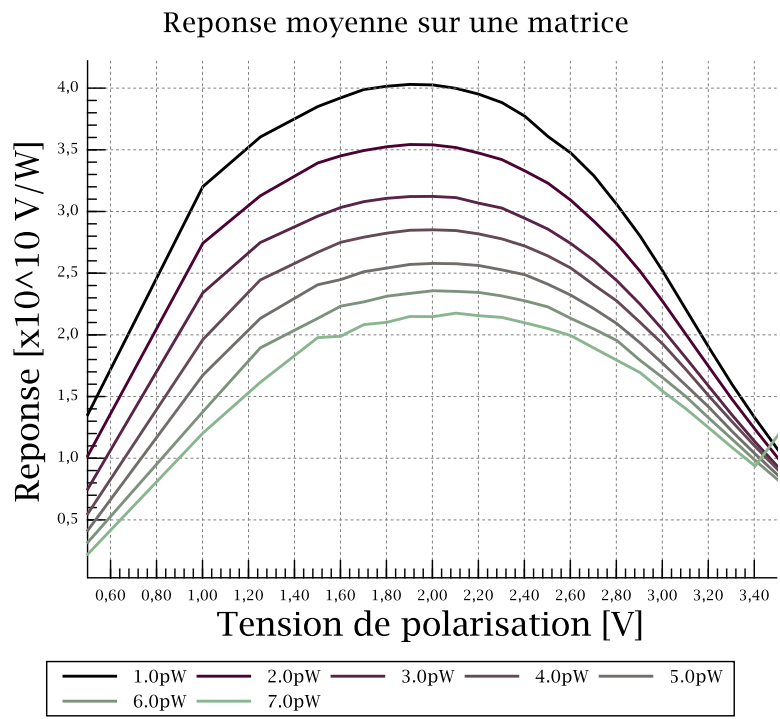}
  \end{center}
  \caption[\'Evolution de la r\'eponse des bolom\`etres]{\'Evolution
  de la r\'eponse des bolom\`etres en fonction de la tension de
  polarisation et du flux incident en mode direct. Chaque point
  correspond \`a la moyenne spatiale des r\'eponses obtenues pour une
  m\^eme matrice du BFP bleu. Les r\'eponses sont mesur\'ees pour une
  modulation en flux de 500~fW/pixel pour un fond de 1~\`a~7~pW/pixel
  comme indiqu\'e dans la l\'egende. Voir le texte pour
  l'interpr\'etation de ces courbes.
  \label{fig:calib_perflabo_sensibilite_description_respDirect}}
\end{figure}

Lors de la campagne d'\'etalonnage du Photom\`etre PACS au MPE, nous
avons effectu\'e des mesures de r\'eponse syst\'ematiques pour
24~valeurs de la tension de polarisation et pour 7~valeurs de flux
incident pour chaque BFP. Chacune de ces tensions a \'et\'e
g\'en\'er\'ee par le programme d\'ecrit dans la
section~\ref{sec:calib_procedure_prediction}. Pour chaque
configuration $(tension, flux)$, nous enregistrons le signal modul\'e
pendant 4~minutes, nous mesurons l'amplitude de la modulation pour
chacun des cycles chopper, et nous gardons la moyenne de ces
amplitudes comme \'etant l'amplitude associ\'ee \`a la configuration
$(tension, flux)$ test\'ee. Notez que sur une dur\'ee de 4~minutes la
d\'erive du gain des d\'etecteurs est n\'egligeable. La
figure~\ref{fig:calib_perflabo_sensibilite_description_respDirect}
montre le r\'esultat de nos mesures pour une matrice du BFP bleu en
mode de lecture direct. Chaque point de la figure est la moyenne
spatiale des amplitudes mesur\'ees sur les 256 pixels d'une m\^eme
matrice.
Les r\'esultats sont donc repr\'esentatifs du comportement global de
la matrice. Chacune des courbes repr\'esente l'\'evolution de la
r\'eponse en fonction de la tension de polarisation pour un flux
donn\'e. \\

\`A tension constante, la temp\'erature du bolom\`etre est une
fonction croissante du flux incident absorb\'e. Il en r\'esulte que la
r\'eponse chute avec le flux incident comme indiqu\'e sur la
figure~\ref{fig:calib_perflabo_sensibilite_description_respDirect}. Ceci
est vrai pour toutes les tensions de polarisation (les courbes ne se
chevauchent pas). \`A flux constant, notez la forme caract\'eristique
des courbes~: la r\'eponse augmente avec la tension de polarisation
appliqu\'ee au bolom\`etre, atteint un maximum, puis retombe aux
fortes polarisations~; et ce pour deux raisons. D'une part, plus le
bolom\`etre est polaris\'e et plus la modulation du signal de sortie
peut \^etre grande. En d'autres termes, l'amplitude des excursions de
point milieu engendr\'ee par une modulation de flux donn\'ee augmente
avec la tension fournie au bolom\`etre~; \cad que la r\'eponse
augmente avec la tension de polarisation (de~0 \`a $\sim$2~V). D'autre
part, la raison pour laquelle les courbes de r\'eponse fl\'echissent
aux fortes tensions de polarisation tient au fait que l'imp\'edance
des thermom\`etres chute avec la temp\'erature et le champs
\'electrique comme indiqu\'e dans l'\'equation~\ref{eq:efros}. La
chute de r\'eponse correspond donc \`a un d\'eplacement du point de
fonctionnement des bolom\`etres sur les courbes de la
figure~\ref{fig:detect_bolocea_fabrication_thermo_R2T} (r\'eponse
$\sim\partial R / \partial T$), ce d\'eplacement \'etant d\^u \`a
l'\'echauffement du thermom\`etre et au fort champ \'electrique
appliqu\'e \`a ses bornes. Par ailleurs, la stabilit\'e du gain des
d\'etecteurs durant les mesures donne des courbes lisses et peu
bruit\'ees. Remarquez toutefois la discontinuit\'e des courbes de
r\'eponse entre~1 et 0.5~V de tension. Cette discontinuit\'e est
certainement due au fait que les points milieux doivent toujours
d\'epasser quelques centaines de mV pour atteindre BOLC sans \^etre
alt\'er\'es (cf sections~\ref{sec:calib_procedure_vrlvhb}
et~\ref{sec:calib_procedure_explore}).

\begin{figure}
  \begin{center}
    \begin{tabular}[t]{ll}
      \includegraphics[width=0.6\textwidth,angle=0]{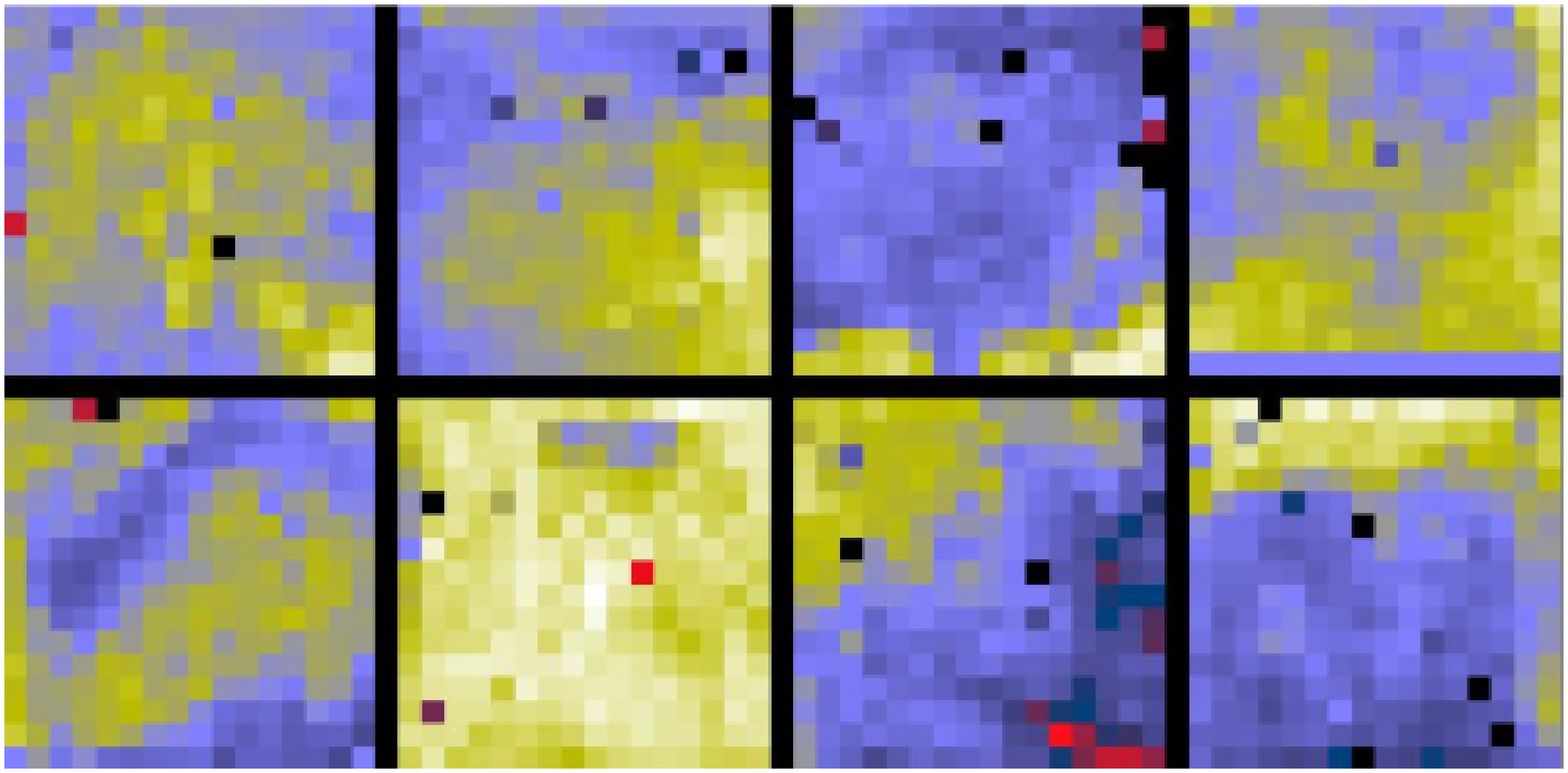} & \includegraphics[width=0.35\textwidth,angle=0]{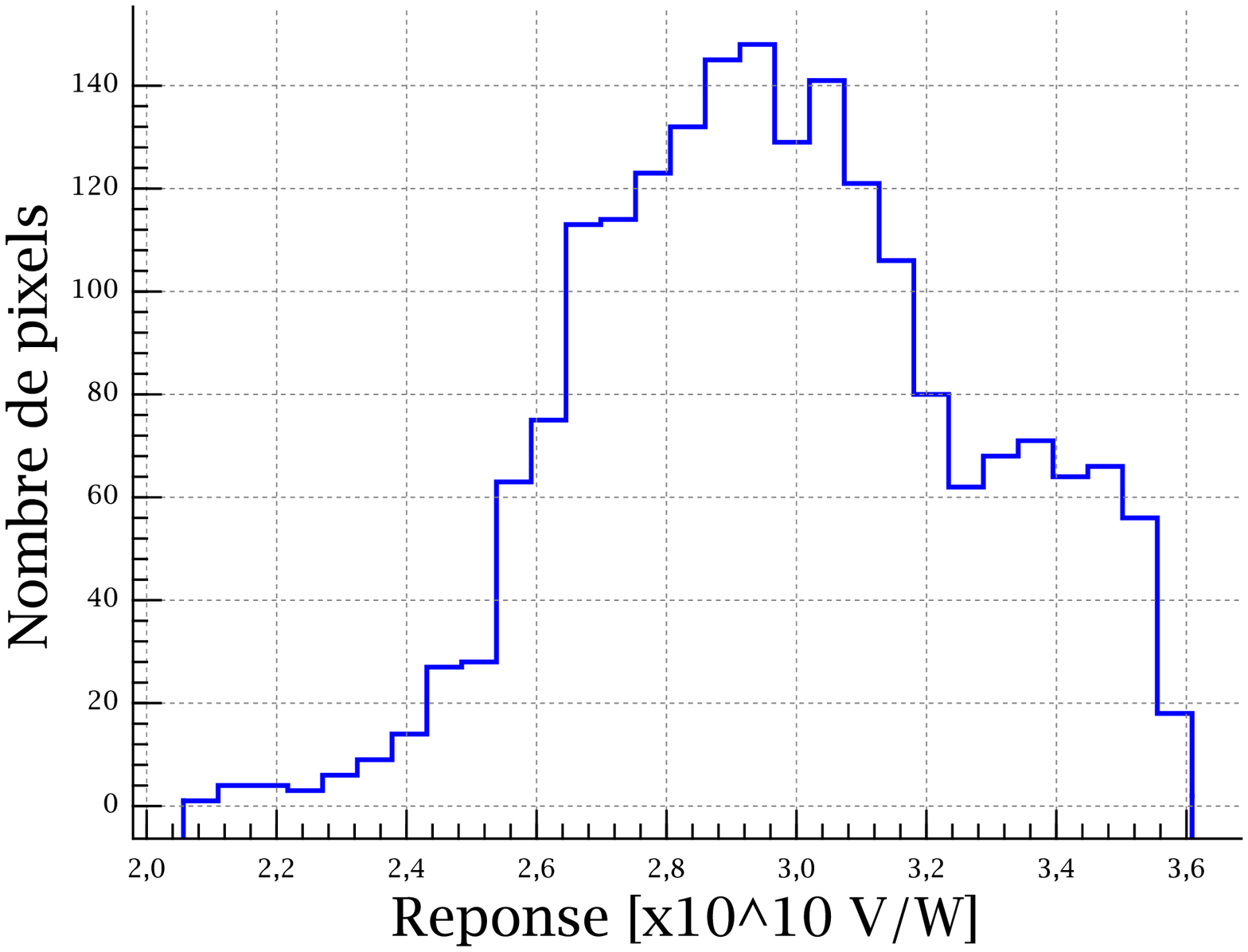} 
    \end{tabular}
  \end{center}
  \caption[Carte et dispersion de la r\'eponse des bolom\`etre sur le
  BFP bleu]{Distribution spatiale de la r\'eponse sur le BFP bleu pour
  une tension de polarisation de 2.7~V, un flux de 2~pW/pixel et une
  modulation de 0.5~pW/pixel. L'histogramme de droite montre la
  dispersion de r\'eponse sur cette carte.
  \label{fig:calib_perflabo_sensibilite_description_respMapHisto}}
\end{figure}

La distribution spatiale de la r\'eponse sur le BFP bleu et sa
dispersion sont pr\'esent\'ees dans la
figure~\ref{fig:calib_perflabo_sensibilite_description_respMapHisto}
pour une tension de polarisation de 2.7~V et un flux incident de
2~pW/pixel. Cette configuration est repr\'esentative des conditions
d'op\'eration de PACS. La r\'eponse moyenne est de
$3\times10^{10}$~V/W, environ deux ordres de grandeur sup\'erieure \`a
la r\'eponse des bolom\`etres r\'esistifs traditionels
\shortcite{turner}. C'est l'utilisation de thermistances \`a tr\`es
haute imp\'edance qui explique la tr\`es grande r\'eponse des
bolom\`etres.

\subsection{Les mesures de bruit}
\label{sec:calib_perflabo_sensibilite_bruit}

\begin{figure}
  \begin{center}
      \includegraphics[width=0.7\textwidth,angle=0]{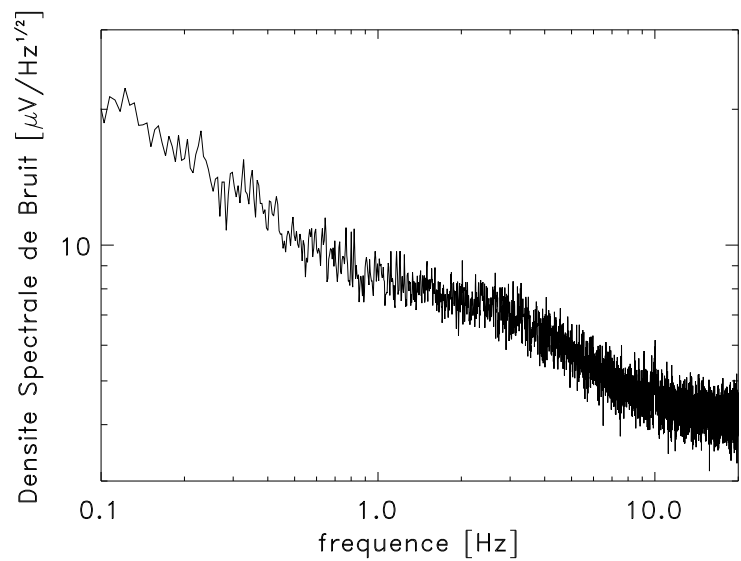}
  \end{center}
  \caption[Spectre de bruit typique d'un bolom\`etre PACS]{Densit\'e
  spectrale de bruit typique d'un pixel du Photom\`etre PACS. Le
  niveau de bruit qui rentre en jeu dans le calcul de la \emph{NEP}
  est mesur\'e autour de 3~Hz dans une bande passante unit\'e. Ce
  spectre contient une composante basse fr\'equence de type 1/f, et
  est att\'enu\'e au-del\`a de quelques Hz par la constante de temps
  du bolom\`etre. Le pic d'\'energie \`a 10~Hz est d\^u \`a un
  battement entre la fr\'equence d'\'echantillonnage (40~Hz) et
  l'environnement \'electromagn\'etique du laboratoire (50~Hz). Une
  analyse d\'etaill\'ee des spectres de bruit est donn\'ee dans les
  sections~\ref{sec:calib_perflabo_tau},
  \ref{sec:calib_perflabo_compare}
  et~\ref{sec:calib_perfobs_oof}. Dans le cas pr\'esent la tension de
  polarisation est de 2.6~V et le flux est de 2~pW/pixel.
  \label{fig:calib_perflabo_sensibilite_description_spectre}}
\end{figure}

Nous appelons \emph{bruit} la valeur moyenne de la densit\'e spectrale
de bruit dans une bande passante unit\'e centr\'ee autour de 3~Hz. La
figure~\ref{fig:calib_perflabo_sensibilite_description_spectre} montre
le spectre de bruit typique d'un pixel \`a partir duquel nous pouvons
mesurer le niveau de bruit utilis\'e dans le calcul de la
\emph{NEP}. Notez que nous \'etudions le bruit des bolom\`etres dans
l'espace de Fourier car cela nous permet de garder l'information
spectrale contenue dans le signal temporel. Nous d\'efinissons la
densit\'e spectrale de bruit comme \'etant la racine carr\'ee de la
densit\'e spectrale de puissance (en [V$^2$/Hz]). L'unit\'e du bruit
que nous mesurons est donc le [V/$\sqrt{\mbox{Hz}}$]. Nous calculons
la densit\'e spectrale de puissance en utilisant une fen\^etre de
Hanning pour apodiser le signal temporel et \'eviter les probl\`emes
d'\emph{aliasing} fr\'equents en traitement du signal. En effet, le
signal temporel \'echantillonn\'e dans l'espace direct a une dur\'ee
finie, c'est-\`a-dire qu'en dehors de l'intervalle de temps
\'echantillonn\'e, le signal vaut~0. Mais l'algorithme qui calcule la
transform\'ee de Fourier duplique le signal et le met bout-\`a-bout
pour simuler un signal infiniment long. Dans la majorit\'e des cas, le
premier point de l'\'echantillon n'a pas la m\^eme valeur que le
dernier point, et dupliquer un tel signal introduit une
discontinuit\'e qui n'est pas initialement pr\'esente dans le
signal. Sans apodisation, le spectre contiendrait de nombreux \og
rebonds \fg et un exc\`es d'\'energie introduit par cette
discontinuit\'e. La solution est de multiplier le signal temporel par
une fen\^etre de Hanning\footnote{Fen\^etre de Hanning utilis\'ee par
IDL~: $f(t)=0.5[1-\cos(\frac{2\pi n}{N})]$ o\`u $n$ est le n\ieme
point et $N$ est le nombre total de points dans l'\'echantillon.}
avant de calculer la transform\'ee de Fourier, ce qui a pour effet
d'amener progressivement \`a~0 les deux extr\'emit\'es du
signal. C'est une mani\`ere astucieuse de pond\'erer le signal et
ainsi d'\'eviter l'introduction d'une discontinuit\'e \og non-physique
\fg dans le signal \`a traiter . Nous devons toutefois normaliser le
spectre obtenu par l'int\'egrale de la transform\'ee de Fourier de la
fonction de Hanning (normalisation par un facteur $\sim$3/8). Par
ailleurs nous avons montr\'e que le th\'eor\`eme de Parseval est
v\'erifi\'e aux erreurs de calcul pr\`es, \cad que la normalisation
n'affecte pas la densit\'e spectrale de bruit.

\begin{figure}
  \begin{center}
      \includegraphics[width=0.7\textwidth,angle=0]{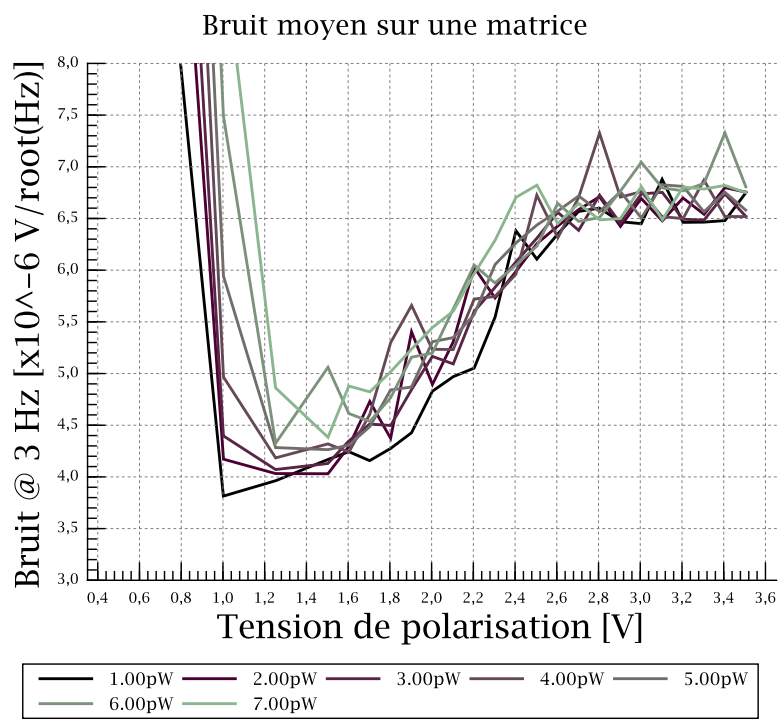}
  \end{center}
  \caption[\'Evolution du bruit des bolom\`etres]{\'Evolution du bruit
  des bolom\`etres en fonction de la tension de polarisation et du
  flux incident en mode direct. Chaque point correspond \`a la moyenne
  spatiale des bruits mesur\'es pour une m\^eme matrice du BFP
  bleu. Les valeurs de bruit pr\'esent\'ees dans cette figure sont
  tir\'ees de densit\'es spectrales de bruit semblables \`a la celle
  de la
  figure~\ref{fig:calib_perflabo_sensibilite_description_spectre} pour
  des fr\'equences centr\'ees autour de 3~Hz. Voir le texte pour
  l'interpr\'etation de ces courbes.
  \label{fig:calib_perflabo_sensibilite_description_noiseDirect}}
\end{figure}

Nous \'etudierons en d\'etail la d\'ependance en fr\'equence des
spectres de bruit dans les sections~\ref{sec:calib_perflabo_tau},
\ref{sec:calib_perflabo_compare} et~\ref{sec:calib_perfobs_oof}, mais
dans la section pr\'esente, nous nous concentrons sur le calcul du
bruit ainsi que sur son \'evolution en fonction du r\'eglage des
bolom\`etres. Pour ce faire, nous utilisons \`a nouveau le programme
d\'ecrit dans la section~\ref{sec:calib_procedure_prediction} pour
g\'en\'erer automatiquement les tensions n\'ecessaires pour polariser
les d\'etecteurs dans chacune des configurations test\'ees. Nous
mesurons syst\'ematiquement le niveau de bruit du Photom\`etre PACS
pour 24~tensions de polarisation et 7~valeurs de flux pour chaque
BFP. Pour chaque configuration $(tension, flux)$, nous enregistrons le
signal pendant 4~minutes, nous calculons la densit\'e spectrale de
bruit et nous extrayons le niveau de bruit \`a 3~Hz. Le r\'esultat de
ces mesures pour une matrice du BFP bleu est pr\'esent\'e dans la
figure~\ref{fig:calib_perflabo_sensibilite_description_noiseDirect}.
Chaque point repr\'esente la moyenne spatiale du bruit mesur\'e sur
les 256 pixels d'une m\^eme matrice. Les r\'esultats sont donc
repr\'esentatifs du comportement global de la matrice. Chacune des
courbes repr\'esente l'\'evolution du bruit en fonction de la tension
de polarisation pour un flux donn\'e.\\

\begin{figure}
  \begin{center}
    \begin{tabular}[t]{ll}
      \includegraphics[width=0.6\textwidth,angle=0]{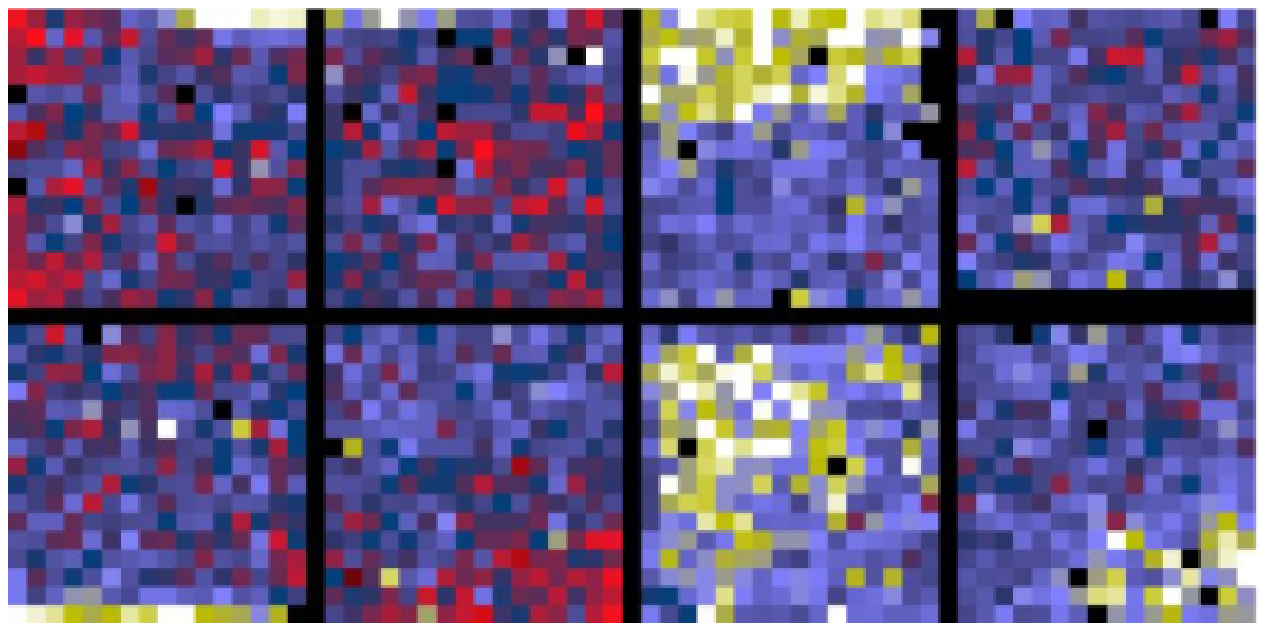} & \includegraphics[width=0.35\textwidth,angle=0]{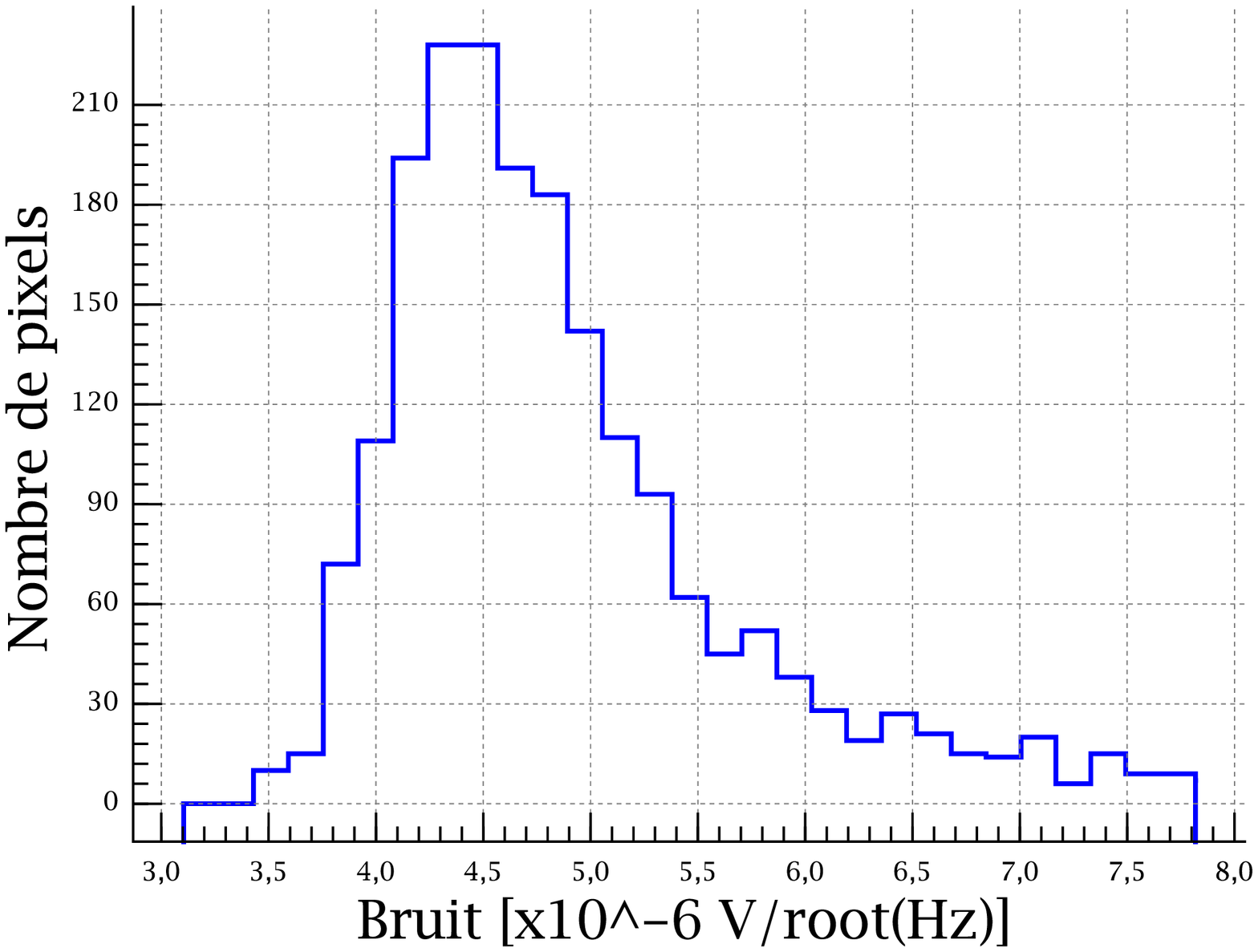} 
    \end{tabular}
  \end{center}
  \caption[Carte et dispersion du bruit des bolom\`etre sur le BFP
  bleu]{Distribution spatiale du bruit \`a 3~Hz sur le BFP bleu pour
  une tension de polarisation de 2.7~V et pour un flux de
  2~pW/pixel. L'histogramme de droite montre la dispersion du bruit
  sur cette carte.
  \label{fig:calib_perflabo_sensibilite_description_noiseMapHisto}}
\end{figure}

Nous pouvons dans un premier temps estimer le niveau de bruit des
bolom\`etres en calculant la contribution des diff\'erentes sources de
bruit que nous avons pr\'esent\'e dans la
section~\ref{sec:intro_bolometrie_thermo_principe_bruit}. \`A partir
de l'\'equation~(\ref{eq:NEPphoton}), nous trouvons une NEP photonique
de l'ordre de $1.5\times 10^{-16}$~W/$\sqrt{\mbox{Hz}}$ pour un flux
incident nominal de $\sim$2~pW/pixel \shortcite{sauvage_note}. En
multipliant cette NEP par la r\'eponse des bolom\`etres ($\sim3\times
10^{10}$~V/W pour une tension de 2~V), nous trouvons un bruit
photonique d'environ 5~$\mu$V/$\sqrt{\mbox{Hz}}$. D'autre part, pour
une r\'esistance \'equivalente de l'ordre de $2\times
10^{11}$~$\Omega$ et une temp\'erature bolom\`etre de 360~mK,
l'\'equation~(\ref{eq:johnson}) donne un bruit Johnson d'environ
2~$\mu$V/$\sqrt{\mbox{Hz}}$. De la m\^eme mani\`ere, nous utilisons
l'\'equation~\ref{eq:phonon} pour calculer le bruit de phonon
g\'en\'er\'e par les bolom\`etres~; nous trouvons un bruit de
$\sim0.2$~$\mu$V/$\sqrt{\mbox{Hz}}$. Notez que les param\`etres
physiques que nous avons choisis d'utiliser pour le calcul des niveaux
de bruit proviennent de r\'esultats r\'ecents pr\'esent\'es dans la
section~\ref{sec:calib_perflabo_ajuste_IV}. La somme quadratique de
ces contributions donne un niveau de bruit de
5.5~$\mu$V/$\sqrt{\mbox{Hz}}$ pour un flux de 2~pW/pixel et une
tension de polarisation de 2~V~; ce qui est en bon accord avec les
mesures de la
figure~\ref{fig:calib_perflabo_sensibilite_description_noiseDirect}.

En ce qui concerne les premiers points de mesure \`a 0.5~V de
polarisation, ils sont encore une fois aberrants. Rappelons simplement
que ce r\'eglage ne permet pas au signal d'\^etre transmis
correctement par l'\'electronique de lecture, le point milieu est trop
faible (\cad inf\'erieur \`a $\sim$300~mV, cf
section~\ref{sec:calib_procedure_vrlvhb}), ce qui engendre un exc\`es
de bruit visible sur la figure. Pour les mesures \`a 1~V de
polarisation, les points milieux se trouvent plus ou moins \`a la
limite de saturation du CL selon le niveau de flux incident. En effet,
pour les faibles flux, les points milieux sont juste au-dessus de
300~mV de sorte que le signal \'electrique est transmis par le CL, et
la mesure de bruit est fiable. Par contre, lorsque le flux augmente,
les points milieux diminuent et passent en-dessous de la limite de
saturation engendrant ainsi un exc\`es de bruit qui s'amplifie au fur
et \`a mesure que le point milieu diminue. Pour les autres
polarisations, l'interpr\'etation est beaucoup plus d\'elicate, et
sans un mod\`ele physique des bolom\`etres il est tr\`es difficile
d'exploiter pr\'ecisemment ces r\'esultats. Je pr\'esente cependant
quelques \'el\'ements de r\'eponse qui pourraient expliquer l'allure
g\'en\'erale des courbes de la
figure~\ref{fig:calib_perflabo_sensibilite_description_noiseDirect}.

L'estimateur que nous avons choisi pour mesurer le bruit est
l\'eg\`erement biais\'e car il d\'epend indirectement de la tension de
polarisation. Nous verrons en effet dans la
section~\ref{sec:calib_perflabo_tau} que le spectre de bruit \'evolue
avec la constante de temps des bolom\`etres (cf
figure~\ref{fig:calib_perflabo_tau_fourier_spec4polar}) et que cette
constante de temps diminue avec la tension de polarisation. Lorsque la
tension est trop faible, la fr\'equence de coupure du bolom\`etre est
inf\'erieure \`a 3~Hz de sorte que notre estimateur mesure le niveau
de bruit dans un r\'egime o\`u le spectre est d\'ej\`a att\'enu\'e par
le filtre. Cependant, ce filtre \'etant du premier ordre, \cad une
att\'enuation de 3~dB par d\'ecade, le bruit mesur\'e \`a 3~Hz n'est
en g\'en\'eral pas significativement att\'enu\'e. Nous pourrions
\'eventuellement estimer le niveau de bruit blanc \`a des fr\'equences
inf\'erieures \`a 3~Hz pour \'eviter les effets de la constante de
temps, mais nous risquerions de le surestimer \`a cause de la
composante en 1/f (cf sections~\ref{sec:calib_perflabo_tau_fourier}
et~\ref{sec:calib_perfobs_oof}). Par contre, lorsque les d\'etecteurs
sont suffisamment aliment\'es, la fr\'equence de coupure se trouve
au-del\`a de 3~Hz et notre estimateur mesure alors correctement le
bruit blanc g\'en\'er\'e par le bolom\`etre.

Par ailleurs, au-del\`a de 1.5~V, le niveau de bruit semble ne pas
d\'ependre du flux incident, ce qui signifie que les bolom\`etres PACS
ne sont pas limit\'es par le bruit de photon mais plut\^ot par leur
bruit intrins\`eque. Cela est surprenant \'etant donn\'e les
estimations de bruit pr\'esent\'ees pr\'ec\'edemment (bruit Johnson et
bruit de phonon inf\'erieurs au bruit photonique). Les courbes de
bruit de la
figure~\ref{fig:calib_perflabo_sensibilite_description_noiseDirect}
sont relativement difficiles \`a analyser et \`a interpr\'eter. Le
lecteur pourra consulter les densit\'es spectrales de bruit ainsi que
la qualit\'e de l'estimateur de bruit dans
l'annexe~\ref{a:spectres}.\\

La distribution spatiale du bruit sur le BFP bleu et sa dispersion
sont pr\'esent\'ees dans la
figure~\ref{fig:calib_perflabo_sensibilite_description_noiseMapHisto}
pour une tension de polarisation de 2.7~V et un flux incident de
2~pW/pixel. Cette configuration est repr\'esentative des conditions
d'op\'eration de PACS. Le bruit moyen est de
4-5~$\mu$V/$\sqrt{\mbox{Hz}}$, environ deux ordres de grandeur
sup\'erieure au bruit des bolom\`etres r\'esistifs traditionnels. C'est
l'utilisation de transistors MOS pour la lecture du signal ainsi que
la tr\`es haute imp\'edance des thermistances qui explique le niveau
de bruit relativement \'elev\'e des bolom\`etres PACS.

\subsection{Le calcul de la \emph{NEP}}
\label{sec:calib_perflabo_sensibilite_NEP}

\begin{figure}
  \begin{center}
      \includegraphics[width=0.7\textwidth,angle=0]{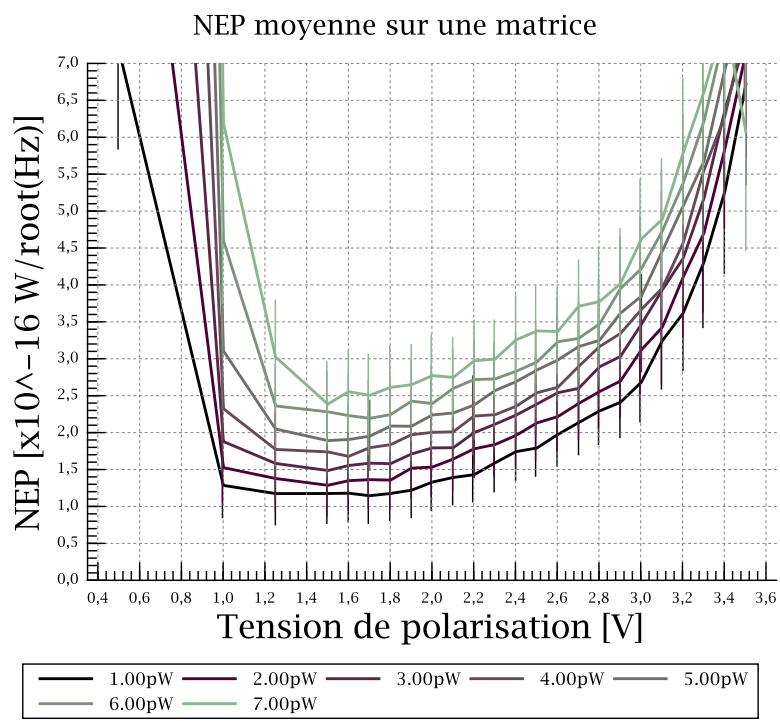}
  \end{center}
  \caption[\'Evolution de la \emph{NEP} des bolom\`etres]{\'Evolution
  de la NEP des bolom\`etres en fonction de la tension de polarisation
  et du flux incident en mode direct. Chaque point correspond \`a la
  moyenne spatiale des NEP mesur\'ees pour une m\^eme matrice du BFP
  bleu. Les valeurs de NEP pr\'esent\'ees dans cette figure sont le
  rapport entre les bruits de la
  figure~\ref{fig:calib_perflabo_sensibilite_description_noiseDirect}
  et la r\'eponse de la
  figure~\ref{fig:calib_perflabo_sensibilite_description_respDirect}. Des
  NEP de l'ordre de 1-3$\times 10^{-16}$~W/$\sqrt{\mbox{Hz}}$ sont
  atteintes suivant le flux et la tension.
  \label{fig:calib_perflabo_sensibilite_description_nepDirect}}
\end{figure}
\begin{figure}
  \begin{center}
    \begin{tabular}[t]{ll}
      \includegraphics[width=0.6\textwidth,angle=0]{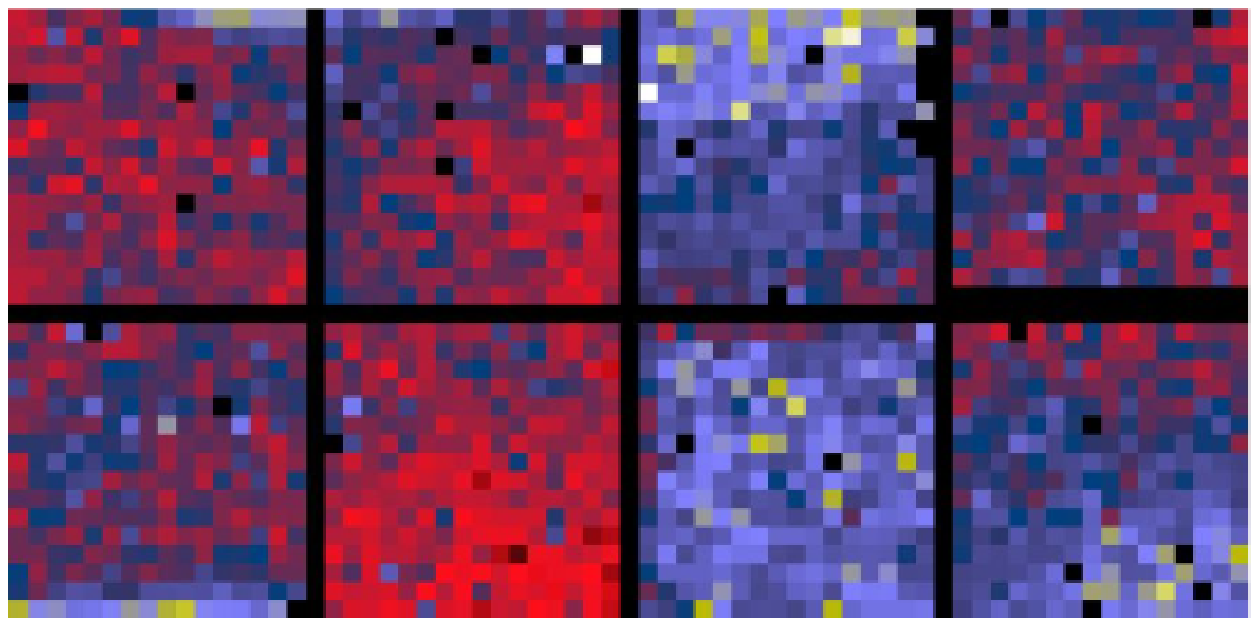} & \includegraphics[width=0.35\textwidth,angle=0]{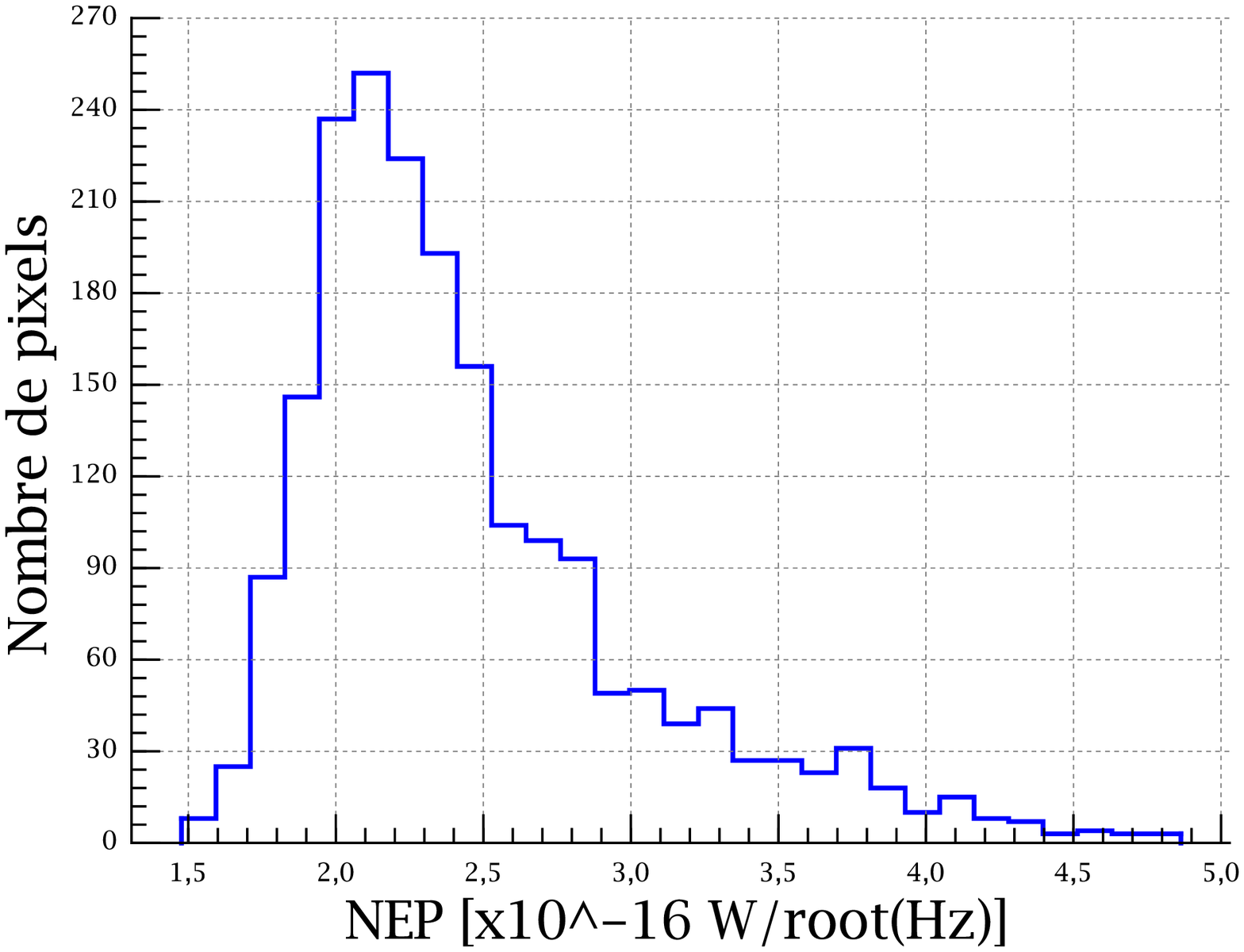} 
    \end{tabular}
  \end{center}
  \caption[Carte et dispersion de la NEP des bolom\`etre sur le BFP
  bleu]{Distribution spatiale de la NEP du BFP bleu pour une tension
  de polarisation de 2.7~V et pour un flux de
  2~pW/pixel. L'histogramme de droite montre la dispersion de NEP sur
  cette carte.
  \label{fig:calib_perflabo_sensibilite_description_nepMapHisto}}
\end{figure}

Une fois les mesures de r\'eponse et de bruit effectu\'ees, le calcul
de la NEP est trivial. \`A partir des
figures~\ref{fig:calib_perflabo_sensibilite_description_noiseDirect}
et~\ref{fig:calib_perflabo_sensibilite_description_respDirect}, nous
faisons le rapport du bruit et de la r\'eponse comme indiqu\'e dans
l'\'equation~(\ref{eq:NEP_br_resp}) et nous obtenons la NEP pour
chacune des configurations $(tension,flux)$ test\'ees. Les r\'esultats
pour le BFP bleu en mode de lecture direct sont pr\'esent\'es dans la
figure~\ref{fig:calib_perflabo_sensibilite_description_nepDirect}. \`A
nouveau, chaque courbe repr\'esente l'\'evolution de la NEP en
fonction de la tension de polarisation pour un flux donn\'e, et chaque
point de la figure est la moyenne spatiale de la NEP calcul\'ee pour
les 256~pixels d'une m\^eme matrice. Les r\'esultats sont donc
repr\'esentatifs du comportement global de la matrice. Pour les tr\`es
basses tensions de polarisation, nous retrouvons les points aberrants
associ\'es \`a la saturation du transistor MOS du circuit de
lecture. Pour les fortes polarisations, la NEP se d\'egrade suite \`a
la chute de r\'eponse, qui est elle-m\^eme due \`a la chute
d'imp\'edance de la thermistance.

La distribution spatiale de la NEP sur le BFP bleu et sa dispersion
sont pr\'esent\'ees dans la
figure~\ref{fig:calib_perflabo_sensibilite_description_nepMapHisto}
pour une tension de polarisation de 2.7~V et un flux incident de
2~pW/pixel. Cette configuration est repr\'esentative des conditions
d'op\'eration de PACS. La NEP typique est de 1-3~$\times
10^{-16}$~W/$\sqrt{\mbox{Hz}}$ suivant le flux et la tension de
polarisation utilis\'es.

Les performances en sensibilit\'e que nous pr\'esentons dans ce
manuscrit sont des NEP totales \og sous flux \fg, \cad qu'elles
contiennent toutes les contributions de bruit (photon, Johnson,
phonon, etc...) pour des conditions normales d'op\'eration de
l'instrument. Nous ne s\'eparons pas la NEP photon de la NEP
d\'etecteur comme c'est parfois le cas dans la litt\'erature
sp\'ecialis\'ee. Pour comparer les performances des matrices de
bolom\`etres du CEA avec celles des autres bolom\`etres, il faut donc
s'assurer que les NEP ont \'et\'e calcul\'ees de la m\^eme fa\c{c}on
et qu'elles sont effectivement comparables. Dans d'autres cas, les
auteurs expriment des NEP d\'etecteurs sans illuminer le plan focal~;
mais les chiffres qu'ils obtiennent ne sont pas vraiment pertinents
puisqu'ils ne sont pas repr\'esentatifs du point de fonctionnement des
bolom\`etres \og sous flux \fg (imp\'edance beaucoup plus grande pour
un flux nul). Quoi qu'il en soit, c'est bien la NEP totale qui est
utilis\'ee lors du calcul de sensibilit\'e d'une cam\'era (cf
section~\ref{sec:calib_perfobs_nep}). L'int\'er\^et de calculer la NEP
d\'etecteur pour un flux donn\'e est que l'on peut la comparer \`a la
NEP photon dans le but de d\'eterminer si le detecteur est BLIP ou
pas.  \`A la fin de ce chapitre, dans la
section~\ref{sec:calib_perflabo_compare}, nous donnerons des
r\'esultats plus quantitatifs sur les performances instrumentales des
matrices de bolom\`etres PACS dans les deux modes de lecture et pour
les deux BFP. Puis nous reprendrons ces r\'esultats dans la
section~\ref{sec:calib_perfobs_nep} pour calculer les performances
observationnelles du Photom\`etre PACS en termes intelligibles par un
astronome.

\section{La non-linearit\'e des bolom\`etres}
\label{sec:calib_perflabo_nonlinear}

Les bolom\`etres sont des d\'etecteurs thermiques intrins\`equement
non-lin\'eaires. En g\'en\'eral l'imp\'edance des senseurs thermiques
varient exponentiellement avec la temp\'erature de sorte que la
r\'eponse des bolom\`etres est une fonction non-lin\'eaire et
d\'ecroissante du flux incident et de la tension appliqu\'ee \`a leurs
bornes. La proc\'edure d'\'etalonnage initialement pr\'evue pour le
photom\`etre PACS contenait un test de non-lin\'earit\'e consistant
\`a mesurer la r\'eponse des bolom\`etres pour diff\'erents flux
incidents et pour des amplitudes de modulation de plus en plus
petites jusqu'\`a atteindre la plus petite modulation de flux
d\'etectable. Ce type de test peut s'av\'erer extr\`emement long \`a
effectuer, voire prohibitif, du fait des longues p\'eriodes de
stabilisation de la temp\'erature des sources d'\'etalonnage (environ
40~minutes pour changer la temp\'erature des sources du banc de test
PACS), et nous allons voir que les r\'esultats de la proc\'edure
d'\'etalonnage contiennent d\'ej\`a l'information utile. La
caract\'erisation de la non-lin\'earit\'e des bolom\`etres est
effectivement n\'ecessaire \`a la pr\'eparation et \`a l'analyse des
futures observations. Elle permet entre autre de calculer la dynamique
effective du Photom\`etre (cf section~\ref{sec:calib_perfobs_nep} pour
les d\'etails), ou encore de corriger le gain des d\'etecteurs
lorsqu'un objet tr\`es brillant, par rapport au fond du t\'elescope,
est observ\'e.

Dans un premier temps nous pr\'esentons et commentons une courbe de
non-lin\'earit\'e extraite de la
figure~\ref{fig:calib_procedure_explore_midpt}, nous proposons une
application de ce type de courbes pour mesurer la temp\'erature et
l'\'emissivit\'e du t\'elescope Herschel, puis nous introduisons le
concept de \emph{r\'eponse statique} qui s'av\`ere \^etre une mesure
relativement fiable de la r\'eponse des bolom\`etres.

\subsection{Bonus de la proc\'edure d'\'etalonnage}
\label{sec:calib_perflabo_nonlinear_courbe}

\begin{figure}
  \begin{center}
      \includegraphics[width=0.7\textwidth,angle=0]{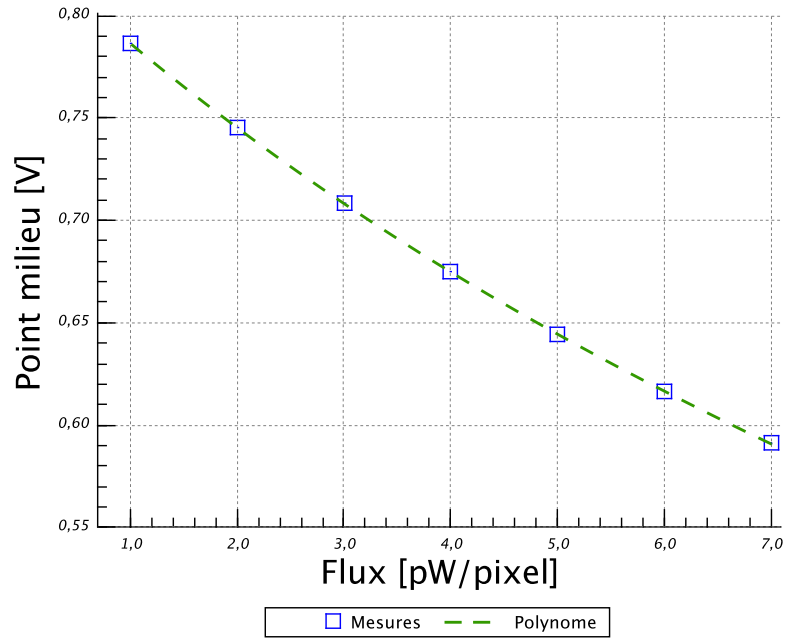}
  \end{center}
  \caption[Courbe d'\'etalonnage de la non-lin\'earit\'e d'un
  bolom\`etre]{\'Evolution du point milieu d'un pixel du BFP bleu en
  fonction du flux incident pour une tension de polarisation de
  2.4~V. Les carr\'es bleus repr\'esentent les points milieux extraits
  de la figure~\ref{fig:calib_procedure_explore_midpt}. La courbe
  verte en pointill\'es est un polyn\^ome du 3\ieme ordre ajust\'e sur
  ces points de mesure. Le bruit RMS associ\'e \`a chacun de ces
  points est de l'ordre de 60~$\mu$V, \cad bien inf\'erieur \`a la
  taille des carr\'es bleus sur cet figure.
  \label{fig:calib_perflabo_nonlinear_linea}}
\end{figure}

L'un des grands avantages de la proc\'edure d'\'etalonnage que nous
avons d\'evelopp\'ee est qu'elle contient bien plus que la seule
information n\'ecessaire au r\'eglage des bolom\`etres. En effet, \`a
partir des mesures en gain faible pr\'esent\'ees dans la
figure~\ref{fig:calib_procedure_explore_midpt}, nous pouvons extraire
l'\'evolution des points milieux de chaque pixel en fonction du flux
incident~; nous pouvons ainsi quantifier la non-lin\'earit\'e du
signal bolom\'etrique. Il n'est donc pas n\'ecessaire de passer des
dizaines d'heures\footnote{Le temps est en effet tr\`es pr\'ecieux
pendant la campagne d'\'etalonnage d'un instrument spatial, et il faut
absolument \'eviter de doubler des tests inutilement.} \`a ex\'ecuter
le test de non-lin\'earit\'e initialement pr\'evu dans le plan de test
puisque l'information que nous cherchons est d\'ej\`a disponible dans
les donn\'ees obtenues en gain faible
(section~\ref{sec:calib_procedure_explore}). La
figure~\ref{fig:calib_perflabo_nonlinear_linea} montre la courbe de
non-lin\'earit\'e d'un bolom\`etre du BFP bleu qui a \'et\'e extraite
de la figure~\ref{fig:calib_procedure_explore_midpt}. Les carr\'es
bleus repr\'esentent les points de mesure et la courbe en pointill\'es
correspond \`a un polyn\^ome du troisi\`eme ordre qui ajuste
parfaitement les donn\'ees. L'allure g\'en\'erale de la courbe est
tr\`es lisse, et son rayon de courbure est tr\`es grand~; ce qui
indique que le comportement des bolom\`etres est ordonn\'e et que le
signal est relativement lin\'eaire dans le domaine de flux que nous
explorons ici. En d'autres termes, pour de petites excursions de flux
autour du point de fonctionnement des bolom\`etres, le signal est
proportionnel au flux incident sur le d\'etecteur~; la r\'eponse est
constante pour de faibles variations de flux. Par exemple, pour un
fond de t\'elescope de 2~pW/pixel et un changement de flux de
100~fW/pixel, ce qui correspondrait \`a un objet relativement brillant
pour l'Observatoire Herschel, nous trouvons un \'ecart inf\'erieur \`a
0.1~\% par rapport \`a un comportement purement lin\'eaire. Cet
\'ecart de point milieu est n\'egligeable, \cad ind\'etectable, car il
est bien inf\'erieur au bruit intrins\`eque des bolom\`etres.

Les courbes de non-lin\'earit\'e sont riches d'informations, et elles
pourraient tr\`es certainement \^etre exploit\'ees lors de la phase de
V\'erifications des Performances du satellite Herschel. Nous pouvons
en effet les utiliser comme des abaques pour mesurer\footnote{La
m\'ethode de calcul est similaire \`a celle utilis\'ee pour obtenir
les points milieux \`a partir des courbes d'\'etalonnage de
l'\'electronique de lecture (cf
section~\ref{sec:calib_procedure_explore_calcul}).} le niveau
d'\'emission du t\'elescope, lorsqu'il aura atteint sa temp\'erature
d'\'equilibre, \`a partir des niveaux de points milieux mesur\'es. Le
flux ainsi mesur\'e dans chacune des trois bandes PACS (et
\'eventuellement les trois bandes de SPIRE) peut nous donner la
temp\'erature et l'\'emissivit\'e du t\'elescope Herschel en
s'appuyant sur les travaux pr\'esent\'es par
\shortciteN{fischer}. Mais ces courbes de non-lin\'earit\'e sont-elles
suffisamment pr\'ecises, en d'autres termes les d\'etecteurs sont-ils
suffisamment stables, pour permettre un calcul fiable du flux \'emis
par le t\'elescope?  Cette question semble tout \`a fait l\'egitime
puisque le spectre de la
figure~\ref{fig:calib_perflabo_sensibilite_description_spectre} montre
une remont\'ee basse fr\'equence qui implique que les points milieux
d\'erivent lentement pendant les mesures. Cela remet en question
l'exactitude et la pertinence des courbes de non-lin\'earit\'e. Il est
donc important d'estimer les incertitudes de mesures li\'ees \`a ces
d\'erives du signal.

Chacun des points de la
figure~\ref{fig:calib_perflabo_nonlinear_linea} repr\'esente la
moyenne temporelle du signal enregistr\'e lors des mesures en gain
faible de la section~\ref{sec:calib_procedure_explore}, et bien que la
d\'eviation standard mesur\'ee sur ces signaux soit de l'ordre de
60~$\mu$V, les v\'eritables barres d'erreurs des courbes de
non-lin\'earit\'e telles que nous les mesurons sont domin\'ees par les
d\'erives basse fr\'equence de l'\'electronique de lecture et des
r\'esistances du bolom\`etre. Notez que le premier et le dernier point
de mesure de la figure~\ref{fig:calib_perflabo_nonlinear_linea} sont
s\'epar\'es de plus de 30~heures~; les gains et offsets ont donc pu
largement d\'eriver au cours du test. Malgr\'e cela, l'\'ecart typique
entre les donn\'ees et le polyn\^ome ajust\'e est de l'ordre de
200~$\mu$V, ce qui correspond \`a une amplitude 30~fois plus petite
que la taille des carr\'es bleus sur la figure! Il serait donc tentant
d'interpr\'eter une si faible dispersion autour du polyn\^ome comme
\'etant la preuve d'une certaine stabilit\'e du syst\`eme. Le
probl\`eme, cependant, est que ce polyn\^ome ne repose sur aucun
mod\`ele physique qui d\'ecrirait le fonctionnement des bolom\`etres,
sa seule justification est qu'il ajuste parfaitement les donn\'ees.
Ce polyn\^ome ne nous permet pas de dissocier la d\'ependance du point
milieu avec le flux (ce que nous cherchons) de la d\'erive induite par
les lentes fluctuations de temp\'erature des d\'etecteurs. Notre
interpr\'etation serait en effet biais\'ee si la temp\'erature
augmentait de fa\c{c}on monotone tout au long du test par exemple, et
l'\'evolution r\'eelle du point milieu en fonction du flux ne
correspondrait alors pas exactement au polyn\^ome ajust\'e. Bien que
l'ajustement soit remarquable et que les points de mesure montrent un
comportement tr\`es ordonn\'e, nous sommes limit\'e dans notre
interpr\'etation et nous ne pouvons que donner une limite inf\'erieure
aux barres d'erreurs de la
figure~\ref{fig:calib_perflabo_nonlinear_linea}. Cette limite
inf\'erieure est de l'ordre de 200~$\mu$V. Pour estimer correctement
les barres d'erreur sur une mesure longue de 30~heures, il faudrait
soit r\'ep\'eter de nombreuses fois ces mesures pour obtenir une
statistique plus riche, mais cela n\'ecessiterait un temps
r\'edhibitoire, soit mod\'eliser le fonctionnement des matrices de
bolom\`etres (r\'esistance + \'electronique) et comparer le mod\`ele
aux points de mesure pour pouvoir dissocier la d\'erive du signal par
rapport aux effets de non-lin\'earit\'e.

Quelle est donc la pertinence d'une courbe de non-lin\'earit\'e telle
que pr\'esent\'ee dans la
figure~\ref{fig:calib_perflabo_nonlinear_linea}? Par exemple, une
variation de quelques centaines de $\mu$V sur une mesure longue de
plus de 30~heures repr\'esente une d\'erive minime, \`a peine un
milli\`eme du signal absolu. D'un point de vue purement instrumental,
une telle stabilit\'e est excellente. Par contre, d'un point de vue
observationnel, une variation de 200~$\mu$V correspondrait \`a la
d\'etection d'une source ponctuelle de pr\`es de 2~Jy (source tr\`es
brillante pour Herschel)! Les courbes de non-lin\'earit\'e ne sont
donc pas assez pr\'ecises pour calculer le flux provenant des objets
c\'elestes. Par contre, le t\'elescope Herschel devrait \'emettre un
flux de l'ordre de 650~Jy dans la bande \`a 110~$\mu$m de PACS, et
nous devrions pouvoir calculer son \'emission avec une pr\'ecision
d'au moins 10~\% \`a partir de courbes comme celle pr\'esent\'ee dans
la figure~\ref{fig:calib_perflabo_nonlinear_linea}.

\subsection{La r\'eponse statique}
\label{sec:calib_perflabo_nonlinear_respStatic}

\begin{figure}
  \begin{center}
    \begin{tabular}{c||c}
      \includegraphics[width=0.48\textwidth,angle=0]{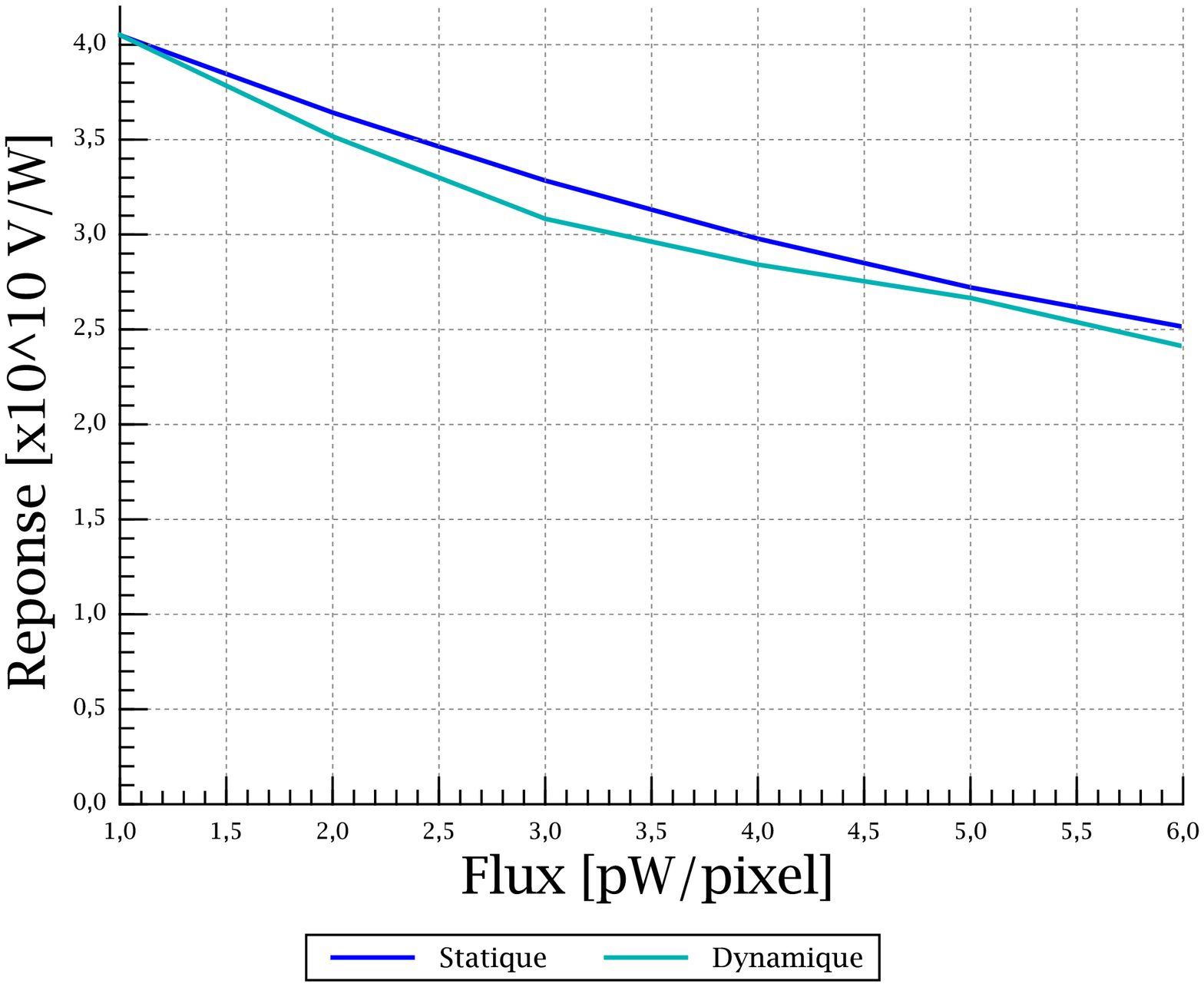} & \includegraphics[width=0.48\textwidth,angle=0]{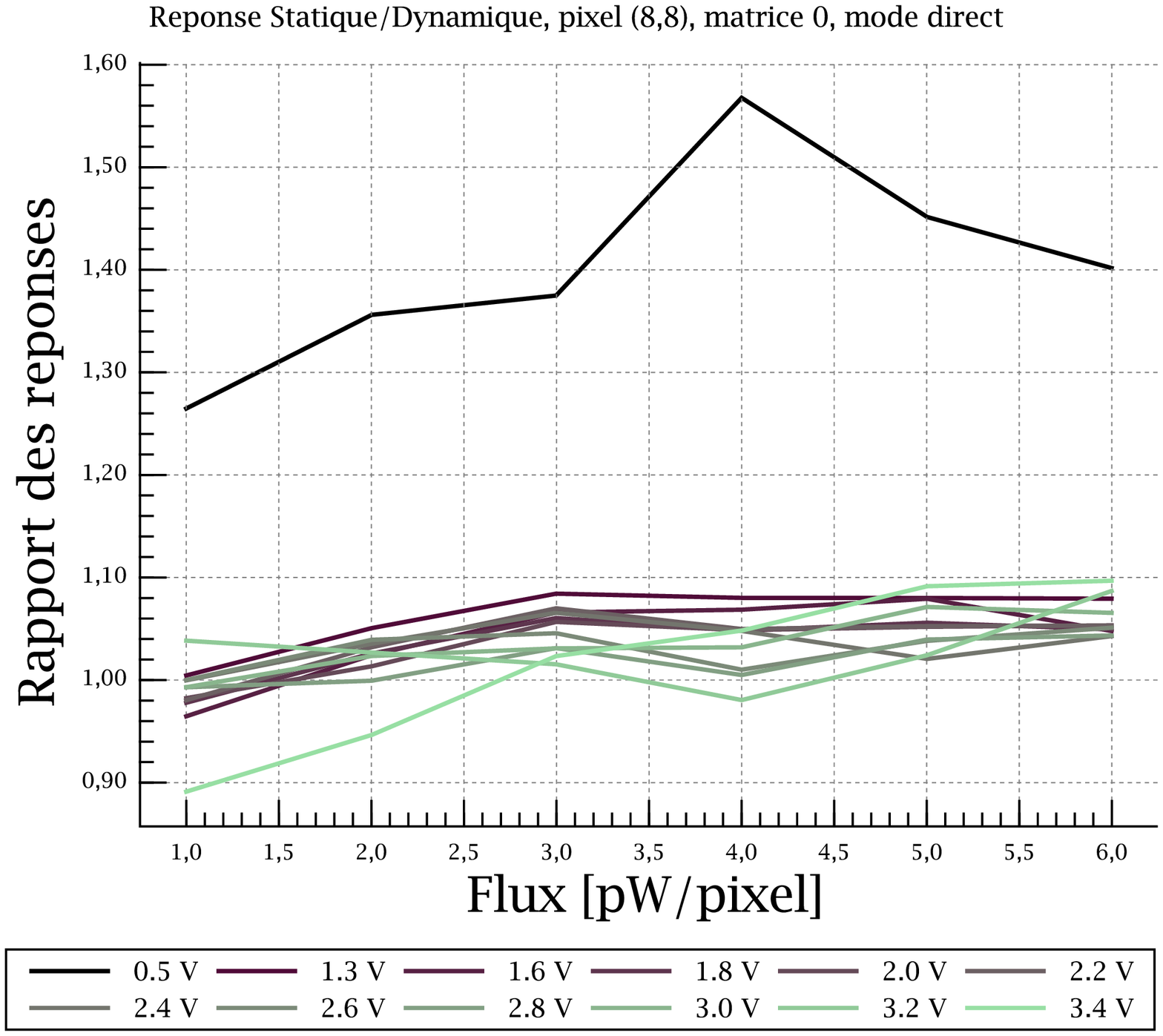} 
    \end{tabular}
  \end{center}
  \caption[Comparaison des r\'eponses statique et
  dynamique]{Comparaison de la r\'eponse statique et de la r\'eponse
  dynamique pour un pixel du BFP bleu. La r\'eponse statique est
  calcul\'ee \`a partir de la
  figure~\ref{fig:calib_perflabo_nonlinear_linea}. \emph{\`A gauche}~:
  R\'eponses statique et dynamique pour une tension de polarisation de
  2.4~V. Les r\'esultats sont tout \`a fait comparables. \emph{\`A
  droite}~: Rapport des r\'eponses statique/dynamique pour plusieurs
  polarisations des bolom\`etres. \`A nouveau, pour la mesure \`a
  0.5~V, les points milieux ne sont pas transmis correctement par le
  CL et la r\'eponse statique est alors erron\'ee. Pour les autres
  tensions, nous trouvons un bon accord ($\sim$10~\%) entre les deux
  mani\`eres de calculer la r\'eponse des bolom\`etres.
  \label{fig:calib_perflabo_nonlinear_resp}}
\end{figure}
Nous avons d\'ej\`a vu que la r\'eponse d'un bolom\`etre repr\'esente
l'efficacit\'e avec laquelle il peut transformer une modulation de
flux en une modulation de signal \'electrique
(section~\ref{sec:calib_perflabo_sensibilite_reponse}). Par ailleurs,
nous venons de voir qu'une courbe de non-lin\'earit\'e donne la
relation entre le signal \'electrique et le flux incident. Ce type de
courbes contient donc en principe toute l'information n\'ecessaire au
calcul de la r\'eponse des bolom\`etres. Nous d\'efinissons la
\emph{r\'eponse statique} d'un pixel \`a une tension de polarisation
donn\'ee comme \'etant la quantit\'e suivante~:
\begin{equation}
    \frac{g_{elec}\,\partial\, V_{ptmil}}{\partial\, flux}
\label{eq:resp_stat}
\end{equation}
o\`u $g_{elec}\sim 0.95$ est le gain total de la cha\^ine
\'electronique\footnote{Il est n\'ecessaire de multiplier le niveau de
point milieu par le gain de l'\'electronique pour se ramener \`a un
signal en sortie de BOLC.} calcul\'e dans la
section~\ref{sec:calib_procedure_vrlvhb}, et $\frac{\partial\,
V_{ptmil}}{\partial\, flux}$ est la pente de la courbe de
non-lin\'earit\'e calcul\'ee pour un flux donn\'e. Remarquez que nous
qualifions cette quantit\'e de \emph{statique} car son calcul ne fait
pas intervenir de modulation du flux
(cf~section~\ref{sec:calib_procedure_explore}). Par opposition, les
r\'eponses pr\'esent\'ees dans la
section~\ref{sec:calib_perflabo_sensibilite_reponse} seront
dor\'enavant appel\'ees \emph{r\'eponses dynamiques} puisqu'elles
n\'ecessitent l'emploi d'un chopper pour moduler le flux. Pour
comparaison, nous pouvons \'ecrire la r\'eponse dynamique sous la
forme $\frac{\Delta\,V_{sortie}}{\Delta\, flux}$
o\`u $\Delta\,V_{sortie}$ est l'amplitude du signal \'electrique en
sortie de l'instrument, et $\Delta\,flux$ est l'amplitude de la
modulation qui a donn\'ee naissance \`a $\Delta\,V_{sortie}$. Le but
de cette section est de comparer les r\'eponses statiques et
dynamiques calcul\'ees \`a partir de deux jeux de donn\'ees
compl\`etement ind\'ependants pour confirmer que le point milieu est
une quantit\'e physique qui repr\'esente effectivement le comportement
des bolom\`etres. Tout d'abord, pour rendre comparable les mesures
statiques et dynamiques, nous utilisons les
figures~\ref{fig:calib_procedure_vrvhb_gainbruit}
et~\ref{fig:calib_perflabo_nonlinear_linea} pour calculer la
quantit\'e $\frac{g_{elec}\,\Delta\,V_{ptmil}}{\Delta\, flux}$ \`a
partir de mesures statiques (nous choisissons $\Delta\, flux=0.5$~pW
pour reproduire \og virtuellement \fg les mesures
chopp\'ees). Remarquez qu'\`a partir de la
figure~\ref{sec:calib_procedure_explore}, nous pouvons en principe
cr\'eer des librairies de cartes de r\'eponse pour n'importe quel fond
de t\'elescope et pour n'importe quel $\Delta\, flux$ en interpolant
les valeurs de point milieu d\'ej\`a mesur\'ees. De telles librairies
ont \'et\'e utilis\'ees lors du stage de Guillaume Willmann pour
d\'evelopper un simulateur de cam\'era sub-millim\'etrique en
antarctique utilisant les matrices de bolom\`etres du CEA
\shortcite{willmann}.
La comparaison des r\'eponses statiques et dynamiques est
pr\'esent\'ee dans la figure~\ref{fig:calib_perflabo_nonlinear_resp}
pour un pixel du BFP bleu en mode de lecture direct. Le graphe de
gauche montre l'\'evolution des deux types de r\'eponses en fonction
du flux incident pour une tension de polarisation donn\'ee
(2.4~V). Nous retrouvons bien le r\'esultat de la
section~\ref{sec:calib_perflabo_sensibilite_reponse} qui montre que la
r\'eponse des bolom\`etres chute d'environ 40~\% entre 1~et
6~pW/pixel.  La courbe bleue, qui repr\'esente la r\'eponse statique,
est parfaitement r\'eguli\`ere car elle correspond \`a la d\'eriv\'ee
d'un polyn\^ome d'ordre~3. Par contre, l'aspect \og irr\'egulier \fg
de la courbe de r\'eponse dynamique est due aux d\'erives basses
fr\'equences du gain des bolom\`etres~; en effet les mesures
chopp\'ees sont modul\'ees avec une p\'eriode plus petite que le \og
temps de coh\'erence \fg du signal (cf
section~\ref{sec:calib_perfobs_oof}), nous nous affranchissons donc
des d\'erives additives (offsets) en soustrayant le signal de deux
plateaux chopper cons\'ecutifs, mais les d\'erives multiplicatives
persistent. Quoi qu'il en soit, les deux courbes co\"incident
remarquablement bien dans cette configuration de l'instrument. Le
graphe de droite donne une vision plus g\'en\'erale et montre que la
r\'eponse statique reproduit de mani\`ere assez fid\`ele l'\'evolution
de la r\'eponse dynamique, avec une pr\'ecision d'environ 10~\%, pour
toutes les tensions de polarisation sauf bien s\^ur celle \`a
0.5~V. En effet, nous savons que pour cette tension le point milieu
n'est pas transmis correctement par l'\'electronique de lecture, et la
r\'eponse statique calcul\'ee \`a partir de ces points milieux n'a
donc aucun sens physique. Notez que l'offset que nous avons \'evoqu\'e
dans la section~\ref{sec:calib_procedure_explore} semble ne pas
affecter le calcul de la r\'eponse statique. Notez que les r\'esultats
pour le BFP rouge sont tout \`a fait similaires \`a ceux pr\'esent\'es
dans la figure~\ref{fig:calib_perflabo_nonlinear_resp} pour le BFP
bleu.


Par ailleurs, il semble que la r\'eponse statique surestime
syst\'ematiquement la r\'eponse dynamique \`a toutes les tensions. Les
r\'eponses dynamiques \'etant mesur\'ees en modulant le flux \`a
0.5~Hz, la fr\'equence de coupure du bolom\`etre ($\sim$5~Hz, cf
section~\ref{sec:calib_perflabo_tau}) n'est certainement pas
responsable de l'att\'enuation du signal modul\'e. Par contre, les
d\'erives basses fr\'equences du d\'etecteur seraient une cause plus
plausible pour expliquer la surestimation que nous observons. En
fait, la diff\'erence fondamentale qu'il existe entre les deux types
de mesure de r\'eponse est que la d\'erive additive (offset) est
corrig\'ee dans les mesures dynamiques alors qu'elle ne l'est pas lors
des mesures statiques. Notez que si la r\'eponse statique surestime
la r\'eponse dynamique, cela signifie que les points milieux diminuent
avec le flux plus rapidement qu'ils ne devraient~; en d'autres termes,
dans le cas id\'eal o\`u il n'y aurait pas de d\'erives basses
fr\'equences, la pente de la
courbe~\ref{fig:calib_perflabo_nonlinear_linea} devrait \^etre \'egale
\`a la r\'eponse dynamique. Nous avons donc une indication que
l'offset du pixel a diminu\'e constamment tout au long du
test. D'autre part, entre le premier et le dernier point de mesure, le
flux incident sur les bolom\`etres a \'et\'e multipli\'e par~7 de
sorte que la temp\'erature du plan focal a d\^u augmenter
l\'eg\`erement (hausse de 1-2~mK au niveau du cryo-r\'efrig\'erateur),
et nous savons par ailleurs que l'offset des transistors de lecture
d\'epend fortement de la temp\'erature. 

L'origine de la surestimation que nous observons pourrait donc se
r\'esumer de la mani\`ere suivante~: la temp\'erature du plan focal
augmente de fa\c{c}on monotone au cours du test, l'offset des
transistors diminue lentement et les points milieux que nous mesurons
sont donc l\'eg\`erement inf\'erieurs \`a ce qu'ils devraient \^etre
sans d\'erive de l'offset. La courbe de non-lin\'earit\'e est alors
sensiblement plus pentue, et par cons\'equent la r\'eponse statique se
retrouve syst\'ematiquement au-dessus de la r\'eponse dynamique.\\
Bien que les points milieux soient \`a m\^eme de repr\'esenter le
comportement physique des bolom\`etres \`a 10~\% pr\`es, nous
continuerons \`a utiliser les r\'eponses dynamiques dans le calcul de
la \emph{NEP} pour garantir la meilleure estimation possible des performances
de la cam\'era.

\section{Le temps de r\'eponse des bolom\`etres}
\label{sec:calib_perflabo_tau}

Tout syst\`eme macroscopique, qu'il soit thermique, m\'ecanique ou
bien \'electrique met un certain temps pour r\'eagir \`a une
excitation donn\'ee, ou stimulus, du syst\`eme. En effet la
propagation de l'information ne peut \^etre instantan\'ee. Le temps
caract\'eristique n\'ecessaire \`a un syst\`eme donn\'e pour
s'\'etablir dans un nouvel \'etat \`a la suite d'une excitation
d\'epend de ses propri\'et\'es physiques. Par exemple, il est
inf\'erieur au milliardi\`eme de seconde pour les processeurs
informatiques actuels mais peut atteindre plusieurs jours voire
plusieurs ann\'ees pour des syst\`emes complexes comme l'atmosph\`ere
terrestre. Les instruments de mesure modernes ne d\'erogent pas \`a la
r\`egle, ils poss\`edent \'egalement un temps caract\'eristique
en-dessous duquel aucune variation temporelle ne peut \^etre
d\'etect\'ee. Il est donc crucial que leur temps de r\'eponse soit
parfaitement adapt\'e \`a leur usage particulier. Par exemple,
l'observatoire Planck va effectuer des balayages du ciel tr\`es
rapides \`a raison de 1~tour/minute, de sorte que la constante de
temps des bolom\`etres de HFI \shortcite{lamarre} doit \^etre
inf\'erieure \`a 4~ms pour pr\'eserver les hautes fr\'equences
spatiales du CMB. Dans le cas de l'observatoire Herschel, la vitesse
maximale de balayage est de 60 secondes d'arc par seconde, not\'ee
[$''$/s], si bien que les matrices de bolom\`etre PACS doivent avoir
des constantes de temps de l'ordre de 30~ms \shortcite{pacs_lutz}.

La pr\'esente section traite exclusivement des mesures en laboratoire
du temps de r\'eponse des bolom\`etres. Son impact sur les futures
observations astronomiques sera pr\'esent\'e dans la
section~\ref{sec:calib_perfobs_scan}. Nous introduisons dans un
premier temps le formalisme n\'ecessaire \`a la bonne interpr\'etation
des r\'esultats.

\subsection{Constante de temps et fr\'equence de coupure}
\label{sec:calib_perflabo_tau_compare}

Il existe plusieurs fa\c{c}ons de mesurer le temps de r\'eponse des
bolom\`etres et nous appelerons indiff\'eremment cette quantit\'e
\emph{constante de temps} si elle est mesur\'ee dans l'espace direct,
ou bien \emph{fr\'equence de coupure} si elle est mesur\'ee dans
l'espace de Fourier. Dans ce paragraphe nous montrerons en particulier
le lien qu'il existe entre ces deux quantit\'es qui d\'ecrivent en
r\'ealit\'e le m\^eme ph\'enom\`ene physique.

\begin{figure}
  \begin{center}
    \begin{tabular}{l}
      \includegraphics[width=0.5\textwidth,angle=0]{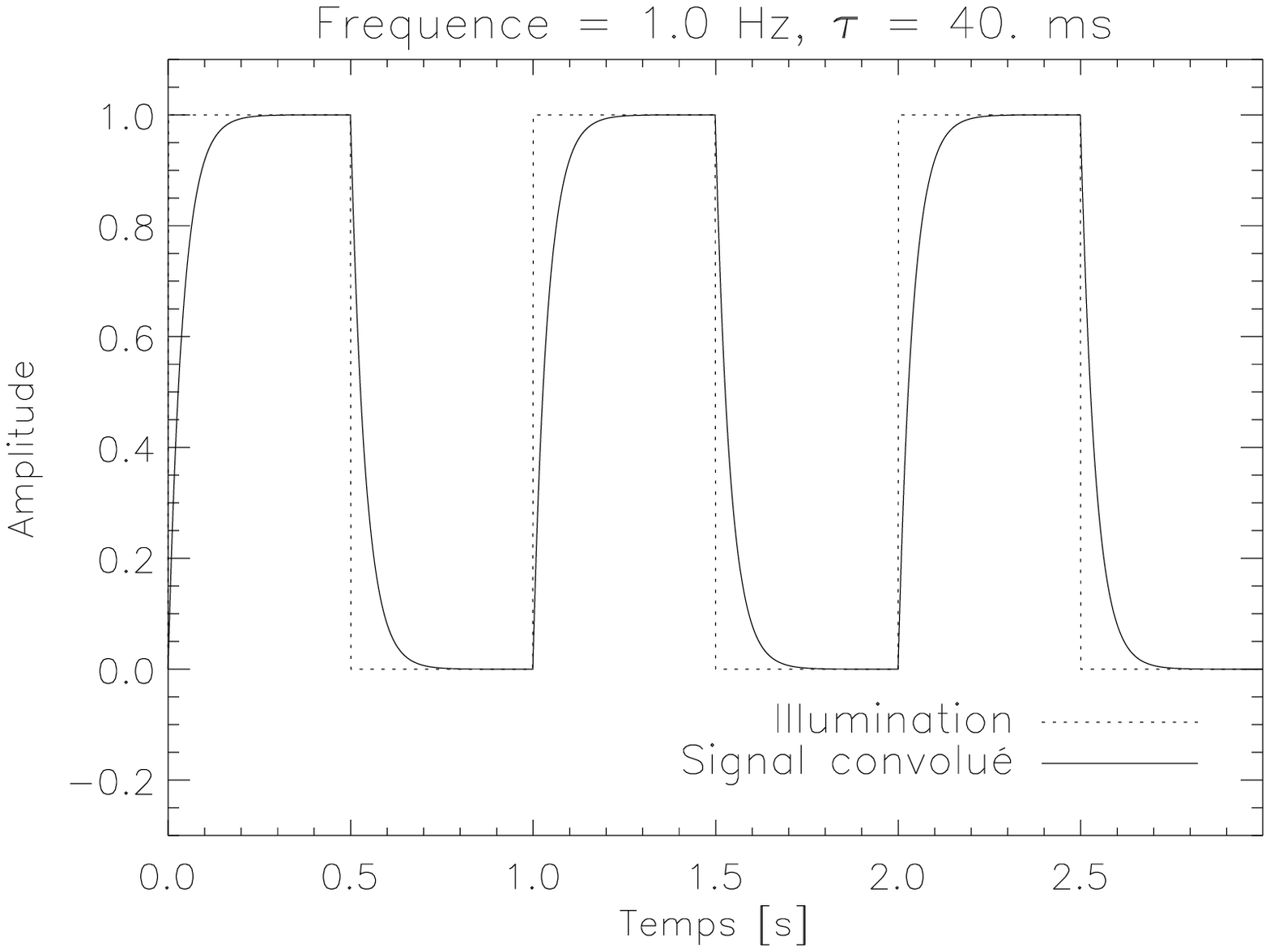}
      \includegraphics[width=0.5\textwidth,angle=0]{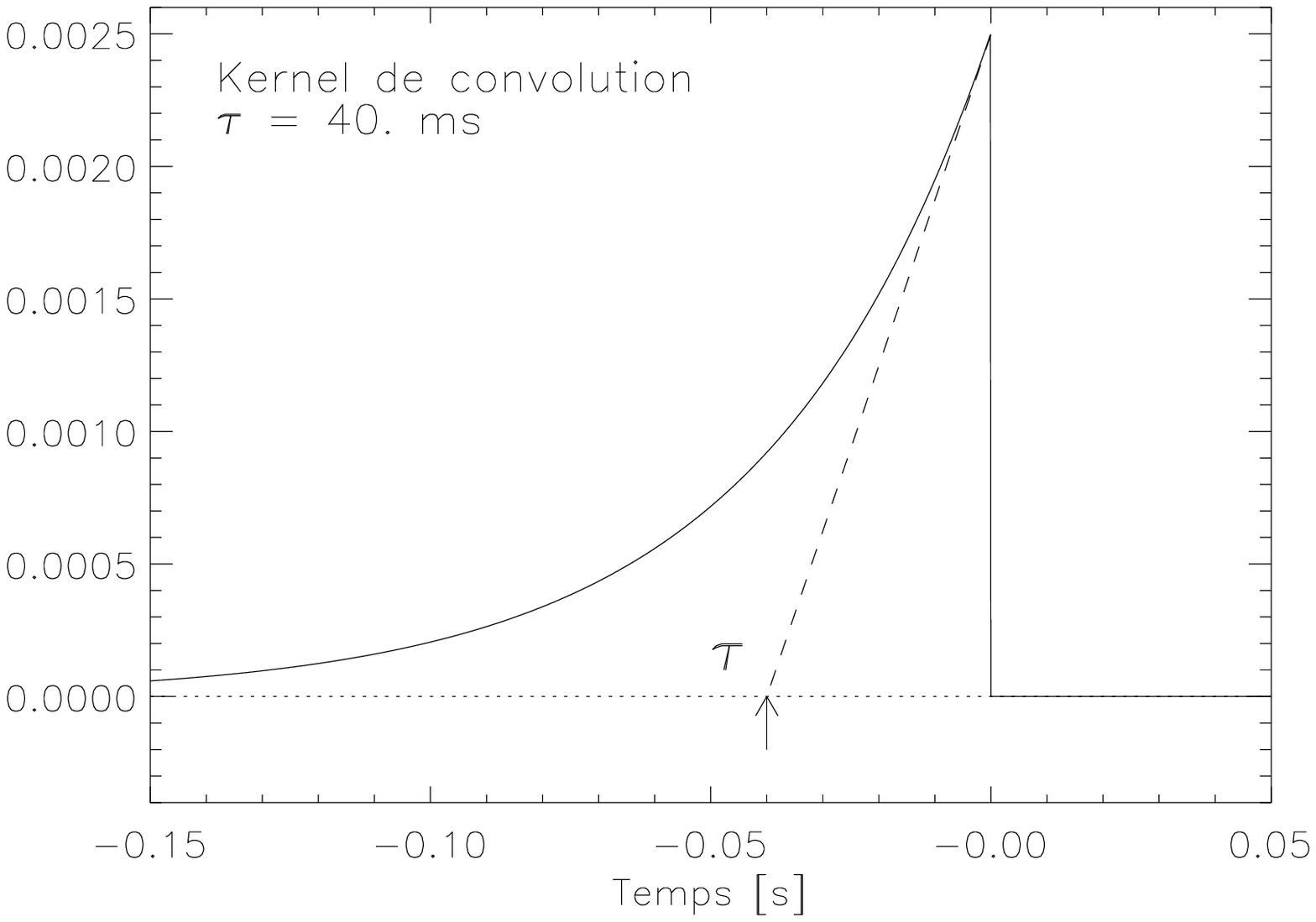}
    \end{tabular}
  \end{center}
  \caption[Simulation du filtre passe-bas dans l'espace
  direct]{\emph{Gauche}: Simulation d'une mesure chopp\'ee \`a 1~Hz
  convolu\'ee par un kernel de la forme $e^{-\frac{t}{\tau}}$ avec
  $\tau=$40~ms. Le signal met environ trois fois la valeur de la
  constante de temps pour se stabiliser \`a la valeur du
  plateau. \emph{Droite}: Kernel de la convolution (initialement de la
  forme $e^{-\frac{t}{\tau}}$ pour $t$>0 puis sym\'etrie par rapport
  \`a l'axe des ordonn\'ees). Il traduit le fait que le signal \`a un
  instant $t$ d\'epend du signal \`a un instant ant\'erieur, mais que
  son influence d\'ecro\^it exponentiellement avec l'\'ecart temporel.
  \label{fig:calib_perflabo_tau_compare_direct}}
\end{figure}

La mani\`ere la plus directe de mesurer le temps de r\'eponse d'un
bolom\`etre est de le soumettre \`a une variation soudaine
d'illumination et de mesurer le temps n\'ecessaire pour qu'il atteigne
son nouvel \'etat d'\'equilibre (en terme de temp\'erature ou de
signal \'electrique). Nous avons vu dans la
section~\ref{sec:intro_bolometrie_thermo_principe} que la r\'eponse
d'un bolom\`etre est soumise \`a un filtre passe-bas du premier
ordre. Nous mod\'elisons donc son comportement en convoluant
l'\'evolution temporelle du flux incident par un kernel de la forme
$e^{-\frac{t}{\tau}}$. Ce kernel est trac\'e sur la
figure~\ref{fig:calib_perflabo_tau_compare_direct} et montre qu'\`a un
instant $t_0$ donn\'e, le signal d\'epend de l'\'etat du bolom\`etre
\`a un instant ant\'erieur, et que plus cet instant est \'eloign\'e de
$t_0$, plus son influence d\'ecro\^it rapidement. Nous pouvons
interpr\'eter ce kernel comme une m\'emoire \`a court terme du
bolom\`etre qui s'estompe exponentiellement avec le temps. La
figure~\ref{fig:calib_perflabo_tau_compare_direct} montre \'egalement
un signal en cr\'eneau qui repr\'esente une modulation de flux, ainsi
que le signal r\'esultant de la convolution par ce kernel. Nous voyons
effectivement que le signal ne s'\'etablit pas instantan\'ement mais
n\'ecessite plusieurs fois la constante de temps $\tau$ avant
d'atteindre un plateau stable.

\begin{figure}
  \begin{center}
    \begin{tabular}{ll}
      \includegraphics[width=0.5\textwidth,angle=0]{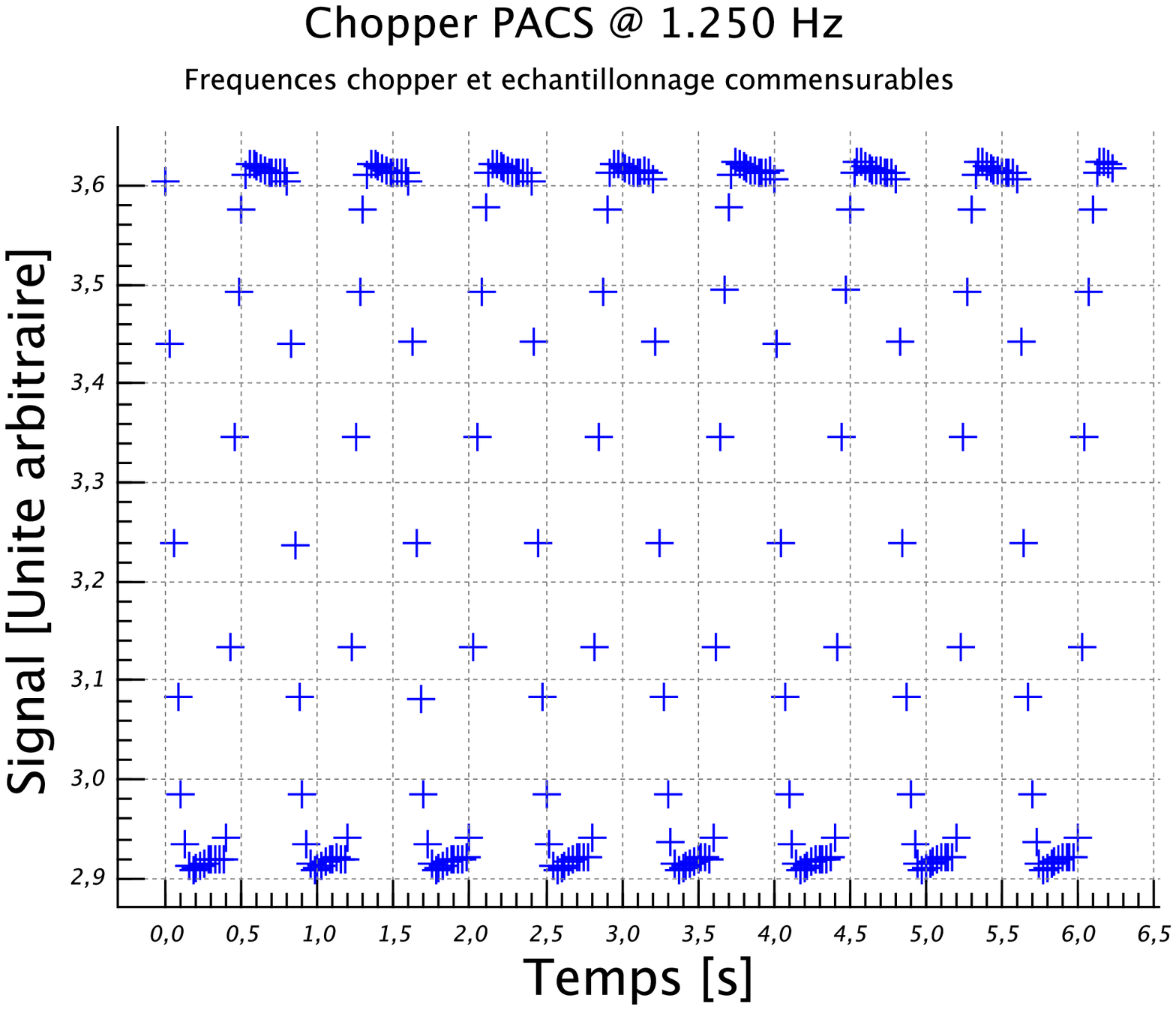}
      \includegraphics[width=0.5\textwidth,angle=0]{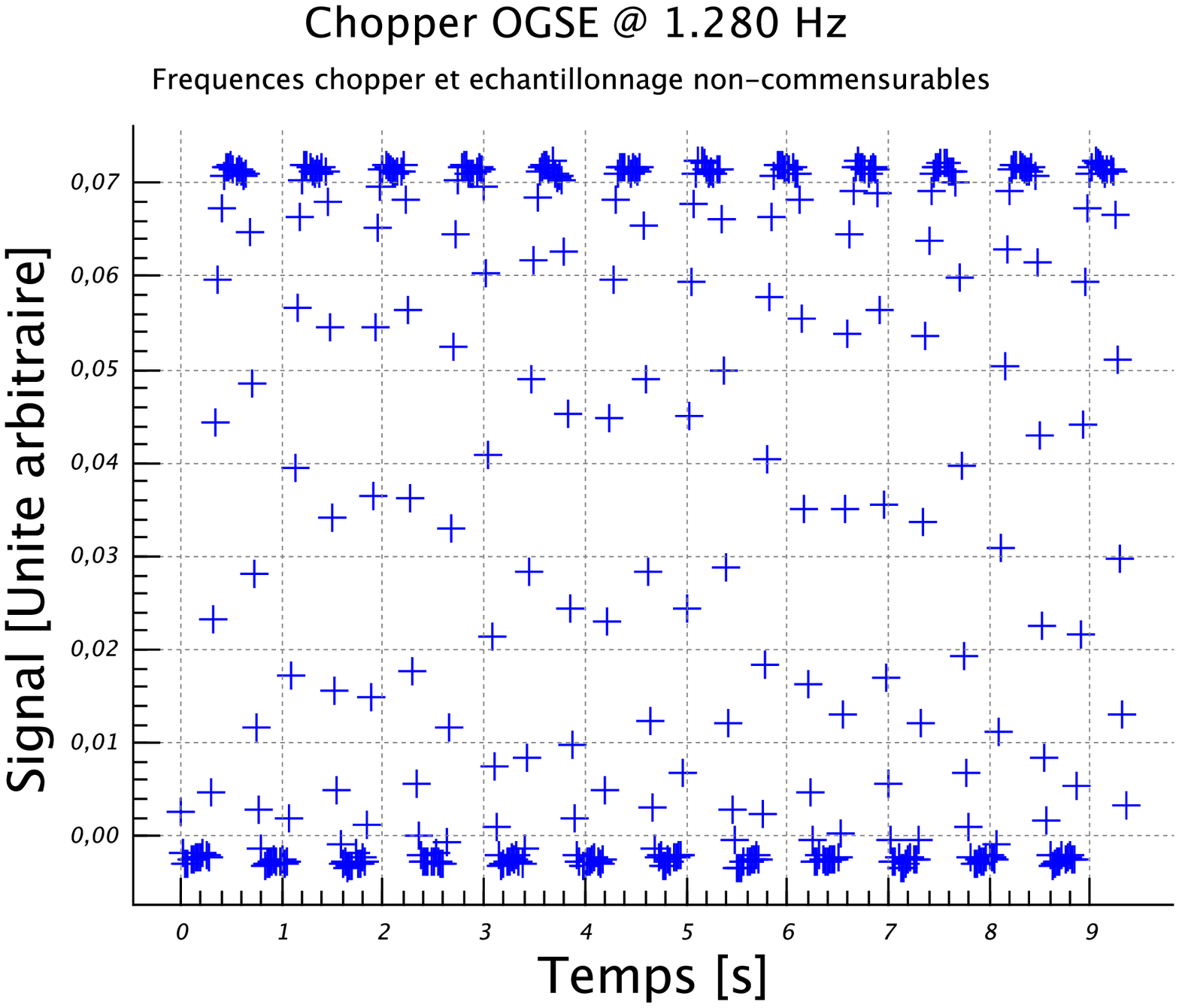} \\
      \includegraphics[width=0.5\textwidth,angle=0]{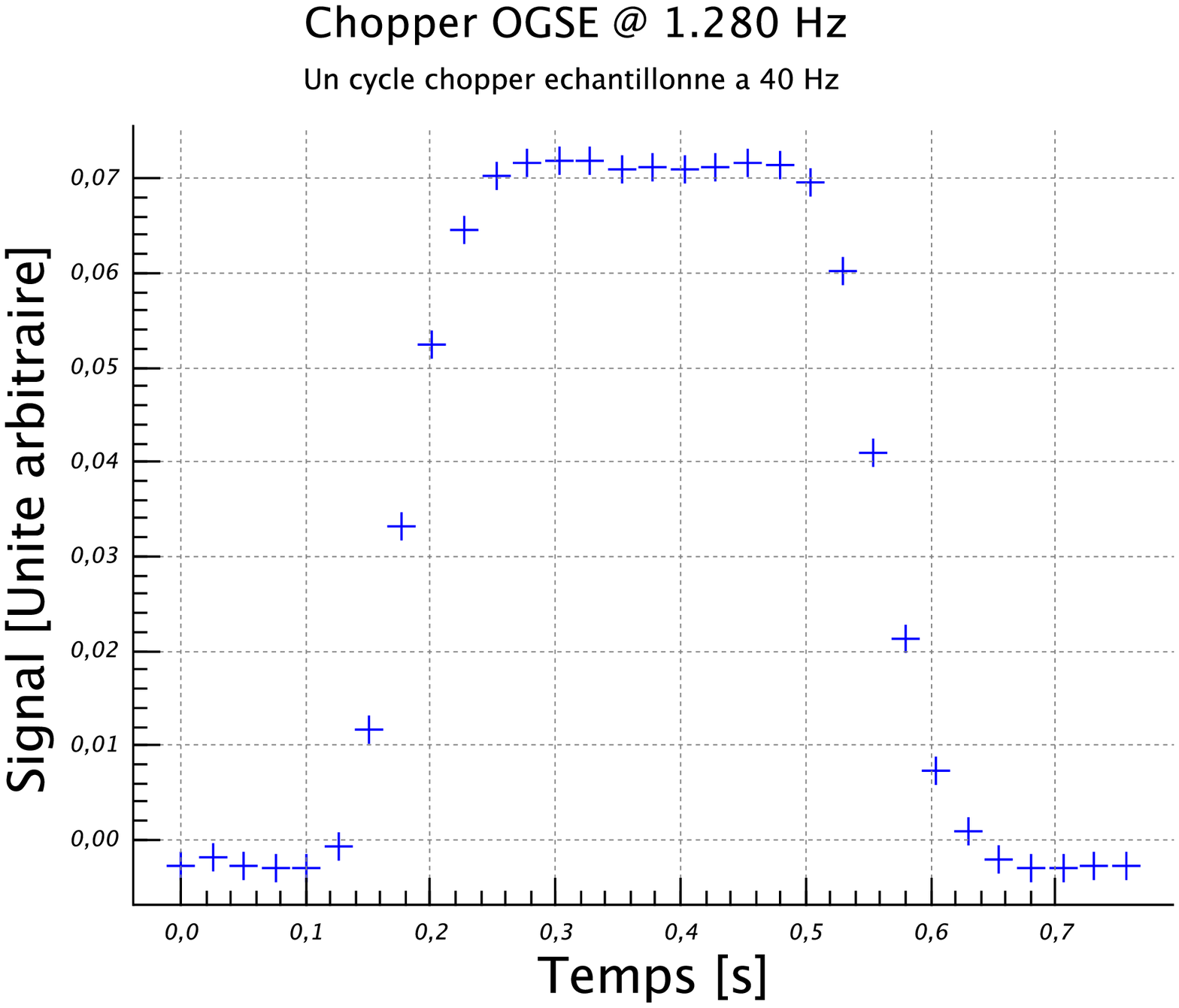}
      \includegraphics[width=0.5\textwidth,angle=0]{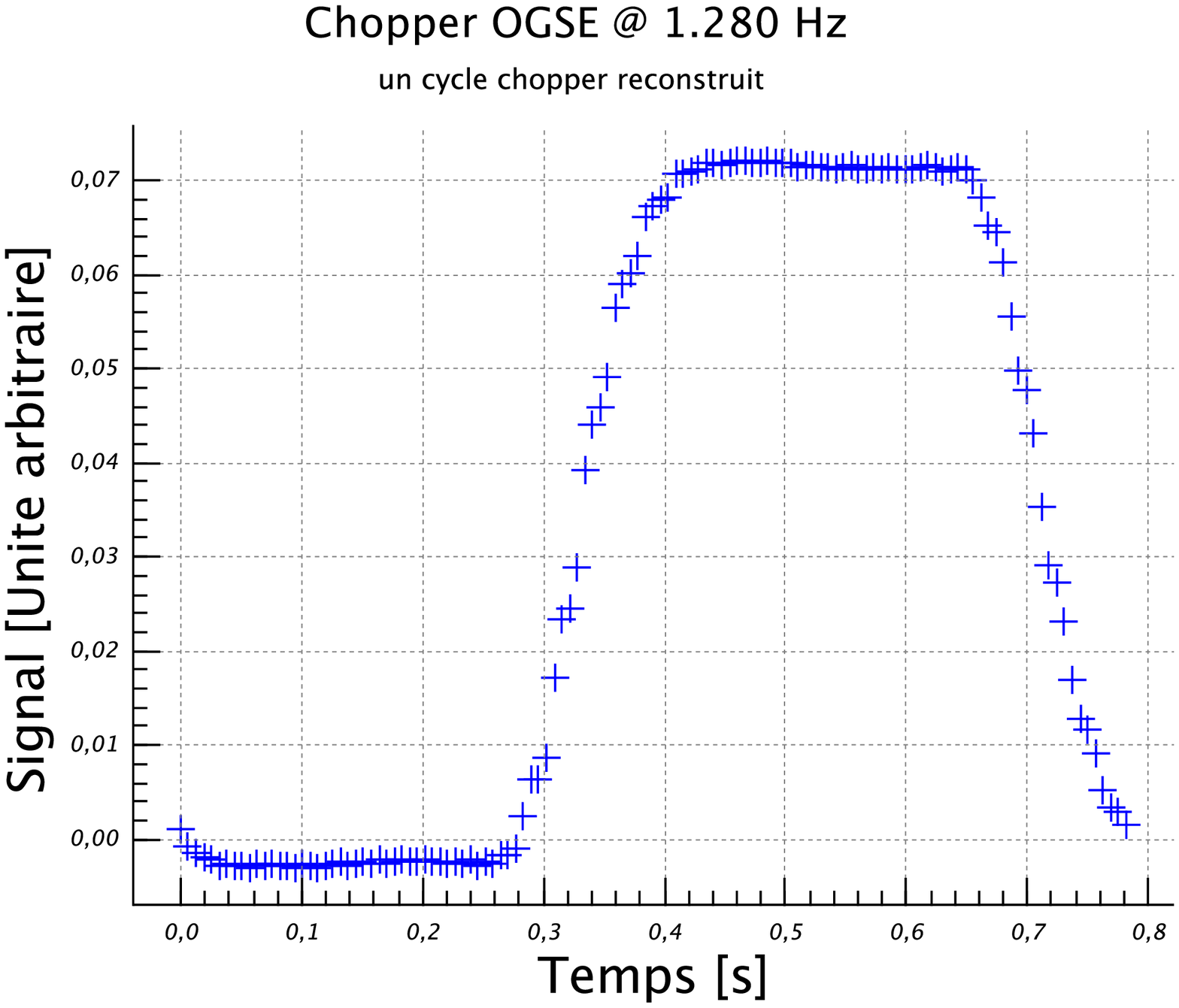}
    \end{tabular}
  \end{center}
  \caption[Commensurabilit\'e fr\'equence chopper/fr\'equence
d'\'echantillonnage]{\emph{Haut gauche}: Signal modul\'e avec le
chopper interne PACS qui est synchronis\'e avec la fr\'equence
d'\'echantillonnage du signal. Chaque cycle chopper contient ici
exactement 32~points ($\frac{40\,Hz}{1.25\,Hz}=32\,points$), et tous
les cycles chopper sont \'echantillonn\'es de la m\^eme fa\c{c}on,
c'est-\`a-dire que les points de mesure tombent toujours au m\^eme
moment dans le cycle chopper quelque soit le cycle. \emph{Haut droit}:
Signal modul\'e avec le chopper du banc de test PACS dont la
fr\'equence de rotation a \'et\'e choisie non-commensurable avec la
fr\'equence d'\'echantillonnage
($\frac{40\,Hz}{1.28\,Hz}=31.25\,points$). Il faut donc 4 cycles
chopper avant qu'un point de mesure ne retombe exactement sur la
m\^eme phase chopper. D'un cycle \`a l'autre tous les points de mesure
sont d\'ecal\'es d'un quart de la p\'eriode
d'\'echantillonnage. \emph{Bas gauche}: Un cycle naturel du chopper du
banc de test PACS contenant 31 points. \emph{Bas droit}: Un cycle du
chopper du banc de test PACS reconstruit \`a partir de 4 cycles et
contenant 125 points au lieu de 31. Le principe de la reconstruction
est d'assigner \`a chaque point de mesure une phase du cycle chopper
entre 0 et 1 puis de les r\'e-ordonner par phase croissante pour
obtenir le cycle reconstruit.
  \label{fig:calib_perflabo_tau_compare_echant}}
\end{figure}

En supposant que le chopper permette une transition franche entre les
deux flux observ\'es, c'est-\`a-dire une transition tr\`es courte
devant le temps de r\'eponse du d\'etecteur, nous devrions pouvoir
ajuster une fonction du type $e^{-\frac{t}{\tau}}$ sur le r\'egime
transitoire pour mesurer la constante de temps $\tau$. Cependant il
est n\'ecessaire que la transition contienne suffisamment de points de
mesure pour permettre un bon ajustement de l'exponentielle. Notez que
nous pouvons affiner l'\'echantillonnage d'un cycle chopper en
choisissant une fr\'equence de modulation non-commensurable avec la
fr\'equence d'\'echantillonnage du signal~; \cad que le nombre de
points de mesure dans un cycle chopper ne doit pas \^etre entier, le
signal est alors \'echantillonn\'e \`a diff\'erentes phases du cycle
chopper. Ceci revient \`a d\'ecaler le premier point de mesure de
chaque cycle chopper d'une fraction de la p\'eriode
d'\'echantillonnage de mani\`ere \`a \'echantillonner plus finement le
signal. Par exemple, pour une fr\'equence d'\'echantillonnage du
signal de 40~Hz et une fr\'equence chopper de 1.28~Hz, il faut
$\frac{40}{1.28}\times 4 = 125$ points, c'est-\`a-dire 4 cycles
chopper entiers, pour qu'un point de mesure se retrouve \`a une m\^eme
phase du cycle chopper. Nous pouvons donc assigner \`a chacun de ces
125 points une phase unique comprise entre 0 et 1 qui indique la
position du chopper au moment o\`u la lecture du signal a lieu. Nous
reconstruisons ensuite un seul cycle chopper avec ces 125 points, au
lieu de 31.25 points initialement, en r\'e-ordonnant les points de
mesure par phase croissante du cycle chopper. La
figure~\ref{fig:calib_perflabo_tau_compare_echant} illustre cette
m\'ethode de reconstruction et montre qu'en choisissant judicieusement
la fr\'equence de modulation du signal, nous pouvons effectivement
affiner l'\'echantillonnage du r\'egime transitoire\footnote{Cette
astuce est utilis\'ee de mani\`ere g\'en\'erale lors des mesures
dynamiques pour s'assurer que les extrema du signal sont effectivement
\'echantillonn\'es et ne se trouvent pas entre deux points de
mesure.}. Les mesures pr\'esent\'ees dans cette figure ont \'et\'e
obtenues avec le chopper du banc de test PACS qui modulait le signal
\`a 1.28~Hz. La figure contient \'egalement un exemple de signal
modul\'e par le chopper interne \`a l'instrument PACS. Les mouvements
de ce dernier sont synchronis\'es sur les acquisitions du signal (via
BOLC puis DECMEC) de sorte qu'il existe toujours un nombre entier de
points par cycle chopper. Dans ce cas, la m\'ethode de reconstruction
d\'ecrite ici n'am\'eliore pas l'\'echantillonnage temporel du
signal~; il est cependant possible de co-additioner les points de
mesures correspondant \`a une m\^eme phase du cycle chopper pour
r\'eduire les incertitudes li\'ees \`a la mesure du signal.

En pratique, la mesure de la constante de temps par ajustement du
r\'egime transitoire n'est pas tr\`es fiable du fait de l'incertitude
sur la forme de la modulation de flux. En effet, ni le banc de test de
Saclay, ni celui de PACS ne produisent une transition suffisamment
franche pour mesurer avec certitude la constante de temps. Faire un
ajustement sur une transition d\'ej\`a liss\'ee ne peut donner qu'une
limite sup\'erieure \`a la v\'eritable valeur de la constante de
temps. Seul le chopper PACS permet de moduler le flux assez rapidement
pour assurer une transition nette. Toutefois, lorsque le chopper PACS
bouge tr\`es vite, le r\'egime transitoire ne contient que deux ou
trois points au maximum, et nous ne pouvons pas ajuster pr\'ecisement
une exponentielle avec aussi peu de points.

Notez toutefois qu'il est possible d'obtenir une d\'ecroissance
exponentielle \og propre \fg en bombardant les matrices de
bolom\`etres avec des particules tr\`es \'energ\'etiques. En effet,
lorsqu'une telle particule interagit avec le bolom\`etre, elle
lib\`ere une fraction de son \'energie directement au c\oe ur du
d\'etecteur, cet apport d'\'energie est cons\'equent et instantan\'e,
la temp\'erature du pixel se relaxe alors en un temps
caract\'eristique $\tau$ et nous pouvons mesurer le nombre d'images
affect\'e par l'impact de la particule. De telles mesures ont \'et\'e
r\'ealis\'ees lors de la campagne de test d'irradiation\footnote{Les
matrices ont re\c{c}u une dose cumul\'ee de 20~et~11~krad de radiation
$\gamma$ (Cobalt~60) \`a des taux de 5~krad/h et 44~rad/h
respectivement sans d\'egradations significatives des
performances. Elles ont \'egalement \'et\'e bombard\'ees avec des
protons (20~MeV) et des particules alpha (30~MeV).} men\'ee \`a
l'Institut de Physique Nucl\'eaire d'Orsay en 2005. Ces travaux ont
fait l'objet d'un poster \`a la conf\'erence RADECS~2006, mais puisque
ma contribution \`a ces tests ne f\^ut que mineure, je renvoie le
lecteur vers l'annexe~\ref{a:nima} de ce manuscrit o\`u se trouve
l'article de \shortciteN{horeau} qui pr\'esente le r\'esultat des
irradiations et des mesures de la constante de
temps. \shortciteANP{horeau} trouvent une constante de temps d'environ
24~ms pour la matrice test\'ee qui est diff\'erente des mod\`eles de
vol PACS. Cette fa\c{c}on de mesurer la constante de temps est en
th\'eorie la plus fiable mais le nombre d'images affect\'ees par
l'impact d\'epend de la quantit\'e d'\'energie effectivement
d\'epos\'ee dans le bolom\`etre et aussi du niveau de d\'etection des
impacts au-dessus du bruit (d\'etection \`a 3$\sigma$). Il reste donc
une incertitude sur le calcul de la constante de temps. Pour affiner
ce type de mesure, il faudrait augmenter consid\'erablement la
fr\'equence d'\'echantillonnage pour bien mesurer la d\'ecroissance
exponentielle.\\

Pour mieux contraindre le temps de r\'eponse d'un bolom\`etre nous
nous tournons plut\^ot vers une m\'ethode qui ne fait pas appel \`a
une modulation de signal. Nous pouvons effectivement mesurer la
fr\'equence de coupure d'un filtre passe-bas en calculant la densit\'e
spectrale de bruit d'un signal temporel. En effet une convolution dans
l'espace direct est \'equivalente \`a une multiplication dans l'espace
de Fourier. Soit $I(t)$ la fonction r\'eelle d\'ecrivant l'\'evolution
temporelle du signal produit par un bolom\`etre\footnote{$I(t)$ inclue
le bruit g\'en\'er\'e par la thermistance du bolom\`etre ainsi que les
fluctuations statistiques du champ de radiation incident.}, et $S(t)$
le signal de sortie mesur\'e. Dans l'espace direct, nous pouvons
\'ecrire:
$$S(t)=I(t)\,\ast\,e^{-t/\tau}$$ 
et dans l'espace de Fourier cette expression se transforme
en:
$$\tilde{S}(\nu)=\tilde{I}(\nu)\times C
\int^{+\infty}_{-\infty}\,e^{-\boldsymbol{i}2\pi\nu t}\,e^{-\frac{t}{\tau}}\,dt$$
o\`u $C$ est une constante de normalisation, $\tilde{S}(\nu)$ et
$\tilde{I}(\nu)$ sont les transform\'ees de Fourier de $S(t)$ et
$I(t)$ respectivement. Nous trouvons la fonction de transfert suivante
qui d\'ecrit l'\'evolution de l'amplitude en fonction de la
fr\'equence:
\begin{center}
  \begin{equation}
    |\tilde{S}(\nu)|=|\tilde{I}(\nu)|\times\frac{1}{\sqrt{\,4\pi^2\nu^2\tau^2
     +1}}
  \label{eq:fonction_transfert}
  \end{equation}
\end{center}
\noindent Aux basses fr\'equences, la fonction de transfert tend vers
$1$, c'est-\`a-dire que les signaux variant lentement sont transmis
sans att\'enuation, alors qu'aux hautes fr\'equences l'amplitude chute
comme $1/\nu$, les signaux rapides se retrouvent donc fortement
att\'enu\'es. Cette fonction de transfert joue bien le r\^ole d'un
filtre passe-bas du premier ordre. La fr\'equence de coupure $\nu_c$
est d\'efinie comme \'etant la fr\'equence \`a laquelle le signal est
att\'enu\'e de 3~dB en puissance. Ceci se traduit pour une densit\'e
spectrale de bruit par:
\begin{center}
  \begin{equation}
    |\tilde{S}(\nu_c)|=\frac{|\tilde{I}(\nu_c)|}{\sqrt{\,2}}\;\;\;\;\;
     \Rightarrow \;\;\;\;\; \nu_c=\frac{1}{2\,\pi\,\tau}
  \label{eq:tau_nu}
  \end{equation}
\end{center}
Nous obtenons donc la formule qui relie la constante de
temps $\tau$, que l'on peut mesurer dans l'espace direct, et la
fr\'equence de coupure $\nu_c$ que l'on mesure dans l'espace de
Fourier.

La constante de temps de la plupart des bolom\`etres est limit\'ee par
le temps n\'ecessaire pour \'evacuer la chaleur de l'absorbeur vers la
source froide. La constante de temps est alors d'origine thermique,
elle vaut $\tau_{th}=C_{th}/G_{th}$ o\`u $C_{th}$ est la capacit\'e
calorifique de l'absorbeur et $G_{th}$ est la conductance thermique
des poutres qui relient l'absorbeur \`a la source froide (cf
section~\ref{sec:intro_bolometrie_thermo_principe}). $\tau_{th}$
d\'epend de la temp\'erature de l'absorbeur et de la source froide. La
situation est un peu diff\'erente pour les matrices de bolom\`etres du
CEA. L'imp\'edance des ponts bolom\'etriques est tellement \'elev\'ee
que la constante de temps \'electrique $\tau_{elec}$ devient
comparable \`a la constante de temps thermique\footnote{Par
construction des matrices PACS, les constantes de temps \'electrique
et thermique sont du m\^eme ordre de grandeur, autour de 10~Hz.}, et
nous devons la prendre en compte dans nos calculs. En effet, le
bolom\`etre poss\`ede une imp\'edance $R\sim$1-10~T$\Omega$ qui,
associ\'ee \`a la capacit\'e \'electrique parasite de l'\'etage haute
imp\'edance du d\'etecteur, forme un filtre passe-bas du premier ordre
o\`u $\tau_{elec}=RC_{elec}$. L'imp\'edance $R$ d\'epend de la
temp\'erature de la thermistance et de la source froide mais aussi du
champ \'electrique aux bornes des r\'esistances formant le pont
bolom\'etrique. Nous nous attendons donc \`a ce que la tension de
polarisation des bolom\`etres soit un param\`etre critique dans la
d\'etermination du temps de r\'eponse des d\'etecteurs. Dans l'espace
de Fourier, la fonction de transfert s'exprime de la fa\c{c}on
suivante~:
\begin{center}
  \begin{equation}
    |\tilde{S}(\nu)|=|\tilde{I}(\nu)|\times\frac{1}{\sqrt{(1-\,4\pi^2\nu^2\,\tau_{th}
\,\tau_{elec})^2\,+\,4\pi^2\nu^2(\tau_{th}+\tau_{elec})^2}}
  \label{eq:fonction_transfert_electrothermic}
  \end{equation}
\end{center}
Le filtre associ\'e correspond au module du produit de la fonction de
transfert complexe de deux filtres passe-bas du premier ordre. Nous
allons maintenant extraire $\tau_{th}$ et $\tau_{elec}$ \`a partir de
deux jeux de donn\'ees ind\'ependants obtenus avec deux proc\'edures
de test tr\`es diff\'erentes. Notez toutefois que l'information utile
pour l'observateur n'est pas la valeur de $\tau_{th}$ ou $\tau_{elec}$
mais plut\^ot la fr\'equence de coupure globale des d\'etecteurs
$\nu_c$.

\subsection{Les mesures dynamiques}
\label{sec:calib_perflabo_tau_direct}

Le travail pr\'esent\'e dans cette section est bas\'e sur la
premi\`ere partie d'un document que j'ai \'ecrit en 2005 dont
l'objectif \'etait de communiquer les performances pr\'eliminaires des
d\'etecteurs de vol aux futurs utilisateurs du Photom\`etre PACS afin
qu'ils puissent pr\'eparer efficacement leurs programmes
d'observations. Le rapport en question se trouve dans
l'annexe~\ref{a:rapport_bp}.

\begin{figure}
  \begin{center}
      \includegraphics[width=0.7\textwidth,angle=0]{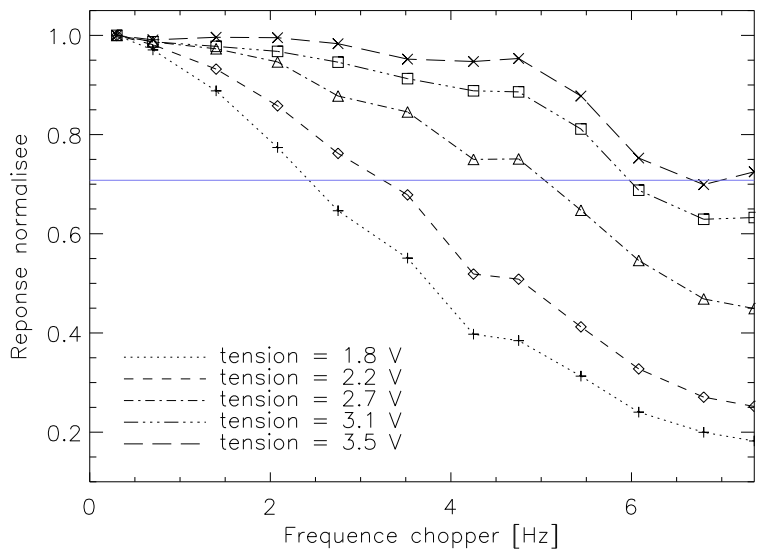}
  \end{center}
  \caption[Mesures dynamiques de la constante de temps]{\'Evolution de
  l'amplitude d'un signal modul\'e en fonction de la fr\'equence de
  modulation. Chaque courbe correspond \`a la moyenne des amplitudes
  mesur\'ees sur une matrice enti\`ere pour une tension de
  polarisation donn\'ee. Le flux incident est de 3~pW/pixel. Les
  courbes sont normalis\'ees par rapport au premier point mesur\'e \`a
  0.3~Hz. L'amplitude des modulations chute avec la fr\'equence comme
  attendu pour un filtre passe-bas, seules les basses fr\'equences ne
  sont pas alt\'er\'ees par le filtre. La ligne horizontale indique
  une att\'enuation de 3~dB et donne en th\'eorie la fr\'equence de
  coupure associ\'ee \`a une tension donn\'ee. Les bolom\`etres sont
  d'autant plus rapides que la tension de polarisation est \'elev\'ee.
  \label{fig:calib_perflabo_tau_direct_bppolar}}
\end{figure}

Le but de la proc\'edure de test que nous d\'ecrivons ici est de
mesurer l'\'evolution de l'amplitude d'un signal modul\'e en fonction
de la fr\'equence de modulation. Nous cherchons \`a mettre en
\'evidence l'att\'enuation de l'amplitude modul\'ee pour en extraire
la valeur des constantes de temps thermique et \'electrique. En
pratique, nous chauffons deux corps noirs \`a l'int\'erieur du
cryostat de test de sorte que l'on obtienne des flux incidents de 3~et
3.5~pW/pixel au niveau du plan focal. Nous utilisons ensuite un
chopper pour moduler le flux \`a une fr\'equence donn\'ee et nous
mesurons l'amplitude du signal \'electrique en sortie du d\'etecteur
pour plusieurs tensions de polarisation (de 1.8~\`a 3.5~V) et pour
plusieurs fr\'equences de modulation (de 0.3~\`a 7.36~Hz). Les
fr\'equences du chopper ont \'et\'e choisies incommensurables avec la
fr\'equence d'\'echantillonnage pour s'assurer que le maximum des
modulations soit bien \'echantillonn\'e (un cycle chopper ne contient
que 5~points de mesure \`a 7.36~Hz). Pour chaque couple de
param\`etres $(tension,frequence)$, nous reconstruisons un seul cycle
chopper en suivant la proc\'edure pr\'esent\'ee pr\'ec\'edemment et
nous calculons l'amplitude de la modulation. Le r\'esultat de ces
mesures est r\'esum\'e dans la
figure~\ref{fig:calib_perflabo_tau_direct_bppolar}. Pour une tension
de polarisation donn\'ee, les amplitudes de modulation sont
normalis\'ees par rapport \`a l'amplitude mesur\'ee \`a 0.3~Hz pour
laquelle le filtre a un effet n\'egligeable. La ligne horizontale est
un indicateur qui correspond \`a une att\'enuation de la modulation de
3~dB ($\sim$71~\%). L'intersection entre les courbes mesur\'ees et
cette ligne horizontale nous donne en th\'eorie la valeur de la
fr\'equence de coupure globale du filtre electro-thermique.

Notez que si la modulation de flux incident \'etait sinuso\"idale,
alors les courbes de la
figure~\ref{fig:calib_perflabo_tau_direct_bppolar} repr\'esenteraient
exactement ce que nous cherchons, \cad le filtre passe-bas dans
l'espace de Fourier. En effet, la transform\'ee de Fourier d'une
sinuso\"ide \'etant un pic de Dirac, mesurer l'amplitude des
modulations pour plusieurs fr\'equences serait alors \'equivalent \`a
multiplier le filtre par un peigne de Dirac. Id\'ealement, cette
proc\'edure reviendrait \`a \'echantillonner le filtre aux
fr\'equences mesur\'ees. Cependant, la modulation fournit par le
chopper ressemble plut\^ot \`a un signal carr\'e liss\'e (cf figure~2
dans l'annexe~\ref{a:rapport_bp}) dont le spectre contient de
nombreuses harmoniques, et la pr\'esence de ces harmoniques entra\^ine
une sous-estimation de la fr\'equence de coupure $\nu_C$. Par exemple,
pour une modulation \`a 0.7~Hz et une fr\'equence de coupure de 5~Hz,
il y a 6~harmoniques qui contribuent \`a l'amplitude associ\'ee \`a
0.7~Hz, alors que pour une modulation \`a 3.52~Hz, seule la
fondamentale contribue \`a l'amplitude des modulations. Plus la
fr\'equence de modulation augmente et plus les harmoniques se
d\'eplacent et sortent de la bande passante du filtre, diminuant ainsi
l'amplitude des modulations.

Le travail men\'e en 2005 repose sur la simulation du filtrage des
modulations dans l'espace direct plut\^ot que dans l'espace de
Fourier. En effet, pour repr\'esenter le plus fid\`element possible le
comportement des bolom\`etres et extraire la valeur des constantes de
temps, nous simulons des signaux carr\'es aux m\^emes fr\'equences que
celles mesur\'ees durant la proc\'edure de test, nous les convoluons
successivement par deux filtres de la forme $e^{-t/\tau}$, puis nous
mesurons l'amplitude du signal associ\'e au couple
$(\tau_{th},\tau_{elec})$ test\'e, et nous reportons le r\'esultat sur
la figure~\ref{fig:calib_perflabo_tau_direct_bppolar} pour
comparaison. Il ressort de ces mesures que le temps de r\'eponse des
bolom\`etres ne d\'epend que faiblement du flux incident. Par contre,
la tension de polarisation a une influence bien plus marqu\'ee sur
leur rapidit\'e. Par exemple, la fr\'equence de coupure moyenne du BFP
bleu est de l'ordre de 2~Hz pour une tension de polarisation de 1.8~V
alors qu'elle vaut 6~Hz pour une tension de 3.5~V (cf
annexe~\ref{a:rapport_bp} pour plus de d\'etails). Pour un flux de
3~pW/pixel et une tension de polarisation nominale de 2.7~V,
l'ajustement des courbes simul\'ees donne des constantes de temps de
58.8~et 20~ms. Nous ne pouvons cependant pas d\'eterminer l'origine
thermique ou \'electrique de chacune de ces constantes. Il faudrait
pour cela mod\'eliser le comportement physique des bolom\`etres du CEA
et pr\'edire leur \'evolution en fonction de chacun des param\`etres
du syst\`eme (temp\'erature du bain, flux incident, tension de
polarisation, $R(T,V)$, $C(T)$, $G(T)$). Notez cependant que la
capacit\'e \'electrique de l'\'etage haute imp\'edance du d\'etecteur
est a priori ind\'ependante de la temp\'erature du bolom\`etre, et que
c'est l'imp\'edance de la thermistance qui d\'efinit enti\`erement
l'\'evolution de la constante de temps \'electrique
($\tau_{elec}=RC_{elec}$). Nous savons par ailleurs que cette
imp\'edance est une fonction fortement d\'ecroissante de la
temp\'erature et du champ \'electrique (cf
\'equation~\ref{eq:efros})~; par cons\'equent, lorsque la dissipation
Joule \'echauffe le bolom\`etre et que la tension aux bornes de la
r\'esistance augmente, l'imp\'edance chute et la constante de temps
s'en trouve alors consid\'erablement raccourcie (cf
figure~\ref{fig:calib_perflabo_tau_direct_bppolar}).



\begin{figure}
  \begin{center}
    \begin{tabular}[t]{ll}
      \includegraphics[width=0.6\textwidth,angle=0]{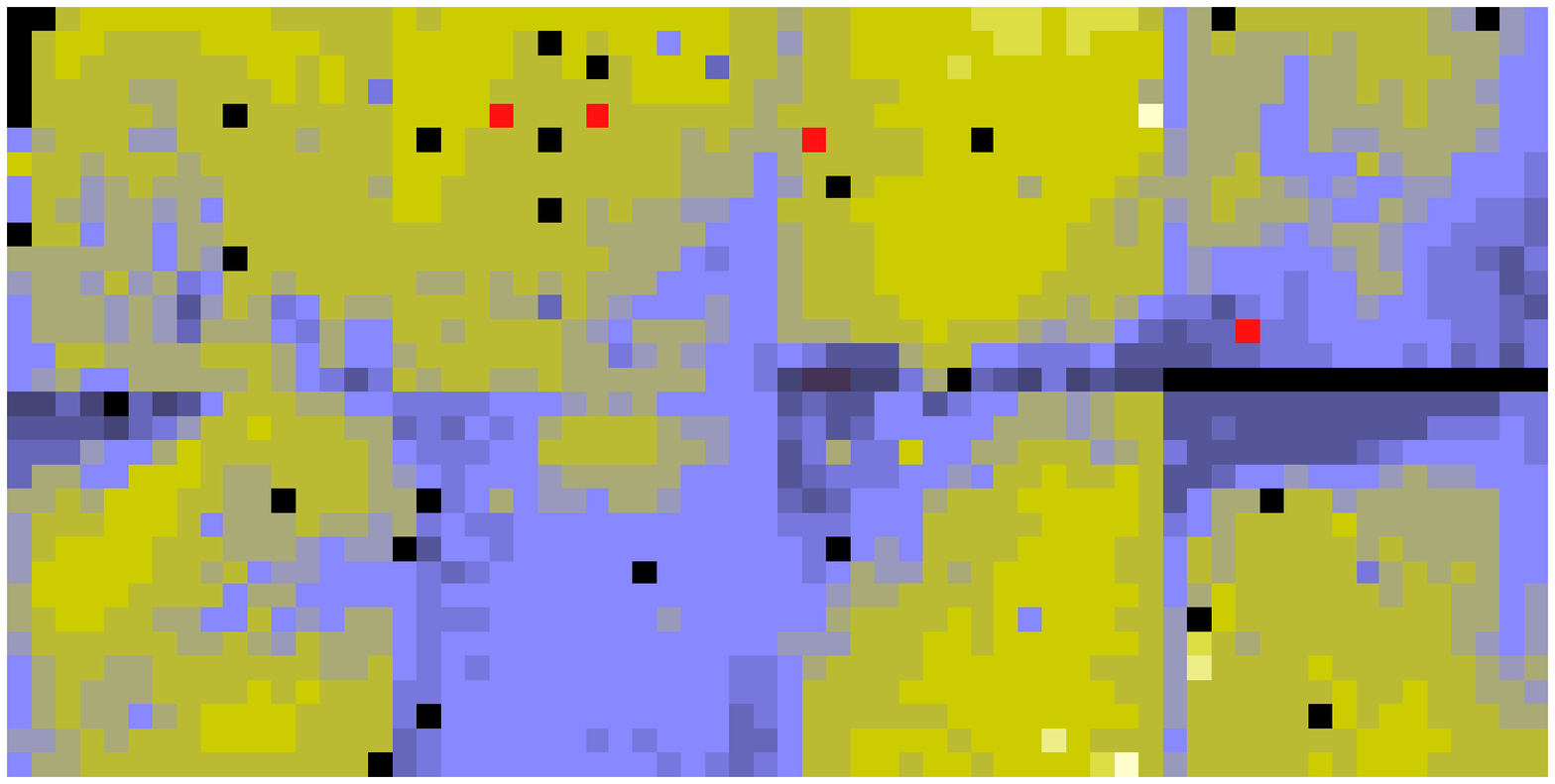} & \includegraphics[width=0.35\textwidth,angle=0]{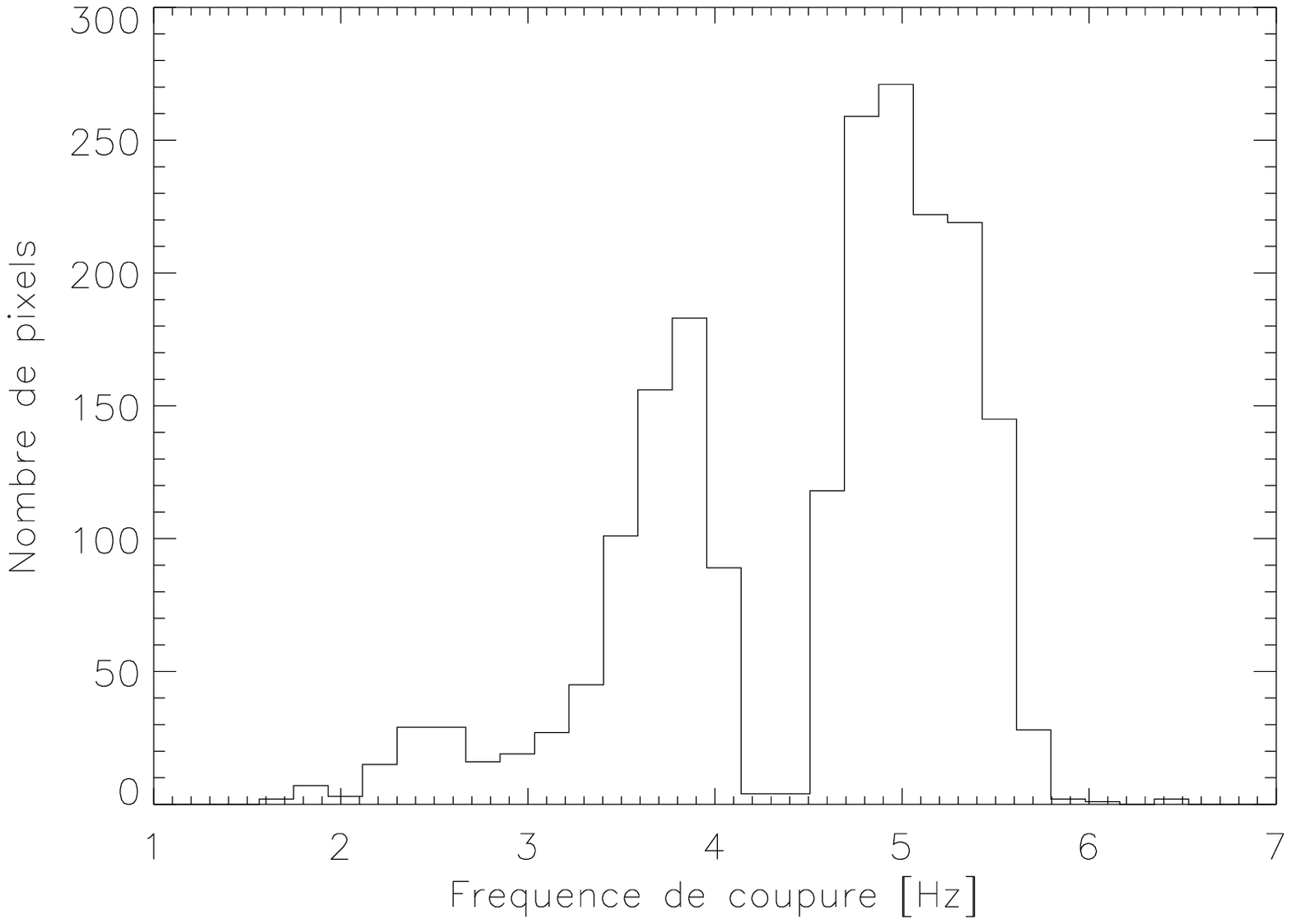} 
    \end{tabular}
  \end{center}
  \caption[Carte et dispersion de la fr\'equence de coupure des
  bolom\`etres du BFP bleu]{Distribution spatiale de la fr\'equence de
  coupure des bolom\`etres du BFP bleu pour un flux de 3~pW/pixel et
  une tension de polarisation de 2.7~V. La carte de r\'eponse de la
  figure~\ref{fig:calib_perflabo_sensibilite_description_respMapHisto}
  est en fait la carte \og compl\'ementaire \fg de
  celle-ci. L'histogramme de droite montre la dispersion de $\nu_C$
  sur cette carte.
  \label{fig:calib_perflabo_tau_direct_bpMapHisto}}
\end{figure}

La distribution spatiale des fr\'equences de coupure mesur\'ees sur le
BFP bleu est pr\'esent\'ee dans la
figure~\ref{fig:calib_perflabo_tau_direct_bpMapHisto} pour un flux
incident de 3~pW/pixel et une tension de polarisation de 2.7~V. Nous
trouvons deux populations de pixels ayant une fr\'equence de coupure
autour de 3.5-4~Hz (en bleu sur la carte) et autour de 5~Hz. Il est
int\'eressant de comparer cette carte avec la carte de r\'eponse de la
figure~\ref{fig:calib_perflabo_sensibilite_description_respMapHisto}. En
effet, les pixels les plus imp\'edants ont une r\'eponse plus
\'elev\'ee (car leur point de fonctionnement sur la
figure~\ref{fig:detect_bolocea_fabrication_thermo_R2T} donne un
coefficient $\alpha=\frac{1}{R}\frac{\partial R}{\partial T}$ plus
grand) et une constante de temps \'electrique plus longue (car
$\tau_{elec}=RC_{elec}$).  Cela explique pourquoi les pixels les plus
r\'epondants sont \'egalement les plus lents, et vice versa~; les
cartes de r\'eponse et de fr\'equence de coupure sont donc \og
compl\'ementaires \fg.

\begin{figure}
  \begin{center}
      \includegraphics[width=1.\textwidth,angle=0]{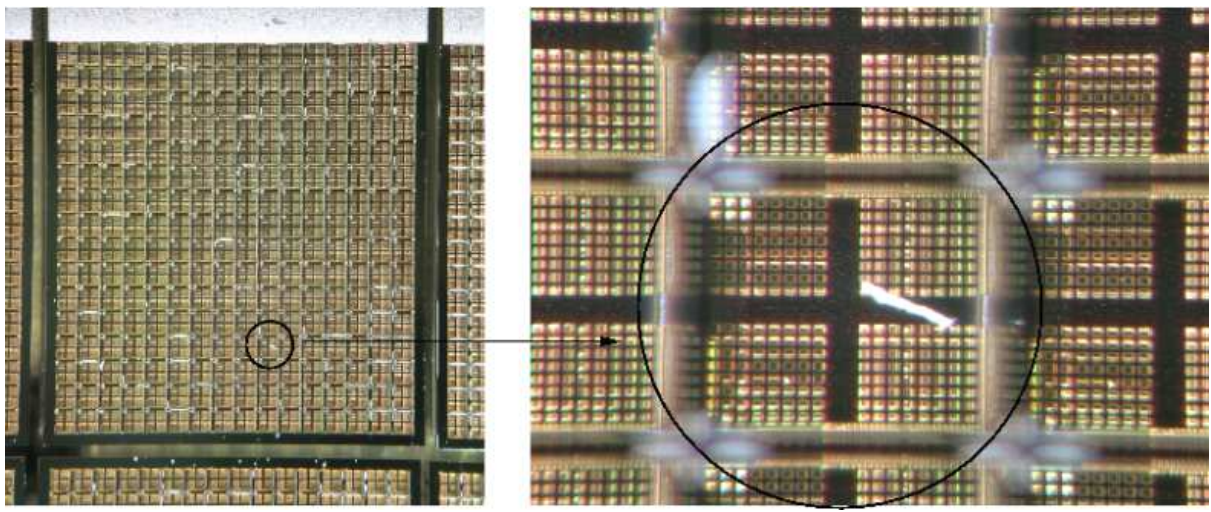}
  \end{center}
  \caption[Corps \'etranger et pixel lent]{Ces photographies ont
  \'et\'e obtenues lors de l'inspection visuelle du BFP bleu avec une
  loupe binoculaire. Celle de gauche montre une matrice sur laquelle
  nous avons rep\'er\'e un corps \'etranger accroch\'e \`a un pixel
  (zoom sur la photographie de droite). Ce r\'esidu augmente
  sensiblement la masse de l'absorbeur de sorte que la constante de
  temps du pixel s'en trouve rallong\'ee (cf matrice du haut
  centre-gauche de la
  figure~\ref{fig:calib_perflabo_tau_direct_bpMapHisto}). Les reflets
  blancs visibles sur les murs inter-pixels sont tr\`es probablement
  des r\'esidus de r\'esine, leur pr\'esence ne change rien \`a la
  constante de temps des bolom\`etres.}
  \label{fig:calib_perflabo_tau_direct_corpsetranger}
\end{figure}

Par ailleurs, lors de l'inspection visuelle des matrices du mod\`ele
de vol, nous avons rep\'er\'e plusieurs corps \'etrangers sur certains
pixels. Ces particules sont probablement maintenues en place par des
forces \'electrostatiques et elles resteront \og accroch\'ees \fg au
pixel tout au long de la mission. Ces r\'esidus sont de taille et de
forme variables, ils proviennent tr\`es certainement de r\'esidus de
r\'esine ou de colle utilis\'ees lors de la fabrication des
matrices. Ces r\'esidus sont relativement petits, quelques dizaines de
microns en g\'en\'eral, mais leur capacit\'e calorifique n'est pas
forc\'ement n\'egligeable et elle s'ajoute quoi qu'il en soit \`a
celle des absorbeurs. Nous nous attendons alors \`a ce que la
constante de temps thermique ($\tau_{th}=C_{th}/G_{th}$) de ces pixels
soit rallong\'ee. La
figure~\ref{fig:calib_perflabo_tau_direct_corpsetranger} montre deux
photographies d'un tel pixel \og alourdi \fg~; nous voyons une vue
rapproch\'ee du corps \'etranger sur celle de droite, et sur l'autre
nous avons une vue plus g\'en\'erale de la matrice qui nous permet de
localiser le pixel\footnote{La matrice en question se trouve en haut
au centre-gauche. En partant du pixel en bas \`a gauche de la matrice,
le pixel lent est le septi\`eme \`a droite et le quatri\`eme en haut.}
sur la carte de bande passante de la
figure~\ref{fig:calib_perflabo_tau_direct_bpMapHisto}~; sa fr\'equence
de coupure est effectivement plus courte et semble aberrante par
rapport \`a celle des pixels voisins.\\

Les mesures de constante de temps se sont av\'er\'ees relativement
difficiles \`a effectuer, et nous devons maintenant prendre un peu de
recul pour estimer la fiabilit\'e de nos r\'esultats. La difficult\'e
majeure que nous avons rencontr\'ee durant ces tests concernait le
contr\^ole de la temp\'erature du plan focal. En effet, nous utilisons
un moteur cryog\'enique pas-\`a-pas pour actionner le chopper. Ce
dispositif fonctionne parfaitement pour les faibles fr\'equences de
modulation, par contre, il est n\'ecessaire de sur-alimenter le moteur
($\sim20$~V) pour pouvoir moduler le signal \`a des fr\'equences
sup\'erieures \`a quelques Hz. Le probl\`eme est que, dans ce cas, le
moteur dissipe beaucoup trop d'\'energie au niveau du plateau
optique. L'\'el\'evation de temp\'erature qui s'ensuit est
significative (plusieurs dixaines de~mK) et perturbe les mesures de la
fa\c{c}on suivante~:
\begin{itemize}
\item L'imp\'edance des bolom\`etres diminue selon
l'\'equation~(\ref{eq:efros}), et la constante de temps
$\tau_{elec}=RC_{elec}$ se trouve alors involontairement
raccourcie. Ceci m\`ene \`a une sur-estimation de la fr\'equence de
coupure.
\item Le point de fonctionnement des r\'esistances se d\'eplace sur la
courbe $R(T)$ de la
figure~\ref{fig:detect_bolocea_fabrication_thermo_R2T} , le
coefficient $\alpha=\frac{1}{R}\frac{\partial R}{\partial T}$
diminue. La r\'eponse des bolom\`etres est donc plus faible lorsque le
moteur \'echauffe le plan focal aux fr\'equences \'elev\'ees. Ceci
m\`ene \`a une sous-estimation de la fr\'equence de coupure.
\item Nous avons aussi observ\'e une \'el\'evation de temp\'erature
d'un des deux corps noirs d'environ 1-2~K lorsque le chopper tourne
trop vite ($\nu_c\gtrsim5$~Hz). Une telle diff\'erence de
temp\'erature changerait significativement le flux incident. Nous
suspectons toutefois que cet \'echauffement ne soit pas r\'eel mais
plut\^ot d\^u \`a un probl\`eme de la sonde de temp\'erature.
\end{itemize}
D'autre part, la d\'etermination des constantes de temps suppose que
la modulation de flux soit carr\'ee. Or, nous savons que ce n'est pas
exactement le cas, elle est plut\^ot lisse et asym\'etrique entre les
deux plateaux chopper (cf figure~2 de
l'annexe~\ref{a:rapport_bp}). Ceci m\`ene encore une fois \`a une
sous-estimation de la fr\'equence de coupure.

Au bout du compte, nous estimons l'erreur des mesures pr\'esent\'ees
dans la figure~\ref{fig:calib_perflabo_tau_direct_bppolar} \`a environ
10-30~\%. Les mesures dynamiques r\'ealis\'ees \`a Saclay n'\'etant
manifestement pas tr\`es fiable, nous nous tournons vers les
densit\'es spectrales de bruit pour en extraire la fr\'equence de
coupure du filtre.

\subsection{Les mesures statiques}
\label{sec:calib_perflabo_tau_fourier}

Calculer la rapidit\'e de r\'eponse d'un bolom\`etre \`a partir de
mesures statiques peut para\^itre pour le moins surprenant, et
pourtant, dans l'espace de Fourier, nous avons acc\`es \`a
l'\'evolution fr\'equentielle du signal sans avoir \`a moduler le flux
incident. En r\'ealit\'e ce sont les bruits blancs g\'en\'er\'es par
le bolom\`etre qui jouent le r\^ole \og d'excitation \fg du signal \`a
toutes les fr\'equences, et nous pouvons alors en th\'eorie voir la
signature du filtre passe-bas dans la densit\'e spectrale de bruit
d'un bolom\`etre. Prenons par exemple le cas du bruit Johnson pour
lequel l'agitation thermique des \'electrons dans la r\'esistance
produit un bruit constant \`a toutes les fr\'equences~; les variations
de tension plus rapides que la constante de temps du bolom\`etre se
trouvent att\'enu\'ees, et nous pouvons interpr\'eter cela comme un
ralentissement des \'electrons les plus rapides. Il en va de m\^eme
pour le bruit thermique auquel cas le d\'eplacement des phonons
rapides est frein\'e par la constante de temps thermique, \cad par les
poutres qui connectent l'absorbeur au puits de chaleur. Nous allons
donc mod\'eliser le comportement des bolom\`etres dans l'espace de
Fourier pour pouvoir en extraire l'information utile, c'est-\`a-dire
la fr\'equence de coupure du filtre passe-bas.

\begin{figure}
  \begin{center}
      \includegraphics[width=0.7\textwidth,angle=0]{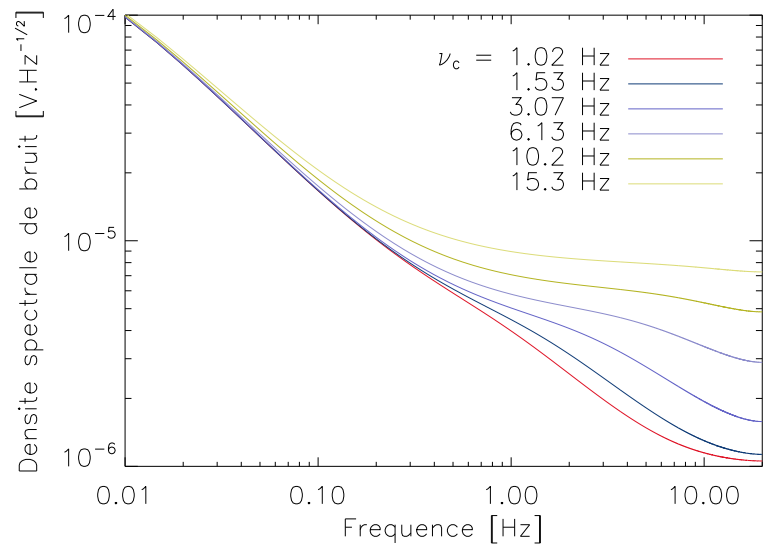}
  \end{center}
  \caption[\'Evolution des spectres simul\'es avec la fr\'equence de
  coupure]{\'Evolution des densit\'es spectrales de bruit simul\'ees
  pour diff\'erentes valeurs de la fr\'equence de coupure. Chaque
  spectre a \'et\'e simul\'e pour des constantes de temps \'electrique
  et thermique identiques, les autres param\`etres qui d\'efinissent
  le niveau de bruit et la remont\'ee basse fr\'equence sont les
  m\^emes pour toutes les courbes. Il semble que le niveau de bruit
  augmente avec la fr\'equence de coupure, ceci est d\^u au repliement
  du spectre~: plus la fr\'equence de coupure augmente, plus
  l'\'energie aux hautes fr\'equences se replie dans la bande
  observ\'ee.
  \label{fig:calib_perflabo_tau_fourier_model_multibp}}
\end{figure}

D'apr\`es l'inventaire des sources de bruit que nous avons dress\'e
dans la section~\ref{sec:intro_bolometrie_thermo_principe_bruit}, nous
devons faire la distinction entre le bruit g\'en\'er\'e par le
bolom\`etre lui-m\^eme et celui g\'en\'er\'e par son \'electronique de
lecture. La bande passante de l'\'electronique basse imp\'edance de
PACS est de 1.5~kHz, nous consid\'erons donc qu'il n'y a aucune
information spectrale au-del\`a de cette fr\'equence. Le bruit de
l'\'electronique est repr\'esent\'e par un bruit blanc de
0~\`a~1500~Hz sur lequel s'ajoute un bruit de la forme
$\beta_{e}/\nu^{\alpha_{e}}$ o\`u le param\`etre $\alpha_{e}$
d\'efinit la pente de la remont\'ee de bruit aux basses fr\'equences
du spectre de l'\'electronique et $\beta_{e}$ donne l'emplacement du
coude de remont\'ee. Le bruit du bolom\`etre est mod\'elis\'e de la
m\^eme fa\c{c}on, \cad un bruit blanc et une remont\'ee basse
fr\'equence de la forme $\beta_{b}/\nu^{\alpha_{b}}$, mais il est en
plus multipli\'e par un filtre passe-bas dont l'amplitude est donn\'ee
par
l'\'equation~(\ref{eq:fonction_transfert_electrothermic}). Puisqu'il
s'agit de bruits non-corr\'el\'es, les spectres de l'\'electronique et
du bolom\`etre sont ajout\'es quadratiquement dans la bande de
1500~Hz. D'autre part, ce spectre simul\'e doit \^etre compar\'e \`a
une densit\'e spectrale de bruit dont la fr\'equence de Nyquist est de
20~Hz, il faut donc le \og replier\footnote{Dans l'espace de Fourier,
l'\'energie pr\'esente dans le signal \`a des fr\'equences
sup\'erieures \`a celle de Nyquist ($1/(2\times t_{mesure})$) est
repli\'ee dans la bande observ\'ee \`a la fa\c{c}on d'un accord\'eon,
\cad que chaque tron\c{c}on de 20~Hz pr\'esent dans le spectre de
1500~Hz est co-additionn\'e, mais un tron\c{c}on sur deux doit \^etre
retourn\'e avant la co-addition.} \fg pour le rendre comparable avec
les spectres mesur\'es. Au total, nous avons besoin de 8~param\`etres
que nous consid\'erons ind\'ependants pour mod\'eliser les spectres de
bruit~: 3~pour le bruit de l'\'electronique, 3~pour le bruit du
bolom\`etre et 2~pour le filtre passe-bas ($\tau_{e}$ et
$\tau_{th}$). La
figure~\ref{fig:calib_perflabo_tau_fourier_model_multibp} montre
l'\'evolution des densit\'es spectrales de bruit obtenues \`a partir
de ce mod\`ele lorsque la fr\'equence de coupure augmente. Ces courbes
ont \'et\'e r\'ealis\'ees avec les m\^emes param\`etres de bruit,
seules les constantes de temps \'evoluent (avec $\tau_{e}=\tau_{th}$
dans le cas pr\'esent). Le fait d'augmenter la fr\'equence de coupure
$\nu_c$ a pour cons\'equence d'ouvrir la bande passante et d'augmenter
le niveau de bruit aux hautes fr\'equences.\\

\begin{figure}
  \begin{center}
    \begin{tabular}[t]{c}
      \includegraphics[width=0.7\textwidth,angle=0]{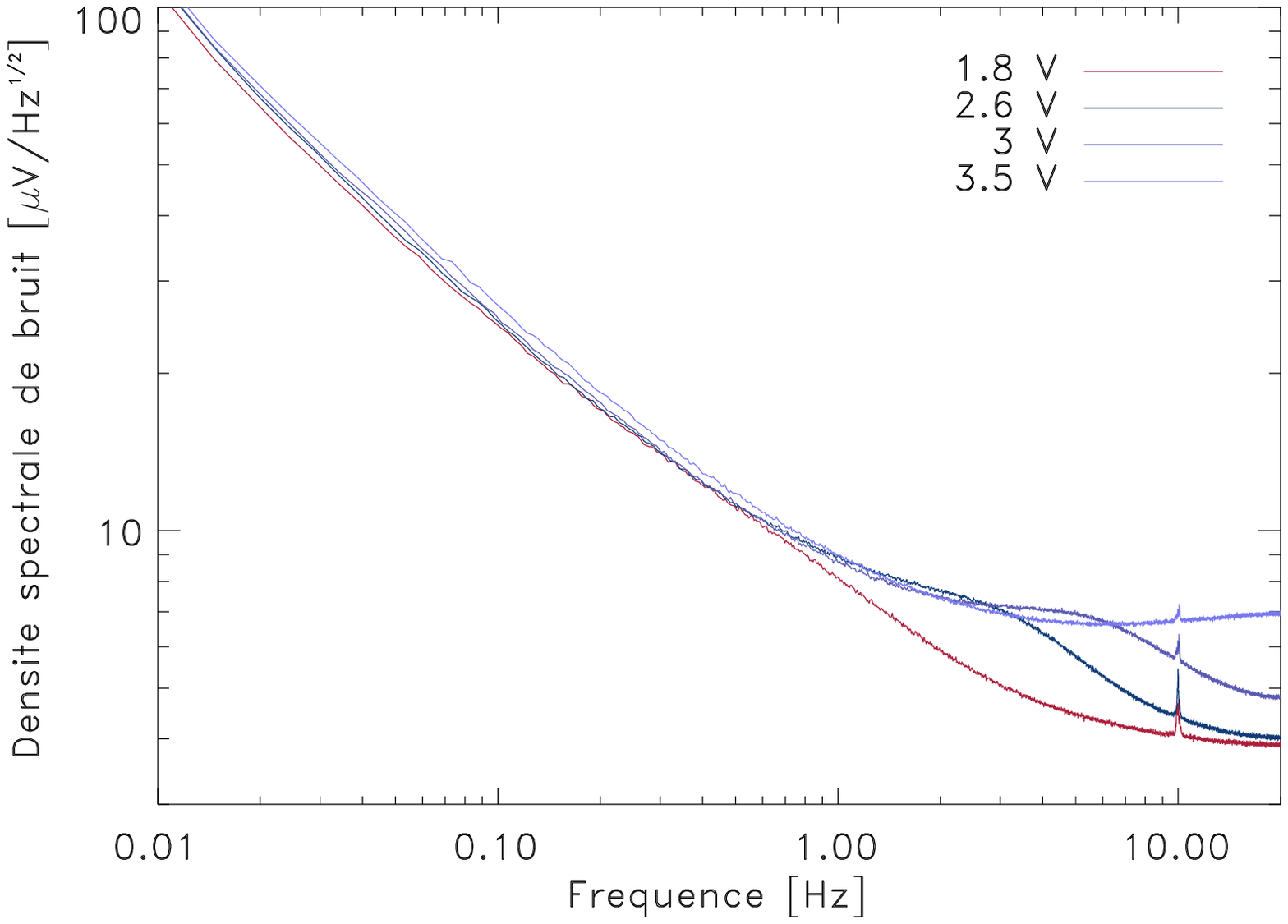}\\
\includegraphics[width=0.7\textwidth,angle=0]{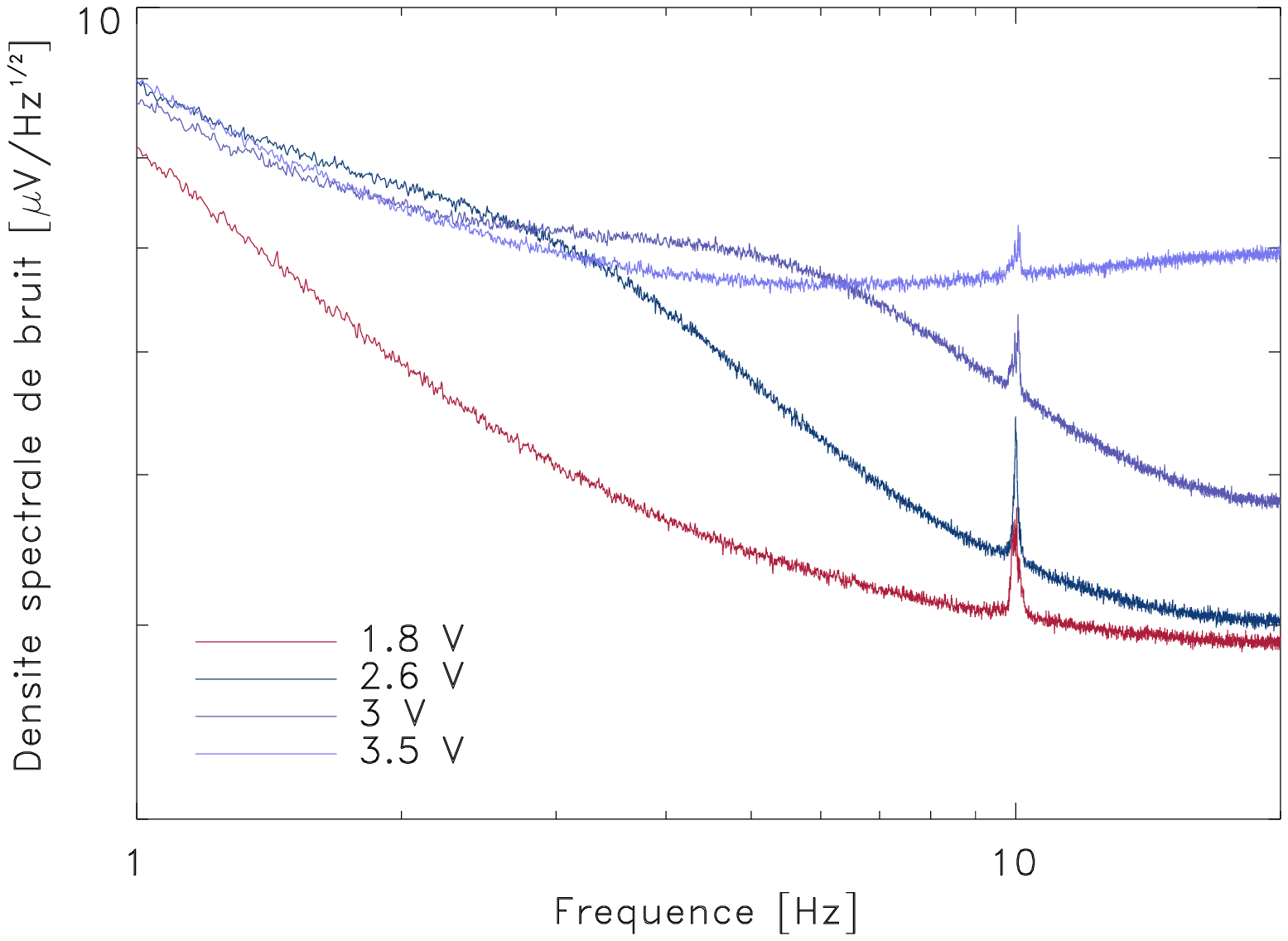}
    \end{tabular}
  \end{center}
  \caption[Mesures de constantes de temps \`a partir de densit\'es
  spectrales de bruit]{\'Evolution de la densit\'e spectrale de bruit
  mesur\'ee en fonction de la tension de polarisation. Ces spectres
  sont obtenus sur des mesures de 3~heures pour un flux incident de
  2~pW/pixel, ils sont de plus moyenn\'es sur un BFP entier pour
  r\'eduire les fluctuations statistiques. Les courbes du haut
  montrent que le bruit basse fr\'equence est ind\'ependant de la
  tension. Le graphe du bas met en \'evidence le d\'eplacement de la
  fr\'equence de coupure du filtre passe-bas des bolom\`etres. Plus la
  fr\'equence de polarisation augmente, plus les d\'etecteurs sont
  rapides.
  \label{fig:calib_perflabo_tau_fourier_spec4polar}}
\end{figure}

D'autre part, nous avons mesur\'e des densit\'es spectrales de bruit
pour diff\'erentes tensions de polarisation dans le but de mettre en
\'evidence l'\'evolution de la constante de temps des bolom\`etres. La
figure~\ref{fig:calib_perflabo_tau_fourier_spec4polar} pr\'esente nos
r\'esultats pour des tensions comprises entre 1.8~et 3.5~V sur le BFP
bleu. Pour obtenir ces spectres, nous avons pris des
sous-\'echantillons de 4~minutes extraits de mesures de 3~heures, nous
avons co-additionn\'e les 45~spectres calcul\'es, puis nous avons
moyenn\'e les spectres des pixels fonctionnels d'une m\^eme matrice du
BFP bleu. Les spectres produits poss\`edent donc des fluctuations
statistiques tr\`es faibles qui laissent apercevoir le d\'eplacement
de la fr\'equence de coupure vers les hautes fr\'equences lorsque la
tension de polarisation augmente (cf
figure~\ref{fig:calib_perflabo_tau_fourier_spec4polar}). Pour
d\'eterminer la valeur de la fr\'equence de coupure de chacun de ces
spectres, nous les avons ajust\'e avec le mod\`ele pr\'esent\'e
pr\'ec\'edemment gr\^ace \`a la m\'ethode du Simplex
\shortcite{nedler}. Nous n'allons pas d\'ecrire le principe de
fonctionnement du Simplex, le lecteur pourra se r\'ef\'erer \`a
l'article de \shortciteN{caceci} pour plus de d\'etails, mais notez
toutefois que le Simplex permet de converger plus rapidement qu'une
simple minimisation de $\chi^2$ qui, elle, s'est av\'er\'ee trop
gourmande en temps de calcul dans notre cas d'\'etude. Le
tableau~\ref{tab:calib_perflabo_tau_fourier_simplex} r\'esume les
r\'esultats d'ajustement obtenus pour les quatres spectres
pr\'esent\'es dans la
figure~\ref{fig:calib_perflabo_tau_fourier_spec4polar}.

\begin{table}
  \begin{center}
    \setlength\extrarowheight{4pt}
    \begin{tabular}[]{p{0.cm}>{\normalsize}p{7.5cm}>{\centering}p{1.2cm}>{\centering}p{1.2cm}>{\centering}p{1.2cm}>{\centering}p{1.2cm}p{0.cm}}
      \toprule
      &\textbf{Tension de polarisation [V]}   & \textbf{1.8} & \textbf{2.6} & \textbf{3.0} & \textbf{3.5} & \\
      \hline \hline
      &Bruit blanc \'electronique [$\mu$V/$\sqrt{\mbox{Hz}}$]   & 0.013 & 0.011 & 0.013 & 0.028 & \\
      &Bruit blanc bolom\`etre [$\mu$V/$\sqrt{\mbox{Hz}}$]   & 0.87 & 3.92 & 3.63 & 0.89 &\\
      &$\beta_e$~: Coude 1/f \'electronique [$\mu$V/$\sqrt{\mbox{Hz}}$]   & 4.36 & 2.39 & 2.35 & 3.38 &\\
      &$\beta_b$~: Coude 1/f bolom\`etre [$\mu$V/$\sqrt{\mbox{Hz}}$]   & 3.02 & 1.85 & 1.62 & 1.73 &\\
      &$\alpha_e$~: Exposant 1/f \'electronique   & 0.47 & 0.78 & 0.81 & 0.87 &\\
      &$\alpha_b$~: Exposant 1/f bolom\`etre   & 0.74 & 0.79 & 0.84 & 0.81 &\\
      &$\tau_e$~: Constante de temps \'electronique [ms]  & 177 & 42 & 10.6 & 1.28 &\\
      &$\tau_{th}$~: Constante de temps bolom\`etre  [ms] & 21 & 1.9 & 11.0 & 4.79 &\\
      \hline
      &$\mathbf{\nu_{c}}$~: \textbf{Fr\'equence de coupure globale [Hz]} & \textbf{0.88} & \textbf{3.78} & \textbf{9.49} & \textbf{31.2} &\\
      \bottomrule
    \end{tabular}
  \caption[Param\`etres d'ajustement des densit\'es spectrales de
  bruit]{R\'esultats des ajustements de densit\'es spectrales de bruit
  en fonction de la tension de polarisation. Nous donnons les
  8~param\`etres n\'ecessaires au mod\`ele pour reproduire les
  spectres, anisi que la valeur calcul\'ee de la fr\'equence de
  coupure globale des bolom\`etres. Voir le texte pour le d\'etail des
  ajustements et l'interpr\'etation de ces chiffres.
  \label{tab:calib_perflabo_tau_fourier_simplex}}
  \end{center}
\end{table}

Les fr\'equences de coupure calcul\'ees \`a partir des valeurs de
$\tau_e$ et $\tau_{th}$ semblent en accord avec les spectres de la
figure~\ref{fig:calib_perflabo_tau_fourier_spec4polar}. Toutefois, il
semble que les 8~param\`etres soient d'une certainement mani\`ere
d\'eg\'en\'er\'es. Par exemple, une baisse de niveau de bruit blanc du
bolom\`etre pourrait \^etre compens\'ee par une l\'eg\`ere hausse du
niveau de bruit blanc de l'\'electronique qui, lui, est repli\'e
75~fois (1500~Hz repli\'e dans une bande de 20~Hz).

Nous avons de plus identifier une raison pour laquelle le mod\`ele ne
peut reproduire exactement les spectres de bruit~: les 8~param\`etres
ne sont pas ind\'ependants comme nous l'avons suppos\'e. Il faudrait
donc affiner ce mod\`ele simpliste en y injectant les relations
physiques qu'il existe entre le niveau de bruit et la valeur de la
constante de temps via les r\'esistances thermique et \'electrique,
par exemple, pour mieux contraindre les ajustements et remonter \`a
des valeurs de bruit plus fiables. Remarquez \'egalement que notre
approche ne peut \^etre appliqu\'ee que pour des spectres dont les
fluctuations statistiques sont tr\`es faibles. En effet il est
difficile d'ajuster des spectres de pixels individuels comme celui
pr\'esent\'e dans la
figure~\ref{fig:calib_perflabo_sensibilite_description_spectre}, cette
approche ne permet donc pas d'obtenir des cartes de constantes de
temps ou de niveaux de bruit comme c'est le cas pour les mesures
dynamiques.

Les mesures de constantes de temps par ajustement de densit\'es
spectrales de bruit repr\'esentent une approche originale et
potentiellement int\'eressante pour extraire des informations
quantitatives globales des matrices de bolom\`etres. Toutefois, dans
son \'etat actuel, le mod\`ele nous permet seulement de confirmer
l'\'evolution de la fr\'equence de coupure avec la tension de
polarisation.

\section{Analyse compar\'ee DDCS/Direct}
\label{sec:calib_perflabo_compare}

Dans les sections pr\'ec\'edentes nous avons pr\'esent\'e le
comportement des matrices de bolom\`etres de fa\c{c}on relativement
g\'en\'erique. Nous proposons maintenant une analyse plus sp\'ecifique
des performances des matrices en comparant les r\'esultats de la
proc\'edure d'\'etalonnage dans les deux modes de lecture ainsi que
pour les deux plans focaux bleu et rouge de PACS.

\subsection{Le point sur les diff\'erents s\'equenceurs}
\label{sec:calib_perflabo_compare_sequenceur}

\begin{figure}
  \begin{center}
      \includegraphics[width=0.9\textwidth,angle=0]{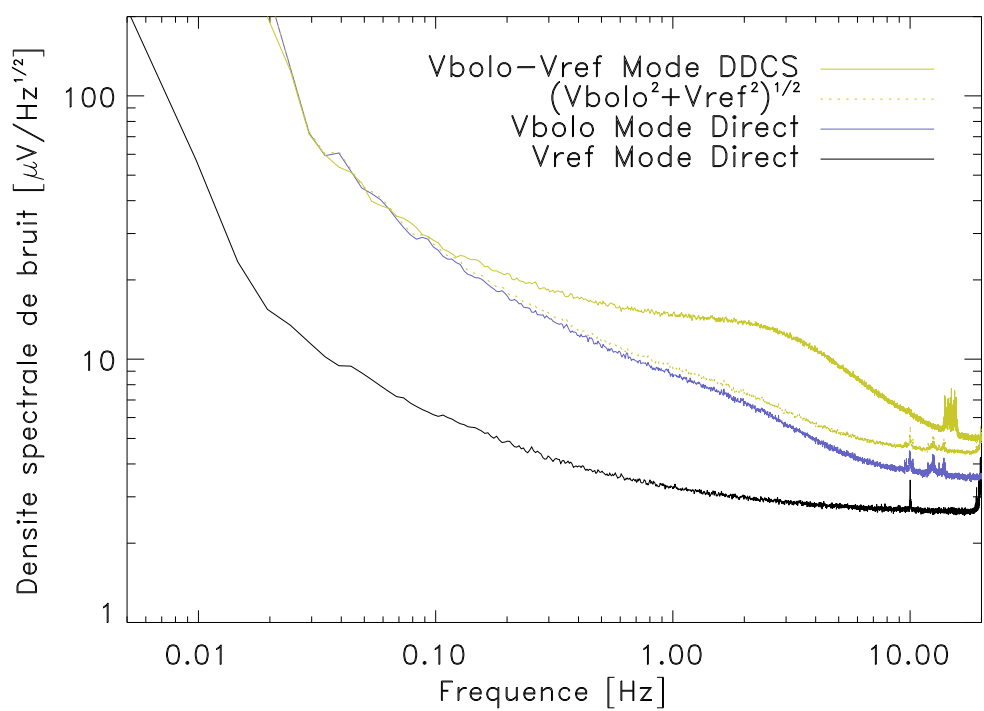}
  \end{center}
  \caption[Densit\'e spectrale de bruit des trois s\'equenceurs
  principaux]{Les trois spectres en trait plein ont \'et\'e calcul\'es
  \`a partir de mesures statiques de 30~minutes dans trois
  s\'equenceurs diff\'erents. Chaque spectre est la moyenne spatiale
  des 256~spectres d'une m\^eme matrice. La courbe du bas repr\'esente
  la densit\'e spectrale de bruit mesur\'ee en mode direct en
  \'echantillonnant seulement la tension $V_{ref}$, c'est le spectre
  de l'\'electronique de lecture. La courbe bleue correspond au
  s\'equenceur que nous appelons mode Direct, seul le signal
  bolom\'etrique $V_{bolo}$ est \'echantillonn\'e. La courbe
  intitul\'ee $\left(V_{bolo}^2+V_{ref}^2\right)^{1/2}$ repr\'esente
  la somme quadratique des contributions $V_{ref}$ et $V_{bolo}$,
  c'est le spectre \og th\'eorique \fg que nous devrions obtenir aux
  hautes fr\'equences en mode de lecture DDCS. La courbe du haut
  montre la densit\'e spectrale de bruit effectivement mesur\'ee en
  mode DDCS.}
  \label{fig:calib_perflabo_compare_seq}
\end{figure}

Comme nous l'avons vu dans la
section~\ref{sec:detect_bolocea_elec_lecture}, la mani\`ere dont le
signal bolom\'etrique est achemin\'e vers BOLC est d\'etermin\'ee par
le s\'equenceur. En mode direct, le PEL de chaque pixel est connect\'e
en permanence au pont bolom\'etrique. En mode DDCS, le PEL d'un m\^eme
pixel voit son potentiel \'electrique alterner entre le point milieu
et la tension de r\'ef\'erence $V_{ref}$ (cf
figure~\ref{fig:detect_bolocea_elec_lecture_seq_schemapix})~; BOLC
produit ensuite la diff\'erence des deux signaux. Il est \'egalement
possible de programmer des s\'equenceurs d\'eriv\'es des modes direct
et DDCS pour \'etudier le comportement du circuit de
lecture. L'annexe~\ref{a:seq} montre les densit\'es spectrales de
bruit que nous avons mesur\'ees pour 9~ s\'equenceurs diff\'erents. De
ces 9~spectres, nous en exploitons 3~qui sont particuli\`erement
pertinents et r\'ev\'elateurs des performances des matrices~: ce sont
les spectres de bruit en mode DDCS, en mode direct sur $V_{bolo}$ et
en mode direct\footnote{Il est en effet possible de bloquer les
transistors $V_{DECX}$ et $CKRL$ de mani\`ere \`a \'echantilloner la
tension de r\'ef\'erence uniquement~; c'est d'ailleurs ce s\'equenceur
que nous avons utilis\'e pour \'etalonner l'\'electronique de lecture
(cf section~\ref{sec:calib_procedure_vrlvhb}).} sur $V_{ref}$. Nous
pr\'esentons la densit\'e spectrale de bruit de ces trois
s\'equenceurs dans la figure~\ref{fig:calib_perflabo_compare_seq}.\\

En mode direct sur $V_{ref}$, nous mesurons en r\'ealit\'e la
densit\'e spectrale de bruit de toute la cha\^ine \'electronique. Le
spectre pr\'esente une remont\'ee basse fr\'equence en-dessous de 1~Hz
due aux variations de gain et d'offset des transistors. Le niveau de
bruit dans le r\'egime de bruit blanc est de l'ordre de
$\sim$3~$\mu$V/$\sqrt{\mbox{Hz}}$, ce qui est en parfait accord avec
les mesures de bruit \emph{r.m.s.} de la
section~\ref{sec:calib_procedure_vrlvhb} (3~$\mu$V/$\sqrt{\mbox{Hz}}$
dans une bande passante de 40~Hz donne environ 19~$\mu$V en
\emph{r.m.s.} par le th\'eor\`eme de Parseval). En mode direct sur
$V_{bolo}$, nous mesurons la densit\'e spectrale de bruit du pont
bolom\'etrique et de l'\'electronique de lecture. Le niveau de bruit
est plus \'elev\'e \`a cause de la contribution des thermistances
(bruit Johnson et bruit de phonon), et le bruit basse fr\'equence est
toujours pr\'esent.

Rappelons que le mode DDCS est une fonctionnalit\'e qui a \'et\'e
introduite dans le circuit de lecture pour r\'eduire le bruit en 1/f
de l'\'electronique froide (cf
section~\ref{sec:detect_bolocea_elec_lecture}). Nous nous attendons
par cons\'equent \`a une diminution du coude de remont\'ee basse
fr\'equence et \`a un niveau de bruit blanc correspondant \`a la somme
quadratique des bruits sur $V_{bolo}$ et sur $V_{ref}$. Toutefois, la
figure~\ref{fig:calib_perflabo_compare_seq} montre que le mode DDCS
pr\'esente un exc\`es de bruit aux hautes fr\'equences (par rapport au
spectre en pointill\'e sur la figure), et que le bruit en 1/f est
sensiblement le m\^eme en mode direct et en mode DDCS. Cela signifie
d'une part qu'il existe une source de bruit suppl\'ementaire
responsable de l'exc\`es aux hautes fr\'equences, et d'autre part que
le bruit en 1/f n'est pas g\'en\'er\'e par l'\'electronique de lecture
mais plut\^ot par les bolom\`etres eux-m\^eme. La modulation
\'electrique du mode DDCS filtre en effet les d\'erives de
l'\'electronique de lecture jusqu'\`a une fr\'equence de 1280~Hz en
r\'ealisant la diff\'erence ($V_{ref}-V_{bolo}$), et les d\'erives
basse fr\'equence qui persistent proviennent forc\'ement des
composants qui se trouvent en amont des transistors $V_{DECX}$ et
$CKRL$, \cad des ponts bolom\'etriques. La
figure~\ref{fig:calib_perflabo_compare_seq} montre que le mode DDCS
est plus \og bruyant \fg que le mode direct \`a toutes les
fr\'equences, et que son int\'er\^et en terme de stabilisation du
signal est nul. Notez toutefois que le mode DDCS est potentiellement
int\'eressant car il permet de filtrer efficacement les perturbations
\'electromagn\'etiques qui seraient capt\'ees au niveau du circuit de
lecture. Par exemple, l'environnement de travail dans le laboratoire
g\'en\`ere un bruit \`a 50~Hz relativement important qui se traduit
par un pic d'\'energie \`a 10~Hz dans les spectres de bruit que nous
mesurons. Nous retrouvons effectivement ce pic en mode direct sur la
figure~\ref{fig:calib_perflabo_compare_seq}, alors qu'il est
totalement absent en mode DDCS.



\subsection{Mesures de bruit}
\label{sec:calib_perflabo_compare_bruit}

\begin{figure}
  \begin{center}
    \begin{tabular}{c||c}
      Bleu & Rouge \\
      \includegraphics[width=0.48\textwidth,angle=0]{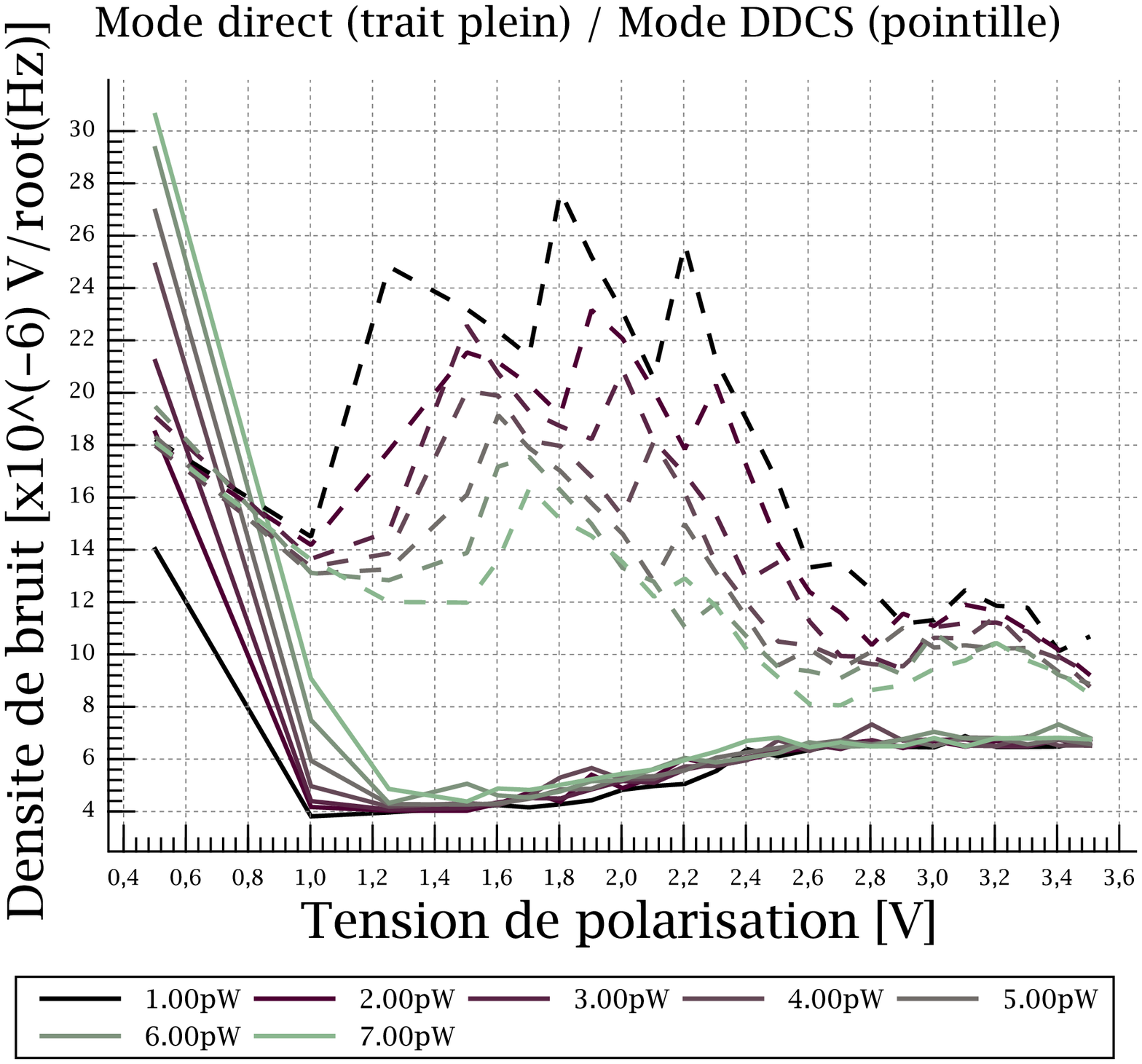} & \includegraphics[width=0.48\textwidth,angle=0]{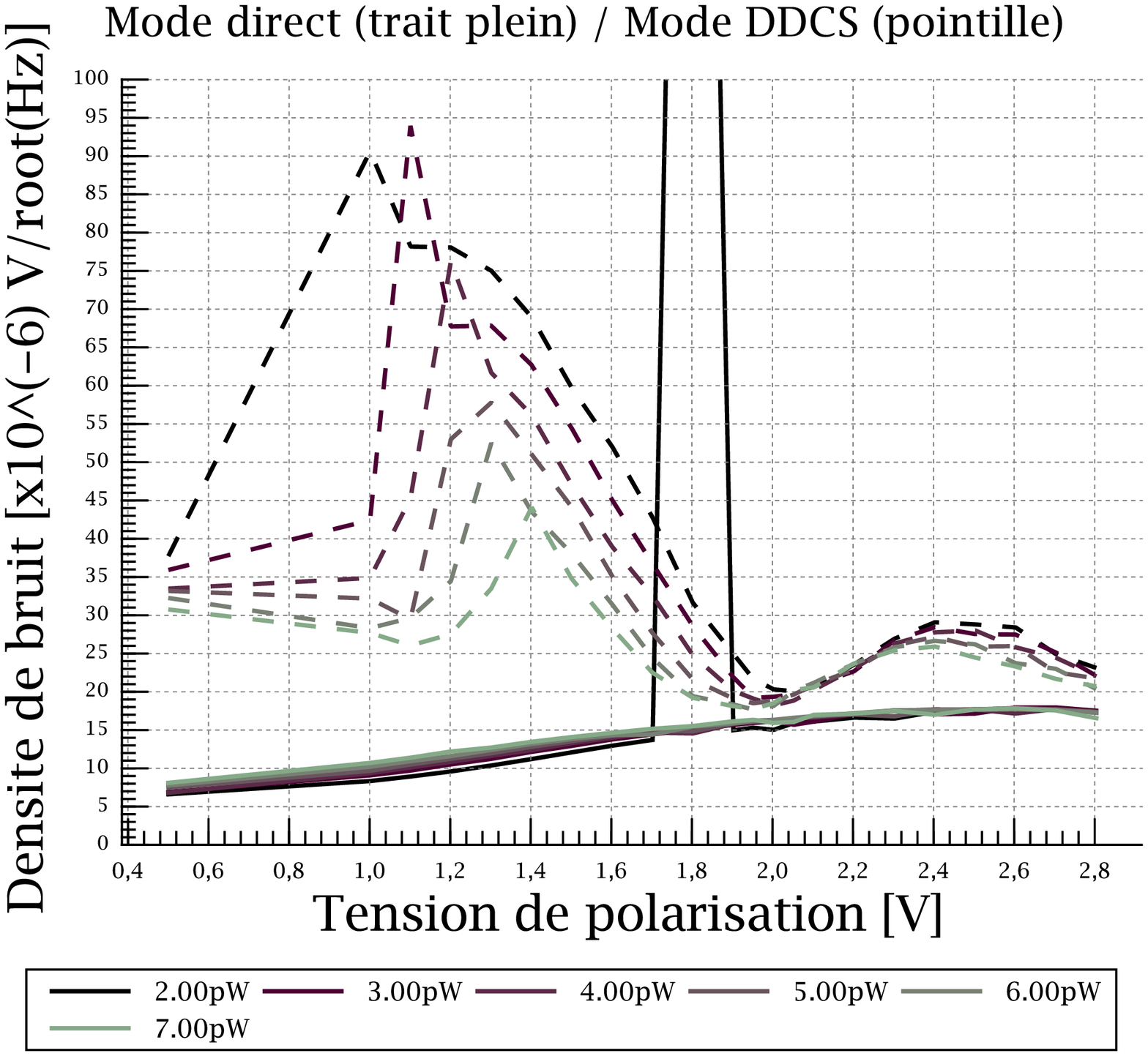} 
    \end{tabular}
  \end{center}
  \caption[Comparaison du bruit en mode DDCS et direct pour les BFP
  bleu et rouge]{Comparaison du niveau de bruit mesur\'e \`a 3~Hz en
  mode DDCS et direct pour les BFP bleu et rouge. \'Evolution du bruit
  moyenn\'e sur une matrice enti\`ere en fonction de la tension de
  polarisation et du flux incident. En mode DDCS, l'\'electronique de
  lecture injecte des charges parasites qui se traduisent par un
  exc\`es de bruit aux basses tensions de polarisation lorsque la
  constante de temps des bolom\`etres est trop courte par rapport \`a
  la fr\'equence d'\'echantillonnage. Le niveau de bruit mesur\'e dans
  les deux modes devient comparable pour des tensions sup\'erieures
  \`a 2.5~V (2~V) sur le BFP bleu (rouge), \cad lorsque les
  bolom\`etres sont suffisamment rapides pour \'ecouler les charges
  parasites avant l'\'echantillonnage du signal.
  \label{fig:calib_perflabo_compare_noise}}
\end{figure}

Pour trouver l'origine de l'exc\`es de bruit du mode DDCS que nous
avons mis en \'evidence dans la
figure~\ref{fig:calib_perflabo_compare_seq}, nous exploitons la
richesse des donn\'ees r\'ecolt\'ees lors de la proc\'edure
d'\'etalonnage comme outil diagnostique. La
figure~\ref{fig:calib_perflabo_compare_noise} pr\'esente l'\'evolution
du bruit en fonction de la tension de polarisation et du flux incident
en mode direct et en mode DDCS pour les deux BFP. Le niveau de bruit
indiqu\'e sur l'axe des ordonn\'ees correspond au bruit mesur\'e \`a
3~Hz dans une bande passante de 1~Hz sur des spectres similaires \`a
ceux de la figure~\ref{fig:calib_perflabo_compare_seq}. Nous avons
d\'ej\`a analys\'e les courbes de bruit en mode direct dans la
section~\ref{sec:calib_perflabo_sensibilite_bruit}, mais l'\'evolution
du bruit en mode DDCS est tr\`es diff\'erente. Nous trouvons un fort
exc\`es de bruit aux basses tensions de polarisation quelque soit la
couleur du BFP~; alors qu'au del\`a de $\sim2.5$~V pour le bleu et de
$\sim1.8$~V pour le rouge, le niveau de bruit en mode DDCS devient
coh\'erent avec celui en mode direct (\cad somme quadratique des
contributions $V_{ref}$ et $V_{bolo}$).

Nous attribuons cette diff\'erence de comportement \`a la pr\'esence
de charges parasites au niveau du circuit de lecture ainsi qu'\`a la
tr\`es haute imp\'edance des bolom\`etres. Nous pensons en effet qu'un
nombre al\'eatoire de charges est inject\'e dans les ponts
bolom\'etriques \`a chaque commutation des transistors $V_{DECX}$ et
$CKRL$ (cf sch\'ema \'electronique,
figure~\ref{fig:detect_bolocea_elec_froide_principe}), et que le temps
n\'ecessaire \`a ces charges pour s'\'ecouler dans les thermistances
d\'epend de l'imp\'edance du circuit et donc de la tension de
polarisation. Aux basses tensions, lorsque les bolom\`etres sont
tr\`es r\'esistifs et donc relativement \og lents \fg, le potentiel du
PEL n'est pas encore \'etabli lors de la conversion de $V_{bolo}$ par
BOLC (cf figure~\ref{fig:detect_bolocea_elec_lecture_seq_schemapix})
de sorte que le signal d\'epend du nombre de charges inject\'ees,
d'o\`u l'exc\`es de bruit pr\'esent aux basses tensions sur la
figure~\ref{fig:calib_perflabo_compare_noise}. Aux fortes tensions,
les bolom\`etres sont moins imp\'edants, les charges s'\'ecoulent plus
rapidement et le signal ne d\'epend plus du nombre de charges
inject\'ees au moment de la conversion par BOLC, l'\'electronique
froide ne perturbe plus la lecture du signal.

Notez par ailleurs qu'un autre ph\'enom\`ene tend \`a r\'eduire les
perturbations lorsque la tension de polarisation augmente. En effet,
pour un nombre donn\'e de charges inject\'ees, la loi d'Ohm implique
que les perturbations en tension diminuent \`a mesure que la
r\'esistance \'equivalente du pont bolom\'etrique diminue. Cependant,
nous ne pouvons pas quantifier ce ph\'enom\`ene sans une
mod\'elisation compl\`ete du syst\`eme, \cad l'injection de charges et
l'\'evolution de la r\'esistance \'equivalente avec la tension de
polarisation.

Notez \'egalement que la tension \`a laquelle le bruit en mode DDCS
rejoint celui en mode direct est inf\'erieure pour le BFP rouge
(1.8~V) que pour le BFP bleu (2.5~V) car il est moins imp\'edant par
construction (cf section~\ref{sec:detect_bolocea_fabrication_thermo}),
et donc plus rapide.

\subsection{Mesures de r\'eponse}
\label{sec:calib_perflabo_compare_reponse}

\begin{figure}
  \begin{center}
    \begin{tabular}{c||c}
      Bleu & Rouge \\
      \includegraphics[width=0.48\textwidth,angle=0]{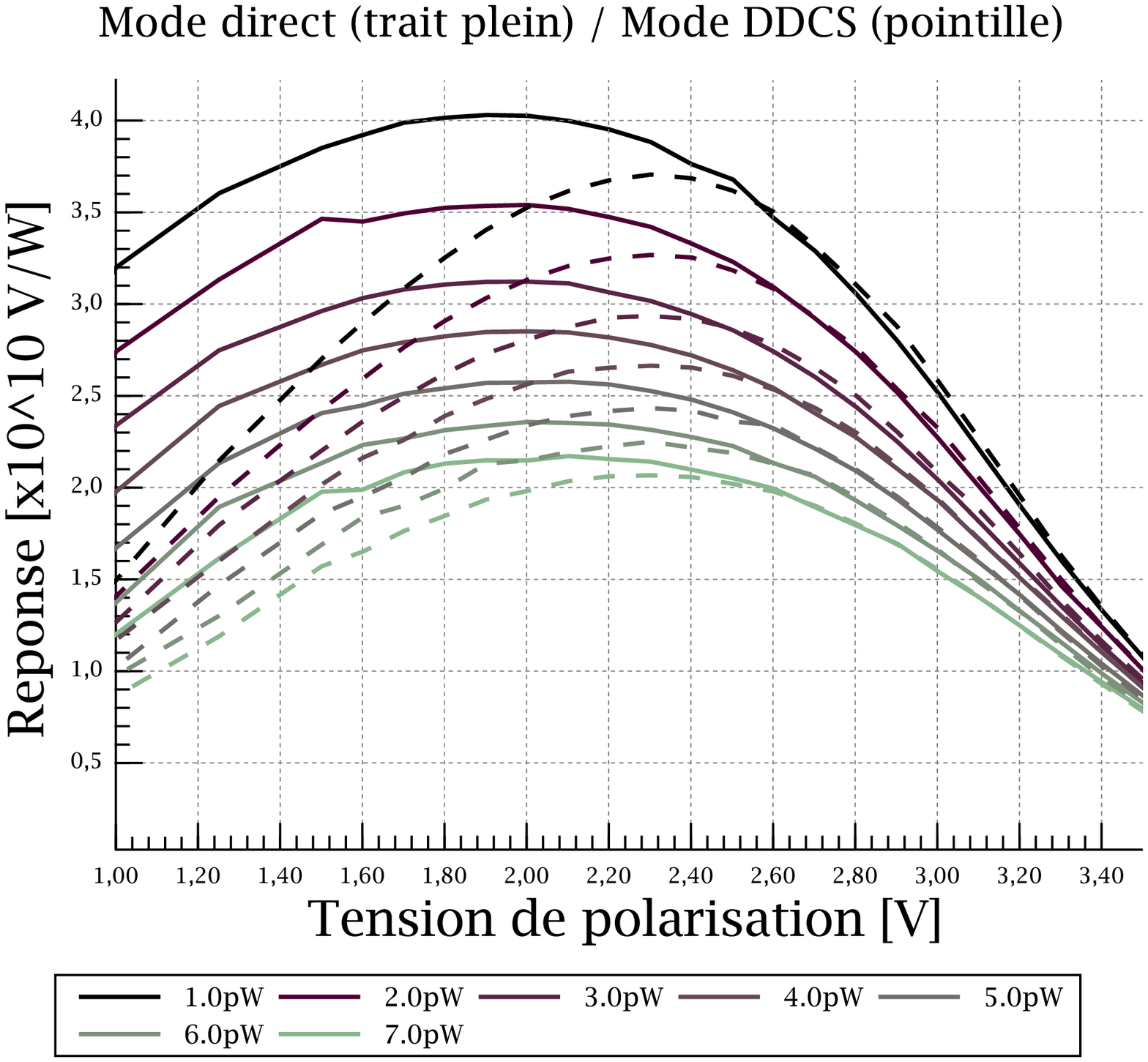} & \includegraphics[width=0.48\textwidth,angle=0]{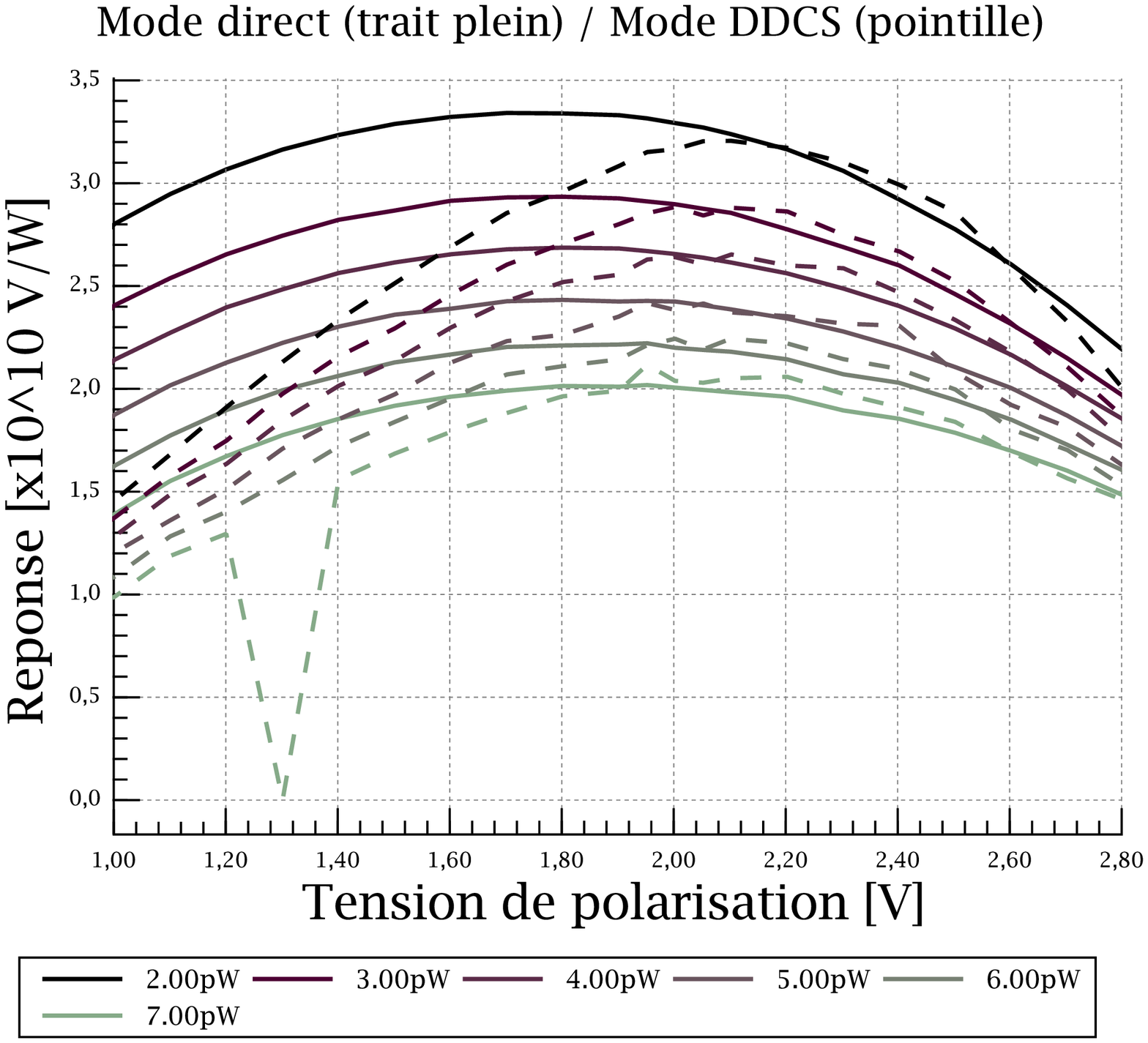} 
    \end{tabular}
  \end{center}
  \caption[Comparaison de la r\'eponse en mode DDCS et direct pour les
  BFP bleu et rouge]{Comparaison de la r\'eponse dynamique en mode
  DDCS et direct pour les BFP bleu et rouge. \'Evolution de la
  r\'eponse moyenn\'ee sur une matrice enti\`ere en fonction de la
  tension de polarisation et du flux incident. En mode DDCS, les
  r\'eponses sont sous-estim\'ees aux basses tensions car le signal
  est \'echantillonn\'e lors d'un r\'egime transitoire entre $V_{ref}$
  et $V_{bolo}$. Comme pour la
  figure~\ref{fig:calib_perflabo_compare_noise}, les deux modes
  pr\'esentent des performances identiques pour des tensions
  sup\'erieures \`a 2.5~V (2~V) sur le BFP bleu (rouge).
  \label{fig:calib_perflabo_compare_resp}}
\end{figure}

La r\'eponse des bolom\`etres est \'egalement affect\'ee par le mode
de lecture DDCS. La figure~\ref{fig:calib_perflabo_compare_resp}
montre l'\'evolution de la r\'eponse des BFP bleu et rouge en fonction
de la tension de polarisation et du flux incident. Les courbes de
r\'eponse en mode DDCS pr\'esentent les m\^emes caract\'eristiques
g\'en\'erales que les courbes en mode direct~: la r\'eponse augmente
avec la tension de polarisation jusqu'\`a ce que l'imp\'edance des
bolom\`etres chute \`a cause des effets de champ et de la dissipation
Joule (cf section~\ref{sec:calib_perflabo_sensibilite_reponse}). Les
deux familles de courbes sont parfaitement superpos\'ees pour les
fortes tensions de polarisation alors que la r\'eponse en mode DDCS
est syst\'ematiquement sous-estim\'ee aux basses tensions. Nous
pensons que la constante de temps des bolom\`etres est encore une fois
responsable des diff\'erences de r\'eponse direct/DDCS.

\begin{figure}
  \begin{center}
      \includegraphics[width=0.9\textwidth,angle=0]{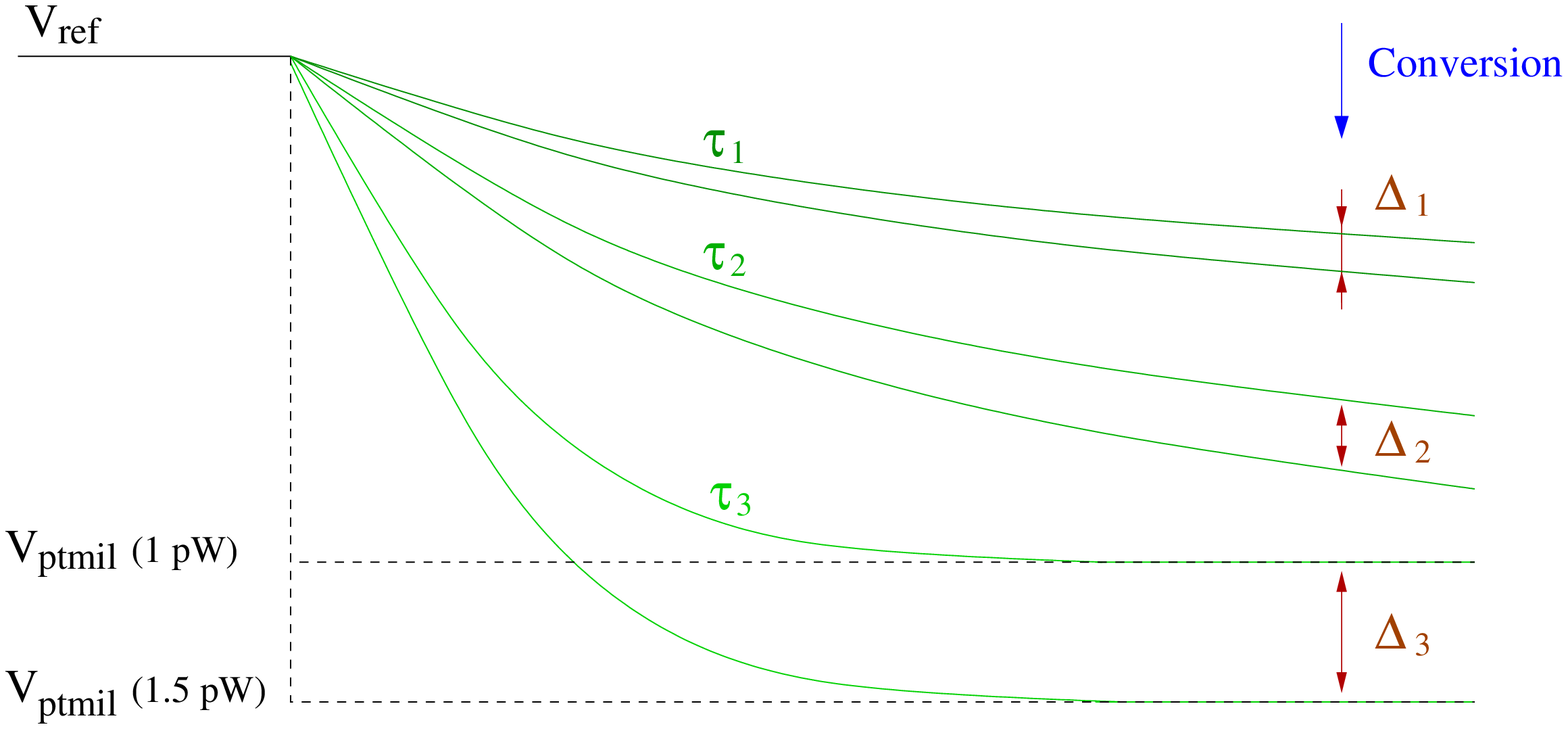}
  \end{center}
  \caption[Sous-estimation de la r\'eponse en mode DDCS]{Sch\'ema
  montrant la sous-estimation de la r\'eponse en mode DDCS pour les
  longues constantes de temps. Le potentiel \'electrique du PEL passe
  de $V_{ref}$ \`a $V_{ptmil}$ avec un temps caract\'eristique $\tau$
  (cf
  figure~\ref{fig:detect_bolocea_elec_lecture_seq_schemapix}). Chaque
  couple de courbes correspond \`a une tension de polarisation
  donn\'ee, \cad \`a une constante de temps donn\'ee o\`u
  $\tau_1>\tau_2>\tau_3$. La courbe sup\'erieure de chaque couple tend
  vers une premi\`ere valeur de point milieu $V_{ptmil}(\mbox{1pW})$,
  l'autre vers $V_{ptmil}(\mbox{1.5pW})$. Au moment de la conversion
  du signal par BOLC, la variation de signal
  $\Delta=|V_{ptmil}(\mbox{1pW})-V_{ptmil}(\mbox{1.5pW})|$
  correspondant \`a la modulation de flux donne
  $\Delta_1<\Delta_2<\Delta_3$. La r\'eponse est sous-estim\'ee
  lorsque la constante de temps est plus longue que la p\'eriode
  d'\'echantillonnage.
  \label{fig:calib_perflabo_compare_resp_PEL}}
\end{figure}

En mode direct, le PEL d'un pixel reste au potentiel du point milieu
en permanence, et tant que la p\'eriode de modulation du flux incident
est inf\'erieure \`a la constante de temps des bolom\`etres, la mesure
de r\'eponse n'est pas biais\'ee. En mode DDCS, le potentiel
\'electrique du PEL alterne entre $V_{ref}$ et $V_{bolo}$ en suivant
des exponentielles d\'ecroissantes (cf sch\'ema de la
figure~\ref{fig:detect_bolocea_elec_lecture_seq_schemapix}).
Si la constante de temps \'electrique des bolom\`etres est plus longue
que la p\'eriode d'\'echantillonnage du signal alors le point milieu
est converti avant qu'il ait atteint sa valeur asymptotique, \cad la
valeur du point milieu qui correspond au flux incident. Dans ce cas la
r\'eponse est sous-estim\'ee. Par contre, aux fortes tensions de
polarisation, lorsque les bolom\`etres sont suffisamment rapides, la
conversion du signal a lieu apr\`es le r\'egime transitoire et la
vraie r\'eponse est alors mesur\'ee. La
figure~\ref{fig:calib_perflabo_compare_resp_PEL} montre l'\'evolution
du PEL d'un pixel plus ou moins rapide et illustre cet effet de
sous-estimation de la r\'eponse.

Comme pour les mesures de bruit pr\'esent\'ees dans la section
pr\'ec\'edente, le r\'egime o\`u les courbes de r\'eponse en mode DDCS
et direct deviennent coh\'erentes se trouve aux alentours de 2.5~V sur
le BFP bleu et de 2.0~V sur le rouge. Notez que cette tension varie
l\'eg\`erement avec la valeur du flux incident~; en effet,
l'imp\'edance des thermom\`etres diminue \`a mesure que le flux
augmente, les bolom\`etres sont alors l\'eg\`erement plus rapides pour
les forts flux.

\subsection{Sensibilit\'e}
\label{sec:calib_perflabo_compare_NEP}

\begin{figure}
  \begin{center}
    \begin{tabular}{c||c}
      Bleu & Rouge \\
      \includegraphics[width=0.48\textwidth,angle=0]{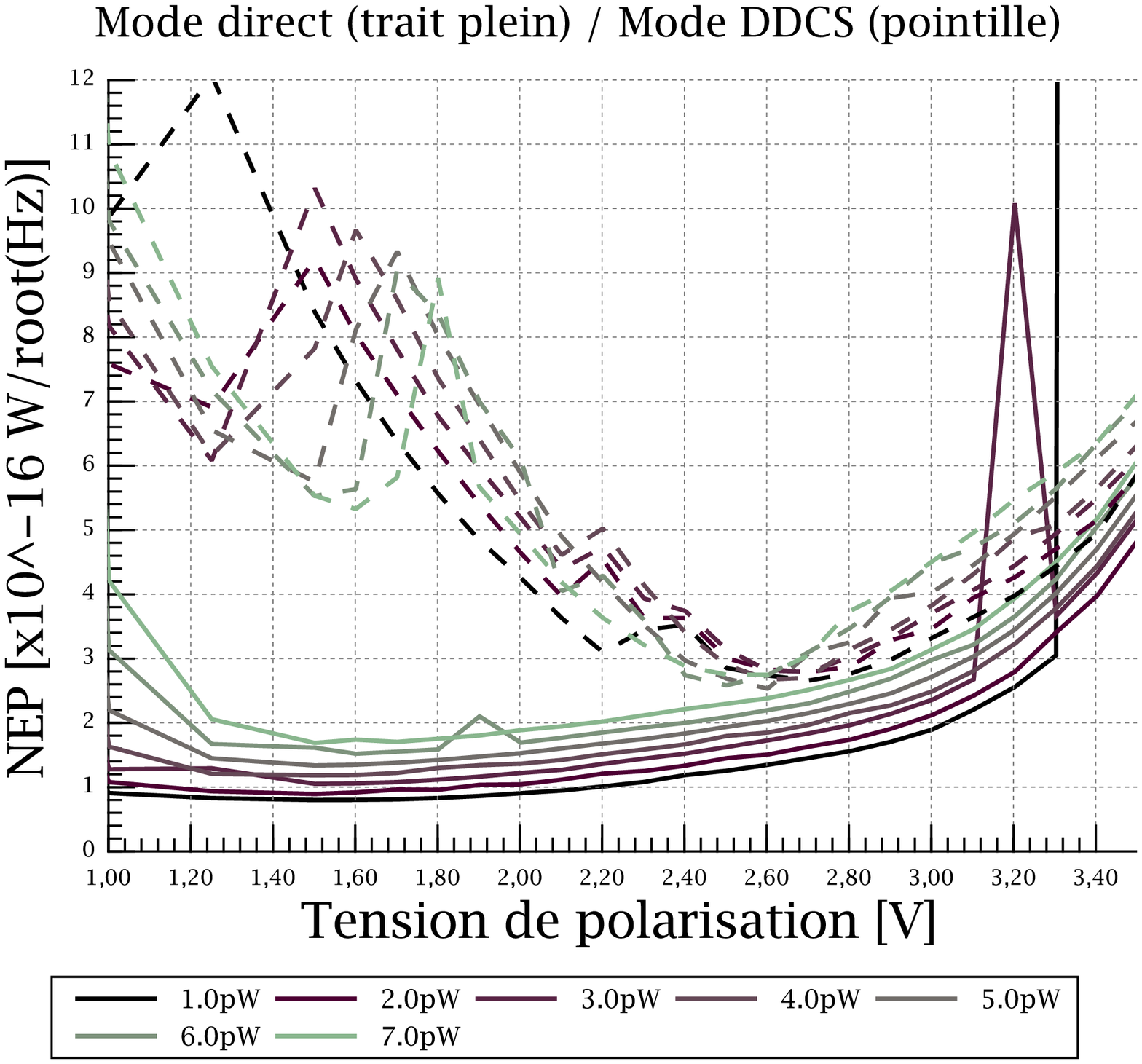} & \includegraphics[width=0.48\textwidth,angle=0]{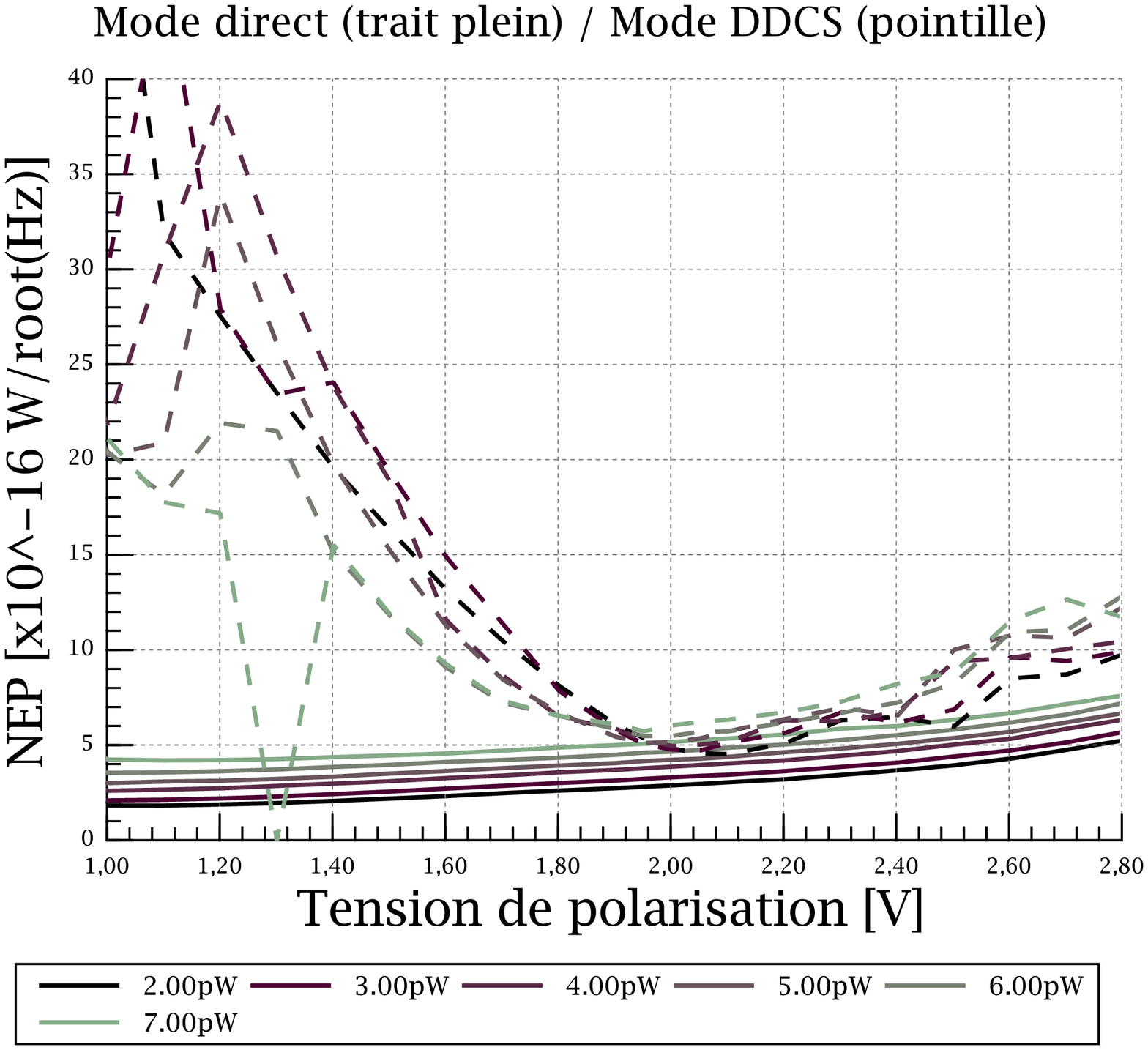} 
    \end{tabular}
  \end{center}
  \caption[Comparaison de la NEP en mode DDCS et direct pour les BFP
  bleu et rouge]{Comparaison de la NEP en mode DDCS et direct pour les
  BFP bleu et rouge. \'Evolution de la NEP moyenn\'ee sur une matrice
  enti\`ere en fonction de la tension de polarisation et du flux
  incident. \`A cause de l'exc\`es de bruit et de la sous-estimation
  de r\'eponse, la NEP optimale en mode DDCS se trouve autour de 2.6~V
  sur le BFP bleu et de 2~V sur le rouge.
  \label{fig:calib_perflabo_compare_nep}}
\end{figure}

La NEP des bolom\`etres \'etant le rapport du bruit et de la r\'eponse
(section~\ref{sec:calib_perflabo_sensibilite_NEP}), nous la calculons
simplement \`a partir des donn\'ees pr\'esent\'ees dans les
figures~\ref{fig:calib_perflabo_compare_noise}
et~\ref{fig:calib_perflabo_compare_resp}. Nous montrons l'\'evolution
de la NEP en fonction de la polarisation, du flux, du mode de lecture
et de la couleur du BFP dans la
figure~\ref{fig:calib_perflabo_compare_nep}. Notez que plus de
150~heures de mesures ont \'et\'e n\'ecessaires pour obtenir ces
courbes, ce qui repr\'esente quelques 1300~configurations
test\'ees. 

En mode direct (DDCS), le minimum de NEP se trouve \`a une
polarisation de $\sim$1.6~V (2.7~V) sur le BFP bleu et de 1~V (2~V)
sur le rouge. Ces tensions sont plus \'elev\'ees en mode DDCS \`a
cause de l'exc\`es de bruit et de la sous-estimation de la r\'eponse
aux basses tensions de polarisation. Le
tableau~\ref{tab:calib_perflabo_compare_nep} r\'esume les valeurs de
NEP obtenues dans les deux modes pour une tension de polarisation de
2.7~V sur le BFP bleu et de 2~V sur le rouge. Nous avons choisi de
donner les performances en mode direct pour les m\^emes tensions qu'en
mode DDCS car la constante de temps des bolom\`etres est trop longue
aux faibles tensions de polarisation (cf
section~\ref{sec:calib_perflabo_tau}). Ces NEP ont \'et\'e extraites
de la figure~\ref{fig:calib_perflabo_compare_nep} pour des flux de
2.75, 1.54 et 2.52~pW/pixel sur les voies bleue, verte et rouge
respectivement.


\begin{table}
  \begin{center}
    \setlength\extrarowheight{4pt}
    \begin{tabular}[]{p{0.cm}p{6cm}>{\centering}p{1.5cm}>{\centering}p{1.5cm}>{\centering}p{1.5cm}p{0.cm}}
      \toprule
      &Bande PACS [$\mu$m]   & 72.5 & 107.5 & 170 & \\
      \hline \hline
      & Mode direct [$\times 10^{-16}$~W/$\sqrt{\mbox{Hz}}$]  & 1.9 & 1.6 & 3.3 & \\
      & Mode DDCS [$\times 10^{-16}$~W/$\sqrt{\mbox{Hz}}$]  & 4.7 & 4.7 & 6.4 & \\
      \bottomrule
    \end{tabular}
  \caption[NEP du Photom\`etre PACS en modes direct et DDCS]{NEP
  mesur\'ee dans les trois bandes spectrales du Photom\`etre PACS pour
  une tension nominale de 2.7~V sur le BFP bleu et de 2.0~V sur le BFP
  rouge.
  \label{tab:calib_perflabo_compare_nep}}
  \end{center}
\end{table}


\section{R\'esultats r\'ecents}
\label{sec:calib_perflabo_ajuste}

Nous sommes maintenant arriv\'es au bout de la proc\'edure
d'\'etalonnage telle que nous l'avions d\'ecrite dans la
section~\ref{sec:detect_outils_concept}. Nous avons mesur\'e les
performances du Photom\`etre PACS, nous avons \'egalement explor\'e de
mani\`ere syst\'ematique le comportement des d\'etecteurs. Notre
compr\'ehension des matrices de bolom\`etres s'est consid\'erablement
accrue au fil de la campagne d'\'etalonnage, et certains r\'esultats
ont d'ailleurs r\'ev\'el\'e de nouveaux probl\`emes li\'es notamment
\`a l'\'electronique de lecture et \`a la grande imp\'edance des
bolom\`etres.

Cette section est consacr\'ee aux derniers ajustements que nous avons
pu r\'ealiser sur le mod\`ele de vol du Photom\`etre PACS juste avant
sa livraison \`a l'ESA. Nous abordons en particulier les probl\`emes
d'injection de charges en mode DDCS, de cross-talk \'electrique et de
mesure de constantes de temps. Nous pr\'esentons \'egalement les
courbes IV que nous avons r\'ecemment mesur\'ees sur le mod\`ele de
rechange PACS gr\^ace \`a une technique astucieuse qui nous a \'et\'e
souffl\'ee par le concepteur des matrices de bolom\`etres, Patrick
Agn\`ese.

\subsection{Le s\'equenceur du mode DDCS}
\label{sec:calib_perflabo_ajuste_DDCS}

Maintenant que nous avons ex\'ecut\'e avec succ\`es la proc\'edure
d'\'etalonnage, nous pouvons revenir sur les r\'eglages secondaires
que nous avions d\^u fixer de fa\c{c}on empirique au d\'ebut de la
proc\'edure (cf section~\ref{sec:detect_outils_concept}). 

Nous avons en particulier test\'e un nouveau s\'equenceur qui permet
d'inhiber les injections de charges dans les ponts bolom\'etriques en
mode DDCS. L'id\'ee est d'ouvrir le transistor $VDECX$, \cad d'isoler
\'electriquement le circuit de d\'etection (cf
figure~\ref{fig:detect_bolocea_elec_froide_principe}), lors des
commutations du transistor $CKRL$ qui ont lieu juste avant et juste
apr\`es la lecture du signal $V_{ref}$~; de cette mani\`ere les
charges g\'en\'er\'ees par $CKRL$ ne peuvent s'\'ecouler dans les
ponts bolom\'etriques. Un rapport de test publi\'e r\'ecemment
\shortcite{sauvage_seq} montre que le nouveau s\'equenceur ne modifie
pas la forme g\'en\'erale des densit\'es spectrales de bruit,
mais que le niveau de bruit est globalement plus bas. Nous obtenons
une diminution maximum de la NEP de 40\% pour le BFP rouge \`a une
tension de 1.6~V. Pour les plus fortes tensions, le gain en
sensibilit\'e du nouveau s\'equenceur est moindre puisque les
bolom\`etres deviennent suffisamment rapides pour \'ecouler les
charges parasites (cf figure~\ref{fig:calib_perflabo_compare_noise}).
Par exemple pour la tension de polarisation optimale (2~V),
l'am\'elioration est de 20\% \og seulement \fg~; ce qui repr\'esente
tout de m\^eme un gain de 35\% sur le temps d'observation pour
atteindre une sensibilit\'e donn\'ee.

Nous avons par ailleurs test\'e diff\'erents r\'eglages pour les
tensions $VDECX$ et $CKRL$ dans le but de r\'eduire le nombre de
charges lib\'er\'ees lors des commutations des transistors. En effet,
le seuil des transistors MOS est de l'ordre de 500~mV \`a 300~mK,
alors que le r\'eglage par d\'efaut que nous avions choisi
correspondait \`a une diff\'erence de potentiel de 2~V. Cette
sur-polarisation assure le fait que tous les transistors sont dans
un \'etat passant m\^eme pour une dispersion des points milieux de
400~mV. Cependant, une telle tension est relativement \'elev\'ee pour
les transistors, et nous pensons diminuer le ph\'enom\`ene d'injection
de charges en abaissant les tensions $VDECX$ et $CKRL$. Nous avons
test\'e 64~couples $(VDECX,CKRL)$ pour la tension de polarisation
optimale, et il appara\^it que le couple $(1.6\mbox{~V},1.6\mbox{~V})$
offre un point de fonctionnement optimum, \cad que les transistors
sont toujours passants et que le niveau de bruit est
consid\'erablement r\'eduit. Koryo Okumura a montr\'e que le gain est
de 20~\`a 50\% sur la NEP suivant la matrice consid\'er\'ee.

Le r\'esultat de ces deux tests confirme notre interpr\'etation selon
laquelle l'exc\`es de bruit aux basses tensions de polarisation est
d\^u \`a des injections de charges parasites par les transistors
$VDECX$ et $CKRL$. Toutefois, les am\'eliorations apport\'ees par ces
nouveaux r\'eglages/s\'equenceur ne sont pas a priori cumulables. En
effet, le s\'equenceur inhibe le transfert de charges vers les ponts
bolom\'etriques alors que les r\'eglages $(VDECX,CKRL)$ brident la
cr\'eation de ces charges. Malgr\'e cela, nous pouvons assurer une
am\'elioration minimum de 20\% sur la sensibilit\'e du mode DDCS.

\subsection{La fr\'equence d'\'echantillonnage}
\label{sec:calib_perflabo_ajuste_echant}

D'apr\`es l'analyse propos\'ee dans la
section~\ref{sec:calib_perflabo_compare}, l'exc\`es de bruit ainsi que
la sous-estimation de la r\'eponse en mode DDCS sont d\^us \`a la trop
longue constante de temps des bolom\`etres aux basses tensions de
polarisation\footnote{Les injections de charges parasites sont \`a
l'origine de l'exc\`es de bruit, mais elles sont r\'ev\'el\'ees par la
constante de temps des bolom\`etres}. Nous avons effectivement
montr\'e que ces effets disparaissaient lorsque les bolom\`etres
devenaient suffisamment rapides.

L'autre alternative qui nous permettrait de tester notre
interpr\'etation consiste \`a d\'ecaler le moment o\`u BOLC convertit
le signal. En effet, pour une constante de temps donn\'ee, diminuer la
fr\'equence d'\'echantillonnage du signal laisse aux points milieux
plus de temps pour s'\'etablir et aux charges parasites pour
s'\'ecouler dans le circuit.

\begin{figure}
  \begin{center}
    \begin{tabular}{ll}
      \includegraphics[width=0.5\textwidth,angle=0]{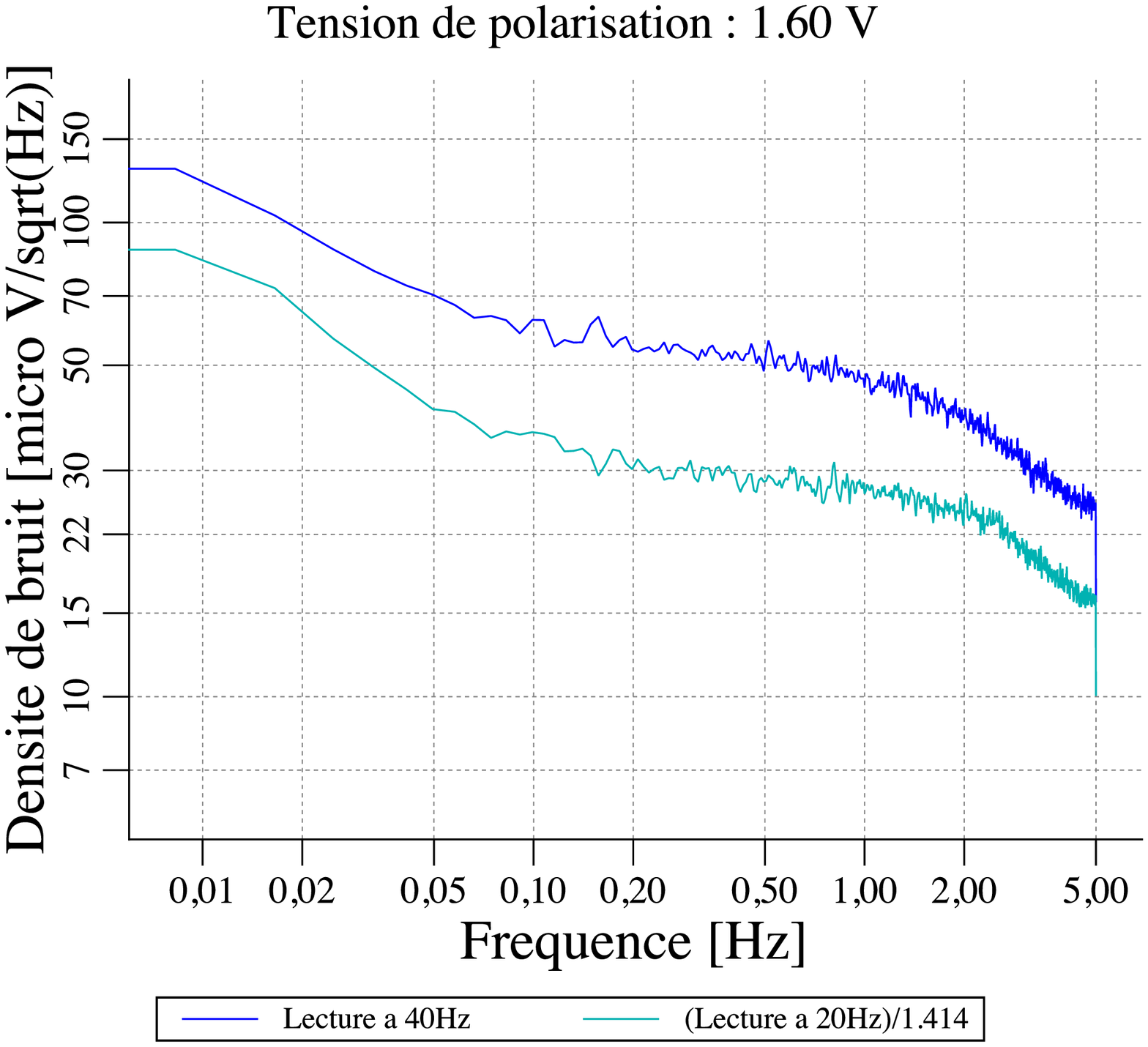}
      \includegraphics[width=0.5\textwidth,angle=0]{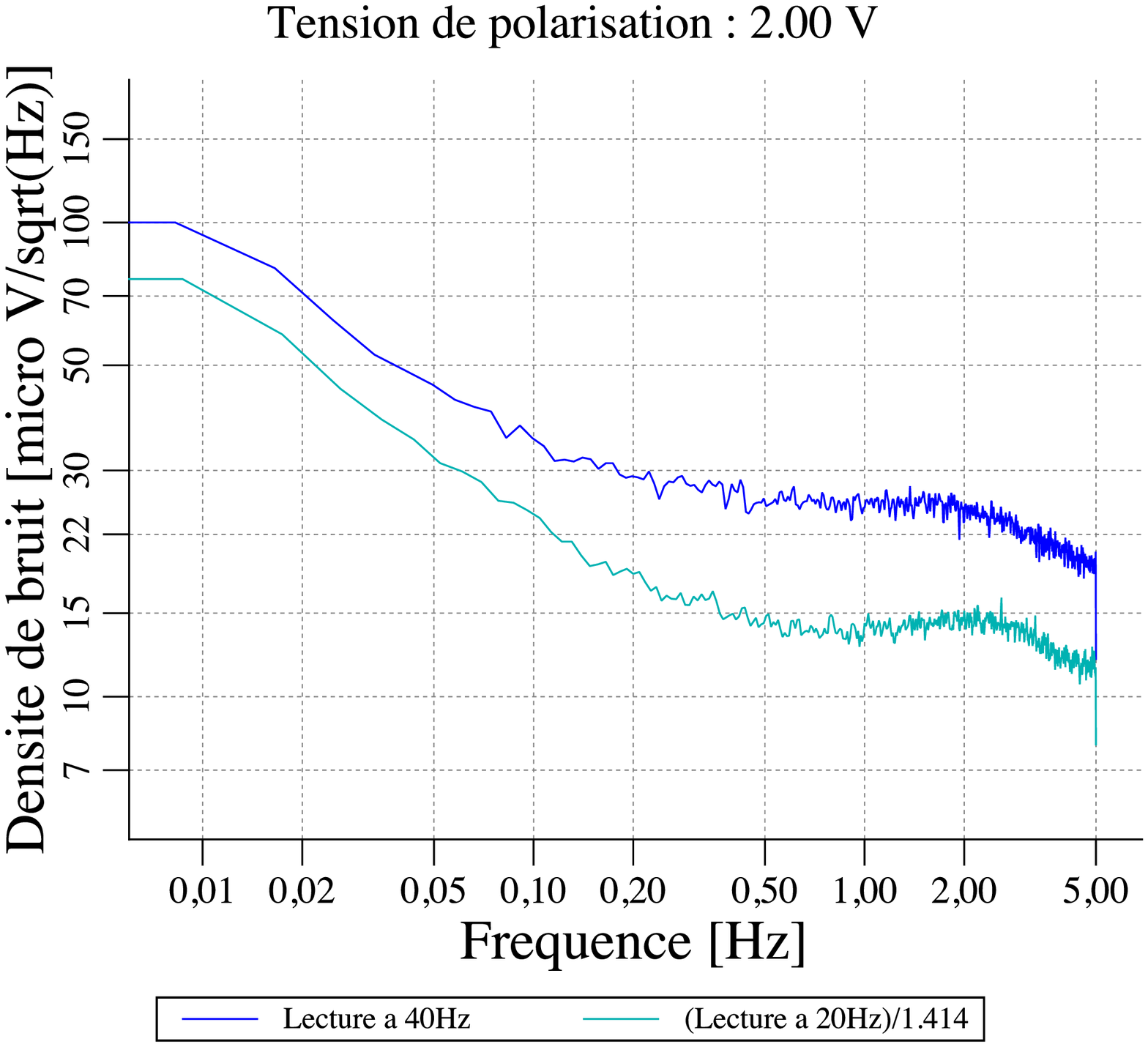} \\
      \includegraphics[width=0.5\textwidth,angle=0]{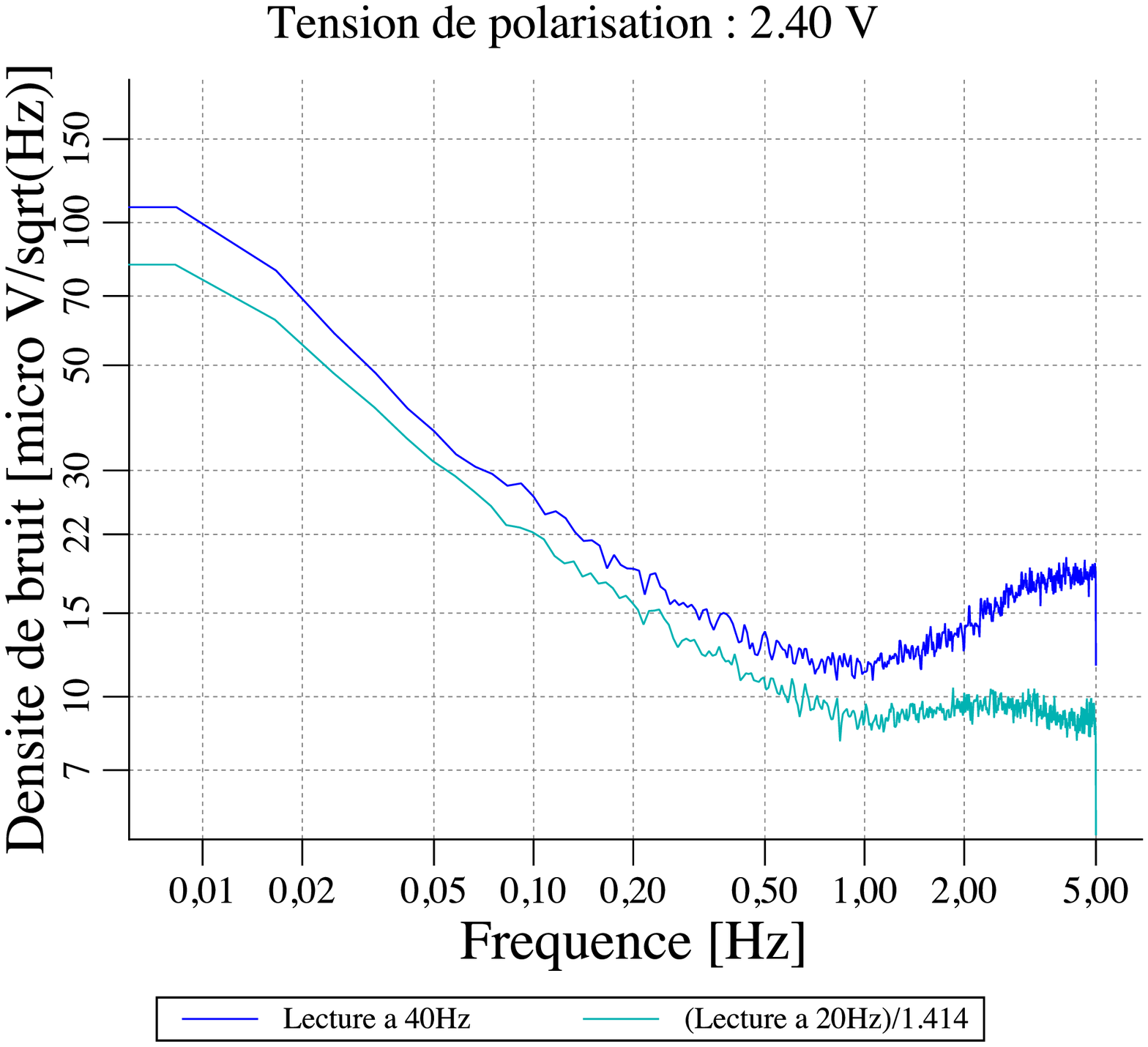}
      \includegraphics[width=0.5\textwidth,angle=0]{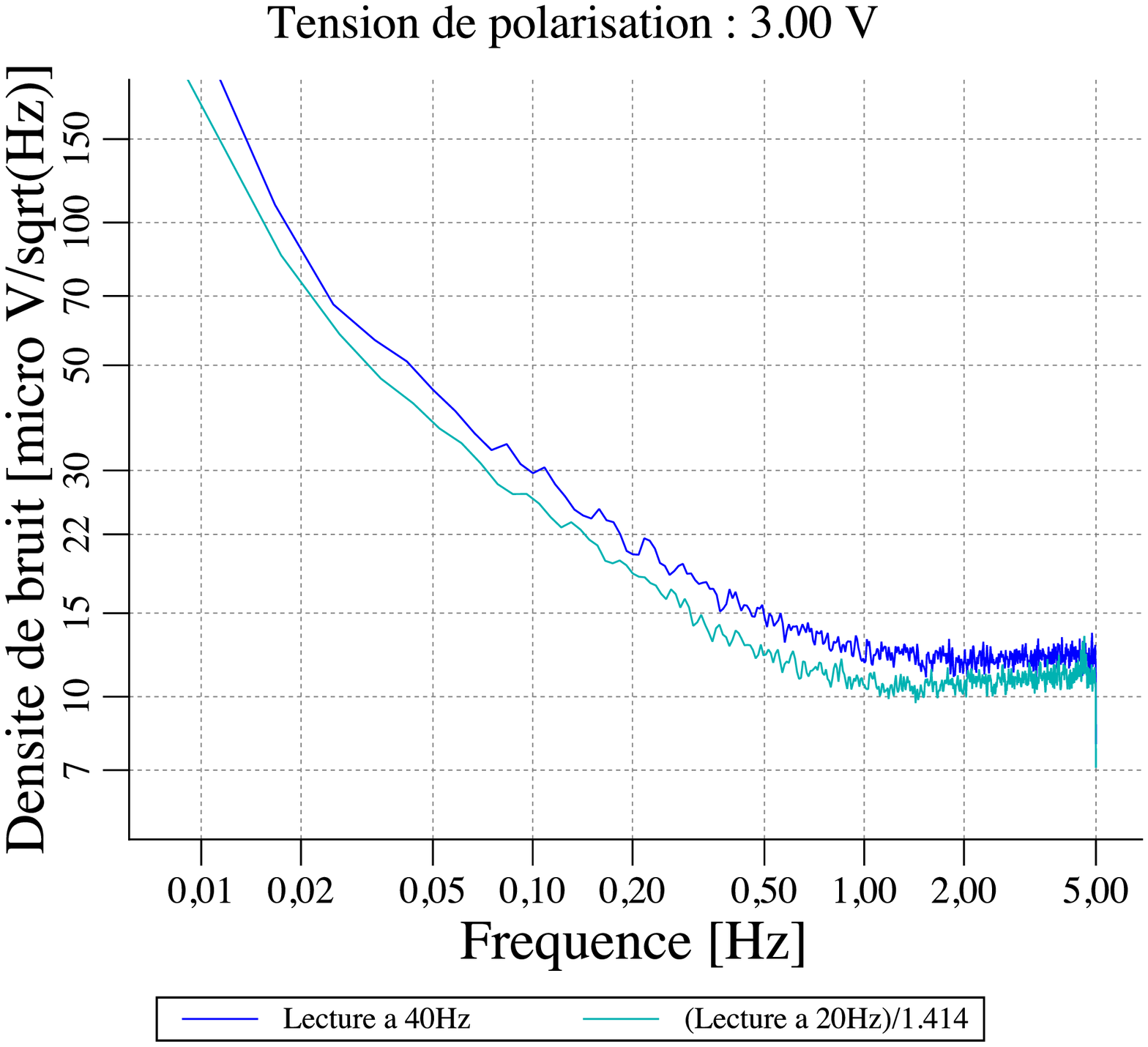}
    \end{tabular}
  \end{center}
  \caption[Densit\'es spectrales de bruit et fr\'equence
  d'\'echantillonnage]{Densit\'es spectrales de bruit moyenn\'ees sur
  une matrice bleue pour deux fr\'equences d'\'echantillonnage et
  quatres tensions de polarisation. Les spectres \'echantillonn\'es
  \`a 20~Hz sont divis\'es par $\sqrt{2}$ pour les rendre comparables
  avec ceux \'echantillonn\'es \`a 40~Hz. Pour les fortes
  polarisations, \cad pour des courtes constantes de temps, les deux
  spectres se superposent. Pour les faibles polarisations, l'exc\`es
  de bruit appara\^it pour les mesures r\'ealis\'ees \`a 40~Hz~: la
  constante de temps est alors plus longue que la p\'eriode
  d'\'echantillonnage (25~ms) et elle r\'ev\`ele les injections de
  charges parasites. Ces spectres ont \'et\'e obtenus en mode DDCS
  pour un flux de 1.5~pW/pixel.
  \label{fig:calib_perflabo_ajuste_echant}}
\end{figure}

Durant la derni\`ere phase de test du Photom\`etre PACS, nous avons
r\'ealis\'e des mesures de bruit pour une fr\'equence
d'\'echantillonnage de 20~Hz. Nous montrons l'\'evolution des
densit\'es spectrales de bruit avec la tension de polarisation dans la
figure~\ref{fig:calib_perflabo_ajuste_echant}. Pour les faibles
tensions, nous constatons une r\'eduction de l'exc\`es de bruit par
rapport aux mesures effectu\'ees \`a 40~Hz. Cela confirme que pour les
faibles tensions et une fr\'equence de 40~Hz, la conversion du signal
par BOLC a lieu dans le r\'egime transitoire de $V_{ref}$ \`a
$V_{ptmil}$, cela signifie que la constante de temps \'electrique du
circuit est plus longue que 25~ms.

D'autres tests de ce type \'etaient pr\'evus lors de la campagne
d\'etalonnage (mesures de bruit et de r\'eponse pour plusieurs
fr\'equences d'\'echantillonnage), mais des contraintes de planning
ainsi que des probl\`emes de compression \`a bord (SPU) nous ont
emp\^ech\'e de poursuivre. Ces tests pourront cependant \^etre
r\'ealis\'es sur le mod\`ele de rechange du photom\`etre.

\subsection{Le cross-talk \'electrique}
\label{sec:calib_perflabo_ajuste_crosstalk}

Durant la campagne d'\'etalonnage de l'instrument PACS, nous avons mis
en \'evidence une forte corr\'elation entre certains pixels du BFP
rouge lorsqu'ils \'etaient fortement illumin\'es, avec entre autre
l'exp\'erience pr\'esent\'ee dans la
figure~\ref{fig:detect_observatoire_phfpu_description_mascot} ou
encore en effectuant des balayages du champ de vue avec fort gradient
d'illumination. Le BFP bleu semble toutefois ne pas souffrir de cette
corr\'elation. Nous pensons que ce cross-talk entre pixels est
d'origine \'electrique plut\^ot qu'optique. En effet, lorsque le
circuit de lecture moyenne imp\'edance est sous-aliment\'e,
l'\'electronique n'est pas suffisamment rapide pour multiplexer le
signal correctement, certains pixels d'une m\^eme colonne de lecture
voient donc leur potentiel \'electrique fortement corr\'el\'e.

Nous avons donc r\'ealis\'e des mesures suppl\'ementaires pour
quantifier la corr\'elation entre le cross-talk et le courant
$I_{VSS}$ qui circule dans le circuit de lecture moyenne
imp\'edance. Koryo Okumura a analys\'e ces tests et a trouv\'e que le
cross-talk disparaissait pour un courant $I_{VSS}$ de 150~nA sur le
BFP rouge. 

Notez que les BU du BFP bleu contiennent chacun deux matrices de
bolom\`etres, et que le courant par d\'efaut fix\'e au d\'ebut de la
campagne \'etait de 300~nA sur le BFP bleu, \cad 150~nA sur chacune
des matrices d'un m\^eme BU, et de 100~nA sur le BFP rouge. Le BFP
bleu \'etait donc correctement aliment\'e, ce qui explique pourquoi
nous n'avions pas observ\'e de cross-talk, alors que le BFP rouge
\'etait sous-aliment\'e.

\subsection{Les mesures de constante de temps}
\label{sec:calib_perflabo_ajuste_tau}

\begin{figure}
  \begin{center}
    \begin{tabular}{ll}
      \includegraphics[width=0.48\textwidth,angle=0]{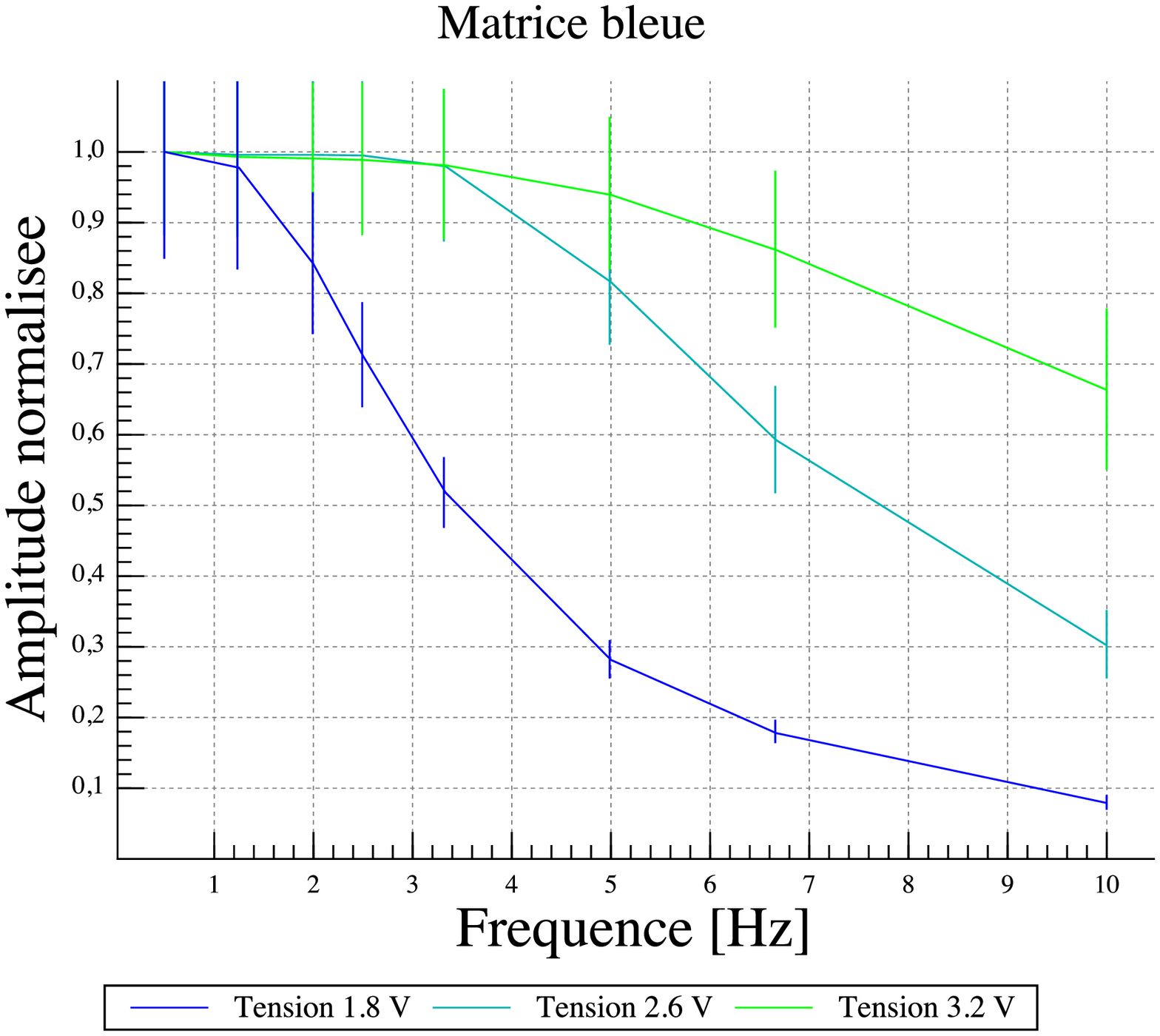} &
      \includegraphics[width=0.48\textwidth,angle=0]{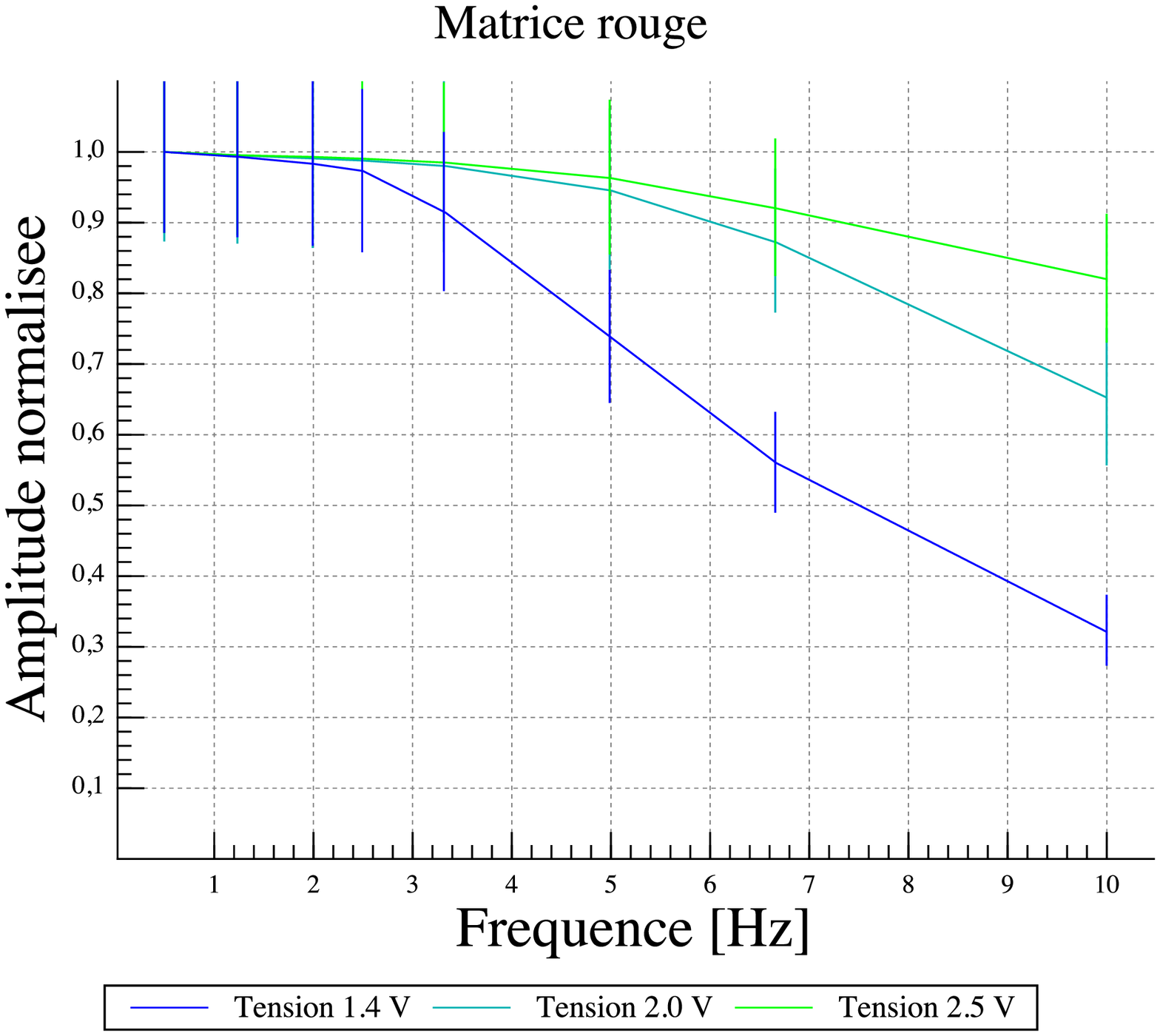}
    \end{tabular}
  \end{center}
  \caption[Mesures dynamiques de la constante de temps r\'ealis\'ees
  au MPE]{Mesures dynamiques de la constante de temps r\'ealis\'ees au
  MPE avec le PhFPU int\'egr\'e dans l'instrument PACS. Le protocole
  de test est le m\^eme que celui pr\'esnt\'e dans la
  section~\ref{sec:calib_perflabo_tau_direct}. Les barres d'erreurs
  indiquent la d\'eviation standard des amplitudes mesur\'ees sur la
  matrice. Les r\'esultats sont plus fiables que ceux pr\'esent\'es
  dans la section~\ref{sec:calib_perflabo_tau_direct} car
  l'environnement thermique du cryostat de test PACS est beaucoup plus
  stable (le chopper interne dissipe tr\`es peu). La fr\'equence de
  coupure augmente avec la tension de polarisation. Les matrices
  rouges \'etant moins imp\'edantes que les bleues par construction,
  leur fr\'equence de coupure est plus \'elev\'ee ($\tau=\frac{1}{2\pi
  RC}$).
  \label{fig:calib_perflabo_ajuste_tau}}
\end{figure}

Dans la section~\ref{sec:calib_perflabo_tau_direct}, nous avons
montr\'e que les mesures dynamiques de la constante de temps
r\'ealis\'ees sur le banc de test de Saclay n'offraient qu'une
pr\'ecision de $\sim$30\% \`a cause de l'environnement thermique et du
mat\'eriel utilis\'e. Nous avons donc r\'ep\'et\'e ces mesures sur le
banc de test\footnote{Le banc de test de l''instrument PACS est
\'egalement appel\'e OGSE pour \emph{Optical Ground Segment
Equipment}} de l'instrument PACS qui est a priori plus stable du fait
de la faible dissipation du chopper interne et de la stabilit\'e des
sources lumineuses internes. Nous pr\'esentons l'\'evolution de
l'amplitude modul\'ee en fonction de la fr\'equence de modulation dans
la figure~\ref{fig:calib_perflabo_ajuste_tau}. Le
tableau~\ref{tab:calib_perflabo_ajuste_tau} donne la valeur de la
fr\'equence de coupure trouv\'ee pour les trois tensions de
polarisation test\'ees. Notez que ces valeurs sont sensiblement
sup\'erieures \`a celles trouv\'ees lors des tests effectu\'es \`a
Saclay (cf figure~\ref{fig:calib_perflabo_tau_direct_bppolar})~; la
cause invoqu\'ee \'etant la surchauffe du plateau optique aux hautes
fr\'equences de modulation (cf
section~\ref{sec:calib_perflabo_tau_direct}).

\begin{table}
  \begin{center}
    \setlength\extrarowheight{4pt}
    \begin{tabular}[]{p{0.cm}p{8cm}>{\centering}p{1.5cm}>{\centering}p{1.5cm}>{\centering}p{1.5cm}p{0.cm}}
      \toprule
      &Tension de polarisation (bleu/rouge) [V]   & 1.8/1.4 & 2.6/2.0 & 3.2/2.5 & \\
      \hline \hline
      & Fr\'equence de coupure matrice bleue [Hz]  & 2.5 & 6 & 9.5 & \\
      & Fr\'equence de coupure matrice rouge [Hz]  & 5.5 & 9 & $>10$ & \\
      \bottomrule
    \end{tabular}
  \caption[Fr\'equence de coupure des BFP bleu et rouge]{Fr\'equence
  de coupure mesur\'ee sur le banc de test PACS pour trois valeurs de
  tension de polarisation.
  \label{tab:calib_perflabo_ajuste_tau}}
  \end{center}
\end{table}

Malgr\'e la stabilit\'e du banc de test PACS, il reste toutefois un
biais dans la mesure de la fr\'equence de coupure. En effet, comme
nous l'avions illustr\'e dans la
figure~\ref{fig:calib_perflabo_tau_compare_echant}, le chopper interne
PACS est synchronis\'e sur l'acquisition du signal de sorte que
l'amplitude de modulation est syst\'ematiquement sous-estim\'ee pour
les hautes fr\'equences de modulation. Par exemple, un signal modul\'e
\`a 10~Hz avec le chopper interne PACS ne contient que 2~points de
mesures par plateau chopper, les extrema du signal risquent donc de ne
pas \^etre \'echantillonn\'es. Les donn\'ees contiennent pourtant
l'information utile, mais il serait n\'ecessaire de mod\'eliser le
comportement du signal modul\'e pour en extraire la fr\'equence de
coupure avec plus de pr\'ecision. Toutefois, les r\'esultats obtenus
lors de ce test sont suffisamment repr\'esentatifs pour pr\'edire les
performances observationnelles du Photom\`etre PACS (cf
section~\ref{sec:calib_perfobs_scan}).



\subsection{Les courbes IV globales}
\label{sec:calib_perflabo_ajuste_IV}

\begin{figure}
  \begin{center}
      \includegraphics[width=0.8\textwidth,angle=0]{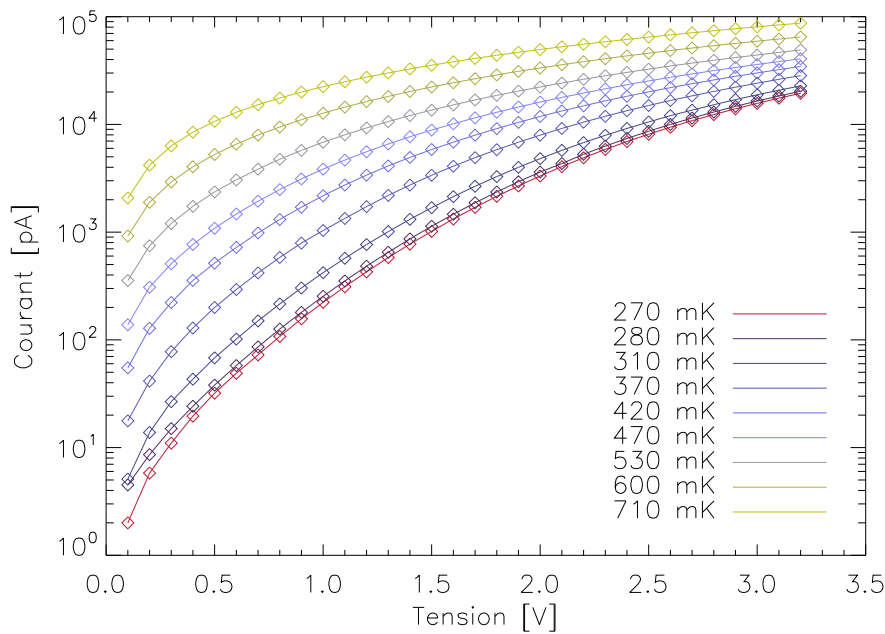}
  \end{center}
  \caption[Courbes~I-V du BFP bleu du mod\`ele de rechange du
  Photom\`etre PACS]{Courbes~I-V globales mesur\'ees sur le BFP bleu
  du mod\`ele de rechange du Photom\`etre PACS, \cad pour
  512~r\'esistances mont\'ees en parall\`ele. Les points de mesure
  inf\'erieurs \`a 10~pA ne sont pas repr\'esentatifs puisqu'ils sont
  du m\^eme ordre de grandeur que la pr\'ecisison de
  l'\'electrom\`etre utilis\'e pour les mesurer.
  \label{fig:calib_perflabo_ajuste_IV}}
\end{figure}

\begin{figure}
  \begin{center}
    \begin{tabular}{ll}
      \includegraphics[width=0.5\textwidth,angle=0]{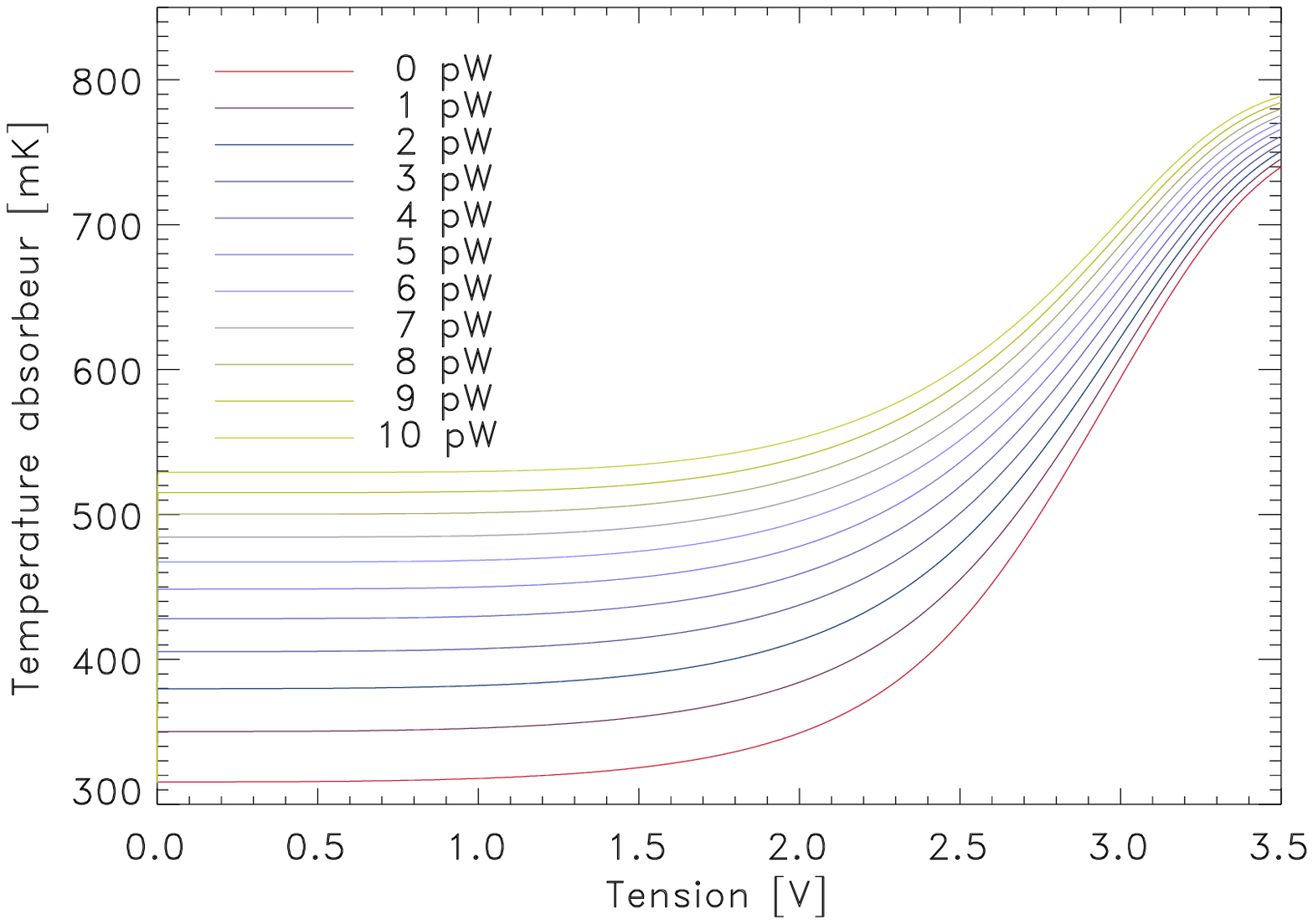}
      \includegraphics[width=0.5\textwidth,angle=0]{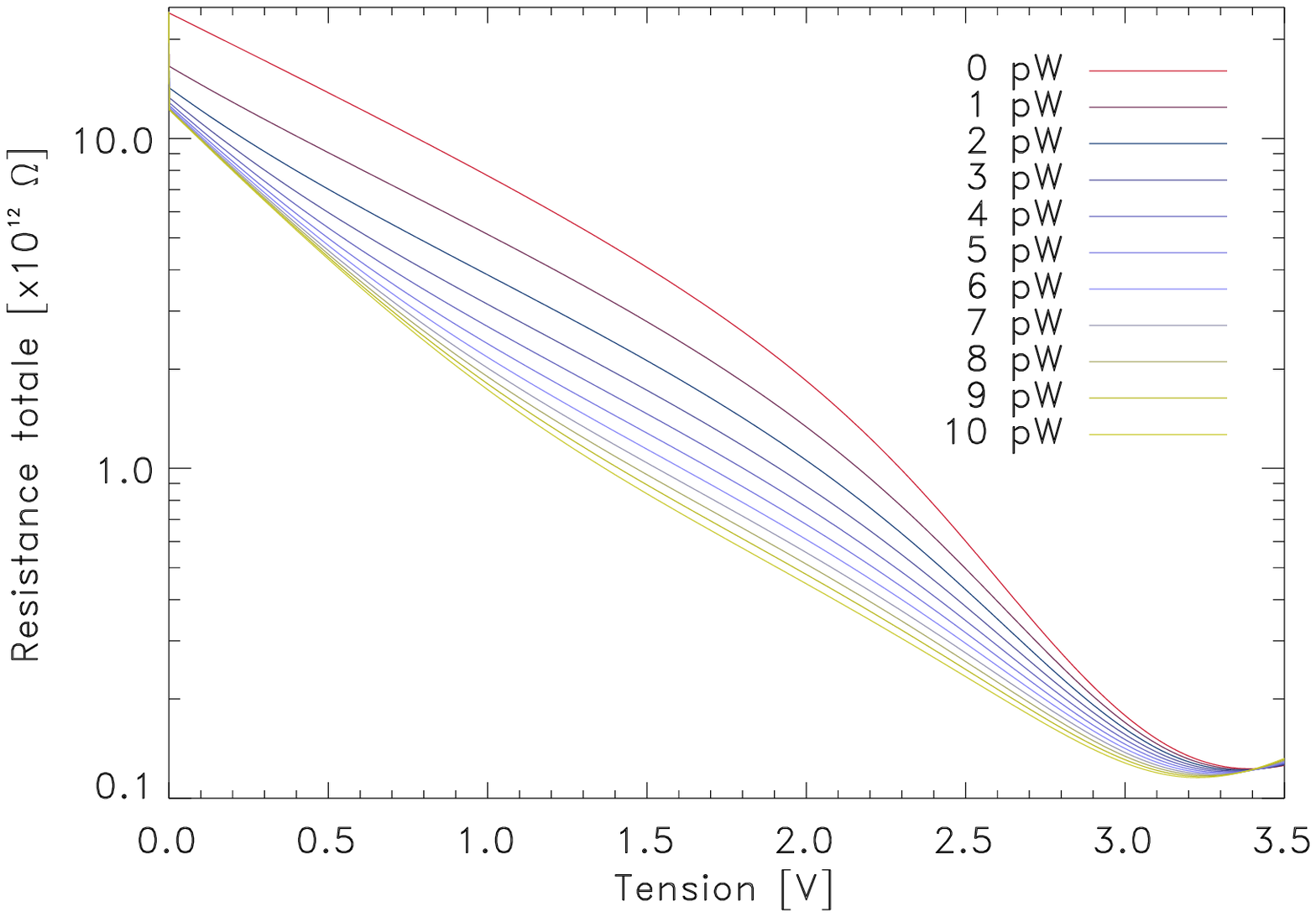} \\
      \includegraphics[width=0.5\textwidth,angle=0]{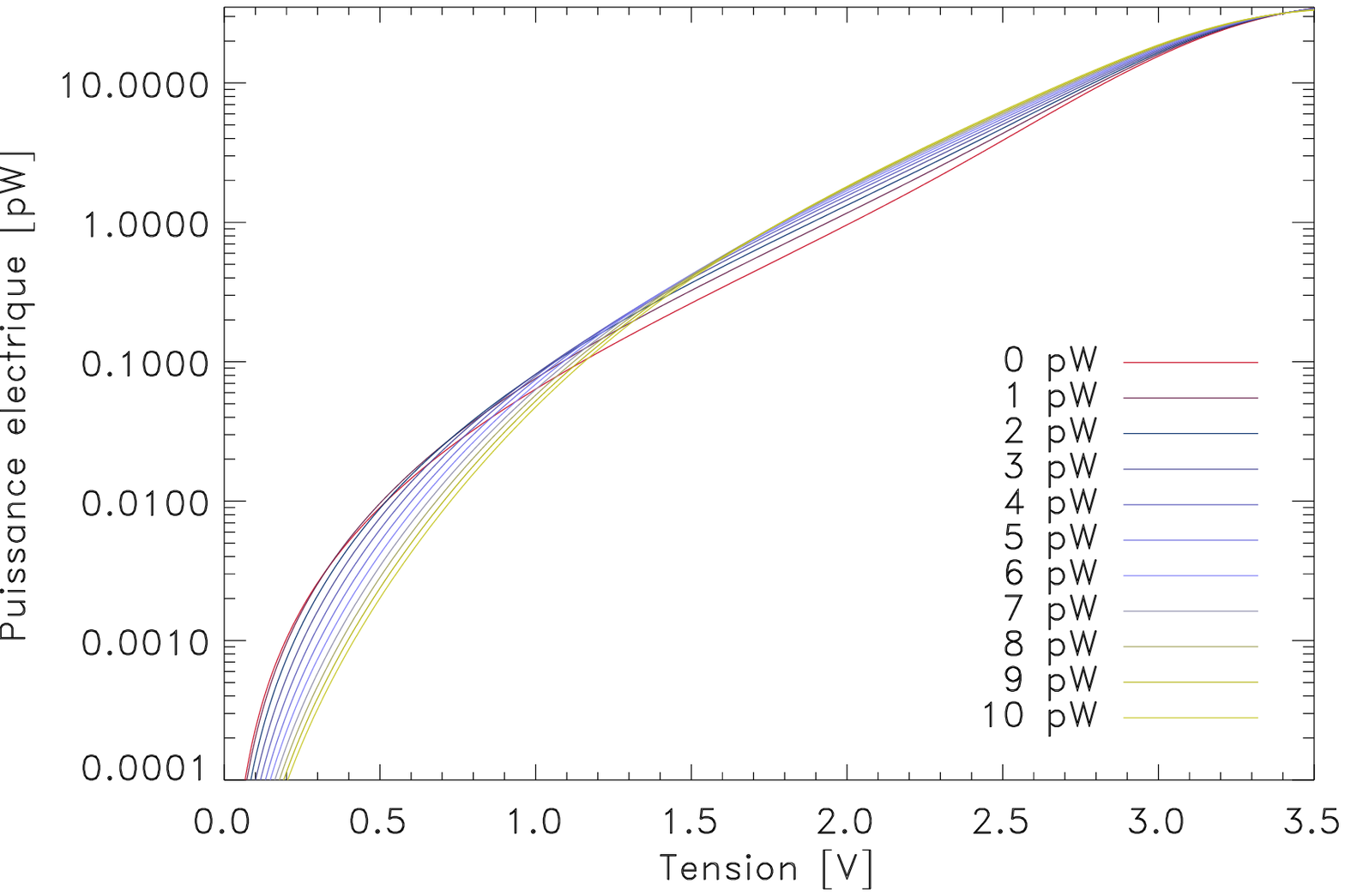}
      \includegraphics[width=0.5\textwidth,angle=0]{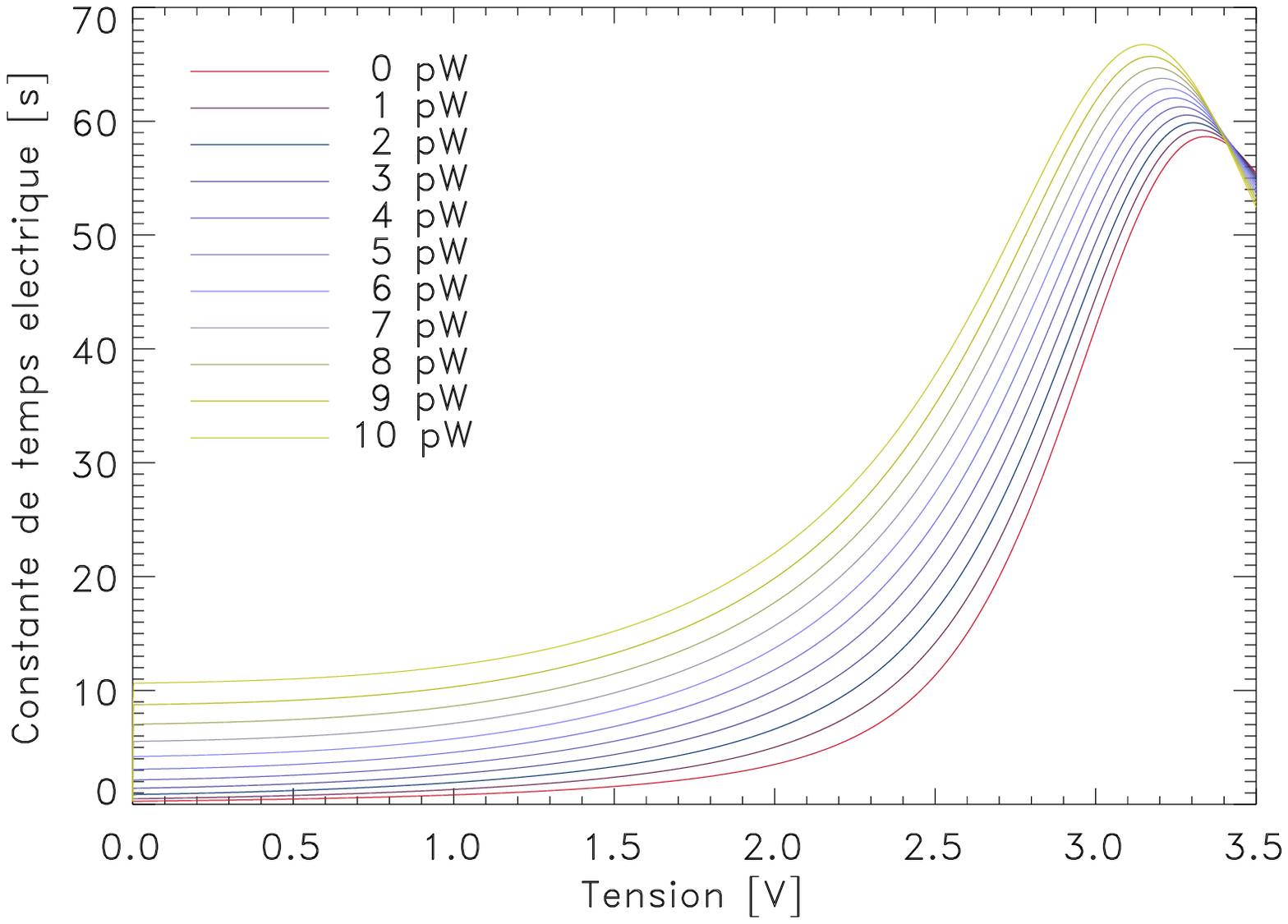}
    \end{tabular}
  \end{center}
  \caption[Mod\'elisation des bolom\`etres du BFP bleu du mod\`ele de
  rechange PACS]{Simulations de l'\'evolution de quelques param\`etres
  d'un bolom\`etre de type PACS en fonction de la tension de
  polarisation et du flux incident. En haut \`a gauche se trouve la
  temp\'erature de l'absorbeur, en haut \`a droite la r\'esistance
  totale d'un pont bolom\'etrique, en bas \`a gauche la puissance
  \'electrique dissip\'ee dans l'absorbeur et en bas \`a droite la
  constante de temps \'electrique du bolom\`etre en supposant une
  capacit\'e \'electrique parasite de 0.1~pF. Ces courbes ont \'et\'e
  obtenues pour une temp\'erature des r\'esistances de charge de
  315~mK~; cette valeur a \'et\'e calcul\'ee \`a partir de
  l'ajustement des points milieux (graphe de droite sur la
  figure~\ref{fig:calib_procedure_explore_midpt}) avec le mod\`ele.
  \label{fig:calib_perflabo_ajuste_model}}
\end{figure}

Dans cette derni\`ere section, nous pr\'esentons des r\'esultats
obtenus r\'ecemment par Louis Rodriguez. Il s'agit de mesures de
courbes~I-V globales r\'ealis\'ees sur le mod\`ele de rechange du BFP
bleu\footnote{Les matrices qui \'equipent le mod\`ele de rechange du
BFP bleu sont du m\^eme type, \cad du m\^eme dopage, que les matrices
du mod\`ele de vol du BFP rouge.}. Dans la
section~\ref{sec:detect_outils_loadcurves}, nous avons expliqu\'e que
le circuit de lecture des matrices ne permet pas de mesurer le courant
qui circule dans chacun des pixels, c'est pour cela que nous avons
opt\'e pour la mesure de points milieux pour explorer le comportement
individuel des bolom\`etres. Cependant, Patrick Agn\`ese nous a
sugg\'er\'e une mani\`ere de mesurer le courant total qui circule dans
un groupe de matrices en connectant un \'electrom\`etre entre les
bornes $V_h$ et $V_{ref}$ (cf
figure~\ref{fig:detect_bolocea_elec_froide_principe}). Nous pouvons
ainsi mesur\'e le courant qui circule dans les 512~r\'esistances de
r\'ef\'erence d'un m\^eme BU. Ces r\'esistances \'etant mont\'ees en
parall\`ele, il est n\'ecessaire de diviser par 512~le courant
mesur\'e pour obtenir le courant moyen circulant dans une seule
r\'esistance, d'o\`u le terme de globales pour qualifier les
courbes~I-V mesur\'ees. La figure~\ref{fig:calib_perflabo_ajuste_IV}
montre ces courbes pour diff\'erentes temp\'eratures de la source
froide\footnote{La temp\'erature de la source froide n'\'etant pas
asservie, l'erreur associ\'ee aux mesures de temp\'erature est de
l'ordre du~mK.}. Notez que les mesures inf\'erieures \`a 10~pA ne sont
pas fiables du fait de l'extr\^eme faiblesse des courants mis en jeu
($\mbox{R}>10^{13}$~$\Omega$).

\`A partir de ces donn\'ees, L.~Rodriguez a extrait les param\`etres
des thermistances, \`a savoir $R_0$, $T_0$ et $L(T)$ (cf
\'equation~\ref{eq:efros}). Il a ensuite inject\'e ces param\`etres
dans le mod\`ele de bolom\`etres d\'evelopp\'e par V.~Rev\'eret
\shortcite{reveret_these} pour reproduire les courbes de rapports
d'imp\'edance mesur\'es durant l'\'etalonnage du Photom\`etre PACS (cf
figure~\ref{fig:calib_procedure_explore_imped}). Il a ainsi pu
calculer d'autres param\`etres physiques des bolom\`etres tels que la
temp\'erature de l'absorbeur, la puissance \'electrique dissip\'ee ou
encore la valeur de la constante de temps \'electrique en fonction de
la tension de polarisation et du flux incident. Quelques r\'esultats
pr\'eliminaires obtenus avec le mod\`ele sont pr\'esent\'es dans la
figure~\ref{fig:calib_perflabo_ajuste_model}. Bien que ces r\'esultats
soient globaux, ils restent tr\`es int\'eressants car ils nous donnent
acc\`es \`a des informations que nous ne pouvons extraire des rapports
d'imp\'edance seuls. Le raffinement du mod\`ele et l'interpr\'etation
de ces courbes sont en cours.

\chapter{Du laboratoire \`a l'observatoire}
\label{chap:calib_perfobs}

\begin{center}
\begin{minipage}{0.85\textwidth}

\small L'objectif de ce chapitre est de traduire les mesures de
performance r\'ealis\'ees en laboratoire en terme de performances
observationnelles. Ceci est une \'etape essentielle qui nous permet
d'\'evaluer les capacit\'es de l'observatoire, mais aussi de d\'efinir
et d'optimiser les modes d'observation de l'instrument avant son
lancement. Nous commencerons par calculer la sensibilit\'e du
Photom\`etre PACS en nous appuyant sur le rapport
de~\shortciteN{sauvage_note}, nous estimerons ensuite la dynamique de
l'instrument, puis nous mettrons en \'evidence le \og temps de
coh\'erence \fg du signal, et enfin nous pr\'esenterons une \'etude
exhaustive du comportement du Photom\`etre en mode d'observation par
balayage du ciel.



% ...
\end{minipage}
\end{center}

\section{De la \emph{NEP} d\'etecteur \` a la sensibilit\'e t\'elescope}
\label{sec:calib_perfobs_nep}

Le terme de \emph{NEP} fait parti du jargon souvent herm\'etique
utilis\'e par les instrumentalistes pour d\'efinir la sensibilit\'e
d'un d\'etecteur IR (cf
section~\ref{sec:calib_perflabo_sensibilite}). Mais la plupart des
futurs utilisateurs de PACS ne seront certainement pas familiers avec
cette terminologie. Et pour cause, la sensibilit\'e d'une cam\'era ne
se r\'esume pas simplement \`a une \emph{NEP}, son calcul fait en plus
intervenir la transmission des \'el\'ements optiques, le mode
d'observation et la technique d'extraction utilis\'ee dans la
r\'eduction des donn\'ees. Les astronomes expriment plut\^ot la
sensibilit\'e en terme de temps d'observation n\'ecessaire pour
d\'etecter une source ponctuelle de flux donn\'e avec un
signal-\`a-bruit donn\'e. Par exemple, la sp\'ecification scientifique
sur la sensibilit\'e du Photom\`etre PACS est de
5$\,$mJy~(5$\,\sigma$~1$\,$hr) \shortcite{pacs_lutz}. Cette section
est d\'edi\'ee au calcul de la sensibilit\'e du Photom\`etre PACS,
l'objectif \'etant de traduire la \emph{NEP} des d\'etecteurs telle
que nous l'avons mesur\'ee en laboratoire en une sensibilit\'e plus
explicite que les astronomes pourront utiliser pour pr\'eparer leurs
futures observations. Ce calcul est inspir\'e de
\shortciteN{sauvage_note}.

La formule g\'en\'erique qui permet d'exprimer le rapport
signal-\`a-bruit $S/N$ d'une observation en fonction de la \emph{NEP}
du d\'etecteur est la suivante~:
\begin{equation}
S/N=\frac{P_{inc}}{NEP}\times\sqrt{2\times T_{obs}}
\label{eq:snr_nep_1}
\end{equation}
o\`u $P_{inc}$ est la puissance incidente en [W] et $T_{obs}$ le temps
d'observation en [s]. Tous les termes de
l'\'equation~(\ref{eq:snr_nep_1}) doivent n\'ecessairement se
rapporter \`a un seul et m\^eme syst\`eme, dans notre cas le
pixel. Nous retrouvons bien que le signal-\`a-bruit est proportionnel
\`a la racine carr\'ee du temps d'observation. Le facteur~2 sous la
racine provient de la th\'eorie de l'\'echantillonnage~; le signal
temporel d'une source observ\'ee durant un temps~$T_{obs}$ est
\'echantillonn\'e selon le crit\`ere de Nyquist jusqu'\`a une
fr\'equence de $1/(2\times T_{obs})$. Nous retrouvons d'ailleurs la
d\'efinition de la \emph{NEP} dans l'\'equation~(\ref{eq:snr_nep_1})~;
elle repr\'esente la puissance incidente que peut d\'etecter un pixel
avec un signal-\`a-bruit de~1 dans une bande passante de 1~Hz, \cad un
temps d'observation d'une demi seconde selon le crit\`ere de Nyquist
($NEP=\frac{P_{inc}}{1}\times\sqrt{1}$).

Cependant, le rapport signal-\`a-bruit mentionn\'e dans la
sp\'ecification de sensibilit\'e s'applique uniquement au flux de la
source, alors que le signal-\`a-bruit de
l'\'equation~(\ref{eq:snr_nep_1}) se r\'ef\`ere au flux total incident
sur un pixel, \cad la somme de $P_{tel}$~et~$P_{source}$ les
puissances incidentes provenant respectivement des sources
d'avant-plan et de la source astrophysique.
Si nous appelons $r.m.s$ le niveau de bruit mesur\'e durant
l'observation, alors nous pouvons \'ecrire le signal-\`a-bruit
global~:
\begin{equation}
S/N=\frac{P_{tel}+P_{source}}{r.m.s}
\label{eq:snr_nep_2}
\end{equation}
Le bruit $r.m.s$ doit ici \^etre exprim\'e en [W/pixel], ceci
n\'ecessite de conna\^itre la r\'eponse du pixel et le bruit mesur\'e
en [V] sur le signal temporel. Ce $r.m.s$ est le bruit qui limite
effectivement la sensibilit\'e de l'observation, et nous pouvons
\'ecrire le signal-\`a-bruit $s/n$ dans le sens astronomique du
terme~:
\begin{equation}
s/n=\frac{P_{source}}{r.m.s.}
\label{eq:snr_nep_3}
\end{equation}
En fait, la valeur de bruit qui doit \^etre utilis\'ee pour calculer
la v\'eritable sensibilit\'e de la cam\'era est celle qui est
mesur\'ee en bout de cha\^ine du processus de r\'eduction des images~;
et celle-ci d\'epend entre autre de la technique d'observation. Par
exemple, pour une observation chopp\'ee, comme c'est souvent le cas
lorsque l'\'emission d'avant-plan domine, le t\'elescope ne passe que
la moiti\'e de son temps \`a pointer la source, l'autre moiti\'e est
utilis\'ee pour mesurer l'\'emission d'avant-plan afin de pouvoir la
soustraire pendant la r\'eduction des donn\'ees. Le temps mentionn\'e
dans l'\'equation~(\ref{eq:snr_nep_1}) est le temps pass\'e sur la
source uniquement, et nous devons re-d\'efinir $T_{obs}$ comme \'etant
le temps total d'observation, \cad qui comprend les \og temps inactifs
\fg (\emph{overhead} en anglais) utilis\'es pour l'\'etalonnage ou le
d\'eplacement du t\'elescope par exemple. Nous introduisons alors le
coefficient $\eta_{obs}$ qui indique la fraction de temps pass\'ee \`a
observer le ciel\footnote{C'est en effet le fond de t\'elescope qui
d\'efinit le niveau de bruit d'une observation, donc m\^eme pour une
observation chopp\'ee o\`u la source est en dehors du champ de vue
plus de la moiti\'e du temps, $\eta_{obs}$ peut quand m\^eme \^etre
sup\'erieur \`a~0.5.}. Notez que le v\'eritable signal-\`a-bruit que
nous cherchons \`a calculer d\'epend \'egalement du type de
photom\'etrie appliqu\'ee durant la r\'eduction. Nous faisons
l'hypoth\`ese d'une photom\'etrie d'ouverture classique pour extraire
le flux des sources ponctuelles, et en sommant le signal de tous les
pixels illumin\'es par une source, nous abaissons le bruit d'un
facteur $\sqrt{N_{pix}}$, o\`u $N_{pix}$ est le nombre de pixels dans
une PSF et la racine carr\'ee provient du fait que le bruit de chacun
des pixels n'est a priori pas corr\'el\'e.  D'autre part, pour la
r\'eduction standard d'une observation chopp\'ee, l'\'emission
d'avant-plan est soustraite pour s'affranchir des d\'erives
d\'etecteurs et des structures spatiales lumineuses dues \`a l'optique
du t\'elescope (ou \`a l'atmosph\`ere dans le cas d'un instrument au
sol). Les bruits s'ajoutant quadratiquement, un autre
facteur~$\sqrt{2}$ s'ajoute alors au calcul de la \emph{NEP}. En
utilisant les trois \'equations pr\'ec\'edentes et les diff\'erents
facteurs correctifs, nous obtenons finalement la formule suivante qui
relie le \og signal-\`a-bruit astronomique \fg $s/n$ \`a la \emph{NEP}
mesur\'ee en laboratoire~:
\begin{equation}
\frac{P_{tel}+P_{source}}{\frac{P_{source}}{s/n}\times\sqrt{\frac{N_{pix}}{2}}}=\frac{P_{tel}+P_{source}}{NEP}\times\sqrt{2\times
T_{obs}\times \eta_{obs}}
\label{eq:snr_nep_4}
\end{equation}
Nous voyons que le terme $P_{tel}+P_{source}$ dispara\^it de la
formule, il semble alors que l'\'emission d'avant plan n'a plus aucune
influence sur la sensibilit\'e du Photom\`etre. Mais cela n'est pas
tout \`a fait vrai, sa contribution est en fait contenue dans le terme
de \emph{NEP} qui lui d\'epend significativement du flux total
incident sur le bolom\`etre (cf
sections~\ref{sec:calib_perflabo_sensibilite_NEP}
et~\ref{sec:calib_perflabo_compare_NEP}). 

Pour une source ponctuelle, le nombre de pixels illumin\'es par la PSF
est approximativement le rapport des angles solides du t\'elescope et
du pixel~: $N_{pix}=\frac{\Omega_{tel}}{\Omega_{pix}}$. Ensuite, pour
calculer la puissance incidente $P_{source}$, nous consid\'erons une
source dont la densit\'e spectrale de flux est~$f_{\nu}$
[W/m$^2$/Hz]. Elle est observ\'ee avec un t\'elescope d'ouverture
effective $A_{eff}$ et \`a travers un syst\`eme optique de bande
spectrale~$\delta\nu$ et de transmission~$t$ (cf
figure~\ref{fig:detect_observatoire_phfpu_description_filtre}). Il
faut \'egalement prendre en compte le param\`etre $\eta_{tel}$ qui
repr\'esente la fraction du flux de la source qui est concentr\'ee
dans le lobe principal du t\'elescope\footnote{L'extraction du flux se
fait dans la PSF uniquement et l'\'energie qui se trouve dans les
lobes d'ordre sup\'erieur n'est en principe pas prise en compte.}. La
puissance collect\'ee dans la PSF \'etant dilu\'ee sur $N_{pix}$
pixels, la puissance incidente par pixel au niveau du plan focal est
alors~:
\begin{equation}
P_{source}=\frac{t\,\delta\nu\, A_{eff}\,\eta_{tel}\,f_{\nu}}{N_{pix}}
\label{eq:snr_nep_5}
\end{equation}
\`A partir des \'equations pr\'ec\'edentes, nous obtenons finalement
la formule analytique qui donne la sensibilit\'e d'une observation,
\cad le flux des sources d\'etect\'ees avec un signal-\`a-bruit $s/n$
en un temps total d'observation $T_{obs}$ connaissant la \emph{NEP} du
d\'etecteur~:
\begin{equation}
f_{\nu}=\frac{NEP}{A_{eff}\,\eta_{tel}\,t\,\delta\nu \,
\sqrt{\frac{\eta_{obs}}{2}}\,\sqrt{\frac{\Omega_{pix}}{\Omega_{tel}}}}\times\frac{1}{\sqrt{2\times
T_{obs}}}\times s/n
\label{eq:snr_nep_6}
\end{equation}
\noindent Pour calculer des valeurs num\'eriques de la sensibilit\'e
du Photom\`etre PACS, nous devons encore expliciter quelques
param\`etres de l'\'equation~\ref{eq:snr_nep_6}. Par exemple, l'angle
solide du pixel s'\'ecrit sous la forme~:
\begin{equation}
\Omega_{pix}=\left(\frac{l_{pix}\,\pi}{3600\times 180}\right)^2
\label{eq:omega_pix}
\end{equation}
o\`u $l_{pix}$ est la taille du pixel en seconde d'arc [$''$]. L'angle
solide du lobe du t\'elescope est quant \`a lui calcul\'e de la
fa\c{c}on suivante~:
\begin{equation}
\Omega_{tel}=\frac{1}{\eta_{ouv}}\times\frac{4\lambda^2}{\pi D^2_{tel}}
\label{eq:omega_tel}
\end{equation}
o\`u $D_{tel}$ est le diam\`etre du t\'elescope et $\eta_{ouv}$
repr\'esente l'efficacit\'e d'ouverture du t\'elescope, \cad que
$\eta_{ouv}\times D_{tel}$ est la taille effective du t\'elescope qui
d\'efinit la forme de la PSF. Cette formule suppose que le t\'elescope
soit utilis\'e \`a sa limite de diffraction
(d'apr\`es~\shortciteNP{rohlfs}).

Le tableau~\ref{tab:calib_perfobs_NEP_param} donne les valeurs
num\'eriques des diff\'erents param\`etres pour chacune des bandes du
Photom\`etre.
\begin{table}
  \begin{center}
    \setlength\extrarowheight{4pt}
    \begin{tabular}[]{p{0.cm}p{7cm}p{2cm}p{2cm}p{2cm}p{0.cm}}
      \toprule
      &\bfseries{Bandes PACS [$\mathbf{\mu}$m]}   & \bfseries{bleue} & \bfseries{Verte} & \bfseries{Rouge} & \\
      \hline \hline
      &Longueur d'onde centrale $\lambda_0$ [$\mu$m]  & 72.5 & 107.5 & 170 & \\
      &Fr\'equence centrale $\nu_0$ [THz]  & 4.14 & 2.79 & 1.76 & \\
      &Largeur spectrale $\delta \nu$ [THz]    & 1.47 & 1.22 & 0.88 &  \\
      \hline \hline
      &Transmission totale $t$  & 0.366 & 0.334 & 0.395 &\\
      \hline \hline
      &Taille pixel $l_{pix}$ [$''$] & 3.2 & 3.2 & 6.4 &\\
      &Angle solide pixel $\Omega_{pix}$ [$\times 10^{-10}$~sr] & 2.40 & 2.40 & 9.63 &\\
      \hline \hline
      &Angle solide t\'elescope $\Omega_{tel}$ [$\times 10^{-10}$~sr] & 9.55 & 13.63 & 34.10 &\\
      &Efficacit\'e t\'elescope $\eta_{tel}$ & 0.64 & 0.73 & 0.77 &\\
      &Surface effective $A_{eff}$ [m$^2$] & 8.48 & 8.48 & 8.48 &\\
      \bottomrule
    \end{tabular}
  \caption[Valeurs num\'eriques utilis\'ees pour le calcul de
  sensibilit\'e]{Synth\`ese des param\`etres utiles au calcul de la
  sensibilit\'e du Photom\`etre PACS. La transmission totale $t$ est
  le produit de la transmission des filtres et dichro\"ique (cf
  figure~\ref{fig:detect_observatoire_phfpu_description_filtre}), de
  la transmission du miroir (87~\%) et de la transmission du diaphragme
  de Lyot (95~\%).
  \label{tab:calib_perfobs_NEP_param}}
  \end{center}
\end{table}
Pour un temps d'observation total $T_{obs}$ de 1~hr, un
signal-\`a-bruit $s/n$ de~5 et une efficacit\'e d'observation
$\eta_{obs}$ de~0.5, l'application num\'erique nous donne la formule
suivante qui relie la sensibilit\'e t\'elescope [W/m$^2$/Hz] \`a la
NEP d\'etecteur [W/$\sqrt{\mbox{Hz}}$]~:

\begin{equation}
f_{\nu}=C_i\times NEP
\label{eq:calib_perfobs_NEP_formule}
\end{equation}
\begin{equation*}
C_{72\,\mu\mbox{m}}=8.05,\,\,\,\, C_{107\,\mu\mbox{m}}=11.13,\,\,\,\,
C_{170\,\mu\mbox{m}}=9.77,\,\,\,\, [\times
10^{-14}\,\,\mbox{m}^{-2}.\mbox{Hz}^{-1/2}]
\end{equation*}

\`A partir de l'\'equation~(\ref{eq:calib_perfobs_NEP_formule}) et du
tableau~\ref{tab:calib_perflabo_compare_nep}, nous pouvons estimer la
sensibilit\'e du Photom\`etre PACS dans ses trois bandes
spectrales. Les r\'esultats\footnote{Les sensibilit\'es donn\'ees ici
ne prennent pas en compte les am\'eliorations de NEP pr\'esent\'ees
dans la section~\ref{sec:calib_perflabo_ajuste} pour le mode DDCS.}
sont pr\'esent\'es dans le tableau~\ref{tab:calib_perfobs_sensib}.\\

\begin{table}
  \begin{center}
    \setlength\extrarowheight{4pt}
    \begin{tabular}[]{p{0.cm}p{6cm}>{\centering}p{1.5cm}>{\centering}p{1.5cm}>{\centering}p{1.5cm}p{0.cm}}
      \toprule
      &Bande PACS [$\mu$m]   & 72.5 & 107.5 & 170 & \\
      \hline \hline
      & Mode direct [mJy (5$\sigma$, 1\,hr)]  & 1.53 & 1.78 & 3.22 & \\
      & Mode DDCS [mJy (5$\sigma$, 1\,hr)]  & 3.78 & 5.23 & 6.25 & \\
      \bottomrule
    \end{tabular}
  \caption[Sensibilit\'e du Photom\`etre PACS en modes direct et
  DDCS]{Estimation des sensibilit\'es du Photom\`etre PACS dans ses
  trois bandes spectrales pour les deux modes de lecture. Ces chiffres
  donnent le flux d'une source ponctuelle en mJy qui serait
  d\'etect\'ee \`a 5~$\sigma$ en une heure d'observation.
  \label{tab:calib_perfobs_sensib}}
  \end{center}
\end{table}




\section{La dynamique du Photom\`etre PACS}
\label{sec:calib_perfobs_dyna}

Dans cette section, nous calculons la dynamique du Photom\`etre PACS,
ce qui permettra aux futurs utilisateurs de l'instrument d'adapter le
mode d'observation et de pr\'evoir d'\'eventuelles saturations lorsque
les champs observ\'es contiennent des objets brillants comme c'est par
exemple le cas pour les observations de condensations pr\'estellaires.

Dans le chapitre~\ref{chap:calib_procedure}, nous avons \'evoqu\'e le
fait que certains r\'eglages des d\'etecteurs pouvaient aboutir \`a de
s\'ev\`eres probl\`emes de saturation. Par exemple, les fortes
tensions de polarisation tendent \`a \og \'etirer \fg les points
milieux, la dispersion du signal devient alors comparable \`a la
dynamique de l'\'electronique chaude, elle peut m\^eme la remplir
enti\`erement pour certaines matrices intrins\`equement dispers\'ees
(cf matrices~7 et~8 sur la
figure~\ref{fig:calib_procedure_prediction_dispersion}). Dans ce cas,
il ne reste aucune marge au signal pour d\'etecter des objets
brillants~; \cad que tout flux incident qui modifierait le niveau de
point milieu conduirait in\'evitablement une fraction de la matrice
\`a saturer les convertisseurs num\'eriques de BOLC.

\begin{figure}
  \begin{center}
    \begin{tabular}{cc}
      \includegraphics[width=0.47\textwidth]{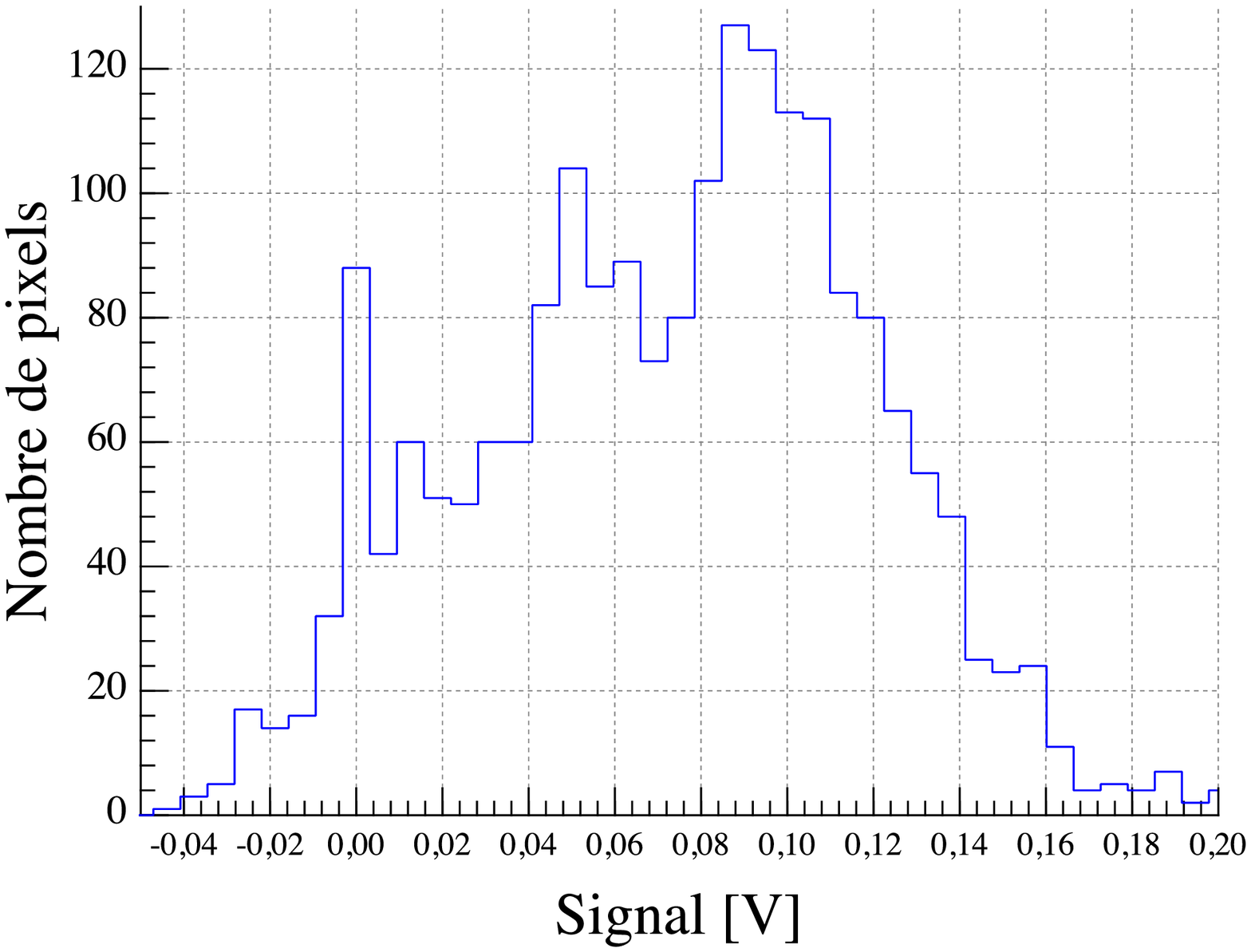}&\includegraphics[width=0.47\textwidth]{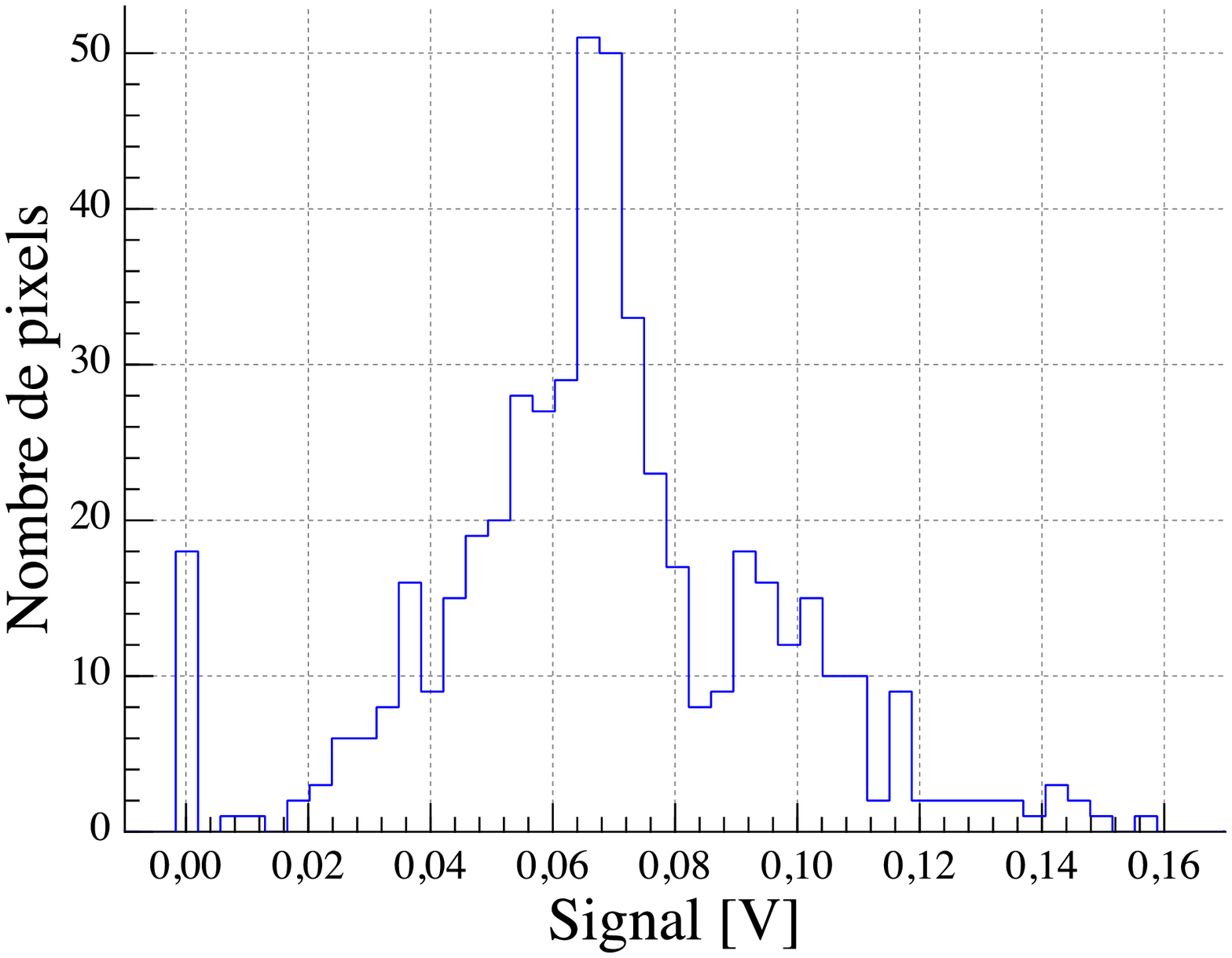}
    \end{tabular}
  \end{center}
  \caption[Histogramme des points milieux des BFP bleu et
  rouge]{Histogrammes des points milieux des BFP bleu (\`a gauche) et
  rouge (\`a droite) obtenus pour la tension de polarisation optimale
  (2.7 et 2.0~V respectivement) et pour un flux incident de 2~pW. La
  dispersion pic-\`a-pic est de~200 et 120~mV respectivement. Les pics
  \`a 0~V correspondent au signal des pixels morts.}
  \label{fig:calib_perfobs_dyna_disp}
\end{figure}

La figure~\ref{fig:calib_perfobs_dyna_disp} pr\'esente l'histogramme
des points milieux des BFP bleu et rouge pour la tension de
polarisation optimale. La dispersion pic-\`a-pic s'\'el\`eve \`a
$\sim$200~mV pour le BFP bleu et \`a $\sim$120~mV pour le rouge. La
dynamique de BOLC \'etant de 330~mV dans le mode nominal d'observation
(cf annexe~\ref{a:dyna_BOLC}), il reste par cons\'equent 130~mV et
210~mV aux BFP bleu et rouge respectivement avant de saturer les ADC
de BOLC. En premi\`ere approximation, nous pouvons diviser ces
tensions par la r\'eponse des bolom\`etres afin d'obtenir le flux
\'equivalent \`a une telle chute de point milieu. Nous trouvons un
flux de 4.3~pW/pixel pour le BFP bleu et de 6.3~pW/pixel sur le
rouge. Notez cependant que ce calcul simpliste sous-estime le flux
\'equivalent puisque la r\'eponse n'est pas constante avec le flux
incident. En effet, la figure~\ref{fig:calib_perflabo_nonlinear_resp}
montre qu'elle chute de 40\% entre~1 et 7~pW/pixel. Un calcul plus
r\'ealiste n\'ecessite d'utiliser des courbes de non-lin\'earit\'e
comme celle pr\'esent\'ee dans la
figure~\ref{fig:calib_perflabo_nonlinear_linea}, et pour la tension de
polarisation optimale, nous trouvons un flux \'equivalent de~$\sim$5
et $\sim$9~pW/pixel pour les BFP bleu et rouge. Nous exprimons ensuite
ces flux en Jansky \`a l'aide de l'\'equation suivante~:
\begin{equation}
P_{source}=t\,\delta\nu\, A_{eff}\,\eta_{tel}\,f_{\nu}\times F
\label{eq:dyna}
\end{equation}
o\`u $F$ est un facteur qui indique la fraction du flux incident dans
le pixel le plus brillant de la PSF. Dans
l'\'equation~(\ref{eq:snr_nep_5}), ce facteur vaut 1/$N_{pix}$~; en
effet nous avions pris la moyenne du flux incident dans une PSF pour
la distribuer sur tous les pixels illumin\'es. Cependant, pour les
calculs de saturation avec source ponctuelle, nous ne pouvons pas
prendre le flux moyen par pixel mais plut\^ot le flux du pixel le plus
brillant. D'apr\`es des simulations r\'ealis\'ees par Koryo Okumura
sur des PSF monochromatiques\footnote{Une estimation plus pr\'ecise du
flux incident sur le pixel le plus brillant n\'ecessiterait
d'int\'egrer les PSF sur la bande spectrale totale en les pod\'erant
par la transmission des filtres.}, nous trouvons $F=0.31$ \`a
70~$\mu$m, 0.165 \`a 100~$\mu$m et 0.25 \`a 160~$\mu$m. 
de sources \'etendues, ce facteur vaut~1.




En supposant que la tension $V_{hb}$ soit r\'egl\'ee de fa\c{c}on \`a
placer les points milieux \og en haut \fg de la dynamique de BOLC, et
que les points milieux les plus bas passent quand m\^eme le CD, nous
obtenons les limites de saturation dans les trois bandes du
photom\`etre PACS qui sont pr\'esent\'ees dans le
tableau~\ref{tab:calib_perfobs_dyna_Jy}.

\begin{table}
  \begin{center}
    \setlength\extrarowheight{4pt}
    \begin{tabular}[]{p{0.cm}p{9cm}>{\centering}p{1.2cm}>{\centering}p{1.2cm}>{\centering}p{1.2cm}p{0.cm}}
      \toprule
      &Bande PACS [$\mu$m]   & 70 & 100 & 160 & \\
      \hline \hline
      & Dispersion points milieux [mV]  & 200 & 200 & 120 & \\
      & Dynamique d\'etecteurs [pW/pixel]  & 5 & 5 & 9 & \\
      & Dynamique instrument source ponctuelle [Jy]  & 550 & 1200 & 1600 & \\
      \bottomrule
    \end{tabular}
  \caption[Dynamique du Photom\`etre PACS]{Dispersion et dynamique du
  Photom\`etre PACS dans ses trois bandes spectrales.
  \label{tab:calib_perfobs_dyna_Jy}}
  \end{center}
\end{table}




\section{La d\'erive du signal}
\label{sec:calib_perfobs_oof}

Comme nous l'avons pr\'esent\'e dans la
section~\ref{sec:intro_bolometrie_thermo_principe_bruit_autre}, tout
syst\`eme \'electronique poss\`ede un bruit basse fr\'equence,
g\'en\'eralement de la forme 1/f, qui se traduit dans l'espace r\'eel
par une lente d\'erive du signal \'electrique. Ce genre d'effet
instrumental est potentiellement n\'efaste pour des observations
astronomiques puisqu'il emp\^eche de distinguer les v\'eritables
variations de flux incident par rapport \`a la d\'erive \'electrique
du d\'etecteur. Il est donc n\'ecessaire de quantifier ce
ph\'enom\`ene afin d'adapter les m\'ethodes d'observation aux
carat\'eristiques des bolom\`etres.

\begin{figure}
  \begin{center}
    \begin{tabular}{cc}
      \multicolumn{2}{c}{\includegraphics[width=0.97\textwidth,angle=0]{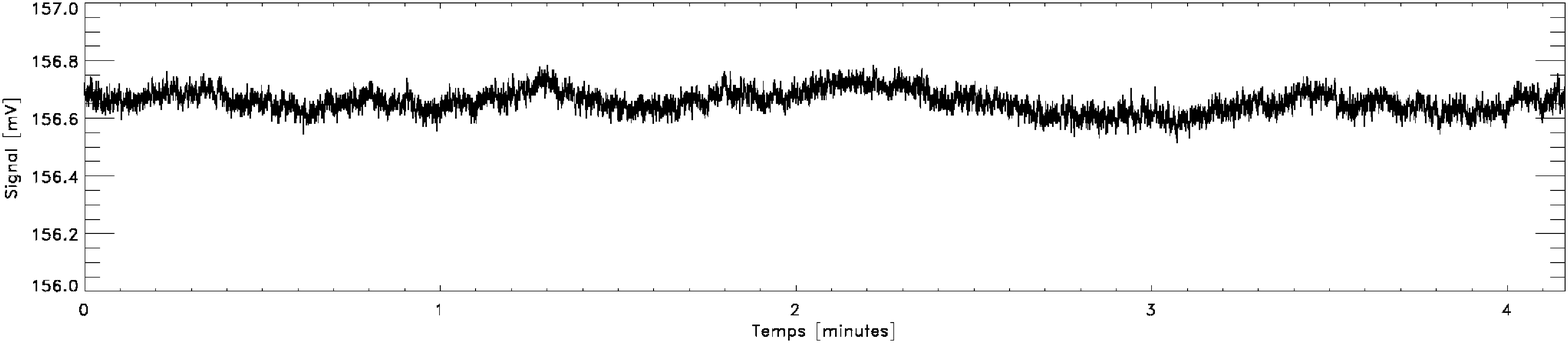}}  \\
      \includegraphics[width=0.47\textwidth,angle=0]{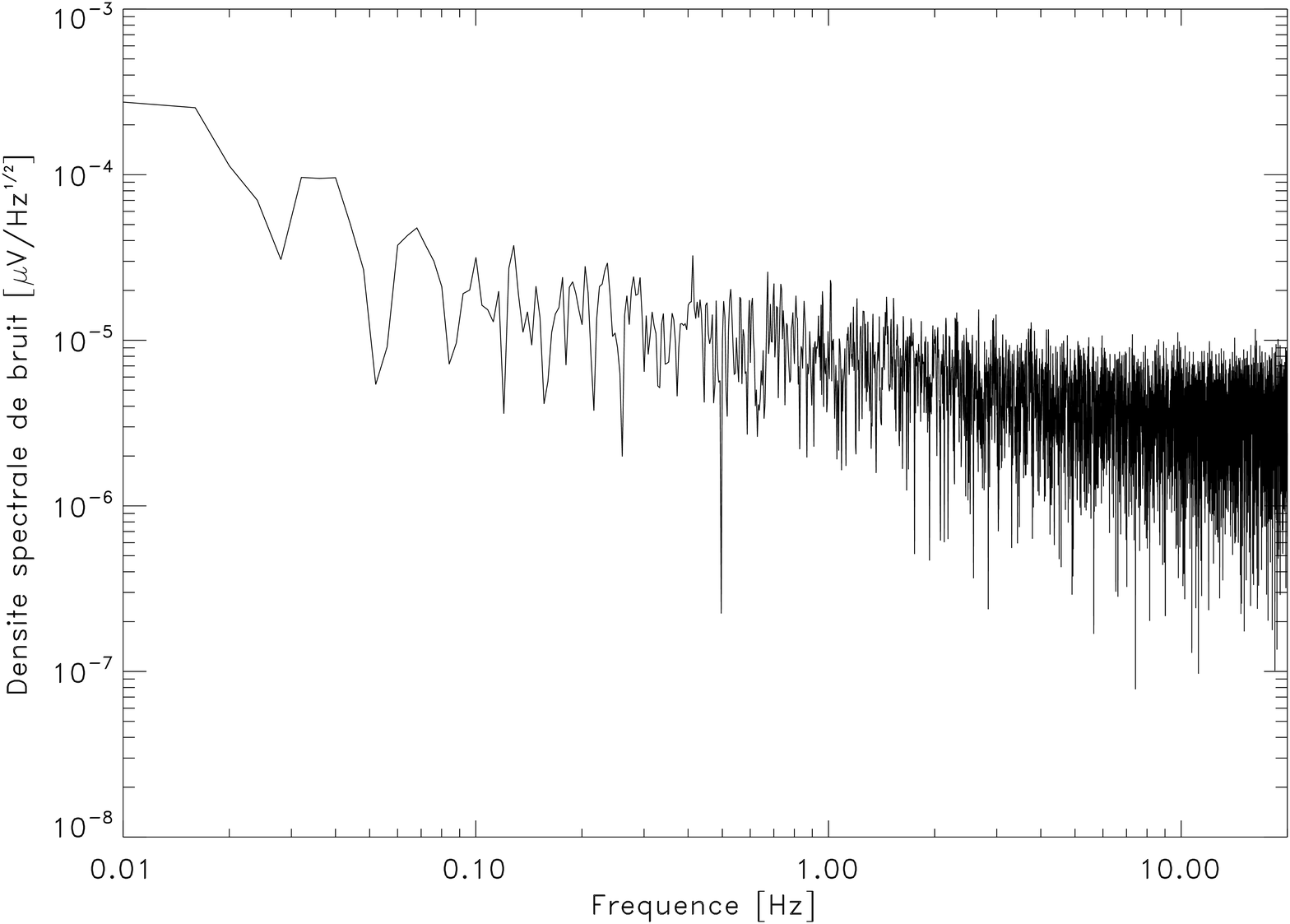}&\includegraphics[width=0.47\textwidth,angle=0]{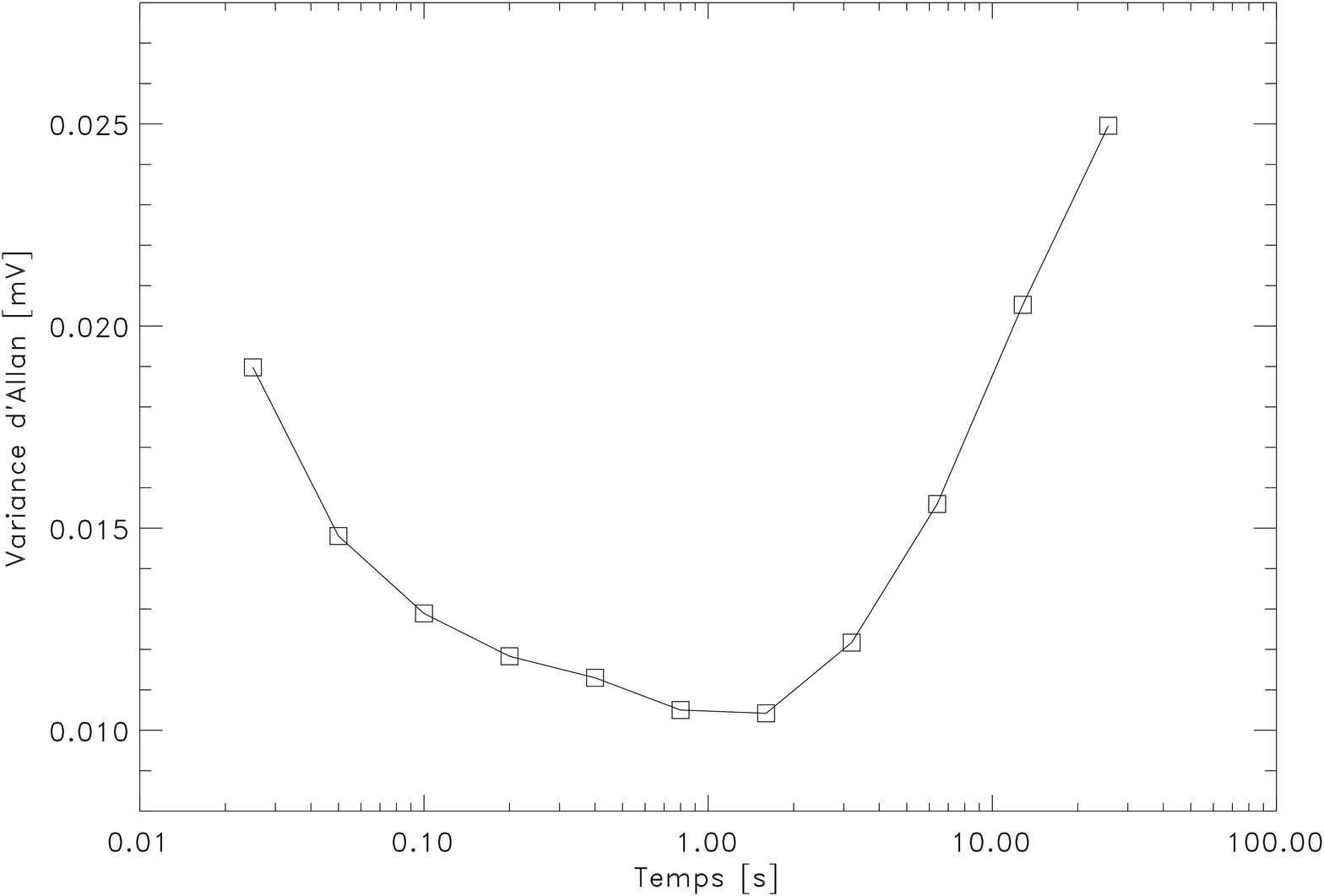}
    \end{tabular}
  \end{center}
  \caption[Stabilit\'e des bolom\`etres~: espace de Fourier Vs
  variance d'Allan]{\'Etude de la stabilit\'e du signal
  bolom\'etrique. La courbe du haut montre l'\'evolution du signal
  d'un pixel bleu sur une dur\'ee de quelques minutes. En bas \`a
  gauche se trouve la densit\'e spectrale de bruit de ce signal. Le
  coude de remont\'ee en 1/f est difficilement rep\'erable. En bas \`a
  droite, nous pr\'esentons la variance d'Allan calcul\'ee pour ce
  m\^eme signal. Le minimum de la courbe donne le \og temps de
  coh\'erence \fg du signal, il nous indique qu'au-del\`a de 1~s la
  d\'erive du signal domine le niveau de bruit blanc des
  bolom\`etres.}
  \label{fig:calib_perfobs_oof_allan}
\end{figure}

Jusqu'\`a pr\'esent, nous avons rencontr\'e ces d\'erives basses
fr\'equences dans l'espace de Fourier uniquement. Par exemple, la
figure~\ref{fig:calib_perflabo_tau_fourier_spec4polar} montre des
densit\'es spectrales de bruit obtenues pour des mesures de 3~heures~;
ces spectres \'etant moyenn\'es sur une matrice enti\`ere, ils
poss\`edent tr\`es peu de fluctuations statistiques et r\'ev\`elent
ainsi une remont\'ee basse fr\'equence autour de 1~Hz.  Cependant, sur
des densit\'es spectrales de bruit mesur\'ees sur des pixels
individuels, le coude de remont\'ee du bruit basse fr\'equence est \og
noy\'e \fg dans les fluctuations statistiques et il est tr\`es
difficile de d\'eterminer clairement la position de ce coude. La
figure~\ref{fig:calib_perfobs_oof_allan} montre par exemple un signal
temporel de 4~minutes et le spectre correspondant sur lequel ni le
coude du bruit en 1/f ni la fr\'equence de coupure du bolom\`etre
n'est clairement visible. Nous nous tournons alors vers une technique
qui est habituellement utilis\'ee pour caract\'eriser la stabilit\'e
des d\'etecteurs h\'et\'erodynes mais qui peut s'appliquer plus
g\'en\'eralement \`a un signal bolom\'etrique comme le notre~: il
s'agit de la variance d'Allan \shortcite{allan}. En termes
simplifi\'es, cette quantit\'e repr\'esente l'\'evolution de la
variance du signal filtr\'e par une moyenne glissante en fonction de
la taille de ce filtre. \shortciteN{ossenkopf} pr\'esente par exemple
un calcul d\'etaill\'e de la variance d'Allan qu'il applique ensuite
aux d\'etecteurs de l'instrument Herschel/HIFI. D'autre part,
l'article de \shortciteN{schieder} montre que pour un signal qui
contient du bruit blanc ainsi que du bruit de d\'erive, comme c'est le
cas pour les matrices de bolom\`etres, alors la variance d'Allan
devrait pr\'esenter un minimum indiquant la p\'eriode de temps \`a
partir de laquelle la d\'erive du signal commence \`a dominer le bruit
blanc du d\'etecteur. Ce temps caract\'eristique donne un \og temps de
coh\'erence \fg du signal qui nous permettra d'optimiser les modes
d'observation de l'instrument. Notez que la d\'erive basse fr\'equence
du signal ne d\'epend pas de la tension de polarisation des
bolom\`etres (cf
figure~\ref{fig:calib_perflabo_tau_fourier_spec4polar}) ni du flux
incident sur les d\'etecteurs (cf annexe~\ref{a:spectres}).

Nous avons donc calcul\'e la variance d'Allan du signal de la
figure~\ref{fig:calib_perfobs_oof_allan} pour comparer le r\'esultat
\`a l'analyse de Fourier. Alors que le coude du bruit basse
fr\'equence n'est pas visible sur le spectre, le minimum de la
variance d'Allan est facilement d\'etectable et donne un temps de
coh\'erence de l'ordre de la seconde, ce qui co\"incide parfaitement
avec la fr\'equence de coude de $\sim$1~Hz que nous avons trouv\'ee
\`a partir de la
figure~\ref{fig:calib_perflabo_tau_fourier_spec4polar}. Puisque les
deux m\'ethodes fournissent des r\'esultats similaires, nous pouvons
affirmer avec confiance que l'\'echelle de temps des d\'erives du
signal est effectivement d'une seconde. De plus, nous avons montr\'e
que cette d\'erive ne s'applique qu'\`a la partie
additive\footnote{Nous d\'ecomposons le signal de la fa\c{c}on
suivante~: $Signal=gain\times flux\,+\,of\!fset$ o\`u l'$of\!fset$ est
ce que nous appelons la partie additive et le $gain$ est la partie
multiplicative.} du signal. En effet, en utilisant les sources
internes d'\'etalonnage du banc de test, nous avons mesur\'e
r\'eguli\`erement l'\'evolution du gain des d\'etecteurs lors de la
campagne d'\'etalonnage, et nous avons montr\'e qu'il fluctue
d'environ 0.1~\% sur des p\'eriodes de plusieurs heures.

Des techniques d'observation permettent de s'affranchir efficacement
de cet effet instrumental. Le principe repose sur la modulation du
signal \`a l'aide du miroir secondaire du t\'elescope,
g\'en\'eralement appel\'e \emph{chopper} ou \emph{wobbler}, pour
observer alternativement le champ o\`u se trouve la source et le fond
de ciel puis un champ vide qui ne contient que le fond de ciel~;
l'objectif \'etant de soustraire les d\'erives additives introduites
par l'instrument ou par les \'emetteurs d'avant-plan dans le cas
d'observations au sol, \cad l'atmosph\`ere. Le t\'elescope est ensuite
g\'en\'eralement \emph{nodd\'e} pour corriger les erreurs introduites
par la diff\'erence de chemin optique entre les deux positions
chopp\'ees. Dans le cas de PACS, le temps de coh\'erence \'etant
d'environ une seconde, il est n\'ecessaire d'effectuer un cycle
chopper toute les secondes de sorte \`a garder l'information sur
l'offset avant qu'il ne d\'erive. La fr\'equence du chopper PACS
devrait par cons\'equent \^etre de $\sim$2~Hz. Notez qu'il est
possible de se dispenser d'un chopper pour moduler le signal. Les
articles de \shortciteN{weferling} et \shortciteN{reichertz} proposent
en effet une technique d'observation qu'ils appellent
\emph{fastscanning} qui consiste \`a balayer le ciel relativement vite
pour que ni les offsets des d\'etecteurs ni l'atmosph\`ere n'aient le
temps de d\'eriver. Ils utilisent entre autre la redondance
d'information sur une matrice de d\'etecteurs pour corriger les
fluctuations de l'atmosph\`ere (plusieurs pixels observent la m\^eme
r\'egion du ciel sur des p\'eriodes plus courtes que le temps de
coh\'erence).

D'autre part, nous avons tent\'e de r\'eduire le bruit basse
fr\'equence des bolom\`etres en cherchant une possible corr\'elation
entre les d\'erives du signal et les d\'erives en temp\'erature des
d\'etecteurs. Nous avons trouv\'e que, pour une mesure longue d'une
heure, la temp\'erature du plan focal d\'erive de fa\c{c}on monotone
sur seulement $\sim$30~$\mu$K~; et qu'une si faible variation
n'affecte pas significativement le signal, \cad que le spectre du
signal brut et celui du signal d\'ecorr\'el\'e se superposent
parfaitement \`a l'exception des premiers points du spectre. En mode
nominal d'observation, le syst\`eme est extr\`emement stable
thermiquement, relativement au niveau de bruit blanc.

\begin{figure}
  \begin{center}
    \begin{tabular}{ll}
      \includegraphics[width=0.48\textwidth,angle=0]{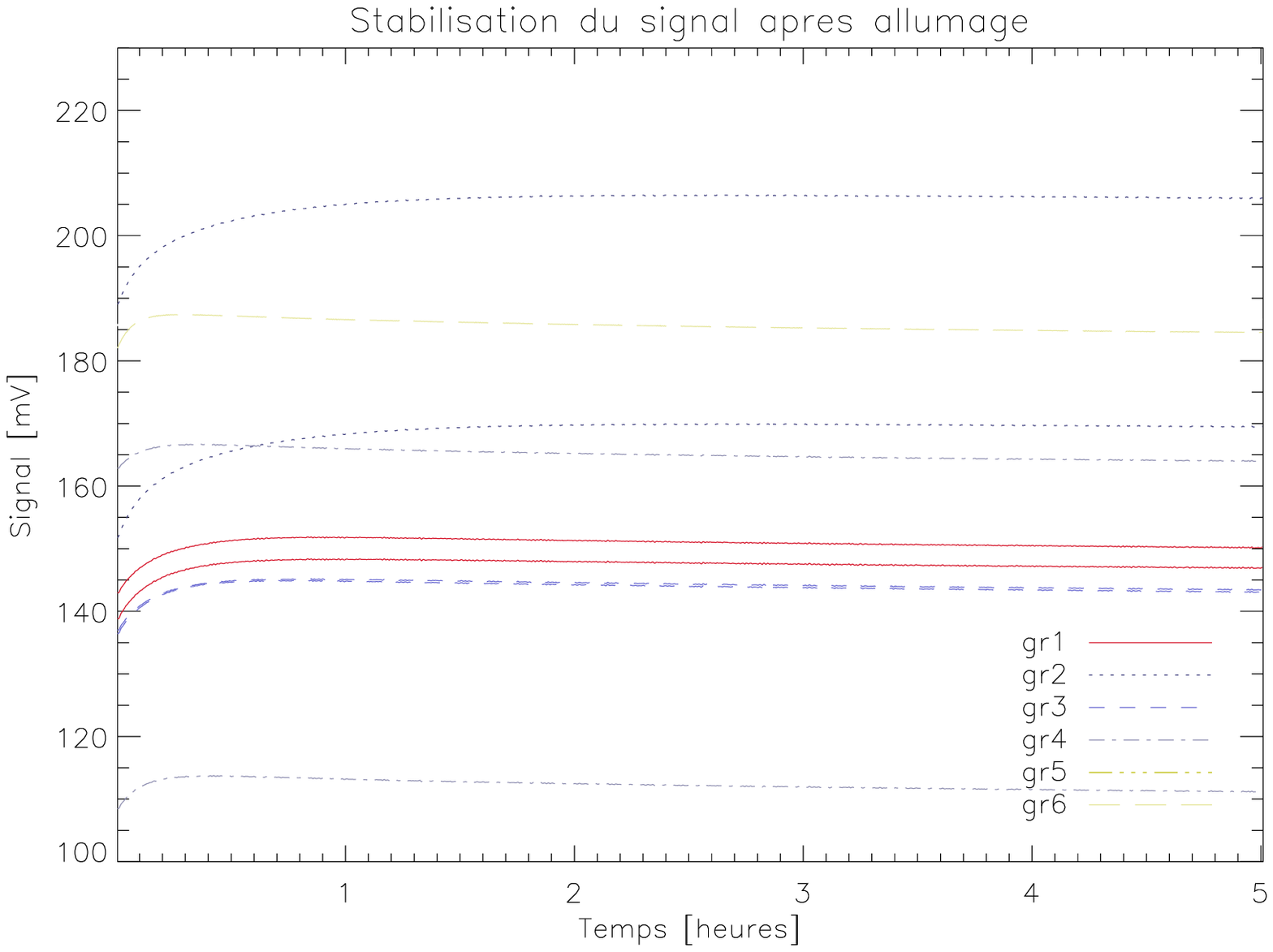}&\includegraphics[width=0.48\textwidth,angle=0]{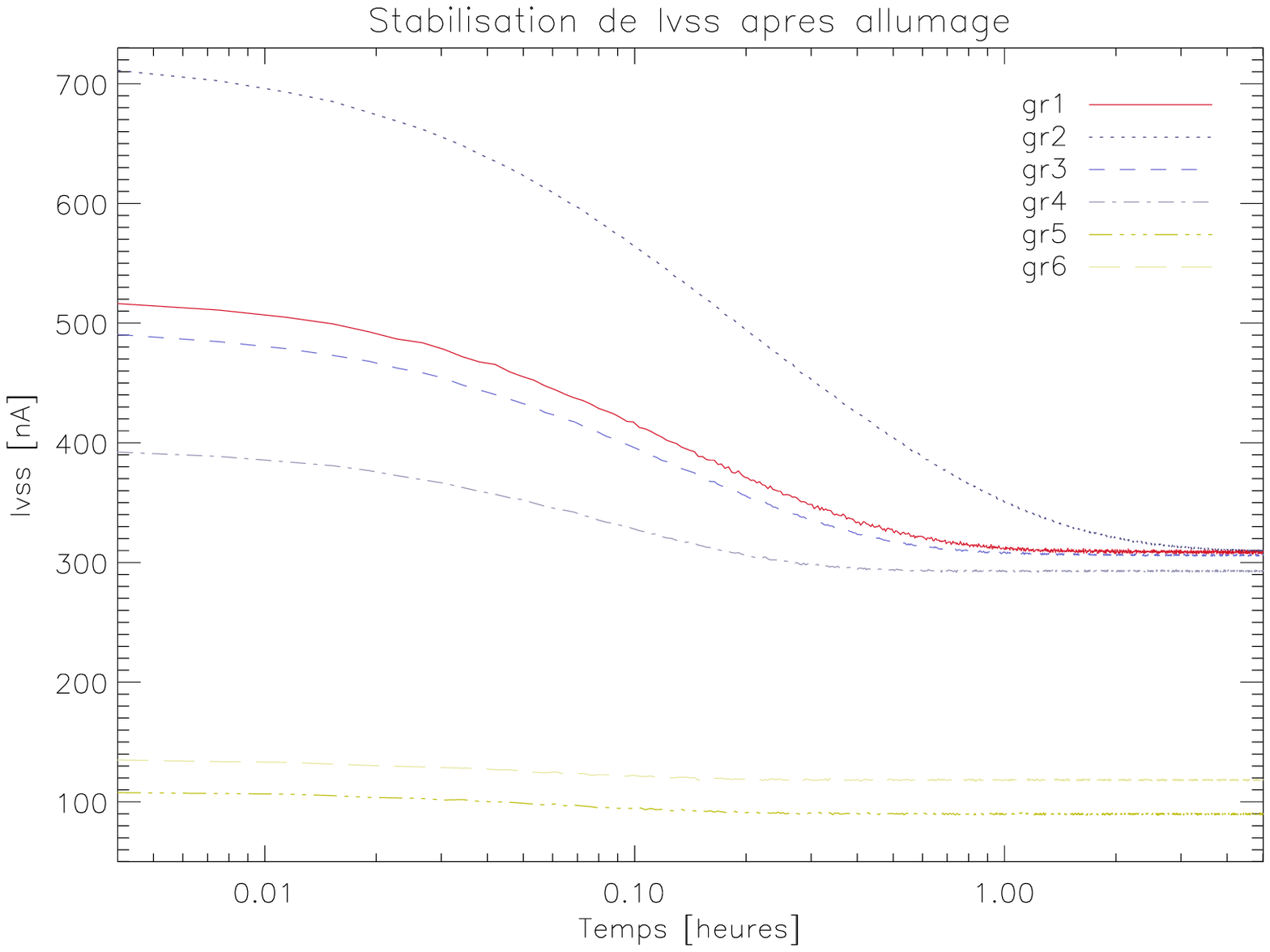}\\
      \includegraphics[width=0.48\textwidth,angle=0]{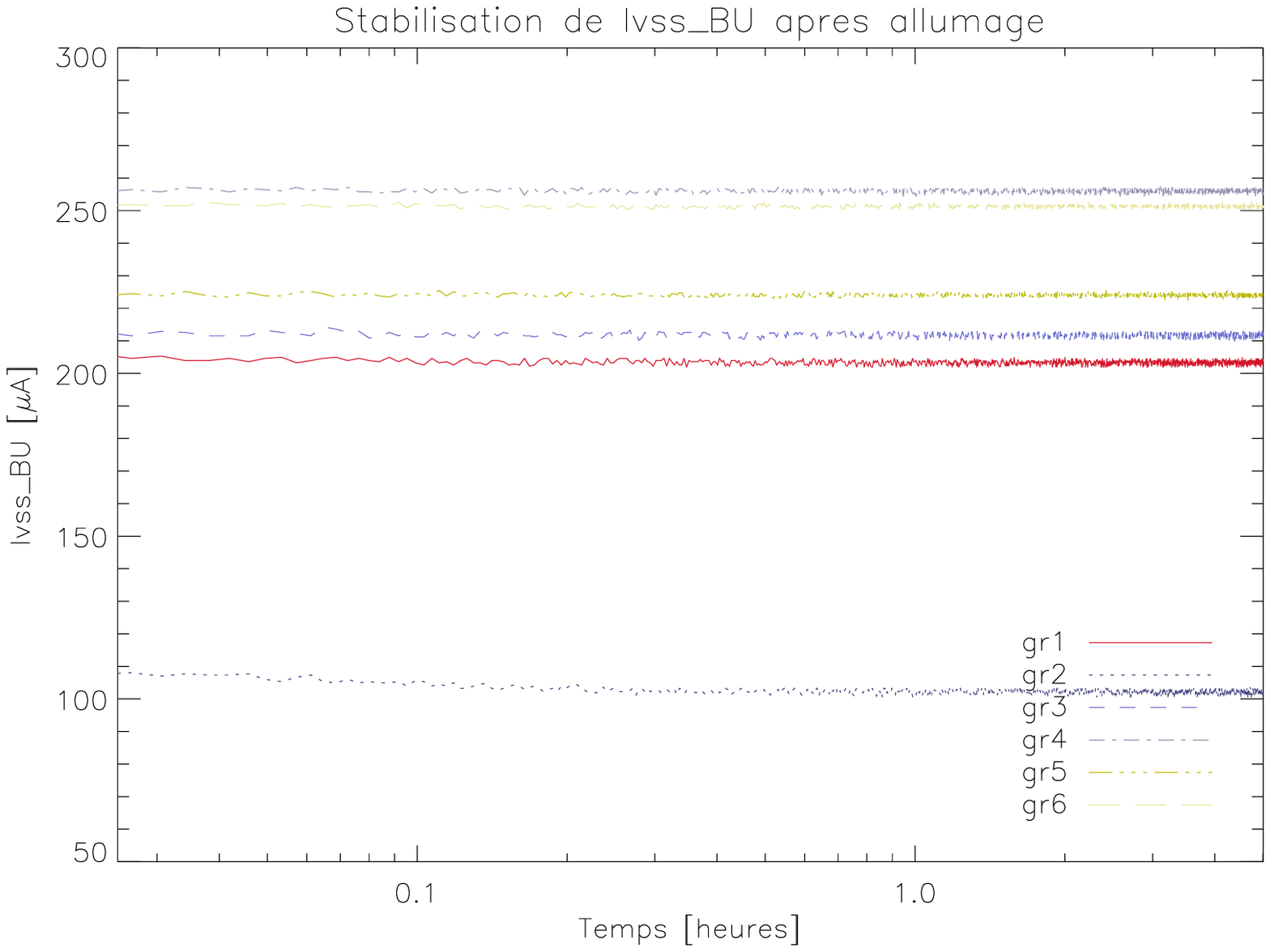}&\includegraphics[width=0.48\textwidth,angle=0]{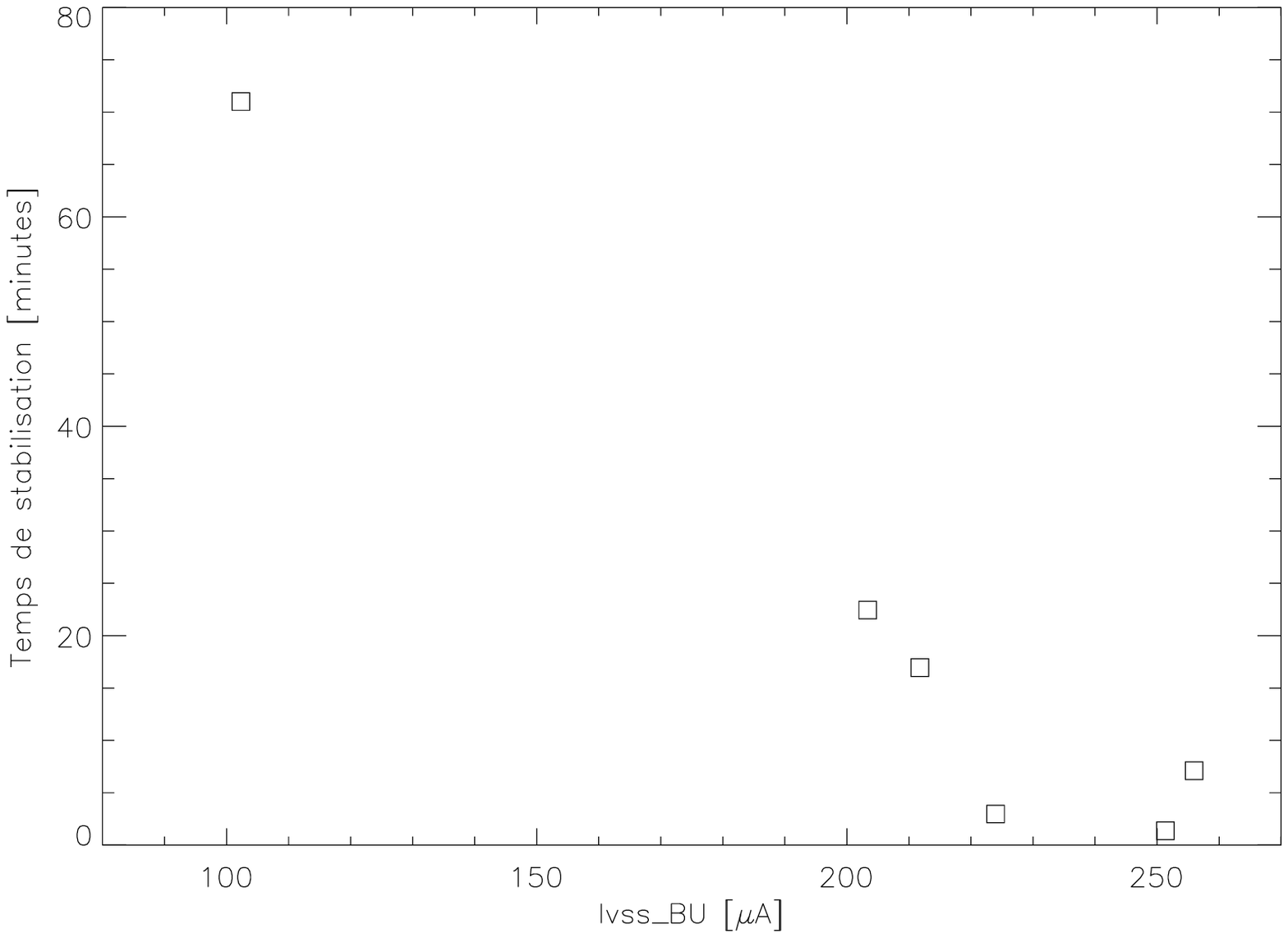}
    \end{tabular}
  \end{center}
  \caption[Stabilisation des d\'etecteurs apr\`es
  allumage]{Corr\'elation entre le courant qui circule dans le BU et
  le temps de stabilisation des d\'etecteurs apr\`es la mise en marche
  de l'instrument. \emph{En haut \`a gauche~:} \'Evolution du signal
  moyen des 6~groupes de d\'etecteurs apr\`es l'allumage de
  PACS. \emph{En haut \`a droite~:} \'Evolution du courant qui circule
  dans le CL, Ivss, de chacun des groupes. Les groupes~5 et~6 (BFP
  rouges) se stabilisent tr\`es rapidement alors que le groupe~2
  n\'ecessite plus d'une heure. \emph{En bas \`a gauche~:} Le courant
  qui circule dans les BU se stabilisent tr\`es rapidement. \emph{En
  bas \`a droite~:} Plus le courant du BU est grand, plus les
  d\'etecteurs se stabilisent rapidement. Le Ivss\_BU du groupe~2 est
  aujourd'hui r\'egl\'e \`a 300~nA comme pour les autres groupes du
  BFP bleu.}
  \label{fig:sec:calib_perfobs_oof_correlSwitchON}
\end{figure}


Toutefois, lors de la mise sous tension des d\'etecteurs, le signal
peut significativement d\'eriver. En effet, l'instrument PACS n'est
pas op\'erationnel d\`es lors qu'il est polaris\'e, il est
n\'ecessaire d'attendre que toutes les charges se repartissent dans le
circuit de lecture et que les d\'etecteurs soient thermalis\'es. Le
temps de stabilisation peut atteindre 80 minutes suivant le r\'eglage
de l'\'electronique de lecture, temps durant lequel aucune observation
n'est r\'ealisable. La
figure~\ref{fig:sec:calib_perfobs_oof_correlSwitchON} montre les
r\'esultats d'une mesure de stabilisation effectu\'ee sur les deux BFP
du mod\`ele de vol du Photom\`etre PACS. Les BFP sont allum\'es au
temps t=0 et re\c{c}oivent un flux constant de 2~pW/pixel. Les
diff\'erents groupes ne se stabilisent pas \`a la m\^eme vitesse. Nous
avons \'egalement trac\'e l'\'evolution du courant qui circule dans le
CL, \emph{Ivss}, qui met un temps similaire pour se stabiliser, et le
courant qui circule dans le BU, \emph{Ivss\_BU}, qui, lui, se
stabilise quasi-instantan\'ement. Notez la corr\'elation que nous
avons trouv\'ee entre Ivss\_BU et le temps de stabilisation du
signal. Dans la figure~\ref{fig:sec:calib_perfobs_oof_correlSwitchON},
ce temps de stabilisation est d\'efini comme \'etant le temps
n\'ecessaire au signal pour atteindre 90~\% de sa valeur
asymptotique. Lorsque le circuit de lecture est sous-aliment\'e comme
c'est le cas du groupe~2 dans ce jeu de donn\'ees, il faut attendre
plus d'une heure avant de pouvoir utiliser les matrices. Le groupe~2
est aujourd'hui aliment\'e comme les autres groupes bleus, \cad avec
un courant Ivss de l'ordre de 300~nA.

\section{Balayage du ciel}
\label{sec:calib_perfobs_scan}


Le t\'elescope Herschel offre un mode de pointage dit de balayage qui
est particuli\`erement adapt\'e aux grands relev\'es du ciel. De
nombreux programmes scientifiques, galactiques et extragalactiques,
utiliseront ce type d'observation. Dans ce paragraphe je commence par
d\'ecrire bri\`evement le mode d'observation par balayage, puis je
pr\'esente le r\'esultat de mes travaux sur la d\'egradation de la PSF
due aux effets instrumentaux tels que la constante de temps des
bolom\`etres et la compression \`a bord des donn\'ees. Pour quantifier
la d\'egradation de la PSF en fonction des r\'eglages et
caract\'eristiques de l'instrument, j'utilise le simulateur PACS qui a
\'et\'e d\'evelopp\'e par le groupe ICC\footnote{\emph{Instrument
Control Center}~: Groupe de scientifiques, dont je fais partie, qui
est responsable de l'utilisation d'un instrument vis-\`a-vis de
l'ESA~; \cad l'\'etalonnage, la d\'efinition des modes d'observation,
etc...}  pour pr\'eparer les op\'erations et l'exploitation de
l'instrument.


\subsection{Le mode d'observation par balayage}
\label{sec:calib_perfobs_scan_scan}

\begin{figure}
  \begin{center}
      \includegraphics[width=0.7\textwidth,angle=0]{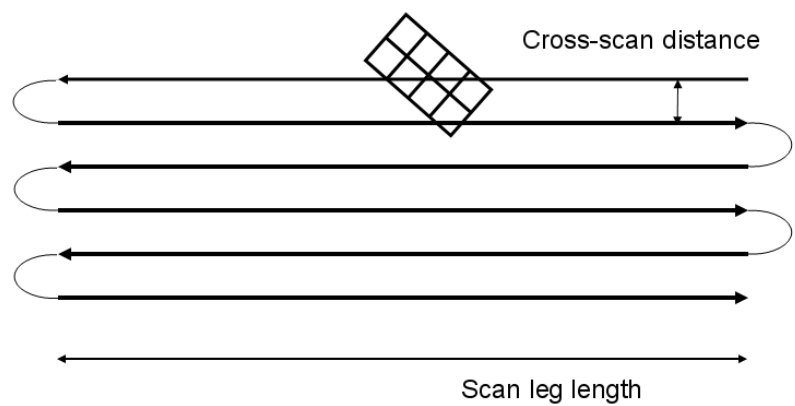}
  \end{center}
  \caption[Principe d'une observation en mode balayage]{Exemple de
  cartographie du ciel en mode balayage avec six lignes (\emph{scan
  legs}). Arriv\'e au bout de la premi\`ere ligne, le satellite
  d\'ec\'el\`ere et repart dans la direction inverse. Lorsqu'il
  atteint la seconde ligne, sa vitesse est d\'ej\`a stabilis\'ee \`a
  la valeur sp\'ecifi\'ee par l'utilisateur, et ainsi de suite
  jusqu'\`a la fin de l'observation. Plus la vitesse de d\'erive du
  satellite est grande, et plus le temps de man\oe uvre n\'ecessaire
  pour changer de ligne est grand. Cette figure est extraite de \og
  PACS Observer's manual \fg disponible sur
  http://herschel.esac.esa.int/Docs/PACS/pdf/pacs\_om.pdf.
  \label{fig:calib_perfobs_scan_scanlegs}}
\end{figure}
Dans ce mode de pointage, le satellite d\'erive \`a vitesse constante
le long de lignes parall\`eles\footnote{Ce sont en fait des portions
de grands cercles que nous pouvons assimiler \`a des lignes
parall\`eles sur de courtes distances. La longueur de ces lignes est
d'ailleurs limit\'ee \`a 20~degr\'es.} et le t\'elescope balaye ainsi
le ciel en suivant une g\'eom\'etrie typique illustr\'ee dans la
figure~\ref{fig:calib_perfobs_scan_scanlegs}. La taille de la r\'egion
observ\'ee est d\'etermin\'ee par le nombre de lignes balay\'ees,
l'espacement entre ces lignes (inf\'erieur au champ de vue de la
cam\'era) et leur longueur. L'observateur d\'efinit \'egalement la
direction du balayage dans le r\'ef\'erentiel du satellite ou bien du
ciel. Et enfin le dernier param\`etre n\'ecessaire pour d\'efinir
enti\`erement une observation est la vitesse angulaire \`a laquelle le
t\'elescope balaye le ciel. Par souci de simplicit\'e, l'ESA a choisi
de n'offrir que trois vitesses de balayage dans ce mode d'observation:
10, 20 et 60~secondes d'arc par seconde (not\'e [$''$/sec]).\\ En mode
balayage le t\'elescope est constamment en mouvement, mais il reste
\`a tout moment \`a l'ombre du Soleil pour garder l'\'equilibre
thermique du miroir. Il pivote \`a vitesse constante autour de l'axe
Soleil-Terre et d\'ecrit des lignes parall\`eles dans le ciel. Au bout
de chacune de ces lignes, le satellite d\'ec\'el\`ere et repart dans
la direction inverse pour balayer la ligne suivante. Mais avec une
masse de pr\`es de 3.5~tonnes, le satellite Herschel poss\`ede une
grande inertie de sorte que le temps de man\oe uvre n\'ecessaire pour
que le t\'elescope fasse demi-tour peut atteindre une fraction
non-n\'egligeable du temps d'observation. Nous d\'efinissons
l'efficacit\'e du mode d'observation par balayage comme \'etant le
rapport $\frac{\mbox{\small temps d'observation}}{\mbox{\small temps
total}}$, o\`u le temps d'observation correspond au temps pass\'e \`a
observer le champ d\'esir\'e, \cad lorsque le t\'elescope a une
vitesse constante, et le temps total est le temps d'observation plus
le temps de man\oe uvre.

\begin{figure}
  \begin{center}
      \includegraphics[width=0.7\textwidth,angle=0]{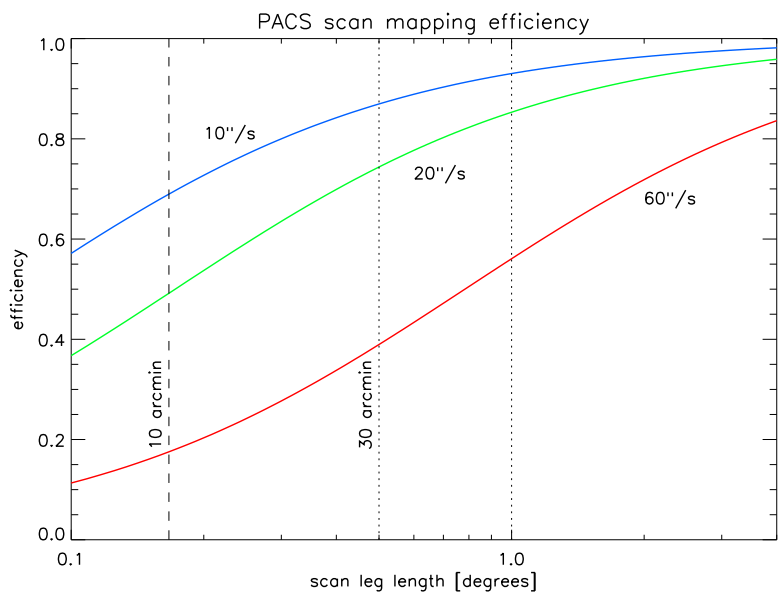}
  \end{center}
  \caption[Efficacit\'e du mode d'observation par
  balayage]{Efficacit\'e compar\'ee du mode d'observation par balayage
  pour trois vitesses en fonction de la longueur de la r\'egion \`a
  cartographier. Le mode d'observation par balayage est
  particuli\`erement bien adapt\'e \`a la cartographie de grandes
  r\'egions du ciel, c'est-\`a-dire sup\'erieures \`a
  $15'\times15'$. Image reproduite avec l'aimable autorisation de
  Bruno Altieri, ESA.
  \label{fig:calib_perfobs_scan_efficiency}}
\end{figure}

Plus la vitesse de balayage est grande, plus le temps pass\'e
en-dehors de la r\'egion \`a cartographier est grande (longue
d\'ec\'el\'eration), et plus l'efficacit\'e diminue. D'autre part,
pour une vitesse donn\'ee, le temps pour faire demi-tour reste
constant et l'efficacit\'e augmente avec la longueur des lignes
balay\'ees, \cad le temps effectif d'observation. La
figure~\ref{fig:calib_perfobs_scan_efficiency} pr\'esente
l'\'evolution de cette efficacit\'e en fonction de la longueur des
lignes pour les trois vitesses de balayage ouvertes \`a la
communaut\'e scientifique. Notez qu'il existe un autre mode
d'observation qui est plus efficace que le mode balayage pour
cartographier les r\'egions du ciel plus petites que $15'\times 15'$~:
c'est le mode raster. Ce mode se rapproche du mode de balayage dans le
sens o\`u le d\'eplacement du t\'elescope d\'ecrit \'egalement un
zigzag (comme sur la figure~\ref{fig:calib_perfobs_scan_scanlegs}),
mais au lieu de d\'eriver \`a vitesse constante, le satellite effectue
des pointages stationnaires. Le mode raster est en quelque sorte un
mode de balayage discr\'etis\'e, la distance entre chaque pointage
\'etant inf\'erieure au champ de vue de la cam\'era. Le temps
n\'ecessaire pour passer d'une position \`a l'autre pouvant atteindre
30~secondes, les observations par raster ne sont pas adapt\'ees aux
grands relev\'es du ciel.\\
Du point de vue de l'observateur, les donn\'ees acquises lors des
man\oe uvres du t\'elescope (c'est-\`a-dire entre chaque lignes
balay\'ees) ne sont pas consid\'er\'ees comme des donn\'ees
scientifiquement exploitables. Il faut donc minimiser ces temps de
man\oe uvre en choisissant un mode d'observation et des param\`etres
adapt\'es \`a la taille du champ \`a cartographier. Cependant, du
point de vue de l'op\'erateur du t\'elescope, les temps de man\oe uvre
repr\'esentent une parfaite occasion pour effectuer des mesures
d'\'etalonnage. En effet, ces \og temps morts\fg s'intercalent \`a
intervalles r\'eguliers durant une m\^eme observation, et il est
tout-\`a-fait possible d'imaginer que les phases
d'acc\'el\'eration-d\'ec\'el\'eration du t\'elescope servent \`a
mesurer le gain des d\'etecteurs r\'eguli\`erement. Une telle
proc\'edure n'a pas encore \'etait clairement d\'efinie par l'ICC
PACS, mais il ne faut pas perdre l'occasion de mettre \`a profit ce
temps pour \'etalonner la cam\'era et potentiellement am\'eliorer la
qualit\'e des observations. Nous pourrions par exemple mesurer le gain
des bolom\`etres \`a intervalle r\'egulier en modulant le signal avec
les deux sources internes de PACS comme nous l'avons d\'ej\`a fait
pour mesurer la r\'eponse des bolom\`etres dans la
section~\ref{sec:calib_perflabo_sensibilite_reponse}.


\subsection{Vitesse de balayage et r\'eglage des bolom\`etres}
\label{sec:calib_perfobs_scan_tau}

Les bolom\`etres poss\`edent une r\'eponse temporelle de la forme
$e^{-\frac{t}{\tau}}$ o\`u $\tau$ est une constante de temps qui
d\'epend entre autre de la tension de polarisation des ponts
bolom\'etriques (cf~sec.~\ref{sec:calib_perflabo_tau_compare}). Dans
l'espace de Fourier l'existence de cette constante de temps se traduit
par un filtre passe-bas du premier ordre dont la fonction de transfert
complexe est:
\begin{center}
  \begin{equation}
    F(\omega) = \frac{1}{1+i\omega\tau}
  \label{eq:filter_freq_bp}
  \end{equation}
\end{center}
o\`u $\omega=2\pi\nu$ est la fr\'equence angulaire, ou pulsation. Par
ailleurs, le mode d'observation par balayage n'utilise pas le chopper
de l'instrument pour moduler le signal, c'est le mouvement du
t\'elescope qui joue ce r\^ole. En effet, l'image du ciel se forme au
niveau du plan focal et d\'efile sur les d\'etecteurs de sorte que les
fr\'equences spatiales pr\'esentes dans la sc\`ene observ\'ee se
retrouvent en fr\'equences temporelles dans le signal
bolom\'etrique. D'apr\`es \shortciteN{hanany}, nous pouvons \'ecrire
$\omega=\vec{k}\cdot\vec{v}$ o\`u $\vec{k}$ est le nombre d'onde
spatial et $\vec{v}$ la vitesse de balayage dans le ciel.
Une carte du ciel obtenue en mode balayage est donc spatialement
filtr\'ee dans l'espace de Fourier par un filtre passe-bas dont la
fonction de transfert est:
\begin{center}
  \begin{equation}
    F(\vec{k}) = \frac{1}{1+ikv\tau cos\,\theta}
  \label{eq:filter_spatial_bp}
  \end{equation}
\end{center}
o\`u $\theta$ est l'angle entre la direction du balayage et la
direction dans le ciel de la structure spatiale \`a
\'etudier. L'amplitude et la phase de ce filtre complexe sont:
\begin{center}
  \begin{equation}
    \|F(\vec{k})\| = \frac{1}{\sqrt{1+(kv\tau \cos\theta)^2}}
    \,\,\,\,\,\,\,\,\,\,\,\,\,\,\,\,\,\,\,\,\,\,\,\,\,\,\,\,\,\,\,\,
    \phi=\tan^{-1}\,(-kv\tau \cos\theta)
  \label{eq:filter_spatial_bp_ampl}
  \end{equation}
\end{center}
La fr\'equence de coupure de ce filtre d\'epend de la direction dans
le ciel et de l'amplitude des structures spatiales, de la vitesse de
d\'erive du t\'elescope et de la valeur de la constante de temps des
bolom\`etres, \cad du r\'eglage des d\'etecteurs. Pour les structures
spatiales qui sont align\'ees avec la direction du balayage, le terme
en $\cos\theta$ vaut~$1$, l'effet du filtre est alors maximum et les
hautes fr\'equences spatiales sont effectivement att\'enu\'ees. Par
contre, les structures perpendiculaires \`a la direction du balayage
ne sont pas du tout filtr\'ees puisque $\cos\theta$ vaut 0. Dans
l'espace direct ceci se traduit par une PSF allong\'ee seulement dans
la direction du balayage, les autres directions \'etant de moins en
moins filtr\'ees \`a mesure que $\cos\theta$ diminue.

\begin{figure}
  \begin{center}
    \begin{tabular}{c}
      \includegraphics[width=0.90\textwidth,angle=0]{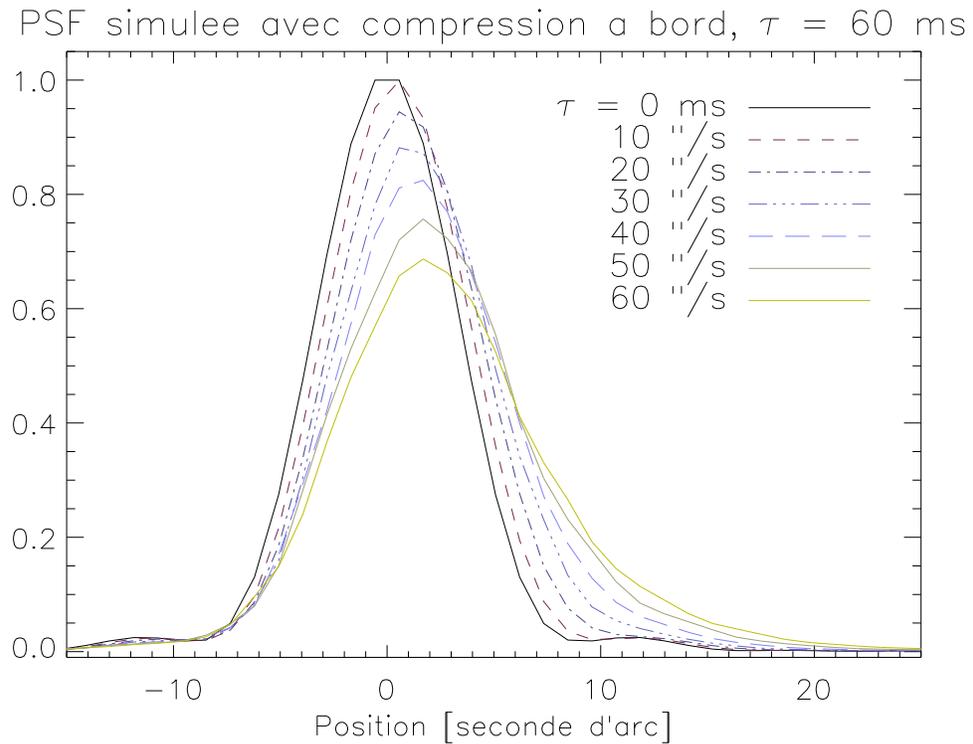}\\
      \includegraphics[width=1.\textwidth,angle=0]{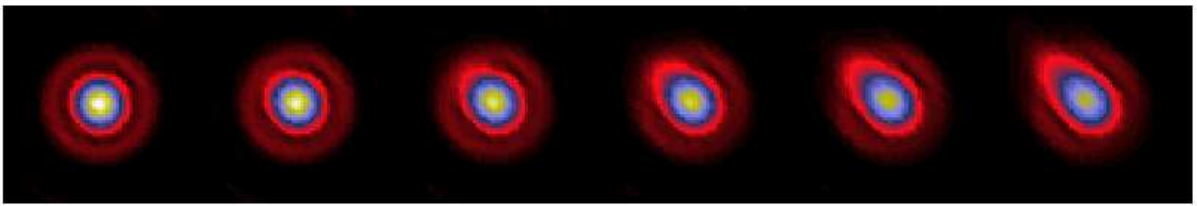}
    \end{tabular}
  \end{center}
  \caption[Simulations de balayage et d\'egradation de PSF (6
  vitesses)]{Les six imagettes (bas de la figure) ont \'et\'e
  g\'en\'er\'ees par le simulateur de PACS. Elles repr\'esentent la
  PSF \`a 110~$\mu$m d'une source balay\'ee \`a 45\textdegree~en un
  seul passage du t\'elescope pour 6 vitesses de balayage
  diff\'erentes. les PSF sont ensuite projet\'ees sur une grille pour
  reconstruire ces imagettes avec un \'echantillonnage de 0.8$''$
  (taille physique du pixel 3.2$''$ sur le BFP bleu et 6.4$''$ sur le
  rouge). De gauche \`a droite les diff\'erentes vitesses sont: 10,
  20, 30, 40, 50 et 60~$''$/sec. La constante de temps utilis\'ee pour
  ces simulations est de 60~ms et la compression du SPU est de 4
  images cons\'ecutives. L'\'echelle de couleur logarithmique met en
  \'evidence l'\'elargissement de la PSF dans la direction du balayage
  et la baisse de contraste. Les courbes du haut pr\'esentent les
  profils de chacune des PSF dans la direction de balayage,
  c'est-\`a-dire dans la direction o\`u le filtrage par la constante
  de temps est maximum.}
  \label{fig:calib_perfobs_scan_tau_6speed}
\end{figure}

Cette analyse dans l'espace de Fourier nous fournit une bonne
description physique du ph\'enom\`ene de filtrage par la constante de
temps. Nous nous int\'eressons maintenant aux r\'esultats des
simulations d'observation par balayage que j'ai effectu\'ees avec le
simulateur PACS dans le but de quantifier la d\'egradation de la PSF
dans ce mode d'observation. Le simulateur PACS est un outil
d\'evelopp\'e par le groupe ICC de Saclay qui inclue les performances
optiques du syst\`eme et les nombreux effets instrumentaux inh\'erents
\`a l'utilisation de bolom\`etres. Par exemple, pour prendre en compte
les effets de la constante de temps, le signal g\'en\'er\'e par le
simulateur \`a un instant $t$ d\'epend du flux incident \`a cet
instant, mais aussi de la valeur du signal \`a l'instant $t-\delta
t$. L'influence du signal ant\'erieur est pond\'er\'ee par une
exponentielle de la forme $e^{\frac{-\delta t}{\tau}}$ ($\delta t$ est
le temps qui s\'epare deux \'echantillonnages successifs). Ce type de
calcul est tout \`a fait similaire \`a une convolution dans l'espace
direct et il exprime l'effet de la constante de temps sur le signal.
La figure~\ref{fig:calib_perfobs_scan_tau_6speed} pr\'esente les
r\'esultats de simulation du balayage d'une source ponctuelle \`a
110~$\mu$m pour six vitesses de scan diff\'erentes. Elle montre un
\'etalement de la PSF dans la direction du mouvement du
t\'elescope. Chacune des six imagettes pr\'esent\'ees est obtenue en
simulant une observation pour laquelle le satellite balaye une source
ponctuelle tr\`es brillante avec un angle de 45\textdegree ~entre la
direction de scan et la grande dimension de la cam\'era.  Le
t\'elescope effectue un seul passage sur la source, d'en bas \`a
droite vers le haut \`a gauche. Les param\`etres du simulateur ont
\'et\'e choisis pour repr\'esenter le mode de fonctionnement nominal
de PACS \`a 110~$\mu$m, c'est-\`a-dire avec une compression \`a bord
de quatre images cons\'ecutives moyenn\'ees et une constante de temps
correspondant au r\'eglage nominal des ponts bolom\'etriques
($V_{polar}=2.7$~V et $\tau=60$~ms). Toutefois, les simulations ne
prennent pas en compte les d\'erives de gain et d'offset puisque nous
voulons mettre en \'evidence les effets de la constante de temps
uniquement. Les imagettes pr\'esent\'ees dans la
figure~\ref{fig:calib_perfobs_scan_tau_6speed} sont des projections
sur une grille de 0.8$''$ des images g\'en\'er\'ees par le simulateur.
Comme pr\'evu par l'analyse de Fourier, nous retrouvons des PSF
\'etal\'ees dans la direction du balayage et intactes dans la
direction orthogonale.

La figure~\ref{fig:calib_perfobs_scan_tau_6speed} pr\'esente
\'egalement des coupes \`a 45\textdegree~des imagettes. Ces profils de
PSF sont ajust\'es par des courbes gaussiennes pour permettre
l'extraction de leurs propri\'et\'es: largeur \`a mi-hauteur, position
et valeur du maximum d'intensit\'e. Par ailleurs, nous d\'efinissons
l'\'elargissement d'une PSF comme \'etant sa largeur \`a mi-hauteur
normalis\'ee par rapport \`a la largeur \`a mi-hauteur d'une PSF
balay\'ee sans constante de temps. De la m\^eme fa\c{c}on le
contraste\footnote{La notion de contraste est ici tr\`es diff\'erente
de celle utilis\'ee habituellement. Les imagettes g\'en\'er\'ees ne
contiennent pas de bruit, nous ne pouvons donc pas normaliser le pic
d'intensit\'e par rapport au niveau de bruit du fond de t\'elescope.}
est le pic d'intensit\'e normalis\'e, et le d\'ephasage est le
d\'ecalage en seconde d'arc du pic d'intensit\'e par rapport \`a la
position de la source dans le ciel. Notez que les profils ne sont pas
sym\'etriques par rapport \`a la position de la source, et ceci est
une cons\'equence du principe de causalit\'e qui implique que la
constante de temps n'affecte que le signal pass\'e. Le
tableau~\ref{tab:calib_perfobs_scan_tau_degradePSF} r\'esume en
quelques chiffres la d\'egradation des PSF simul\'ees \`a 110~$\mu$m
pour une constante de temps de 60~ms et une compression \`a bord
nominale de 4~images cons\'ecutives.\\

\begin{table}
  \begin{center}
    \setlength\extrarowheight{4pt}
    \begin{tabular}[]{p{0.cm}>{\bfseries}p{4cm}>{\centering}p{1cm}>{\centering}p{1cm}>{\centering}p{1cm}>{\centering}p{1cm}>{\centering}p{1cm}>{\centering}p{1cm}>{\centering}p{1cm}p{0.cm}}
      \toprule
      &Vitesse [$''$/sec]   & \bfseries{0} & \bfseries{10} & \bfseries{20} & \bfseries{30} & \bfseries{40} & \bfseries{50} & \bfseries{60} &   \\
      \hline \hline
      &Contraste          & 1 & 0.97 & 0.92 & 0.86 & 0.80 & 0.72 & 0.65 & \\
      &\'Elargissement    & 1 & 1.02 & 1.07 & 1.14 & 1.23 & 1.33 & 1.41 & \\
      &D\'ephasage [$''$]   & 0 & 0.47 & 0.89 & 1.28 & 1.60 & 1.91 & 2.19 & \\
      \bottomrule
    \end{tabular}
  \caption[D\'egradation de la PSF en fonction de la vitesse de
  balayage]{Synth\`ese des r\'esultats sur l'\'etude de la
  d\'egradation de la PSF en fonction de la vitesse de balayage. Ces
  chiffres ont \'et\'e extraits des simulations pr\'esent\'ees dans la
  figure~\ref{fig:calib_perfobs_scan_tau_6speed} o\`u la constante de
  temps est de 60~ms. Voir le texte pour plus de d\'etails.
  \label{tab:calib_perfobs_scan_tau_degradePSF}}
  \end{center}
\end{table}

Pour des vitesses de balayage de 10 et 20$''$/sec, la d\'egradation de
la PSF est relativement faible (inf\'erieure \`a 10~\%) de sorte que
la qualit\'e des observations n'est pas significativement affect\'ee
par la constante de temps des bolom\`etres. Par contre, pour une
vitesse de 60$''$/sec, la PSF est substantiellement alt\'er\'ee: la
limite de d\'etection chute de 35~\% sur les sources ponctuelles (la
photom\'etrie reste peu affect\'ee puisque nous pouvons int\'egrer le
flux sur les pixels voisins qui ont r\'ecup\'er\'e une partie du
flux), le pic d'intensit\'e est d\'ecal\'e de plus de 2 secondes
d'arc, et pour les sources \'etendues la r\'esolution spatiale diminue
de plus de 40~\%.\\

\begin{figure}
  \begin{center}
    \begin{tabular}{c}
      \includegraphics[width=0.90\textwidth,angle=0]{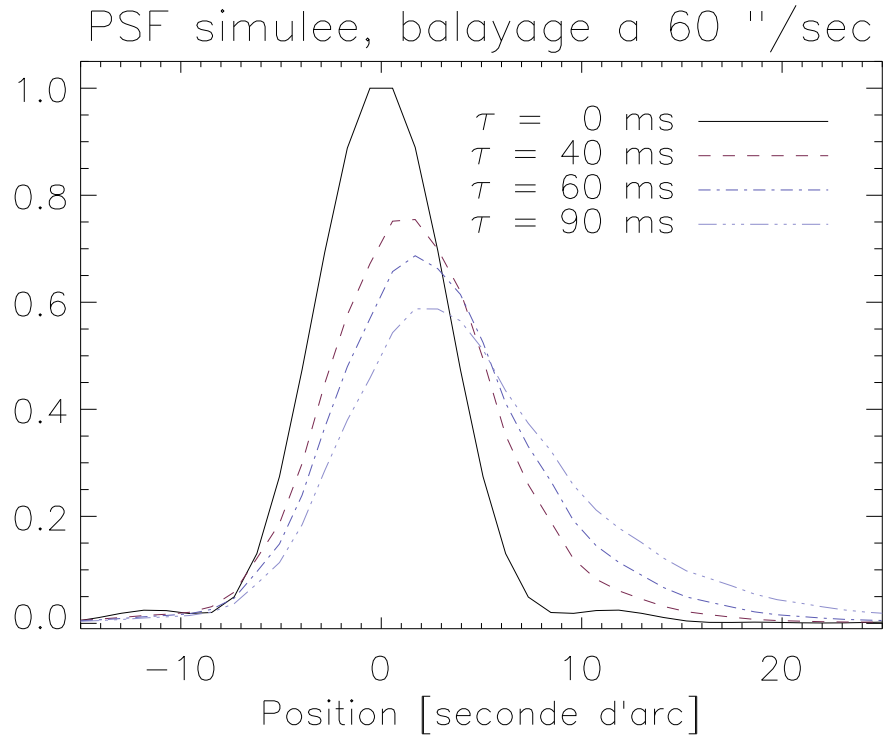}\\
      \includegraphics[width=0.9\textwidth,angle=0]{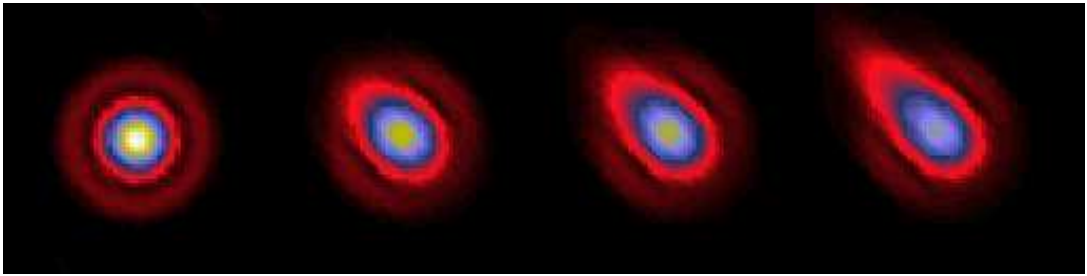}
    \end{tabular}
  \end{center}
  \caption[Simulations de balayage pour 4 constantes de
  temps]{Simulations de balayage de sources ponctuelles \`a 110~$\mu$m
  avec un angle de 45\textdegree ~et une vitesse de 60$''$/sec.  Sur les
  quatre imagettes (bas de la figure), la constante de temps augmente
  de gauche \`a droite: 0, 40, 60 et 90~ms. La valeur de $\tau$
  augmente effectivement lorsque la tension de polarisation des
  bolom\`etres diminue. Pour ces simulations la compression \`a bord
  est de 4 images successives. Le graphe (haut) montre l'\'evolution
  des profils de PSF avec la constante de temps.}
  \label{fig:calib_perfobs_scan_tau_3polar}
\end{figure}

Il est toutefois possible de limiter la d\'egradation de la PSF en
changeant le r\'eglage des d\'etecteurs. En effet, d'apr\`es
l'\'equation~\ref{eq:filter_spatial_bp_ampl}, la fr\'equence de
coupure du filtre d\'epend de la valeur de la constante de temps
$\tau$, qui elle-m\^eme d\'epend de la tension de polarisation des
bolom\`etres (section~\ref{sec:calib_perflabo_tau}). Il est donc
potentiellement int\'eressant d'augmenter la tension aux bornes des
d\'etecteurs pour diminuer leur temps de r\'eponse et ainsi minimiser
l'\'elargissement de la PSF. \`A nouveau le simulateur PACS s'av\`ere
\^etre un outil tr\`es appropri\'e pour examiner ce genre de
probl\`eme. La figure~\ref{fig:calib_perfobs_scan_tau_3polar} montre
l'\'evolution de la PSF projet\'ee sur une grille de 0.8$''$ pour
quatre valeurs de constante de temps (de gauche \`a droite dans la
figure: 0, 40, 60 et 90~ms). La vitesse de balayage est la m\^eme pour
toutes les simulations, c'est-\`a-dire 60$''$/sec, et la compression
\`a bord est nominale. Nous retrouvons comme pr\'evu des PSF plus
piqu\'ees (moins floues) pour les courtes constantes de temps. La
courbe en trait plein ($\tau=0$~ms) repr\'esente le comportement d'un
bolom\`etre infiniment rapide et sert de r\'ef\'erence pour la
comparaison des autres PSF. La courbe o\`u $\tau=40$~ms correspond \`a
une tension de polarisation de 3.5~V (cf
section~\ref{sec:calib_perflabo_tau}) et offre les meilleures
performances en terme de d\'egradation de PSF. La courbe o\`u
$\tau=60$~ms correspond \`a la polarisation nominale qui est un
compromis entre sensibilit\'e et rapidit\'e. Et enfin, la courbe o\`u
$\tau=90$~ms ne correspond pas pr\'ecis\'ement \`a une polarisation
que nous avons test\'ee mais elle illustre le comportement d'un
d\'etecteur sous-polaris\'e. Le
tableau~\ref{tab:calib_perfobs_scan_tau_degradePSFtau} donne les
caract\'eristiques des profils de PSF de la
figure~\ref{fig:calib_perfobs_scan_tau_3polar} en fonction de la
valeur de $\tau$ pour une vitesse de balayage de 60$''$/sec.
\begin{table}
  \begin{center}
    \setlength\extrarowheight{4pt}
    \begin{tabular}[]{p{0.cm}>{\bfseries}p{6cm}>{\centering}p{1cm}>{\centering}p{1cm}>{\centering}p{1cm}>{\centering}p{1cm}p{0.cm}}
      \toprule
      &Constante de temps [ms]   & \bfseries{0} &\bfseries{40} & \bfseries{60} & \bfseries{90} & \\
      \hline \hline
      &Contraste                 & 1 & 0.73 & 0.65 & 0.55& \\
      &\'Elargissement           & 1 & 1.29 & 1.41 & 1.61 &\\
      &D\'ephasage [$''$]          & 0 & 1.4 & 2.20 & 3.20 &\\
      \bottomrule
    \end{tabular}
  \caption[D\'egradation de la PSF en fonction de la constante de
  temps]{Param\`etres des PSF simul\'ees pr\'esent\'ees dans la
  figure~\ref{fig:calib_perfobs_scan_tau_3polar}. La vitesse de
  balayage est de 60$''$/sec et la compression \`a bord est de 4 images
  cons\'ecutives.
  \label{tab:calib_perfobs_scan_tau_degradePSFtau}}
  \end{center}
\end{table}
Pour les d\'etecteurs du Photom\`etre PACS, augmenter la tension de
polarisation \`a 3.5~V permet d'am\'eliorer la qualit\'e de la PSF
d'environ 10~\% par rapport \`a la polarisation nominale. Selon
l'objectif scientifique de l'observation, ces 10~\% peuvent
repr\'esenter un gain appr\'eciable sur la r\'esolution spatiale de la
carte reconstruite. Il faut toutefois garder \`a l'esprit que les
performances de la cam\'era sont un compromis entre rapidit\'e et
sensibilit\'e. En effet, si les d\'etecteurs sont fortement
polaris\'es alors la sensibilit\'e peut chuter d'un facteur 2 (cf
section~\ref{sec:calib_perflabo_compare_NEP}), ce qui est prohibitif
pour la plupart des observations Herschel qui visent \`a observer des
objets relativement peu brillants. Cependant, un tel r\'eglage
pourrait \^etre acceptable pour les programmes d'observation o\`u la
sensibilit\'e n'est pas une contrainte forte et dont le but est de
cartographier de tr\`es grandes r\'egions du ciel avec une bonne
r\'esolution spatiale. Quoiqu'il en soit, la sur-polarisation des
bolom\`etres ne pourrait \^etre pertinente que pour une vitesse de
balayage de 60$''$/sec. En effet, \`a 10 ou 20$''$/sec, le gain sur la
taille de la PSF est marginal et ne pourrait justifier une telle perte
de sensibilit\'e. Notez par ailleurs que les r\'esultats pr\'esent\'es
dans cette section sont en bon accord avec les \'etudes
pr\'eliminaires que j'avais conduites en 2005 (cf
annexe~\ref{a:rapport_bp}) avant que le simulateur PACS ne soit
capable de g\'en\'erer des observations par balayage qui incluent les
effets instrumentaux tels que le temps de r\'eponse des d\'etecteurs.

\subsection{Taux de compression et vitesse de balayage}
\label{sec:calib_perfobs_scan_compression}

Le potentiel observationnel du Photom\`etre PACS d\'epend bien entendu
des performances des d\'etecteurs, mais \'egalement de la technique
d'observation utilis\'ee. Dans la section pr\'ec\'edente nous avons
mis en avant l'influence crois\'ee d'un effet instrumental, le temps
de r\'eponse des d\'etecteurs, et d'un param\`etre observationnel, la
vitesse de balayage du t\'elescope. Nous nous int\'eressons maintenant
\`a un param\`etre totalement ind\'ependant des performances
intrins\`eques des bolom\`etres mais qui est cependant susceptible
d'engendrer une d\'egradation de la PSF: la compression des donn\'ees
par le SPU \`a bord du satellite. En effet, le d\'ebit maximal de
donn\'ees que l'observatoire peut transmettre vers la Terre est de
l'ordre de 130~kbits/s alors que le d\'ebit g\'en\'er\'e par le
Photom\`etre PACS est d'environ 1600~kbits/s (cf
section~\ref{sec:detect_observatoire_phfpu_description}). Le SPU
utilise des algorithmes de compression sans perte pour r\'eduire le
d\'ebit de donn\'ees d'un facteur~4 mais il reste encore un facteur 4
de compression \`a atteindre. La solution qui a \'et\'e retenue
consiste \`a moyenner 4~images successives ramenant ainsi le d\'ebit
de donn\'ees dans les limites impos\'ees par l'antenne \`a haut gain
du satellite. De plus, dans le mode d'observation parall\`ele,
\emph{PMode}, de Herschel dans lequel les photom\`etres de PACS et de
SPIRE sont utilis\'es simultan\'ement, il est n\'ecessaire de
compresser les donn\'ees PACS d'un facteur~2 suppl\'ementaire amenant
\`a 8 le nombre d'images moyenn\'ees par le SPU.  Ce type de \og
compression par moyennage\fg requiert une puissance de calcul
relativement faible qui reste compatible avec le CPU embarqu\'e sur
Herschel. D'autre part, le bruit contenu dans le signal temporel
n'\'etant pas corr\'el\'e, le calcul de la moyenne de $N$~images a
l'avantage de r\'eduire d'un facteur~$\sqrt{N}$ le bruit du signal
transmis. Cependant, lorsque le t\'elescope observe en mode de
balayage, l'information spatiale contenue dans chacune des images
moyenn\'ees se retrouve \og m\'elang\'ee\fg et l'image r\'esultante
risque d'\^etre d\'egrad\'ee. En effet le Photom\`etre PACS
\'echantillonne enti\`erement le champ de vue du t\'elescope toutes
les 25~ms, et lorsque le champ change entre chaque mesure, chacune des
images enregistr\'ees repr\'esente une r\'egion diff\'erente du
ciel. Plus le t\'elescope balaye vite le ciel, et plus les
diff\'erences entre les images prises dans la moyenne sont
importantes. Plus le nombre d'images moyenn\'ees est grand, et plus
l'image r\'esultante sera d\'egrad\'ee par la \og compression
\fg. Pour quantifier l'impact de la compression \`a bord sur la
qualit\'e des observations par balayage, je me tourne une nouvelle
fois vers le simulateur PACS.

\begin{figure}
  \begin{center}
      \includegraphics[width=1.\textwidth,angle=0]{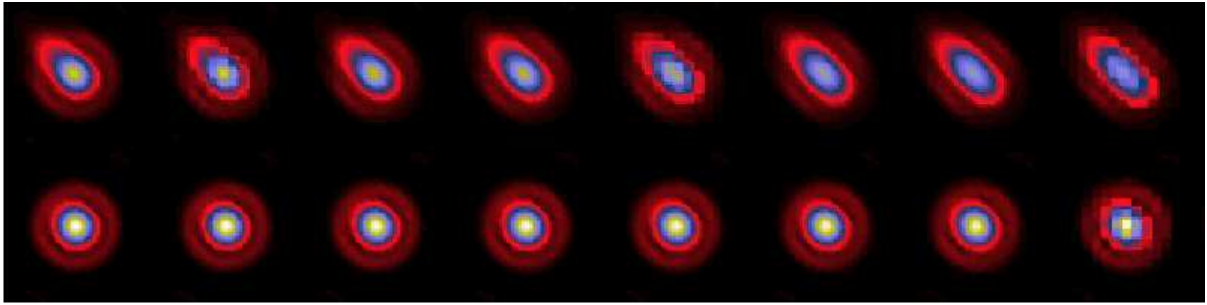}
  \end{center}
  \caption[Simulation de PSF balay\'ees et compression \`a
  bord]{Chaque imagette est la projection sur une grille de 0.8$''$
  d'une source ponctuelle observ\'ee en mode balayage avec le
  simulateur PACS. Le t\'elescope effectue un seul passage sur la
  source d'en-bas \`a droite vers le haut \`a gauche. La constante de
  temps des bolom\`etres est de 60~ms. La fr\'equence
  d'\'echantillonnage est 40~Hz. Les vitesses de balayage sont de
  60$''$/s (ligne du haut) et de 20$''$/s (ligne du bas). De gauche \`a
  droite sur la figure, le taux de compression par moyennage passe
  d'un facteur~2 \`a un facteur~9. Les PSF balay\'ees \`a 60$''$/s sont
  significativement alt\'er\'ees pour les forts taux de compression
  \`a bord. D'autre part, la r\'esolution spatiale de certaines
  imagettes projet\'ees est limit\'ee par la taille physique du
  pixel. Voir dans le texte pour les explications.
  \label{fig:calib_perfobs_scan_compression_ima}}
\end{figure}

Pour chaque simulation pr\'esent\'ee dans la
figure~\ref{fig:calib_perfobs_scan_compression_ima}, nous simulons
plusieurs observations par balayage d'une source ponctuelle en ne
changeant qu'un seul param\`etre \`a la fois. La fr\'equence
d'\'echantillonnage de la cam\'era est toujours de 40~Hz, la constante
de temps est fix\'ee \`a 60~ms, et les PSF sont reprojet\'ees sur une
grille de 0.8$''$ de c\^ot\'e. Nous explorons le taux de compression
de 2 \`a 9~images cons\'ecutives moyenn\'ees pour deux vitesses de
balayage (20~et~60$''$/s). Au premier coup d'oeil sur la
figure~\ref{fig:calib_perfobs_scan_compression_ima} nous remarquons
qu'au moins~4 configurations donnent des PSF \`a l'aspect douteux (par
exemple pour un nombre de moyenne $N_{SPU}$=9 et une vitesse de
balayage de $V_{scan}$=20$''$/s, imagette en-bas \`a droite).  Pour
mettre hors de cause un \'eventuel bug du simulateur, je me suis
int\'eress\'e \`a ce r\'esultat inattendu et pourtant
pr\'evisible. J'en donnerai l'expliquation d\'etaill\'ee dans la
section~\ref{sec:calib_perfobs_scan_angle}.

\begin{figure}
  \begin{center}
    \begin{tabular}{ll}
      \includegraphics[width=0.5\textwidth,angle=0]{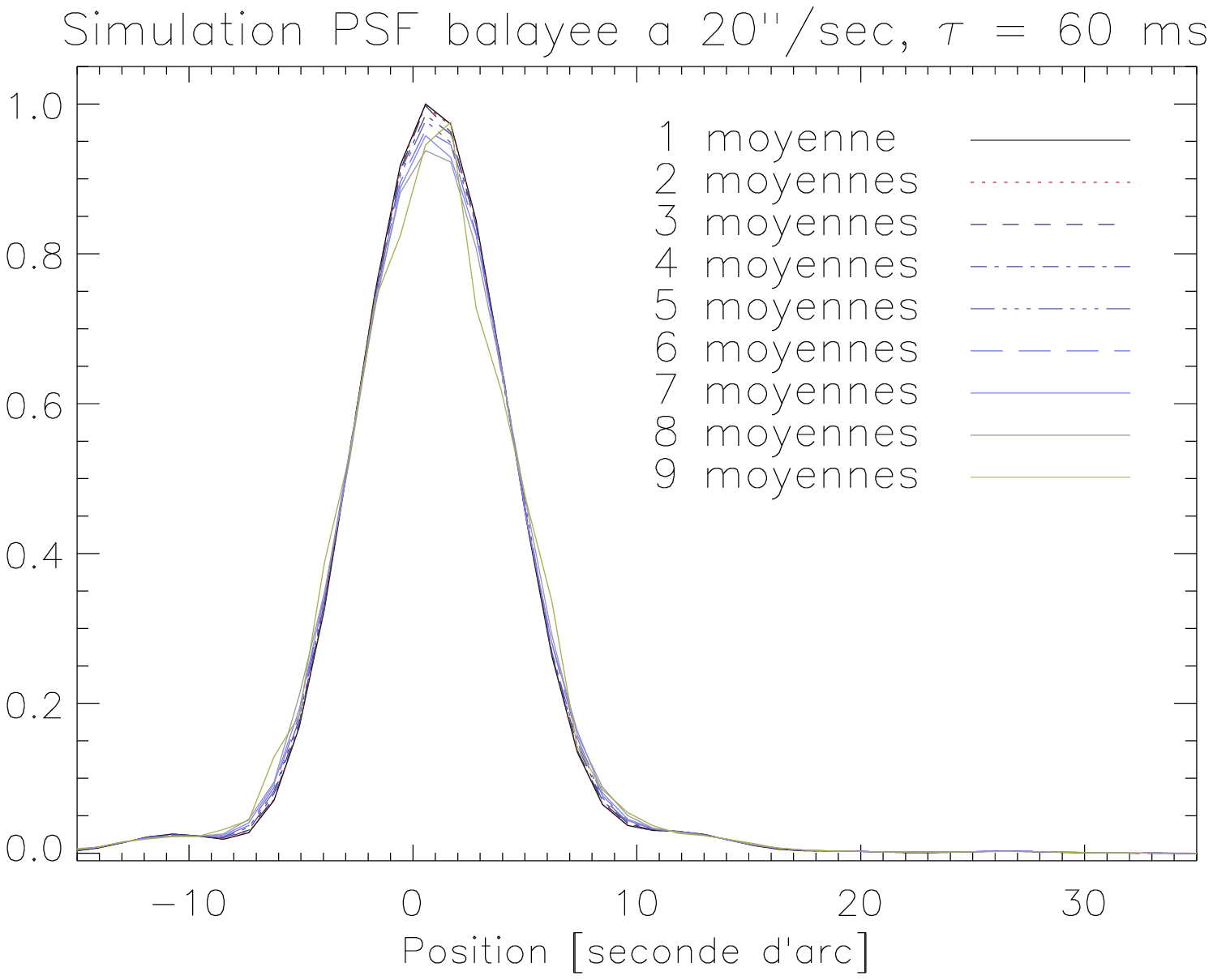}
      &
      \includegraphics[width=0.5\textwidth,angle=0]{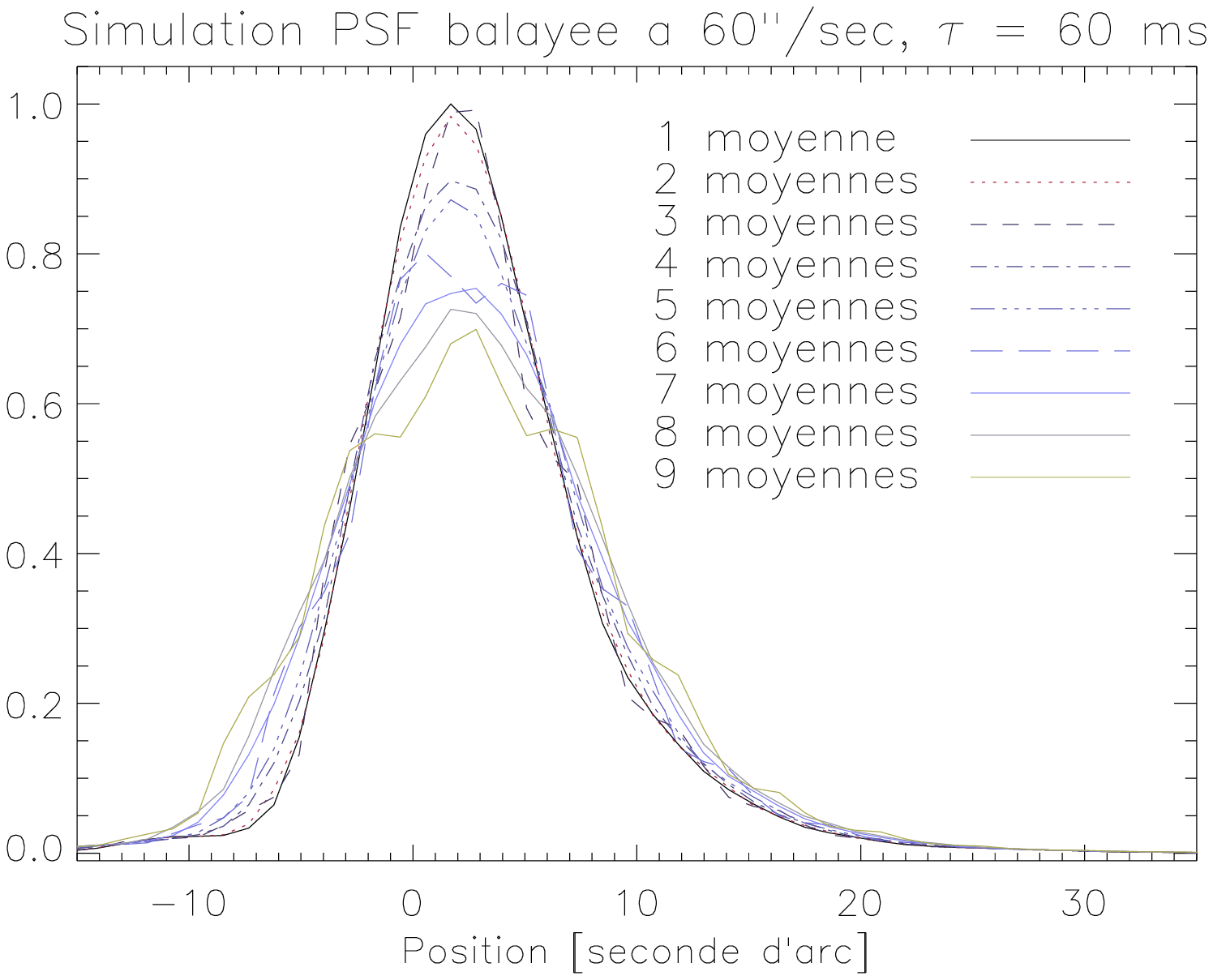}
    \end{tabular}
  \end{center}
  \caption[Profils des PSF simul\'ees et compression \`a bord]{Profils
  des PSF simul\'ees de la
  figure~\ref{fig:calib_perfobs_scan_compression_ima}. La
  d\'egradation de la PSF par le moyennage \`a bord est n\'egligeable
  \`a 20$''$/s mais devient significative pour un balayage du ciel \`a
  60$''$/s. Notez que la compression \`a bord r\'ealis\'ee par le
  simulateur ne d\'ecale pas le pic d'intensit\'e car la position
  associ\'ee \`a l'image moyenn\'ee est la moyenne des positions de
  chacune des images prises dans la moyenne. Notez \'egalement que
  l'\'elargissement des PSF est sym\'etrique contrairement \`a
  l'\'elargissement produit par la constante de temps (cf
  figures~\ref{fig:calib_perfobs_scan_tau_6speed}
  et~\ref{fig:calib_perfobs_scan_tau_3polar})~; cette derni\`ere ne
  pouvant influencer le signal dans le futur.}
  \label{fig:calib_perfobs_scan_compression_psf}
\end{figure}

En ce qui concerne les PSF simul\'ees, nous pouvons extraire leurs
profils et les ajuster avec des gaussiennes pour quantifier leurs
propri\'et\'es, de la m\^eme mani\`ere que dans la section
pr\'ec\'edente. La figure~\ref{fig:calib_perfobs_scan_compression_psf}
pr\'esente le profil des PSF simul\'ees et montre que l'impact de la
compression \`a bord est minimis\'e pour les faibles vitesses de
balayage et pour les petits nombres de moyenne, \cad lorsque les
images prises dans la moyenne ne diff\`erent pas trop les unes des
autres. Nous voyons que pour une vitesse de balayage du ciel de
20$''$/s la PSF est peu affect\'ee par le nombre d'images
moyenn\'ees. Par contre, les sources balay\'ees \`a 60$''$/s se
trouvent significativement \'elargies et att\'enu\'ees dans la
direction du balayage lorsque le nombre de moyenne augmente. Par
exemple pour $N_{SPU}$=8, le pic d'intensit\'e baisse de
30~\%. D'apr\`es les figures~\ref{fig:calib_perfobs_scan_tau_6speed}
et~\ref{fig:calib_perfobs_scan_compression_psf} la PSF d'une source
ponctuelle balay\'ee \`a 60$''$/s en \emph{PMode} ($N_{SPU}$=8) est
att\'enu\'ee et \'elargie d'un facteur~2 par rapport \`a une source
balay\'ee \`a 10$''$/s, ce qui repr\'esente une d\'egradation
consid\'erable des performances de l'Observatoire. Le \emph{PMode}
semble \^etre assez populaire parmi les futurs utilisateurs
d'Herschel puisqu'il fournira des donn\'ees PACS et SPIRE
simultan\'ement, mais il est imp\'eratif d'avertir les observateurs
d\'esirant balayer le ciel \`a grande vitesse que la sensibilit\'e
effective ne sera pas celle annonc\'ee comme nominale. Pour ne citer
qu'un seul programme n\'ecessitant l'utilisation du \emph{PMode} \`a
grande vitesse de balayage, je citerais le programme clef de temps
ouvert de Sergio Molinari \shortcite{molinari} qui consiste \`a
cartographier enti\`erement le plan galactique de -0.5 \`a +0.5
degr\'es de latitude de part et d'autre de l'\'equateur galactique.



\subsection{Compression \`a bord, vitesse et angle de balayage}
\label{sec:calib_perfobs_scan_angle}

Dans cette section je m'int\'eresse \`a l'\'echantillonnage du ciel
par le Photom\`etre PACS, avec le but avou\'e d'expliquer les \og
aberrations\fg observ\'ees lors des simulations pr\'esent\'ees dans la
figure~\ref{fig:calib_perfobs_scan_compression_ima}. Car il s'agit
bien d'un effet d'\'echantillonnage qui donne \`a certaines PSF cet
aspect \og pixelis\'e \fg. Il faut rappeler que les images
g\'en\'er\'ees par le simulateur PACS ont \'et\'e projet\'ees sur une
grille de 0.8$''$ de c\^ot\'e pour reconstruire la PSF, alors que la
taille physique des pixels est de 3.2$''$ dans le cas pr\'esent du BFP
bleu. Toutes les PSF de la
figure~\ref{fig:calib_perfobs_scan_compression_ima} sont donc
\'echantillonn\'ees \`a 0.8$''$, mais l'\'echantillonnage de certaines
est effectivement limit\'e \`a 3.2$''$. Ceci est d\^u \`a un effet
g\'eom\'etrique qui se produit pour certaines combinaisons des
param\`etres d'observation.

Prenons l'exemple de l'imagette en-bas \`a droite de la
figure~\ref{fig:calib_perfobs_scan_compression_ima} qui pr\'esente
tous les sympt\^omes d'un \'echantillonnage grossier par rapport aux
autres imagettes. La fr\'equence de lecture est de 40~Hz et le nombre
de moyennes \`a bord est de~9. Les images transmises vers la Terre
sont donc \'echantillonn\'ees \`a une fr\'equence effective
$\nu_{eff}=\frac{40}{9}=4.44$~Hz. D'autre part le t\'elescope d\'erive
\`a raison de 20$''$ par seconde de sorte que deux images successives
se trouvent d\'ecal\'ees de 4.5$''$ l'une de l'autre. Dans le cas des
simulations pr\'esent\'ees, le balayage s'effectue suivant la
diagonale d'une matrice (angle de 45\textdegree~avec la grande
longueur de la cam\'era), la distance qui s\'epare le centre de deux
pixels dans cette direction est de $3.2''/\cos (45)=4.5''$. Le
t\'elescope se d\'eplace donc d'un pixel exactement entre chaque
\'echantillonnage du champ de vue. Pour les images simul\'ees avec
$N_{SPU}=9$ et $V_{scan}=20''$/s, quelque soit la finesse de la grille
de projection utilis\'ee, l'\'echantillonnage de l'image
reconstruite reste limit\'ee par la taille physique du pixel.

\begin{figure}
  \begin{center}
      \includegraphics[width=0.7\textwidth,angle=0]{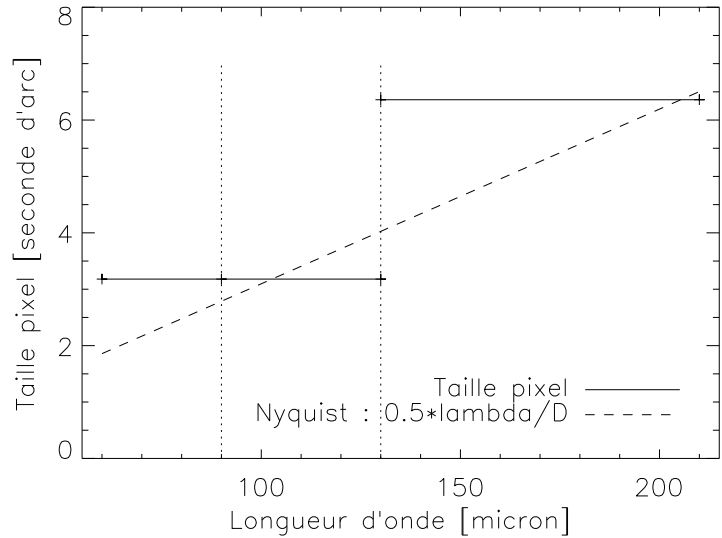}
  \end{center}
  \caption[\'Echantillonnage Nyquist sur les deux plans
  focaux]{Comparaison entre la taille physique des pixels et
  l'\'echantillonnage optimum des images selon le crit\`ere de Nyquist
  (0.5F$\lambda$). La cam\'era n'echantillonne pas \`a Nyquist pour
  les bandes \`a 75 et 170~$\mu$m.
  \label{fig:calib_perfobs_scan_compression_nyquist}}
\end{figure}

\`A l'inverse, il est possible de choisir une combinaison
($N_{SPU},V_{scan},\theta_{scan}$) telle que l'\'echantillonnage
spatial de la carte reconstruite soit sup\'erieur \`a
l'\'echantillonnage nominal de la cam\'era d\'efini par la taille du
pixel. Ceci peut s'av\'erer extr\^emement int\'eressant pour
\'echantillonner le ciel plus pr\'ecis\'ement, c'est-\`a-dire affiner
l'\'echantillonnage jusqu'\`a atteindre la r\'esolution spatiale
maximale donn\'ee par le th\'eor\`eme d'\'echantillonnage de
\emph{Nyquist-Shannon} (\'echantillonnage maximum \`a 0.5F$\lambda$
pour une PSF filtrant \`a F$\lambda$, cf
section~\ref{sec:intro_bolometrie_bolo_matrice}). La
figure~\ref{fig:calib_perfobs_scan_compression_nyquist} montre la
taille physique d'un pixel du Photom\`etre PACS en fonction de la
longueur d'onde et la compare au crit\`ere de Nyquist. La figure
r\'ev\`ele que les voies centr\'ees \`a 75~et 170~$\mu$m
sous-\'echantillonnent la PSF au sens de Nyquist. Je vais donc
g\'en\'eraliser l'analyse faite pr\'ec\'edemment pour le cas
particulier de la figure~\ref{fig:calib_perfobs_scan_compression_ima}
dans le but de d\'eterminer les combinaisons de
($N_{SPU},V_{scan},\theta_{scan}$) qui permettent d'atteindre la
r\'esolution maximale pour une observation par balayage avec
couverture homog\`ene\footnote{\og Couverture homog\`ene\fg signifie
que le t\'elescope n'effectue qu'un seul passage sur une m\^eme zone
du ciel, d'apr\`es le manuel d'utilisateur de HSPOT, tout comme les
simulations pr\'esent\'ees dans ce chapitre.}.

\begin{figure}
  \begin{center}
    \begin{tabular}{l}
      \includegraphics[width=0.55\textwidth,angle=0]{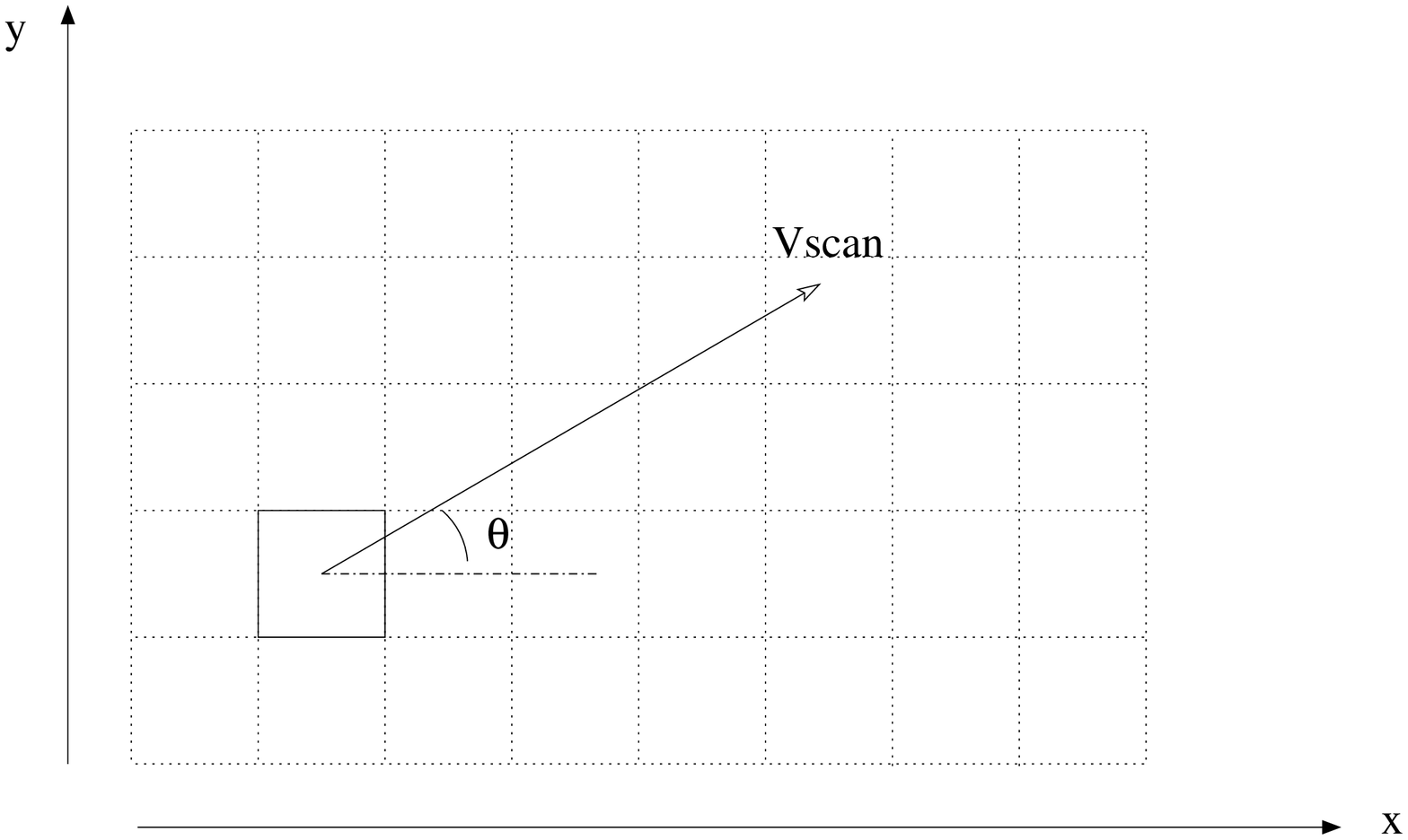}
      \includegraphics[width=0.40\textwidth,angle=0]{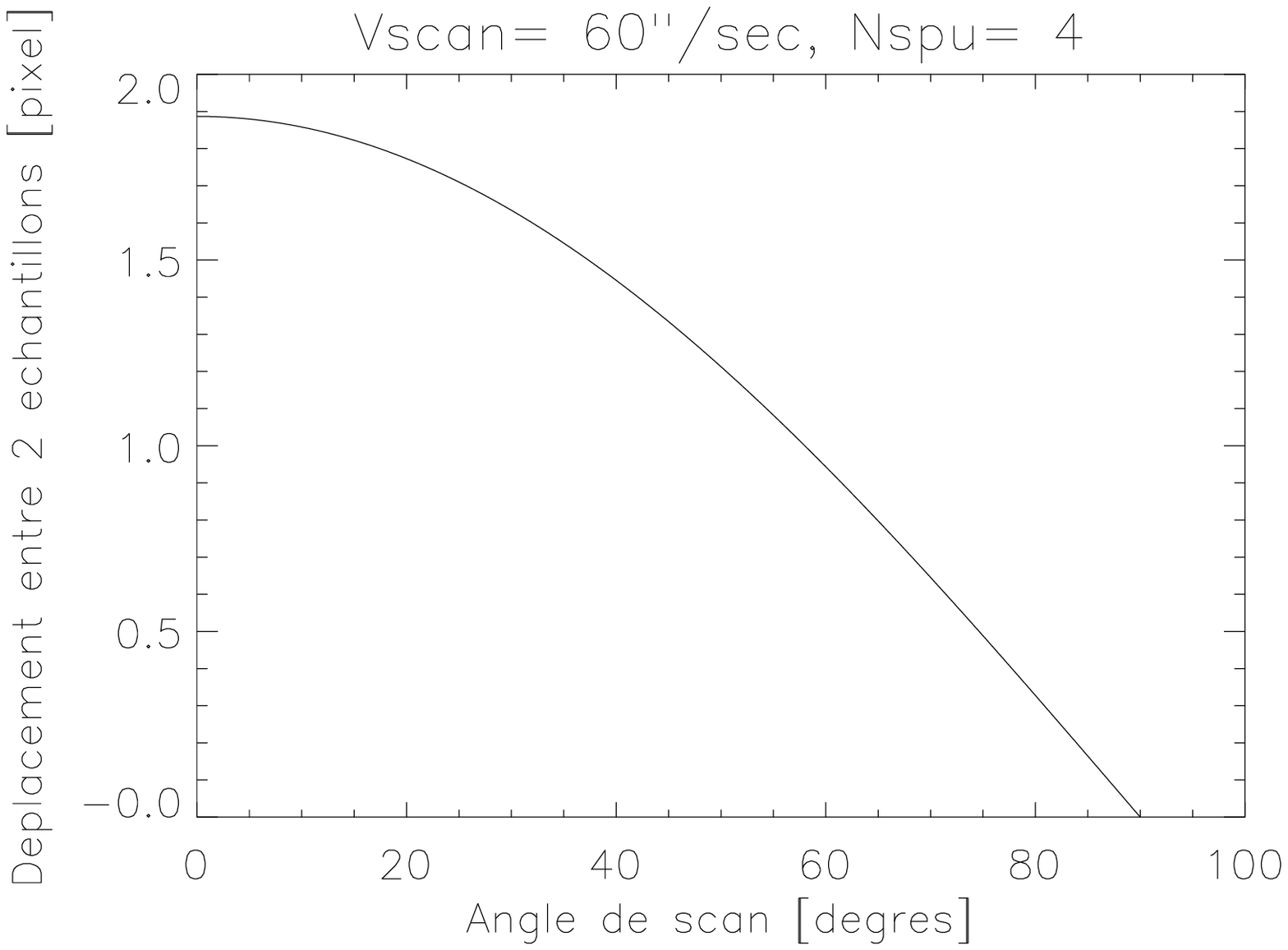}
    \end{tabular}
  \end{center}
  \caption[D\'eplacement d'un pixel en mode de balayage]{\emph{\`A
  gauche}: Sch\'ema du plan focal PACS projet\'e sur le ciel.  Le
  t\'elescope d\'erive \`a la vitesse $V_{scan}$. L'angle
  $\theta_{scan}$ d\'efinit la direction de balayage du t\'elescope
  par rapport \`a l'axe des $x$ (grande longueur de la
  cam\'era). \emph{\`A droite}: D\'eplacement, suivant le grand axe de
  la cam\'era, d'un point du ciel entre deux \'echantillons en
  fonction de l'angle de balayage. La fr\'equence effective est
  $\nu_{eff}=\frac{40}{N_{SPU}}$.
  \label{fig:calib_perfobs_scan_compression_pix}}
\end{figure}

\begin{figure}
  \begin{center}
      \includegraphics[width=0.9\textwidth,angle=0]{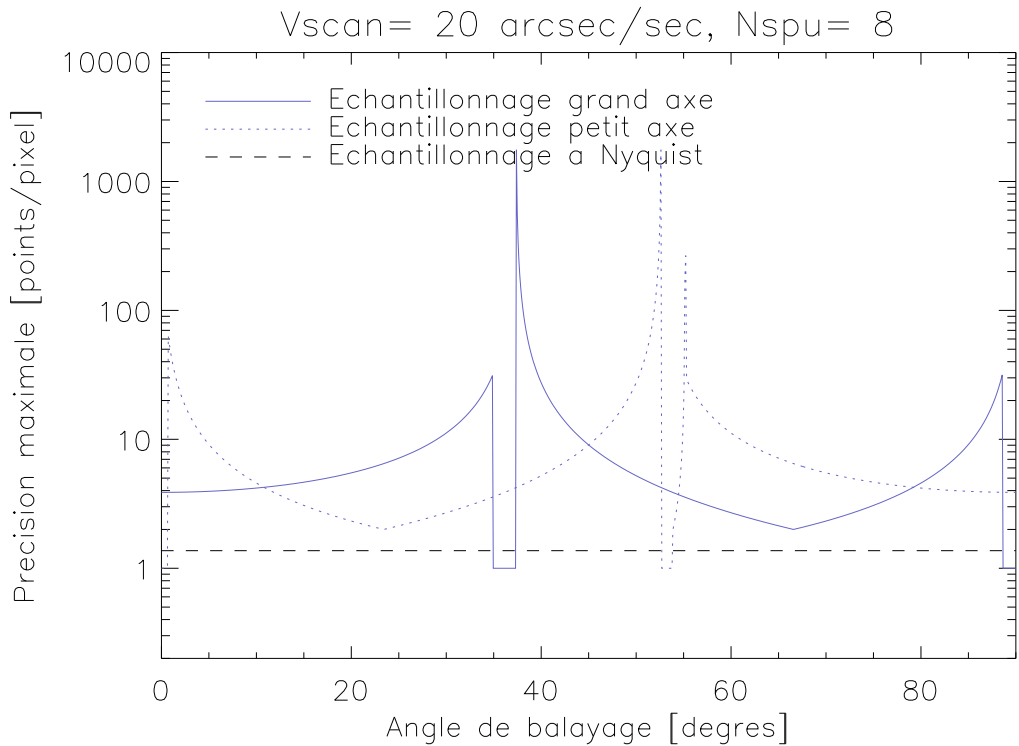}
  \end{center}
  \caption[Pr\'ecision de reconstruction de cartes
  balay\'ees]{\'Echantillonnage spatial maximal avec laquelle nous
  pouvons reconstruire une carte balay\'ee \`a 75~$\mu$m pour
  $N_{SPU}=8$ et $V_{scan}=20''$/s. Cette configuration correspond au
  mode parall\`ele de l'Observatoire Herschel.  Les courbes en trait
  plein et en pointill\'es repr\'esentent respectivement la
  pr\'ecision d'\'echantillonnage suivant l'axe des $x$ et suivant
  l'axe des $y$ selon la notation de la
  figure~\ref{fig:calib_perfobs_scan_compression_pix}. Ce sont des
  portions d'hyperbole qui d\'ependent de la taille du pixel, de la
  g\'eom\'etrie du plan focal et de la distance parcourue entre deux
  images successives. La droite en tiret donne le crit\`ere de Nyquist
  \`a la longueur d'onde consid\'er\'ee. La plupart des angles de
  balayage autorise un \'echantillonnage optimum. Cependant, pour
  quelques angles, l'\'echantillonnage est inf\'erieur au crit\`ere de
  Nyquist.
  \label{fig:calib_perfobs_scan_compression_echantScan}}
\end{figure}

La figure~\ref{fig:calib_perfobs_scan_compression_pix} montre le
sch\'ema de quelques pixels d'une matrice PACS ainsi que les
param\`etres utilis\'es pour le calcul de l'\'echantillonnage
spatial. La vitesse de balayage est $V_{scan}$, la direction du
balayage fait un angle $\theta_{scan}$ avec la grande longueur de la
camera (64 pixels pour le plan bleu et 32 pour le plan rouge). La
grande longueur correspond aux abscisses sur la figure. Le graphe de
droite montre la projection sur l'axe des $x$ de l'\'ecart angulaire
entre deux images cons\'ecutives en fonction de l'angle de
balayage. La fr\'equence d'\'echantillonnage temporelle effective est
$\nu_{eff}=\frac{40}{N_{SPU}}$. Si le t\'elescope se d\'eplace d'un
angle~$\alpha$ dans le ciel entre deux \'echantillonnages du champ de
vue, alors l'\'echantillonnage spatial de l'image reconstruite
sera~$\frac{1}{\alpha}$. Plus le d\'eplacement du t\'elescope est
petit entre deux images successives, et plus l'\'echantillonnage
spatial est fin. Cependant le v\'eritable calcul est un peu plus
subtil car il prend en compte la taille physique du pixel et la
longueur du plan focal.  Pour simplifier la g\'eom\'etrie du
probl\`eme, je consid\`ere un point du ciel qui balaye le champ de vue
du Photom\`etre en passant par le centre du plan focal. Avant de
projeter et de co-additioner chacune des images individuelles sur une
grille, j'utilise l'astrom\'etrie de l'observation afin de les
superposer de sorte que les toutes les structures du ciel
co\"{i}ncident. Je calcule ensuite la distance qui s\'epare le centre
des pixels ayant \'echantillonn\'e un m\^eme point du ciel pendant
l'observation. La pr\'ecision maximale de la projection est l'inverse
de la distance ainsi calcul\'ee. Les courbes sont donc des portions
d'hyperbole qui d\'ependent de la taille du pixel, de la g\'eom\'etrie
du plan focal et de la distance parcourue entre deux images
successives. La
figure~\ref{fig:calib_perfobs_scan_compression_echantScan} pr\'esente
le r\'esultat de mes calculs pour le cas ($N_{SPU}=8$,
$V_{scan}=20''$/s), et le compare au crit\`ere d'\'echantillonnage de
Nyquist. Nous voyons que la plupart des angles authorise un
\'echantillonnage spatial sup\'erieur au crit\`ere de Nyquist. Notez
toutefois que le sur-\'echantillonnage de la PSF n'apporte aucune
information spatiale suppl\'ementaire sur la sc\`ene observ\'ee. Par
contre, pour certains angles ($\sim$37\textdegree~pour l'axe des $x$
et $\sim$53\textdegree~pour l'axe des $y$) la
figure~\ref{fig:calib_perfobs_scan_compression_echantScan} montre que
la pr\'ecision est limit\'ee par la taille du pixel, \cad que
l'\'ecart entre deux images successives est approximativement d'un
pixel. Pour une observation \`a~75 ou 170~$\mu$m, la carte du ciel
reconstruite ne sera pas \'echantillonn\'ee de fa\c{c}on optimale (cf
figure~\ref{fig:calib_perfobs_scan_compression_nyquist}). De plus, si
la grille de projection est plus fine que la taille du pixel, alors
l'image reconstruite appara\^itra mieux \'echantillonn\'ee dans une
direction que dans l'autre.

L'aspect discontinu des courbes de la
figure~\ref{fig:calib_perfobs_scan_compression_echantScan} est d\^u au
calcul th\'eorique qui ne prend pas en compte les aberrations optiques
et autres alt\'erations du champ de vue du t\'elescope. Inclure ces
aberrations aurait tendance \`a \og adoucir \fg les courbes et \`a
amortir les variations brutales de r\'esolution. Le but de mon analyse
est tout de m\^eme atteint puisque nous avons maintenant un moyen de
pr\'edire les angles pour lesquels une observation par balayage
sous-\'echantillonne le ciel. Dans le
tableau~\ref{tab:calib_perfobs_scan_compression_table}, je recense
toutes les combinaisons des param\`etres d'observation ($N_{SPU}$,
$V_{scan}$, $\theta_{scan}$) qui sont propos\'ees par l'Observatoire
Herschel et qui ne permettent pas un \'echantillonnage optimum au sens
de Nyquist. Notez que le tableau ne contient que des angles compris
entre 0 et 90\textdegree, pour les angles de balayage sup\'erieurs \`a
90\textdegree~il suffit de prendre les angles suppl\'ementaires
(sym\'etrie axiale). Nous voyons que la contrainte sur les angles de
balayage est relativement faible, il n'y a qu'une petite fraction des
angles possibles qui sont \`a bannir, il sera donc facile pour les
utilisateurs d'Herschel de s'assurer du bon \'echantillonnage de leurs
observations en \'evitant les quelques configurations cit\'ees dans le
tableau~\ref{tab:calib_perfobs_scan_compression_table}.\\


\begin{table}
  \begin{center}
    \setlength\extrarowheight{2pt}
    \begin{tabular}[]{p{0.cm}>{\bfseries}p{3cm}>{\centering}p{2cm}>{\centering}p{2cm}>{\centering}p{2cm}>{\centering}p{2cm}p{0.cm}}
      \toprule
      & &\multicolumn{2}{c}{\bfseries{BFP bleu}}&\multicolumn{2}{c}{\bfseries{BFP rouge}}&\\
      \cmidrule(r){3-4}\cmidrule(l){5-6}
      & & Axe des $x$ & Axe des $y$ & Axe des $x$ & Axe des $y$&\\
      \hline \hline
      & & 36 - 37$\,^\circ$ & 0$\,^\circ$ & & & \\
      & $\mathbf{N_{SPU}=8}$ & 57 - 58$\,^\circ$ & 15$\,^\circ$ & 55 - 57$\,^\circ$  & 0$\,^\circ$ & \\
      & $\mathbf{V_{scan}=60''/s}$ & 74$\,^\circ$ & 31$\,^\circ$ & 88 - 90$\,^\circ$ & 33$\,^\circ$ & \\
      & & 89-90$\,^\circ$ & 52$\,^\circ$ & & & \\
      \hline
      & $\mathbf{N_{SPU}=8}$ & 35 - 37$\,^\circ$ & 0$\,^\circ$ &  &  & \\
      & $\mathbf{V_{scan}=20''/s}$ & 88 - 90$\,^\circ$ & 52 - 53$\,^\circ$ & \raisebox{1.5ex}[0cm][0cm]{84 - 90$\,^\circ$} & \raisebox{1.5ex}[0cm][0cm]{0 - 3$\,^\circ$} & \\
      \hline
      & $\mathbf{N_{SPU}=4}$ & 57$\,^\circ$ & 0 - 3$\,^\circ$ &  &  & \\
      & $\mathbf{V_{scan}=60''/s}$ & 89 - 90$\,^\circ$ & 32$\,^\circ$ & \raisebox{1.5ex}[0cm][0cm]{86 - 90$\,^\circ$} & \raisebox{1.5ex}[0cm][0cm]{0 - 1$\,^\circ$} & \\
      \hline
      & $\mathbf{N_{SPU}=4}$ & &  &  &  & \\
      & $\mathbf{V_{scan}=20''/s}$ & \raisebox{1.5ex}[0cm][0cm]{87 - 90$\,^\circ$} & \raisebox{1.5ex}[0cm][0cm]{0 - 1$\,^\circ$} & \raisebox{1.5ex}[0cm][0cm]{78 - 90$\,^\circ$} & \raisebox{1.5ex}[0cm][0cm]{0 - 16$\,^\circ$} & \\
      \hline
      & $\mathbf{N_{SPU}=4}$ & &  &  &  & \\
      & $\mathbf{V_{scan}=10''/s}$ & \raisebox{1.5ex}[0cm][0cm]{84 - 90$\,^\circ$} & \raisebox{1.5ex}[0cm][0cm]{0 - 3$\,^\circ$} & \raisebox{1.5ex}[0cm][0cm]{66 - 90$\,^\circ$} & \raisebox{1.5ex}[0cm][0cm]{0 - 12$\,^\circ$} & \\
      \bottomrule
    \end{tabular}
  \caption[Angles de balayage \`a \'eviter pour \'echantillonner \`a
  Nyquist]{Liste exhaustive des angles de balayage qui n'autorisent
  pas l'\'echantillonnage du ciel suivant le crit\`ere de
  Nyquist. Toutes les combinaisons possibles des param\`etres
  d'observation de Herschel sont recens\'ees dans le tableau. Les deux
  cas ($N_{SPU}=8,V_{scan}=60$) et ($N_{SPU}=8,V_{scan}=20$)
  correspondent au \emph{Pmode} de l'Observatoire. Le calcul a \'et\'e
  effectu\'e pour des angles compris entre~0 et 90\textdegree,
  au-del\`a il suffit de prendre les angles suppl\'ementaires. Notez
  que les angles suivant les axes $x$ et $y$ ne sont pas
  compl\'ementaires, la raison \'etant que le plan focal n'est pas
  sym\'etrique par rapport \`a sa diagonale. Les angles qui
  n'apparaissent pas dans le tableau permettent donc de reconstruire
  des cartes du ciel \`a la r\'esolution spatiale optimale. Le calcul
  de ces angles d\'epend de $N_{SPU}$, $V_{scan}$, $\theta_{scan}$, de
  la taille du pixel et de la g\'eom\'etrie du plan focal~; par contre
  il ne prend pas en compte les aberrations optiques du syst\`eme.
  \label{tab:calib_perfobs_scan_compression_table}}
  \end{center}
\end{table}

Remarquez que l'\'etude pr\'esent\'ee ici est bas\'ee sur un
raisonnement semblable \`a celui pr\'esent\'e dans la
section~\ref{sec:calib_perflabo_tau_compare} \`a propos de
l'\'echantillonnage temporel d'un signal chopp\'e. De la m\^eme
fa\c{c}on nous comparons la fr\'equence d'\'echantillonnage effective
du signal $\nu_{eff}$ avec la fr\'equence de modulation g\'en\'er\'ee
par le mouvement du t\'elescope
($\omega=\vec{k}\cdot\vec{v}=kV_{scan}\cos{\theta_{scan}}$, cf
sec.~\ref{sec:calib_perfobs_scan_tau}). Lorsque ces deux fr\'equences
sont commensurables, le t\'elescope se d\'eplace d'un nombre entier de
fois la taille du pixel entre deux \'echantillons successifs, ce qui
revient \`a \'echantillonner le ciel toujours sur la m\^eme
grille. Par contre lorsqu'elles sont non-commensurables, le champ de
vue se d\'ecale d'une fraction de pixel entre deux images et le ciel
est alors \'echantillonn\'e plus finement (voir la
figure~\ref{fig:calib_perflabo_tau_compare_echant} pour faire le
parall\`ele avec un signal chopp\'e). Notez \'egalement que le
photom\`etre SPIRE, et les autres bolom\`etres \`a cornet de mani\`ere
g\'en\'erale, poss\`edent des angles privil\'egi\'es qui permettent
d'\'echantillonner le ciel au sens de Nyquist. Par contre, les
contraintes sont beaucoup plus importantes pour ces types de
d\'etecteurs (cornets \`a 1~ou~2F$\lambda$) que pour les bolom\`etres
PACS~; par exemple SPIRE n'offre que deux \og angles magiques \fg
\shortcite{sibthorpe}, tous les autres angles de balayage
sous-\'echantillonnent le ciel.




\chapter{Conclusions}
\label{sec:conclu}

\section{Pour les futurs d\'eveloppements} 
\label{sec:conclusion_recom}

En 1997, la conception de matrices de bolom\`etres monolithiques
multiplex\'es \`a froid est apparue comme un projet audacieux pour
certains, p\'erilleux pour d'autres. Et pourtant, le photom\`etre PACS
est aujourd'hui livr\'e \`a l'ESA avec les performances
attendues. Pour capitaliser l'investissement du CEA, le LETI et le
DAPNIA poursuivent leurs efforts et d\'eveloppent une fili\`ere
bolom\`etre infrarouge. Trois projets sont actuellement en
cours. ARTEMIS \shortcite{talvard} est un instrument
sub-millim\'etrique grand champ qui poss\`ede 16 matrices de
bolom\`etres dans le plan focal et qui devrait \^etre install\'e sur
le t\'elescope APEX dans les Andes chiliennes. PILOT
\shortcite{bernard} est un projet ballon dont l'objectif est de
mesurer la polarisation du milieu interstellaire dans l'infrarouge
lointain~; la plateforme sera \'equip\'ee d'une matrice de type
PACS. CAMISTIC \shortcite{minier_camistic} est une cam\'era qui
devrait \^etre install\'ee sur le t\'elescope IRAIT en Antarctique
pour r\'ealiser des mesures de transparence de l'atmosph\`ere dans
l'infrarouge lointain.

Fort de l'exp\'erience acquise lors de la caract\'erisation et de
l'\'etalonnage des matrices de bolom\`etres PACS, je propose
maintenant quelques commentaires et recommendations pour faciliter les
futurs tests et \'eventuellement am\'eliorer les performances et
l'exploitation des instruments \`a venir~:

\begin{itemize}
\item La possibilit\'e d'injecter une tension de r\'ef\'erence \`a
proximit\'e des bolom\`etres a jou\'e un r\^ole central dans la
proc\'edure d'\'etalonnage des d\'etecteurs~; il est donc n\'ecessaire
de conserver cette fonctionnalit\'e pour les futurs d\'eveloppements.
\item Pour le Photom\`etre PACS, la sonde de temp\'erature la plus
proche des matrices se trouve sur le doigt froid qui connecte les BFP
au cryo-r\'efrig\'erateur
(figure~\ref{fig:detect_observatoire_phfpu_description_dichro}). Or,
plusieurs interfaces thermiques s\'eparent la sonde de temp\'erature
et les bolom\`etres, les mesures absolues de la temp\'erature des
d\'etecteurs sont donc potentiellement biais\'ees. Il est capital pour
les futurs instruments d'ajouter une sonde de temp\'erature proche des
matrices, sur le BU et/ou le CL. Cela permettrait d'\'etudier
quantitativement la corr\'elation qu'il existe entre le bruit basse
fr\'equence du signal et les lentes d\'erives de temp\'erature du plan
focal.
\item Le s\'equenceur des futures matrices devrait \^etre plus
flexible que celui de PACS.  En effet, l'\'electronique de lecture
pouvant fonctionner \`a 1280~Hz en mode DDCS
(40~Hz~$\times$~16~pixels~$\times$~($V_{ptmil}$ et $V_{ref}$)), il
serait int\'eressant de proposer un mode direct qui
\'echantillonnerait le signal \`a 80~Hz plut\^ot qu'\`a 40~Hz, une
telle modification de l'\'electronique chaude serait
mineure\footnote{La modification consisterait \`a modifier le mode
DDCS en sommant, plut\^ot qu'en soustrayant, les deux conversions
successives d'une m\^eme phase pixel (cf
section~\ref{sec:detect_bolocea_elec_lecture}).} et nous gagnerions un
facteur $\sqrt{2}$ sur le rapport signal-\`a-bruit. De plus, si le
s\'equenceur \'etait programmable sur des p\'eriodes plus longues que
la phase pixel, nous pourrions adapter le mode DDCS pour prendre en
compte les mesures de d\'erive basse fr\'equence de la
section~\ref{sec:calib_perfobs_oof}~; \cad que pour minimiser les
perturbations induites par le mode DDCS, il serait pr\'ef\'erable
d'\'echantillonner $V_{ptmil}$ pendant le temps de coh\'erence du
signal ($\sim$1~s) puis de commuter sur $V_{ref}$ pour effectuer la
mesure de r\'ef\'erence pendant une fraction de seconde seulement
avant de retourner sur $V_{ptmil}$. Cette mani\`ere de proc\'eder est
certainement la plus efficace pour \'eliminer les d\'erives basse
fr\'equence dues au circuit de lecture en perturbant le moins possible
le niveau de point milieu. Enfin, avec un s\'equenceur plus flexible,
nous pourrions bloquer l'adressage sur un pixel unique et
\'echantillonner le signal \`a des fr\'equences allant de quelques~Hz
\`a 1280~Hz~; nous pourrions ainsi contraindre plus pr\'ecis\'ement la
constante de temps de l'\'electronique de lecture et
confirmer/infirmer l'existence de l'offset mis en \'evidence dans la
section~\ref{sec:calib_procedure_explore_imped}.
\item La dynamique de l'\'electronique chaude devrait \^etre
augment\'ee soit en utilisant des ADC 32-bits au lieu de 16-bits soit
en r\'eduisant la dispersion de signal avant num\'erisation, en
soustrayant analogiquement des cartes de points milieux par exemple.
\item \'Etant donn\'e la saturation des transistors du circuit de
lecture lorsque la tension d'entr\'ee du CL est inf\'erieure \`a
$\sim$300~mV (section~\ref{sec:calib_procedure_vrlvhb}), il faudrait
modifier l'\'electronique chaude pour autoriser les valeurs positives
de $V_l$, cela permettrait aux points milieux de d\'epasser la limite
de 300~mV pour les faibles tensions de polarisation, et nous pourrions
de surcro\^it r\'ealiser un \'etalonnage plus fiable de
l'\'electronique de lecture (mesures $V_h=V_l$ et $V_{hb}$ au lieu de
$V_{ref}$ et $V_{hb}$ comme sugg\'er\'e dans la
section~\ref{sec:calib_procedure_explore_imped}).
\item Lorsque deux matrices sont connect\'ees sur un m\^eme BU, il
faut veiller \`a bien les appairer, \cad s'assurer que le niveau de
points milieux des deux matrices est bien comparable. En effet, le
groupe~4 du BFP bleu (les deux matrices en haut \`a droite sur la
figure~\ref{fig:calib_procedure_explore_dispersion}) comporte deux
matrices dont les offsets sont d\'ecal\'es d'environ 100~mV, leur
r\'eglage a donc \'et\'e assez d\'elicat puisque soit l'une soit
l'autre \'etait \`a la limite de saturation de l'ADC.
\item Il serait \'egalement utile d'impl\'ementer un contr\^ole actif
de la temp\'erature du plan focal, les mesures de courbes~I-V seraient
alors plus pr\'ecises, et nous pourrions r\'eit\'erer la proc\'edure
de caract\'erisation pour diff\'erentes temp\'eratures de la source
froide pour mesurer l'\'evolution du bruit et de la r\'eponse des
bolom\`etres avec la temp\'erature.
\item Il faudrait poursuivre l'exploitation du mod\`ele de bolom\`etre
de \shortciteN{reveret_these} avec les courbes~I-V r\'ecemment
mesur\'ees et les mesures de points milieux. Il serait \'egalement
utile d'\'etendre ce mod\`ele en y incluant l'\'electronique de
lecture, nous pourrions ainsi \'etudier de mani\`ere plus quantitative
l'impact des injections de charges parasites sur les mesures de bruit.
\item D'apr\`es des mesures r\'ecentes de courbes~I-V r\'ealis\'ees
sur des matrices extr\^emement imp\'edantes, \cad avec des
r\'esistances de l'ordre de $10^{15}$~$\Omega$, il semble que la
r\'eponse de ces d\'etecteurs ne soient que deux fois sup\'erieures
\`a celle des matrices PACS alors que celles-ci sont trois ordres de
grandeur moins imp\'edantes. Cela indique que la r\'eponse des
bolom\`etres n'est pas limit\'ee par la r\'eponse \'electrique
($\alpha\propto dR/dT$) mais plut\^ot par la r\'eponse thermique
(\'equation~\ref{eq:fig:intro_bolometrie_thermo_principe_capacitecalorific}). Une
am\'elioration de la NEP, pour le projet SPICA par exemple, dont
l'objectif est d'atteindre $10^{-18}$~W/$\sqrt{\mbox{Hz}}$,
n\'ecessitera une diminution de la capacit\'e calorifique des grilles
absorbantes, soit en abaissant la temp\'erature de fonctionnement \`a
100~mK soit en r\'eduisant la masse de la grille (1~$\mu$m
d'\'epaisseur au lieu de 5~$\mu$m). Cependant, la r\'eduction de la
capacit\'e calorifique risque de rallonger consid\'erablement la
constante de temps thermique des bolom\`etres. Les \'etudes de
faisabilit\'e pour les d\'etecteurs de SPICA sont en cours au LETI et
au DAPNIA.
\end{itemize}

\section{Conclusions et perspectives} 
\label{sec:conclusion_conclu}

Les appels \`a proposition de l'ESA pour les projets clefs de temps
garanti et de temps ouvert ont r\'ecemment montr\'e que le
Photom\`etre PACS a \'et\'e tr\`es largement solicit\'e par la
communaut\'e astronomique. En effet, si l'on consid\`ere que
l'Observatoire Spatial Herschel poss\`ede cinq voies de d\'etection,
c'est-\`a-dire deux photom\`etres avec SPIRE et PACS et trois
spectrom\`etres avec SPIRE, PACS et HIFI, le Photom\`etre PACS \`a lui
seul repr\'esente 25\% des demandes de temps garanti et plus de 50\%
des demandes de temps ouvert (34\% en instrument principal et 20\% en
mode parall\`ele).

Dans cette th\`ese, j'ai montr\'e que la sensibilit\'e du Photom\`etre
PACS est conforme \`a la sp\'ecification scientifique du PhFPU qui est
de 5~mJy (5$\,\sigma$, 1~hr) sur source ponctuelle. Dans
l'\'eventualit\'e o\`u l'environnement \'electromagn\'etique de
l'instrument soit trop instable pour le mode direct, j'ai propos\'e
plusieurs options pour am\'eliorer les performances des matrices en
mode DDCS. J'ai \'egalement d\'etermin\'e la bande d'utilisation des
bolom\`etres, \cad de 1~\`a 5~Hz environ, ce qui a permis d'adapter
les diff\'erents modes d'observation aux caract\'eristiques des
d\'etecteurs afin d'assurer une utilisation optimale de l'instrument
en vol.

Mais l'aventure n'est pas finie, le lancement de Herschel est pr\'evu
pour la fin de l'ann\'ee~2008, et les six premiers mois de voyage
seront consacr\'es \`a la v\'erification des performances des
instruments. Cette p\'eriode sera sans doute tr\`es intense pour les
membres du groupe ICC. Pour ma part, je reste impliqu\'e dans le
projet Herschel et je devrai pouvoir apporter mon aide pour affiner le
r\'eglage des bolom\`etres puisque je vais effectuer mon postdoc au
NASA Herschel Science Center \`a Caltech/IPAC, Pasadena.

Du c\^ot\'e astrophysique, j'ai eu l'opportunit\'e au cours de ma
th\`ese d'aller observer avec le VLT dans le cadre du temps garanti de
l'instrument VISIR \shortcite{lagage}. L'objectif de ces observations
\'etait de r\'ealiser des mesures spectroscopiques basse r\'esolution
en infrarouge moyen (8-13~$\mu$m) d'une galaxie naine \`a flamb\'ee
d'\'etoiles, \`a savoir He~2-10. Cette galaxie est particuli\`erement
int\'eressante car elle contient de jeunes amas stellaires encore
profond\'ement enfouis dans leur cocon de gaz et de poussi\`ere. Ces
amas contiendraient plusieurs milliers d'\'etoiles OB, ce qui justifie
leur qualification de super-amas stellaires. Il est vraisemblable que
ces super-amas \'evoluent en amas globulaires tels que ceux observ\'es
aujourd'hui dans notre Galaxie. Toutefois, compte tenu des fortes
pressions et des maintes d\'erives de la date de livraison du PhFPU,
et par cons\'equent du temps suppl\'ementaire qui nous a \'et\'e
octroy\'e, j'ai donn\'e la priorit\'e \`a l'\'etalonnage des matrices
plut\^ot qu'\`a l'analyse des spectres de He~2-10. J'ai cependant
trait\'e les donn\'ees VISIR mais l'interpr\'etation des spectres
obtenus n'a pas encore commenc\'e. Je compte finir ce travail lors de
mon postdoc \`a IPAC.

\newpage

\dspaceoff

\bibliography{biblio}

\dspaceon

\begin{appendix}

   \chapter{Publications}
\label{a:publi}

\section{Astronomy \& Astrophysics}
\label{a:a&a}
L'article pr\'esent\'e dans cette annexe a \'et\'e soumis au Journal
A\&A. Nous avons r\'ecemment re\c{c}u le rapport du \emph{referee} qui
exige une version plus courte et plus claire. Nous avons donc
d\'ecid\'e de condenser l'information en nous focalisant sur les
points innovants, nous voulons \'egalement ajouter les r\'esultats
r\'ecents qui concernent les simulations du comportement des
bolom\`etres.\\

\noindent Une version compl\`ete de la th\`ese, avec annexes, peut \^etre
t\'el\'echarg\'ee \`a l'adresse suivante~:\\
http://tel.archives-ouvertes.fr/docs/00/29/20/30/PDF/dapnia\_07-12-T.pdf
 
\vfill

\section{SPIE}
\label{a:spie}
Cet article a \'et\'e publi\'e dans les \emph{proceedings} de la
conf\'erence SPIE \og Astronomical Telescopes and Instrumentation \fg
qui s'est tenue \`a Orlando, Floride, en~2006.  J'ai pr\'esent\'e ce
travail \`a l'oral dans la session \og Space Telescopes and
Instrumentation~: Optical, Infrared and Millimeter \fg.\\

\noindent Une version compl\`ete de la th\`ese, avec annexes, peut \^etre
t\'el\'echarg\'ee \`a l'adresse suivante~:\\
http://tel.archives-ouvertes.fr/docs/00/29/20/30/PDF/dapnia\_07-12-T.pdf
 
\vfill

\section{NIM-A}
\label{a:nima}
Cet article a \'et\'e publi\'e dans le journal \og Nuclear Instruments
and Methods in Physics Research Section A \fg. J'ai pr\'esent\'e ce
travail \`a l'oral lors de la conf\'erence internationale \og New
Developments in Photodetection \fg qui a eu lieu \`a Beaune en 2005.\\

\noindent Une version compl\`ete de la th\`ese, avec annexes, peut \^etre
t\'el\'echarg\'ee \`a l'adresse suivante~:\\
http://tel.archives-ouvertes.fr/docs/00/29/20/30/PDF/dapnia\_07-12-T.pdf
 
\vfill

\section{RADECS 2006}
\label{a:radec_horeau}

Cet article a \'et\'e \'ecrit par Beno\^it Horeau et devrait \^etre
publi\'e sous peu dans \og IEEE Transactions on Nuclear Science
\fg. Ce travail a \'et\'e pr\'esent\'e dans un poster lors de la
conf\'erence RADECS~2006 qui a eu lieu \`a Ath\`enes.\\

\noindent Une version compl\`ete de la th\`ese, avec annexes, peut \^etre
t\'el\'echarg\'ee \`a l'adresse suivante~:\\
http://tel.archives-ouvertes.fr/docs/00/29/20/30/PDF/dapnia\_07-12-T.pdf

\vfill

\section{Rapport sur les bandes passantes}
\label{a:rapport_bp}

\noindent Une version compl\`ete de la th\`ese, avec annexes, peut \^etre
t\'el\'echarg\'ee \`a l'adresse suivante~:\\
http://tel.archives-ouvertes.fr/docs/00/29/20/30/PDF/dapnia\_07-12-T.pdf

\vfill

   \chapter{Densit\'es spectrales des 9~combinaisons du s\'equenceur PACS}
\label{a:seq}

\noindent Cette figure contient les densit\'es spectrales de bruit
mesur\'ees pour chacun des 9~s\'equenceurs possible pour les matrices
du Photom\`etre PACS. La description du s\'equenceur est donn\'ee dans
la section~\ref{sec:detect_bolocea_elec_lecture}, et les densit\'es
spetrales de bruit des principaux s\'equenceurs du Photom\`etre sont
discut\'ees dans la
section~\ref{sec:calib_perflabo_compare_sequenceur}.\\

\begin{itemize}
\item La ligne du haut correspond au mode de lecture DDCS
($CKRL_{h-l}$ et $VDECX_{h-l}$ commutent \`a 1280~Hz pour
\'echantillonner alternativement $V_{bolo}$ et $V_{ref}$).
\item La ligne du milieu correspond au mode de lecture Direct en
\'echantillonnant seulement le signal bolom\'etrique $V_{bolo}$
($CKRL_{h-l}=0$~V, $VDECX_{h-l}=2$~V).
\item La ligne du bas correspond au mode de lecture Direct en
\'echantillonnant seulement le signal de r\'ef\'erence $V_{ref}$.
\item La colonne de gauche correspond \`a la premi\`ere conversion
r\'ealis\'ee par BOLC.
\item La colonne du milieu correspond \`a la deuxi\`eme conversion
r\'ealis\'ee par BOLC.
\item La colonne de droite correspond \`a la diff\'erence des deux
converisons r\'ealis\'ees par BOLC.\\
\end{itemize}

\noindent La \og bosse \fg pr\'esente dans les spectres lorsque le
signal bolom\'etrique est \'echantillonn\'e correspond \`a une
p\'eriode de 1~minute, c'est la p\'eriode d'asservissement de la
temp\'erature des corps noir et de la structure 4~K du PhFPU, elle
n'est donc pas reli\'ee aux performances des d\'etecteurs.

\vfill

\begin{center}
\includegraphics[width=1.4\textwidth,angle=270]{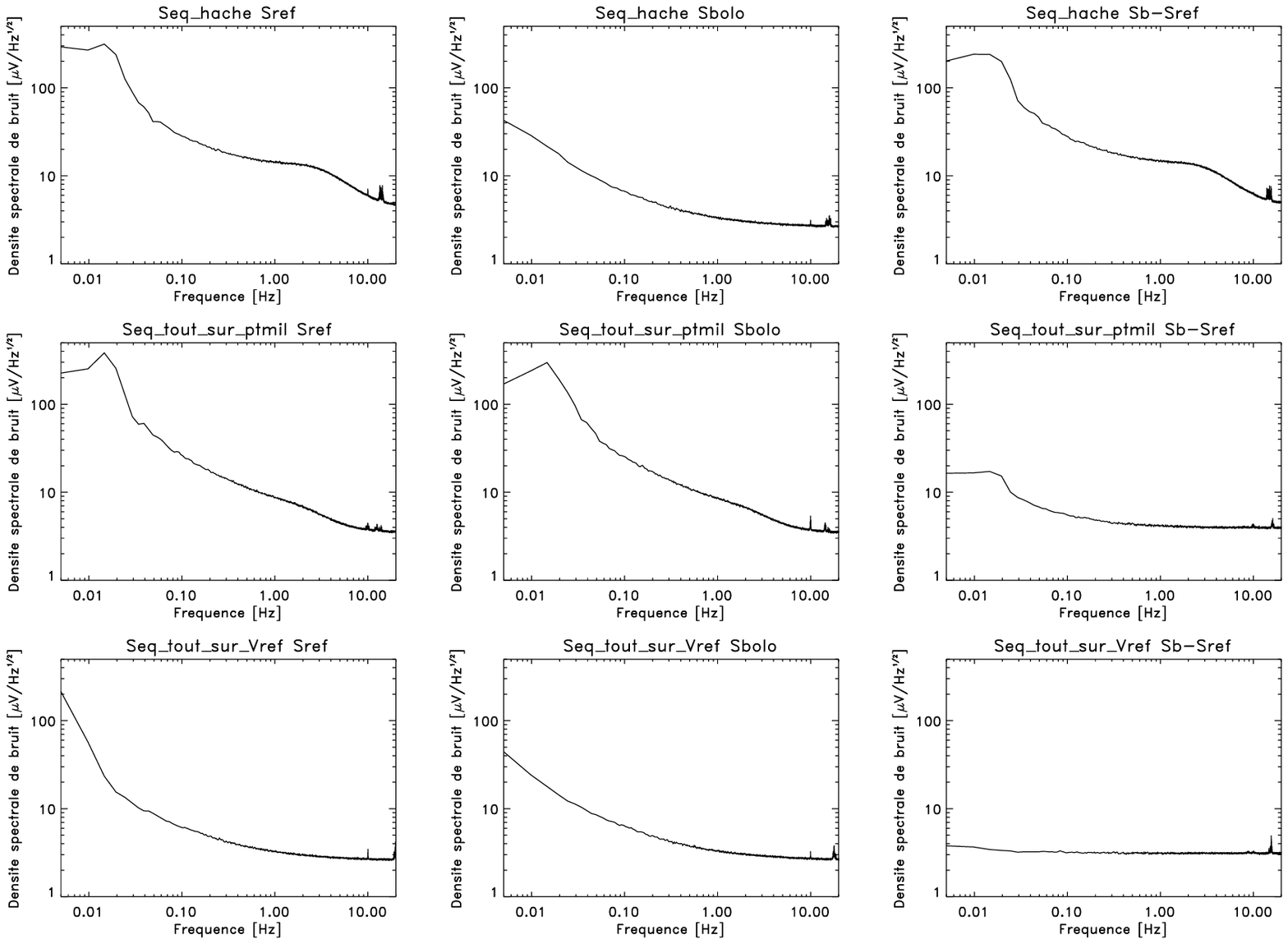}
\end{center}

\vfill

   \chapter{La dynamique de BOLC}
\label{a:dyna_BOLC}

\noindent Ce document pr\'esente la dynamique des convertisseurs
num\'eriques (ADC 16 bits) de l'\'electronique chaude dans les trois
gains disponibles et pour les trois modes de lecture offerts par BOLC.
Ce document a \'et\'e n\'ecessaire pour automatiser le r\'eglage des
bolom\`etres en donnant des contraintes sur les tensions d'entr\'ee de
BOLC.

\vfill

\begin{center}
\includegraphics[scale=0.6]{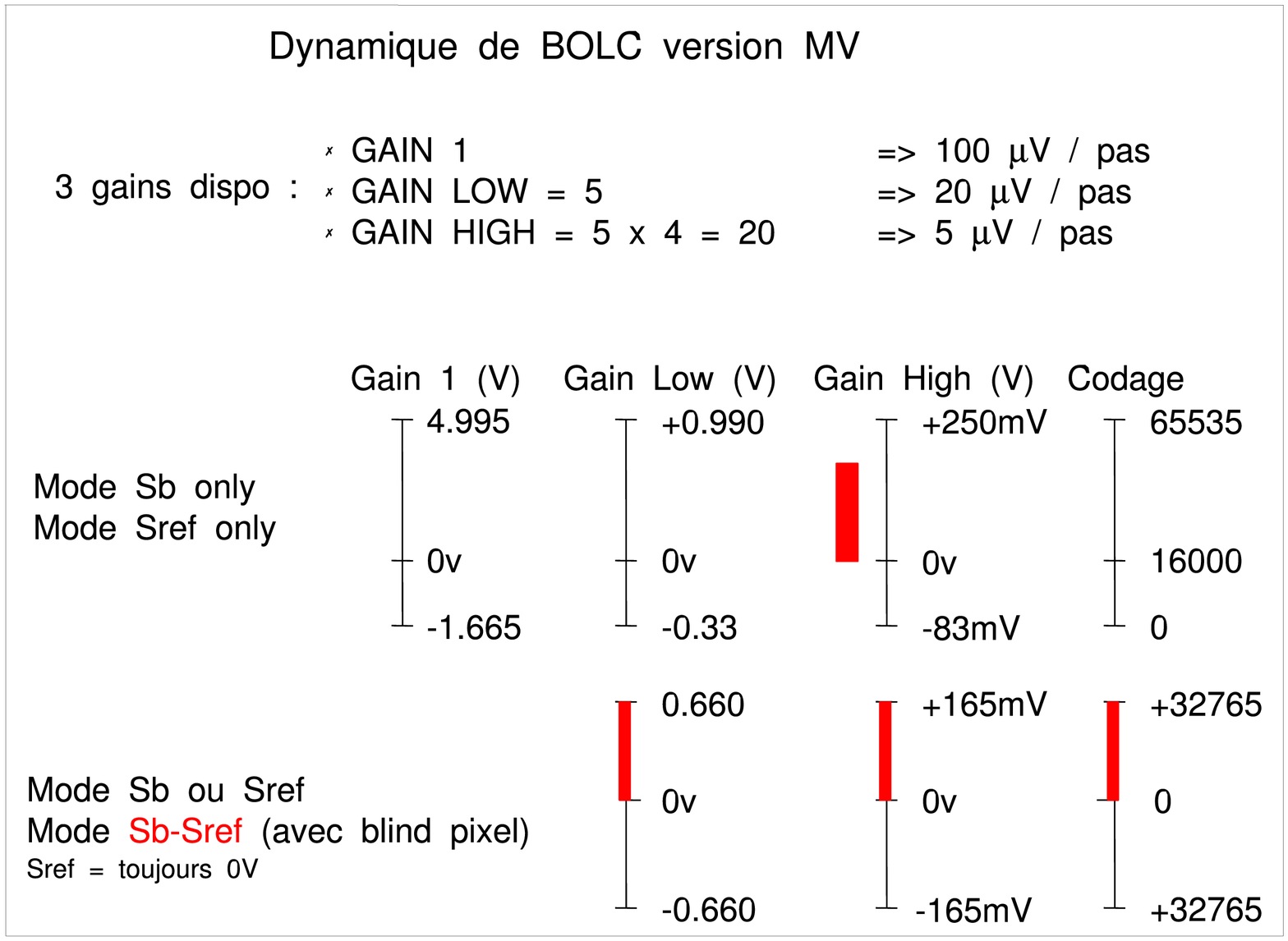}
\end{center}

\vfill
   \chapter{Configuration des plans focaux de vol du Photom\`etre PACS}
\label{a:pacs_bfp}

\noindent Les deux documents de cette annexe montrent la configuration
des plans focaux de vol du Photom\`etre PACS et permettent
l'identification des diff\'erents sous-syst\`emes. Nous y trouvons des
informations sur~:
\begin{itemize}
\item La g\'eom\'etrie des plans focaux, \cad la position relative des
matrices, et l'emplacement du doigt froid connect\'e au
cryo-r\'efrig\'erateur.
\item Le num\'ero des limandes utilis\'ees pour connecter les
diff\'erents \'etages \'electroniques (2~K-4~K et 2~K-0.3~K),
\item le num\'ero des Sous-Module Buffer (SMB) qui contiennent chacun
un BU (\'etage 2~K). Les SMB bleus contr\^olent deux matrices alors
que les SMB rouges contr\^olent une seule matrice.
\item Le num\'ero et le type de dopage des Sous-Module D\'etection
(SMD). Les SMD bleus contiennent deux matrices de type~2.5 et les
rouges en contiennent une seule de type~2 (plus imp\'edante que les
type~2.5).
\item Le num\'ero et la provenance des CL et CD de chaque matrice
constituant les SMD.
\end{itemize}

\vfill

\begin{center}
\includegraphics[scale=0.8]{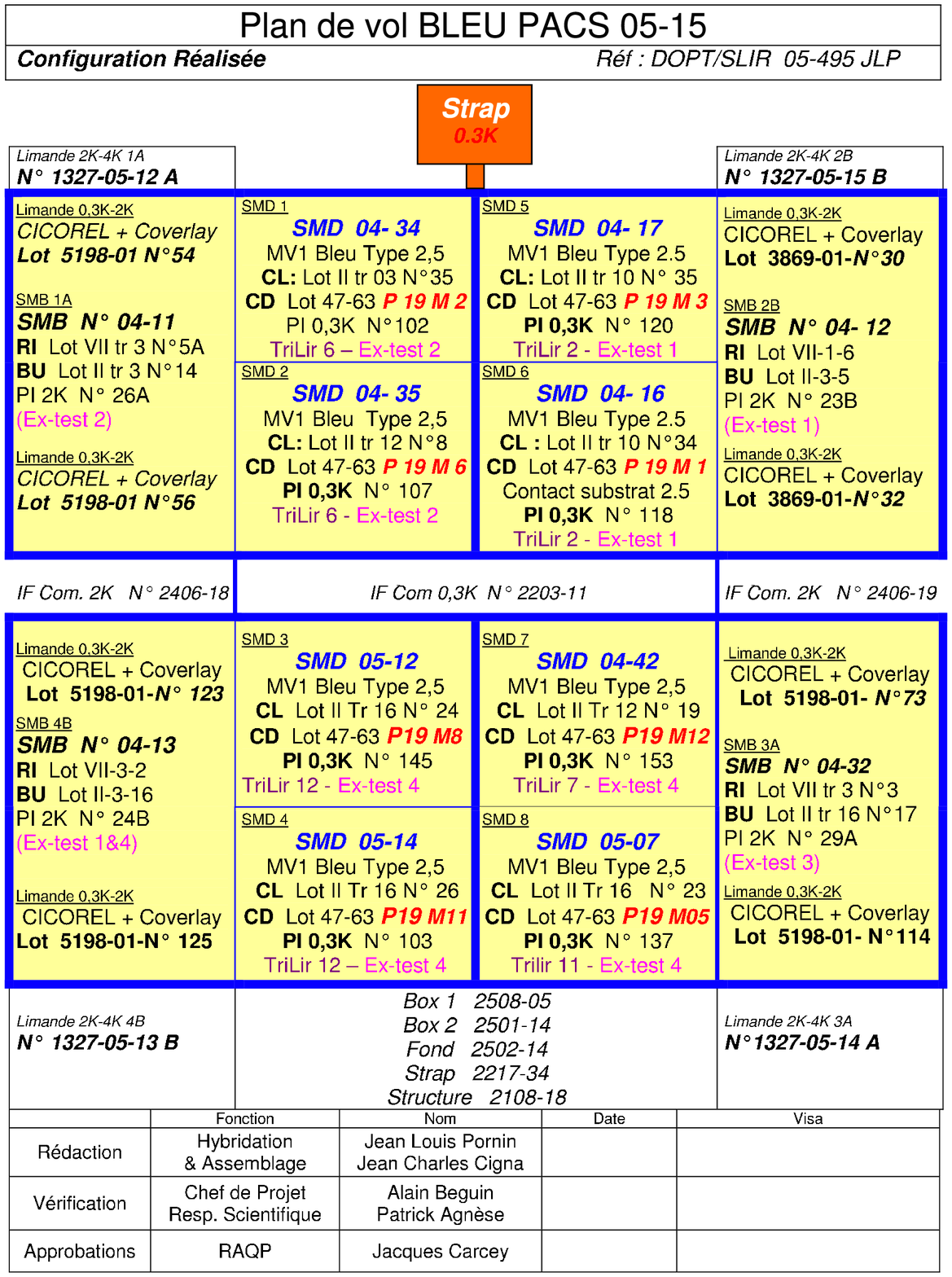}
\end{center}

\begin{center}
\includegraphics[scale=0.8]{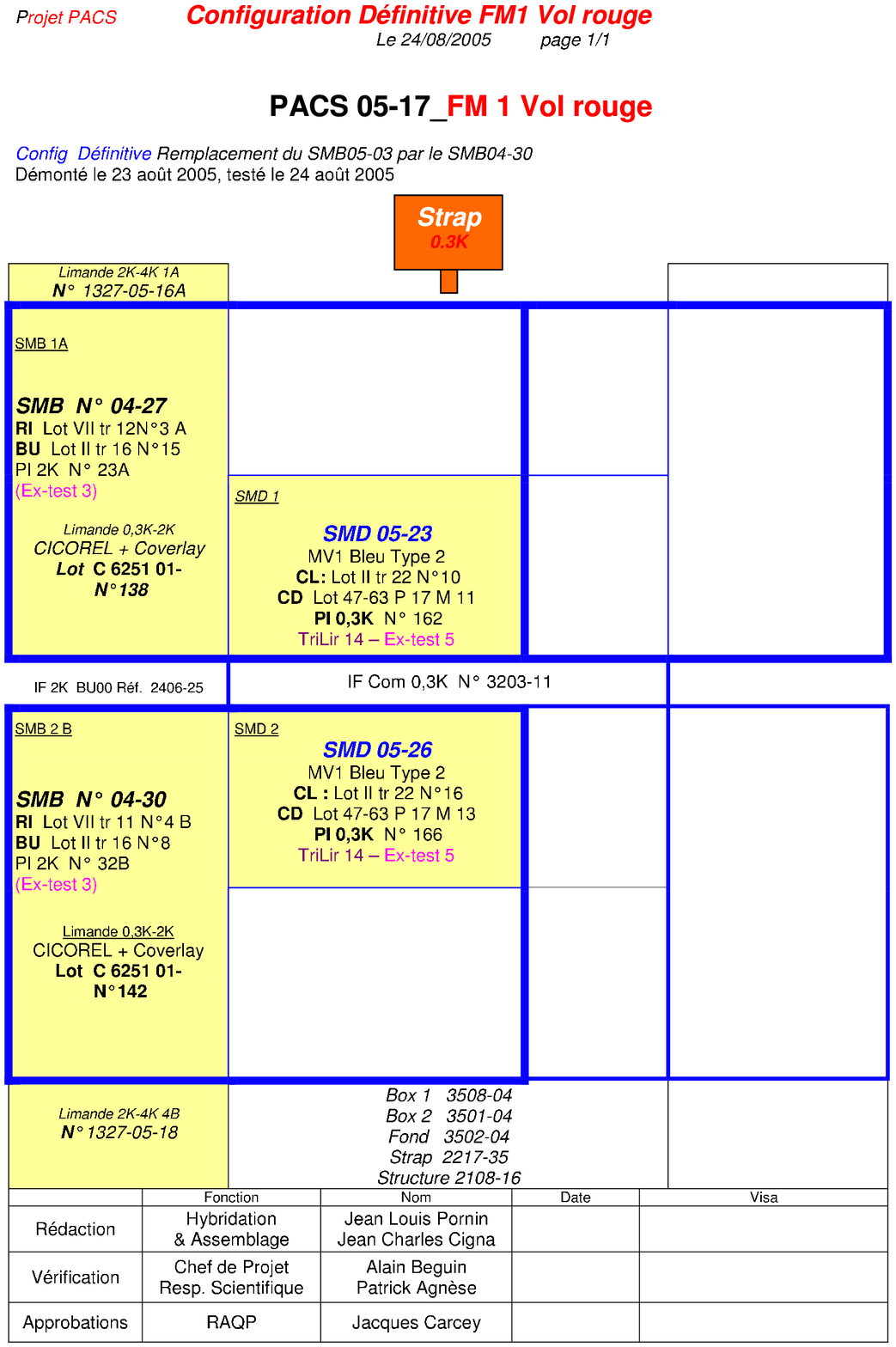}
\end{center}

\cleardoublepage
   \chapter{Estimateur et densit\'es spectrales de bruit}
\label{a:spectres}

\noindent Les deux premi\`eres figures pr\'esent\'ees dans cette
annexe montrent l'\'evolution de l'estimateur du niveau de bruit en
fonction de la tension de polarisation, du flux incident et du mode de
lecture. Les deux courbes suivantes montrent les densit\'es spectrales
de bruit correspondantes. Ces courbes ont \'et\'e calcul\'ees sur des
\'echantillons de 4~minutes obtenus lors des mesures de bruit en gain
fort durant la campagne d'\'etalonnage de PACS au MPE, Garching.

\noindent Une version compl\`ete de la th\`ese, avec annexes, peut \^etre
t\'el\'echarg\'ee \`a l'adresse suivante~:\\
http://tel.archives-ouvertes.fr/docs/00/29/20/30/PDF/dapnia\_07-12-T.pdf

\vfill

\end{appendix}

\end{document}